\documentclass[a4paper,12pt,titlepage,twosided]{report}

\usepackage{graphics}
\usepackage{graphicx}
\usepackage{color}
\usepackage{amsmath}
\usepackage{amsfonts}
\usepackage{amssymb}
\usepackage{psfrag}
\usepackage{psfig}
\usepackage{epsfig}
\usepackage{ifthen}
\usepackage{fancyheadings}
\usepackage{calc}
\usepackage{graphpap}
\usepackage{eufrak}
\usepackage{mathbbol}
\usepackage{supertabular}
\usepackage{longtable}

\numberwithin{equation}{section}

\newcommand{\manifold}{\mathcal}

\newcommand{\define}{\textsl}
\renewcommand{\emph}{\textsl}

\newcommand{\program}{\textsc}

\renewcommand{\d}{\mathrm{d}}
\newcommand{\diff}[2]{
  \ifthenelse{\equal{#1}{}}%
{\frac{\mathrm{d}\hphantom{#2}}{\mathrm{d}#2}}%
{\frac{\mathrm{d}#1}{\mathrm{d}#2}}}
\newcommand{\ddiff}[2]{
  \ifthenelse{\equal{#1}{}}%
{\frac{\mathrm{d}^2\hphantom{#2}}{\mathrm{d}#2^2}}
{\frac{\mathrm{d}^2#1}{\mathrm{d}#2^2}}}
\newcommand{\pardiff}[2]{\frac{\partial#1}{\partial#2}}
\newcommand{\parddiff}[2]{\frac{\partial^2#1}{\partial#2^2}}
\newcommand{\pparddiff}[3]{\frac{\partial^2#1}{\partial#2\,\partial#3}}

\newcommand{\cosec}{\mathop{\mathrm{cosec}}}

\newcommand{\tensor}[1]{\boldsymbol{#1}}
\renewcommand{\vec}[1]{\boldsymbol{#1}}

\newcommand{\expct}[3][]{\left\langle#3\left|#2\right|#3\right\rangle_{\mathrm{#1}}}
\newcommand{\bra}[1]{\left\langle\right.#1\left|\right.}
\newcommand{\ket}[1]{\left.\left|#1\right.\right\rangle}

\newcommand{\vac}[3][]{\left\langle#2\right\rangle^{#3}_{\mathrm{#1}}}   
\newcommand{\abs}[1]{\left|#1\right|}

\newlength{\arrowlength}
\renewcommand{\nearrow}[1]{
\setlength{\unitlength}{\arrowlength}
\begin{picture}(20,10)
\put(0,0){\line(2,1){20}}
\put(0,0){\vector(-2,-1){0}}
\put(20,10){\vector(2,1){0}}
\put(11,4){\makebox(0,0)[tl]{\tiny$#1$}}
\end{picture}}
\renewcommand{\swarrow}[1]{
\setlength{\unitlength}{\arrowlength}
\begin{picture}(20,10)
\put(0,10){\line(2,-1){20}}
\put(0,10){\vector(-2,1){0}}
\put(20,0){\vector(2,-1){0}}
\put(11,6){\makebox(0,0)[bl]{\tiny$#1$}}
\end{picture}}
\newcommand{\ddiagdots}{
\setlength{\unitlength}{\arrowlength}
\begin{picture}(20,10)
\put(0,10){\circle*{1}}
\put(10,5){\circle*{1}}
\put(20,0){\circle*{1}}
\end{picture}}
\newcommand{\udiagdots}{
\setlength{\unitlength}{\arrowlength}
\begin{picture}(20,10)
\put(0,0){\circle*{1}}
\put(10,5){\circle*{1}}
\put(20,10){\circle*{1}}
\end{picture}}
\newcommand{\downdots}{
\setlength{\unitlength}{\arrowlength}
\begin{picture}(2,10)
\put(0,0){\circle*{1}}
\put(0,5){\circle*{1}}
\put(0,10){\circle*{1}}
\end{picture}}

\newcommand{\topbott}[2]{\left\{ #1 \atop #2 \right\}}
\newcommand{\tpit}{[\theta\to\pi-\theta]}
\newcommand{\NPadj}{\varphi}
\newcommand{\indhel}{h}
\newcommand{\stepx}{\Delta x}
\newtheorem{theorem}{Theorem}
\newcommand{\routine}{\emph}
\newcommand{\bDelta}{\mathbb{\Delta}}
\newcommand{\draft}[1]{}
\newcommand{\ddraft}[1]{}
\newcommand{\catdraft}[1]{}

\begin{document}

\pagenumbering{roman}
\begin{titlepage}
\begin{center}
\textbf{\huge Electromagnetic Quantum }\\
\hspace{\fill}\\
\textbf{\huge Field Theory on}\\
\hspace{\fill}\\
\textbf{\huge Kerr-Newman Black Holes}\\
\hspace{\fill}\\
by\\
\hspace{\fill}\\
\large Marc Casals i Casanellas\\
\hspace{\fill}\\
\hspace{\fill}\\
A dissertation submitted to\\
the National University of Ireland\\
in partial fulfilment of the requirements\\
of the degree of Philosophiae Doctor\\
\hspace{\fill}\\
\hspace{\fill}\\
Department of Mathematical Physics\\
Faculty of Science\\
University College Dublin\\
\hspace{\fill}\\
\hspace{\fill}\\
February 23, 2004
\end{center}
\hspace{\fill}\\
\begin{tabular}{rl}
\textbf{Supervisor:}&Professor Adrian C.~Ottewill\\
\textbf{Head of Department:}&Professor Adrian C.~Ottewill
\end{tabular}
\end{titlepage}

\renewcommand{\abstractname}{\vphantom{.}\vspace{-2truecm}Abstract}
\begin{abstract}

We study classical and quantum aspects of electromagnetic
perturbations on black hole space-times. We develop an elegant
formalism introduced by Wald, which sets up the theory of linear
perturbations in a Type-D background in a compact and transparent manner. 
We derive natural expressions for the electromagnetic potential 
in the ingoing and upgoing gauges in
terms of the single Newman-Penrose (NP) scalar $\phi_0$. 
This enables the formulation of the quantum theory of the electromagnetic field
as that of a complex scalar field. Unfortunately, the field
equations for $\phi_0$ in the Kerr-Newman background are
non-separable, except in the Reissner-Nordstr\"{o}m limit.

We study the separable spin-1, classical field equations obeyed by the
NP scalars $\phi_{\pm1}$ 
in the Kerr-Newman background
and find, for various limits, 
the asymptotic behaviour of the radial and angular solutions. 
We correct and build on a study by Breuer, Ryan and
Waller to find a uniformly valid asymptotic behaviour for large frequency of the
angular solutions 
and the eigenvalues. 
We complement our asymptotic analysis with the numerically obtained solution of the 
radial and angular differential equations.

We follow Candelas, Chrzanowski and Howard (CCH) in their
canonical quantization of the electromagnetic potential and field.
We study the form of the renormalized stress-energy tensor (RSET) in the 
past Boulware vacuum
close to the horizon. 
In contrast with a calculation in CCH, its leading order behaviour close to the horizon 
corresponds to minus the stress tensor of a thermal distribution at the Hawking temperature 
rigidly rotating with the horizon.
We prove that expressions given by CCH for the expectation value of the stress tensor in the 
past Boulware, past Unruh and $\ket{CCH^-}$ states
lead to a lack of symmetry under parity. 
We show that the origin of this asymmetry is the non-symmetrization of
the quantum operators in the derivation of the expressions in CCH. 
We give the correct expressions, and present a detailed analysis of the resulting RSETs.

\end{abstract}

\vspace*{5 cm}
\begin{flushright}
{\bf \it \large Als meus avis
}
\end{flushright}

\chapter*{\large Agra\"{i}ments/Acknowledgements\markboth{Acknowledgements}{Acknowledgements}}


I wish to thank the following people, who have helped me 
illuminate
black holes.

My supervisor, Professor Adrian Ottewill, for offering me
the invaluable opportunity
to take up a job that I relish doing -research in physics, and also for his 
ingenious advice
throughout.
All the other members of staff and postgraduates in the Mathematical Physics 
and the Mathematics departments.
Gavin Duffy deserves 
an
outstanding mention for his 
endless, unselfish help.
I am also particularly indebted to Michael Mackey and Thomas Unger for their 
assisstance in computer-related issues.
I wish to thank Professor Valeri P. Frolov for useful comments.

All my friends, with whom I have enjoyed many 
dreamy
conversations on merry
nights.
In particular, my girlfriend Siobh\'{a}n, 
who has illuminated me on many sunny nights.

Aquesta tesi est\`{a} dedicada
a tots els meus avis, 
que amb el seu amor m'han fet veure les estrelles.
Estic agra\"{i}t 
als meus pares per haver-me transm\`{e}s
la curiositat d'entrar dins de forats negres i l'habilitat de saber-ne sortir.
A tots els meus amics, amb qui m'he estrellat moltes nits.

This research was financially supported in part by Enterprise Ireland.



\tableofcontents

\clearpage
\pagenumbering{arabic}
\chapter*{Overview}\label{intro}%
\addcontentsline{toc}{chapter}{Overview}
\markboth{}{Overview}

Quantum field theory in a curved background is a semiclassical
approximation to a quantum theory of gravity in which the matter
fields are quantized but the gravitational field is described by a
classical background space-time. The theory may be taken to
include one-loop quantum gravitational effect by considering the
linear perturbation of the gravitational field as a massless
spin-2 field on the background space-time.  In the semiclassical
theory, the classical stress-energy tensor is replaced in
Einstein's field equation by the renormalized expectation value of
the stress-energy tensor in a suitable state $\ket{\Psi}$ of the
system:
$$R_{\mu\nu}-\frac{1}{2}Rg_{\mu\nu}+\Lambda g_{\mu\nu}=+\frac{8\pi G}{c^4}
\vac[ren]{\hat{T}_{\mu\nu}}{\Psi}.
$$

Quantum field theory in a curved background
is the framework of this thesis. Even though this theory does not
provide a full account of quantum gravity, it does provide an
understanding of the influence of the gravitational field on
quantum field theoretic results, which a complete quantum theory
of gravitation should account for. Indeed, it is expected that
this semiclassical approach is valid in the limit that the length
and time scales of the quantum processes are much larger than the
Planck length
($\left(G\hbar/c^3\right)^{1/2}=1.62\times10^{-33}$cm) and the
Planck time ($\left(G\hbar/c^5\right)^{1/2}=5.39\times10^{-44}$s). 

The first important results that quantum field theory in a curved background achieved related
to the discovery of particle production by rapidly varying gravitational fields. In particular,                     
the pivotal result discovered by Hawking in 1975
~\cite{ar:Hawking'75} was that black holes radiate as black bodies
due to particle creation.
Many related investigations have followed since then. Most of
them, however, relate to a spherically-symmetric (Schwarzschild)
black hole. Not so many relate to the more realistic case of a
rotating, axially-symmetric (Kerr) black hole. The difficulty of
dealing with the latter rather than the former is apparent already
from noting that a spherically-symmetric black hole possesses an
infinite-number of rotational symmetries whereas the
axially-symmetric black hole only possesses one.


The most common field theory studied in the literature is scalar (spin-0)
field theory. The main focus of this thesis is on electromagnetic (spin-1) 
perturbations in an electrically-charged, axially-symmetric (Kerr-Newman) background.
Some of the results, however, are more general and apply to
general spin theory in backgrounds of a wider type. 
Some other results apply to the Kerr background.
We will indicate throughout this thesis what spin theory
and what background the results apply to.

The algebra of general spin theory is more complicated
than that for spin-0, and the Newman-Penrose formalism which
exploits the underlying symmetries of the space-time is introduced
in order to simplify it. It was with the use of the Newman-Penrose
formalism that Teukolsky ~\cite{ar:Teuk'73} decoupled the field
equations for different spin fields in the Kerr background and
expressed the decoupled equations into a single ``master''
differential equation, where the spin appeared as a
parameter. More importantly, he managed to separate this equation
thereby making it possible to express the general solution as a
sum over generalised Fourier modes.


The content of this thesis is organized as follows:

In Chapter \ref{ch:intro} we give an overview of the Newman-Penrose formalism
and the Kerr-Newman space-time.

In Chapter \ref{ch:field eqs.} we study the solutions to the
Maxwell equations in a curved background. We do so by developing a
particularly elegant formalism, which was presented by Wald
~\cite{ar:Wald'78}. We also suggest an approach to quantizing the
electromagnetic field, different from the one commonly used in the
literature, which allows the reduction of the quantum theory for
the electromagnetic field to that of a complex scalar field.

Chapters \ref{ch:radial sln.} and \ref{ch:SWSH} provide an analysis of the radial and angular differential
Teukolsky equations respectively.
Both chapters suggest various approaches for solving the equations and describe the methodology 
used to find the numerical solutions for each case.
The numerical results are presented and discussed.
We study the behaviour of the solutions in different asymptotic regions.
In each chapter we also provide a review background of the related existing results in the literature.

In Chapter \ref{ch:high freq. spher} we present the results of
an analysis of the angular solutions in the limit of high
frequency. This analysis is based on a previous one done by
Breuer, Ryan and Waller ~\cite{ar:BRW}. Their work, however, was
flawed and incomplete. This chapter corrects and completes their
work and thus gives a complete account of the behaviour for large
frequency of the angular functions and the eigenvalues for general integral spin,
which has not been presented in the literature before.
The analytic work is fully complemented with a numerical study.

The last chapter focuses on quantum
field theory for the electromagnetic field in the Kerr and Kerr-Newman backgrounds. 
The method used follows the method of canonical quantization of the electromagnetic field used by Candelas,
Chrzanowski and Howard ~\cite{ar:CCH}. 
We give an overview of the results achieved thus far in the literature in the particular problem of
defining a vacuum state with certain desirable properties in the Kerr background.
We present numerical results obtained for the difference of the expectation value 
of the stress-energy tensor between two states for spin-1.
In the last three sections we deal with three separate issues.
In one section, we remark on some of the differences between the spin-1 and the spin-0 cases in the calculation of flux of
energy and angular momentum of a black hole, and present calculations in the cases that the field is in the past Unruh and past Boulware states.
In the following section, we rederive and review a calculation in ~\cite{ar:CCH} on the renormalized stress energy
tensor for the past Boulware state close to the horizon. 
We show numerically that it corresponds to that of a rigidly rotating thermal reservoir of particles.
In the last section we thoroughly discuss the apparent lack of symmetry that the expectation value
of the stress tensor in different states exhibits based on the formulae of Candelas, Chrzanowski and Howard.
Finally, we give for the first time in the literature the correct form of the equations for the calculation of these quantities, 
and give a physical interpretation of the results.

In the remainder of this thesis, we use  geometrized Planck units:
$c=G=\hbar=1$,
and follow the sign conventions of Misner, Thorne and Wheeler ~\cite{bk:M&T&W}.       
We follow the tetrad and Newman-Penrose formalism conventions of Chandrasekhar ~\cite{bk:Chandr}.
A Greek-letter index indicates a tensor index whereas a Latin-letter index in parentheses indicates a tetrad index. 

\include{chap0}
\chapter{Introduction}  \label{ch:intro}

\draft{1) titles of all sections/chapters will be changed, 2) should titles have capital letters at start of all words?
3) specify what is valid for Kerr and what for Kerr-Newman, 4) don't think refs. ~\cite{bk:Carter-DeWittDeWitt} and ~\cite{bk:Carter-Cargese'86}
are correctly specified}

\catdraft{posar en algun lloc:''(left as an exercise for the smart reader)''}


\section{Tetrad/Newman-Penrose formalism} \label{sec:NP formalism}

Consider a 4-dimensional Riemannian space with a signature $+2$. 
A \define{tetrad basis} consists of four contravariant vectors $\{\vec{e}_{(a)}, a=1,2,3,4\}$ such that
\begin{equation}
e_{(a)}{}^{\alpha}e_{(b)}{}_{\alpha}=\eta_{(a)(b)}
\end{equation}
where $\left(\eta_{(a)(b)}\right)$ is a constant, symmetric matrix. The inverse to the matrix $\left(e_{(a)}{}^{\alpha}\right)$ is
$\left(e^{(a)}{}_{\alpha}\right)$ in the sense that: $e_{(a)}{}^{\alpha}e^{(b)}{}_{\alpha}=\delta^{(b)}{}_{(a)}$ and 
$e_{(a)}{}^{\alpha}e^{(a)}{}_{\beta}=\delta^{\alpha}{}_{\beta}$. Similarly, the inverse of the constant, symmetric matrix is
given by $\eta^{(a)(b)}\eta_{(b)(c)}=\delta^{(a)}{}_{(c)}$.

The tetrad components of a tensor field are the projections of the tensor field onto the tetrad frame. They are 
obtained by contracting the tensor components of the field with those of the vectors in the tetrad basis:
\begin{equation}
\begin{aligned}
A_{(a_1)\dots(a_r)}&=e_{(a_1)}{}^{\alpha_1}\dots e_{(a_r)}{}^{\alpha_r}A_{\alpha_1\dots\alpha_r} & \text{if} \quad \tensor{A}\in T^0{}_r\\
A^{(a_1)\dots(a_r)}&=
e^{(a_1)}{}^{\alpha_1}\dots e^{(a_r)}{}^{\alpha_r}A_{\alpha_1\dots\alpha_r}=
\eta^{(a_1)(b_1)}\dots\eta^{(a_r)(b_r)}A_{(b_1)\dots(b_r)}  & \text{if} \quad \tensor{A}\in T^r{}_0
\end{aligned}
\end{equation}
where $T^r{}_s$ is a tensor field of type $(r,s)$ on the Riemannian space. 
It is clear that tetrad indices are raised and lowered with $\eta^{(a)(b)}$ and $\eta_{(a)(b)}$ respectively.

The contravariant vectors $\vec{e}_{(a)}=e_{(a)}{}^{\alpha}\pardiff{}{x^{\alpha}}$, considered as tangent vectors, define
the \define{directional derivatives}: $A_{(a_1)\dots(a_r),(b)}=e_{(b)}{}^{\alpha}\pardiff{}{x^{\alpha}}A_{(a_1)\dots(a_r)}$.
The \define{intrinsic derivative} of $A_{(a_1)\dots(a_r)}$ in the direction $\vec{e}_{(b)}$ is defined in relation to the directional 
derivative as
\begin{equation}
\begin{aligned}
& A_{(a_1)\dots(a_r)|(b)}\equiv A_{\alpha_1\dots\alpha_r;\beta} e_{(a_1)}{}^{\alpha_1}\dots e_{(a_r)}{}^{\alpha_r}e_{(b)}{}^{\beta}= \\
& =A_{(a_1)\dots(a_r),(b)}-\eta^{(n)(m)}
\left(\gamma_{(n)(a_1)(b)}A_{(m)(a_2)\dots(a_r)}+\gamma_{(n)(a_2)(b)}A_{(a_1)(m)(a_3)\dots(a_r)}+\dots + 
\right. \\ & \left.
\qquad \qquad \qquad \qquad \qquad +\gamma_{(n)(a_r)(b)}A_{(a_1)\dots (a_{r-1})(m)}  \right)
\end{aligned}
\end{equation}
where we have used the following definition of the \define{Ricci rotation-coefficients}: 
\begin{equation} \label{eq.def. of Ricci rot.-coeffs.}
\gamma_{(c)(a)(b)}\equiv e_{(a)\alpha;\beta}e_{(c)}{}^{\alpha}e_{(b)}{}^{\beta} 
\end{equation}
The Ricci rotation-coefficients are anti-symmetric in the first pair of indices: $\gamma_{(c)(a)(b)}=-\gamma_{(a)(c)(b)}$, 
due to the fact that $\eta_{(a)(b)}$ are constant.

The \define{Lie bracket} $\left[\vec{e}_{(a)},\vec{e}_{(b)}\right]$ is itself a vector and can therefore be expressed as a linear
combination of the basis vectors
\begin{equation}
\left[\vec{e}_{(a)},\vec{e}_{(b)}\right]=C^{(c)}{}_{(a)(b)}\vec{e}_{(c)}
\end{equation}
where the coefficients
$C^{(c)}{}_{(a)(b)}$ are called the \define{structure constants}. These constants can be readily expressed in terms of the rotation coefficients
as $C^{(c)}{}_{(a)(b)}=\gamma^{(c)}{}_{(b)(a)}-\gamma^{(c)}{}_{(a)(b)}$ and we therefore have the \define{commutation relations}:
\begin{equation} \label{eq:commutation rlns.}
\left[\vec{e}_{(a)},\vec{e}_{(b)}\right]=\left(\gamma^{(c)}{}_{(b)(a)}-\gamma^{(c)}{}_{(a)(b)} \right)\vec{e}_{(c)}
\end{equation}

The \define{Newman-Penrose (NP) formalism} ~\cite{ar:N&P'62} is a particular case of the tetrad formalism whereby the four vectors in the 
basis are chosen to be null, with two of the basis vectors being real and the other two complex-conjugates of each other. 
All four basis vectors are also chosen to be orthogonal by pairs. The remaining normalization conditions are usually chosen so that
\begin{equation}  \label{eq:metric for NP tetrad}
\left(\eta_{(a)(b)}\right)=\left( \eta^{(a)(b)}\right)=
\left( 
\begin{array}{cccc}
0 & 1 &  0 &  0 \\
1 & 0 &  0 &  0 \\
0 & 0 &  0 & -1 \\
0 & 0 & -1 &  0
\end{array}
\right)
\end{equation}
Each vector in the NP basis is designated by a specific symbol: 
\begin{equation}
\vec{l}\equiv \vec{e}_{(1)}=\vec{e}^{(2)},\vec{n}\equiv \vec{e}_{(2)}=\vec{e}^{(1)}, \vec{m}\equiv \vec{e}_{(3)}=-\vec{e}^{(4)}, \vec{m}^*\equiv \vec{e}_{(4)}=-\vec{e}^{(3)}
\end{equation}
These vectors define directional derivatives which are denoted by the following symbols:
\begin{equation}
\begin{aligned}
D&\equiv e_{(1)}^{\mu}\nabla_{\mu}=e^{(2) {\mu}}\nabla_{\mu}       \qquad       & \bDelta &\equiv e_{(2)}^{\mu}\nabla_{\mu}=e^{(1) \mu}\nabla_{\mu}  \\
\delta&\equiv e_{(3)}^{\mu}\nabla_{\mu}=-e^{(4) {\mu}}\nabla_{\mu} \qquad       & \delta^*&\equiv e_{(4)}^{\mu}\nabla_{\mu}=-e^{(3) {\mu}}\nabla_{\mu}
\end{aligned}
\end{equation}
\catdraft{com es que no es $D\equiv e_{(1)}^{\mu}\pardiff{}{x^{\mu}}$?? en un Ott. hi posa $\nabla$ i en l'altre $\pardiff{}{x^{\mu}}$? sembla que per
def. hauria de ser $\pardiff{}{x^{\mu}}$ pero en pas de $l^{\mu;\nu}=...$ a $Dl^{\mu}=...$ sembla que cal que sigui $\nabla$??}

The Ricci rotation-coefficients, called \define{spin coefficients} within the NP formalism, are also given specific symbols:
\begin{equation}  \label{eq:def. spin coeffs.}
\begin{aligned}
\kappa  &\equiv\gamma_{(3)(1)(1)} \qquad &\rho &\equiv\gamma_{(3)(1)(4)}  \qquad &\epsilon &\equiv\frac{1}{2}\left(\gamma_{(2)(1)(1)}+\gamma_{(3)(4)(1)}\right) \\ 
\sigma  &\equiv\gamma_{(3)(1)(3)} \qquad &\mu  &\equiv\gamma_{(2)(4)(3)}  \qquad &\gamma   &\equiv\frac{1}{2}\left(\gamma_{(2)(1)(2)}+\gamma_{(3)(4)(2)}\right) \\
\lambda &\equiv\gamma_{(2)(4)(4)} \qquad &\tau &\equiv\gamma_{(3)(1)(2)}  \qquad &\alpha   &\equiv\frac{1}{2}\left(\gamma_{(2)(1)(4)}+\gamma_{(3)(4)(4)}\right) \\
\nu     &\equiv\gamma_{(2)(4)(2)} \qquad &\pi  &\equiv\gamma_{(2)(4)(1)}  \qquad &\beta    &\equiv\frac{1}{2}\left(\gamma_{(2)(1)(3)}+\gamma_{(3)(4)(3)}\right)
\end{aligned}
\end{equation}

The ten independent components of the Weyl tensor $C_{\alpha\beta\mu\nu}$ are represented by the five complex \define{Weyl scalars}:
\begin{equation}  \label{eq:def. Weyl scalars}
\begin{aligned}
\psi_{-2}&\equiv -C_{(1)(3)(1)(3)}  \qquad & \psi_{-1}&\equiv -C_{(1)(2)(1)(3)} \\
\psi_{0}&\equiv -C_{(1)(3)(4)(2)}   \qquad &          &                 \\ 
\psi_{+1}&\equiv -C_{(1)(2)(4)(2)}  \qquad & \psi_{+2}&\equiv -C_{(2)(4)(2)(4)} 
\end{aligned}
\end{equation}
The other tetrad components of the Weyl tensor can be obtained from these five scalars by either using any of the 
symmetries it possesses ($C_{(\alpha\beta)\gamma\delta}=0=C_{\alpha\beta(\gamma\delta)}$ and 
$C_{\alpha\beta\gamma\delta}=C_{\gamma\delta\alpha\beta}$; we preferred to use tensor instead of tetrad
indices for these symmetries in order to avoid mixing brackets that indicate a tetrad index with brackets that indicate symmetrization) 
or its trace-freeness $C^{(a)}{}_{(b)(c)(a)}=0$
or else the \define{cyclic condition}: $C_{(1)(2)(3)(4)}+C_{(1)(3)(4)(2)}+C_{(1)(4)(2)(3)}=0$. Of course, in flat space-time all 
Weyl scalars are zero: $\psi_{\indhel}=0, \forall \indhel$.
\catdraft{Adrian diu que cyclic condition es Ricci id. i que ho esmenti pero a mi em sembla que es Bianchi id.?}
Notice that the NP Weyl scalars 
$\{\psi_{-2},\psi_{-1},\psi_0,\psi_{+1},\psi_{+2}\}$ are respectively named 
$\{\psi_{0},\psi_{+1},\psi_{+2},\psi_{+3},\psi_{+4}\}$ in most of the bibliography cited in this thesis.
The index notation we use in this thesis for all NP scalars follows that of Carter ~\cite{bk:Carter-Cargese'86}
(a similar notation was originally used by Price ~\cite{ar:Price'72}).

Similarly, the ten components of the Ricci tensor can be represented by four real and three complex scalars.

The anti-symmetric Maxwell tensor $F_{\alpha \beta}$ is expressed in terms of the three complex scalars:
\begin{subequations} \label{eq:def. of phi's}
\begin{align}
&\phi_{-1} \equiv F_{(1)(3)} \\
&\phi_{0} \equiv \frac{1}{2}\left(F_{(1)(2)}+F_{(4)(3)}\right) \\
&\phi_{+1} \equiv F_{(4)(2)}
\end{align}
\end{subequations}
The inverse relations are given by
\begin{equation}
F_{(a)(b)}=
\left(
\begin{array}{cccc}
0 & \phi_0+\phi_0^* & \phi_{-1} & \phi_{-1}^*     \\
-\phi_0-\phi_0^* & 0 & -\phi_{+1}^* & -\phi_{+1}  \\
-\phi_{-1} & \phi_{+1}^* & 0 &-\phi_0+\phi_0^*    \\
-\phi_{-1}^* & \phi_{+1} & \phi_0-\phi_0^* & 0 
\end{array}
\right)
\end{equation}

The commutation relations (\ref{eq:commutation rlns.}) are very useful and we therefore show their explicit
form in the NP formalism:
\begin{subequations} \label{eq:NP commutation rlns.}
\begin{align}
\bDelta D-D\bDelta&=(\gamma+\gamma^*)D+(\epsilon+\epsilon^*)\bDelta-(\pi+\tau^*)\delta-(\tau+\pi^*)\delta^* \\
\delta D-D\delta&=(\alpha^*+\beta-\pi^*)D+\kappa\bDelta-(\rho^*+\epsilon-\epsilon^*)\delta-\sigma\delta^* \\
\delta\bDelta-\bDelta\delta&=-\nu^*D+(\tau-\alpha^*-\beta)\bDelta+(\mu-\gamma+\gamma^*)\delta+\lambda^*\delta^* \\
\delta^*\delta-\delta\delta^*&=(\mu^*-\mu)D+(\rho^*-\rho)\bDelta+(\alpha-\beta^*)\delta+(\beta-\alpha^*)\delta^*
\end{align}
\end{subequations}

The Ricci and the Bianchi identities in the NP formalism can be found in ~\cite{bk:Chandr}.

The first-order change in a basis vector $\vec{e}_{(a)}$ when it suffers an infinitesimal displacement $\vec{\zeta}$ is 
given by $\delta e_{(a)\alpha}(\vec{\zeta})=e_{(a)\alpha;\beta}\zeta^{\beta}=-\gamma_{(a)(b)(c)}e^{(b)}{}_{\alpha}\zeta^{(c)}$.
Several consequences may be derived by studying the effect of such a displacement on the vector $\vec{l}$.
First of all, it is easy to see that
\begin{equation}
\delta l_{\alpha}(\vec{l})=l_{\alpha;\beta}l^{\beta}=(\epsilon+\epsilon^*)l_{\alpha}-\kappa m^*_{\alpha}-\kappa^* m_{\alpha} \label{eq:covar.deriv. of l in l-direction}
\end{equation}

It is clear from (\ref{eq:covar.deriv. of l in l-direction}) that the $\vec{l}$-vectors form a congruence of null geodesics
if, and only if, $\kappa=0$. If that is the case, they are affinely parametrized if, and only if, in addition $\Re(\epsilon)=0$.
The symbols $\Re$ and $\Im$ represent the real and imaginary parts respectively.

Furthermore, from equation (\ref{eq.def. of Ricci rot.-coeffs.}) together with the definition (\ref{eq:def. spin coeffs.})
of the spin coefficients, it is straight-forward to prove that
\begin{subequations}
\begin{align}
\frac{1}{2}l^{\alpha}{}_{;\alpha}&=-\Re(\rho) \label{eq:expansion of  l-bundle} \\
\frac{1}{2}l_{[\alpha;\beta]}l^{\alpha;\beta}&=\left(\Im(\rho)\right)^2 \label{eq:rotation of  l-bundle} \\
\frac{1}{2}l_{(\alpha;\beta)}l^{\alpha;\beta}&=\left(\Re(\rho)\right)^2+|\sigma|^2  \label{eq:shear of  l-bundle}
\end{align}
\end{subequations}

If we consider at each point of a null $\vec{l}$-ray a small circle orthogonal to $\vec{l}$ with that point as its centre,
and we then follow into the future null-direction the rays of the congruence $\vec{l}$ which intersect the circle, the circle may
become contracted or expanded, rotated or sheared. From equation (\ref{eq:expansion of  l-bundle}), the quantity $-\Re(\rho)$ measures the
possible expansion or contraction and, from (\ref{eq:rotation of  l-bundle}), $\Im(\rho)$ measures the possible rotation of the circle.
Finally, equation (\ref{eq:shear of  l-bundle}) shows that $|\sigma|$ is the shear of the bundle of $\vec{l}$-rays, as it measures the 
extent to which neighbouring $\vec{l}$-rays are sliding past each other.

Note that the full set of NP equations is invariant under the interchange 
$\{\vec{l}\leftrightarrow\vec{n}, \vec{m}\leftrightarrow\vec{m^*}\}$.   
\catdraft{only for Type D with $\vec{l}$ and $\vec{n}$ being pral.null.dirs.?Adrian:no, valid in gral.}
Such an interchange results in the following transformations:
\begin{equation} \label{eq:effect of l<->n,m<-m^* on NP scalars and spin coeffs.}
\begin{aligned}
\psi_{-2}&\leftrightarrow\psi_{+2} & \psi_{-1}&\leftrightarrow\psi_{+1} & \psi_{0}&\leftrightarrow\psi_{0} \\
\phi_{-1}&\leftrightarrow -\phi_{+1} & \phi_{0}&\leftrightarrow -\phi_{0} & \\
\kappa &\leftrightarrow -\nu & \rho &\leftrightarrow -\mu & \sigma&\leftrightarrow-\lambda \\
\alpha&\leftrightarrow-\beta & \epsilon&\leftrightarrow-\gamma & \pi&\leftrightarrow-\tau
\end{aligned}
\end{equation}
It is also clear that complex-conjugation of a gravitational or electromagnetic NP quantity is equivalent 
to swopping the tetrad indices 3 and 4.  

We can subject the NP frame to a Lorentz transformation at some point and extend it continuously through all of 
space-time. We have six degrees of freedom to rotate the frame while keeping $\eta_{(a)(b)}$ unchanged, 
corresponding to the six parameters of the
group of Lorentz transformations, and we can view such a general rotation as composed of the following 
three classes of rotations where $a$ and $b$ are complex fields and $A$ and $\varphi$ real fields:
\begin{itemize}

\item[a)] \textsl{rotation of class I}: 
\begin{equation}
\begin{aligned}
\vec{l}&\rightarrow\vec{l}           & \qquad \vec{n}&\rightarrow\vec{n}+a^*\vec{m}+a\vec{m^*}+aa^*\vec{l}\\
\vec{m}&\rightarrow \vec{m}+a\vec{l} & \qquad \vec{m^*}&\rightarrow \vec{m^*}+a^*\vec{l}
\end{aligned}
\end{equation}

The effect of such a rotation on the Maxwell scalars is: 
\begin{equation}
\begin{aligned}
\phi_{-1}&\rightarrow \phi_{-1} \qquad & \phi_{0}&\rightarrow \phi_{0}+a^*\phi_{-1} \qquad & \phi_{+1}&\rightarrow \phi_{+1}+2a^*\phi_0+a^{* 2}\phi_{-1}
\end{aligned}
\end{equation}
and on the Weyl scalars: 
\begin{equation}
\begin{aligned}
\psi_{-2}&\rightarrow \psi_{-2}                                           \quad               &\psi_{-1}&\rightarrow \psi_{-1}+a^*\psi_{-2} \\
\psi_{0}&\rightarrow \psi_{0}+2a^*\psi_{-1}+a^{* 2}\psi_{-2}              \quad               &\psi_{+1}&\rightarrow\psi_{+1}+3a^*\psi_0+3a^{* 2}\psi_{-1}+a^{* 3}\psi_{-2}\\
\psi_{+2}&\rightarrow\makebox[0pt][l]{$\displaystyle\psi_{+2}+4a^*\psi_{+1}+6a^{* 2}\psi_0+4a^{* 3}\psi_{-1}+a^{* 4}\psi_{-2}$}
\end{aligned}
\end{equation}

\item[b)] \textsl{rotation of class II}:
\begin{equation}
\begin{aligned}
\vec{l}&\rightarrow\vec{l}+b^*\vec{m}+b\vec{m^*}+bb^*\vec{n} &  \qquad\vec{n}&\rightarrow\vec{n} \\
\vec{m}&\rightarrow \vec{m}+b\vec{n}                          & \qquad\vec{m^*}&\rightarrow \vec{m^*}+b^*\vec{n}
\end{aligned}
\end{equation}

The effect of a rotation of class II on the Maxwell scalars can be derived from that of a rotation of class I using the transformations
(\ref{eq:effect of l<->n,m<-m^* on NP scalars and spin coeffs.}) and including a complex-conjugation since the exchange $
\vec{m}\leftrightarrow\vec{m^*}$ is not being performed now. The result is: 
\begin{equation}
\begin{aligned}
\phi_{-1}&\rightarrow \phi_{-1}+2b\phi_0+b^2\phi_{+1} \qquad & \phi_{0}&\rightarrow \phi_{0}+b\phi_{+1} \qquad & \phi_{+1}&\rightarrow \phi_{+1}
\end{aligned}
\end{equation}
and
\begin{equation}
\begin{aligned}
\psi_{-2}&\rightarrow \makebox[0pt][l]{$\displaystyle\psi_{-2}+4b\psi_{-1}+6b^2\psi_0+4b^3\psi_{+1}+b^4\psi_{+2}$} \\
\psi_{-1}&\rightarrow \psi_{-1}+3b\psi_0+3b^2\psi_{+1}+b^3\psi_{+2}   \qquad    &  \psi_{0}&\rightarrow \psi_{0}+2b\psi_{+1}+b^2\psi_{+2} \\
\psi_{+1}&\rightarrow \psi_{+1}+b\psi_{+2}                            \qquad    & \psi_{+2}&\rightarrow \psi_{+2}
\end{aligned}
\end{equation}

\item[c)] \textsl{rotation of class III}:
\begin{equation}
\begin{aligned}
\vec{l}&\rightarrow A^{-1}\vec{l}             & \qquad\vec{n}&\rightarrow A\vec{n} \\
\vec{m}&\rightarrow e^{i\varphi}\vec{m}       & \qquad\vec{m^*}&\rightarrow e^{-i\varphi}\vec{m^*}
\end{aligned}
\end{equation}

The Maxwell and Weyl scalars are respectively changed as follows under this rotation
\begin{equation}
\begin{aligned}
\phi_{-1}&\rightarrow A^{-1}e^{i\varphi}\phi_{-1} \qquad & \phi_{0}&\rightarrow \phi_{0} \qquad & \phi_{+1}&\rightarrow Ae^{-i\varphi}\phi_{+1}
\end{aligned}
\end{equation}
and
\begin{equation}
\begin{aligned}
\psi_{-2}&\rightarrow A^{-2}e^{2i\varphi}\psi_{-2} \qquad & \psi_{-1}&\rightarrow A^{-1}e^{i\varphi}\psi_{-1} \\
\psi_{0}&\rightarrow \psi_{0}                      &          &                                        \\
\psi_{+1}&\rightarrow Ae^{-i\varphi}\psi_{+1}      \qquad & \psi_{+2}&\rightarrow A^2e^{-2i\varphi}\psi_{+2}
\end{aligned}
\end{equation}

\end{itemize}

These three different classes of rotation allow us to classify the Weyl tensor in four different types depending
on how many of the Weyl scalars we can make zero by subjecting them to Lorentz transformations. Considering $\psi_{+2}\neq 0$
(if it were zero we could make it non-zero with a rotation of class I unless all Weyl scalars vanished or space were conformally flat),
we can make $\psi_{-2}$ vanish by applying a rotation of class II with a parameter $b$ satisfying:
\begin{equation} \label{eq:eq. for pnd's of Weyl tensor}
\psi_{-2}+4\psi_{-1}b+6\psi_0b^2+4\psi_{+1}b^3+\psi_{+2}b^4=0
\end{equation}
This equation for $b$ can have from 1 to 4 distinct solutions and the corresponding new direction(s) of the vector $\vec{l}$ (given by
$\vec{l}+b^*\vec{m}+b\vec{m^*}+bb^*\vec{n}$) is(are) called the \define{principal null direction(s) or Debever-Penrose direction(s)} of the Weyl tensor.
The Weyl tensor is called \define{algebraically general} if there are four distinct roots and otherwise it is called 
\define{algebraically special}. The \define{Petrov classification} further classifies the Weyl tensor depending on how many
distinct roots equation (\ref{eq:eq. for pnd's of Weyl tensor}) possesses:

\begin{itemize}
\item[a)] \define{Petrov Type I}: four distinct roots. By successively applying different
classes of rotation, both $\psi_{-2}$ and $\psi_{+2}$ can be made to vanish but $\psi_{\pm1}$ and $\psi_{0}$ cannot.

\item[b)] \define{Petrov Type II}: one double and two single roots. $\psi_{\pm2}$ and $\psi_{-1}$ can be made to vanish but $\psi_{+1}$ and $\psi_{0}$ cannot.

\item[c)] \define{Petrov Type D}: two distinct double roots. $\psi_{\pm2}$ and $\psi_{\pm1}$ can be made to vanish but $\psi_{0}$ cannot.

\item[d)] \define{Petrov Type III}: one triple and one single root. $\psi_{\pm2}$, $\psi_{-1}$ and $\psi_{0}$ can be made to vanish but $\psi_{+1}$ cannot.

\item[e)] \define{Petrov Type N}: only one distinct root. $\psi_{-2}$, $\psi_{\pm1}$ and $\psi_{0}$ can be made to vanish but $\psi_{+2}$ cannot.

\end{itemize}

Finally, the next theorem establishes a direct relationship between the values of certain spin coefficients and the Weyl tensor type. 
This theorem is restricted to the vacuum.
\\
\define{Goldberg-Sachs theorem}: If the Riemann tensor is of Type II and a null 
basis is so chosen that $\vec{l}$ is the repeated null direction and $\psi_{-2}=\psi_{-1}=0$, then $\kappa=\sigma=0$; and, 
conversely, if $\kappa=\sigma=0$, then $\psi_{-2}=\psi_{-1}=0$ and the Riemann tensor is of Type II.
\\
A corollary to this theorem results from the interchange $\vec{l}\leftrightarrow\vec{n}$ applied to the theorem. The corollary states that
if the Riemann tensor is of Petrov Type D, then the congruences formed by the two principal null-directions, $\vec{l}$
and $\vec{n}$, must both be geodesic (from (\ref{eq:covar.deriv. of l in l-direction})) and shear-free 
(from (\ref{eq:shear of  l-bundle})), i.e., $\kappa=\sigma=\nu=\lambda=0$ when $\psi_{\pm2}=\psi_{\pm1}=0$; and
conversely.

Kundt and Thompson ~\cite{ar:Kundt&Thomp'62} and Robinson and Schild ~\cite{ar:Robin&Schild'63} 
gave a generalization of the Goldberg-Sachs theorem which is not restricted to vacuum solutions.
Kundt and Tr\"{u}mper ~\cite{ar:Kundt&Trump'62} gave a generalization of the same theorem to
a particular type of Einstein-Maxwell fields that includes the Kerr-Newman solution described
in the following section.

\draft{'Exact slns. of Einsteins eqs.' seems to indicate that Kundt and Tr\"{u}mper's generalization imply that for tyep D 
s-t's (e.g.,K-N) it is either $\kappa=0$ or $\sigma=0$ but not necessarily both zero, but Carter'68 indicates that
this generalization implies that they are both zero for K-N?}

All black hole solutions of general relativity are of Petrov Type D and it is therefore possible to
choose a null tetrad such that the four spin coefficients $\kappa$, $\sigma$, $\nu$ and $\lambda$ and all Weyl scalars
except for $\psi_{0}$ are zero. Such is the case of both the Kinnersley and the Carter null tetrads, which we have 
chosen to use in all our calculations.

For the physically important, Petrov Type D backgrounds, the corresponding Weyl tensor and its dual satisfy
the equations
\begin{equation}  \label{eq:Weyl tensor in Petrov Type D s-t}
\begin{aligned}
C_{\alpha\beta\gamma[\delta} l_{\epsilon]}l^{\beta}l^{\gamma}&=0 \quad & \quad {}^*C_{\alpha\beta\gamma[\delta} l_{\epsilon]}l^{\beta}l^{\gamma}&=0
\end{aligned}
\end{equation}
where $\vec{l}$ represents any one of the two principal null congruences.

\catdraft{1) no se veure com (\ref{eq:Weyl tensor in Petrov Type D s-t}) son equivalents a $\psi_{\pm2}=\psi_{\pm1}=0$?,
2) a 'Exact sln. of Einstein eqs. no hi posa el dual a (\ref{eq:Weyl tensor in Petrov Type D s-t})?}


\section{The Kerr-Newman space-time} \label{sec:K-N}

\draft{generalize everything to Kerr-Newman}

The action integral
\begin{equation} \label{eq:Einstein-Maxwell action integral}
S=\int_{\mathcal{D}}\left(\frac{1}{16\pi}R+L_{\text{emag}}\right)\sqrt{-g}\d{x}^4
\end{equation}
corresponds to interacting gravitational and electromagnetic fields, 
where the integration is performed over the interior of a four-dimensional region $\mathcal{D}$ and 
$L_{\text{emag}}$ is the Lagrangian of the electromagnetic field.
The result of extremizing this action integral for interacting gravitational and electromagnetic fields
created by a mass $M$, intrinsic angular 
momentum per unit mass $a$ and charge $Q$ as seen at radial infinity, and subject to the existence
of a physically nonsingular horizon is the \define{Kerr-Newman geometry} and its associated electromagnetic field. 
These gravitational and electromagnetic fields were first found by Newman, Couch, Chinnapared, Exton, Prakash and Torrence ~\cite{ar:Newmanetal'65}
by applying a transformation to the charged, spherical solution of Reissner-Nordstr\"{o}m. 
The Kerr-Newman metric in the \define{Boyer-Lindquist ~\cite{ar:Boyer&Lind'67} co-ordinate system}  $\{t,r,\theta,\phi\}$
is 
\begin{equation} \label{eq:K-N metric}
\begin{aligned}
\d{s}^2
&=-\left(\frac{\Delta-a^2\sin^2\theta}{\Sigma}\right)\d{t}^2-\frac{2a\sin^2\theta(r^2+a^2-\Delta)}{\Sigma}\d{\phi}\d{t}+  \\
&+\left[\frac{(r^2+a^2)^2-a^2\Delta\sin^2\theta}{\Sigma}\right]\sin^2\theta\d{\phi}^2+\frac{\Sigma}{\Delta}\d{r}^2+\Sigma\d{\theta}^2=\\
&=-\frac{\Delta}{\Sigma}\left(\d{t}-a\sin^2\theta\d{\phi}\right)^2+\frac{\sin^2\theta}{\Sigma}\left[(r^2+a^2)\d{\phi}-a\d{t}\right]^2+
\frac{\Sigma}{\Delta}\d{r}^2+\Sigma\d{\theta}^2
\end{aligned}
\end{equation}
where
\begin{equation}\label{eq:def. of Sigma}
\Sigma\equiv r^2+a^2\cos^2\theta
\end{equation}
\begin{equation}
\Delta \equiv r^2-2Mr+a^2+Q^2
\end{equation}
The associated electromagnetic field is
\begin{equation} \label{eq:associated K-N emag. field}
\begin{aligned}
\vec{F}&=\frac{Q}{\Sigma^4}(r^2-a^2\cos^2\theta)\d{r}\wedge\left[\d{t}-a\sin^2\theta\d{\phi}\right]+ \\
&+2\frac{Q}{\Sigma^4}ar\cos\theta\sin\theta\d{\theta}\wedge\left[(r^2+a^2)\d{\phi}-a\d{t}\right]
\end{aligned}
\end{equation}

The co-ordinate $\phi$ is required to be periodic with period $2\pi$ so that the Kerr-Newman metric is
asymptotically flat for large $r$. 
The metric coefficients in Boyer-Lindquist co-ordinates are independent of $t$ and $\phi$. The spacetime
is therefore stationary and axially symmetric.
The Kerr-Newman metric possesses two Killing vectors associated with these two symmetries:   
\catdraft{com sabem que no n'hi ha mes??}
\begin{equation}
\vec{\xi}\equiv \pardiff{}{t}
\end{equation}
and
\begin{equation}
\vec{\psi}\equiv \pardiff{}{\phi}
\end{equation}

The Kerr-Newman metric is singular at 
\begin{equation} \label{eq:coord. sing. of K-N}
\Delta=0
\end{equation}
and also at
\begin{equation} \label{eq:curvature sing. of K-N}
\Sigma=0
\end{equation}
The evaluation of the curvature invariants shows that (\ref{eq:coord. sing. of K-N}) is a co-ordinate singularity 
while (\ref{eq:curvature sing. of K-N}) is a true, curvature singularity. 
\catdraft{p.159Hawk\&Ellis per R-N pero tambe sembla ser aixi per K-N->check;compte:a p.163Hawk\&Ellis es veu que per Kerr per $a^2>m^2$ timelike geods. hit sing.?!tot i
que cita Carter'68, que es per K-N?}
The physical singularity is not spacelike, as it is in the Schwarzschild case, but timelike, 
so that the singularity may be avoided by timelike and null curves.
In consequence, given any spacelike surface, it is always possible to find timelike and null curves that hit the singularity
and do not cross the spacelike surface. Therefore, Cauchy surfaces do not exist for the full space-time. 
They do exist for the exist for the exterior region, however. 
The physical singularity is at $\{r=0, \theta=\pi/2\}$, in Boyer-Lindquist co-ordinates. These co-ordinates, however, are not
to be treated as the usual polar co-ordinates in flat space. Indeed, in the limit $Q=M=0$ 
\catdraft{tambe caldria $a=0$ pero? pero llavors $\{\tilde{r},\tilde{\theta},\tilde{\phi}\}={r,\theta,\phi}$ i per tant ${r,\theta,\phi}$ si
serien polar coords. en flat s-t??}
the metric (\ref{eq:K-N metric}) would not be equal to the flat space metric if the Boyer-Lindquist co-ordinates were to be 
equated to the usual polar co-ordinates. 
Newman and Janis ~\cite{ar:Newman&Janis'65} have shown that it is instead the following set of co-ordinates $\{\tilde{r},\tilde{\theta},\tilde{\phi}\}$
that correspond to polar co-ordinates in flat space-time:
\begin{equation} \label{eq:polar coords. in Kerr}
\begin{aligned}
\tilde{r}^2&=r^2+a^2\sin^2\theta \\
\tan\tilde{\phi}&=\frac{\tan\phi-a/r}{1+(a/r)\tan\phi} \\
\cos\tilde{\theta}&=\frac{r\cos\theta}{(r^2+a^2\sin^2\theta)^{1/2}} 
\end{aligned}
\end{equation}
The singularity at $\{r=0, \theta=\pi/2\}$ corresponds to the circle $\{\tilde{r}=a, \tilde{\theta}=\pi/2\}$
and it is therefore a \define{ring singularity}. Carter ~\cite{ar:Carter'68a} has shown that the number of geodesics that may
reach the singularity is more restricted than in the Kerr case. For Kerr-Newman, no timelike geodesics can reach the singularity
and null geodesics may only reach it if they lie in the equator and have a uniquely determined angular momentum.   
In the Kerr background, on the other hand, both timelike and null geodesics may reach the singularity if they lie in the 
equator and their angular momentum lies within a finite range.

The other type of singularity corresponds to the solutions of (\ref{eq:coord. sing. of K-N}), which are 
\begin{equation} \label{eq:horizons}
r_{\pm}=M\pm\sqrt{M^2-a^2-Q^2}
\end{equation}
The two hypersurfaces $r=r_{\pm}$ are event horizons, as we shall now prove. 
If the surface $f(t,r,\theta,\phi)=0$ is to be a null hypersurface containing the Killing vectors $\vec{\xi}$ and $\vec{\psi}$
then it must be: $f(r,\theta)=0$ and $\d{f}_{\alpha}\d{f}^{\alpha}=0$. These two conditions imply
\begin{equation}
\frac{\Delta}{\Sigma}\left(\pardiff{f}{r}\right)^2+\frac{1}{\Sigma}\left(\pardiff{f}{\theta}\right)^2=0
\end{equation}
The only solutions of this equation which are periodic in $\theta$ are the null hypersurfaces $r=r_{\pm}$. 
The hypersurface $r=r_-$ is called the \define{inner event horizon} and the hypersurface $r=r_+$, the \define{outer event horizon}.
In the case $Q^2+a^2> M^2$ there is no co-ordinate singularity and therefore the curvature singularity
(\ref{eq:curvature sing. of K-N}) is a naked singularity. This case does not describe black holes.  
The \define{extreme case} denotes the case $Q^2+a^2=M^2$. The inner and outer horizons then coincide at $r_+=r_-=M$.
For the rest of this thesis we will restrict ourselves to the case $a^2+Q^2<M^2$, unless specified otherwise.
The radius of the outer horizon is related to the surface area $\mathcal{A}$ of a Kerr-Newman black hole by 
$\displaystyle \mathcal{A}=4\pi (r_+^2+a^2)$.  

The Killing vector $\vec{\xi}$ becomes a null vector 
where the equation $0=g_{\mu\nu}\xi^{\mu}\xi^{\nu}=g_{tt}$ is satisfied, which has two roots: 
\begin{equation}
r_{\topbott{1}{2}}=M\pm\sqrt{M^2-Q^2-a^2\cos^2\theta}
\end{equation} 
The surface defined by the Boyer-Lindquist radius $r_1$
is called the \define{stationary limit surface}. The stationary limit surface is timelike everywhere except at the axis, where it
is null and it coincides with the surface $r=r_+$. 
The region between the event horizon and the stationary limit surface, i.e., 
for $r_+<r<r_1$ is called the \define{ergosphere}. Within the ergosphere, the vector $\vec{\xi}$ is spacelike and therefore observers will not
be able to remain at rest with respect to radial infinity.
Outside the stationary limit surface, i.e., for $r\in(r_1,+\infty)$, $\vec{\xi}$ is a timelike vector.

The Killing vector $\vec{\psi}$ becomes timelike close to the ring singularity for negative values of $r$. Because 
$\phi$ is a periodic co-ordinate, the orbits of $\vec{\psi}$ are closed and therefore there exist closed timelike curves in
a neighborhood of the ring singularity. 

\draft{mention Killing tensor or conformal Killing spinor (p.893M,T\&W;p.321Wald)? and Killing-Yano tensor (p.10Ott\&Winst'00)?}

We can construct another Killing vector as
\begin{equation}
\vec{\chi}\equiv \vec{\xi}+\Omega_+\vec{\psi}
\end{equation}
with
\begin{equation} \label{eq:def. Omega_+}
\Omega_+\equiv \frac{a}{r_+^2+a^2}
\end{equation}
We shall see later on that $\Omega_+$ represents the angular velocity of the horizon.
The Killing vector $\vec{\chi}$ becomes null wherever $0=g_{\mu\nu}\chi^{\mu}\chi^{\nu}=g_{tt}+2\Omega_+g_{t\phi}+\Omega_+^2g_{\phi\phi}$ is satisfied. 
Its only real root corresponds to the \define{speed-of-light surface} and is given by
\begin{equation}  \label{eq:def. r_SOL}
r_{\text{SOL}}(\theta)=2\sqrt{-C}\cos\left(\frac{\Theta}{3}\right)-\frac{r_+}{3}
\end{equation}
where 
\begin{equation}
\begin{aligned}
A&=r_+^2+a^2(2-\sin^2\theta)-\frac{(r_+^2+a^2)^2}{a^2\sin^2\theta};   \quad & \quad  C&=\frac{3A-r_+^2}{9}\\    
B&=r_+A+2M\left[a^2\sin^2\theta-2(r_+^2+a^2)+\frac{(r_+^2+a^2)^2}{a^2\sin^2\theta}\right]  &&\\
D&=\frac{9r_+A-27B-2r_+^3}{54}; \quad & \quad \Theta&=\cos^{-1}\left(\frac{D}{\sqrt{-C^3}}\right)
\end{aligned}
\end{equation}
The vector $\vec{\chi}$ is timelike between the event horizon and the speed-of-light surface, and it is spacelike 
for a radius larger than $r_{\text{SOL}}$.

\begin{figure}[p]
\centering
\includegraphics*[width=90mm]{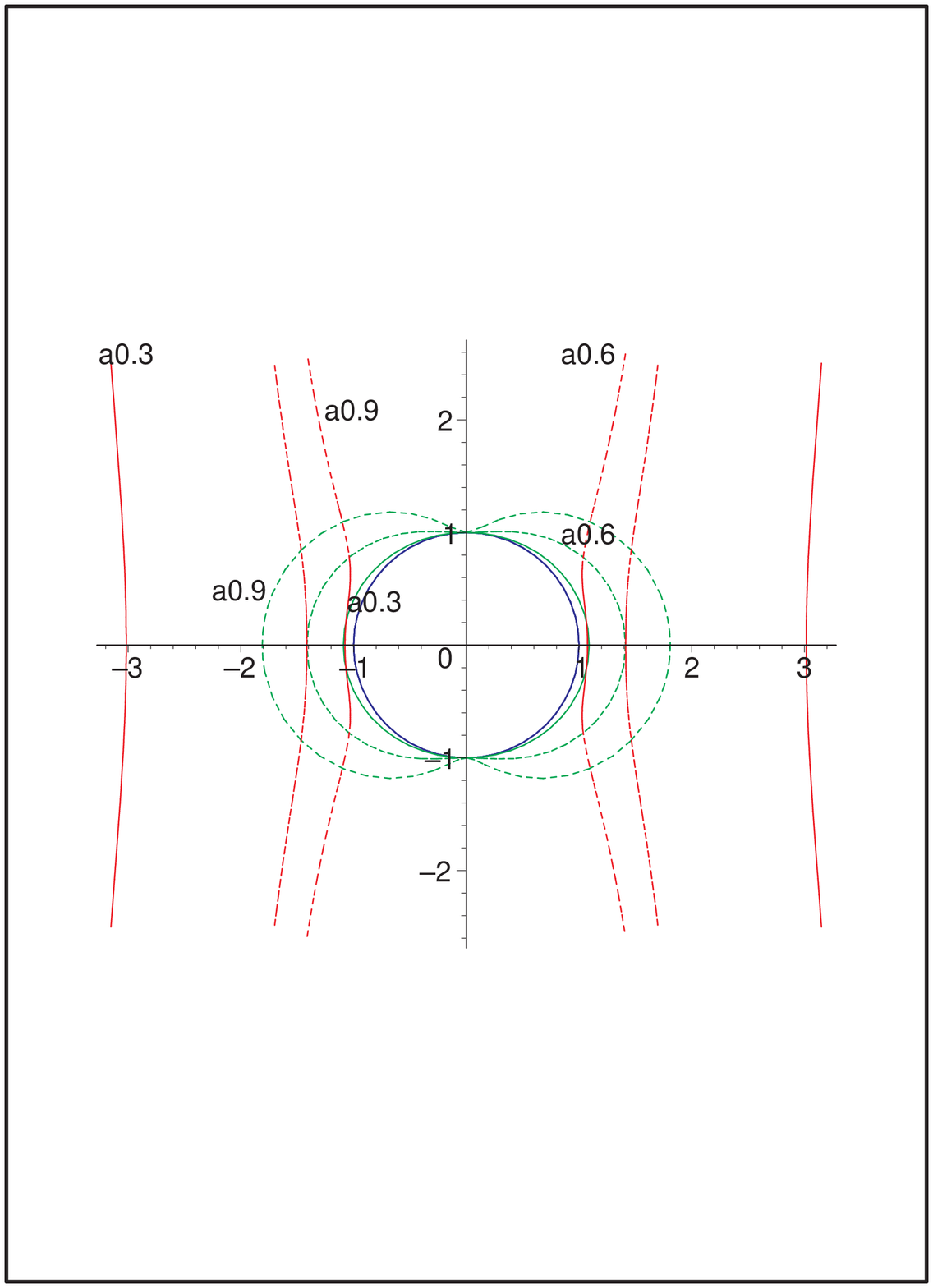}
\caption{Values of the radii $r_+$ (blue), $r_1$ (green) and $r_{\text{SOL}}$ (red) as functions of the angle $\theta$, 
corresponding to hypersurfaces of constant $t$ and $\phi$. 
The parameters of the black hole are: $Q=0$ and $a=0.3$ (straight line), $a=0.6$ (dashed line) and $a=0.9$ (lighter, dashed line).
Values have been normalized so that $r_+=1$. With this normalization, the radius $r_{\text{SOL}}$ would not vary with $Q$.}
\label{fig:plot_KNsurfaces_Q0}
\end{figure}

Figure \ref{fig:plot_KNsurfaces_Q0} shows the radii of the outer event horizon, the static limit surface and
the speed-of-light surface as functions of the angle $\theta$ for various values of the intrinsic angular momentum per unit mass $a$.
The value of the mass has been chosen to be $M=(1+a^2+Q^2)/2$ so that $r_+=1$. Since $r_{\text{SOL}}$ only depends on $Q$ via $r_+$,
with the chosen normalization this radius does not vary with $Q$. As the intrinsic angular momentum per unit mass increases, 
$r_{\text{SOL}}$ diminishes whereas $r_1$ increases, as expected. The radius of the speed-of-light surface 
becomes infinite at the axis of symmetry and it reaches a minimum value at the equator.
The cusp observed for the speed-of-light surface at the axis of symmetry is merely a manifestation of the choice
of co-ordinate system, rather than the geometry of the space-time.

Apart from the Killing isometries, the Kerr-Newman solution 
is also invariant under the discrete symmetry
\begin{equation}  \label{eq:symm. (t,phi)->(-t,-phi)}
(t,\phi)\to (-t,-\phi)
\end{equation}
as expected from a rotational source, and under the \define{parity operation $\mathcal{P}$}:
\begin{equation}  \label{eq:parity op.}
\mathcal{P} \equiv (\vec{r}\to-\vec{r})=(\theta \to \pi-\theta,\phi \to \phi+\pi)
\end{equation}
We say that $M$, $a$ and $Q$ are respectively the mass, intrinsic angular momentum per unit mass and charge of the black
hole in the sense that
\begin{subequations}
\begin{align}
M&=-\frac{1}{8\pi} \int_S\eta_{\alpha\beta\gamma\delta}\nabla^{\gamma}\xi^{\delta}\d{x}^{\alpha}\wedge\d{x}^{\beta} \\
J\equiv Ma&=\frac{1}{16\pi}\int_S\eta_{\alpha\beta\gamma\delta}\nabla^{\gamma}\psi^{\delta}\d{x}^{\alpha}\wedge\d{x}^{\beta} \\
4\pi Q&=\frac{1}{2}\int_S\eta_{\alpha\beta\gamma\delta}F^{\gamma\delta}\d{x}^{\alpha}\wedge\d{x}^{\beta}
\end{align}
\end{subequations}
where $\eta_{\alpha\beta\gamma\delta}$ are the components of the space-time's volume 4-form and $S$ is any spacelike 
2-surface which has the topology of a 2-sphere, completely surrounds the source and lies entirely in the vacuum region. 
By comparison with the Schwarzchild and the Reissner-Nordstr\"{o}m solutions it can be easily seen from 
(\ref{eq:K-N metric}) and (\ref{eq:associated K-N emag. field}) that $M$ and $Q$ represent the mass and the
electric charge in the limit of large positive $r$. Similarly, in the limit of large negative $r$, the mass and the charge are respectively
$-M$ and $-Q$. On the other hand, the parameter $a$ is the cause for Coriolis-type forces which in the limit $r\to+\infty$ are identical to the ones
created by a rotating body with angular momentum $Ma$ in the weak-field limit. 

\catdraft{em manca veure que el camp asympt. seguent es el que correspon al de un ?? en flat space, de manera que Q sigui carrega,etc:
Note that in the limit of large $r$, the dominant components of the electric and magnetic fields (\ref{eq:associated K-N emag. field})
are
\begin{equation} \label{eq:associated K-N emag. field;large r}
\begin{aligned}
E_{\hat{r}}&=\frac{Q}{r^2} & B_{\hat{r}}&=\frac{2Qa}{r^3}\cos\theta & B_{\hat{\theta}}&=\frac{Qa}{r^3}\sin\theta 
\end{aligned}
\end{equation}
where the hat on an index indicates that they are components with respect to the basis 
$\{\vec{\omega}^{t}=\d{t},\vec{\omega}^{r}=\d{r},\vec{\omega}^{\theta}=r\d{\theta},\vec{\omega}^{\phi}=r\sin\theta\d{\phi}\}$.}

The \define{surface gravity on the outer[inner] horizon} is $\kappa_+[\kappa_-]$ 
defined as
\begin{equation} \label{eq:surface grav.}
\kappa_{\pm}\equiv\frac{r_{\pm}-r_{\mp}}{2(r_{\pm}^2+a^2)}
\end{equation}
For a Schwarzchild black hole, $\kappa_+$ is the value of the force that must be exerted at radial infinity 
to hold a unit test mass in place in the limit of it lying on the horizon. 
\catdraft{que vol dir ``in place''?ja que anul.lar l'accelerac. no vol dir que la veloc. sigui zero i que la part. no
es mogui. pq. es ``at radial infinity''?}

The radial co-ordinate $r_*$ defined by
\begin{equation} \label{eq: def. dr_*/dr}
\diff{r_*}{r}=\frac{(r^2+a^2)}{\Delta}
\end{equation}
is commonly known as the \define{tortoise} co-ordinate. We choose the constant of integration in 
(\ref{eq: def. dr_*/dr}) so that $r_*=r_*(r)$ coincides with Chandrasekhar's ~\cite{bk:Chandr}, i.e.,
\begin{equation} \label{eq: r_*}
r_*=r+\frac{1}{2\kappa_+}\ln(r-r_+)+\frac{1}{2\kappa_-}\ln(r-r_-)
\end{equation}

\draft{give metric as maximal extension in null coords. (p.157Hawk\&Ellis for R-N)}

\begin{figure}[p]
\setlength{\unitlength}{1.5pt}
\begin{center}
\begin{picture}(142,284)
\put(71,142){
\rotatebox{45}{\makebox(0,0){
\begin{picture}(250,250)
\put(45,0){\line(1,0){55}}
\put(0,50){\line(1,0){150}}
\put(0,100){\line(1,0){200}}
\put(50,150){\line(1,0){200}}
\put(100,200){\line(1,0){150}}
\put(150,250){\line(1,0){55}}
\put(0,45){\line(0,1){55}}
\put(50,0){\line(0,1){150}}
\put(100,0){\line(0,1){200}}
\put(150,50){\line(0,1){200}}
\put(200,100){\line(0,1){150}}
\put(250,150){\line(0,1){55}}
\put(100,50){\circle*{2}}
\put(150,50){\circle*{2}}
\put(150,100){\circle*{2}}
\qbezier[50](50,0)(75,25)(100,50)
\qbezier[50](50,0)(100,0)(100,50)
\qbezier[50](50,0)(50,50)(100,50)
\qbezier[50](50,0)(87,13)(100,50)
\thicklines
\qbezier(50,0)(63,37)(100,50)
\thinlines
\qbezier[50](0,50)(25,75)(50,100) 
\qbezier[50](0,50)(50,50)(50,100)
\qbezier[50](0,50)(0,100)(50,100)
\qbezier[50](0,50)(13,87)(50,100)
\thicklines
\qbezier(0,50)(37,63)(50,100)
\thinlines
\qbezier[50](100,50)(125,75)(150,100) 
\qbezier[50](100,50)(150,50)(150,100)
\qbezier[50](100,50)(100,100)(150,100)
\qbezier[50](100,50)(137,63)(150,100)
\qbezier[50](100,50)(113,87)(150,100)
\qbezier[50](50,100)(75,125)(100,150) 
\qbezier[50](50,100)(100,100)(100,150)
\qbezier[50](50,100)(50,150)(100,150)
\qbezier[50](50,100)(87,113)(100,150)
\qbezier[50](50,100)(63,137)(100,150)
\qbezier[50](150,100)(175,125)(200,150) 
\qbezier[50](150,100)(200,100)(200,150)
\qbezier[50](150,100)(150,150)(200,150)
\qbezier[50](150,100)(187,113)(200,150)
\thicklines
\qbezier(150,100)(163,137)(200,150)
\thinlines
\qbezier[50](100,150)(125,175)(150,200)  
\qbezier[50](100,150)(150,150)(150,200)
\qbezier[50](100,150)(100,200)(150,200)
\qbezier[50](100,150)(113,187)(150,200)
\thicklines
\qbezier(100,150)(137,163)(150,200)
\thinlines
\qbezier[50](150,200)(175,225)(200,250)
\qbezier[50](150,200)(200,200)(200,250)
\qbezier[50](150,200)(150,250)(200,250)
\qbezier[50](150,200)(187,213)(200,250)
\qbezier[50](150,200)(163,237)(200,250)
\qbezier[50](200,150)(225,175)(250,200)
\qbezier[50](200,150)(250,150)(250,200)
\qbezier[50](200,150)(200,200)(250,200)
\qbezier[50](200,150)(237,163)(250,200)
\qbezier[50](200,150)(213,187)(250,200)
\qbezier[50](100,50)(75,75)(50,100)
\qbezier[50](100,50)(50,50)(50,100)
\qbezier[50](100,50)(100,100)(50,100)
\qbezier[50](100,50)(63,63)(50,100)
\qbezier[50](100,50)(87,87)(50,100)
\qbezier[50](150,100)(125,125)(100,150)
\qbezier[50](150,100)(100,100)(100,150) 
\qbezier[50](150,100)(150,150)(100,150)
\qbezier[50](150,100)(113,113)(100,150)
\qbezier[50](150,100)(137,137)(100,150)
\qbezier[50](200,150)(175,175)(150,200)
\qbezier[50](200,150)(150,150)(150,200)
\qbezier[50](200,150)(200,200)(150,200)
\qbezier[50](200,150)(163,163)(150,200)
\qbezier[50](200,150)(187,187)(150,200)
\put(125,42){\rotatebox{-45}{\makebox(0,0){$\manifold{I}^-$}}}
\put(158,75){\rotatebox{-45}{\makebox(0,0){$\manifold{I}^+$}}}
\put(92,75){\rotatebox{-45}{\makebox(0,0){$\manifold{H}^-$}}}
\put(125,108){\rotatebox{-45}{\makebox(0,0){$\manifold{H}^+$}}}
\put(155,45){\rotatebox{-45}{\makebox(0,0){$i^0$}}}
\put(142,92){\rotatebox{-45}{\makebox(0,0){$i^+$}}}
\put(107,57){\rotatebox{-45}{\makebox(0,0){$i^-$}}}
\put(75,25){\rotatebox{-45}{\makebox(0,0){$-III^*$}}}
\put(25,75){\rotatebox{-45}{\makebox(0,0){$-III$}}}
\put(125,75){\rotatebox{-45}{\makebox(0,0){$I=-I$}}}
\put(125,175){\rotatebox{-45}{\makebox(0,0){$III$}}}
\put(175,125){\rotatebox{-45}{\makebox(0,0){$III^*$}}}
\put(75,125){\rotatebox{-45}{\makebox(0,0){$I^*$}}}
\put(75,75){\rotatebox{-45}{\makebox(0,0){$-II$}}}
\put(125,125){\rotatebox{-45}{\makebox(0,0){$II=II^*$}}}
\definecolor{light}{gray}{.5}
\put(175,175){\rotatebox{-45}{\makebox(0,0){\textcolor{light}{$II$}}}}
\put(225,175){\rotatebox{-45}{\makebox(0,0){\textcolor{light}{$I$}}}}
\put(175,225){\rotatebox{-45}{\makebox(0,0){\textcolor{light}{$I$}}}}
\put(220,90){\vector(-3,4){45}}
\put(229,75){\rotatebox{-45}{\makebox(0,0){$r=r_-$}}}
\put(170,40){\vector(-3,4){45}}
\put(179,25){\rotatebox{-45}{\makebox(0,0){$r=r_+$}}}
\put(195,65){\vector(-3,4){38}}
\put(204,50){\rotatebox{-45}{\makebox(0,0){$r=0$}}}
\end{picture}}}}
\end{picture}
\end{center}
\caption{
Conformal diagram (see ~\cite{bk:Wald'84}) for the extended Kerr-Newman spacetime along the axis of 
symmetry for the case $a^2+Q^2<M^2$. 
The dotted lines indicate surfaces of constant $r$. 
The bold lines indicate the ring singularities, which lie on $\theta=\pi/2$.
}
\label{fig:PenroseKerr}
\end{figure}
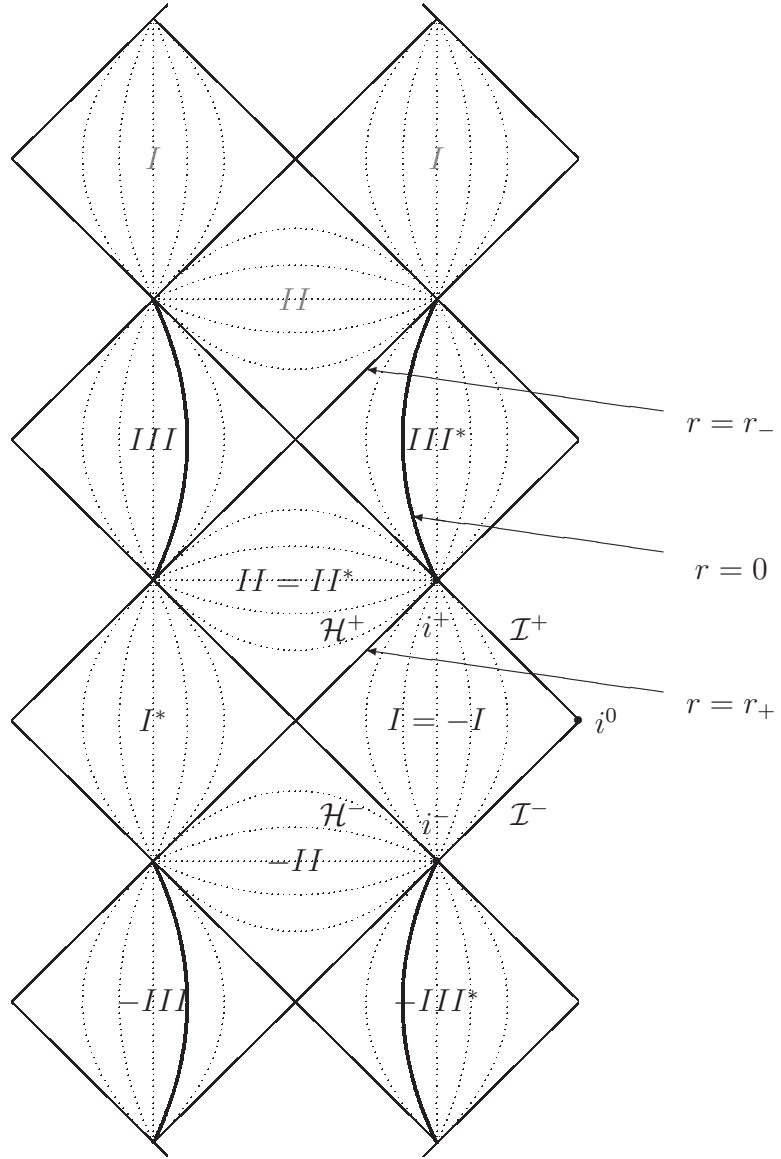

The \define{retarded $u$ and advanced $v$ time co-ordinates} are defined via
\begin{equation}
\begin{aligned}
\d{u}&\equiv \d{t}-\d{r_*} \\
\d{v}&\equiv \d{t}+\d{r_*}
\end{aligned}
\end{equation}
we can also define a new pair of angular co-ordinates by
\begin{equation}
\begin{aligned}
\d{\bar{\phi}}&\equiv\d{\phi}+\frac{a}{\Delta}\d{r} \\
\d{\bar{\phi'}}&\equiv \d{\phi}-\frac{a}{\Delta}\d{r}
\end{aligned}
\end{equation}
\draft{p.423Chandr. calls $\{v,r,\theta,\bar{\phi}\}$ Kerr-Schild?}
The system of co-ordinates $\{v,r,\theta,\bar{\phi}\}$ is called \define{Kerr system} of co-ordinates (~\cite{bk:M&T&W}), and it is a generalization
of the ingoing Eddington-Finkelstein system in the Schwarzchild space-time. 
The Kerr-Newman metric can be analytically extended across the horizons $r_{\pm}$ by transforming to the Kerr system.
Both the metric and the associated electromagnetic field are indeed analytic at $r=r_{\pm}$ when expressed in these co-ordinates. 
In this set of co-ordinates, we denote the regions $r_+<r<+\infty$, $r_-<r<r_+$ and $-\infty<r<r_-$ by $I$, $II$ and $III$ respectively.
Analogously, the metric can also be extended across the horizons by transforming to the co-ordinates $\{u,r,\theta,\bar{\phi'}\}$.
In this set of co-ordinates, we denote the regions $r_+<r<+\infty$, $r_-<r<r_+$ and $-\infty<r<r_-$ by $-I$, $-II$ and $-III$ respectively.
We have assumed that the common region to both co-ordinate systems is the one for $r>r_+$ and therefore regions $I$ and $-I$ are the
same. We can now take another patch formed with the co-ordinates $\{u,r,\theta,\bar{\phi'}\}$ containing regions $I^*$, $II^*$ and 
$III^*$, such that regions $II$ and $II^*$ coincide. Boyer and Lindquist found a system of co-ordinates, analogous to Kruskal's
for the Schwarzchild solution, which spans the regions $\{I,II,I^*,-II\}$ and such that the metric is regular throughout these regions.
Similarly, they found another system of co-ordinates that spans, and is regular throughout (except at the ring singularity),
regions $II$, $III$, $III^*$ plus another region which, like $II$, is bounded by two pairs of horizons. 
This extension procedure can be continued indefinitely both upward and downward.
The result, which is the \define{maximal analytic extension} of the space-time for the case $a^2+Q^2<M^2$, is depicted in figure \ref{fig:PenroseKerr}. 
The surfaces $\mathcal{H}^{\pm}$ are the future and past horizons respectively,
and the surfaces $\mathcal{I}^{\pm}$ are the future and past null infinity respectively, 
as shown in figure \ref{fig:PenroseKerr}. 
The point $i^0$ is spacelike infinity and the points $i^\pm$ are future and past null infinity respectively.

Regions $I$ and $I^*$ are asymptotically flat regions where $r_+<r<\infty$. 
Regions $II$ and $-II$, where $r_-<r<r_+$, represent a black hole and 
a white hole respectively. Unlike the Schwarzschild space-time, they do not contain the curvature singularity, but the
inner horizon instead. In these regions, the surfaces $r=const.$ are spacelike and therefore these regions contain 
closed trapped surfaces. 

If the inner horizon is crossed from region $II$, then regions $III$ and $III^*$, where $-\infty<r<r_-$ and which are identical in structure to each other,
are encountered. These regions contain the curvature singularity. Since it is a ring singularity, it is possible to pass through it
and enter another asymptotically flat region 
for $r\to-\infty$. Alternatively, since the singularity is timelike,
it may be avoided by entering a region which is identical in structure to $-II$.  

The global structure of the space-time in the extreme case is similar to the one for the non-extreme case, but differs
in that the regions $r_-<r<r_+$ do not exist.


\draft{p.1Ott\&Winst'00: $r=r_-$ is a Cauchy horizon}

A new azimuthal angular variable $\phi_+$ may be defined as
\begin{equation}
\phi_+\equiv\phi-\Omega_+t
\end{equation}
It is found that in the co-ordinate system $\{u,v,\theta,\phi_+\}$ and close to the past and future horizons:
\begin{equation}
\vec{\chi}=
\begin{cases}
\pardiff{}{u} & \text{at} \quad \mathcal{H^-} \\
\pardiff{}{v} & \text{at} \quad \mathcal{H^+}
\end{cases}
\end{equation}

The \define{rigidly rotating co-ordinate system} is given by $\left\{t_+,r,\theta,\phi_+\right\}$, where $t_+\equiv t$. 
In this system, the Killing vector $\vec{\chi}$ takes up the form
\begin{equation}
\vec{\chi}=\pardiff{}{t_+}
\end{equation}
Accordingly, we can construct two hamiltonians associated one with the Killing vector $\vec{\xi}$ 
and the other one with the Killing vector $\vec{\chi}$:
\begin{equation}
\begin{aligned}
\hat{H}_{\vec{\xi}}&\equiv i\vec{\xi}=i\pardiff{}{t} \\
\hat{H}_{\vec{\chi}}&\equiv i\vec{\chi}=i\pardiff{}{t_+}
\end{aligned}
\end{equation}

\catdraft{com es que arreu (e.g.,p.906MT\&W) es energia$=-\vec{p}*$Killing vector??}

\section{Orthonormal and null tetrads} \label{sec:particular tetrads}

An observer who moves along a world line of constant $r$ and $\theta$ with an angular velocity 
$\omega\equiv\d{\phi}/\d{t}$
relative to the asymptotic rest frame has a tetrad $\{\vec{e}_{(t)},\vec{e}_{(r)},\vec{e}_{(\theta)},\vec{e}_{(\phi)}\}$ 
associated with him. Such an observer sees no local change in the geometry and is therefore considered a \define{stationary
observer} relative to the local geometry. If his angular velocity is zero, and therefore he moves along a world line of 
constant $r$, $\phi$ and $\theta$, he is a \define{static observer (SO)} relative to radial infinity. A SO moves along the integral
curves of $\vec{\xi}$.         
If we require $\vec{e}_{(r)}$ and $\vec{e}_{(\theta)}$ to be parallel to 
$\partial/\partial r$ and $\partial/\partial \theta$ 
respectively, we then find that the 
vectors in the tetrad of a stationary observer are given by        
\begin{equation} \label{eq:tetrad of stationary obs.}
\begin{aligned}
\vec{e}_{(t)}&=\frac{1}{\sqrt{\left|g_{tt}+2\omega g_{t\phi}+\omega^2g_{\phi\phi}\right|}}\left(\pardiff{}{t}+\omega\pardiff{}{\phi}\right) \\
\vec{e}_{(r)}&=\sqrt{\frac{\Delta}{\Sigma}}\pardiff{}{r}   \\
\vec{e}_{(\theta)}&=\sqrt{\frac{1}{\Sigma}}\pardiff{}{\theta} \\
\vec{e}_{(\phi)}&=\frac{1}{\sqrt{\left|g_{tt}+2\omega g_{t\phi}+\omega^2g_{\phi\phi}\right|}}\frac{1}{\sqrt{g^2_{t\phi}-g_{tt}g_{\phi\phi}}}
\left[-(g_{t\phi}+\omega g_{\phi\phi})\pardiff{}{t}+(g_{tt}+\omega g_{t\phi})\pardiff{}{\phi}\right]
\end{aligned}
\end{equation}
Only the angular velocities $\omega$ such that $\vec{e}_{(t)}$ is a timelike vector are valid. The corresponding range of validity for $\omega$
is $\omega_{\text{min}}<\omega<\omega_{\text{max}}$ where
\begin{subequations}
\begin{align}
\omega_{\topbott{\text{min}}{\text{max}}}&=\Omega\mp\sqrt{\Omega^2-\frac{g_{tt}}{g_{\phi\phi}}} \\
\Omega&\equiv\frac{1}{2}\left(\omega_{\text{min}}+\omega_{\text{max}}\right)=-\frac{g_{t\phi}}{g_{\phi\phi}}  \label{eq:def. of Omega}
\end{align}
\end{subequations}
Note that at asymptotic radial infinity it is: $\displaystyle r\omega_{\topbott{\text{min}}{\text{max}}}=\mp 1$, as it should be.
As the Boyer-Lindquist radius $r$ decreases, the value of $\omega_{\text{min}}$ increases until it reaches the value
zero at the stationary limit surface. If $r$ decreases below the stationary limit surface then both $\omega_{\text{min}}$
and $\omega_{\text{max}}$ are positive and all stationary observers must orbit around the black hole with positive angular
velocity. Static observers cannot exist inside the ergosphere and hence the name of static limit surface for the region
outside the ergosphere. As $r$ decreases below the stationary limit surface, the two positive quantities $\omega_{\text{min}}$
and $\omega_{\text{max}}$ become larger and closer to each other until they coincide 
with value $+\sqrt{g_{tt}/g_{\phi\phi}}$ 
at the event horizon. All timelike 
curves 
at the horizon point inside the black hole.
\define{Dragging of inertial frames} refers to the fact that as a stationary observer approaches the horizon, 
the range of validity for its angular velocity as seen by an observer at infinity becomes narrower and contains
ever larger values.            

The quantity $p_{\phi}\equiv \vec{p}\vec{\psi}=p_{\alpha}\psi^{\alpha}$, where $\vec{p}$ is the 4-momentum of a certain observer, is the component
of angular momentum of that observer along the black hole's spin axis.
This quantity is conserved for geodesic observers.
The only stationary observers for whom this quantity is zero are those with an angular velocity $\Omega$ (\ref{eq:def. of Omega}). 
These observers are the closest analogue to the static observers in Schwarzschild space-time on the Kerr-Newman space-time in the
sense that their 4-velocity is orthogonal to $\vec{\psi}$, the hypersurfaces of constant $t$. 
These observers are called 
\define{zero angular momentum observers (ZAMO)}, or alternatively, \define{locally non-rotating observers (LNRO)}. 
The angular velocity $\omega_{ZAMO}$($=\Omega$) of the ZAMOs as they approach the horizon tends to $\omega_{ZAMO}(r=r_+)=\Omega_+$, which can therefore be
interpreted as the angular velocity of the horizon. The stationary observers whose angular velocity is constant and equal to $\Omega_+$ are observers
that follow integral curves of $\vec{\chi}$ and are called \define{rigidly rotating observers (RRO)}.        
Their 4-velocity becomes null at the speed-of-light surface $r=r_{\text{SOL}}$ and is spacelike for $r>r_{\text{SOL}}$.

The last orthonormal tetrad we wish to present is the \define{Carter orthonormal tetrad} ~\cite{ar:Carter'68b}, 
\catdraft{aquesta es la ref. a ``grav. in astrophysics'' i a p.34GavPhD, tot i que jo no trobo aquesta tetrada en aquesta ref. i CCH i Ott\&Winst'00Lett coinc. a donar
la mateixa ref., diff. de ~\cite{ar:Carter'68b}?.......}
which corresponds to a stationary observer (\ref{eq:tetrad of stationary obs.}) with angular velocity 
\begin{equation}
\omega_{c}=\frac{a}{r^2+a^2}
\end{equation}
The Carter orthonormal tetrad is given by
\begin{equation} \label{eq:def. Carter ortho. tetrad}
\begin{aligned}
\vec{e}_{(t)}&=+\frac{r^2+a^2}{\sqrt{\Delta\Sigma}}\left[\pardiff{}{t}+\frac{a}{r^2+a^2}\pardiff{}{\phi}\right] \\
\vec{e}_{(\phi)}&=\frac{1}{\sqrt{\Sigma}}\left[a\sin\theta\pardiff{}{t}+\frac{1}{\sin\theta}\pardiff{}{\phi}\right]
\end{aligned}
\end{equation}
together with $\vec{e}_{(r)}$ and $\vec{e}_{(\theta)}$ of equation (\ref{eq:tetrad of stationary obs.}).

All the above tetrads are orthonormal, consisting of one timelike and three spacelike vectors, and they can therefore be associated to observers
is some regions of exterior Kerr-Newman.
The Newman-Penrose formalism, however, is constructed with a basis of four null vectors. We will next give the two null tetrads that
are most commonly used, which are the only two we have made use of in this thesis.

Kerr and Schild ~\cite{co:Kerr&Schild'65} have shown that the Kerr-Newman metric can be expressed as
\begin{equation} \label{eq:K-N metric in terms of flat space metric}
g_{\alpha\beta}=\eta_{\alpha\beta}+2Hk_{\alpha}k_{\beta}
\end{equation}
where $H$ is a scalar field and $k_{\alpha}$ are the covariant components of a null vector field, then
the null congruence $k_{\alpha}$ is geodesic and $H$ can be chosen so that $k_{\alpha}$ is an affinely-parametrized geodesic.
They further showed that $k_{\alpha}$ is then a principal null direction.  

The Weyl tensor for the Kerr-Newman metric has two principal null directions;  
these correspond to the two different null vectors $\vec{k}$ such that equation (\ref{eq:K-N metric in terms of flat space metric})
for the Kerr-Newman metric is satisfied.
The Kerr-Newman solution is therefore of Petrov Type D.

The \define{advanced Eddington-Finkelstein co-ordinate system} is defined by
the set of co-ordinates $\{\bar{t},r,\theta,\bar{\phi}\}$ where
\begin{equation}
\d{\bar{t}}\equiv\d{t}+\frac{2Mr}{\Delta}\d{r} \\
\end{equation}
If the Kerr-Newman metric is expressed in the advanced Eddington-Finkelstein co-ordinates, we obtain in contravariant form:
\begin{equation} \label{eq:contravar. K-N metric in t_bar,r,phi_bar,theta coords.}
\parddiff{}{s}=-\parddiff{}{\bar{t}}+\frac{1}{\Sigma}\left[(r^2+a^2)\parddiff{}{r}+2a\pardiff{}{r}\pardiff{}{\bar{\phi}}
+\frac{1}{\sin^2\theta}\parddiff{}{\bar{\phi}}+\parddiff{}{\theta}\right]-\frac{2Mr}{\Sigma}\left(\pardiff{}{\bar{t}}-\pardiff{}{r}\right)^2
\end{equation}
By comparing equations (\ref{eq:K-N metric in terms of flat space metric}) and (\ref{eq:contravar. K-N metric in t_bar,r,phi_bar,theta coords.}),
we can immediately see that the vector for one of the principal null congruences must be:
\begin{equation} \label{eq:l in t_bar,r,phi_bar,theta coords}
\vec{n}
=
\frac{\Delta}{2\Sigma}\left(
\pardiff{}{\bar{t}}-\pardiff{}{r}
\right)
\end{equation}
in the advanced Eddington-Finkelstein co-ordinate system.
\draft{1) give proportionality factor, 2) specify horizons in text below}
\catdraft{hauria d'aclarir pq. es tria aquesta normalitzac.->explicac. deu ser a Kinnersley's paper-llegir-lo! o sino, p.902MT\&W}
The other principal null direction $\vec{l}$ may be obtained by using the fact that the Kerr-Newman metric is invariant under 
the symmetry $(t,\phi)\to(-t,-\phi)$, 
It may alternatively be obtained by performing the transformation $\d{r}\to -\d{r}$, which interchanges ingoing and outgoing rays. 
The vectors $\vec{l}$ and $\vec{n}$ are respectively outgoing and ingoing. 
Their normalization is chosen so that they are regular at the inner and outer horizons 
when expressed in the advanced Eddington-Finkelstein system. 
The Boyer-Lindquist co-ordinates are singular at the horizon. 
For a particle or photon falling inward through the horizon not only the Boyer-Lindquist co-ordinate $t$
will go to infinity as the horizon is approached, but also $\phi$ will go to infinity, due to the dragging
of inertial frames. A set of co-ordinates that remain finite as a particle or photon fall inward through the horizon
will therefore need to perform an infinite untwisting of $\phi$ as well as an infinite compression of $t$ 
in the neighborhood of the horizon. The Kerr co-ordinate system 
achieves both objectives. The ingoing principal null congruence expressed in the Kerr co-ordinates is simply
\begin{equation}
\vec{n}=-\pardiff{}{r}
\end{equation}
\draft{must find exact equation}
It can be shown that the only particle that can remain forever at the horizon is massless and is the one that follows
the curves of the outgoing principal null congruence $\vec{l}$. We therefore say that $\vec{l}$ 
is the generator of the horizon. These particles have angular velocity $\Omega_+$. 
The fact that their angular velocity is non-zero is another consequence of the dragging of inertial frames. 

The congruences given by $\vec{l}$ and the ones given by $\vec{n}$ are both geodesic and are the two principal null directions
of the Kerr-Newman space-time.
The \define{Kinnersley tetrad} ~\cite{ar:Kinnersley'69} consists in the null vectors $\vec{l}$ and $\vec{n}$
together with another null vector, $\vec{m}$, and its complex conjugate, chosen so that the
normalization conditions correspond to those of the Newman-Penrose tetrad, i.e., (\ref{eq:metric for NP tetrad}). 
The Kinnersley tetrad is therefore formed with the vectors
\begin{subequations} \label{eq:def. null Kinnersley tetrad}
\begin{align}
\vec{l}&=\frac{1}{\Delta}\left[(r^2+a^2)\pardiff{}{t}+\Delta\pardiff{}{r}+a\pardiff{}{\phi}\right]\\
\vec{n}&=\frac{1}{2\Sigma}\left[(r^2+a^2)\pardiff{}{t}-\Delta\pardiff{}{r}+a\pardiff{}{\phi}\right]\\
\vec{m}&=\frac{-\rho^*}{\sqrt 2}\left[ia\sin\theta\pardiff{}{t}+\pardiff{}{\theta}+i\cosec\theta\pardiff{}{\phi}\right]
\end{align}
\end{subequations}

\draft{1) say that $\vec{l}=\partial{}{u}$,etc (eqs.4,7Newman\&Jannis'65,eqs.1.13Newman\&Jannis'65,eq.3.3N\&P'68)-only for flat s-t?; 
2) describe properly properties,etc and idem for Carter tetrad,
3)say that $\vec{l}$ is principal null vect. of Maxwell tensor (~\cite{ar:Newmanetal'65}) and define what this means}

The spin coefficients with respect to this tetrad of the Kerr-Newman metric are given by (\ref{eq:def. spin coeffs.}):
\begin{equation}\label{eq:spin coeffs. in Kinnersley in Kerr}
\begin{aligned}
\rho&=-\frac{1}{r-ia\cos\theta};& \beta&=-\frac{\rho^*\cot\theta}{2\sqrt2};&
 \pi&=\frac{ia\rho^2\sin\theta}{\sqrt2};\\
\tau&=\frac{-ia\sin\theta}{\sqrt2\Sigma};& \mu&=\frac{\rho\Delta}{2\Sigma};&
 \gamma&=\mu+\frac{(r-M)}{2\Sigma};\\
\alpha&=\pi-\beta^*; &
\kappa&=\sigma=\lambda=\nu=\epsilon=0 &
\end{aligned}
\end{equation}
In terms of the Newman-Penrose formalism with the Kinnersley tetrad, the Weyl tensor and the electromagnetic field corresponding to the Kerr-Newman
solution of the Einstein-Maxwell equations are given by
\begin{subequations}  \label{eq:NP scalars in K-N}
\begin{align}
\psi_{\pm2}&=\psi_{\pm1}=0 & \psi_{0}&=M\rho^3+Q^2\rho^3\rho^*  \label{eq:NP Weyl scalars in K-N} \\
\phi_{\pm1}&=0 & \phi_0&=-\frac{1}{2}iQ\rho^2                   \label{eq:NP Maxwell scalars in K-N}
\end{align}
\end{subequations}

It is easy to see that the effect of the parity operator (\ref{eq:parity op.}) on the NP Kinnersley tetrad and spin coefficients is
\begin{equation} \label{eq:parity op. on NP objs.}
\begin{array}{lll}
\mathcal{P}X=X \qquad\quad  & \mathcal{P}Y=Y^* \qquad\quad  &  \mathcal{P}Z=-Z^*
\\
X=\vec{l},\vec{n},D,\Delta
\qquad\quad  &
Y=\epsilon,\rho,\mu,\gamma
\qquad\quad  &
Z=\vec{m},\delta,\tau,\pi,\alpha,\beta
\end{array}
\end{equation}

Under the symmetry transformation (\ref{eq:symm. (t,phi)->(-t,-phi)}), the Kinnersley tetrad transforms in the 
following manner:
\begin{equation} \label{Kinn. tetrad under t,phi->-t,-phi}
\begin{aligned}
\vec{l}&\to -\frac{\Sigma}{\Delta}\vec{n},&
\vec{n}&\to -\frac{\Delta}{\Sigma}\vec{l},&
\vec{m}&\to \frac{\rho^*}{\rho}\vec{m^*} & \text{under} \quad (t,\phi)\to (-t,-\phi)
\end{aligned}
\end{equation}
From their definition (\ref{eq:def. of phi's}), the NP Maxwell scalars will accordingly
transform under these symmetries as
\begin{subequations} 
\begin{align}
\phi_{\indhel}
\to \phi^*_{\indhel} \quad \forall \indhel
\qquad \quad
& 
\text{under} \quad (\theta,\phi)\to (\pi-\theta,\phi+\pi) \label{NP scalars under parity} \\
\left.
\begin{array}{ll}
\phi_{-1}&\to \Delta^{-1}\rho^{-2}\phi_{+1}
\\
\phi_{0}&\to -\phi_{0}
\\
\phi_{+1}&\to \Delta\rho^2\phi_{-1}
\end{array}
\right\}
 & \text{under} \quad (t,\phi)\to (-t,-\phi) \label{NP scalars under t,phi->-t,-phi}
\end{align}
\end{subequations}
subject to the boundary conditions being the same.

We have decided to mainly use the Kinnersley tetrad when working with the Newman-Penrose formalism because of the 
fact that this the null tetrad used by the overwhelming majority of the literature. The reason for this popular choice is probably
that the decoupling and separability of the linear perturbations of general spin in the Kerr background 
discovered by Teukolsky (~\cite{ar:Teuk'72}, ~\cite{ar:Teuk'73}) was originally obtained using the Kinnersley tetrad. Nevertheless, the tetrad 
(\ref{eq:def. null Kinnersley tetrad}) does not fully exploit the symmetries of the space-time. 
The Carter null tetrad ~\cite{bk:Carter-Cargese'86}, on the other hand, does exploit the symmetries of the space-time as it does not violate
the symmetry (\ref{eq:symm. (t,phi)->(-t,-phi)}).             
The \define{Carter null tetrad} in the co-ordinate system $\{r,q,\tilde{\phi},\tilde{t}\}$ where $q\equiv a\cos\theta$,
$\tilde{\phi}\equiv a^{-1}\phi$ and $\tilde{t}\equiv t-a\phi$ is
\begin{subequations} \label{eq:def. Carter null tetrad}
\begin{align}
\vec{{}_cl}&=\frac{1}{\sqrt{2\Delta\Sigma}}\left[-\Delta\pardiff{}{r}+\pardiff{}{\tilde{\phi}}+r^2\pardiff{}{\tilde{t}}\right]\\
\vec{{}_cn}&=\frac{1}{\sqrt{2\Delta\Sigma}}\left[+\Delta\pardiff{}{r}+\pardiff{}{\tilde{\phi}}+r^2\pardiff{}{\tilde{t}}\right]\\
\vec{{}_cm}&=\frac{1}{\sqrt{2(a^2-q^2)\Sigma}}\left[-(a^2-q^2)\pardiff{}{q}-i\pardiff{}{\tilde{\phi}}+iq^2\pardiff{}{\tilde{t}}\right]
\end{align}
\end{subequations}
We keep the same values for $\Sigma$ and $\rho$ as the ones we have been using so far, which in the new co-ordinates
can be written as $\Sigma=r^2+q^2$ and $\displaystyle \rho=\frac{-1}{r-iq}$.
The spin coefficients in the Carter null tetrad are
\begin{equation}\label{eq:spin coeffs. in Carter null in Kerr}
\begin{aligned}
{}_c\mu&=-\sqrt{\frac{\Delta}{2\Sigma}}\rho;& \qquad\qquad {}_c\rho&={}_c\mu; \\
{}_c\epsilon&=-\frac{\Delta\rho+(r-M)}{2\sqrt{2\Delta\Sigma}};& {}_c\gamma&={}_c\epsilon; \\
{}_c\tau&=\sqrt{\frac{(a^2-q^2)}{2\Sigma}}i\rho; & {}_c\pi&=-{}_c\tau; \\
{}_c\alpha&=-\frac{i(a^2-q^2)\rho+q}{2\sqrt{2(a^2-q^2)\Sigma}}; & {}_c\beta&=-{}_c\alpha; \\
{}_c\kappa&={}_c\sigma={}_c\lambda={}_c\nu=0 &
\end{aligned}
\end{equation}




\section{Black hole thermodynamics} \label{sec:b-h thermodynamics}

Hawking ~\cite{ar:Hawking'75} showed in 1975 that particle creation could occur at late times as a result 
of the collapse of a star to a Schwarzschild black hole.  
He also found that the nature of this radiation is thermal, black body radiation at the \define{Hawking temperature} 
\begin{equation} \label{eq:Hawking temp.}
T_H=\frac{\kappa_+}{2\pi}
\end{equation}
Consider the collapse of a spherical, uncharged star and the vacuum state that corresponds to quantizing the field
with respect to the standard incoming exponential modes at $\mathcal{I^-}$. 
If the field is in this state, then an inertial partial detector at $\mathcal{I^-}$ will detect no particles.
As the radius of the collapsing star becomes close to the radius of its event horizon, 
the waves coming in from $\mathcal{I^-}$ suffer a blue-shift as they approach the surface of the star which is much smaller than
the red-shift they suffer as they emerge through the star.       
\catdraft{The null rays $u=const$, for large values of $u$, when traced backwards in time and reflected out to $\mathcal{I^-}$,
pile up densely along the latest incoming wave that managed to escape through the star before the star reached
its Schwarzschild radius.}
\catdraft{i.e., totes les outgoing waves que arriben a l'infinit per late times son waves
que de fet eren incoming waves i que han atravessat l'estrella ABANS que aquesta esdevingues un b-h??! i.e., la Hawking
rad. no correspon a parts. que va creant el b-h a tothora del no-res (p.264B\&D) sino a parts. 'creades'?/vingudes de $\mathcal{I^-}$?
d'abans que es formes el b-h (totes elles)????????!!!!!!!!!!} 
The Bogolubov transformation between these outgoing waves as they reach $\mathcal{I^+}$ and the standard 
exponential modes at $\mathcal{I^+}$ can be calculated.
The result is that an inertial partial detector at $\mathcal{I^+}$ will detect a flux of outgoing particles corresponding
to a thermal spectrum of a black body at the Hawking temperature. This flux of particles is the \define{Hawking radiation}.
The mass of the black hole will decrease in time due to the emission of this flux, that is, the black hole evaporates.   
\catdraft{pero si aquestes parts. han estat emeses ABANS que es convertis en b-h llavors la massa del b-h en si no hauria de variar???}  
If the black hole is rotating, more particles will be emitted by its gravitational field with angular momentum 
of the same sign as that of the black hole's.
Similarly, if the black hole is charged, more particles will be emitted by its electromagnetic field with charge 
of the same sign as that of the black hole's. 
Both the angular momentum and charge of the black hole will also decrease in time. 

Heuristically, the Hawking radiation can be explained in the following manner. Virtual particle-antiparticle pairs created 
with wavelength $\lambda$ separate temporarily to a distance $\sim\lambda$ before they are reannihilated. For those
pairs of wavelength of the order of the size of the black hole, $\lambda\sim M$, the tidal forces
between the particle and the antiparticle become so large that they separate before reannihilation. 
One particle escapes to infinity, thus contributing to the Hawking radiation, whilst the other particle
enters the black hole following a timelike path of negative energy relative to infinity.
\catdraft{1)com s'explica en termes de annihilation/creation ops.? quines son les Bogolubov transf. entre boulware i Unruh?,
2) si es te en compte back-reaction llavors el b-h ja no es troba en eq. amb rad. (p.271B\&D)=>H-H vac. state
no te cap mena de sentit fis. ja que no es estable i=> no existira mai? que te de bo el H-H vac.? (p.280B\&D?)}
In a strict sense, however, it is not possible to talk about particles in curved space-time, except at an asymptotically flat region.
It is instead the stress energy tensor the right tool to use in order to describe the natural processes. 

The above results suggest a relationship between black hole processes and thermodynamics. This relationship 
is at a much more essential level than the mere existence of thermal radiation emitted by black holes. Indeed, a correlation
has been found between certain properties of black holes and the four laws of thermodynamics. This correlation is the following:

\underline{0th law}\\
The zeroth law of thermodynamics states that the temperature is constant throughout a system in thermal equilibrium.\\
The black hole analogue of this law is the theorem ~\cite{bk:Carter-DeWittDeWitt} that the surface gravity parameter $\kappa_+$, 
and therefore the Hawking temperature, is constant over the event horizon of a stationary black hole.

\underline{1st law}\\
The first law of thermodynamics states that for a process that only involves infinitesimal variations 
of the thermodynamic co-ordinates, it must be:  
\begin{equation}
\delta E=T\delta S+P\delta V
\end{equation}
Bardeen, Carter and Hawking ~\cite{ar:B&Carter&Hawk'73} have found that for variations in the 
metric of a stationary black hole that do not alter its stationarity, it must be:
\begin{equation}
\delta M=\frac{1}{8\pi}\kappa_+\delta \mathcal{A}+\Omega_+\delta J
\end{equation}
From equation (\ref{eq:Hawking temp.}) it follows that
the two laws are analogous if a correspondence is established between $E$ and $M$,
between the entropy $S$ and $\displaystyle \frac{\mathcal{A}}{4}$, and finally, between the ``work term'' $P\delta V$ and $\Omega_+\delta J$.

\underline{2nd law}\\
\define{Hawking's area theorem} ~\cite{ar:Hawking'71b} 
requires the total event horizon area $\mathcal{A}_T$ of black hole space-times to be non-decreasing, i.e.,  
$\d{\mathcal{A}_T}\ge0$, in all black hole processes for which the weak energy condition is satisfied. 
Since the entropy of the black hole is $S=\frac{\mathcal{A}}{4}$ from the first law, 
Hawking's area theorem is just a special case of the second law of thermodynamics: $\d{S_T}\ge0$.
Black hole evaporation does not violate the 2nd law of thermodynamics since, even though the mass, and therefore also the event
horizon area, decrease in time, account must be taken of the increase in entropy outside the black hole due to the
Hawking radiation.    

\underline{3rd law}\\
The third law of thermodynamics states that it is not possible to reach the absolute zero of temperature
through a finite series of processes.    
An analogue of this formulation of the third law exists  in black hole physics. 
Israel ~\cite{ar:Israel'86} formulated it as: 
a non-extremal black hole ($T_H>0$) cannot become extremal ($T_H=0$) at a finite advanced time $v$ in any continuous
process in which the stress-energy tensor of accreted matter stays bounded and satisfies the weak energy condition
in a neighbourhood of the outer horizon.  
The \define{cosmic censorship hypothesis} states that naked singularities cannot form from gravitational collapse.
Injection of matter whose energy density is or becomes negative in a neighbourhood of the outer horizon of a non-extreme
black hole can not only result in the formation of an extreme black hole but also violate cosmic censorship. 


\section{Physical phenomena associated with the ergosphere} \label{sec:superrad.}

The existence of the ergosphere has important physical consequences, both on a classical and on a quantum level.

Penrose ~\cite{ar:Penrose'69} showed in 1969 that it is possible to extract energy from black holes that possess an ergosphere, 
in what is referred to as the \define{Penrose process}. 
The reason being that the change in the mass of a black hole when a test-particle falls into it is the energy
of the particle as measured by a SO. This energy is equal to $-\vec{p}\vec{\xi}$ for an uncharged particle. 
We know that the Killing vector $\vec{\xi}$ is timelike outside the ergosphere but spacelike inside it, 
and therefore for certain timelike orbits of the uncharged test-particle, its energy may be negative as viewed from radial infinity by a SO.
These orbits of negative energy are, however, confined entirely within the ergosphere. 
Therefore, in order to send in a test-particle from outside the ergosphere that carries negative energy into the black hole, its orbit must be changed.
Extraction of energy from the black hole can be achieved by dividing a geodesic particle that is coming in from radial infinity into two other particles, 
once inside the ergosphere. One of the particles falls into the black hole following a negative energy orbit as seen by a SO, 
and the other one is retrieved at radial infinity possessing an energy larger than that of the initially incoming particle.

If both the test-particle and the black hole are charged, then the region within which the energy of the particle may be negative as viewed
from infinity by a SO is not exactly the ergosphere, but a region called the \define{effective ergosphere} ~\cite{bk:M&T&W}. 
Indeed, if the charges of the particle and of the black hole have opposite sign then the energy of the particle as viewed from infinity by a SO
will be more negative and therefore the effective ergosphere will be larger than the ergosphere. 
Analogously, if the charges have the same sign then the effective ergosphere will be smaller than the ergosphere. 

The second law of black hole dynamics shows that there is a limit to the decrease in the mass of the black hole which may
be achieved via the Penrose process. This limit was found by Christodoulou ~\cite{ar:Christo'70} and 
Christodoulou and Ruffini ~\cite{ar:Christo&Ruff'71}, independently
of Hawking's result ~\cite{ar:Hawking'71b}, and is the \define{irreducible mass} of the black hole: 
\begin{equation}
M_{\text{irred}}=\frac{1}{2}\sqrt{r_+^2+a^2}
\end{equation}

There is a corresponding effect in classical waves to the Penrose process, which was discovered by Zel'dovich ~\cite{ar:Zel'71} and
Starobinski\u{\i} ~\cite{ar:Starob'73a} for spin-0 and shortly after by Starobinski\u{\i} and Churilov ~\cite{ar:Starob'73b} for spin-1 and spin-2.
We shall see in subsequent chapters that certain modes that are part of the Fourier mode decomposition of field perturbations
of black holes, may be reflected back by a black hole with an ergosphere with an amplitude larger than that of the incident
wave mode. 
These modes are called \define{superradiant modes} and the amplification effect of these modes by the
rotating black hole is called \define{superradiance}.
As for the Penrose process, the part of the incident superradiant wave mode that has been transmitted through 
to the future horizon carries negative energy into the black hole.
\catdraft{hauria de ser negat. E mesurat per local obs. ja que te $\tilde{w}<0$ (i $w>0$) pero de fet te negat. E mesurada per obs.
llunya i pos. E mesurada per obs. local. l'Adrian no sap veure el pq., diu de mirar Wald's book}
Similarly, certain field wave modes that emerge from the past horizon may be reflected by the black hole back to the future horizon with an
increased amplitude. These modes are also superradiant modes.
The phenomenon of superradiance occurs for the spin-0, spin-1 and spin-2 fields but it does not for the neutrino case.

Finally, there is a quantum counterpart to classical supperradiance, which was discovered by Unruh ~\cite{ar:Unruh'74} in 1974.
He showed that a black hole possessing an ergosphere emits out to $\mathcal{I}^+$ positive fluxes of energy and angular momentum 
when the field is in a quantum state which is empty at both $\mathcal{H}^-$ and $\mathcal{I}^-$. 
The only contribution to these fluxes is from superradiant field modes.
As a consequence of the emission of these positive fluxes, the mass and the angular momentum of the rotating black hole decrease.
This process is called the \define{Starobinski\u{\i}-Unruh effect}. In Chapter \ref{stress-energy tensor} we will describe this effect further 
and calculate its radial flux for the spin-1 case.

\draft{p.264B\&D:$\tilde{\omega}$ seems to be defined differently for charged black holes so that superradiant cond. includes the charge??}


\chapter{Field equations}  \label{ch:field eqs.}


\section{Introduction} \label{sec:Intro. in field eqs.}

\draft{throughout, when I say `arbitrary Type D background' might not be correct and should be restricted
to vacuum Type D backgrounds which is what the Goldberg-Sachs theorem applies to, 
also including K-N (and others?) because of its generalization, but not `arbitrary'?}

\draft{1) diff. sign for $\eta_{\mu\nu}$ in McL\&Ott than in Chrzan's; 
2) there's a floating $(-1)^m$ and possibly $2\pi$? 3) $\NPadj_j$ as a sourceless vs. non-sourceless eqn. sln.
4) move some eqs. to an Appendix?}
\catdraft{1) ja he vist clar que cap de les 2 possibles eqs. a solucionar son separables (i.e., una per $P^{\dagger}_0$
i una per $P_0$; les dues opcions de cadascun son identiques) tant per B-L tetrad com per Carter tetrad =>4 possibles eqs.??! apart de veure 
si en eq.$P_0 \phi_0=0$ puc trobar una func. f t.q. $P_0 \phi_0=Op.nou(f*\phi_0)=0$ ; 2) que son les eqs.6.8,6.9Chrzan.?}

We consider the field equations satisfied by the electromagnetic field in a general Type D background.
The background is considered fixed and the electromagnetic field appearing in the field equations 
is the total electromagnetic field, regardless of its origin, whether background or perturbation. 
In the case of an uncharged background, however, the electromagnetic field appearing in the field equations
may alternatively be interpreted as the one corresponding to the first order of a linear perturbation only.

By making use of the NP formalism, Teukolsky (~\cite{ar:Teuk'72}, ~\cite{ar:Teuk'73}) 
showed that the equations describing linear electromagnetic, neutrino and gravitational 
perturbations of a general Type D background can be decoupled.
He further showed that some of the decoupled equations can be solved by separation of variables 
in the Boyer-Lindquist co-ordinate system in the Kerr background.
Carter ~\cite{ar:Carter'68b} had previously shown the separability in the scalar case.
Cohen and Kegeles ~\cite{ar:Coh&Keg'74} showed in 1974 that in a Type D background 
all the electromagnetic field components may be derived by double differentiation from one
single, complex Maxwell scalar, thereby acting as a Debye potential. 
This scalar then carries the two dynamical degrees of freedom of the perturbed field.
Chrzanowski ~\cite{ar:Chrzan'75} was the first author to give, using Teukolsky's results, 
analytic expressions for the linear electromagnetic and gravitational perturbation potentials in the Kerr background 
in the homogeneous case. 
He showed that the electromagnetic potential may be derived from the Debye potential of Cohen and Kegeles's.
He obtained the expressions for the potential from
a conjecture about the form of the Green function for the uncoupled field equations, which he proved to be
correct in the Kerr background for the particular case of spin-1 perturbations for high frequency and also for spin-0 perturbations.
However, it was Wald ~\cite{ar:Wald'78} in 1978 who developed a very elegant and general formalism
which was underlying Teukolsky's and Chrzanowski's results. 
Wald's formalism, which is valid in an arbitrary background, proved in a very simple manner that
Chrzanowski's results (and also his conjecture about the Green function) were correct.
This formalism also gives a much better understanding of the potential and field solutions and the relationships between the
different quantities. 

In this chapter we are going to develop Wald's formalism for the electromagnetic case
and therefore derive Chrzanowski's expressions for the electromagnetic potential in the Kerr background.
The clarity of this formalism will allow us to explain some features
of the field equations, the origin of which remained so far unclear. This formalism will also enable us
to produce the various expressions for the potential and the NP Maxwell scalars in a very compact and simple way. 
This derivation makes clear the origin of these expressions as well as the relationships between them.
We will establish, when appropriate, the parallelism between our expressions and the analogous equations 
in the literature as we unravel Wald's formalism for the electromagnetic case.
Our initial results are valid in an arbitrary background, are then specialized to a Type D background 
where $\kappa=\sigma=\nu=\lambda=0$
and finally to the Kerr-Newman background. In particular, we generalize Teukolsky's results to the Kerr-Newman background.

Finally, we derive formal and simple expressions for the potential in the ingoing and upgoing gauges, both in terms
of one single, complex Maxwell scalar. This means that quantum field theory could be constructed from these expressions
as if it were a complex scalar field theory. This is indeed the case in the Reissner-Nordstr\"{o}m background.
Unfortunately, though, we shall show that this is not possible in the Kerr background since the uncoupled
equation for this Maxwell scalar cannot be solved by separation of variables.


\section{Wald's formalism} \label{sec:Wald in arbitrary type D s-t}



In this section we will describe Wald's ~\cite{ar:Wald'78} formalism for finding an analytic expression for the solution
of a coupled equation in terms of a solution of a related decoupled equation.
We will later use the latter solution to find the electromagnetic potential and field components. 
Everything in this section is valid in any smooth manifold with a smooth metric. 

Let $\mathcal{E}$ be a linear, partial differential operator and $f$ a tensor field of the type on which $\mathcal{E}$ acts. 
The general field equation we want to solve is given by
\begin{equation}\label{eq:Wald, field eq.}
\mathcal{E}(f)=0
\end{equation}
Suppose that we have been able to derive a decoupled equation for a new variable $\phi\equiv \mathcal{T}(f)$ 
by applying the operator $\mathcal{S}$ on (\ref{eq:Wald, field eq.}):
\begin{equation}\label{eq:Wald, decoupled eq.}
\mathcal{S}\mathcal{E}(f)=\mathcal{O}\mathcal{T}(f)=\mathcal{O}(\phi)=0
\end{equation}
where the operator $\mathcal{O}$ is defined by the first equality in (\ref{eq:Wald, decoupled eq.}), 
and $\mathcal{O}(\phi)=0$ is a decoupled equation.

Throughout this chapter we shall use the following definition of the \define{adjoint} of a differential operator: 
if $Q_{\mu_1 \dots \mu_m}{}^{\nu_1 \dots \nu_n}$
is a linear, partial differential operator mapping components $f_{\nu_1 \dots \nu_n}$ of tensor fields of type
$(0,n)$ to components $Q_{\mu_1 \dots \mu_m}{}^{\nu_1 \dots \nu_n}f_{\nu_1 \dots \nu_n}$ of tensor fields of type
$(0,m)$, then the adjoint of $Q_{\mu_1 \dots \mu_m}{}^{\nu_1 \dots \nu_n}$, denoted by $Q^{\dagger^{\nu_1 \dots \nu_n}}{}_{\mu_1 \dots \mu_m}$,
is defined to be the (unique) linear, partial differential operator mapping components $g^{\mu_1 \dots \mu_m}$ of
tensor fields of type $(m,0)$ to components $Q^{\dagger^{\nu_1 \dots \nu_n}}{}_{\mu_1 \dots \mu_m}g^{\mu_1 \dots \mu_m}$
of tensor fields of type $(n,0)$ such that $\forall f_{\nu_1 \dots \nu_n}$,$g^{\mu_1 \dots \mu_m}$,
\begin{equation} \label{eq:def. adj. op., no cc}
g^{\mu_1 \dots \mu_m}\left(Q_{\mu_1 \dots \mu_m}{}^{\nu_1 \dots \nu_n}f_{\nu_1 \dots \nu_n}\right)-
\left(Q^{\dagger^{\nu_1 \dots \nu_n}}{}_{\mu_1 \dots \mu_m}g^{\mu_1 \dots \mu_m}\right)f_{\nu_1 \dots \nu_n}=t^{\alpha}{}_{; \alpha}
\end{equation}

Wald showed in the following manner that by direct differentiation of a solution of the adjoint of the decoupled equation 
a solution of the initial field equation is obtained, if its operator is self-adjoint.
Let $\NPadj$ be the solution of the adjoint of the decoupled equation, i.e.,
\begin{equation}\label{eq:Wald, adj. of decoupled eq.}
\mathcal{O}^{\dagger}(\NPadj)=0
\end{equation}
Then, if we apply the adjoint of (\ref{eq:Wald, decoupled eq.}) to $\NPadj$ we obtain
\begin{equation}\label{eq:Wald, adj. of field eq.}
\mathcal{E}^{\dagger}\mathcal{S}^{\dagger}(\NPadj)=\mathcal{T}^{\dagger}\mathcal{O}^{\dagger}(\NPadj)=0
\end{equation}
Therefore, if $\mathcal{E}=\mathcal{E}^{\dagger}$, then $f=\mathcal{S}^{\dagger}(\NPadj)$ is a solution of 
the field equation (\ref{eq:Wald, field eq.}). The variable $\phi$ satisfying 
(\ref{eq:Wald, decoupled eq.}) is then given by $\phi=\mathcal{T}\mathcal{S}^{\dagger}(\NPadj)$.


\section{Maxwell equations and gauge invariance} \label{sec:gauge inv.}

The Maxwell equations can be written in terms of the potential $\tensor{A}$ and the 
source terms $\tensor{J}$ as
\begin{equation} \label{eq:Maxwell eqs. with potential}
D_{\alpha}{}^{\beta}A_{\beta}=J_{\alpha}
\end{equation}
where
\begin{equation} \label{eq:def. op. D}
D_{\alpha}{}^{\beta}\equiv \delta_{\alpha}{}^{\beta}\nabla^{\gamma}\nabla_{\gamma}-
\delta_{\gamma}{}^{\beta}\nabla^{\gamma}\nabla_{\alpha}
\end{equation}
\catdraft{pq. (\ref{eq:def. op. D}) coinc. amb eq.5.6Mandl\&Shaw;eq.1.41Itzyk\&Zub caldria que 
$\delta_{\gamma}{}^{\beta}\nabla^{\gamma}\nabla_{\alpha}=\nabla^{\alpha}\nabla_{\gamma}\delta_{\gamma}{}^{\beta}$-es cert?}
We are using rationalized units in the sense that Maxwell equations are written as in (\ref{eq:Maxwell eqs. with potential}),
whereas Teukolsky ~\cite{ar:Teuk'73} uses unrationalized units which include a factor $4\pi$ in these equations.

Alternatively, in terms of the field components
\begin{equation}\label{eq:def. of F}
F_{\alpha\beta}=-F_{\beta\alpha}\equiv A_{\beta ;\alpha}-A_{\alpha;\beta}
\end{equation}
the Maxwell equations become
\begin{subequations}\label{eq:Maxwell eqs. with F}
\begin{align}
g^{\alpha\gamma}F_{\alpha\beta;\gamma}&=J_{\beta} \label{eq:Maxwell eq. with F and J} \\
F_{[\alpha\beta;\gamma]}&=0
\end{align}
\end{subequations}
The \define{law of current conservation} 
\begin{equation} \label{eq:law of current conserv.}
\nabla_{\alpha}J^{\alpha}=0
\end{equation}
follows directly from the field equation (\ref{eq:Maxwell eq. with F and J}) 
and the antisymmetry of the electromagnetic tensor $F_{\alpha\beta}$. 

In terms of the NP Maxwell scalars (\ref{eq:def. of phi's}) the Maxwell equations (\ref{eq:Maxwell eqs. with F})
take the form:
\begin{subequations} \label{eq:Maxwell eqs. with phi's}
\begin{align}
&\phi_{0|(1)}-\phi_{-1|(4)}=\frac{1}{2}J_{(1)}\label{eq:Maxwell eq. with phi0(1),phi_1(4)} \\
&\phi_{+1|(3)}-\phi_{0|(2)}=\frac{1}{2}J_{(2)}\label{eq:Maxwell eq. with phi0(2),phi1(3)}\\
&\phi_{0|(3)}-\phi_{-1|(2)}=\frac{1}{2}J_{(3)}\label{eq:Maxwell eq. with phi0(3),phi_1(2)}\\
&\phi_{+1|(1)}-\phi_{0|(4)}=\frac{1}{2}J_{(4)}\label{eq:Maxwell eq. with phi0(4),phi1(1)}
\end{align}
\end{subequations}

It is well known that there is a certain freedom in the choice of the potentials that satisfy the 
Maxwell equations (\ref{eq:Maxwell eqs. with potential}) and yield the same electromagnetic field via equations (\ref{eq:def. of F}). 
Indeed, if a certain potential satisfies equations (\ref{eq:Maxwell eqs. with potential}) then it is always possible to apply 
the \define{gauge transformation}
\begin{equation} \label{eq:gauge transf.}
A_{\alpha} \rightarrow A'_{\alpha}=A_{\alpha}+\pardiff{\Phi}{x_{\alpha}}
\end{equation}
and the transformed potential $A'_{\alpha}$ will also satisfy the Maxwell equations and it will yield the same electromagnetic
field as the potential $A_{\alpha}$. The invariance of the field under these transformations is called \define{gauge invariance}.


\catdraft{No veig com (\ref{eq:homog.Maxwell eqs. with potential in Lorenz gauge}) 
pot ser valid per K-N donada eq.22.19dMTW i ja que el Ricci tensor de K-N no es zero ja que 
no es vacuum s-t. Ara be, nomes l'aplico a K-N per radial inf., on es flat s-t i el Ricci tensor es zero (?)}
The homogeneous Maxwell equations (\ref{eq:Maxwell eqs. with potential}) in the Lorentz gauge $\nabla_{\alpha}A^{\alpha}=0$ become,
in a vacuum space-time,
\begin{equation} \label{eq:homog.Maxwell eqs. with potential in Lorenz gauge}
\nabla^{\gamma}\nabla_{\gamma} A_{\alpha}=0
\end{equation}
A plane-wave solution is of the form
\begin{equation} \label{eq:plane wave potential}
A_{\alpha}=e_{\alpha}e^{ik_{\beta}x^{\beta}}+e_{\alpha}^*e^{-ik_{\beta}x^{\beta}}
\end{equation}
where the following conditions
\begin{subequations}
\begin{align}
k_{\alpha}k^{\alpha}&=0 \\
k_{\alpha}e^{\alpha}&=0 \label{eq:polarization perp. to dir.of propagation}
\end{align}
\end{subequations}
must be satisfied as a consequence of (\ref{eq:homog.Maxwell eqs. with potential in Lorenz gauge}) 
and the Lorentz condition respectively.
The tensor $\vec{e}$ is called the \define{polarization} tensor 
and $\vec{k}$ is the direction of propagation. 
The Lorentz gauge describes a transverse wave via (\ref{eq:polarization perp. to dir.of propagation}).
Because of the condition (\ref{eq:polarization perp. to dir.of propagation}) out of the four components
of $\vec{e}$ only three are independent.
Lorentz gauge still leaves a certain freedom in the choice of the electromagnetic potential: we can perform 
the gauge transformation (\ref{eq:gauge transf.}) with
\begin{equation}
\Phi(x)=i\epsilon e^{ik_{\beta}x^{\beta}}-i\epsilon^* e^{-ik_{\beta}x^{\beta}}
\end{equation}
The transformed potential is then given by
\begin{equation}
A'_{\alpha}=e'_{\alpha}e^{ik_{\beta}x^{\beta}}+e_{\alpha}^{' *}e^{-ik_{\beta}x^{\beta}}
\end{equation}
with
\begin{equation} \label{eq:gauge transf. on polar.tensor}
e'_{\alpha}=e_{\alpha}-\epsilon k_{\alpha} 
\end{equation}
and $\epsilon$ an arbitrary parameter. The transformed potential $A'_{\alpha}$ will also satisfy the homogeneous Maxwell
equations as well as the Lorenz gauge condition. This means that of the three independent components of $\vec{e}$
only two of them are physically significant. 

In an asymptotically flat space-time, like Kerr-Newman, the above discussion about the plane wave (\ref{eq:plane wave potential})
is valid at radial infinity for a wave travelling in the direction of $\vec{l}$, i.e., 
such that $\vec{k}=\vec{l}$. 
\catdraft{1) why? at infinity it should behave as a radial wave, not a plane wave, which differe by for ex. factor 1/r, 
which interferes in derivation and conclusions from Lorentz gauge? 
2) our waves later are in gauges such that at infinity they coincide with the Lorentz gauge?}
In particular, the condition (\ref{eq:polarization perp. to dir.of propagation})
implies that $e^{n}=0=e_{l}$ at radial infinity. 
The gauge transformation (\ref{eq:gauge transf. on polar.tensor}) then means that, asymptotically, $e_m$ is left invariant as well
as $e'_{l}=e_{l}=0$ whereas $e'_{n}=e_{n}-\epsilon$. We can therefore make $e'_{n}$ equal to zero by chosing
$\epsilon=e_{n}$ so that it is only $e_m$ and $e_{m^*}$ that carry physical significance at radial infinity.

Any plane wave $\psi$, which is transformed by a rotation of an angle $\vartheta$ about the direction
of propagation into
\begin{equation} \label{eq:def. helicity}
\psi'=e^{ih\vartheta}\psi
\end{equation}
is said to have \define{helicity} $h$. 
When subjecting the null vector $\vec{l}$ to a rotation of an angle $\vartheta$, 
then $\vec{l}$ and $\vec{n}$ are left unchanged whereas $m^{\mu}\rightarrow e^{i\vartheta}m^{\mu}$.
\ddraft{why?}
Thus, an electromagnetic wave propagating in the direction of $\vec{l}$ in an asymptotically flat background
can be decomposed into parts which, at radial infinity, have helicity $-1$, $0$ and $+1$. 
However, the physically significant helicities at radial infinity are $\pm1$, not $0$.

The conclusions we have reached for an electromagnetic wave travelling in the direction of $\vec{l}$ in
an asymptotically flat background are equally valid for an electromagnetic wave travelling in the 
direction of $\vec{n}$ in the same background.



\section{Wald's formalism for spin-1} \label{sec:Wald for spin-1}

In this section we will unfold for the spin-1 case Wald's formalism described in Section \ref{sec:Wald in arbitrary type D s-t}. 
The following are the correspondences between the general operators and other objects used in that section 
and the ones we are going to use for the spin-1 case:

\begin{equation} \label{eq: correspondence Wald-Ott formalism}
\begin{aligned}
f &\leftrightarrow \tensor{A};& \phi &\leftrightarrow \phi;& \NPadj &\leftrightarrow \NPadj && \\
\mathcal{E} &\leftrightarrow D=2K^{\dagger}K;\quad& \mathcal{T} &\leftrightarrow K;\quad& \mathcal{S}&\leftrightarrow 2\Pi;\quad& \mathcal{O} &\leftrightarrow P
\end{aligned}
\end{equation}
Clearly, since $D=2K^{\dagger}K$, the spin-1 field equations are self-adjoint.  

The gravitational quantities (i.e., spin coefficients and null tetrad) appearing in the Maxwell equations 
in NP form are the ones of the background that we are considering.
However, as mentioned in the introduction, the interpretation of the electromagnetic quantities in the Maxwell
equations can differ if the background considered is uncharged. 
If the background is charged, then it is considered fixed and the NP Maxwell scalars include the electromagnetic field
associated to the background (e.g., (\ref{eq:NP Maxwell scalars in K-N}) for Kerr-Newman) as well as an electromagnetic perturbation. 
On the other hand, if the background is uncharged, like Kerr, we may alternatively consider the Maxwell scalars in the 
Maxwell equations to correspond only to the first order of a linear electromagnetic perturbation.
Since the electromagnetic stress-energy tensor is second order in the electromagnetic field, the change in an uncharged background
caused by the first order perturbation will be of second order. 
The change in the gravitational quantities can then be neglected to first order in the Maxwell equations  
and we still consider the gravitational quantities in the equations to be background quantities.
Unfortunately, this second interpretation is not possible in the case of the Kerr-Newman background 
(although it is in Reissner-Nordstr\"{o}m's): all efforts in the literature to decouple the equations 
for the coupled electromagnetic-gravitational perturbations in this background have been unsuccessful so far. 
See Chandrasekhar ~\cite{bk:Chandr} for a description of this treatment.

We thus choose to follow, even in uncharged backgrounds, the first interpretation out of the two described above.
This reflects on the fact that we will always use the same symbol, $\phi_{\indhel}$, 
to refer to the Maxwell scalars both in charged and uncharged backgrounds.
That is, the symbol $\phi_{\indhel}$ will always represent an electromagnetic perturbation together with the electromagnetic field, 
if there is one, associated to the background. 
The symbols for the gravitational quantities always represent background quantities.
\catdraft{aixo que dic no es gens clar: 1) grav.pertorb. no te pq. ser interpretat com a alterac. del background
sino com una grav.wave externa?, 2) en vista de p.1117Teyk'72,p.2457Fack\&Ips'72,Ipser'71??!->p.ex., sln. que
es $\phi_0\propto \rho^2$, i que correspon a K-N, sembla continuar essent una pertorbac.?}
Ipser ~\cite{ar:Ipser'71} has proven that the only time-independent electromagnetic perturbation of the Kerr metric that is physically acceptable, 
is axisymmetric and is given by (\ref{eq:NP Maxwell scalars in K-N}), which corresponds to the addition of charge to the black hole,
that is, to the passage from the Kerr metric to the Kerr-Newman metric.   

We now define the operator $K_{\indhel}{}^{(a)}$ which maps the potential onto the NP scalars, i.e., 
\begin{equation}\label{eq:phi as func. of potential with K}
\phi_{\indhel}=K_{\indhel}{}^{(a)}A_{(a)}
\end{equation}
We can easily calculate $K_{\indhel}{}^{(a)}$ from (\ref{eq:def. of phi's}) by expanding out the intrinsic 
derivatives in (\ref{eq:def. of F}) after writing this equation in tetrad form. In this manner we obtain:
\begin{subequations} \label{eq:op. K}
\begin{align}
K_{-1}{}^{(a)}&=-\delta_{(1)}^{(a)}\delta+\delta_{(2)}^{(a)}\kappa+\delta_{(3)}^{(a)}(D+\epsilon^*-\epsilon-\rho^*)-\delta_{(4)}^{(a)}\sigma \\
K_{0}{}^{(a)}&=\frac{1}{2}
\begin{aligned}[t] \Big[ &
-\delta_{(1)}^{(a)}(\bDelta-\gamma-\gamma^*+\mu^*-\mu)+\delta_{(2)}^{(a)}(D+\epsilon+\epsilon^*+\rho-\rho^*)+ 
\\ &
+\delta_{(3)}^{(a)}(\delta^*-\pi-\tau^*+\beta^*-\alpha)-\delta_{(4)}^{(a)}(\delta+\pi^*+\tau+\beta-\alpha)
\Big] \end{aligned}
\\
K_{+1}{}^{(a)}&=\delta_{(1)}^{(a)}\nu+\delta_{(2)}^{(a)}(\delta^*+\alpha+\beta^*-\tau^*)-\delta_{(3)}^{(a)}\lambda-\delta_{(4)}^{(a)}(\bDelta+\mu^*+\gamma-\gamma^*)
\end{align}
\end{subequations}

Similarly, expanding out the intrinsic derivatives in (\ref{eq:Maxwell eqs. with phi's}) we can express 
the Maxwell equations in operator form as
\begin{equation}\label{eq:Maxwell eqs. with K^dagger}
2K^{\dagger \indhel}{}_{(a)}\phi_{\indhel}=J_{(a)}
\end{equation}
where we raise and lower the Maxwell scalar index by applying
\begin{equation} \label{eq:raise/lower index i}
\begin{array} {cc}
(\epsilon^{ij})=
\left(
\begin{array}{ccc}
0 & 0  & -1  \\
0 & 2  &  0 \\
-1 & 0 & 0
\end{array}
\right)
& \quad
(\epsilon_{ij})=
\left(
\begin{array}{ccc}
0 & 0  & -1  \\
0 & 1/2  &  0 \\
-1 & 0 & 0
\end{array}
\right)
\end{array}
\end{equation}
respectively.

The adjoint of $K_{\indhel}{}^{(a)}$ can be easily calculated from the definition (\ref{eq:def. adj. op., no cc}) of adjoint to give
\begin{subequations} \label{eq:op. K^dagger}
\begin{align}
K^{\dagger}_{-1}{}^{(a)}&=\delta_{(1)}^{(a)}(\delta+2\beta-\tau)+\delta_{(2)}^{(a)}\kappa-\delta_{(3)}^{(a)}(D+2\epsilon-\rho)-\delta_{(4)}^{(a)}\sigma \\
K^{\dagger}_{0}{}^{(a)}&=\frac{1}{2}
\begin{aligned}[t] \Big[&+
\delta_{(1)}^{(a)}(\bDelta+2\mu)-\delta_{(2)}^{(a)}(D-2\rho)- 
\\ &
-\delta_{(3)}^{(a)}(\delta^*+2\pi)+\delta_{(4)}^{(a)}(\delta-2\tau)
\Big] \end{aligned}
\\
K^{\dagger}_{+1}{}^{(a)}&=\delta_{(1)}^{(a)}\nu-\delta_{(2)}^{(a)}(\delta^*-2\alpha+\pi)-\delta_{(3)}^{(a)}\lambda+\delta_{(4)}^{(a)}(\bDelta+\mu-2\gamma)
\end{align}
\end{subequations}

Combining equations (\ref{eq:Maxwell eqs. with K^dagger}) and (\ref{eq:phi as func. of potential with K}) together and comparing 
with (\ref{eq:Maxwell eqs. with potential}) it follows that
\begin{equation}\label{eq:op.D in terms of op.K}
D_{(a)}{}^{(b)}=2K^{\dagger \indhel}{}_{(a)}K_{\indhel}{}^{(b)}
\end{equation}
as indicated at the beginning of the section.

All equations given so far in this section are valid for an arbitrary background.
We now specialize to a Type D background such that $\kappa=\sigma=\nu=\lambda=0$
and a tetrad for which $\vec{l}$ and $\vec{n}$ correspond to the two principal null-directions. 
From the corollary of the Goldberg-Sachs theorem, we have that $\kappa=\sigma=\nu=\lambda=0$
for an empty Type D background. 
Even though the Goldberg-Sachs theorem only applies to empty space-times, its
generalization by Kundt and Tr\"{u}mper means that what follows also applies to the Kerr-Newman background.

We will now proceed to decouple the NP equations (\ref{eq:Maxwell eqs. with K^dagger})
by applying a new operator $\Pi_j{}^{(a)}$ onto them in the manner of (\ref{eq:Wald, decoupled eq.}) for the general case.
We therefore need to find an operator
$\Pi_j{}^{(a)}$ such that $\Pi_j{}^{(a)}K^{\dagger^{\indhel}}{}_{(a)}$ vanishes unless $\indhel=j$.
We make the ansatz $\Pi_j{}^{(a)}=K_{j}{}^{(a)}+\xi_j{}^{(a)}$, then it can be easily
checked that the operators
\begin{subequations}\label{eq:op. Pi in terms of K}
\begin{align}
\Pi_{-1}{}^{(a)}=&\delta_{(1)}^{(a)}\left(K_{-1}{}^{(1)}+2\tau\right)+\delta_{(3)}^{(a)}\left(K_{-1}{}^{(3)}-2\rho\right)\\
\begin{split}
\Pi_{0}{}^{(a)}=&\delta_{(1)}^{(a)}\left(2K_{0}{}^{(1)}-2\mu\right)+\delta_{(3)}^{(a)}\left(2K_{0}{}^{(3)}+2\pi\right) 
\quad \text{or} \quad \\
&\delta_{(2)}^{(a)}\left(2K_{0}{}^{(2)}-2\rho\right)+\delta_{(4)}^{(a)}\left(2K_{0}{}^{(4)}+2\tau\right)
\end{split} \\
\Pi_{+1}{}^{(a)}=&\delta_{(2)}^{(a)}\left(K_{+1}{}^{(2)}+2\pi\right)+\delta_{(4)}^{(a)}\left(K_{+1}{}^{(4)}-2\mu\right)
\end{align}
\end{subequations}
which take the explicit form
\begin{subequations}\label{eq:op. Pi}
\begin{align}
\Pi_{-1}{}^{(a)}=&-\delta_{(1)}^{(a)}\left(\delta+\pi^*-\alpha^*-\beta-2\tau\right)+\delta_{(3)}^{(a)}\left(D+\epsilon^*-\epsilon-\rho^*-2\rho\right)\label{eq:a)op. Pi}\\
\begin{split}
\Pi_{0}{}^{(a)}=&-\delta_{(1)}^{(a)}\left(\bDelta-\gamma-\gamma^*+\mu^*+\mu\right)+\delta_{(3)}^{(a)}\left(\delta^*+\pi-\tau^*+\beta^*-\alpha\right)
\quad \text{or} \quad \\
&+\delta_{(2)}^{(a)}\left(D+\epsilon+\epsilon^*-\rho-\rho^*\right)-\delta_{(4)}^{(a)}\left(\delta+\pi^*-\tau+\beta-\alpha^*\right) 
\end{split}\label{eq:b)op. Pi} \\
\Pi_{+1}{}^{(a)}=&\delta_{(2)}^{(a)}\left(\delta^*+\alpha+\beta^*-\tau^*+2\pi\right)-\delta_{(4)}^{(a)}\left(\bDelta+\mu^*+\gamma-\gamma^*+2\mu\right)\label{eq:c)op. Pi}
\end{align}
\end{subequations}
are the ones we are looking for. 
Note that the two expressions for $\Pi_{0}{}^{(a)}$ are not equal but they both achieve the desired decoupling
and they are physically equivalent.
The decoupling operators satisfy
\begin{equation} \label{eq:decoupling eq.}
2\Pi_j{}^{(a)}K^{\dagger^{\indhel}}{}_{(a)}=\delta^{\indhel}_jP_j
\end{equation}
where the operators $P_j$ are given by
\begin{subequations} \label{eq:op. P}
\begin{align}
\begin{split}
P_{-1}=&-2\left[(D-2\rho-\rho^*-\epsilon+\epsilon^*)(\bDelta-2\gamma+\mu)-  \right. \\
&\left.-(\delta-2\tau-\alpha^*-\beta+\pi^*)(\delta^*-2\alpha+\pi)\right] \end{split} \label{eq:op. Pa}\\
\begin{split}
P_0=&-2\left[(\bDelta-\gamma-\gamma^*+\mu+\mu^*)(D-2\rho)-(\delta^*+\pi-\tau^*+\beta^*-\alpha)(\delta-2\tau)\right]
\quad \text{or} \quad \\
&-2\left[(D-\rho-\rho^*+\epsilon+\epsilon^*)(\bDelta+2\mu)-(\delta-\tau+\beta-\alpha^*+\pi^*)(\delta^*+2\pi)\right] 
\end{split} \label{eq:op. Pb}\\
\begin{split}
P_{+1}=&-2\left[(\bDelta+2\mu+\mu^*-\gamma^*+\gamma)(D-\rho+2\epsilon)- \right.\\
&\left. -(\delta^*+2\pi-\tau^*+\alpha+\beta^*)(\delta-\tau+2\beta)\right]\end{split} \label{eq:op. Pc}
\end{align}
\end{subequations}
Thus, applying $\Pi_j{}^{(a)}$ onto the coupled NP Maxwell equations (\ref{eq:Maxwell eqs. with K^dagger})
will decouple them to
\begin{equation} \label{eq:decoupled eq.} 
P_j\phi_j=\Pi_j{}^{(a)}J_{(a)}
\end{equation}
which, in the absence of sources, is the equivalent of $\mathcal{O}(\phi)=0$ of 
Section ~\ref{sec:Wald in arbitrary type D s-t} for the spin-1 case in a Type D background.
There is no summation over the index $j$ in equation (\ref{eq:decoupled eq.}).

Note that operators $P_{+1}$ and $P_{-1}$ are obtainable one from the other under the interchange 
$\{\vec{l}\leftrightarrow\vec{n}, \vec{m}\leftrightarrow\vec{m^*}\}$, which results in (\ref{eq:effect of l<->n,m<-m^* on NP scalars and spin coeffs.}).
Likewise for the operators $\Pi_{+1}{}^{(a)}J_{(a)}$ and $-\Pi_{-1}{}^{(a)}J_{(a)}$. 
This means that equations (\ref{eq:decoupled eq.}) for $\phi_{+1}$ and $\phi_{-1}$ can be obtained one 
from the other under the interchange $\{\vec{l}\leftrightarrow\vec{n}, \vec{m}\leftrightarrow\vec{m^*}\}$, 
as expected. 

The two expressions for the operator $P_0$ in (\ref{eq:op. Pb}), formed 
from either of the two expressions for $\Pi_{0}{}^{(a)}$, are identical, as can be checked by using the commutation relations 
(\ref{eq:NP commutation rlns.}). 
Not only the left hand side of (\ref{eq:decoupled eq.}) remains the same whichever expression
for $\Pi_0{}^{(a)}$ we use, but so also does the right hand side when we make use of the law (\ref{eq:law of current conserv.}) of current conservation.
The NP scalar $\phi_0$ resulting 
from the use of either expression for $\Pi_{0}{}^{(a)}$ will be the same. Indeed, the two expressions 
correspond to different gauge choices, as we see specifically below.

By substituting $\phi_j$ from (\ref{eq:phi as func. of potential with K}) and $J_{(a)}$ from 
the tetrad form of (\ref{eq:Maxwell eqs. with potential}) into the decoupled NP equations (\ref{eq:decoupled eq.}), we obtain
the operator relation
\begin{equation}
P_jK_j{}^{(b)}=\Pi_j^{(a)}D_{(a)}{}^{(b)}
\end{equation}
The adjoint of this equation results in (\ref{eq:Wald, adj. of field eq.}) for the electromagnetic
case in a Type D background:
\begin{equation} \label{eq:adj. of field eq.}
K^{\dagger}_j{}^{(b)}P^{\dagger}{}_j=D^{(b)}{}_{(a)}\Pi^{\dagger}_j{}^{(a)}
\end{equation}
where
\begin{subequations}\label{eq:op. P^dagger}
\begin{align}
P^{\dagger}_{-1}&=-2\left[(\bDelta+\gamma-\gamma^*+\mu^*)(D+\rho+2\epsilon)-(\delta^*-\tau^*+\alpha+
\beta^*)(\delta+\tau+2\beta)\right]\\
\begin{split}
P^{\dagger}_{0}&=-2\left[(D+\rho-\rho^*+\epsilon+\epsilon^*)\bDelta-
(\delta+\tau+\beta-\alpha^*+\pi^*)\delta^*\right] \text{ or }\\
&=-2\left[(\bDelta-\gamma-\gamma^*-\mu+\mu^*)D-(\delta^*-\pi-\tau^*+\beta^*-\alpha)\delta\right]
\end{split} \label{eq:op. P^dagger_0} \\
P^{\dagger}_{+1}&=-2\left[(D-\rho^*-\epsilon+\epsilon^*)(\bDelta-2\gamma-\mu)-
(\delta-\beta-\alpha^*+\pi^*)(\delta^*-\pi-2\alpha)\right]
\end{align}
\end{subequations}
and
\begin{subequations}\label{eq:op. Pi^dagger}
\begin{align}
&\Pi^{\dagger}_{-1}{}^{(a)}=\delta_{(1)}^{(a)}\left(\delta+2\beta+\tau \right)-\delta_{(3)}^{(a)}\left(D+2\epsilon+\rho\right)\\
&\Pi^{\dagger}_{0}{}^{(a)}=\delta_{(1)}^{(a)}\bDelta-\delta_{(3)}^{(a)}\delta^* \qquad\text{or} \qquad
-\delta_{(2)}^{(a)}D+\delta_{(4)}^{(a)}\delta\\
&\Pi^{\dagger}_{+1}{}^{(a)}=-\delta_{(2)}^{(a)}\left(\delta^*-2\alpha-\pi\right)+\delta_{(4)}^{(a)}\left(\bDelta-2\gamma-\mu\right)
\end{align}
\end{subequations}

As a consequence of (\ref{eq:adj. of field eq.}), if $\NPadj_j$ is a solution of the 
adjoint of the decoupled NP equation (\ref{eq:decoupled eq.}) 
for the the Maxwell scalars with $J_{(a)}$=0, i.e.,
\begin{equation} \label{eq:adj. of decoupled eq.}
P^{\dagger}{}_j\NPadj_j=0
\end{equation}
then we can obtain a potential, which is a solution of the sourceless Maxwell equations (\ref{eq:Maxwell eqs. with potential}), 
by direct differentiation of $\NPadj_j$:
\begin{equation} \label{eq:potential as a func. of psi}
A_j{}^{(a)}=\Pi^{\dagger}_j{}^{(a)}\NPadj_j
\end{equation}
\catdraft{pero eqs. Teuk. valen tambe pel cas en que hi ha sources, i.e., valen no nomes per sourceless case com semblo indicar aqui?}

This equation is the one corresponding to $f=\mathcal{S}^{\dagger}(\NPadj)$ for the spin-1 case in a Type D background.
There is a complication, which is that this expression for the potential turns out to be pure gauge:
\begin{equation} \label{eq:pure gauge potential}
K^{\indhel}{}_{(a)}A_j{}^{(a)}=K^{\indhel}{}_{(a)}\Pi^{\dagger}_j{}^{(a)}\NPadj_j=\frac{1}{2}\delta^{\indhel}_jP^{\dagger}{}_j\NPadj_j=0
\end{equation}
\draft{is the expression `pure gauge' correct, when we prove in Section \ref{sec:polarization} that even though 
it does not contribute to the NP scalars it does contribute to the electric and mag. fields? how is such a thing possible??}
However, if we impose that the potential components $A_j{}^{\alpha}$ must be real, we can express them as

\begin{equation} \label{eq:make potential real}
A_j{}^{\alpha}=\left(\Pi^{\dagger}_j{}^{\alpha}\NPadj_j\right)^* \pm \Pi^{\dagger}_j{}^{\alpha}\NPadj_j
\end{equation}
As we have just seen in (\ref{eq:pure gauge potential}) the second term is pure gauge so it does not contribute
to $\phi_j$, but the first term is non-trivial, i.e., $K^{\indhel}{}_{\alpha}\left(\Pi^{\dagger}_j{}^{\alpha}\NPadj_j\right)^*\neq0$.
In what follows we are only 
going to include the non-trivial term in the potential components $A_j{}^{\alpha}$.
We can therefore obtain four different expressions for the potential depending on what scalar $\NPadj_j$
and what expression for $\Pi^{\dagger}_0{}^{(a)}$ we choose to use:
\begin{subequations} \label{eq:potential as a func. of psi_j^*}
\begin{align}
&A_{-1}{}^{\alpha}=\left(\Pi^{\dagger}_{-1}{}^{\alpha}\NPadj_{-1}\right)^*=
\left[l^{\alpha}(\delta^*+2\beta^*+\tau^*)-m^{* \alpha}(D+2\epsilon^*+\rho^*)\right]\NPadj^*_{-1} 
\label{eq:A_1 as a func. of psi_j^*} \\
&A_{0}{}^{\alpha}=\left(\Pi^{\dagger}_{0}{}^{\alpha}\NPadj_{0}\right)^*=
\left[l^{\alpha}\bDelta-m^{* \alpha}\delta\right]\NPadj^*_{0} \quad \text{or} \quad
\left[-n^{\alpha}D+m^{\alpha}\delta^*\right]\NPadj^*_{0} 
\label{eq:A0 as a func. of psi_j^*} \\
&A_{+1}{}^{\alpha}=\left(\Pi^{\dagger}_{+1}{}^{\alpha}\NPadj_{+1}\right)^*=
\left[-n^{\alpha}(\delta-2\alpha^*-\pi^*)+m^{\alpha}(\bDelta-2\gamma^*-\mu^*)\right]\NPadj^*_{+1} 
\label{eq:A1 as a func. of psi_j^*}
\end{align}
\end{subequations}
All different expressions are gauge-related. In particular, the two expressions for $A_{0}{}^{\alpha}$ are
related by
\begin{equation} \label{eq:rln. between the two A_0^alpha expressions}
\left[l^{\alpha}\bDelta-m^{* \alpha}\delta\right]\NPadj^*_{0}+\NPadj^{*;\alpha}_0=\left[-n^{\alpha}D+m^{\alpha}\delta^*\right]\NPadj^*_{0}
\end{equation}
We can finally calculate the NP Maxwell scalars using (\ref{eq:phi as func. of potential with K}) and choose for each
$\phi_j$ any one of the expressions (\ref{eq:potential as a func. of psi_j^*}) for the potential that we wish. 
In particular, if we choose either of the two expressions we have for $A_{0}{}^{\alpha}$ we obtain
\begin{subequations} \label{eq:phi_j as funcs. of psi_0^*}
\begin{align}
\phi_{-1}&=K_{-1 \alpha}A_0{}^{\alpha}=-(D+\epsilon^*-\epsilon-\rho^*)\delta\NPadj_0^* \\
\phi_{0}&=K_{0 \alpha}A_0{}^{\alpha}=(\rho^*-\rho)\bDelta\NPadj_0^*-(\delta^*+\beta^*-\alpha)\delta\NPadj_0^*\\
\phi_{+1}&=K_{+1 \alpha}A_0{}^{\alpha}=-(\bDelta+\mu^*+\gamma-\gamma^*)\delta^*\NPadj_0^*
\end{align}
\end{subequations}
If instead, we choose to use $A_{-1}{}^{\alpha}$ or $A_{+1}{}^{\alpha}$, we obtain
\begin{subequations} \label{eq:phi_j as funcs. of psi_{-1}^*}
\begin{align}
\phi_{-1}&=K_{-1 \alpha}A_{-1}{}^{\alpha}=-(D-\epsilon+\epsilon^*-\rho^*)(D+2\epsilon^*+\rho^*)\NPadj_{-1}^* \label{eq:phi_{-1} as funcs. of psi_{-1}^*}\\
\phi_{0}&=K_{0 \alpha}A_{-1}{}^{\alpha}=\left[-(D+\epsilon^*+\epsilon)(\delta^*+2\beta^*+\tau^*)+(\pi+\tau^*)(D+2\epsilon^*+\rho^*)\right]\NPadj_{-1}^*\\
\phi_{+1}&=K_{+1 \alpha}A_{-1}{}^{\alpha}=-(\delta^*+\alpha+\beta^*-\tau^*)(\delta^*+2\beta^*+\tau^*)\NPadj_{-1}^*
\end{align}
\end{subequations}
or
\begin{subequations} \label{eq:phi_j as funcs. of psi_{+1}^*}
\begin{align}
&\phi_{-1}=K_{-1 \alpha}A_{+1}{}^{\alpha}=-\delta(\delta-2\alpha^*-\pi^*)\NPadj_{+1}^* \label{eq:phi_{-1} as funcs. of psi_{+1}^*}\\
\begin{split}
&\phi_{0}=K_{0 \alpha}A_{+1}{}^{\alpha}= \\
&=\frac{1}{2}\left[-(\bDelta-\gamma-\gamma^*+\mu^*-\mu)(\delta-\alpha^*-\pi^*)-
(\bDelta+\mu^*+\gamma-\gamma^*)(\bDelta-2\gamma-\mu^*)\right]\NPadj_{+1}^*
\end{split} \\
&\phi_{+1}=K_{+1 \alpha}A_{+1}{}^{\alpha}=-(\bDelta+\mu^*+\gamma-\gamma^*)(\bDelta-2\gamma^*-\mu^*)\NPadj_{+1}^*
\end{align}
\end{subequations}
respectively. Expressions (\ref{eq:phi_j as funcs. of psi_0^*}), (\ref{eq:phi_j as funcs. of psi_{-1}^*}) 
and (\ref{eq:phi_j as funcs. of psi_{+1}^*}) were originally obtained by Cohen and Kegeles ~\cite{ar:Coh&Keg'74}.
\catdraft{pero no acaben de coinc. ben be??}

Finally, consider the equations
\begin{equation} \label{eq:particular rln. among psi_j} 
2\tilde{\Pi}^{\dagger -i}{}_{(a)}\NPadj_{\indhel}=\tilde{J}_{(a)}
\end{equation}
where $\tilde{\Pi}_{\pm1}{}^{(a)}\equiv \Pi_{\pm1}{}^{(a)}$ and $\tilde{\Pi}_{0}{}^{(a)}$ is the arithmetic
average of the two expressions for $\Pi_{0}{}^{(a)}$ given in (\ref{eq:b)op. Pi}). 
The tetrad components $\tilde{J}_{(a)}$ correspond to new `source terms' which are so far arbitrary.
Analogously to (\ref{eq:Maxwell eqs. with K^dagger}) these new equations can be decoupled by applying the operator
$K_{-j}{}^{(a)}$ to give
\begin{equation}
P^{\dagger}_{j}\NPadj_j=K_{-j}{}^{(a)}\tilde{J}_{(a)}
\end{equation}
That is, the solutions of equations (\ref{eq:particular rln. among psi_j}) with $\tilde{J}_{(a)}=0$ are also 
solutions of equations (\ref{eq:adj. of decoupled eq.}).
We will see in the next section 
that, in the Kerr-Newman space-time, equations (\ref{eq:particular rln. among psi_j}) are satisfied 
for some new source terms $\tilde{J}_{(a)}$ proportional to $J_{(a)}$.
From equations (\ref{eq:particular rln. among psi_j}) when $\tilde{J}_{(a)}=0$,
we can express the different terms with $\NPadj_{\pm1}^*$ 
appearing in (\ref{eq:A_1 as a func. of psi_j^*}) and (\ref{eq:A1 as a func. of psi_j^*}) as operators on $\NPadj_0^*$ 
to yield the simple expressions
\begin{subequations} \label{eq:A_pm1 as a func. of psi_0^*}
\begin{align}
&A_{-1}{}^{\alpha}=\left[l^{\alpha}\bDelta-m^{* \alpha}\delta\right]\NPadj^*_{0} \\
&A_{+1}{}^{\alpha}=\left[-n^{\alpha}D+m^{\alpha}\delta^*\right]\NPadj^*_{0}
\end{align}
\end{subequations}
which coincide precisely with the two expressions we had obtained for $A_{0}{}^{\alpha}$ in
(\ref{eq:A0 as a func. of psi_j^*}).

\draft{quina es la importancia/que aporta aquest darrer corol.lari?? (sota p.5Ott)}


\section{Wald's formalism for spin-1 in the Kerr-Newman background} \label{sec:Wald in K-N}

In order to separate the differential equations for the Maxwell scalars in the Kinnersley
tetrad (\ref{eq:def. null Kinnersley tetrad}) in Boyer-Lindquist coordinates in the Kerr background, 
Teukolsky ~\cite{ar:Teuk'73} started off from the decoupled equations (\ref{eq:decoupled eq.}) for $\phi_{\pm1}$.
However, the final expressions he gave were in terms of operators acting on $\phi_{-1}$ and on
$\rho^{-2}\phi_{+1}$, instead of $\phi_{+1}$. 
As Wald ~\cite{ar:Wald'78} remarked, in the Kerr background the quantity $(\psi_{0})^{-2/3}\phi_{+1}\propto \rho^{-2}\phi_{+1}$ 
satisfies the adjoint of the decoupled field equation for $\phi_{-1}$, i.e., (\ref{eq:adj. of decoupled eq.}) for $j=-1$. 
In other words, $\NPadj_{-1}=\rho^{-2}\phi_{+1}$ in the Kerr background. 
We are next going to show why this is so by finding a simple relationship between the solutions $\phi_{\indhel}$ of the decoupled equations
and the solutions $\NPadj_{\indhel}$ of the adjoint of the decoupled equations in a general Type D background.

Suppose there is a scalar function $v$ such that
\begin{equation}\label{eq:K as a func. of Pi via v}
K_j{}^{(a)}=v^{-1}\tilde{\Pi}_j{}^{(a)}v
\end{equation}
By using its adjoint 
\begin{equation}\label{eq:K^dagger as a func. of Pi^dagger via v}
K^{\dagger}_j{}^{(a)}=v\tilde{\Pi}^{\dagger}_j{}^{(a)}v^{-1}
\end{equation}
together with the adjoint of equation (\ref{eq:decoupling eq.}), it follows that
\begin{equation}\label{eq:P as a func. of P^dagger via v}
P_j=vP^{\dagger}{}_{-j}v^{-1}
\end{equation}
where we used the fact that, because the two expressions for $P_0$ in (\ref{eq:op. Pb}) are
identical, then $\delta^{\indhel}_jP_j=2\tilde{\Pi}_j{}^{(a)}K^{\dagger^{\indhel}}{}_{(a)}$.
When substituting the form (\ref{eq:P as a func. of P^dagger via v}) 
of $P_j$ into the decoupled equations (\ref{eq:decoupled eq.}) we obtain
\begin{equation}\label{eq:adj. of decoupled inhomogeneous eq.}
P^{\dagger}_{-j}\left(v^{-1}\phi_{j}\right)=v^{-1}\tilde{\Pi}_j{}^{(a)}J_{(a)}
\end{equation}
where we have used the property mentioned earlier that $\tilde{\Pi}_j{}^{(a)}J_{(a)}=\Pi_j{}^{(a)}J_{(a)}$.
Comparing equations (\ref{eq:adj. of decoupled inhomogeneous eq.}) with (\ref{eq:adj. of decoupled eq.}), we then have
a simple relationship between the solutions of the decoupled equations and the solutions 
of the adjoint of the decoupled equations:
\begin{equation}\label{eq:psi_j as a func. of phi_{-j}}
\NPadj_j=v^{-1}\phi_{-j}
\end{equation}
up to a factor of proportionality, which we choose to be one.
This relationship does not involve operators, unlike equations (\ref{eq:phi_j as funcs. of psi_0^*}), 
(\ref{eq:phi_j as funcs. of psi_{-1}^*}) or (\ref{eq:phi_j as funcs. of psi_{+1}^*}).
Condition (\ref{eq:K as a func. of Pi via v}) for the scalar function $v$
is equivalent to it satisfying the relations
\begin{subequations} \label{eq:conds. on scalar func. v}
\begin{align}
v^{-1}Dv&=D+2\rho\\
v^{-1}\bDelta v&=\bDelta-2\mu\\
v^{-1}\delta v&=\delta+2\tau\\
v^{-1}\delta^*v&=\delta^*-2\pi
\end{align}
\end{subequations}
as can be seen by comparing (\ref{eq:op. K}) and (\ref{eq:op. Pi}).

It is easy that if the property (\ref{eq:psi_j as a func. of phi_{-j}}) is satisfied by a certain background, then
the relations (\ref{eq:particular rln. among psi_j}) are satisfied with $\tilde{J}_{(a)}=v^{-1}J_{(a)}$.
Indeed, we first replace $\NPadj_j$ with $v^{-1}\phi_{-j}$ in the coupled NP equations 
(\ref{eq:Maxwell eqs. with K^dagger}), which relate $\phi_0$ to $\phi_{\pm1}$.
We then make use of (\ref{eq:K^dagger as a func. of Pi^dagger via v}) and we immediately
find that equations (\ref{eq:particular rln. among psi_j}), which relate 
$\NPadj_0$ to $\NPadj_{\pm1}$, are satisfied with $\tilde{J}_{(a)}=v^{-1}J_{(a)}$.
Equations (\ref{eq:A_pm1 as a func. of psi_0^*}) would then be valid in the sourceless case $J_{(a)}=0$.

In the Kerr-Newman background, the scalar function that satisfies condition (\ref{eq:K as a func. of Pi via v})
exists and it can be easily checked that it is $v=\rho^2$.
We therefore have that
\begin{equation} \label{eq:psi_j as a func. of phi_{-j} in K-N}
\NPadj_j=\rho^{-2}\phi_{-j}
\end{equation}
in the Kerr-Newman background.

We can then make use of the simple relation (\ref{eq:psi_j as a func. of phi_{-j} in K-N})
and then equations (\ref{eq:potential as a func. of psi_j^*}) yield
\begin{subequations} \label{eq:potential as a func. of phi_j^*}
\begin{align}
A_{-1}{}^{\alpha}&=\left(\Pi^{\dagger}_{-1}{}^{\alpha}\rho^{-2}\phi_{+1}\right)^*=
\left[l^{\alpha}(\delta^*+2\beta^*+\tau^*)-m^{* \alpha}(D+2\epsilon^*+\rho^*)\right]
\rho^{* -2}\phi^*_{+1} \label{eq:A_1 as a func. of phi_j^*} \\
A_{0}{}^{\alpha}&=\left(\Pi^{\dagger}_{0}{}^{\alpha}\rho^{-2}\phi_{0}\right)^*=
\left[l^{\alpha}\bDelta-m^{* \alpha}\delta\right]\rho^{* -2}\phi^*_{0} 
\text{     or     } 
\left[-n^{\alpha}D+m^{\alpha}\delta^*\right]\rho^{* -2}\phi^*_{0} 
\label{eq:A0 as a func. of phi_j^*} \\
A_{+1}{}^{\alpha}&=\left(\Pi^{\dagger}_{+1}{}^{\alpha}\rho^{-2}\phi_{-1}\right)^*=
\left[-n^{\alpha}(\delta-2\alpha^*-\pi^*)+m^{\alpha}(\bDelta-2\gamma^*-\mu^*)\right]
\rho^{* -2}\phi^*_{-1} \label{eq:A1 as a func. of phi_j^*}
\end{align}
\end{subequations}
where $\phi_j$ are any solutions of sourceless (\ref{eq:decoupled eq.}). 

We know from (\ref{eq:A_pm1 as a func. of psi_0^*}) and the discussion above 
that equations (\ref{eq:potential as a func. of phi_j^*}) reduce to the simple expressions
\begin{subequations} \label{eq:A_pm1 as a func. of phi_0^*}
\begin{align}
&A_{-1}{}^{\alpha}=\left[l^{\alpha}\bDelta-m^{* \alpha}\delta\right]\rho^{* -2}\phi^*_{0} \\
&A_{+1}{}^{\alpha}=\left[-n^{\alpha}D+m^{\alpha}\delta^*\right]\rho^{* -2}\phi^*_{0}
\end{align}
\end{subequations}
in the sourceless case $J_{(a)}=0$ in the Kerr-Newman background.


\section{Equation for $\phi_0$} \label{sec:eq. for phi_0 in Kerr}


It is clear from the relation (\ref{eq:psi_j as a func. of phi_{-j} in K-N}) 
that by using one set of equations among (\ref{eq:phi_j as funcs. of psi_0^*}),
(\ref{eq:phi_j as funcs. of psi_{-1}^*}) and (\ref{eq:phi_j as funcs. of psi_{-1}^*}) 
the knowledge of any one NP Maxwell scalar of our choice suffices to obtain all the components of the electromagnetic field tensor.
This NP Maxwell scalar is therefore a Debye potential for electromagnetic perturbations.
This complex Maxwell scalar carries all the information of the theory, that is, 
its real and imaginary parts represent the two dynamical degrees of theory of the perturbed field.
Furthermore, the potential can be readily derived from that same Maxwell scalar via the appropriate 
equation in (\ref{eq:potential as a func. of phi_j^*}).

\catdraft{pero a intro. Chrzan. sembla excloure $\phi_0$ pel fet que no es inv. sota gauge transfs. i inf. tetrad rotations??}

When quantizing the theory in the Kerr-Newman space-time, we shall see that a complete set of mode solutions
requires two sets of solutions with different boundary conditions. Two possible sets are solutions with
`ingoing' and `upgoing' boundary conditions. We shall see in the next section that the potentials 
$A_{-1}{}^{\alpha}$ and $A_{+1}{}^{\alpha}$ adapt themselves in a natural way to the 
former and latter type of boundary conditions respectively. 
It would therefore be very useful to quantize the theory by using the simple expressions (\ref{eq:A_pm1 as a func. of phi_0^*}), 
which yield the potentials naturally adapted to a set of complete solutions from one single Maxwell scalar.

This is indeed the case
in the Reissner-Nordstr\"{o}m space-time where the equation for $\phi_{0}$ can be separated. 
This is the procedure that ultimately underlies in the calculation in ~\cite{ar:J&McL&Ott'91} of simple, 
elegant expressions for the NP Maxwell scalars and expectation values of the stress-energy tensor. 
Unfortunately, as we shall now see, we have not been able to separate the equation for $\phi_{0}$ 
in the Kerr-Newman or, indeed, Kerr backgrounds.

Teukolsky showed that the differential equations for $\phi_{-1}$ and for $\rho^{-2}\phi_{+1}$
are separable in the Kinnersley tetrad in Boyer-Lindquist co-ordinates in the Kerr background.
He indicated that these equations are actually separable in any co-ordinates related to Boyer-Lindquist's
by: $t\to t+f_1(r)+f_2(\theta)$, $\phi\to \phi+g_1(r)+g_2(\theta)$, $r\to h(r)$ and $\theta\to j(\theta)$.  
However, he does not find a decoupled equation in relation to $\phi_{0}$.

\catdraft{hauria d'esmentar aqui Fack\&Ips'72}

Using the Kinnersley tetrad, we calculated $P^{\dagger}_0(\NPadj_0)$ explicitly in Boyer-Lindquist co-ordinates 
in the Kerr-Newman background where $\NPadj_0=\rho^{-2}\phi_0$ and $P^{\dagger}_0$ is given by (\ref{eq:op. P^dagger_0})
(remember that both expressions for $P^{\dagger}_0$ are identical).
We found that $2\rho^{* -1}P^{\dagger}_0(\rho^{-2}\phi_0)$ differs only slightly from
the equation that Teukolsky wrote down explicitly. 
We write it later in (\ref{eq:Teuk.eq.}) with $\indhel=0$, and where the function $\Omega_\indhel$ to solve for is $\rho^{-1}\phi_0$. 
The only difference being an extra term $2\rho^2\left(M/\rho^*+Q^2\right)\rho^{-1}\phi_0$. 
That is, writing the equation $P^{\dagger}_0(\NPadj_0)=0$ explicitly gives
\begin{equation} \label{eq:eq. for rho^-1phi0}
\begin{aligned} 
&\left\{\left[\frac{(r^2+a^2)^2}{\Delta}-a^2\sin^2\theta\right]\parddiff{}{t}+
\frac{4Mar}{\Delta}\pparddiff{}{t}{\phi}+
\left[\frac{a^2}{\Delta}-\frac{1}{\sin^2\theta}\right]\parddiff{}{\phi}- \right.\\
&\left. -\frac{\partial}{\partial r}\left(\Delta\pardiff{}{r}\right)-
\frac{1}{\sin\theta}\frac{\partial}{\partial \theta}\left(\sin\theta\pardiff{}{\theta}\right)+
2\rho^2\left(\frac{M}{\rho^*}+Q^2\right)\right\}(\rho^{-1}\phi_0)=0
\end{aligned}
\end{equation}
The differential equation for $\rho^{-1}\phi_0$ is therefore surprisingly similar to the one for the scalar field,
where the only difference is the extra term indicated.
It is the form of this extra term:
\begin{equation}
\frac{2}{(r-ia\cos\theta)^2}\left[-M(r+ia\cos\theta)+Q^2\right]
\end{equation}
that stops (\ref{eq:eq. for rho^-1phi0}) from being separable, even in the Kerr background where $Q=0$.
This extra term does not vanish in the Reissner-Nordstr\"{o}m background where $a=0$, however it reduces to just 
$2(-Mr+Q^2)/r^2$ so that (\ref{eq:eq. for rho^-1phi0}) does indeed become separable in this background.

\catdraft{1) posar p.24K(sota) altra justificac. pq. no es sep.by vars. pero cal donar expressio per $\phi_0$ que no he donat encara?
2) caldria demostrar que no hi ha cap f(r,theta) t.q. eq. per $f*\phi_0$ sigui sep.by vars. pero no he fet (p.25K(bis2))! }

After seeing that the differential equation for $\phi_0$ is not separable in the Kinnersley tetrad in Kerr one might wonder whether
the inherent symmetry in the Carter null tetrad renders the differential equation for the Maxwell scalars separable.

We define ${}_cP_{0}$ as either of the two identical expressions in (\ref{eq:op. P}) 
where the directional derivatives and the spin coefficients in the Carter null tetrad are given by 
(\ref{eq:def. Carter null tetrad}) and (\ref{eq:spin coeffs. in Carter null in Kerr}) respectively.
In the Kerr background, the equation ${}_cP_{0}{}_c\phi_{0}=0$ 
can be explicitly expressed as an operator acting on $\rho^{-1}{}_c\phi_{0}$, where the subscript $c$ in 
the NP Maxwell scalar indicates the use of the Carter null tetrad. 
This equation is
\begin{equation} \label{eq:eq. for rho^-1cphi0}
\begin{aligned}
&\left[\Delta{}_c\mathcal{D}^{\dagger}_2{}_c\mathcal{D}_0-2(r-M)\rho-2\Delta\rho^2+
(a^2-q^2){}_c\mathcal{L}^{\dagger}_2{}_c\mathcal{L}_0+2i\rho(\sigma-q)+   
\right. \\ &\left.
\qquad \qquad \qquad \qquad \qquad  \qquad  +2(a^2-q^2)\rho^2\right]
(\rho^{-1}{}_c\phi_{0})=0
\end{aligned}
\end{equation}
where we have already written out the $\tilde{t}$ and $\tilde{\phi}$ dependence as $e^{-i(\sigma\tilde{t}-\tilde{m}\tilde{\phi})}$ already.
The operators in (\ref{eq:eq. for rho^-1cphi0}) are defined as
\begin{subequations} \label{eq:def. cD_n,cL_n}
\begin{align}
{}_c\mathcal{D}_{n}^
{ \topbott{}{\dagger}} &\equiv \partial_{r} \mp \frac{i(r^2\sigma-\tilde{m})}{\Delta}+n\frac{r-M}{\Delta} \\
{}_c\mathcal{L}_{n}^
{ \topbott{}{\dagger}} &\equiv \partial_{q} \pm \frac{\tilde{m}+q^2\sigma}{a^2-q^2}-n\frac{q}{a^2-q^2}
\end{align}
\end{subequations}
The equation (\ref{eq:eq. for rho^-1cphi0}) is clearly not separable in the variables $r$ and $q$ due to the various terms
containing $\rho$ or $\rho^2$.
                                                                                                             

\section{The Teukolsky equation and the homogeneous potential solution} \label{sec:Wald in Kerr}

As mentioned earlier, Teukolsky (~\cite{ar:Teuk'72}, ~\cite{ar:Teuk'73}) wrote the differential equations 
for spin-1 perturbations in the Kerr background as operators acting
on $\phi_{-1}$ and $\rho^{-2}\phi_{+1}$ (rather than $\phi_{+1}$ itself).
Since the latter turns out to be $\NPadj_{-1}$ in this background, 
the actual equations that Teukolsky wrote down explicitly correspond to
\begin{subequations} \label{eq:Teuk.eqs. in op. form}
\begin{align}
&P_{-1}\phi_{-1}=-\Pi_{-1}{}^{\alpha}J_{\alpha}\\
&P_{-1}^{\dagger}\NPadj_{-1}=P_{-1}^{\dagger}(\rho^{-2}\phi_{+1})=-\rho^{-2}\Pi_{+1}{}^{\alpha}J_{\alpha}
\end{align}
\end{subequations}
which are particular cases of (\ref{eq:decoupled eq.}) and (\ref{eq:adj. of decoupled inhomogeneous eq.}) 
(with $v=\rho^2$ and $\tilde{\Pi}_{+1}{}^{\alpha}=\Pi_{+1}{}^{\alpha}$) respectively.
The reason why these are the equations
that he wrote out explicitly (rather than $P_{+1}\phi_{+1}=0$ instead of $P_{-1}^{\dagger}\NPadj_{-1}=0$,
or $P_{+1}^{\dagger}\NPadj_{+1}=0$ instead of $P_{-1}\phi_{-1}=0$) is that these are 
the equations that turn out to be separable. 
A similar situation holds for the NP Weyl scalars and the corresponding differential equations that Teukolsky gave.

Teukolsky wrote equations (\ref{eq:Teuk.eqs. in op. form}) using the Kinnersley
tetrad in Boyer-Lindquist coordinates.
Since Teukolsky's equations are the ones that we are are going to solve, we will write them out
explicitly. 
Teukolsky presented the results for the electromagnetic and gravitational perturbations in ~\cite{ar:Teuk'72} and 
he proved them and extended them to the neutrino case in ~\cite{ar:Teuk'73}. Carter ~\cite{ar:Carter'68b} had
previously shown the separability for the scalar case.
Teukolsky wrote the field equations in the Kerr background in compact form for the various spin fields, 
as one single `master' equation where the parameter $\indhel=0,\pm 1/2, \pm1,\pm2$ refers to the helicity of the field.
An analogous equation can be derived in the Kerr-Newman background.
In this background, we derived the equation for spin-1 whereas for spin-1/2 and spin-2 it is given in ~\cite{ar:Bose'75}.
We will still refer to the original Teukolsky equation with the inclusion of 
the modifications so that it is valid in the Kerr-Newman background
as the Teukolsky equation. 
This equation, valid in Kerr-Newman background, is:
\begin{equation} \label{eq:Teuk.eq.}
\begin{aligned} 
&\left[\frac{(r^2+a^2)^2}{\Delta}-a^2\sin^2\theta\right]\parddiff{\Omega_{\indhel}}{t}+
\frac{2(2Mr-Q^2)a}{\Delta}\pparddiff{\Omega_{\indhel}}{t}{\phi}+
\left[\frac{a^2}{\Delta}-\frac{1}{\sin^2\theta}\right]\parddiff{\Omega_{\indhel}}{\phi}- \\
&-\Delta^{-h}\frac{\partial}{\partial r}\left(\Delta^{h+1}\pardiff{\Omega_{\indhel}}{r}\right)-
\frac{1}{\sin\theta}\frac{\partial}{\partial \theta}\left(\sin\theta\pardiff{\Omega_{\indhel}}{\theta}\right)-
2h\left[\frac{a(r-M)}{\Delta}+\frac{i\cos\theta}{\sin^2\theta}\right]\pardiff{\Omega_{\indhel}}{\phi}- \\
&-2h\left[\frac{(Mr-Q^2)r-Ma^2)}{\Delta}-r-ia\cos\theta\right]\pardiff{\Omega_{\indhel}}{t}+
\\ & +\left(h^2\cos^2\theta-h-\frac{2Q^2}{\Sigma}\delta_{2,|h|}\right)\Omega_{\indhel}=
\Sigma T_{\indhel}
\end{aligned}
\end{equation}

\begin{table}
\begin{center}
\begin{tabular}{c|c|c}
$h$&$\Omega_{\indhel}$&$T_{\indhel}$ \\
\hline
\hline
0&$\Phi$&
$4\pi T^{\alpha}{}_{\alpha}$\\
\hline
$+\frac{1}{2}$&$\chi_{-1/2}$& $X_{-1/2}$\\
$-\frac{1}{2}$&$\rho^{-1}\chi_{1/2}$&$X_{+1/2}$\\
\hline
+1&$\phi_{-1}$&$J_{-1}=-\Pi_{-1}{}^{\alpha}J_{\alpha}$\\
-1&$\rho^{-2}\phi_{+1}$&$\rho^{-2}J_{+1}=-\rho^{-2}\Pi_{+1}{}^{\alpha}J_{\alpha}$\\
\hline
+2&$\psi_{-2}$&$8\pi T_{-2}$\\
-2&$\rho^{-4}\psi_{+2}$&$8\pi\rho^{-4}T_{+2}$\\
\end{tabular}
\end{center}
\caption{Field quantities $\Omega_{\indhel}$ and source terms $T_{\indhel}$ in the Teukolsky equation (\ref{eq:Teuk.eq.}). 
The quantities $T_{\pm 2}$ and $X_{\pm1/2}$ are the result of the decoupling operators acting on the sources
for the gravitational and neutrino cases respectively (see ~\cite{ar:Bose'75}, ~\cite{ar:Teuk'73}).} \label{table:quantities in Teuk.eq.}
\end{table}
where the field $\Omega_{\indhel}$ and the source term $T_{\indhel}$ denote different quantities depending on the value of the
helicity $\indhel$ as indicated in Table \ref{table:quantities in Teuk.eq.}.
Clearly the Teukolsky equation (\ref{eq:Teuk.eq.}) is separable for any value 
of the helicity $\indhel$. 
Its solution can therefore be written as a sum over the Fourier modes
\begin{equation} \label{eq:Fourier expansion for Omega_h}
\begin{aligned}
\Omega_{\indhel}(t,r,\theta,\phi)&=\int_{-\infty}^{+\infty}\d{\omega}\sum_{l=|\indhel|}^{+\infty}\sum_{m=-l}^{+l}{}_{lm\omega}\Omega_{\indhel}(t,r,\theta,\phi)  \\
{}_{lm\omega}\Omega_{\indhel}(t,r,\theta,\phi)&=\frac{1}{\sqrt{2\pi}}{}_{\indhel}R_{lm\omega}(r){}_{\indhel}S_{lm\omega}(\theta)e^{-i\omega t}e^{+im\phi}=\\
&={}_{\indhel}R_{lm\omega}(r){}_{\indhel}Z_{lm\omega}(\theta,\phi)e^{-i\omega t}
\end{aligned}
\end{equation}
\catdraft{should sum over $P$ be included in (\ref{eq:Fourier expansion for Omega_h})?!}
where we have made the obvious definition
\begin{equation}
{}_{\indhel}Z_{lm\omega}(\theta,\phi)\equiv \frac{(-1)^{m+1}}{\sqrt{2\pi}}{}_{\indhel}S_{lm\omega}(\theta)e^{+im\phi}
\end{equation}
We impose that the angular function ${}_{\indhel}S_{lm\omega}$ is normalized to one:
\begin{equation} \label{eq:normalization SWSH}
\int_0^{\pi}\d{\theta}\sin\theta{}_{\indhel}S_{lm\omega}^2=1
\end{equation}
The $t$- and $\phi$- dependences of the modes ${}_{lm\omega}\Omega_{\indhel}$ are a consequence of the fact that the 
Kerr-Newman background is stationary and axially symmetric. 
The parameter $l$ labels the eigenvalues of the angular differential equation for ${}_{\indhel}S_{lm\omega}$.
The sign factor $(-1)^{m+1}$ appearing in the definition of ${}_{\indhel}Z_{lm\omega}$ differs from that
of Chrzanowski ~\cite{ar:Chrzan'75} because of a difference in the normalization of the angular function 
${}_{\indhel}S_{lm\omega}$. The inclusion of this sign factor simplifies the equations for the field and the potential
that we derive later, given our normalization of the spherical functions.

\catdraft{1) justify limits of integration and sums, 2) pq. ha de tenir aquestes$t-$ and $\phi-$ dependencies? }

Specifically for the electromagnetic case:
\begin{equation}\label{eq:sep.vars. for phi}
{}_{lm\omega}\phi_{\indhel}(t,r,\theta,\phi)=
\rho^{h+1}{}_{-\indhel}R_{lm\omega}(r){}_{-\indhel}Z_{lm\omega}(\theta,\phi)e^{-i\omega t}
\end{equation}
i.e.,
\begin{subequations} \label{eq:sep.vars. for Omega}
\begin{align}
{}_{lm\omega}\Omega_{-1}={}_{lm\omega}\NPadj_{-1}(t,r,\theta,\phi)=\rho^{-2}{}_{lm\omega}\phi_{+1}(t,r,\theta,\phi)=
&{}_{-1}R_{lm\omega}(r){}_{-1}Z_{lm\omega}(\theta,\phi)e^{-i\omega t}
\\
{}_{lm\omega}\Omega_{+1}={}_{lm\omega}\phi_{-1}(t,r,\theta,\phi)=
&{}_{+1}R_{lm\omega}(r){}_{+1}Z_{lm\omega}(\theta,\phi)e^{-i\omega t}
\end{align}
\end{subequations}

The radial and angular Teukolsky equations into which the sourceless Teukolsky equation (\ref{eq:Teuk.eq.}) 
separates for the electromagnetic case can be written in a compact manner as follows:
\draft{check que es exactament aquesta $\lambda$ de Chandr.}
\begin{subequations} \label{eq:op.eq. for R}
\begin{align}
\left(\Delta\mathcal{D}_1\mathcal{D}^{\dagger}_1+2i\omega r\right){}_{+1}R_{lm\omega}&={}_{-1}\lambda_{lm\omega}{}_{+1}R_{lm\omega} \label{eq:op.eq. for R1} \\
\left(\Delta\mathcal{D}^{\dagger}_0\mathcal{D}_0-2i\omega r\right){}_{-1}R_{lm\omega}&={}_{-1}\lambda_{lm\omega}{}_{-1}R_{lm\omega} \label{eq:op.eq. for R_1}
\end{align}
\end{subequations}
and
\begin{subequations} \label{eq:op.eq. for S}
\begin{align}
\left(\mathcal{L}^{\dagger}_0\mathcal{L}_1-2a\omega\cos\theta\right){}_{+1}S_{lm\omega}&=-{}_{-1}\lambda_{lm\omega}{}_{+1}S_{lm\omega} \label{eq:op.eq. for S1} \\
\left(\mathcal{L}_0\mathcal{L}^{\dagger}_1+2a\omega\cos\theta\right){}_{-1}S_{lm\omega}&=-{}_{-1}\lambda_{lm\omega}{}_{-1}S_{lm\omega} \label{eq:op.eq. for S_1}
\end{align}
\end{subequations}
where the constant of separation ${}_{\indhel}\lambda_{lm\omega}$ is an eigenvalue of the angular equation.
We are using the definitions of the operators
\begin{subequations}
\begin{align}
\mathcal{L}_{n}^
{ \topbott{}{\dagger}} &\equiv \partial_{\theta} \pm \mathcal{Q}+n\cot\theta  \label{eq:def. L_n} \\   
\mathcal{D}_{n}^
{ \topbott{}{\dagger}} &\equiv \partial_{r} \mp \frac{iK}{\Delta}+2n\frac{r-M}{\Delta}    \label{eq:def. D_n}
\end{align}
\end{subequations}
where
\begin{subequations}
\begin{align}
\mathcal{Q} &\equiv -a\omega\sin\theta+\frac{m}{\sin\theta}  \label{eq:def. Q}   \\
K &\equiv (r^2+a^2)\omega-am             \label{eq:def. K}
\end{align}
\end{subequations}
Throughout this thesis, we use the convention that the upper and lower symbols inside braces go 
with the upper and lower signs in the equation.
We will analyze the radial and angular equations in the following chapters.

We can now substitute expressions (\ref{eq:sep.vars. for phi}) for ${}_{lm\omega}\phi_{\pm 1}$ 
into (\ref{eq:potential as a func. of phi_j^*}) and use
the symmetries (\ref{eq:R symm.->cc,-m,-w}) and (\ref{eq:S symm.->-s,-m,-w}) of the radial and angular functions:
\begin{equation} \label{eq:R symm.->cc,-m,-w and S symm.->-s,-m,-w}
{}_{\indhel}R_{lm\omega}(r)={}_{\indhel}R_{l-m-\omega}^*(r) 
\qquad  \text{and} \qquad
{}_{\indhel}S_{lm\omega}(\theta)=(-1)^{\indhel+m}{}_{-\indhel}S_{l-m-\omega}(\theta)
\end{equation} 
which we shall show in the next two chapters.  We then obtain for the homogeneous potential:
\begin{subequations} \label{eq:potential as a func. of RS}
\begin{align}
\begin{split}
{}_{lm\omega}A_{-1}{}^{\alpha}&=\left(\Pi^{\dagger}_{-1}{}^{\alpha}\rho^{-2}{}_{lm\omega}\phi_{+1}\right)^*=  \\
&=\begin{gathered}[t]
\left[l^{\alpha}(\delta^*+2\beta^*+\tau^*)-m^{* \alpha}(D+2\epsilon^*+\rho^*)\right] \\
{}_{-1}R_{l-m-\omega}(r){}_{+1}Z_{l-m-\omega}(\theta,\phi)e^{+i\omega t}
\end{gathered}
\end{split} \label{eq:potential_1 as a func. of RS} \\
\begin{split}
{}_{lm\omega}A_{+1}{}^{\alpha}&=\left(\Pi^{\dagger}_{+1}{}^{\alpha}\rho^{-2}{}_{lm\omega}\phi_{-1}\right)^*= \\
&=
\left[-n^{\alpha}(\delta-2\alpha^*-\pi^*)+m^{\alpha}(\bDelta-2\gamma^*-\mu^*)\right] \\
&\qquad \qquad \qquad \qquad \qquad  \rho^{* -2}{}_{+1}R_{l-m-\omega}(r){}_{-1}Z_{l-m-\omega}(\theta,\phi)e^{+i\omega t}= \\
&=
\rho^{* -2}\left[-n^{\alpha}(\delta-2\alpha^*+\pi^*)+m^{\alpha}(\bDelta-2\gamma^*+\mu^*)\right] \\
&\qquad \qquad \qquad \qquad \qquad  {}_{+1}R_{l-m-\omega}(r){}_{-1}Z_{l-m-\omega}(\theta,\phi)e^{+i\omega t}
\end{split}
\label{eq:potential1 as a func. of RS}
\end{align}
\end{subequations}

The real potential (\ref{eq:make potential real}) can be expressed as a Fourier mode sum as
\begin{subequations}
\begin{align}
A_j{}^{\alpha}&=\int_{-\infty}^{+\infty}\d{\omega}\sum_{l=|\indhel|}^{+\infty}\sum_{m=-l}^{+l}\sum_{P=\pm 1}{}_{lm\omega P}a{}_{lm\omega P}A_j{}^{\alpha}
\label{eq:Fourier expansion for A} \\
{}_{lm\omega P}A_j{}^{\alpha}&\equiv {}_{lm\omega}A_j{}^{\alpha}+P(-1)^{l+m}{}_{l-m-\omega}A_j{}^{* \alpha} \label{eq:def. of lmwPA}
\end{align}
\end{subequations}
\draft{include sum over $P$ and coeff. $a$ in (\ref{eq:Fourier expansion for Omega_h}) so that it corresponds with (\ref{eq:Fourier expansion for A})?}
where the potential modes are obtained from (\ref{eq:potential as a func. of RS}):
\begin{subequations} \label{eq:tableIChrzan.}
\begin{align}
\begin{split}
&{}_{l-m-\omega P}A_{-1}{}^{\alpha}
=
\\
&=\Big\{
\left[l^{\alpha}(\delta^*+2\beta^*+\tau^*)-m^{* \alpha}(D+2\epsilon^*+\rho^*)\right]
{}_{-1}R_{lm\omega }(r){}_{+1}Z_{lm\omega }(\theta,\phi)e^{-i\omega t}
+\\&+P
\left[l^{\alpha}(\delta+2\beta+\tau)-m^{\alpha}(D+2\epsilon+\rho)\right]
{}_{-1}R_{lm\omega }(r){}_{-1}Z_{lm\omega }(\theta,\phi)e^{-i\omega t}
\Big\}
\end{split} \label{eq:A_1, tableIChrzan.} \\
\begin{split}
&{}_{l-m-\omega P}A_{+1}{}^{\alpha}
=
\\
&=\Big\{
\rho^{* -2}\left[-n^{\alpha}(\delta-2\alpha^*+\pi^*)+m^{\alpha}(\bDelta-2\gamma^*+\mu^*)\right] 
{}_{+1}R_{lm\omega }(r){}_{-1}Z_{lm\omega }(\theta,\phi)e^{-i\omega t}
+\\&+P
\rho^{* -2}\left[-n^{\alpha}(\delta^*-2\alpha+\pi)+m^{* \alpha}(\bDelta-2\gamma+\mu)\right] 
{}_{+1}R_{lm\omega }(r){}_{+1}Z_{lm\omega }(\theta,\phi)e^{-i\omega t}
\Big\}
\end{split}
\label{eq:A1, tableIChrzan.}
\end{align}
\end{subequations}

Expressions (\ref{eq:tableIChrzan.}), valid in the Kerr-Newman background, 
were originally obtained by Chrzanowski ~\cite{ar:Chrzan'75} in the Kerr background.
The only difference with his expressions 
is an overall change in the sign of $m$ and $\omega$, justified because of the sum over $m$ and integration over $\omega$.
The parameter $P$ is summed in (\ref{eq:Fourier expansion for A}) over the values $+1$ and $-1$, corresponding 
to two linearly independent polarization states for the potential, as we shall see in the last chapter.
This was indeed Chrzanowski's justification for the inclusion of this sum.
The two linearly independent solutions are actually related via the parity operation $\mathcal{P}$, as we shall also see
in the last chapter. Chrzanowski made use of this relationship to calculate one linearly independent solution from the other.
We have chosen the sign factor and the change in the signs of $m$ and $\omega$ 
in equation (\ref{eq:def. of lmwPA}) so that this
relationship between the two independent solutions holds like in Chrzanowski's.
The Fourier coefficients must satisfy the following condition so that the potential remains real:
\begin{equation} \label{eq: c.c. of lmwPa}
{}_{lm\omega P}a^{*}=(-1)^{l+m}P{}_{l-m-\omega P}a
\end{equation}
This condition becomes immediately clear by making use of the symmetries of the radial and angular functions and 
the normalization we have chosen for them, which we shall give in the following chapters.

We will label the electromagnetic potential with the superscript `in', `up', `out' or `down' to indicate that
the radial function used has the corresponding boundary conditions, which are made explicit
in the next chapter.

For completeness and so that we can establish the appropriate comparison with his results, we will next 
briefly outline the method that Chrzanowski ~\cite{ar:Chrzan'75} uses to calculate the homogeneous 
electromagnetic potential prior to adaptation to our operator notation.


Chrzanowski's starting point is a conjecture made previously by Chrzanowski and Misner ~\cite{ar:Chrzan&Misner'75}. 
Knowing what the retarded Green function for the radial Teukolsky equation looks like for the scalar case in Kerr, 
they conjecture that the form of the retarded Green function for the spin-1 case in Kerr is
\begin{equation} \label{eq:conjecture Green func. for Teuk.eq. for spin-1}
G_{\mu\alpha}(x,x')=
\begin{cases}
\int_{-\infty}^{+\infty}\d{\omega}\sum_{lmP}
\frac{i\omega }{|\omega|}g^{PP'}{}_{lm\omega P}A^{\text{up}}_{\mu}(x){}_{lm\omega P'}A^{\text{out} *}_{\alpha}(x'), & r>r'\\
\int_{-\infty}^{+\infty}\d{\omega}
\sum_{lmP}\frac{i\omega }{|\omega|}g^{PP'}{}_{lm\omega P}A^{\text{in}}_{\mu}(x){}_{lm\omega P'}A^{\text{dn} *}_{\alpha}(x'), & r<r'\\
\end{cases}
\end{equation}
which Chrzanowski proves to be valid for high frequency. 
As a corollary of Wald's results in Section \ref{sec:Wald in arbitrary type D s-t}, this form of the Green
function is valid for all frequency, not just high frequency. 
The quantity $g^{PP'}$ is the reciprocal of $g_{PP'}$, which is a 2-dimensional metric for the polarization states;
it is $g_{PP'}=-\delta_{PP'}$ if the states are orthogonal.
It immediately follows that
\begin{equation} \label{eq:A from Green func. conjecture}
A_{\alpha}=\int_{-\infty}^{+\infty}\d{\omega}\sum_{lmP}
\frac{i\omega }{|\omega|}g^{PP'}{}_{lm\omega P}A^{\text{up}}_{\alpha}\left\langle {}_{lm\omega P'}A^{\text{out}}_{\beta},J^{\beta}\right\rangle
\end{equation}
where the inner product is defined as $\left\langle A,B\right\rangle \equiv \int\d^4x\sqrt{-g}A^*B$ 
and the modes have been normalized so that 
\begin{equation} \label{eq:norm Aup/out/in/dn}
\left\langle {}_{lm\omega P}A^{\text{up}}_{\alpha},{}_{l'm'w'P'}A^{\text{out} \alpha}\right\rangle_{\mathcal{I}^+}=
\left\langle {}_{lm\omega P}A^{\text{in}}_{\alpha},{}_{l'm'w'P'}A^{\text{dn} \alpha}\right\rangle_{\mathcal{H}^+}=
\frac{\omega}{|\omega|}g_{PP'}\delta_{ll'}\delta_{mm'}\delta(w-w')
\end{equation}
When applying the operators $K_{-1}{}^{\alpha}$ and 
$\rho^{-2}K_{+1}{}^{\alpha}$ on (\ref{eq:A from Green func. conjecture}) we obtain
\begin{equation} \label{eq:phi from Green func. conjecture}
\Omega_{\pm 1}=
\int_{-\infty}^{+\infty}\d{\omega}\sum_{lmP}\frac{i\omega }{|\omega|}{}_{lm\omega}\Omega^{\text{up}}_{\pm 1}\left\langle {}_{lm\omega}A^{\text{out}}_{\beta},J^{\beta}\right\rangle
\end{equation}
where
\begin{equation}
{}_{lm\omega}A^{\text{out}}_{\beta}\equiv\sum_{P,P'}g^{PP'}{}_{lm\omega P}A^{\text{out}}_{\beta}
\end{equation}

On the other hand, the radial Teukolsky equation can be solved by the method of radial Green's functions, so that the
solution of the Teukolsky equation can be expressed as
\begin{equation} \label{eq:Omega_h from Green func.}
\begin{aligned}
\Omega_{\indhel}
&=\int_{-\infty}^{+\infty}\d{\omega}\sum_{lmP} {}_{\indhel}R_{lm\omega}{}_{\indhel}Z_{lm\omega}e^{-i\omega t}= \\
&=\int_{-\infty}^{+\infty}\d{\omega}\sum_{lmP}
\frac{i\omega }{|\omega|}{}_{lm\omega}\Omega^{\text{up}}_{\indhel}\left\langle {}_{-\indhel}R^{\text{out}}_{lm\omega}\ {}_{\indhel}Z_{lm\omega}e^{-i\omega t},T_{\indhel}\right\rangle
\end{aligned}
\end{equation}
where use has been made of the symmetry (\ref{eq:R symm.->cc,-s}) of the radial equation. 
Since, from Table \ref{table:quantities in Teuk.eq.}, it is 
$T_{\indhel}=-\rho^{h-1}\Pi_{-h \beta}J^{\beta}$ for $\indhel=\pm1$, 
integrating by parts gives
\begin{equation} \label{eq:int. by parts Omega_h from Green func.}
\Omega_{\pm 1}=\int_{-\infty}^{+\infty}\d{\omega}\sum_{lmP}\frac{i\omega }{|\omega|}{}_{lm\omega}\Omega^{\text{up}}_{\pm 1}
\left\langle -\Pi_{\mp 1 \beta}{}^{\dagger *}\rho^{* (\pm 1-1)}{}_{\mp 1}R^{\text{out}}_{lm\omega}\ {}_{\pm 1}Z_{lm\omega}e^{-i\omega t},J^{\beta}\right\rangle
\end{equation}
The complex conjugation of $\Pi^{\dagger}$ is due to the slightly different definitions of adjoint used in 
(\ref{eq:def. adj. op., no cc}) and implicit in Chrzanowski's inner product.
Comparing (\ref{eq:phi from Green func. conjecture}) with (\ref{eq:int. by parts Omega_h from Green func.}) we have
\begin{equation} \label{eq:Chrzan. potential as a func. of RS}
{}_{lm\omega}A_{\beta}=-\Pi_{\mp 1 \beta}{}^{\dagger *}\rho^{* (h-1)}{}_{\mp 1}R_{lm\omega}\ {}_{\pm 1}Z_{lm\omega}e^{-i\omega t}
\end{equation}
where the label `out' has been dropped since the same argument could have carried through
with the advanced Green function rather than the retarded one, and the result obtained is
thus independent of the boundary condition. 
Chrzanowski also obtained equivalent results for the spin-2 case.
The expression (\ref{eq:Chrzan. potential as a func. of RS}) coincides
with our result (\ref{eq:potential as a func. of RS}) except for having the opposite sign for $m$ and $\omega$.

As mentioned earlier, $A_{+1}{}^{\alpha}$ and $A_{-1}{}^{\alpha}$ correspond to two different gauge choices, 
neither of which is the Lorentz gauge. Clearly from (\ref{eq:potential as a func. of RS}), the potential
with helicity $-1$ corresponds to the `ingoing gauge', i.e., 
\begin{equation}
l^{\alpha}A_{-1 \alpha}=A_{-1 l}=0
\end{equation}
 is the gauge condition.
This potential is transverse at the future horizon and at past infinity and will thus be used in calculations in these asymptotic
regions. The potential with helicity $+1$ corresponds to the `upgoing gauge':
\begin{equation}
n^{\alpha}A_{+1 \alpha}=A_{+1 n}=0
\end{equation}
This potential is transverse at the past horizon and at future infinity and will be used in calculations in these regions.

Since Chrzanowski obtains the NP scalar $\phi_{-1}$ from the ingoing gauge potential via equation (\ref{eq:int. by parts Omega_h from Green func.})
he calls $\phi_{-1}$ the `ingoing' field component. 
This is a consequence of the fact
that $T_{+1}$, which contains $\Pi^{-1}{}^{\alpha}$ and thus does not involve $J_n$, 
is the source term in the differential equation for $\phi_{-1}$. 
The `ingoing' potential $A_{-1 \alpha}$ was calculated by integrating $T_{+1}$ by parts. 
Equivalently, he calls $\phi_{+1}$ the `upgoing' field component.
This notation is in agreement with the asymptotics for large $r$. Indeed, as we shall see in the next chapter, 
the asymptotic behaviour of the solution of the Teukolsky equation separately for outgoing and ingoing waves in the limit $r\to +\infty$ is 
\begin{equation} \label{eq:peeling th}
\begin{aligned}
\Omega_{\indhel}&\sim r^{-(2h+1)}e^{+i\omega r}, &\quad r^{-1}e^{-i\omega r}&\quad (r\rightarrow +\infty) \\
\phi_{+1}&\sim r^{-1}e^{+i\omega r},     &\quad r^{-3}e^{-i\omega r}&\quad (r\rightarrow +\infty) \\
\phi_{0}&\sim r^{-2}e^{+i\omega r},     &\quad r^{-2}e^{-i\omega r}&\quad (r\rightarrow +\infty) \\
\phi_{-1}&\sim r^{-3}e^{+i\omega r},     &\quad r^{-1}e^{-i\omega r}&\quad (r\rightarrow +\infty)
\end{aligned}
\end{equation}
It is therefore the `upgoing'[`ingoing'] scalar $\phi_{+1[-1]}$ the one with the asymptotically dominant behaviour 
for the upgoing[ingoing] waves. The above asymptotic behaviour (\ref{eq:peeling th}) was originally obtained by
Newman and Penrose ~\cite{ar:N&P'62} and is commonly referred to as the \define{peeling off theorem}. 

\catdraft{no se si posar text seguent pq. no ho acabo d'entendre i pq. tampoc he def. emag.pral. null dirs.?:
``The reason for this name is that as we move backwards from infinity along a suitable null geodesic the two
electromagnetic principal null directions ``peel off'' from the outgoing radial direction. The same behaviour is shown
by the principal null directions of the Weyl tensor''}

\catdraft{1) no l'anomena 'upgoing' sino 'outgoing' pero canvio nomenclatura.teuk'72 tambe li diu 'outgoing'-segur que faig be de canviar
de nomenclatura??, 2) include asympts. close to horizon?}

\catdraft{1) even though $A_{\pm 1}{}^{\alpha}$ `naturally' adapt themselves to `in' and `up' b.c., it is still possible to
calculate only one of them by using the corresponding NP scalar first with `in' b.c. and then with `up' b.c. 
We would then still have `in' and `up' b.c. potentials calculated from one single NP scalar. 
Is this not possible? or it is possible but calculations with potential that uses `unnatural' b.c. would end up being difficult?,
Adrian:correct, could be done but calculations would be messy
2) is it not possible to do QFT directly with NP scalars instead of potential?-this is really what we end up doing, no?Adrian:yes}


\section{Field components and Teukolsky-Starobinski\u{\i} identities} \label{sec:Teuk-Starob. ids.}

\draft{sign $(-1)^m$ floating everywhere->check!}

In order to obtain an expression for the NP scalars $\phi_{\indhel}$ we
substitute the quantities $\phi_{-1}$ and $\NPadj_{-1}$ in their mode 
expressions (\ref{eq:sep.vars. for Omega}) into equations (\ref{eq:phi_j as funcs. of psi_{-1}^*})and use the symmetries 
(\ref{eq:R symm.->cc,-m,-w and S symm.->-s,-m,-w})
for the radial and angular functions. The result is
\begin{subequations} \label{eq:phi_j as funcs. of R_1S1}
\begin{align}
{}_{l-m-w}\phi_{-1}&=K_{-1 \alpha}{}_{l-m-w}A_{-1}{}^{\alpha}=-\mathcal{D}_0\mathcal{D}_0{}_{-1}R_{lm\omega}(r){}_{+1}Z_{lm\omega}(\theta,\phi)e^{-i\omega t}\\
\begin{split}
{}_{l-m-w}\phi_{0}&=K_{0 \alpha}{}_{l-m-w}A_{-1}{}^{\alpha}=   \\
&=\frac{\rho^2}{\sqrt{2}}\left[\left(\rho^{-1}\mathcal{D}_0+1\right)
\mathcal{L}_{1}+ia\sin\theta\mathcal{D}_0\right]{}_{-1}R_{lm\omega}(r){}_{+1}Z_{lm\omega}(\theta,\phi)e^{-i\omega t}
\end{split}     \\
{}_{l-m-w}\phi_{+1}&=K_{0 \alpha}{}_{l-m-w}A_{-1}{}^{\alpha}=-\frac{\rho^2}{2}\mathcal{L}_0\mathcal{L}_1{}_{-1}R_{lm\omega}(r){}_{+1}Z_{lm\omega}(\theta,\phi)e^{-i\omega t}
\end{align}
\end{subequations}

Similarly, we can simplify (\ref{eq:phi_j as funcs. of psi_{+1}^*}) to
\begin{subequations} \label{eq:phi_j as funcs. of R1S_1}
\begin{align}
&{}_{l-m-w}\phi_{-1}=K_{-1 \alpha}{}_{l-m-w}A_{+1}{}^{\alpha}=-\frac{1}{2}\mathcal{L}_0^{\dagger}\mathcal{L}_1^{\dagger}
{}_{+1}R_{lm\omega}(r){}_{-1}Z_{lm\omega}(\theta,\phi)e^{-i\omega t} \\
\begin{split}
&{}_{l-m-w}\phi_{0}=K_{0 \alpha}{}_{l-m-w}A_{+1}{}^{\alpha}=  \\
&=-\frac{\rho^2}{2\sqrt{2}}\left[\left(\rho^{-1}\mathcal{D}_0^{\dagger}+1\right)
\mathcal{L}^{\dagger}_{1}+ia\sin\theta\mathcal{D}_0^{\dagger}\right]\Delta{}_{+1}R_{lm\omega}(r){}_{-1}Z_{lm\omega}(\theta,\phi)e^{-i\omega t}
\end{split} \\
&{}_{l-m-w}\phi_{+1}=K_{+1 \alpha}{}_{l-m-w}A_{+1}{}^{\alpha}=-\frac{\rho^2}{4}\Delta\mathcal{D}^{\dagger}_0\mathcal{D}^{\dagger}_0\Delta
{}_{+1}R_{lm\omega}(r){}_{-1}Z_{lm\omega}(\theta,\phi)e^{-i\omega t}
\end{align}
\end{subequations}


The factor of proportionality in the relation (\ref{eq:psi_j as a func. of phi_{-j}}) 
has arbitrarily been chosen to be one. This means that the scalars ${}_{lm\omega}\phi_{\pm 1}$
derived from (\ref{eq:psi_j as a func. of phi_{-j}}) and (\ref{eq:sep.vars. for Omega}) 
will have an arbitrary normalization which will not necessarily have to coincide with the one 
of the scalars ${}_{lm\omega}\phi_{\pm1}$ obtained from 
(\ref{eq:phi_j as funcs. of R_1S1}) or (\ref{eq:phi_j as funcs. of R1S_1}), as we shall see.
Of course, we could always change the normalization of the potentials in 
(\ref{eq:potential as a func. of RS}) so that expressions (\ref{eq:sep.vars. for phi}) held, 
but instead we will keep the potentials and the NP scalars resulting from 
(\ref{eq:potential as a func. of RS}) and (\ref{eq:phi_j as funcs. of R_1S1}) 
or (\ref{eq:phi_j as funcs. of R1S_1}).

The only difference in the calculation of expressions (\ref{eq:phi_j as funcs. of R_1S1}) and (\ref{eq:phi_j as funcs. of R1S_1})
is that the potentials used correspond to two different gauge choices and therefore the NP scalars ${}_{lm\omega}\phi_{\indhel}$ 
should be the same whichever set of expressions we choose to use.
However, so far the solutions ${}_{\indhel}R_{lm\omega}$ and ${}_{\indhel}S_{lm\omega}$ of the radial and angular Teukolsky equations 
have arbitrary normalizations and boundary conditions (in the radial case).
This means that the normalizations and boundary conditions of ${}_{-1}R_{lm\omega}$ and ${}_{+1}S_{lm\omega}$ 
used in (\ref{eq:phi_j as funcs. of R_1S1}) are independent of those of ${}_{+1}R_{lm\omega}$ and ${}_{-1}S_{lm\omega}$ 
used in (\ref{eq:phi_j as funcs. of R1S_1}). Therefore the normalization and boundary conditions of the
scalars ${}_{lm\omega}\phi_{\indhel}$ which are derived from (\ref{eq:phi_j as funcs. of R_1S1}) do not necessarily 
coincide with those of the ${}_{lm\omega}\phi_{\indhel}$ which are derived from (\ref{eq:phi_j as funcs. of R1S_1}). 
Since expressions (\ref{eq:phi_j as funcs. of R_1S1})[(\ref{eq:phi_j as funcs. of R1S_1})] contain the potential
in the `ingoing[upgoing] gauge', ${}_{lm\omega}A_{-[+]1}{}^{\alpha}$, it will be natural to use ingoing[upgoing] 
boundary conditions for the radial solution when using these expressions and name the resulting scalars ${}_{lm\omega}\phi_{\indhel}^{\text{in[up]}}$.

\catdraft{no se si explicac. a p.3KK(V) de pq. ${}_{lm\omega}\phi_{\indhel}$ no encaixen amb expressions originals l'he expressat prou be aqui ni si es
del tot correcta?}


Apart from a normalization factor, the expressions for each one
of the NP scalars ${}_{lm\omega}\phi_j$ in (\ref{eq:phi_j as funcs. of R_1S1}) and in (\ref{eq:phi_j as funcs. of R1S_1}) must be equal as long as 
the same boundary conditions are used for ${}_{-1}R_{lm\omega}$ and ${}_{+1}R_{lm\omega}$. By equating
them we find the \define{Teukolsky-Starobinski\u{\i} identities}:
\begin{subequations} \label{eq:Teuk-Starob. ids.,indet.consts.}
\begin{align}
\mathcal{D}_0\mathcal{D}_0{}_{-1}R_{lm\omega}&=C{}_{+1}R_{lm\omega} & \mathcal{L}_0\mathcal{L}_1{}_{+1}S_{lm\omega}&=D{}_{-1}S_{lm\omega} 
\label{eq:a)Teuk-Starob. ids.,indet.consts.} \\
\Delta\mathcal{D}^{\dagger}_0\mathcal{D}^{\dagger}_0\Delta{}_{+1}R_{lm\omega}&=C'{}_{-1}R_{lm\omega} & 
\mathcal{L}^{\dagger}_0\mathcal{L}^{\dagger}_1{}_{-1}S_{lm\omega}&=D'{}_{+1}S_{lm\omega} \label{eq:b)Teuk-Starob. ids.,indet.consts.}
\end{align}
\end{subequations}
where, for clarity, we drop the subindices $\{lm\omega\}$ in the constants of proportionality $\{C,C',D,D'\}$. 
There are three restrictions on these constants. 
Firstly, by applying the operator $\Delta\mathcal{D}^{\dagger}_0\mathcal{D}^{\dagger}_0$
on the radial equation (\ref{eq:a)Teuk-Starob. ids.,indet.consts.}) and using the radial relation
(\ref{eq:b)Teuk-Starob. ids.,indet.consts.}) and (\ref{eq:op.eq. for R_1}) the condition 
\begin{equation} \label{eq:op. equal to B^2}
\Delta\mathcal{D}^{\dagger}_0\mathcal{D}^{\dagger}_0\Delta\mathcal{D}_0\mathcal{D}_0={}_1B_{lm\omega}^2=CC'
\end{equation}
follows. The first equality in (\ref{eq:op. equal to B^2}) is only valid when operating on ${}_{-1}R_{lm\omega}$. 
We are using the definition
\begin{equation} \label{eq:def. B}
{}_1B_{lm\omega}^2\equiv {}_{-1}\lambda_{lm\omega}^2+4ma\omega-4a^2\omega^2
\end{equation}
The subindex $1$ in ${}_1B_{lm\omega}$ indicates spin-1 case.
Similarly, when applying the operator $\mathcal{L}^{\dagger}_0\mathcal{L}^{\dagger}_1$
on the angular equation (\ref{eq:a)Teuk-Starob. ids.,indet.consts.}) 
and using (\ref{eq:op.eq. for S1}) and the angular (\ref{eq:b)Teuk-Starob. ids.,indet.consts.}),
it follows that $\mathcal{L}^{\dagger}_0\mathcal{L}^{\dagger}_1\mathcal{L}_0\mathcal{L}_1={}_1B_{lm\omega}^2=DD'$, with the first equality
being valid only when operating on ${}_{+1}S_{lm\omega}$.

\catdraft{com es que enlloc no poso ortogonalitat dels SWSH???! (e.g., com eq.2.5Ott\&Winst'00 pels spherical) ho satisfan o no??}

The normalization (\ref{eq:normalization SWSH}) of the spherical functions immediately implies that $D=D'$. 
We therefore have $D$ and $D'$ determined: $D=D'={}_1B_{lm\omega}$, and
freedom in the choice of $C$ and $C'$ subject to the restriction $CC'={}_1B_{lm\omega}^2$. 
Traditionally (~\cite{ar:CCH},~\cite{ar:CMSR},~\cite{ar:Teuk&Press'74}),
the choice $C=1/2$ and $C'=2{}_1B_{lm\omega}^2$ has been made and we will be faithful to tradition by making the same choice.
The Teukolsky-Starobinski\u{\i} identities (\ref{eq:Teuk-Starob. ids.,indet.consts.}) then become
\begin{subequations} \label{eq:Teuk-Starob. ids.}
\begin{align}
\mathcal{D}_0\mathcal{D}_0{}_{-1}R_{lm\omega}&=
\frac{1}{2}{}_{+1}R_{lm\omega} & \mathcal{L}_0\mathcal{L}_1{}_{+1}S_{lm\omega}&={}_1B_{lm\omega}{}_{-1}S_{lm\omega} \label{eq:a)Teuk-Starob. ids.} \\
\Delta\mathcal{D}^{\dagger}_0\mathcal{D}^{\dagger}_0\Delta{}_{+1}R_{lm\omega}&=2{}_1B_{lm\omega}^2{}_{-1}R_{lm\omega} & 
\mathcal{L}^{\dagger}_0\mathcal{L}^{\dagger}_1{}_{-1}S_{lm\omega}&={}_1B_{lm\omega}{}_{+1}S_{lm\omega} \label{eq:b)Teuk-Starob. ids.}
\end{align}
\end{subequations}

We will also include here the Teukolsky-Starobinski\u{\i} identities for the angular functions for the spin-2 case (~\cite{bk:Chandr}), since
we will need them in Chapter \ref{ch:high freq. spher}. They are:
\begin{equation} \label{eq:Teuk-Starob ids. for spher.s=2}
\begin{aligned}
\mathcal{L}_{-1}\mathcal{L}_{0}\mathcal{L}_{1}\mathcal{L}_{2} {}_{+2}S_{lm\omega}&={}_{2}B_{lm\omega} {}_{-2}S_{lm\omega} \\
\mathcal{L}^{\dagger}_{-1}\mathcal{L}^{\dagger}_{0}\mathcal{L}^{\dagger}_{1}\mathcal{L}^{\dagger}_{2} {}_{-2}S_{lm\omega}&={}_{2}B_{lm\omega} {}_{+2}S_{lm\omega}
\end{aligned}
\end{equation}
where
\begin{equation}
\begin{aligned}
{}_{2}B_{lm\omega}^{2}&\equiv
{}_{-2}\lambda_{lm\omega}^{2}({}_{-2}\lambda_{lm\omega}+2)^{2}
-8(a\omega)^{2}{}_{-2}\lambda_{lm\omega}\left\{\left(1-\frac{m}{a\omega}\right)\left[5{}_{-2}\lambda_{lm\omega}+6\right]-12\right\}+ \\
&+144(a\omega)^{4}\left(1-\frac{m}{a\omega}\right)^{2}
\end{aligned}
\end{equation}
The signs of ${}_sB_{lm\omega}$ for spin $s=1$ and $2$ are arbitrary, but we will take them to be both positive.
It can be shown that if they are taken to be positive, 
then (\ref{eq:Teuk-Starob. ids.}) and (\ref{eq:Teuk-Starob ids. for spher.s=2}) agree with the sign in 
the symmetry (\ref{eq:S symm.->pi-t,-s}) of the angular function.

We finally use the Teukolsky-Starobinski\u{\i} identities to simplify 
(\ref{eq:phi_j as funcs. of R_1S1}) and (\ref{eq:phi_j as funcs. of R1S_1}). 
The result is a set of very simple expressions for the NP scalar modes ${}_{lm\omega}\phi_{\pm 1}^{\text{in/up}}$:
\begin{equation} \label{eq:phi_0/2(in/up)}
\begin{aligned}
{}_{l-m-\omega}\phi_{-1}^{\text{in}}&=-\frac{1}{2}{}_{+1}R^{\text{in}}_{lm\omega}{}_{+1}Z_{lm\omega}e^{-i\omega t} \\ 
{}_{l-m-\omega}\phi_{+1}^{\text{in}}&=-\frac{{}_{1}B_{lm\omega}}{2}\rho^2{}_{-1}R^{\text{in}}_{lm\omega}{}_{-1}Z_{lm\omega}e^{-i\omega t} \\
{}_{l-m-\omega}\phi_{-1}^{\text{up}}&=-\frac{{}_{1}B_{lm\omega}}{2}{}_{+1}R^{\text{up}}_{lm\omega}{}_{+1}Z_{lm\omega}e^{-i\omega t}      \\
{}_{l-m-\omega}\phi_{+1}^{\text{up}}&=-\frac{{}_{1}B_{lm\omega}^2}{2}\rho^2{}_{-1}R^{\text{up}}_{lm\omega}{}_{-1}Z_{lm\omega}e^{-i\omega t}
\end{aligned}
\end{equation}

Chandrasekhar ~\cite{bk:Chandr} obtained an expression for the other NP scalar by comparing Maxwell equations 
(\ref{eq:Maxwell eq. with phi0(1),phi_1(4)}) and (\ref{eq:Maxwell eq. with phi0(4),phi1(1)}) 
and using the Teukolsky-Starobinski\u{\i} identities.
He also found another, equivalent expression for the same NP scalar using Maxwell equations 
(\ref{eq:Maxwell eq. with phi0(2),phi1(3)}) and (\ref{eq:Maxwell eq. with phi0(3),phi_1(2)}) instead. These expressions are:
\begin{equation} \label{eq:phi0(ch)}
\begin{aligned}
{}_{l-m-\omega}\phi_{0}^{\text{in}}&=-\frac{\rho^2}{2^{3/2}{}_{1}B_{lm\omega}}\left[\left(\rho^{-1}\mathcal{D}_0^{\dagger}+1\right)
\mathcal{L}^{\dagger}_{1}+ia\sin\theta\mathcal{D}_0^{\dagger}\right]\Delta{}_{+1}R_{lm\omega}^{\text{in}}{}_{-1}Z_{lm\omega}e^{-i\omega t}= \\
&=\frac{\rho^2}{\sqrt{2}}\left[\left(\rho^{-1}\mathcal{D}_0+1\right)
\mathcal{L}_{1}+ia\sin\theta\mathcal{D}_0\right]{}_{-1}R_{lm\omega}^{\text{in}}{}_{+1}Z_{lm\omega}e^{-i\omega t} \\
{}_{l-m-\omega}\phi_{0}^{\text{up}}&=-\frac{\rho^2}{2^{3/2}}\left[\left(\rho^{-1}\mathcal{D}_0^{\dagger}+1\right)
\mathcal{L}^{\dagger}_{1}+ia\sin\theta\mathcal{D}_0^{\dagger}\right]\Delta{}_{+1}R_{lm\omega}^{\text{up}}{}_{-1}Z_{lm\omega}e^{-i\omega t}= \\
&=\frac{\rho^2{}_{1}B_{lm\omega}}{\sqrt{2}}\left[\left(\rho^{-1}\mathcal{D}_0+1\right)
\mathcal{L}_{1}+ia\sin\theta\mathcal{D}_0\right]{}_{-1}R_{lm\omega}^{\text{up}}{}_{+1}Z_{lm\omega}e^{-i\omega t}
\end{aligned}
\end{equation}

Making use of the relations (\ref{eq:DR1 as func. of R_1,DR_1}) and (\ref{eq:LSs as func. of S_s,LS_s}) it can be immediately
checked that the two different expressions for ${}_{lm\omega}\phi_{0}^{\text{in}}$ are indeed equivalent.   
Likewise for the two expressions for ${}_{lm\omega}\phi_{0}^{\text{up}}$.

In practise we are only going to numerically calculate one radial function (${}_{-1}R$) and its derivative with both `in' and `up'
boundary conditions and one angular function (${}_{-1}S$) and its derivative. 
We will then calculate the other radial and angular functions and their derivatives from linear expressions derived from the 
Teukolsky-Starobinski\u{\i} identities. We will obtain the NP Maxwell scalars from the simple expressions (\ref{eq:phi_0/2(in/up)})
and (\ref{eq:phi0(ch)}).

\chapter{Radial solution} \label{ch:radial sln.}

\draft{include more graphs for radial func.?}

\section{Introduction}

The Teukolsky equation (\ref{eq:Teuk.eq.}) for the field perturbation $\Omega_\indhel$ in the Kerr-Newman background
\ddraft{at least for spin-1...}
is separable in Boyer-Lindquist coordinates.
The resulting radial differential equation in the vacuum case is
\begin{equation} \label{eq:radial teuk. eq.}
\Delta^{-\indhel }\diff{}{r}\left(\Delta^{\indhel+1}\diff{{}_{\indhel}R_{lm\omega}}{r}\right)-{}_{\indhel}V{}_{\indhel}R_{lm\omega}=0
\end{equation}
where the potential is given by
\begin{equation} \label{eq:radial teuk. potential}
{}_{\indhel}V=\frac{2i\indhel (r-M)K-K^2}{\Delta}-4i\indhel \omega r+{}_{\indhel}\lambda_{lm\omega}
\end{equation}
where ${}_{\indhel}\lambda_{lm\omega}$ is the eigenvalue of the angular equation, which we deal with in the next chapter.
It is immediate from the radial equation (\ref{eq:radial teuk. eq.}) that the following symmetries are satisfied:
\begin{subequations} \label{eq:R symms.}
\begin{align}
&{}_{\indhel}R_{lm\omega}(r)=\Delta^{-\indhel }{}_{-\indhel }R_{lm\omega}^*(r)   \label{eq:R symm.->cc,-s} \\
&{}_{\indhel}R_{lm\omega}(r)={}_{\indhel}R_{l-m-\omega}^*(r)  \label{eq:R symm.->cc,-m,-w}
\end{align}
\end{subequations}
These symmetries, however, is only satisfied subject to particular boundary conditions, which we will explore
in the next section.

Except for the case $\omega=0$, the radial equation (\ref{eq:radial teuk. eq.}), has regular singular points
at the two roots of $\Delta$, i.e., at the inner and outer horizons, and an irregular singular point at infinity.
This equation is therefore not soluble in terms of standard functions and we do not know an integral representation
of its solutions. We are forced to solve it numerically.

The radial potential (\ref{eq:radial teuk. potential}) is a complex, long-range potential.
In the next section we are going to see two possible transformations, one derived by Detweiler ~\cite{ar:Detw'76} 
and the other one by Sasaki and Nakamura ~\cite{ar:Sasa&Naka'82}, that convert the radial potential
into a short-range one. Detweiler's main interest was in solving the homogeneous radial equation 
for spin-1 whereas Sasaki and Nakamura's was in solving the inhomogeneous radial equation for spin-2. The two
approaches are, as a matter of fact, particular cases of a general-spin method that we present in the next section.
This method is valid in the Kerr-Newman background whereas both Detweiler's and Sasaki and Nakamura's results
were restricted to the Kerr background.
In the same section we give the full set of transformations between the coefficients of the different radial solutions.
We also study a particularly symmetric solution of the radial equation.

In the two subsequent sections we describe the numerical method we have used to integrate the homogeneous radial equation
for spin-1. The numerical results are compared against the literature.

In Section \ref{sec:asympts. close to r_+} we calculate the asymptotic behaviour of the radial solution close to the
horizon, which is needed in Chapter \ref{stress-energy tensor}. We follow the method used by Candelas  ~\cite{ar:Candelas'80}.
He, however, developed the method for spin-0 and we extend it to general-spin and specialize to spin-1 only at the very end.

In the last section, we find the asymptotic behaviour for small frequency 
of the radial function, based on a method used by Page ~\cite{ar:PageI'76}, which we extend and generalize to the Kerr-Newman background.


\section{Short-range potentials} \label{sec:short-range potentials}

\catdraft{1) should $b_{\pm}$ below be real??, 2) i que?pq. cal que es compleixi (\ref{eq:asympt. form of short-range sln.})?}
A second-order differential equation 
\begin{equation}
\ddiff{Y(x)}{x}+A(x)\diff{Y(x)}{x}+B(x)Y(x)=0
\end{equation}
is said to be \define{short-range} if, and only if, $A(x)=O(x^{-n})$ and 
$B(x)=b_{\pm}^2+O(x^{-n})$ 
when $x\to \pm \infty$ with $n\geq 2$ and where $b_{\pm}$ are constants.
If this condition is guaranteed, then the asymptotic form of the solution is
\begin{equation} \label{eq:asympt. form of short-range sln.}
Y(x) \sim 
\begin{cases}
e^{\pm ib_+x}   & (x\to +\infty) \\
e^{\pm ib_-x}   & (x\to -\infty)
\end{cases}
\end{equation}
The potential (\ref{eq:radial teuk. potential}) in the radial Teukolsky equation is a long-range potential.
We will therefore not solve numerically this equation but we will instead solve one derived from it, which is short-range.
Detweiler ~\cite{ar:Detw'76} on the one hand and Sasaki and Nakamura ~\cite{ar:Sasa&Naka'82} on the other have
independently derived from the radial Teukolsky equation two different differential equations which are
short-range and valid in the Kerr background. 

Both derivations impose for the resulting differential equations to be short-range, but there are two main 
differences between the two derivations. 
One difference is that Detweiler requires the potential to be real and the differential equation to have the same form as the
radial Teukolsky equation, whereas Sasaki and Nakamura require
the differential equation to become the Regge-Wheeler equation in the limit $a \rightarrow 0$ for $\indhel=-2$.
The Regge-Wheeler equation is the differential equation that governs the odd parity gravitational perturbations
of the Schwarzschild space-time. 
The other key difference between the two derivations is that Detweiler's main interest was in 
solving the homogeneous differential equation whereas Sasaki and Nakamura's was in solving the inhomogeneous one. 
Detweiler's new source term behaves actually worse than the original source for $r \rightarrow +\infty$
whereas Sasaki and Nakamura's new source term is short-range. 

In ~\cite{ar:Detw'76}, Detweiler derived for general spin the set of equations that the new, real potential and the new 
radial function should satisfy in order to meet his requirements.
However, he only solved them and showed the explicit form of the new potential and radial function for the case of $\indhel=-1$. 
In ~\cite{ar:Detw'77}, he further wrote the required general form for any spin for the new potential and
radial function in terms of the potential ${}_{\indhel}V$ and the variables that define the new radial
function. These variables were left undetermined satisfying certain general-spin equations. Sasaki and Nakamura ~\cite{ar:Sasa&Naka'82}
were mainly interested in gravitational perturbations and their whole derivation was restricted to $\indhel=-2$.

Both derivations are, in fact, particular cases of a more general derivation which we have calculated for 
the homogeneous case and will present next. 
This derivation is valid in the Kerr-Newman background.
We also show when and how the two approaches differ and justify why we chose to pursue Detweiler's approach rather 
than Sasaki and Nakamura's.

First note that the solution ${}_{-\indhel }R_{lm\omega}$ can be expressed in terms of the solution ${}_{\indhel}R_{lm\omega}$ and
its derivative. We only need to use one of the radial Teukolsky-Starobinski\u{\i} identities (\ref{eq:Teuk-Starob. ids.}) and express
the second derivative of the function appearing in the identity in terms of the function and its first derivative by using the 
radial Teukolsky equation. The result is
\begin{equation} \label{eq:R_s as func of Rs and Rs'}
{}_{-\indhel }R_{lm\omega}=a_D{}_{\indhel}R_{lm\omega}+b_D\Delta^{\indhel+1}\diff{{}_{\indhel}R_{lm\omega}}{r}
\end{equation}
where, for $\indhel =-1$,
\begin{equation} \label{eq:def a_D,b_D}
\begin{aligned}
a_D&=-\frac{2}{\Delta}\left[2K^2+\Delta(iK'-{}_{-1}\lambda_{lm\omega})\right] \\
b_D&=\frac{4iK}{\Delta}
\end{aligned}
\end{equation}
where a primed function denotes differentiation with respect to its only argument, $r$ in this case.
The general transformation of the radial function ${}_{\indhel}R_{lm\omega}$ which preserves the form of the linear wave equation 
(\ref{eq:radial teuk. eq.}) is
\begin{equation} \label{eq:chi as func of Rs and Rs'}
\chi_{lm\omega}=\alpha(r){}_{\indhel}R_{lm\omega}+\beta(r)\Delta^{\indhel+1}\diff{{}_{\indhel}R_{lm\omega}}{r}
\end{equation}
or equivalently, using (\ref{eq:R_s as func of Rs and Rs'}),
\begin{equation} \label{eq:chi as func of Rs and R_s}
\chi_{lm\omega}=p(r){}_{\indhel}R_{lm\omega}+q(r){}_{-\indhel }R_{lm\omega}
\end{equation}
with 
\begin{equation} \label{eq:def alpha,beta}
\begin{aligned}
\alpha&=p+a_Dq \\
\beta&=b_Dq
\end{aligned} 
\end{equation}
Detweiler's and Sasaki and Nakamura's derivations differ in the choices of the conditions on
$\alpha$ and $\beta$ (or equivalently, $p$ and $q$).
Transformation (\ref{eq:chi as func of Rs and Rs'}) can be inverted to give
\begin{equation} \label{eq:Rs,Rs' as funcs. of chi and chi'}
\begin{aligned}
\gamma {}_{\indhel}R_{lm\omega}&=\left(\alpha+\beta'\Delta^{\indhel+1}\right)\chi_{lm\omega}-\beta\Delta^{\indhel+1}\diff{\chi_{lm\omega}}{r} \\
\gamma \diff{{}_{\indhel}R_{lm\omega}}{r}&=-\left(\alpha'+\beta\Delta^{\indhel}{}_{\indhel}V\right)\chi_{lm\omega}+\alpha\diff{\chi_{lm\omega}}{r}
\end{aligned}
\end{equation}
where
\begin{equation}
\gamma=\alpha\left(\alpha+\beta'\Delta^{\indhel+1}\right)-\beta\Delta^{\indhel+1}\left(\alpha'+\beta\Delta^{\indhel}{}_{\indhel}V\right)
\end{equation}
If we take the first and second derivatives of $\chi_{lm\omega}$ in (\ref{eq:chi as func of Rs and Rs'}) 
with respect to $r$ and use (\ref{eq:radial teuk. eq.}),
we find that the differential equation satisfied by $\chi_{lm\omega}$ is
\begin{equation} \label{eq:diff eq chi}
\Delta^{-\indhel }\diff{}{r}\left(\Delta^{\indhel+1}\diff{\chi_{lm\omega}}{r}\right)-\Delta F\diff{\chi_{lm\omega}}{r}-{}_{\indhel}U(r)\chi_{lm\omega}=0
\end{equation}
with
\begin{subequations}
\begin{align}
F &\equiv \frac{\gamma'}{\gamma} \label{eq:def. F,S&N'82}  \\
{}_{\indhel}U & \equiv 
{}_{\indhel}V+\frac{\Delta^{-\indhel }}{\beta}\left[\left(2\alpha+\beta'\Delta^{\indhel+1}\right)'-F\left(\alpha+\beta'\Delta^{\indhel+1}\right)\right] 
\label{eq:sU in terms of sV}
\end{align}
\end{subequations}
It is then useful to define a new dependent variable
\begin{equation} \label{eq:def. of X}
X_{lm\omega}\equiv(r^2+a^2)^{1/2}\Delta^{\indhel/2}\chi_{lm\omega}
\end{equation}
From the differential equation (\ref{eq:diff eq chi}) for $\chi_{lm\omega}$ we then find that $X_{lm\omega}$ satisfies
\begin{equation} \label{eq:diff. eq. for X}
\ddiff{X_{lm\omega}}{r_*}-\mathcal{F}\diff{X_{lm\omega}}{r_*}-{}_{\indhel}\mathcal{U}X_{lm\omega}=0
\end{equation}
with
\begin{subequations}
\begin{align}
\mathcal{F} &\equiv \frac{\Delta F}{(r^2+a^2)} \\
G & \equiv \frac{s\Delta'}{2(r^2+a^2)}+\frac{r\Delta}{(r^2+a^2)^2} \\
{}_{\indhel}\mathcal{U} &\equiv \frac{\Delta {}_{\indhel}U}{(r^2+a^2)^2}+G^2+\diff{G}{r_*}-\frac{\Delta FG}{(r^2+a^2)}
\end{align}
\end{subequations}
In order to obtain now Detweiler's derivation as a particular case of the above, we impose that the differential equation 
(\ref{eq:diff eq chi}) satisfied by $\chi_{lm\omega}$ is of the same form as the radial Teukolsky equation 
(\ref{eq:radial teuk. eq.}).
Detweiler makes this requirement as a starting point. 
That is, we require that $F=0$, and therefore, from (\ref{eq:def. F,S&N'82}), that $\gamma=const\equiv \kappa$. 
The other requirement Detweiler makes is for the potential ${}_{\indhel}U$ to be real.
This requirement implies, using
equations (\ref{eq:radial teuk. eq.}), (\ref{eq:chi as func of Rs and R_s}) and (\ref{eq:diff eq chi}), 
that the constant $\kappa$ and the functions $\alpha$ and $\beta$ (or equivalently 
$p$ and $q$ via (\ref{eq:def alpha,beta}) where $a_D$ and $b_D$ are assumed to be known, as
we do for the spin-1 case) must satisfy the equations
\begin{subequations} \label{eq:eqs. 33,46,50,51Detw76}
\begin{align}
\Delta^{2\indhel}\kappa\kappa^*&=a_D^2-a'_Db_D\Delta^{\indhel+1}+a_Db'_D\Delta^{\indhel+1}-b_D^2\Delta^{2\indhel+1}{}_{\indhel}V  \\
\kappa^*q&=\Delta^{-\indhel }p^* \label{eq:eq50Detw76} \\
\kappa\kappa^*&=\kappa^*p^2+\kappa p^{*2}+(a_D+a_D^*)\Delta^{-\indhel }pp^*+b_D\Delta(pp^{*'}-p'p^*)  \label{eq:eq51Detw76}  
\end{align}
\end{subequations}
The simplest choices for $\kappa$ and $p$ are made by assuming that they are real. 
From equations (\ref{eq:eqs. 33,46,50,51Detw76}), where we now specialise to the $\indhel =-1$ case 
and therefore have $a_D$ and $b_D$ given by (\ref{eq:def a_D,b_D}), it follows that
\begin{subequations} 
\begin{align}
\kappa&=\kappa^*=\left(4{}_{-1}\lambda_{lm\omega}^2-16a^2\omega^2+16a\omega m\right)^{1/2}=2{}_1B_{lm\omega} \label{eq:val. for kappa} \\
p&=\frac{\kappa}{\sqrt{2}}\left(\frac{4K^2}{\Delta}-2{}_{-1}\lambda_{lm\omega}+\kappa\right)^{-1/2} \label{eq:val. for p} \\
q&=\frac{p\Delta}{\kappa}  \label{eq:val. for q} 
\end{align}
\end{subequations}
Detweiler ~\cite{ar:Detw'76} shows that the term inside the square root
in (\ref{eq:val. for p}) is strictly positive as long as
\begin{equation}
{}_{-1}\lambda_{lm\omega}-a^2\omega^2+2a\omega m<\frac{\sqrt{5}}{4} \qquad \text{for} \qquad a\omega >-\frac{1}{4}
\end{equation}
The numerical results in ~\cite{ar:Teuk&Press'74} show that this condition necessarily holds.
\catdraft{com pot ser que es demostri que es compleix per a TOT $c>-1/4$?}

The transformation of the radial function from ${}_{-1}R_{lm\omega}$ to $X_{lm\omega}$ is now fully determined 
via equations (\ref{eq:chi as func of Rs and Rs'}) and (\ref{eq:def. of X}), 
since we know $\alpha$ and $\beta$ from (\ref{eq:def alpha,beta})
and $p$ and $q$ from (\ref{eq:val. for p}) and (\ref{eq:val. for q}) respectively.

From equations (\ref{eq:radial teuk. potential}), (\ref{eq:def alpha,beta}),
(\ref{eq:eq50Detw76}) and (\ref{eq:sU in terms of sV}) 
where now $F=0$, we can obtain the potential in the differential equation for $\chi_{lm\omega}$:
\begin{equation} \label{eq:val. for sU}
{}_{-1}U(r)=-\frac{K^2}{\Delta}+{}_{-1}\lambda_{lm\omega}+\frac{\Delta(Kp')'}{Kp}
\end{equation}
Finally, with (\ref{eq:val. for sU}) and $F=0$ the differential equation (\ref{eq:diff. eq. for X}) for the dependent variable
$X_{lm\omega}$ becomes
\begin{equation} \label{eq:diff. eq. for X,s=-1}
\ddiff{X_{lm\omega}}{r_*}-{}_{-1}\mathcal{U}(r)X_{lm\omega}=0
\end{equation}
with
\begin{equation} \label{eq:potential -1mathcalU}
\begin{aligned}
{}_{-1}\mathcal{U}&=\frac{\Delta{}_{-1}\lambda_{lm\omega}-K^2}{(r^2+a^2)^2}+\frac{\Delta^2(Kp')'}{(r^2+a^2)^2Kp}-
\left(\frac{\Delta}{r^2+a^2}\right)^{3/2}\left[\frac{\Delta^{1/2}}{(r^2+a^2)^{1/2}}\right]''=  \\
&=\frac{\left[-\omega (r^2+a^2)+am\right]^2}{(r^2+a^2)^2}+
\frac{\Delta{}_{-1}\lambda_{lm\omega}}{(r^2+a^2)^2}-
\frac{\Delta(\Delta r^2+4Ma^2r-Q^2(a^2-r^2))}{(r^2+a^2)^4}-                                     \\
&-\frac{\Delta\left[\Delta(10r^2+2\nu^2)-(r^2+\nu^2)(11r^2-10rM+\nu^2)\right]}{(r^2+a^2)^2\left[(r^2+\nu^2)^2+\eta\Delta\right]}+ \\
&+\frac{12\Delta r(r^2+\nu^2)^2\left[\Delta r-(r^2+\nu^2)(r-M)\right]}{(r^2+a^2)^2\left[(r^2+\nu^2)^2+\eta\Delta\right]^2}
-\frac{\Delta(r-M)^2\eta \left[2(r^2+\nu^2)^2-\eta\Delta \right]}{(r^2+a^2)^2\left[(r^2+\nu^2)^2+\eta\Delta\right]^2}
\end{aligned}
\end{equation}
where 
\begin{equation}
\begin{aligned}
\nu^2 &\equiv a^2-am/\omega  \\
\eta&\equiv \frac{\kappa-2{}_{-1}\lambda_{lm\omega}}{4\omega^2}
\end{aligned}
\end{equation}

Sasaki and Nakamura ~\cite{ar:Sasa&Naka'82} derive the 
differential equation (\ref{eq:diff. eq. for X}) for $X_{lm\omega}$ but only for the case $\indhel =-2$. They require 
(\ref{eq:diff. eq. for X}) for $\indhel =-2$ to be short-range and to reduce to the
Regge-Wheeler equation in the limit $a \rightarrow 0$. They show that the transformation
\begin{equation} \label{eq:chi as func of Rs,Rs',Rs'', S&N'82}
\chi_{lm\omega}=\frac{f\Delta(r^2+a^2)}{gj}\mathcal{D}_0\left[j\mathcal{D}_0\left(\frac{g{}_{\indhel}R_{lm\omega}}{r^2+a^2}\right)\right]
\end{equation}
where $f$, $g$ and $j$ are undetermined functions of $r$ guarantees a short-range potential as long as $f$, $g$ and $j$
are regular functions with no zero-points and $(1)\ f=const+O(r^{-1})$, $(2)\ g=const+O(r^{-2})$, $(3)\ h=const+O(r^{-2})$
for $r \rightarrow +\infty$ and $(4)$ they all are $O(1)$ for $r \rightarrow r_+$. The differential
equation (\ref{eq:diff. eq. for X}) for $\indhel =-2$ becomes the Regge-Wheeler equation for $a=0$ if $f$, $g$ and $j$ 
are constant in that case.

Even though Sasaki and Nakamura's derivation is purely limited to the case of gravitational perturbations, 
a similar transformation for the spin-1 case could be found. Such a transformation
would possibly deliver a short-range source in the inhomogeneous case, but in the homogeneous case in principle
it would not have any advantage over Detweiler's derivation. 
Nevertheless, we still
tried to obtain a similar transformation to (\ref{eq:chi as func of Rs,Rs',Rs'', S&N'82}) such that for 
$\indhel =-1$ the new, radial differential equation is short-range, in case it turned out to be simpler than the one, 
(\ref{eq:diff. eq. for X,s=-1}), given by Detweiler. The generalized version of (\ref{eq:chi as func of Rs,Rs',Rs'', S&N'82})
we used is
\begin{equation} \label{eq:gralized. chi as func of Rs,Rs',Rs'', S&N'82}
\chi_{lm\omega}=\frac{f\Delta^n(r^2+a^2)^p}{gj}\mathcal{D}_0\left[j\mathcal{D}_0\left(\frac{g{}_{\indhel}R_{lm\omega}}{(r^2+a^2)^q}\right)\right]
\end{equation}
but there was no set of values $\{n,p,q\}$ and set of functions $\{f,g,j\}$ such that the resulting
equation (\ref{eq:diff. eq. for X}) for $X_{lm\omega}$ in the case $\indhel =-1$ is short-range.

\catdraft{curious: we don't calculate the deriv. of ${}_{+1}R$ at any stage...}

We therefore decided to follow Detweiler's derivation and solve numerically the differential
equation (\ref{eq:diff. eq. for X,s=-1}). 
We thus find the radial function ${}_{-1}R_{lm\omega}$ and its derivative from the solution $X_{lm\omega}$ and its derivative
with (\ref{eq:Rs,Rs' as funcs. of chi and chi'}). 
The radial function ${}_{+1}R_{lm\omega}$ can then be obtained with (\ref{eq:R_s as func of Rs and Rs'}).
The first term in the potential (\ref{eq:potential -1mathcalU}) tends to $-\omega^2$ at infinity ($r\to +\infty$) and to 
$-\tilde{\omega}^2$ at the horizon ($r\to r_+$), whereas all the other terms go as $O(r^{-2})$ at infinity 
and vanish at the horizon.
We can therefore define two sets of solutions with the following asymptotic behaviours:
\begin{subequations} \label{eq:X_in/up}
\begin{align}  
X^{\text{in}}_{lm\omega} & \sim
\begin{cases} 
B^{\text{in}}_{lm\omega}e^{-i\tilde{\omega}r_*} & (r\rightarrow r_+) \\
e^{-i\omega r_*}+A^{\text{in}}_{lm\omega}e^{+i\omega r_*} & (r\rightarrow +\infty)
\end{cases} \label{eq:X_in}
\\
X^{\text{up}}_{lm\omega} & \sim
\begin{cases}
e^{+i\tilde{\omega}r_*}+A^{\text{up}}_{lm\omega}e^{-i\tilde{\omega}r_*} & 
(r\rightarrow r_+) \\
B^{\text{up}}_{lm\omega}e^{+i\omega r_*} & (r\rightarrow +\infty)
\end{cases} \label{eq:X_up}
\end{align}
\end{subequations}
When the behaviour of the solution modes in terms of the time $t$ and the angle $\phi$
is included, we can find the asymptotic behaviour of the solution modes in terms
of the advanced and retarded time co-ordinates:
\begin{subequations} \label{eq:X_in/up as func. of u,v}
\begin{align}  
X^{\text{in}}_{lm\omega}e^{+im\phi-i\omega t} & \sim
\begin{cases} 
e^{-i\omega v+im\phi} & \text{at}\ \mathcal{I}^- \\
A^{\text{in}}_{lm\omega}e^{-i\omega u+im\phi} & \text{at}\ \mathcal{I}^+ \\
0 & \text{at}\ \mathcal{H}^- \\
B^{\text{in}}_{lm\omega}e^{-i\tilde{\omega}v+im\phi_+} & \text{at}\ \mathcal{H}^+
\end{cases} \label{eq:X_in as func. of u,v}
\\
X^{\text{up}}_{lm\omega}e^{+im\phi-i\omega t} & \sim
\begin{cases} 
0 & \text{at}\ \mathcal{I}^- \\
B^{\text{up}}_{lm\omega}e^{-i\omega u+im\phi} & \text{at}\ \mathcal{I}^+ \\
e^{-i\tilde{\omega}u+im\phi_+} & \text{at}\ \mathcal{H}^- \\
A^{\text{up}}_{lm\omega}e^{-i\tilde{\omega}v+im\phi_+} & \text{at}\ \mathcal{H}^+
\end{cases} \label{eq:X_up as func. of u,v}
\end{align}
\end{subequations}
Equation (\ref{eq:X_in as func. of u,v}) represents a wave emerging from $\mathcal{I}^-$, being partially
scattered back to $\mathcal{I}^+$ and partially transmitted through to $\mathcal{H}^+$. 
Similarly, (\ref{eq:X_up as func. of u,v}) represents a wave emerging from $\mathcal{H}^-$, being partially
scattered back to $\mathcal{H}^+$ and partially transmitted through to $\mathcal{I}^+$. 

\catdraft{should this interpretation not depend on the signs of $\omega$ and $\tilde{\omega}$?Adrian:no it's $u$,$v$ that matters}

Both sets of modes are eigenfunctions of the hamiltonians $\hat{H}_{\vec{\xi}}$ and $\hat{H}_{\vec{\chi}}$ with 
eigenvalues $\omega$ and $\tilde{\omega}$ respectively:
\begin{equation} \label{eq:hamiltonians on in/up modes}
\begin{aligned}
\hat{H}_{\vec{\xi}}X^{\bullet}_{lm\omega}e^{+im\phi-i\omega t} &=\omega X^{\bullet}_{lm\omega}e^{+im\phi-i\omega t}\\
\hat{H}_{\vec{\chi}}X^{\bullet}_{lm\omega}e^{+im\phi-i\omega t} &=\tilde{\omega}X^{\bullet}_{lm\omega}e^{+im\phi-i\omega t}
\end{aligned}
\end{equation}
where the symbol $\bullet$ indicates either `in' or `up'.
We will restrict the definition of the `in' and `up' modes to those modes with positive $\omega$ and 
positive $\tilde{\omega}$ respectively. 
It then follows that the `in' and `up' modes are positive frequency with respect to 
the Killing vectors $\vec{\xi}$ and $\vec{\chi}$ respectively.

Analogously, it is possible to find the asymptotic behaviour of solutions ${}_{\indhel}R^{\text{in/up}}_{lm\omega}$ 
$\forall h=0, \pm 1/2, \pm 1, \pm3/2, \pm2$ of the radial Teukolsky equation:
\begin{subequations} \label{eq:R_in/up}
\begin{align}
{}_{\indhel}R^{\text{in}}_{lm\omega} & \sim
\begin{cases} \label{eq:R_in}
{}_{\indhel}R^{\text{in,tra}}_{lm\omega}\Delta^{-\indhel }e^{-i\tilde{\omega}r_*} & (r\rightarrow r_+) \\
{}_{\indhel}R^{\text{in,inc}}_{lm\omega}r^{-1}e^{-i\omega r_*}+{}_{\indhel}R^{\text{in,ref}}_{lm\omega}r^{-1-2\indhel}e^{+i\omega r_*} & 
(r\rightarrow +\infty)
\end{cases}
\\
{}_{\indhel}R^{\text{up}}_{lm\omega} & \sim
\begin{cases}\label{eq:R_up}
{}_{\indhel}R^{\text{up,inc}}_{lm\omega}e^{+i\tilde{\omega}r_*}+{}_{\indhel}R^{\text{up,ref}}_{lm\omega}\Delta^{-\indhel }e^{-i\tilde{\omega}r_*} & 
\qquad (r\rightarrow r_+) \\
{}_{\indhel}R^{\text{up,tra}}_{lm\omega}r^{-1-2\indhel}e^{+i\omega r_*} & 
\qquad (r\rightarrow +\infty)
\end{cases}
\end{align}
\end{subequations}

From equations (\ref{eq:Rs,Rs' as funcs. of chi and chi'}), (\ref{eq:def. of X}), (\ref{eq:X_in}) and (\ref{eq:X_up})
\catdraft{em sembla que em cal expressio per X no a 1r ordre en expansio asympt. sino a algun ordre mes=>incloure'ls!!}
we can find the asymptotic coefficients of ${}_{-1}R^{\bullet}_{lm\omega}$ from those of $X^{\bullet}_{lm\omega}$:
\begin{equation} \label{eq:R_1 coeffs from X's}
\begin{aligned}
\frac{{}_{-1}R^{\text{in,ref}}_{lm\omega}}{{}_{-1}R^{\text{in,inc}}_{lm\omega}A^{\text{in}}_{lm\omega}}&=\frac{4\omega^2}{{}_1B_{lm\omega}} &\qquad\quad
\frac{{}_{-1}R^{\text{in,tra}}_{lm\omega}}{{}_{-1}R^{\text{in,inc}}_{lm\omega}B^{\text{in}}_{lm\omega}}&=
\frac{-sgn(\tilde{\omega})|\omega|i}{(r_+^2+a^2)^{1/2}\EuFrak{N}^*} \\
\frac{{}_{-1}R^{\text{up,ref}}_{lm\omega}}{{}_{-1}R^{\text{up,inc}}_{lm\omega}A^{\text{up}}_{lm\omega}}&=\frac{-i{}_1B_{lm\omega}}{4K_+\EuFrak{N}^*} &\quad
\frac{{}_{-1}R^{\text{up,tra}}_{lm\omega}}{{}_{-1}R^{\text{up,inc}}_{lm\omega}B^{\text{up}}_{lm\omega}}&=\frac{|\omega|(r_+^2+a^2)^{1/2}}{|K_+|} \\
{}_{-1}R^{\text{in,inc}}_{lm\omega}&=\frac{1}{2^{3/2}|\omega|} &\quad {}_{-1}R^{\text{up,inc}}_{lm\omega}&=\frac{-2^{1/2}(r_+^2+a^2)^{1/2}\tilde{\omega}}{{}_1B_{lm\omega}}
\end{aligned}
\end{equation}
where $K_+\equiv K(r_+)$ and we have also defined the new variable
\begin{equation} \label{eq: def. EuFrak{N}_s}
\EuFrak{N} \equiv iK_++\frac{(r_+-r_-)}{2}  
\end{equation}

In the calculation of ${}_{-1}R^{\text{in,tra}}_{lm\omega}/{}_{-1}R^{\text{in,inc}}_{lm\omega}$ in (\ref{eq:R_1 coeffs from X's}) 
we needed an extra term in the asymptotic
expansion of the ingoing part (the outgoing part is simply obtained by complex conjugation since the potential ${}_{-1}\mathcal{U}$ is real)
of $X_{lm\omega}$ close to the horizon. By introducing the asymptotic expansion
\begin{equation} \label{eq:asympt. for r=r_+ of X_in}
\frac{X^{\text{in}}_{lm\omega}}{B^{\text{in}}_{lm\omega}}=\left[1+\alpha_1(r-r_+)+O((r-r_+)^2)\right]e^{-i\tilde{\omega}r_*}
\end{equation}
in the differential equation (\ref{eq:diff. eq. for X,s=-1}) and performing a Taylor series expansion around $r_+$ of the 
potential (\ref{eq:potential -1mathcalU}), we find from the second order term that
\begin{equation} \label{eq:val. of alpha_1}
\begin{aligned}
&\alpha_1=\\
&=\frac{-1}{2\EuFrak{N}^*}\left[{}_{-1}\lambda_{lm\omega}-\frac{4Ma^2r_+-Q^2(a^2-r_+^2)}{(r_+^2+a^2)^2}+\frac{a^2+Q^2}{r_+^2+\nu^2}-
\frac{4amr_+\tilde{\omega}}{r_+-r_-}-\frac{2(r_+-M)^2\eta}{(r_+^2+\nu^2)^2}\right]
\end{aligned}
\end{equation}
In the calculation of ${}_{-1}R^{\text{in,inc}}_{lm\omega}$, an extra term is also needed in the asymptotic
expansion of the ingoing part of $X_{lm\omega}$ for large $r$:
\begin{equation}
X_{lm\omega}=\left[1+\frac{\beta_1}{r}+O(r^{-2})\right]e^{-i\omega r_*}
\end{equation}
with
\begin{equation}
\beta_1=-\frac{({}_{-1}\lambda_{lm\omega}+2a\omega m)i}{2\omega}
\end{equation}

After obtaining the asymptotic coefficients of ${}_{-1}R^{\bullet}_{lm\omega}$ from those of $X^{\bullet}_{lm\omega}$,
we just need to derive those of ${}_{+1}R^{\bullet}_{lm\omega}$ to complete the asymptotic picture of the solutions to the
radial Teukolsky equation for spin-1. This is achieved by using the transformation
(\ref{eq:R_s as func of Rs and Rs'}) together with the asymptotic behaviour in (\ref{eq:R_in/up}):
\begin{equation} \label{eq:R1 coeffs from R_1's}
\begin{aligned}
\frac{{}_{+1}R^{\text{in,inc}}_{lm\omega}}{{}_{-1}R^{\text{in,inc}}}&=-2^3\omega^2; & \qquad
\frac{{}_{+1}R^{\text{in,ref}}_{lm\omega}}{{}_{-1}R^{\text{in,ref}}_{lm\omega}}&=
\frac{{}_{+1}R^{\text{up,tra}}_{lm\omega}}{{}_{-1}R^{\text{up,tra}}_{lm\omega}}=-\frac{{}_1B_{lm\omega}^2}{2\omega^2}     \\
\frac{{}_{+1}R^{\text{in,tra}}_{lm\omega}}{{}_{-1}R^{\text{in,tra}}_{lm\omega}}&=
\frac{{}_{+1}R^{\text{up,ref}}_{lm\omega}}{{}_{-1}R^{\text{up,ref}}_{lm\omega}}=-2^3K_+ \EuFrak{N}^*i; &
\frac{{}_{+1}R^{\text{up,inc}}_{lm\omega}}{{}_{-1}R^{\text{up,inc}}_{lm\omega}}&=\frac{-i{}_1B_{lm\omega}^2}{2K_+\EuFrak{N}} 
\end{aligned}
\end{equation}

As mentioned in Section \ref{sec:eq. for phi_0 in Kerr}, the differential equation (\ref{eq:eq. for rho^-1phi0}) 
for $\rho^{-1}\phi_0$ is separable when $a=0$. We can therefore write 
\begin{equation} \label{eq:phi0 in R-N}
r{}_{lm\omega}\phi_0={}_0R_{lm\omega}(r){}_0S_{lm\omega}(\theta)e^{-i\omega t}e^{+im\phi}
\end{equation}
in the Reissner-Nordstr\"{o}m background where ${}_0R_{lm\omega}(r)$ and ${}_0S_{lm\omega}(\theta)$ are, respectively,
solutions of the radial and angular differential equations resulting from such separation.
The differential equation for $r{}_0R_{lm\omega}$ coincides with the 
differential equation (\ref{eq:diff. eq. for X,s=-1}) for $X_{lm\omega}$ with $a=0$.
That is, $r{}_0R_{lm\omega}=X_{lm\omega}$ when $a=0$. 
In the Reissner-Nordstr\"{o}m background 
not only ${}_{lm\omega}\phi_0$ has the neat radial funcionality of (\ref{eq:phi0 in R-N}),
but also the expressions for ${}_{\pm1}R_{lm\omega}$ in terms of $r{}_0R_{lm\omega}$ and its derivative are very simple ones.
It therefore seems reasonable to hope that in the Kerr-Newman background the expressions for ${}_{\pm 1}R_{lm\omega}$ 
in terms of $X_{lm\omega}$ and its derivative are also simple ones. 
More importantly, one would also hope that the expression for ${}_{lm\omega}\phi_0$ in terms of $X_{lm\omega}$ 
and its derivative is simpler than in terms of ${}_{\pm 1}R_{lm\omega}$ and its derivative (see (\ref{eq:phi0(ch)})). 
Unfortunately, the expressions we obtained for ${}_{\pm 1}R_{lm\omega}$ and 
${}_{lm\omega}\phi_0$
in terms of $X_{lm\omega}$ and its derivative 
are actually much more complicated than the ones we already have for ${}_{+1}R_{lm\omega}$ and ${}_{lm\omega}\phi_0$
in terms of ${}_{-1}R_{lm\omega}$ and its derivative.
\catdraft{no estic posant p.6F(IV)->(XVI) pq. no crec que sigui prou interessant}

It is clear from their asymptotic behaviour that neither ${}_{\indhel}R^{\text{in}}_{lm\omega}$ nor 
${}_{\indhel}R^{\text{up}}_{lm\omega}$ satisfy the symmetry (\ref{eq:R symm.->cc,-s}). 
As a matter of fact, we shall now show that, under this symmetry, the functions ${}_{\indhel}R^{\bullet}_{lm\omega}$
transform to the radial funcions that are derived from the solution $X_{lm\omega}^{\bullet *}$ and its derivative.
We construct a new radial function ${}_{-1}\bar{R}_{lm\omega}^{\bullet}$ 
derived from $X_{lm\omega}^{\bullet *}$ in the same manner that ${}_{-1}R_{lm\omega}^{\bullet}$ is derived from
$X_{lm\omega}^{\bullet}$:
\begin{equation} \label{eq:bar{R}_1}
\kappa {}_{-1}\bar{R}_{lm\omega}=
\left(\alpha+\beta'\right)\left[\Delta^{1/2}(r^2+a^2)^{-1/2}X_{lm\omega}^* \right]-\beta\diff{}{r}\left[\Delta^{1/2}(r^2+a^2)^{-1/2}X_{lm\omega}^* \right]
\end{equation}
It can then be checked using equations (\ref{eq:R_1 coeffs from X's}) and (\ref{eq:R1 coeffs from R_1's}) that the symmetries
\begin{equation}
\begin{aligned}
{}_{+1}R^{\bullet}_{lm\omega}&=2{}_1B_{lm\omega}\Delta^{-1}{}_{-1}\bar{R}^{\bullet *}_{lm\omega} \\
{}_{+1}\bar{R}^{\bullet}_{lm\omega}&=2{}_1B_{lm\omega}\Delta^{-1}{}_{-1}R^{\bullet *}_{lm\omega} 
\end{aligned}
\end{equation}
are satisfied, where ${}_{+1}\bar{R}^{\bullet}_{lm\omega}$ 
is calculated by applying the operator in the radial equation (\ref{eq:a)Teuk-Starob. ids.}) to
${}_{-1}\bar{R}^{\bullet}_{lm\omega}$.
Renaming ${}_{\pm 1}\bar{R}^{\text{in}}_{lm\omega}$ and ${}_{\pm 1}\bar{R}^{\text{up}}_{lm\omega}$
by ${}_{\pm 1}R^{\text{out}}_{lm\omega}$ and ${}_{\pm 1}R^{\text{down}}_{lm\omega}$
respectively, we have the following two sets of modes:
\begin{subequations} \label{eq:R_out/down}
\begin{align}
{}_{\pm 1}R^{\text{out}}_{lm\omega}&\equiv (2{}_1B_{lm\omega})^{\pm 1}\Delta^{\mp 1}{}_{\mp 1}R^{\text{in} *}_{lm\omega} 
 \sim \\ & \sim
\begin{cases} \label{eq:R_out}
{}_{\pm 1}R^{\text{out,tra}}_{lm\omega}e^{+i\tilde{\omega}r_*} & (r\rightarrow r_+) \\
{}_{\pm 1}R^{\text{out,inc}}_{lm\omega}r^{-1\mp 2}e^{+i\omega r_*}+
{}_{\pm 1}R^{\text{out,ref}}_{lm\omega}r^{-1}e^{-i\omega r_*} & 
(r\rightarrow +\infty)
\end{cases}
\\
{}_{\pm 1}R^{\text{down}}_{lm\omega}&\equiv (2{}_1B_{lm\omega})^{\pm 1}\Delta^{\mp 1}{}_{\mp 1}R^{\text{up} *}_{lm\omega} 
 \sim \\ & \sim
\begin{cases}\label{eq:R_down}
{}_{\pm 1}R^{\text{down,inc}}_{lm\omega}\Delta^{\mp 1}e^{-i\tilde{\omega}r_*}+
{}_{\pm 1}R^{\text{down,ref}}_{lm\omega}e^{+i\tilde{\omega}r_*} & 
(r\rightarrow r_+) \\
{}_{\pm 1}R^{\text{down,tra}}_{lm\omega}r^{-1}e^{-i\omega r_*} & (r\rightarrow +\infty)
\end{cases}
\end{align}
\end{subequations}
where
\begin{equation} \label{eq:def. coeffs. Rs_out/down}
\begin{aligned}
{}_{\pm 1}R^{\text{out,inc}}_{lm\omega}&\equiv (2{}_1B_{lm\omega})^{\pm 1}{}_{\mp 1}R^{\text{in,inc} *}_{lm\omega}, &
{}_{\pm 1}R^{\text{down,inc}}_{lm\omega}&\equiv (2{}_1B_{lm\omega})^{\pm 1}{}_{\mp 1}R^{\text{up,inc} *}_{lm\omega} \\
{}_{\pm 1}R^{\text{out,ref}}_{lm\omega}&\equiv (2{}_1B_{lm\omega})^{\pm 1}{}_{\mp 1}R^{\text{in,ref} *}_{lm\omega}, &
{}_{\pm 1}R^{\text{down,ref}}_{lm\omega}&\equiv (2{}_1B_{lm\omega})^{\pm 1}{}_{\mp 1}R^{\text{up,ref} *}_{lm\omega} \\
{}_{\pm 1}R^{\text{out,tra}}_{lm\omega}&\equiv (2{}_1B_{lm\omega})^{\pm 1}{}_{\mp 1}R^{\text{in,tra} *}_{lm\omega}, &
{}_{\pm 1}R^{\text{down,tra}}_{lm\omega}&\equiv (2{}_1B_{lm\omega})^{\pm 1}{}_{\mp 1}R^{\text{up,tra} *}_{lm\omega} 
\end{aligned}
\end{equation}
Note that the factor $(2{}_1B_{lm\omega})^{\pm 1}$ is needed so that the Teukolsky-Starobinski\u{\i} identities
are satisfied.
Similarly, since the radial modes ${}_{\pm 1}R^{\text{out}}_{lm\omega}$ and ${}_{\pm 1}R^{\text{down}}_{lm\omega}$
are obtained from $X^{\text{in} *}_{lm\omega}$ and $X^{\text{up} *}_{lm\omega}$ respectively,
we rename the latter as
\begin{subequations} \label{eq:X_out/down}
\begin{align}  
X^{\text{out}}_{lm\omega}\equiv X^{\text{in} *}_{lm\omega}   & \sim
\begin{cases} 
B^{\text{out}}_{lm\omega}e^{+i\tilde{\omega}r_*} & (r\rightarrow r_+) \\
e^{+i\omega r_*}+A^{\text{out}}_{lm\omega}e^{-i\omega r_*} & (r\rightarrow +\infty)
\end{cases} \label{eq:X_out}
\\
X^{\text{down}}_{lm\omega}\equiv X^{\text{up} *}_{lm\omega} & \sim
\begin{cases}
e^{-i\tilde{\omega}r_*}+A^{\text{down}}_{lm\omega}e^{+i\tilde{\omega}r_*} & 
(r\rightarrow r_+) \\
B^{\text{down}}_{lm\omega}e^{-i\omega r_*} & (r\rightarrow +\infty)
\end{cases} \label{eq:X_down}
\end{align}
\end{subequations}
with
\begin{equation} \label{eq:def. coeffs. X_out/down}
\begin{aligned}
A^{\text{out}}_{lm\omega}&\equiv A^{\text{in} *}_{lm\omega}, & \qquad \qquad 
A^{\text{down}}_{lm\omega}&\equiv A^{\text{up} *}_{lm\omega} \\
B^{\text{out}}_{lm\omega}&\equiv B^{\text{in} *}_{lm\omega}, &
B^{\text{down}}_{lm\omega}&\equiv B^{\text{up} *}_{lm\omega} &
\end{aligned}
\end{equation}
We follow the same positive-frequency convention for the `out' and `down' modes
as that for the `in' and `up' modes respectively, namely, their definition is restricted
to modes with positive $\omega$ and positive $\tilde{\omega}$ respectively.
The asymptotic behaviour in terms of the advanced and retarded time co-ordinates of these
two new sets of functions is
\begin{subequations} \label{eq:X_out/down as func. of u,v}
\begin{align}  
X^{\text{out}}_{lm\omega}e^{+im\phi-i\omega t} & \sim
\begin{cases} 
A^{\text{out}}_{lm\omega}e^{-i\omega v+im\phi} & \text{at}\ \mathcal{I}^- \\
e^{-i\omega u+im\phi} & \text{at}\ \mathcal{I}^+ \\
B^{\text{out}}_{lm\omega}e^{-i\tilde{\omega}u+im\phi_+} & \text{at}\ \mathcal{H}^- \\
0 & \text{at}\ \mathcal{H}^+
\end{cases} \label{eq:X_out as func. of u,v}
\\
X^{\text{down}}_{lm\omega}e^{+im\phi-i\omega t} & \sim
\begin{cases} 
B^{\text{down}}_{lm\omega}e^{-i\omega v+im\phi} & \text{at}\ \mathcal{I}^- \\
0 & \text{at}\ \mathcal{I}^+ \\
A^{\text{down}}_{lm\omega}e^{-i\tilde{\omega}u+im\phi_+} & \text{at}\ \mathcal{H}^- \\
e^{-i\tilde{\omega}v+im\phi_+} & \text{at}\ \mathcal{H}^+
\end{cases} \label{eq:X_down as func. of u,v}
\end{align}
\end{subequations}
Modes (\ref{eq:X_out as func. of u,v}) describe a wave going out to $\mathcal{I}^+$
whereas modes (\ref{eq:X_down as func. of u,v}) describe a wave going down $\mathcal{H}^+$. 

It is easy to check that the `out' NP scalars are precisely the ones that relate to the `in' NP scalars 
under the symmetry transformation $(t,\phi)\to (-t,-\phi)$ as in (\ref{NP scalars under t,phi->-t,-phi}).
Likewise for the `down' NP scalars with respect to the `up' NP scalars.
Indeed, using the radial symmetry (\ref{eq:R symm.->cc,-m,-w}) and the angular symmetry (\ref{eq:S symm.->-s,-m,-w}),
which we shall see in the next chapter, it immediately follows that
\begin{equation} \label{NP scalars in/up->out/down}
\left.
\begin{array}{ll}
{}_{lm\omega}\phi_{-1}^{\text{in/up}}&\to (-1)^{m+1}\Delta^{-1}\rho^{-2}{}_{l-m-\omega}\phi_{+1}^{\text{out/down}} \\
{}_{lm\omega}\phi_{0}^{\text{in/up}}&\to -(-1)^{m+1}{}_{l-m-\omega}\phi_{0}^{\text{out/down}}  \\
{}_{lm\omega}\phi_{+1}^{\text{in/up}}&\to (-1)^{m+1}\Delta\rho^2{}_{l-m-\omega}\phi_{-1}^{\text{out/down}}  
\end{array}
\right\} \text{under} \quad (t,\phi)\to (-t,-\phi)
\end{equation}
\draft{must check factor 2 and sign for $\phi_0$}

We can also find the pair of radial functions ${}_{\pm1}R^{\text{sym}}_{lm\omega}$ that satisfies the
symmetry (\ref{eq:R symm.->cc,-s}). Imposing the condition
\begin{equation} \label{eq:cond. for R symm.}
C\Delta^{-1}{}_{-1}R^{\text{sym} *}_{lm\omega}={}_{+1}R^{\text{sym}}_{lm\omega}(=2\mathcal{D}_0\mathcal{D}_0 {}_{-1}R^{\text{sym}}_{lm\omega})
\end{equation}
where $C$ is a factor of proportionality, 
we find that this new, symmetric radial function can be expressed in terms of ${}_{-1}R^{\text{in}}_{lm\omega}$ and ${}_{-1}R^{\text{up}}_{lm\omega}$ as
\begin{equation} \label{eq:R_1(sym) en func. de R_1(up/in)}
\begin{aligned}
{}_{-1}R^{\text{sym}}_{lm\omega}&=
\alpha\left[
A^{\text{sym}}
{}_{-1}R^{\text{in}}_{lm\omega}+{}_{-1}R^{\text{up}}_{lm\omega}
\right]   \\
A^{\text{sym}}&\equiv 
\frac{1}{{}_{-1}R^{\text{in,tra}}_{lm\omega}}
\left[\frac{iC}{8K_+\EuFrak{N}^*}\left(\frac{\alpha^*}{\alpha}\right)-{}_{-1}R^{\text{up,ref}}_{lm\omega}\right]
\end{aligned}
\end{equation}
with $|C|=2{}_1B_{lm\omega}$ and where $\alpha$ is an arbitrary complex number. 
Another consequence of imposing the symmetry (\ref{eq:cond. for R symm.}) are the following
new relations:
\begin{subequations}
\begin{align}
&{}_{+1}R^{\text{in}}_{lm\omega}=\frac{8iK_+\EuFrak{N}^*{}_{-1}R^{\text{in,tra}}_{lm\omega}}{\Delta}
\left[\frac{{}_{-1}R^{\text{up,ref} *}_{lm\omega}}{{}_{-1}R^{\text{in,tra} *}_{lm\omega}}{}_{-1}R^{\text{in} *}_{lm\omega}-
{}_{-1}R^{\text{up} *}_{lm\omega}\right]  \\
\begin{split}
&{}_{+1}R^{\text{up}}_{lm\omega}=\\
&=\frac{8iK_+\EuFrak{N}^*{}_{-1}R^{\text{up,ref}}_{lm\omega}}{\Delta}
\left[\left(1-\frac{2^4\omega^4}{{}_1B_{lm\omega}^2\left|{}_{-1}R^{\text{in,ref}}_{lm\omega}\right|^2}\right)
\frac{{}_{-1}R^{\text{up,ref} *}_{lm\omega}}{{}_{-1}R^{\text{in,tra} *}_{lm\omega}}{}_{-1}R^{\text{in} *}_{lm\omega}-
{}_{-1}R^{\text{up} *}_{lm\omega}\right]
\end{split}
\end{align}
\end{subequations}
It immediately follows from (\ref{eq:R_1(sym) en func. de R_1(up/in)}) 
that the asymptotic form for ${}_{-1}R^{\text{sym}}_{lm\omega}$ is
\begin{align} 
\frac{{}_{-1}R^{\text{sym}}_{lm\omega}}{\alpha} & =
\begin{cases} \label{eq:asympt. R_1(sym)}
\displaystyle
e^{+i\tilde{\omega}r_*}+
\frac{iC}{8K_+\EuFrak{N}^*}\left(\frac{\alpha^*}{\alpha}\right)
\Delta e^{-i\tilde{\omega}r_*} & (r\rightarrow r_+) \\
\displaystyle
-\frac{4\omega^2}{{}_1B_{lm\omega}}
\left(\frac{\alpha^*}{\alpha}\right)
A^{\text{sym} *}re^{+i\omega r_*}+A^{\text{sym}}\frac{1}{r}e^{-i\omega r_*} & (r\rightarrow +\infty)
\end{cases}
\end{align}


\catdraft{1) explicar pq. descarto finite els. method d'App.DSasa\&Naka'82!!
2)explicar com resulta que Detw's potential es short-range, 3)explicar que vol dir short-range
4)explicar quin metode fa servir Chandr i com es relaciona amb els que explico aqui!}

An alternative transformation of the radial equation is given by Teukolsky and Press ~\cite{ar:Teuk&Press'74}.
They perform the change of variable ${}_{\indhel}Y_{lm\omega}\equiv\Delta^{\indhel/2}(r^2+a^2)^{1/2}{}_{\indhel}R_{lm\omega}$, and the radial equation
transforms to
\begin{equation}
\ddiff{{}_{\indhel}Y_{lm\omega}}{r_*}+V_{TP}\ {}_{\indhel}Y_{lm\omega}=0
\end{equation}
where the potential is
\begin{equation}
\begin{aligned}
V_{TP}&=\frac{K^2-2i\indhel K(r-M)+\Delta(4ir\omega \indhel-{}_{-1}\lambda_{lm\omega})-\indhel ^2(M^2-a^2)}{(r^2+a^2)^2}- \\
&-\frac{\Delta(2Mr^3+a^2r^2-4Mra^2+a^4)}{(r^2+a^2)^4}
\end{aligned}
\end{equation}
The potential is invariant under a change in the sign of the helicity parameter $\indhel $ together with complex conjugation. 
In consequence, the wronskian formed with two solutions ${}_{\indhel}Y_{lm\omega}$ and ${}_{-\indhel }Y_{lm\omega}^*$ is constant, 
where we are using the following definition of wronskian
\begin{equation} \label{eq:def. radial wronsk.}
W[f(r),g(r)]\equiv 
\diff{f(r)}{r_*}g(r)-f(r)\diff{g(r)}{r_*}
\end{equation}
As Detweiler's radial potential ${}_{-1}\mathcal{U}$ in (\ref{eq:potential -1mathcalU}) is real, the wronskian
of a solution $X_{lm\omega}$ and its complex conjugate is also constant. The various possible wronskians for spin-1 constructed with the 
`in' and `up' solutions are shown in Table \ref{table:radial wronsks}. Each one of the wronskians for the solutions ${}_{\indhel}Y_{lm\omega}$ 
can be derived from a particular one of the wronskians for the solutions $X_{lm\omega}$, and viceversa. The correspondence between
the wronskians of the different types of solutions is indicated by the same letter on the right margin 
of the table. It can be easily shown that the wronskians for the solution ${}_{\indhel}Y_{lm\omega}$ can be expressed in terms
of the solutions ${}_{\indhel}R_{lm\omega}$ as
\begin{equation} \label{eq:gral. radial wronsk.,R_+/-1}
W[{}_{+1}Y_{lm\omega},{}_{-1}Y_{lm\omega}^*]= 
{}_{-1}R^*_{lm\omega}\mathcal{D}^{\dagger}_0\left(\Delta{}_{+1}R_{lm\omega}\right)-
\Delta{}_{+1}R_{lm\omega}\mathcal{D}^{\dagger}_0{}_{-1}R^*_{lm\omega}=Ci
\end{equation}
where $C$ is a real constant, and in particular,
\begin{equation} \label{eq:gral. radial wronsk.,R_+/-1 symm.}
W[{}_{+1}Y_{lm\omega}^{\text{sym}},{}_{-1}Y_{lm\omega}^{\text{sym}*}]=0 
\end{equation}
It is also useful to note the following two relations between the wronskians of the `in'
and `up' solutions once the normalization constants, which are determined later in (\ref{eq:normalization consts.}), have been included:
\begin{equation} \label{eq: wronskian in=-up}  
\begin{aligned}
W[{}_{+1}Y_{lm\omega}^{\text{up}},{}_{-1}Y_{lm\omega}^{\text{up} *}]&=-W[{}_{+1}Y_{lm\omega}^{\text{in}},{}_{-1}Y_{lm\omega}^{\text{in} *}]     \\
W[{}_{+1}Y_{lm\omega}^{\bullet},{}_{-1}Y_{lm\omega}^{\bullet *}]&=
+W[{}_{+1}Y_{l-m-\omega}^{\bullet},{}_{-1}Y_{l-m-\omega}^{\bullet *}]
\end{aligned}
\end{equation}

\begin{table} 
\begin{tabular}{|rclc|}
\hline
$r \rightarrow r_+$  & & $r \rightarrow +\infty$ &  \\
\hline
\hline
$2i\tilde{\omega}\left(1-\abs{A^\text{up}}^2\right)$   
&$=W[X^{\text{up}},X^{\text{up} *}]=$& 
$2i\omega \abs{B^{\text{up}}}^2$  & (a) \\
$-2i\tilde{\omega}A^{\text{up} *}B^{\text{in}}$ 
&$=W[X^{\text{in}},X^{\text{up} *}]$=& 
$2i\omega A^{\text{in}}B^{\text{up} *}$  & (b)\\
$2i\tilde{\omega}B^{\text{in}}$ 
&$=W[X^{\text{in}},X^{\text{up}}]=$& 
$2i\omega B^{\text{up}}$  & (c)\\
$-2i\tilde{\omega}\abs{B^{\text{in}}}^2$
&$=W[X^{\text{in}},X^{\text{in} *}]=$& 
$-2i\omega \left(1-\abs{A^{\text{in}}}^2\right)$  & (d)  \\
\hline
$\frac{-i{}_1B^2}{K_+}+2^4K_+\abs{\EuFrak{N}}^2i\abs{{}_{-1}R^{\text{up,ref}}}^2$
&$=W[{}_{+1}Y^{\text{up}},{}_{-1}Y^{\text{up} *}]=$&
$\frac{-i{}_1B^2}{\omega}\abs{{}_{-1}R^{\text{up,tra}}}^2$ & (a) \\
$2^4iK_+\abs{\EuFrak{N}}^2{}_{-1}R^{\text{in,tra}}{}_{-1}R^{\text{up,ref *}}$
&$=W[{}_{+1}Y^{\text{in}},{}_{-1}Y^{\text{up} *}]=$&
$\frac{-i{}_1B^2}{\omega}{}_{-1}R^{\text{up,tra *}}{}_{-1}R^{\text{in,ref}}$ & (b)\\
$2\EuFrak{N}^*{}_{-1}R^{\text{in,tra}}$
&$=W[{}_{-1}Y^{\text{in}},{}_{-1}Y^{\text{up}}]=$&
$-2i\omega {}_{-1}R^{\text{up,tra}}$ & (c)\\
$2^4iK_+\abs{\EuFrak{N}}^2\abs{{}_{-1}R^{\text{in,tra}}}^2$
&$=W[{}_{+1}Y^{\text{in}},{}_{-1}Y^{\text{in} *}]=$&
$2^4i\omega^3-\frac{i{}_1B^2}{\omega}\abs{{}_{-1}R^{\text{in,ref}}}^2$ & (d)
\\ \hline
\end{tabular}
\caption{Wronskians for the radial solutions. For clarity purposes, the subindices $\{lm\omega\}$ have been dropped.
The values of the constants at the left and right columns have been obtained with the asymptotics for the radial 
solutions at the horizon and at infinity respectively.} \label{table:radial wronsks}
\end{table}

\draft{PROBLEM: (c) probably has wrong sign but can't prove it (p.60K(IV))?}


We give two well-known, useful expressions
which are valid for any spin-1 radial solutions satisfying the Teukolsky-Starobinski\u{\i} identities as given in (\ref{eq:Teuk-Starob. ids.}):
\begin{subequations} \label{eq:R1,DR1 as func. of R_1,DR_1}
\begin{align}
\Delta{}_{+1}R_{lm\omega}&=2\left[({}_{-1}\lambda_{lm\omega}+2i\omega r)-2iK\mathcal{D}_0\right]{}_{-1}R_{lm\omega} 
\label{eq:R1 as func. of R_1,DR_1} \\
2{}_1B_{lm\omega}^2\mathcal{D}^{\dagger}_0 \left(\Delta{}_{+1}R_{lm\omega}\right)&=
\left[({}_{-1}\lambda_{lm\omega}-2ir\omega)\mathcal{D}_0+2i\omega \right]{}_{-1}R_{lm\omega} 
\label{eq:DR1 as func. of R_1,DR_1}
\end{align}
\end{subequations}

We will finalize this section by referring to the phenomenon of superradiance which we described in
Section \ref{sec:superrad.}. This phenomenon is manifest in the wronskian relations in Table \ref{table:radial wronsks}.
We shall see in Chapter \ref{stress-energy tensor} that the squared modulus of the reflection coefficient 
$A^{\text{in}}_{lm\omega}$ is equal to the 
fractional gain or loss of energy in a scattered wave mode ${}_{lm\omega}\phi^{\text{in}}_{\indhel}$. 
Wronskian relation (d) in Table \ref{table:radial wronsks} shows that for the modes such that $\tilde{\omega}\omega<0$ 
the squared modulus of this reflection coefficient must be greater than one, and thus the wave mode is reflected back
with a gain of energy. Since the `in' modes are only defined for positive $\omega$, superradiance occurs for these
modes for negative $\tilde{\omega}$ only. 
The transmitted part of the superradiant wave falls into the rotating black hole carrying in negative energy.
Similarly, from wronskian relation (a), when $\tilde{\omega}\omega<0$ the squared modulus of the reflection coefficient 
$A^{\text{up}}_{lm\omega}$ for the `up' modes must be greater than one. Therefore, the `up' modes, which are defined
for positive $\tilde{\omega}$, that experience superradiance are those for which $\omega<0$.
The condition $\tilde{\omega}\omega<0$ for superradiance, which is the same for scalar and gravitational perturbations, 
clearly shows that this phenomenon is only possible if $a\neq 0$ and therefore it only occurs if the black hole
possesses an ergosphere.


\section{Numerical method} \label{sec:num. method; radial func.}

\catdraft{al prog: 1) que son c1....c6? quina es la seva diff. amb $x_1$? 2) com determino cutoff lower i upper?}
We wrote the Fortran90 program \program{raddrv2KN.f} that solves the short-range differential equation (\ref{eq:diff. eq. for X,s=-1}) with the
real potential (\ref{eq:potential -1mathcalU}).
The program then uses equation (\ref{eq:Rs,Rs' as funcs. of chi and chi'}) to find ${}_{-1}R^{\bullet}_{lm\omega}$ and its derivative.
The variable of numerical integration is $r_*$. 
In this section we will describe the various methods used by this program as well as its structure.

We cannot set the initial condition for the radial function in the program at $r_*=-\infty$ ($r=r_+$) and therefore we set it instead 
slightly away from the horizon, at $r_*=r_{* 0}$ ($r=r_0\gtrsim r_+$). 
The value of the function $X^{\text{in}}_{lm\omega}$ (\ref{eq:X_in})
at $r_{* 0}$ is accurately given by the first order expansion (\ref{eq:asympt. for r=r_+ of X_in}). 
The differential equation (\ref{eq:diff. eq. for X,s=-1}) is solved with the driver 
routine \routine{odeint} described below so that 
$G\equiv X^{\text{in}}_{lm\omega}(r)/B^{\text{in}}_{lm\omega}$ and its derivative with respect to $r_*$ 
are obtained at a finite series of values of $r_*$ ranging from $r_{* 0}$ to a final value $r_{* f} \gg r_+$.

The reflection and transmission coefficients $A^{\bullet}_{lm\omega}$ and $B^{\bullet}_{lm\omega}$ are obtained 
by using the wronskian relations in Table \ref{table:radial wronsks} where the wronskians are evaluated at the last point $r_{* f}$ of the integration. 
The value of $X^{\text{up}}_{lm\omega}$ at $r_{* f}$ used to calculate the coefficients
is obtained by inserting the asymptotic expansion
\begin{equation} \label{eq:X_up/B_up, r->inf}
\begin{aligned}
S\equiv \frac{X^{\text{up} *}_{lm\omega}}{B^{\text{up} *}_{lm\omega}} & \rightarrow 
\exp{\left(-i\omega r_*+\sum_{i=1}^6 \frac{{}_ic_{lm\omega}}{r^i}\right)} & (r_* \rightarrow +\infty)
\end{aligned}
\end{equation}
into the differential equation and finding the values ${}_ic_{lm\omega}$, which we include in Appendix \ref{ch:App.A}. 
We initially calculated all the various wronskian relations numerically.
However, we found a numerical problem when calculating $W\left[G,G^*\right]$
for large $r_*$ for modes for which $\tilde{\omega}/\omega|B^{\text{in}}_{lm\omega}|^2$ 
is of the order of the precision of the calculations, $10^{-32}$ in our case.
Since $\left(1-|A^{\text{in}}_{lm\omega}|^2\right)=\tilde{\omega}/\omega {|B^{\text{in}}_{lm\omega}|^2}$, for those modes $|A^{\text{in}}_{lm\omega}|$
must be equal to $1$ within the first $32$ digits. 
But since that is the precision of the calculations, the next digits are round-off error and therefore the value of
$W\left[G,G^*\right]=-2i\omega \left(1-|A^{\text{in}}_{lm\omega}|^2\right)/|B^{\text{in}}_{lm\omega}|^2$ is all round-off error.      
We avoid this problem by setting $W\left[G,G^*\right]$ and $W\left[S,S^*\right]$
directly in the program equal to $-2i\tilde{\omega}$ and $2i\omega $ respectively. 
All four coefficients can then be found from these analytical values of the wronskians together with
the numerical calculations of $W\left[G,S\right]$ and $W\left[G,S^*\right]$
at $r_{* f}$, which do not pose any numerical problem. However, any subsequent evaluations of 
$\left(1-|A^{\text{in}}_{lm\omega}|^2\right)$ for the mentioned modes will obviously carry along large round-off error.
We found those modes to be the ones with either large $l$, small $m$ or small $\omega $. 
For example, $(1)$ for $l=8$, $\omega=0.3$: when $m=3$ the error in is in the 2nd digit already while for smaller $m$ all digits are wrong, 
$(2)$ for $l=8$, $m=1$: when $\omega<0.6$ all digits are wrong,  
and $(3)$ for $m=1$, $\omega=0.3$: when $l \geq 7$ all digits are wrong.

Once the reflection and transmission coefficients are calculated we can obtain the 
`up' solution at all points in the interval $[r_{* 0},r_{* f}]$ where $X^{\text{in}}_{lm\omega}$ 
has been calculated. For this purpose we may use the expression
\begin{equation}
X^{\text{up}}_{lm\omega}=\frac{1}{B^{\text{in} *}_{lm\omega}}\left(X^{\text{in *}}_{lm\omega}-A^{\text{in}}_{lm\omega}X^{\text{in}}_{lm\omega}\right)
\end{equation}
which follows from the wronskian relations in Table \ref{table:radial wronsks}.
The radial functions ${}_{-1}R^{\bullet}_{lm\omega}$ and their derivative can then be obtained via (\ref{eq:Rs,Rs' as funcs. of chi and chi'})
with $\indhel =-1$ and their coefficients via (\ref{eq:R_1 coeffs from X's}). We wish to normalize ${}_{-1}R^{\bullet}_{lm\omega}$ with 
${}_{-1}R^{\bullet \text{,inc}}_{lm\omega}$ set equal to $1$. We thus divide both coefficients and radial functions across by
${}_{-1}R^{\bullet \text{,inc}}_{lm\omega}$, which is given in (\ref{eq:R_1 coeffs from X's}).

\draft{describe splining?}

The second-order differential equation is rewritten as two coupled first-order differential equations in the usual way with a change
of variable $z\equiv\d{X}/\d{r_*}$. That is, equation (\ref{eq:diff. eq. for X,s=-1}) is solved as
\begin{equation} \label{eq:radial prog. derivs.}
\begin{aligned}
\diff{\Re X}{r_*}&=\Re z & \diff{\Im X}{r_*}&=\Im z \\
\diff{\Re z}{r_*}&={}_{-1}\mathcal{U}\Re X & \diff{\Im z}{r_*}&={}_{-1}\mathcal{U}\Im X 
\end{aligned}
\end{equation}
The notation we will use within the routines described below is the following.
The independent variable $r_*$ is going to be called $x$.
The two dependent complex variables $X$ and $\d{X}/\d{r_*}$ evaluated at a given point $x_n$ 
are going to be represented by $Y_n$. For clarity of notation the index 
that refers to one particular differential equation out of the set of four has been eliminated,
as all the following equations and descriptions are straight-forwardly generalizable to a set
of equations. 
The subindices ${lm\omega}$ have also been dropped in the radial functions for clarity.
Finally, the functions on the right hand side of equations (\ref{eq:radial prog. derivs.}) are going to be denoted by $f$.

The actual integration of the differential equation (\ref{eq:radial prog. derivs.}) is done
with the routines \routine{odeint}(driver)$\rightarrow$\routine{bsstep}(stepper)$\rightarrow$
$\topbott{\text{\routine{mmid}(algorithm)}}{\text{\routine{pzextr}(extrapolation)}}$, 
where the arrow indicates a routine call to another one.

The driver routine sets up the quantities that determine the desired accuracy for the numerical solution.
It then calls the stepper routine
with the present values of $x$, $Y$, $f$ and a suggested stepsize and receives back and stores the values of the actual stepsize $\stepx$ used and
the calculated value $y(x+\stepx)$. It then starts again at the new point $x+\stepx$ until it reaches the final point $r_{* f}$. 
The stepper routine sets up 
the number of subintervals to divide $[x,x+\stepx]$ in and calls the algorithm routine to perform the integration from $x$ to $x+\stepx$ with this number of subintervals.
It then extrapolates the results obtained 
with different numbers of subintervals
in order to improve on the accuracy of the final result. It changes the present stepsize 
if needed and performs again the above steps until the result $y(x+\stepx)$ is found within the desired accuracy. 
It finally estimates the most efficient stepsize to be taken in the next integration. 
The algorithm routine integrates the solution from $x$ to $x+\stepx$ for a certain stepsize and 
a certain number of subintervals of $[x,x+\stepx]$.
Finally, the extrapolation routine, called by the stepper, extrapolates 
various values of $y(x+\stepx)$ obtained with increasing number of subintervals
to the value that would be obtained if an infinite number of subintervals were used. It also gives an estimate for the error of the method.

We have used the forms of these routines as given in ~\cite{bk:NumRec} and have adapted them to solve the particular
problem (\ref{eq:radial prog. derivs.}). We give below a description of these routines in order to show
how the integration of the differential equation is performed.

\textbf{\routine{odeint} driver routine}

\routine{odeint} contains a loop that calculates the value $Y_n$ of the solution at the point $x_n$ whose value 
is increased at each iteration of the loop by a stepsize $\stepx$ from an initial value $x_0$ until a final point $x_f$.
The stepsize $\stepx$ may vary from one iteration to the next.
Within each iteration, \routine{odeint} first calculates the value $f$ of the derivatives 
(\ref{eq:radial prog. derivs.}) at the present point $x_n$.
It then sets the quantity whose fractional error will be compared against the error of the method to decide whether
the method has converged or not.
Since the value of the solution may
change a lot in magnitude from one point $x_n$ to the other, the error may be determined by $\epsilon Y_n$ where $\epsilon$
is the desired fractional error, i.e., the solution will be good to one part in $\epsilon$. 
However, if the solution goes through a zero value this quantity would not be a good indicator of the
error there. Another situation we must look out for is the accumulation of round-off error: the smaller the stepsize $\stepx$ is, the
higher the number of times we will have to evaluate the derivatives in (\ref{eq:radial prog. derivs.}) and therefore
the larger the accumulated error might become. If the fractional error were given in terms of $\stepx\d{Y_n}/\d{x}$,
both of the mentioned situations would be taken care of. 
A good method for assessing whether the desired accuracy is met or not is therefore achieved by comparing the absolute 
error of the solution against $\epsilon Y^{\text{scal}}_n$ where
\begin{equation} \label{eq:def. yscal,odeint}
Y^{\text{scal}}_n=|Y_n|+\left|\stepx\diff{Y_n}{x}\right|+\delta
\end{equation}
and $\delta$ is of the order of the precision of the machine as a safeguard in case the other two terms in $Y^{\text{scal}}_n$
are effectively zero.
\routine{odeint} then stores the values of $x_n$ and $Y_n$ calculated in the previous iteration and calls the stepper routine. 
The iteration finishes and \routine{odeint} stops whenever the final point $x_f$ is reached
or else when a certain maximum number of iterations has been performed.

\textbf{\routine{bsstep} stepper routine}

\draft{mention that this is the Burlisch-Stoer method!}

This routine makes use of the modified midpoint method included in the routine \routine{mmid} to find the value of the solution of the differential equation
at the point $x+\stepx$ from the knowledge of the value of the solution and the derivatives in (\ref{eq:radial prog. derivs.}) 
at the point $x$. This method requires splitting the interval $[x,x+\stepx]$ into a certain number $N$ of substeps. By calling \routine{mmid} the stepper
routine \routine{bsstep} obtains the value of $Y(x+\stepx)$ using a series of values of $N$ and then uses the extrapolation routine 
\routine{pzextr} to find the value of $Y(x+\stepx)$ for $\stepx/N\rightarrow 0$.
This idea for obtaining the value of the solution as though an infinite number of substeps were used is known as Richardson's deferred
approach to the limit. The most efficient series of values of $N$ that is known is the one given by 
$N_l=2l$. 
In practise, the number
of terms in the series is limited to 
$l_{max}=8$. 
The reason is that for $N>N_8$ little more efficiency is gained whereas roundoff error can become a problem.

We will denote by ${}_NY_{n+1}$ the value of the solution at the point $x=x_{n+1}$ calculated by \routine{mmid} using $N$ substeps
and will denote  by $Y_{n+1}^{(k)}$ the result of the extrapolation for $\stepx/N\rightarrow 0$ when up to $k$ terms in the series
${}_{N_l}Y_{n+1}$ ($l=1\dots k$) 
have been used. A loop will keep incrementing the number of terms in the series to do the extrapolation with, i.e., it will calculate 
${}_{N_l}Y_{n+1}$ ($l=1\dots k$)
and the corresponding $Y_{n+1}^{(k)}$ for increasing $k$, and finish when either 
$k=l_{max}$
or convergence has been achieved. Convergence is achieved whenever the relative error $\epsilon_k$ of the method at this 
$k$th iteration is smaller than the desired tolerance $\epsilon$. 
We will denote this value of $k$ by $k_f$. 
The relative error is calculated as
\begin{equation}
\epsilon_k=\frac{1}{S}\left| \frac{Y^{\text{err}}}{Y^{\text{scal}}} \right|
\end{equation}
where $Y^{\text{err}}$ is the absolute error of the extrapolation (given by \routine{pzextr} as $D_{k-1,1}$, defined in 
(\ref{eq:def.C,D})) and $Y^{\text{scal}}$ is provided by \routine{odeint} as (\ref{eq:def. yscal,odeint}). 
$S$ is a safety factor that we set equal to $1/4$ since the estimate of the error is not exact. 

After calculating $Y(x+\stepx)$ up to the desired relative error and before returning 
to the driver routine in order to calculate the solution at a new point, \routine{bsstep} estimates the
number $N_q$ of substeps such that the calculation of the series ${}_{N_l}Y_{n+1}$ ($l=1\dots q$) is most efficient.
It then calculates the corresponding stepsize $\stepx_q$ that would yield a value for the solution
within the desired accuracy; 
this is the stepsize that should be attempted in the next step. 
In general, we denote by $\stepx_i$ the stepsize that provides convergence
when series with a final number $N_i$ of substeps is used.
By `most efficient' is meant the one that requires the smallest amount of work per unit step, where 
the amount of work is given by the number of 
times that we need to evaluate the right hand side of (\ref{eq:radial prog. derivs.}).
It can be checked from (\ref{eq:modif. midpt. method}) that for the series ${}_{N_l}Y_{n+1}$ ($l=1\dots k$) this number of evaluations
is given by the recursive relation
\begin{equation} \label{eq:work A_k in mmid}
\begin{aligned}
A_1&=N_1+1 \\
A_{k}&=A_{k-1}+N_k
\end{aligned}
\end{equation}
On the other hand, 
the error in the extrapolation for the series ${}_{N_l}Y_{n+1}$ ($l=1\dots k$) is calculated by the routine \routine{pzextr} and is of order $(\stepx)^{2k-1}$. 
Therefore, if $\epsilon_k$ is the relative error 
in the extrapolation for this series using
a stepsize of $\stepx$, then the stepsize $\stepx_k$ required 
to obtain a relative error of order of the desired tolerance $\epsilon$ when using the same 
series
of substeps is estimated to be
\begin{equation} \label{eq:estimated new stepsize}
\stepx_k=\stepx\left(\frac{\epsilon}{\epsilon_k}\right)^{1/(2k-1)}
\end{equation}
The amount of work per unit step when using 
the series $N_l$ ($l=1\dots k$) and the
stepsize $\stepx_k$ is therefore equal to
\begin{equation} \label{eq:work per unit step W_k in mmid}
W_k=\frac{A_k}{\stepx_k}\stepx=A_k\left(\frac{\epsilon_k}{\epsilon}\right)^{1/(2k-1)}
\end{equation}
which has been nondimensionalized by multiplying by $\stepx$.
The optimal number $N_q$ of substeps is then given by the term $q$ in the series 
$N_l$ ($l=1\dots k$)
such that $W_q=\min_{k=1,\dots ,k_f} W_k$
\catdraft{probl. ja que sembla dir que com mes gran es k mes gran es l'error i per tant despres de $k_f$ ja no es menor
que epsilon, pero aixo no te sentit, nomes si H>1 pero llavors com mes punts tenim per fer extrapolac. mes gran es l'error??!->prendre nomes 1 punt?!}
) and
the corresponding stepsize $\stepx_q$ that provides convergence is obtained from (\ref{eq:estimated new stepsize}).

The following factor $\alpha(i,j)$, given in ~\cite{bk:NumRec}, is the factor by which $\stepx_i$ 
is to be multiplied so that the resulting stepsize
$\alpha(i,j)\stepx_i$ provides convergence 
when series with a final number $N_j$ of substeps is used:
\begin{equation} \label{eq:def. alpha,bsstep}
\begin{aligned}
\alpha(i,j)=\epsilon^{\frac{A_{i+2}-A_{j+2}}{(2i+3)(A_{j+2}-A_1+1)}} & \qquad \text{for} & i<j
\end{aligned}
\end{equation}
This factor helps to improve the routine in two particular circumstances.
The first one is in the case that the current stepsize $\stepx$ being used is too small, which will be indicated by the fact that 
$q=k_f$. In that case, increasing the stepsize to $\stepx_{q+1}$ might be a better choice. We do not know the value $\stepx_{q+1}$ but it can
be calculated from $\stepx_q$ with (\ref{eq:def. alpha,bsstep}). If using $\stepx_{q+1}$ is more efficient than using
$\stepx_q$ then we choose $\stepx_{q+1}$ over $\stepx_q$. This check follows from the definition (\ref{eq:work per unit step W_k in mmid}) 
of work per unit stepsize and turns out to be $A_{q+1}\alpha(q,q+1)>A_{q+2}$. 
The second circumstance in which $\alpha(i,j)$ is useful is in the case that the current stepsize $\stepx$ is too large
to achieve convergence with. 
This situation is detected by the condition $\stepx_k \alpha(k,q+1)<\stepx$. 
If this situation occurs then the current stepsize $\stepx$ is abandoned and the stepsize given by
$\stepx_k \alpha(k,q)$ is attempted instead. 

\textbf{\routine{mmid} algorithm routine}

The routine \routine{mmid} uses an algorithm called the modified midpoint method, which is 
based on a variation of Euler's method: $Y_{n+1}=Y_n+\stepx f(x_n,Y_n)$. If instead of evaluating
the derivative $f$ at the point $x_n$ it is evaluated at a middle point between $x_n$ and $x_{n+1}$ we then have the second-order
Runge-Kutta method. The modified midpoint method splits the interval $[x_n,x_{n+1}]$ into a
sequence of $N$ intervals equally spaced by $\stepx/N$ and then uses the second-order Runge-Kutta method at the end points of 
the intervals except at the very first and last points. The result is
\begin{equation} \label{eq:modif. midpt. method}
\begin{aligned}
z_0&\equiv Y_n \\
z_1&\equiv z_0+\frac{\stepx}{N}f(x_n,z_0) \\
z_{m+1}&\equiv z_{m-1}+\frac{2\stepx}{N}f\left(x_n+\frac{m\stepx}{N},z_m\right) \quad \text{for} \quad m=1,2,\dots,N-1 \\
{}_NY_{n+1}&=\frac{1}{2}\left[z_N+z_{N-1}+\frac{\stepx}{N}f(x_n+\stepx,z_N)\right]
\end{aligned}
\end{equation}
This algorithm is also second-order but it has the nice feature that its truncation error contains only
even powers of $\stepx/N$. As a consequence, if we combine the result ${}_NY_{n+1}$ obtained using a sequence of $N$ intervals
with the one ${}_{N/2}Y_{n+1}$ obtained using half as many intervals in the manner $Y_{n+1}=(4{}_NY_{n+1}-{}_{N/2}Y_{n+1})/3$, 
then the approximation $Y_{n+1}$ is fourth-order accurate even though is uses approximately (for large $N$) only 1.5 times 
as many derivative evaluations. The routine \routine{mmid} directly implements the algorithm (\ref{eq:modif. midpt. method}).

\textbf{\routine{pzextr} extrapolation routine}

Given a set of sample values $\{Z_1,Z_2,...,Z_k\}$ of a function at the sample points $\{w_1,w_2,...,w_k\}$, polynomial inter- or extrapolation 
consists in approximating the value of the function at a certain point $w$ by evaluating at $w$ the unique $(k-1)$-degree
polynomial such that its value at each one of the points $w_l$ coincides with the sample values $Z_l$.
This polynomial is given by Lagrange's formula:
\begin{equation} \label{eq:Lagrange interp}
\begin{aligned}
P(w)&=\frac{(w-w_2)(w-w_3)\dots(w-w_k)}{(w_1-w_2)(w_1-w_3)\dots(w_1-w_k)}Z_1+ \\
&+\frac{(w-w_1)(w-w_3)\dots(w-w_k)}{(w_2-w_1)(w_2-w_3)\dots(w_2-w_k)}Z_2+ \\
+\dots &+\frac{(w-w_1)(w-w_2)\dots(w-w_{k-1})}{(w_k-w_1)(w_k-w_2)\dots(w_k-w_{k-1})}Z_k
\end{aligned}
\end{equation}
Neville's algorithm implements (\ref{eq:Lagrange interp}) in a way that not only gives an error estimate but it also makes it easy to
calculate the polynomial when an extra point $w_{k+1}$ is added instead of having to evaluate the awkward formula (\ref{eq:Lagrange interp}) 
back from scratch. If we define $P_{l(l+1)\dots (l+m)}$ as the
unique $m$-degree polynomial passing through the points $\{w_l,w_{l+1},\dots w_{l+m}\}$ and define the differences
\begin{subequations} \label{eq:def.C,D}
\begin{align}
C_{m,l}&\equiv P_{l\dots(l+m)}-P_{l\dots(l+m-1)} \\
D_{m,l}&\equiv P_{l\dots(l+m)}-P_{(l+1)\dots(l+m)}
\end{align}
\end{subequations}
we then have from (\ref{eq:Lagrange interp}) the recursive relations
\begin{subequations} \label{eq:recursive C,D}
\begin{align}
C_{m+1,l}&=\frac{(w_{l+m+1}-w)(C_{m,l+1}-D_{m,l})}{w_l-w_{l+m+1}} \\
D_{m+1,l}&=\frac{(w_l-w)(C_{m,l+1}-D_{m,l})}{w_l-w_{l+m+1}}
\end{align}
\end{subequations}
The Tableau 
below 
schematizes how a polynomial $P_{l\dots (l+m)}$ of degree $m$ can be obtained from a polynomial 
(either $P_{(l+1)\dots (l+m)}$ or $P_{l\dots (l+m-1)}$) of degree $(m-1)$ that interpolates the same points except for one of the two 
ends (either $w_l$ or $w_{l+m}$) with the
knowledge of either $D_{m,l}$ or $C_{m,l}$ which can be obtained
using the recursive relations (\ref{eq:recursive C,D}).

\scalebox{0.78}{%
$\displaystyle\label{eq:Neville Tableau}
\begin{array}{cccccccccc}
w_{1}&Z_{1}=P_{1}\\
&&\swarrow{C_{1,1}}\\
&&&P_{12}\\
&&\nearrow{D_{1,1}}&&\swarrow{C_{2,1}}\\
w_{2}&Z_{2}=P_{2}&&&&\ddiagdots\\
\downdots&\downdots&&&&&\swarrow{C_{k-2,1}}\\
\downdots&\downdots&&&&&&P_{12\ldots (k-1)}\\
\downdots&\downdots&&&&&\nearrow{D_{k-2,1}}&&\swarrow{C_{k-1,1}}\\
\downdots&\downdots&&&&\udiagdots&&&&P_{12\ldots k}\\
\downdots&\downdots&&&\nearrow{D_{2,k-3}}&&&&\nearrow{D_{k-1,1}}\\
\downdots&\downdots&&P_{(k-2)(k-1)}&&&&\udiagdots\\
\downdots&\downdots&\nearrow{D_{1,k-2}}&&\swarrow{D_{2,k-2}}&&\nearrow{D_{3,k-3}}\\
w_{k-1}&Z_{k-1}=P_{k-1}&&&&P_{(k-2)(k-1)k}\\
&&\swarrow{C_{1,k-1}}&&\nearrow{D_{2,k-2}}\\
&&&P_{(k-1)k}\\
&&\nearrow{D_{1,k-1}}\\
w_{k}&Z_{k}=P_{k}\\
\end{array}%
$}

\draft{number for the tableau does not appear}

Apart from the values of $w_l$ and $Z_l$ ($l=1,\dots ,k$), the routine \routine{mmid} assumes initial knowledge of the values 
$D_{l,k-1-l}$ ($l=1,\dots,k-2$), which have in practise been obtained from the previous call
to this routine with the first $k-1$ sample points.
The routine \routine{pzextr} calculates $D_{l,k-l}$ ($l=1,\dots,k-1$) via (\ref{eq:recursive C,D}) with $w=0$
and stores these values for the next time it is called. It finally evaluates at $w=0$ the polynomial of degree $k$ by adding these values
to the initial starting point $P_k$, i.e.,
\begin{equation}
P_{1\dots k}=P_k+\sum_{l=1}^{k-1}D_{l,k-l}
\end{equation}
It finally returns $D_{k-1,1}$ as an estimate of the error of the extrapolation.

The correspondence of notation with the routines \routine{bsstep} and \routine{odeint} is established by 
letting $w_l=h/N_l$ and $Z_l={}_{N_l}Y_{n+1}$ where the subindex $n+1$ has been dropped as all the values 
of $Z_l$ refer to the same point $x_{n+1}$.


\section{Numerical results} \label{sec:num. results; radial func.}

The numerical reflection coefficients $|A^{\text{in}}_{lm\omega}|^2$ we obtained agree with Chandrasekhar's ~\cite{bk:Chandr}
TableIX, Chapter 8 as shown in Table \ref{table:Chandr's tableIX,ch.8}.

\begin{table}
\begin{tabular}{ccc}
$\omega $ & Chandrasekhar's & numerical result  \\
\hline                                                                   
-0.35 & 1.01565 & 1.015649                         \\
-0.455593 & 0.11332 & 0.11332896
\end{tabular}
\caption{The coefficient $|A^{\text{in}}_{lm\omega}|^2$ for $M=1$, $Q=0$, $a=0.95$, $l=1$, $m=-1$ as given by Chandrasekhar's (~\cite{bk:Chandr})
TableIX, Chapter 8 and numerically calculated with the program \program{raddrv2KN.f}.
}\label{table:Chandr's tableIX,ch.8}
\end{table}

In Table \ref{table:data R_1 a=0.95,l=2,m=-2,w=-0.5;tableVChandr}, which we include in Appendix \ref{ch:App.A} because of its large size,
we check the results for the radial solution
for the particular case $\indhel =-1$, $Q=0$, $a=0.95$, $l=2$, $m=-2$, $\omega=-0.5$ against the ones given by Chandrasekhar ~\cite{bk:Chandr}
in Table V of his Appendix. 
The radius of the event horizon corresponding to these values is: $r_+/M \simeq 1.3122$.
The normalization taken by Chandrasekhar is the same as that of ${}_{-1}R^{\text{sym}}_{lm\omega}$ and is thus related
to the numerical `in'/`up' solutions that we obtained via the relation (\ref{eq:R_1(sym) en func. de R_1(up/in)}). 
Since we do not know the value of $\alpha$ that Chandrasekhar used, we matched at the point $r/M=4$ Chandrasekhar's value
${}_{-1}R^{\text{chandr}}_{lm\omega}$ with the value of ${}_{-1}R^{\text{sym,num}}_{lm\omega}$,
which is calculated from the numerical values ${}_{-1}R^{\text{in/up,num}}_{lm\omega}$ via (\ref{eq:R_1(sym) en func. de R_1(up/in)}).
The resulting value for $\alpha$ is 
$-0.68514+1.6271i$ 
and the values of ${}_{-1}R^{\text{sym,num}}_{lm\omega}$
at other points $r$ where calculated with it. Since $\alpha$ is actually obtained with Chandrasekhar's values, only 
its first 5 digits are actually valid, and simililary for ${}_{-1}R^{\text{sym,num}}_{lm\omega}$. These values are plotted
in Figure \ref{fig:R_1 a=0.95,l=2,m=-2,w=-0.5;tableVChandr}, where an erroneous glitch is observed for the real part of
$\d{{}_{-1}R^{\text{chandr}}_{lm\omega}}/d{r}$ but not in our numerical results.

\begin{figure}[p]
\centering
\begin{tabular}{cc}
\includegraphics*[width=70mm]{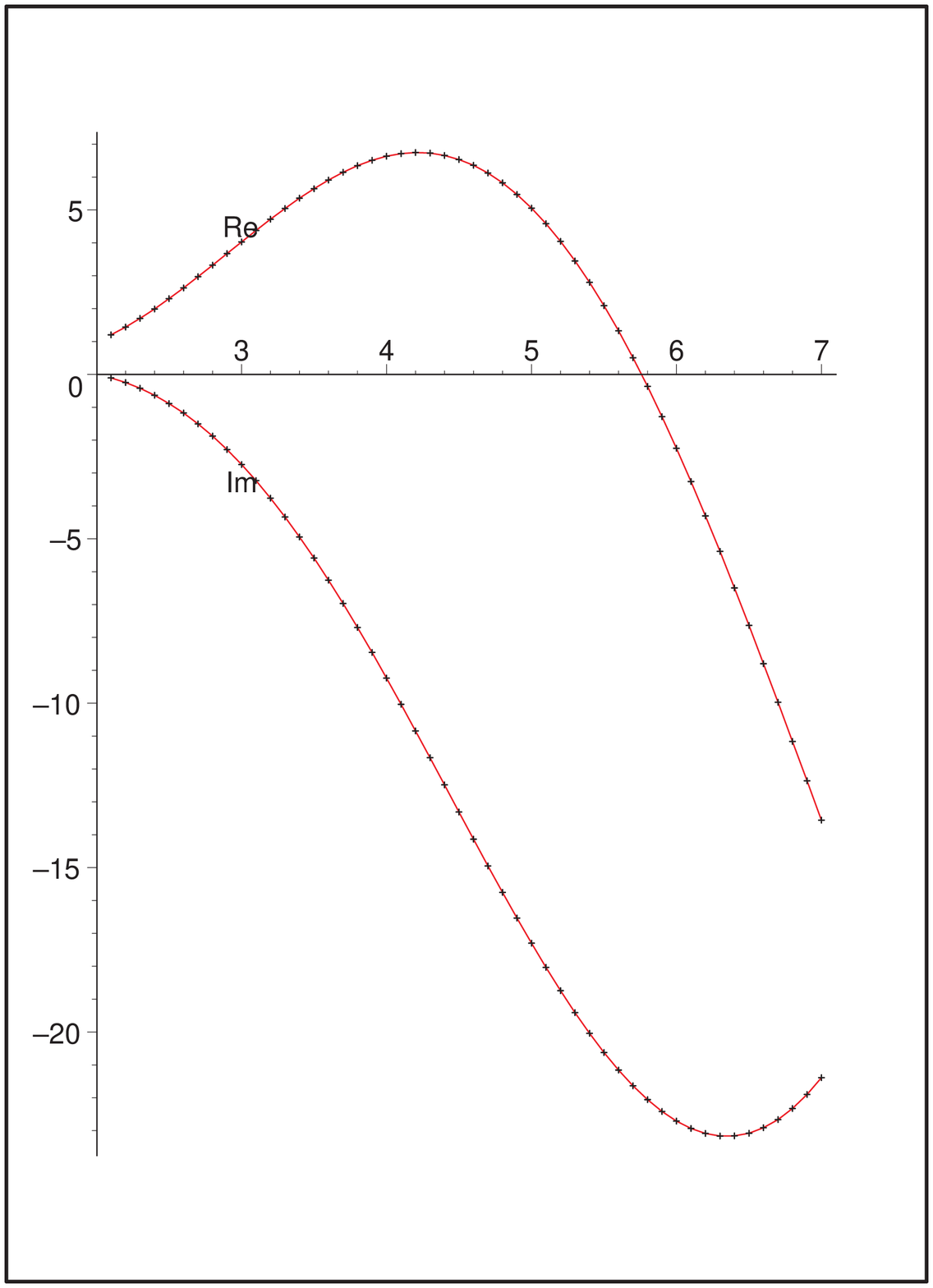} &
\includegraphics*[width=70mm]{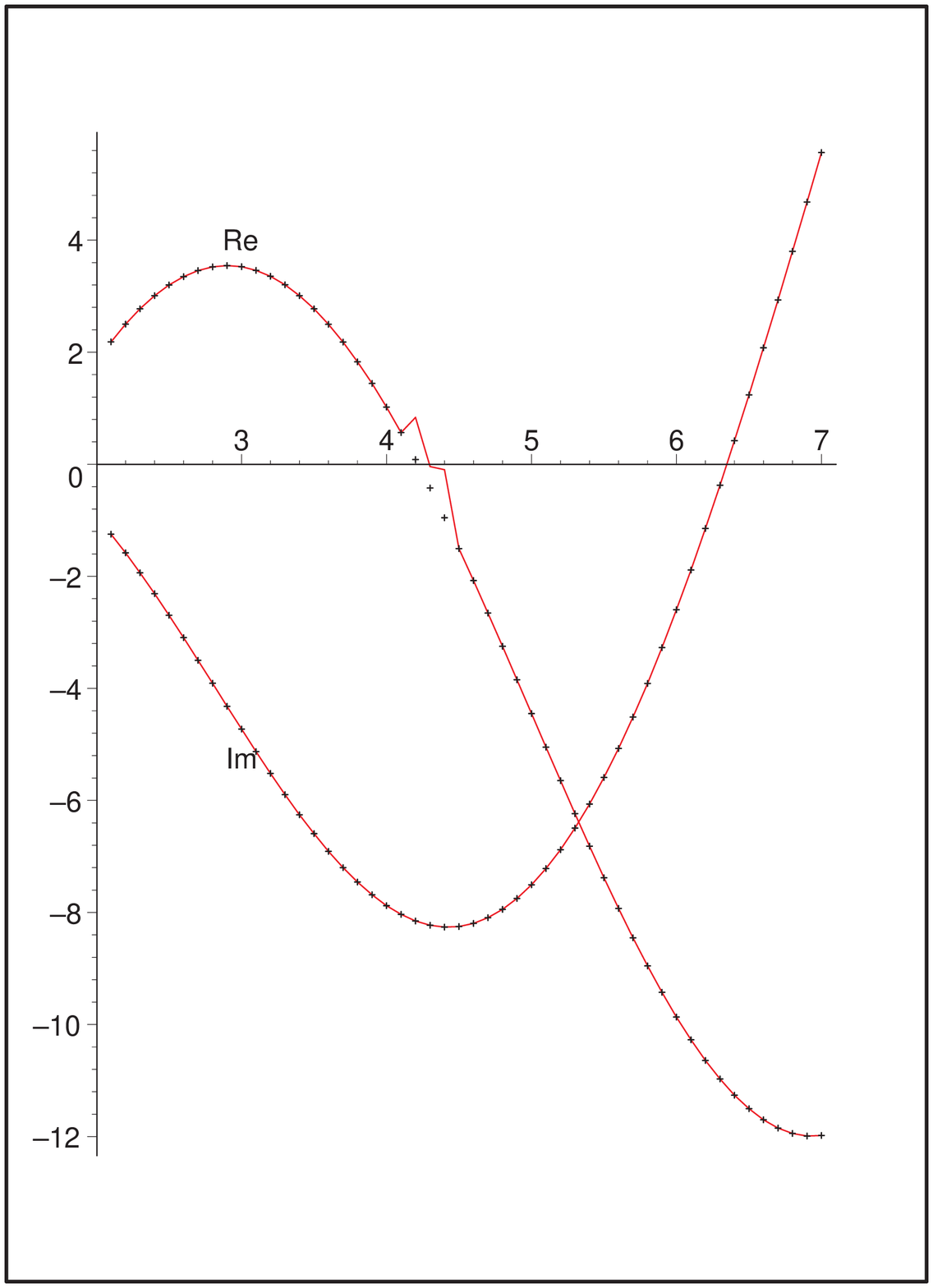} \\
${}_{-1}R_{2,-2,-0.5}$ & $\diff{{}_{-1}R_{2,-2,-0.5}}{r}$
\end{tabular}
\caption{Real and Imaginary parts of ${}_{\indhel}R^{\text{chandr}}_{lm\omega}$ (in bold red line), ${}_{-1}R^{\text{sym,num}}_{lm\omega}$ 
(in black crosses) and their derivatives 
for $\indhel =-1$, $Q=0$, $a=0.95$, $l=2$, $m=-2$, $\omega=-0.5$.}
\label{fig:R_1 a=0.95,l=2,m=-2,w=-0.5;tableVChandr}
\end{figure}


\section{Asymptotics close to the horizon} \label{sec:asympts. close to r_+}

In this section we are interested in finding the asymptotic behaviour of the radial
function close to the horizon. 
The main objective for this study is its particular application in Section \ref{sec:RRO} in the last chapter.
We calculate in that section the behaviour close to the horizon of the renormalized expectation value of
the stress-energy tensor when the field is in the past Boulware state.
As we shall show in that section, only the `up' modes are of interest for that calculation. 
That calculation involves a factor in front of the `up' radial functions that decreases exponentially with $\tilde{\omega}$.
We will therefore not consider the case of large $\tilde{\omega}$. 

This study is based on the one performed by Candelas  ~\cite{ar:Candelas'80}.
Even though Candelas started the calculation for general spin, he soon confined it to the scalar case. It is our
intention to complete his asymptotic calculation for general spin and only specialize to the spin-1 case at the very end.

We perform a Taylor series expansion around $r=r_+$ of the coefficients of ${}_{\indhel}R_{lm\omega}$ and its
derivatives appearing in the radial Teukolsky equation (\ref{eq:radial teuk. eq.}). We only keep the first order terms
in the expansion and also terms that involve parameters which might become very large. We obtain:
\begin{equation} \label{eq:non-approx eqA4Cand'80}
(r-r_+)\ddiff{{}_{\indhel}R_{lm\omega}}{r}+(\indhel+1)\diff{{}_{\indhel}R_{lm\omega}}{r}- \left[\frac{{}_{\indhel}\lambda_{lm\omega}
-4i\omega \indhel r}{r_+-r_-}-
\frac{q(q-2i\indhel)}{4(r-r_+)}\right]
{}_{\indhel}R_{lm\omega}=0
\end{equation}
where $q$ is defined as
\begin{equation} \label{eq:def. of Q_p}
q \equiv \frac{2K_+}{r_+-r_-}    
\end{equation}
The parameters in (\ref{eq:non-approx eqA4Cand'80}) that might become very large independently of the limit
$r \rightarrow r_+$ are ${}_{\indhel}\lambda_{lm\omega}$, $\omega $ and $\tilde{\omega}$. 
As mentioned, we are going to discard the possibility $\tilde{\omega} \rightarrow \infty$.
Keeping $\tilde{\omega}$ bounded means that either both $m$ and $\omega $ are bounded or else that
$\displaystyle \omega \rightarrow \infty$ and $m \sim \omega/\Omega_+$. 
We are going to restrict ourselves to the first possibility (i.e., $m$ and $\omega $ bounded) since it is only 
for this case that we are able to find the behaviour of the angular solutions, needed in the calculation in Section \ref{sec:RRO}. 
This is a crucial point, as we will see later. We thus have that the only
term in (\ref{eq:non-approx eqA4Cand'80}) that might become very large independently of $r \rightarrow
r_+$ is the one with ${}_{\indhel}\lambda_{lm\omega}$. Since we are keeping $m$ and $\omega $ bounded, ${}_{\indhel}\lambda_{lm\omega}$ can
only become large if we let $l \rightarrow +\infty$.   \catdraft{pq? justificar}
Because of factors possessing positive powers of $l$ for large $l$ multiplying the `up' radial functions
in the calculations that we will perform in Section \ref{sec:RRO}, we are only interested in the asymptotics close
to the horizon in the modes with $l \rightarrow +\infty$.


Even though we only show the angular equation resulting from the separation of variables in
the Teukolsky equation valid $\forall \indhel$ in the next chapter, we will give here a straight-forward result 
regarding this equation that we need in order to pursue the present calculations.
When letting $l \rightarrow +\infty$ and keeping $\omega $ and $m$ bounded in the Teukolsky angular 
equation (\ref{eq:ang. teuk. eq.}), all the terms in the coefficient of the angular function ${}_{\indhel}S_{lm\omega}(\theta)$ can be ignored
except for ${}_{\indhel}\lambda_{lm\omega}$ and those with a $1/\sin \theta$ in
them. This is equivalent to setting $a\omega=0$ in the angular equation. This means that in the limit $l \rightarrow +\infty$
(with $m$ and $\omega $ bounded) the angular solution reduces to the spin-weighted spheroidal harmonics: 
${}_{\indhel}S_{lm} \rightarrow {}_{\indhel}Y_{lm}$ and that ${}_{\indhel}\lambda_{lm\omega} \rightarrow (l-\indhel )(l+\indhel+1)\rightarrow l^2$.
The latter expression implies that ${}_1B_{lm\omega}\rightarrow l^2$ in the same limit.
We refer the reader to the next chapter for a description of the different angular solutions and eigenvalues.

We can now approximate equation (\ref{eq:non-approx eqA4Cand'80}) in the limit 
$l\rightarrow +\infty$ (with $\omega $ and $m$ bounded):
\begin{equation} \label{eq:eqA4Cand'80}
(r-r_+)\ddiff{{}_{\indhel}R_{lm\omega}}{r}+(\indhel+1)\diff{{}_{\indhel}R_{lm\omega}}{r}-\left[\frac{l^2}{r_+-r_-}-
\frac{q(q-2i\indhel)}{4(r-r_+)}\right]
{}_{\indhel}R_{lm\omega}=0
\end{equation}
Note that $m$ and $\omega$ do not appear explicitly anymore in the differential equation in this limit.
This differential equation can be rewritten as
\begin{equation} \label{eq:rewriting eqA4Cand'80}
z^2\ddiff{{}_{\indhel}W_{lm\omega}}{z}+z\diff{{}_{\indhel}W_{lm\omega}}{z}-\left[z^2+\indhel ^2-q(q-2i\indhel)\right]{}_{\indhel}W_{lm\omega}=0
\end{equation}
after the change of variables 
\begin{subequations}
\begin{align}
x &\equiv \frac{r-r_+}{2(r_+-M)}           \\
z &\equiv 2lx^{1/2}   \\
{}_{\indhel}W_{lm\omega} &\equiv x^{\indhel/2}{}_{\indhel}R_{lm\omega}
\end{align}
\end{subequations}
The solutions of the differential equation (\ref{eq:rewriting eqA4Cand'80}) are the modified Bessel functions: 
\begin{equation}
{}_{\indhel}W_{lm\omega}=I_{\pm(\indhel+iq)}(2lx^{1/2}), K_{\indhel+iq}(2lx^{1/2})
\end{equation}
It is at this point that Candelas' analysis specializes to the scalar case. We pursue it for general spin.
The asymptotic behaviour close to the horizon of the `up' radial functions with $l \rightarrow +\infty$
is given by
\begin{equation} \label{eq:eqA6Cand'80}
{}_{\indhel}R_{lm\omega}^{\text{up}} \rightarrow
{}_{\indhel}a_lx^{-\indhel /2}K_{\indhel+iq}(2lx^{1/2})+{}_{\indhel}b_lx^{-\indhel /2}I_{-(\indhel+iq)}(2lx^{1/2})
\qquad (l\rightarrow +\infty,r \rightarrow r_+)
\end{equation}
which is uniformly valid in $l$. The factors ${}_{\indhel}a_l$ and ${}_{\indhel}b_l$ are the coefficients of the two independent solutions.

\catdraft{em manca especificar aqui a (\ref{eq:eqA6Cand'80}) limit comu entre $l \rightarrow +\infty$ i $r \rightarrow r_+$, quin mes rapid que quin?!}

The following asymptotic formulae for the modified Bessel functions are well known (~\cite{bk:AS}):
\begin{equation} \label{eq:eqs9.7.1,9.7.2A&S}
\begin{aligned}
I_{\nu}(z)&\rightarrow\frac{e^{z}}{\sqrt{2\pi z}}&
(\abs{z} &\rightarrow +\infty,\ \nu\ \text{fixed})\quad\text{if}\quad \abs{\arg z} < \pi/2\\
K_{\nu}(z)&\rightarrow\sqrt{\frac{\pi}{2z}}e^{-z}&
(\abs{z} &\rightarrow +\infty,\ \nu\ \text{fixed})\quad\text{if}\quad \abs{\arg z} < 3\pi/2\\
\end{aligned}
\end{equation}
and
\begin{equation} \label{eq:eqs9.6.7,9.6.9A&S}
\begin{aligned}
I_{\nu}(z)&\rightarrow\frac{1}{\Gamma (\nu+1)}\left(\frac{z}{2}\right)^{\nu}&
(z &\rightarrow 0,\ \nu\ \text{fixed})\quad\text{if}\quad \nu\neq -1,-2,\dots\\
K_{\nu}(z)&\rightarrow 
\begin{cases}
\frac{\Gamma (\nu)}{2}\left(\frac{z}{2}\right)^{-\nu}\quad \text{if}\ \Re{\nu}>0 \\
\frac{\Gamma (-\nu)}{2}\left(\frac{z}{2}\right)^{\nu}\quad \text{if}\ \Re{\nu}<0 
\end{cases}
&(z &\rightarrow 0,\ \nu\ \text{fixed})
\end{aligned}
\end{equation}
If we fix $r$ close to the horizon in (\ref{eq:eqA6Cand'80}) and let $l \rightarrow +\infty$, the value of the radial
potential at that fixed value of $r$ will go to infinity and thus it must be
${}_{\indhel}R^{\text{up}}_{lm\omega} \rightarrow 0$.
From the formulae (\ref{eq:eqs9.7.1,9.7.2A&S}), it is only possible that ${}_{\indhel}R^{\text{up}}_{lm\omega} \rightarrow 0$ 
in the limit $l \rightarrow +\infty$ with $r$ fixed in (\ref{eq:eqA6Cand'80}) 
if the coefficient ${}_{\indhel}b_l$ decreases exponentially with $l$. 
It follows from this result together with the formulae (\ref{eq:eqs9.6.7,9.6.9A&S})
that in the limits $r \rightarrow r_+$ and $l \rightarrow +\infty$, while keeping $lx^{1/2}$ finite,
the second term in the asymptotic expression (\ref{eq:eqA6Cand'80}) can be neglected with
respect to the first one.   
In this last statement we have made the implicit assumption that if the coefficient ${}_{\indhel}a_l$ 
decreases for large $l$, then it does slower than the coefficient ${}_{\indhel}b_l$. 
This assumption is proved to be correct in what follows. We have from the above discussion that
\begin{equation} \label{eq:approx eqA6Cand'80}
{}_{\indhel}R^{\text{up}}_{lm\omega} \rightarrow {}_{\indhel}a_lx^{-\indhel /2}K_{\indhel+iq}\left(2lx^{1/2}\right)
\qquad (l\rightarrow +\infty,r \rightarrow r_+,lx^{1/2}\ \text{finite})
\end{equation}

\draft{there does not seem to be any reason why any expressions so far are equally valid for `in' modes....?}

We can determine the coefficient ${}_{\indhel}a_l$ by comparison with the WKB
approximation (\ref{eq:R_up}). By taking the limit $lx^{1/2}\rightarrow 0$ on the 
solution (\ref{eq:approx eqA6Cand'80}) we obtain
\begin{equation} \label{eq:approx eqA6Cand'80,r->rplus}
\begin{aligned}
{}_{\indhel}R^{\text{up}}_{lm\omega} \rightarrow 
&
\frac{{}_{\indhel}a_l\pi
(r_+-r_-)^{2\indhel}l^{-\indhel-iq}I_{\tilde{\omega}}^*}{2\sin [(\indhel+iq)\pi]
\Gamma(1-\indhel-iq)}\Delta^{-\indhel }e^{-i\tilde{\omega}r_*}-
\\&-
\frac{{}_{\indhel}a_l\pi
l^{\indhel+iq}I_{\tilde{\omega}}}{2\sin [(\indhel+iq)\pi]
\Gamma(1+\indhel+iq)}e^{+i\tilde{\omega}r_*}  
\qquad 
(l\rightarrow \infty,r \rightarrow r_+,lx^{1/2}\rightarrow 0)
\end{aligned}
\end{equation}
where
\begin{equation} \label{eq:I_wtilde}
I_{\tilde{\omega}}\equiv e^{-\tilde{w}r_+}\left(4M\kappa_+\right)^{-\frac{i\tilde{w}}{2\kappa_+}}\left(-4M\kappa_-\right)^{-\frac{-i\tilde{w}}{2\kappa_-}}
\end{equation}
Comparing this asymptotic expression with the WKB approximation (\ref{eq:R_up}) it follows that
\begin{equation} \label{eq:al and Rup,ref}
\begin{aligned}
{}_{\indhel}a_l&=- \frac{2\sin [(\indhel+iq)\pi]\Gamma(1+\indhel+iq)I_{\tilde{\omega}}^*}{\pi}l^{-\indhel-iq}{}_{\indhel}R^{\text{up,inc}}_{lm\omega}   \\
{}_{\indhel}R^{\text{up,ref}}_{lm\omega}&=
-\frac{\Gamma(1+\indhel+iq)}{\Gamma(1-\indhel-iq)}(r_+-r_-)^{2\indhel}I_{\tilde{\omega}}^{* 2}l^{-2(\indhel+iq)}{}_{\indhel}R^{\text{up,inc}}_{lm\omega}
\end{aligned}
\end{equation}

We now specialize to the spin-1 case. Combining equations
(\ref{eq:approx eqA6Cand'80}) and (\ref{eq:al and Rup,ref}), and using
the same normalization as the one used in the numerical results of the preceding section (i.e., setting         
${}_{-1}R^{\text{up,inc}}_{lm\omega}=1$ and  ${}_{+1}R^{\text{up,inc}}_{lm\omega}=-iB^2/\left(2\EuFrak{N}K_+\right)$), we have
\begin{equation} \label{eq:R1 `up' approx l->inf,r->rplus}
\begin{aligned}
{}_{+1}R^{\text{up}}_{lm\omega} &\rightarrow \frac{-2I_{\tilde{\omega}}^*l^{3-iq}}{(r_+-r_-)K_+\Gamma(-iq)}
x^{-1/2}K_{1+iq}(2lx^{1/2}) 
\\ &\qquad \qquad \qquad \qquad\qquad (l\rightarrow +\infty,r \rightarrow r_+,lx^{1/2}\ \text{finite}) \\
{}_{-1}R^{\text{up}}_{lm\omega} &\rightarrow \frac{i(r_+-r_-)I_{\tilde{\omega}}^*l^{+1-iq}}{K_+\Gamma(-iq)}
x^{1/2}K_{-1+iq}(2lx^{1/2}) 
\\ &\qquad \qquad \qquad \qquad\qquad (l\rightarrow +\infty,r \rightarrow r_+,lx^{1/2}\ \text{finite})
\end{aligned}
\end{equation}
\catdraft{he fet hipotesi raonable de com ha de ser (\ref{eq:R1 `up' approx l->inf,r->rplus}) per K-N pero encara l'he de comprovar!}
\draft{there is no reason why asymptotics should work for case ${}_{+1}R^{\text{up,inc}}_{lm\omega}=-iB^2/\left(2\EuFrak{N}K_+\right)$
since we have started off with assumption that wave goes to zero as the potential goes as $l^2$, but in this case wave goes as $l^4$?}

It is also useful to give the expressions that the `up' radial functions (\ref{eq:R1 `up' approx l->inf,r->rplus})
adopt in this limit whenever the constants of normalization (\ref{eq:normalization consts.}) 
that we shall give in the last chapter are included. These expressions are, in compact form:
\begin{equation} \label{eq:R1 `up' approx l->inf,r->rplus;compact version}
|N^{\text{up}}|{}_{\indhel}R^{\text{up}}_{lm\omega} \rightarrow 
A_{\indhel}Nx^{-1/2}K_{\indhel+iq}(2lx^{1/2}) \qquad (l\rightarrow +\infty,r \rightarrow r_+,lx^{1/2}\ \text{finite})
\end{equation}
for $\indhel =\pm 1$, where
\begin{equation} \label{eq:def.A_s}
A_{\indhel} \equiv
\left \{
\begin{array}{ll}
\displaystyle -4 &, \indhel=+1 \\
\displaystyle \frac{2i}{l^4} &, \indhel=-1
\end{array}
\right \}
\left[l(r_+-r_-)\right]^{-\indhel }
\end{equation}
and
\begin{equation} \label{eq:def.D}
N \equiv \frac{I_{\tilde{\omega}}^*l^{-iq}}{\sqrt{2^3\pi K_+}\Gamma(-iq)}
\end{equation}
\catdraft{he fet hipotesi raonable de com ha de ser (\ref{eq:def.D}) per K-N pero encara l'he de comprovar!}
We therefore have finally found the asymptotic behaviour of the `up' radial modes with $l\rightarrow +\infty$ 
close to the horizon with $\tilde{\omega}$, and both $m$ and $\omega$, bounded. We believe that this is the method behind 
the approximation given by Candelas, Chrzanowski and Howard ~\cite{ar:CCH} in their
TableII. Their result, however, does not exactly coincide with either (\ref{eq:R1 `up' approx l->inf,r->rplus}) or
(\ref{eq:R1 `up' approx l->inf,r->rplus;compact version})
(as a matter of fact, in their table there is a quantity $\rho$ that they have not defined and it cannot be the spin coefficient
as it cannot have a $\theta$-dependency). 

Note that there is no reason why the `up' radial modes in (\ref{eq:R1 `up' approx l->inf,r->rplus}), or including the normalization
constant in (\ref{eq:R1 `up' approx l->inf,r->rplus;compact version}) should diverge in the stated limits. 
In fact, they clearly do not in the case of helicity $-1$. 
It is only when other factors (as in (\ref{eq:phi_0/2(in/up)})) are included that the resulting expressions we will deal with
in Section \ref{sec:RRO} diverge in this limit.

We are also interested in finding the result of applying the operator $\mathcal{D}_0^{\dagger}\Delta$ on the asymptotic
solution (\ref{eq:R1 `up' approx l->inf,r->rplus;compact version}), as we will need this result in later
calculations.
It immediately follows from (\ref{eq:R1 `up' approx l->inf,r->rplus;compact version}) and (\ref{eq:def.A_s}) that
\begin{equation} \label{eq:Ddagger Delta R1 `up' approx l->inf,r->rplus}
\begin{aligned}
&\frac{1}{A_{\indhel}N}\mathcal{D}_0^{\dagger}\left(\Delta{}_{\indhel}R^{\text{up}}_{lm\omega}\right) \rightarrow \\
&\rightarrow (r_+-r_-)x^{-\indhel /2} \left[\left(-\frac{\indhel}{2}+1+i\frac{q}{2}\right)K_{\indhel+iq}(2lx^{1/2})+
lx^{1/2}K'_{\indhel+iq}(2lx^{1/2})
\right]  \\
& \qquad \qquad \qquad  \qquad \qquad\qquad \qquad\qquad(l\rightarrow +\infty,r \rightarrow r_+,lx^{1/2}\ \text{finite})
\end{aligned}
\end{equation}
An important simplification for $\indhel =+1$ happens when using the recurrence relation (~\cite{bk:AS})
\begin{equation} \label{eq:eq.9.6.26A&S}
K'_{\nu}(z)=-K_{\nu-1}(z)-\frac{\nu}{z}K_{\nu}(z)
\end{equation}
for the modified Bessel function. 
Expression (\ref{eq:Ddagger Delta R1 `up' approx l->inf,r->rplus}) then reduces to
\begin{equation} \label{eq:Ddagger Delta R+1 `up' approx l->inf,r->rplus}
\frac{1}{A_{\indhel}N}\mathcal{D}^{\dagger}_0\left(\Delta{}_{+1}R^{\text{up}}_{lm\omega}\right) \rightarrow 
-\frac{(r_+-r_-)}{2x^{1/2}} K_{iq} \qquad (l\rightarrow +\infty,r \rightarrow r_+)
\end{equation}
\catdraft{hauria de ser $(l\rightarrow +\infty,r \rightarrow r_+,lx^{1/2}\ \text{finite})$ a (\ref{eq:Ddagger Delta R+1 `up' approx l->inf,r->rplus})?}
We used the program \program{raddrv2KN.f} described in Section \ref{sec:num. method; radial func.}
to compare the numerical solution with the analytic
asymptotic approximation (\ref{eq:R1 `up' approx l->inf,r->rplus}) we have found.
Graphs \ref{fig:plot candelas'80 radial approx,w=0.5}--\ref{fig:relat.err. plot candelas'80 radial approx,w=0.01}
show that this approximation indeed tends to the non-approximated (numerical) solution and that as 
$r \rightarrow r_+$ the approximation is better for the higher values of $l$, as predicted.


\begin{figure}[p]
\centering
\includegraphics*[width=70mm]{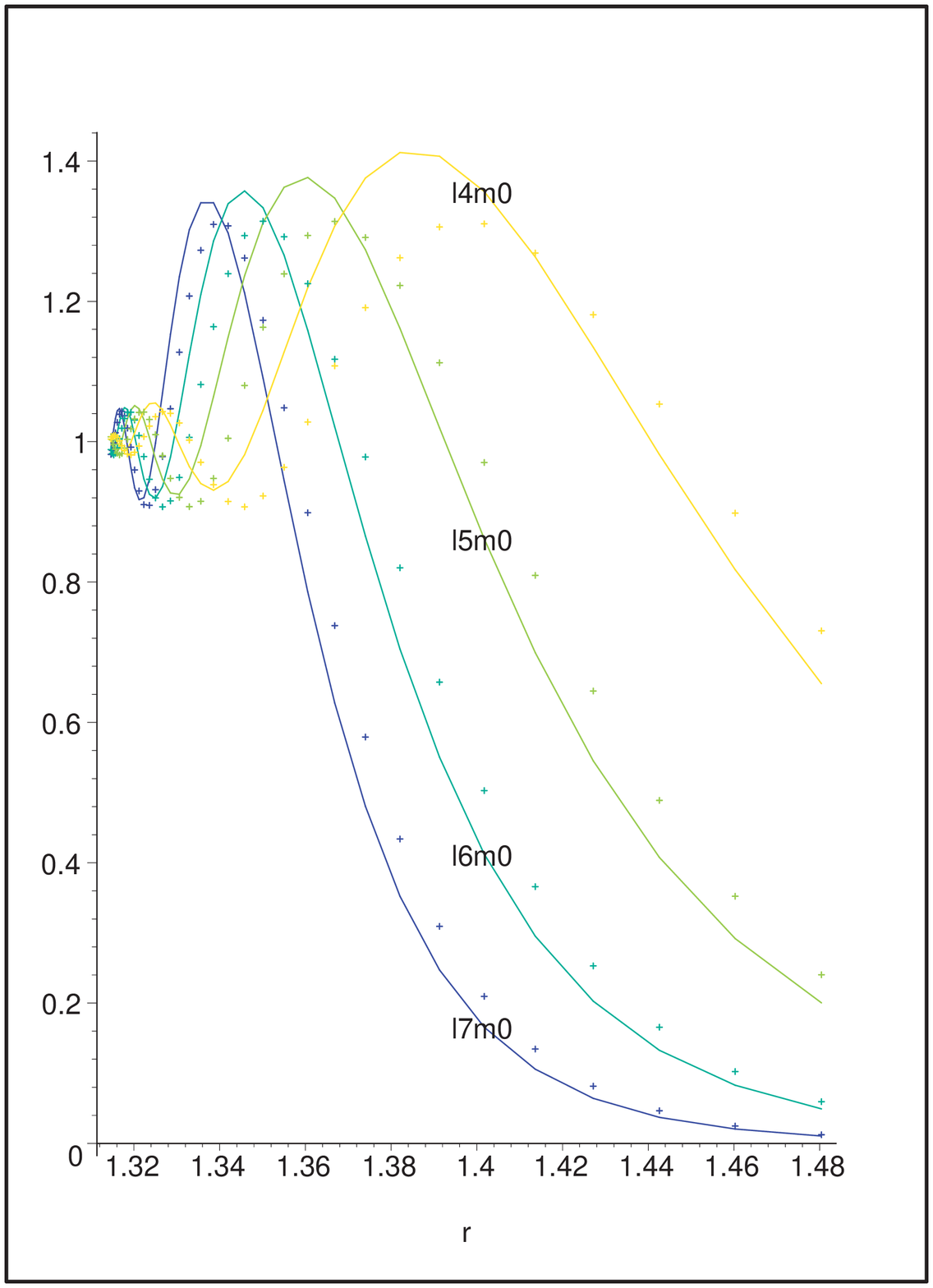}
\caption{$\abs{{}_{-1}R^{\text{up}}_{lm\omega}}^2$, $l=4\dots 7$, $m=0,\omega=0.5$ (dots are the numerical solution and straight lines are the 
approximation from (\ref{eq:R1 `up' approx l->inf,r->rplus})).}
\label{fig:plot candelas'80 radial approx,w=0.5}
\includegraphics*[width=70mm]
{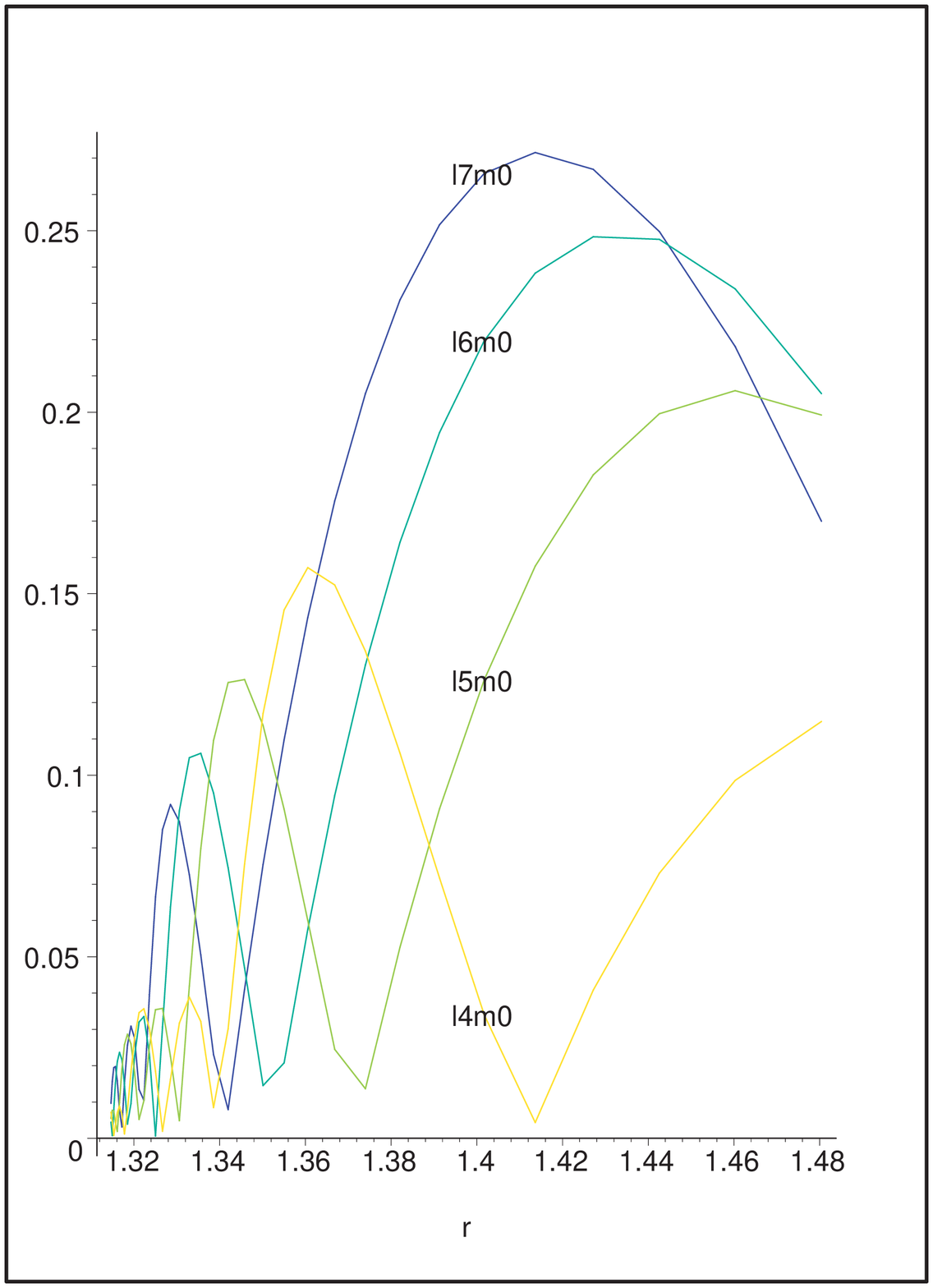}  
\caption{
Relative error $\frac{\abs{\abs{{}_{-1}R^{\text{up,num}}_{lm\omega}}^2-\abs{{}_{-1}R^{\text{up,approx}}_{lm\omega}}^2}}
{\abs{{}_{-1}R^{\text{up,num}}_{lm\omega}}^2}$, $l=4\dots 7$, $m=0$, $\omega=0.5$.}
\label{fig:relat.err. plot candelas'80 radial approx,w=0.5}
\end{figure}

\begin{figure}[p]
\centering
\includegraphics*[width=70mm]{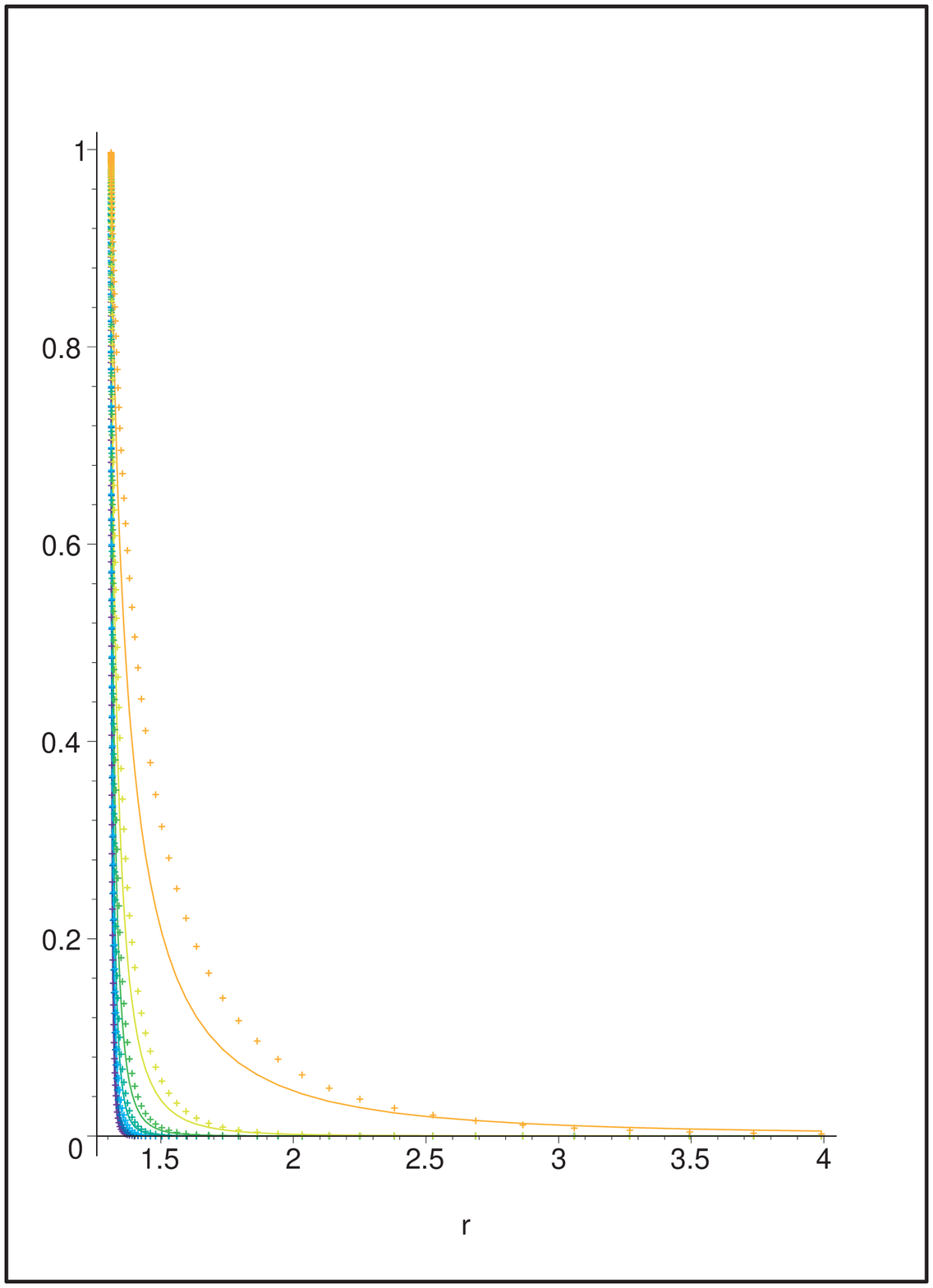}
\caption{$\abs{{}_{-1}R^{\text{up}}_{lm\omega}}^2$, $l=4\dots 7$, $m=0$, $\omega=0.01$.
Correspondence between colours and modes is the same as in Figure \ref{fig:plot candelas'80 radial approx,w=0.5}.
(dots are the numerical solution and straight lines are the approximation from (\ref{eq:R1 `up' approx l->inf,r->rplus})).}
\label{fig:plot candelas'80 radial approx,w=0.01}
\includegraphics*[width=70mm]{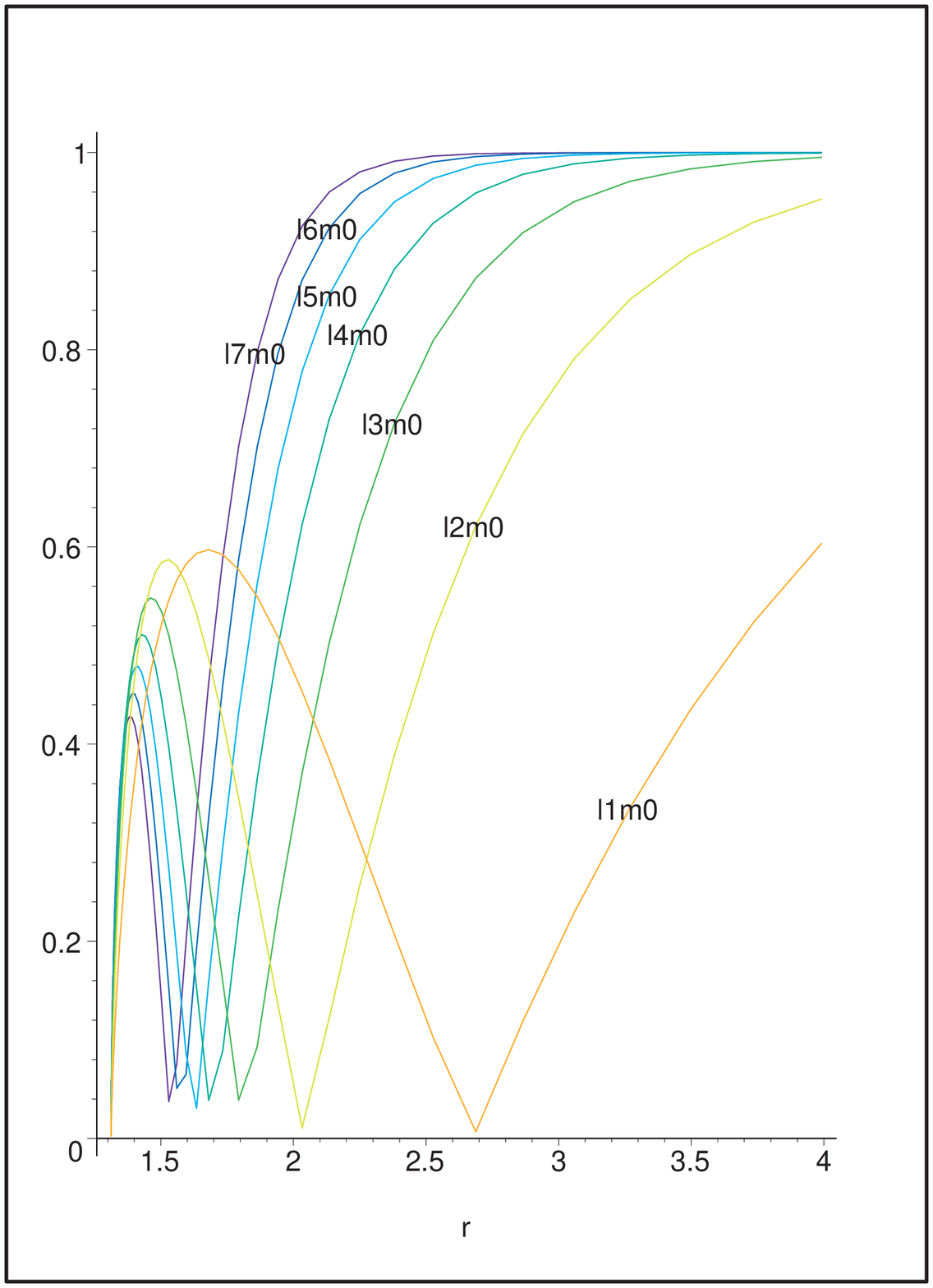} 
\caption{Relative error $\frac{\abs{\abs{{}_{-1}R^{\text{up,num}}_{lm\omega}}^2-\abs{{}_{-1}R^{\text{up,approx}}_{lm\omega}}^2}}
{\abs{{}_{-1}R^{\text{up,num}}_{lm\omega}}^2}$, $l=1\dots 7$, $m=0$, $\omega=0.01$.}
\label{fig:relat.err. plot candelas'80 radial approx,w=0.01}
\end{figure}


\section{Asymptotics for small frequency}

We conclude this chapter with an asymptotic analysis of the radial functions for $M\omega \ll 1$.
This analysis follows that of Page ~\cite{ar:PageI'76} for general spin `in' perturbations in the Kerr background.
We complete it by generalizing it to the Kerr-Newman space-time and obtaining the asymptotics for the `up' solutions as well.
This analysis is useful for studying what are the contributions to the Fourier sum of modes with $\omega \sim 0$.

The radial Teukolsky equation can be rewritten as
\begin{equation} \label{eq:sln approx eqA7Page'76} 
\begin{aligned}
&
x^2(x+1)^2\ddiff{{}_{\indhel}R_{lm\omega}}{x}+(\indhel+1)x(x+1)(2x+1)\diff{{}_{\indhel}R_{lm\omega}}{x}+
\\ & +
\Bigg\{k_p^2x^4+
2k_px^3\left[\alpha_pk_p+i\indhel\right]
+x^2\left[-{}_{\indhel}\lambda_{lm\omega}+(3i\indhel+q)k_p+\alpha_p^2k_p^2\right]+
\\&+
x\left[-{}_{\indhel}\lambda_{lm\omega}-i\indhel q+(i\indhel \alpha_p+q\alpha_p)k_p\right]
-i\indhel \frac{q}{2}+\frac{q^2}{4}\Bigg\}{}_{\indhel}R_{lm\omega}=0
\end{aligned}
\end{equation}

where we have defined
\begin{subequations} \label{eq:def.x,Q_p,k_p,alpha_p}
\begin{align}
k_p &\equiv \omega (r_+-r_-)                        \\
\alpha_p &\equiv \frac{2r_+}{r_+-r_-}
\end{align}
\end{subequations}
Approximating this equation for $k_p<<1$ by keeping the term with the lowest order in $k_p$ in the 
coefficient of each power of $x$ we obtain
\begin{equation} \label{eq:eqA7Page'76}
\begin{aligned}
&x^2(x+1)^2\ddiff{{}_{\indhel}R_{lm\omega}}{x}+(\indhel+1)x(x+1)(2x+1)\diff{{}_{\indhel}R_{lm\omega}}{x}+ \\
&\qquad \qquad +\left[k_p^2x^4+2i\indhel k_px^3-{}_{\indhel}\lambda_{lm\omega} x(x+1)-\frac{i\indhel q}{2}(2x+1)+\frac{q^2}{4}\right]{}_{\indhel}R_{lm\omega}=0
\end{aligned}
\end{equation}

In what follows we take ${}_{\indhel}\lambda_{lm\omega} =(l-\indhel )(l+\indhel+1)$, which is exact for $a\omega=0$ and it is an 
approximation for $a\omega <<1$. Alternatively, we will consider it to be exact for $a\omega <<1$ and use it as the redefinition 
of the parameter $l$. This implies that now $l$ is nearly, but not an exactly, an integer. This will 
avoid possible singularities in the $\Gamma$-functions appearing in the solutions obtained below. 

We find now approximations to the solution of the differential equation (\ref{eq:eqA7Page'76}) in two different limits, which have
an overlapping region. We will then proceed to match the two approximations in this common region of validity.

\begin{itemize}

\item Approximation for $k_px<<l+1$

The approximation for $k_px<<l+1$ is obtained by keeping only the lowest term in $k_px$ in 
equation (\ref{eq:eqA7Page'76}) and re-writing it as

\begin{equation} \label{eq:approx low k_px eqA7Page'76}
\begin{aligned}
\ddiff{{}_{\indhel}R_{lm\omega}}{z}
&+(\indhel+1)\left[\frac{\indhel }{z}+\frac{1}{z-1}\right]\diff{{}_{\indhel}R_{lm\omega}}{z}-\\
& \qquad \qquad -\frac{{}_{\indhel}R_{lm\omega}}{z(z-1)}
\left[{}_{\indhel}\lambda_{lm\omega}-\frac{q^2-2i\indhel q}{4z}+\frac{q^2+2i\indhel q}{4(z-1)}\right]=0
\end{aligned}
\end{equation}
where $z \equiv -x$.
The general solution of the differential equation (\ref{eq:approx low k_px eqA7Page'76}), which we denote by $R_1$, is
given in terms of the hypergeometric functions as
\begin{equation} \label{eq:sln approx low k_px eqA7Page'76} 
\begin{aligned}
&R_1=C_1x^{-\indhel-iq/2}(x+1)^{-\indhel+iq/2}{}_2F_{1}(-l-\indhel ,l-\indhel+1;1-\indhel-iq;-x)+ 
\\ &
+C_2(-1)^{\indhel}x^{+iq/2}(x+1)^{-\indhel+iq/2}{}_2F_1(-l+iq,1+l+iq;1+\indhel+iq;-x)
\end{aligned}
\end{equation}
where $C_1$ and $C_2$ are constants of integration.
The asymptotic behaviour of this solution close to the horizon and for large $r$ is
\begin{subequations} 
\begin{align}
&R_1 \sim
C_1(r_{+}-r_{\_})^{2\indhel}I_{\tilde{\omega}}^*\Delta^{-\indhel }e^{-i\tilde{\omega}r_{*}}+
C_2I_{\tilde{\omega}}e^{+i\tilde{\omega}r_{*}}, \qquad (x\rightarrow 0)
\label{eq:sln approx low k_px eqA7Page'76,r->rplus} \\
\begin{split}
&R_1 \sim \\
&\sim      C_1 \left[\frac{\Gamma(1-\indhel-iq)\Gamma(1+2l)}{\Gamma(1-\indhel+l)\Gamma(1+l-iq)}x^{l-\indhel }+
\right.\\&\left.\qquad \qquad \qquad 
          +\frac{\Gamma(1-\indhel-iq)\Gamma(-1-2l)}{\Gamma(-\indhel-l)\Gamma(-l-iq)}x^{-l-\indhel-1}\right]+
\\&
+(-1)^{\indhel}C_2 \left[\frac{\Gamma(1+\indhel+iq)\Gamma(1+2l)}{\Gamma(1+l+iq)\Gamma(1+l+\indhel )}x^{l-\indhel }+
\right.\\&\left.\qquad \qquad \qquad 
          +\frac{\Gamma(1+\indhel+iq)\Gamma(-1-2l)}{\Gamma(-l+iq)\Gamma(\indhel-l)}x^{-l-\indhel-1}\right],
 \qquad (x>>\abs{q}/2+1)
\end{split}
\end{align}
\end{subequations}

\item Approximation for $x>>\abs{q}/2+1$

In order to obtain a solution of the differential equation (\ref{eq:eqA7Page'76}) which is valid in the region $x>>|q|/2+1$ 
we only keep the terms in the equation with higher powers in $x$:
\begin{equation} \label{eq:approx large x eqA7Page'76}
\ddiff{{}_{\indhel}R_{lm\omega}}{x}+\frac{2(\indhel+1)}{x}\diff{{}_{\indhel}R_{lm\omega}}{x}+
\left[k_p^2+\frac{2i\indhel k_p}{x}-\frac{{}_{\indhel}\lambda_{lm\omega}}{x^2}\right]{}_{\indhel}R_{lm\omega}=0
\end{equation}
Its general solution, which we denote by $R_2$ is given in terms of the confluent hypergeometric functions as
\begin{equation} \label{eq:sln approx large x eqA7Page'76} 
\begin{aligned}
R_2&=D_1x^{-\indhel+l}e^{-ik_px}{}_1F_{1}(1-\indhel+l,2l+2;2ik_px)+
\\&+
D_2x^{-1-\indhel-l}e^{-ik_px}{}_1F_1(-l-\indhel ,-2l;2ik_px)
\end{aligned}
\end{equation}
where $D_1$ and $D_2$ are constants of integration.
The behaviour of this solution in the region of overlap with $R_1$ and for large $r$ is
\begin{subequations}
\begin{align}
&R_2 \sim
D_1x^{l-\indhel }+D_2x^{-1-\indhel-l}, (k_px<<l+1)
\label{eq:sln approx large x eqA7Page'76,kx<<l+1} \\
\begin{split}
&R_2\sim \\
&\sim 
      x^{-1}e^{-ik_px} \left[\frac{\Gamma(2l+2)}{\Gamma(1+\indhel+l)}D_1(-2ik_p)^{-1+\indhel-l}+
                             \frac{\Gamma(-2l)}{\Gamma(+\indhel-l)}D_2(-2ik_p)^{\indhel+l}\right]+
\\&+
      x^{-2\indhel-1}e^{+ik_px} \left[\frac{\Gamma(2l+2)}{\Gamma(1-\indhel+l)}D_1(2ik_p)^{-1-\indhel-l}+
                                \frac{\Gamma(-2l)}{\Gamma(-\indhel-l)}D_2(2ik_p)^{-\indhel+l}\right],
\\ &\qquad \qquad \qquad \qquad \qquad\qquad \qquad\qquad \qquad\qquad\qquad \qquad (x \rightarrow +\infty)
\end{split}
\end{align}
\end{subequations}

\end{itemize}

We can now match the two solutions $R_1$ and $R_2$ in the region of overlap, which is given by 
$\abs{q}/2+1<<x<<(l+1)/k_p$. We obtain:

\newcommand{\newcommandtest}{\frac{\Gamma(1-\indhel-iq)\Gamma(1+2l)}{\Gamma(1-\indhel+l)\Gamma(1+l-iq)}}

\begin{equation} \label{eq:rlns between C and D coeffs}
\begin{aligned}
D_1&=  C_1 \newcommandtest+
(-1)^{\indhel}C_2 \frac{\Gamma(1+\indhel+iq)\Gamma(1+2l)}{\Gamma(1+l+iq)\Gamma(1+l+\indhel )}                 \\
D_2&=  C_1 \frac{\Gamma(1-\indhel-iq)\Gamma(-1-2l)}{\Gamma(-\indhel-l)\Gamma(-l-iq)}+
(-1)^{\indhel}C_2 \frac{\Gamma(1+\indhel+iq)\Gamma(-1-2l)}{\Gamma(-l+iq)\Gamma(\indhel-l)}
\end{aligned}
\end{equation}

Relations (\ref{eq:rlns between C and D coeffs}) coincide with the equivalent ones in ~\cite{ar:J&McL&Ott'95} in the limit $Q=a=0$ and 
with the equivalent ones in ~\cite{th:GavPhD} in the limit $Q=\indhel=0$.

We finally proceed to obtain the behaviour for small frequency of the `in' and `up' radial functions and coefficients:

\begin{itemize}

\item `in' solution:

Comparing the behaviour of $R_1$ for $r \rightarrow r_+$ in 
equation (\ref{eq:sln approx low k_px eqA7Page'76,r->rplus}) with the `in' WKB approximation (\ref{eq:R_in}) in the same 
region we find
\begin{equation} \label{eq:C_in in terms of R_in coeffs}
\begin{aligned}
C^{\text{in}}_1&=\frac{{}_{\indhel}R^{\text{in,tra}}_{lm\omega}I_{\tilde{\omega}}}{(r_{+}-r_{\_})^{2\indhel}}                   \\
C^{\text{in}}_2&=0
\end{aligned}
\end{equation}

Similarly, comparing $R_2$ and the `in' WKB approximation solution for $r \rightarrow +\infty$ we find
\begin{equation} \label{eq:match R_2 and R_in for r->inf}
\begin{aligned}
\frac{\Gamma(2l+2)}{\Gamma(1+\indhel+l)}D^{\text{in}}_1(-2ik_p)^{-1+\indhel-l}+
&\frac{\Gamma(-2l)}{\Gamma(+\indhel-l)}D^{\text{in}}_2(-2ik_p)^{\indhel+l}=  \\
                  & \qquad \qquad =\frac{{}_{\indhel}R^{\text{in,inc}}_{lm\omega}}{(r_{+}-r_{\_})}                                         \\
\frac{\Gamma(2l+2)}{\Gamma(1-\indhel+l)}D^{\text{in}}_1(2ik_p)^{-1-\indhel-l}+
&\frac{\Gamma(-2l)}{\Gamma(-\indhel-l)}D^{\text{in}}_2(2ik_p)^{-\indhel+l}=  \\
                  & \qquad \qquad =\frac{{}_{\indhel}R^{\text{in,ref}}_{lm\omega}}{(r_{+}-r_{\_})^(1+2\indhel)}
\end{aligned}
\end{equation}

Combining together equations (\ref{eq:rlns between C and D coeffs}), (\ref{eq:C_in in terms of R_in coeffs})
and (\ref{eq:match R_2 and R_in for r->inf}) we can
find asymptotic expressions for small frequency for the coefficients of the `in' radial solutions:
\begin{equation} \label{eq:R_in coeffs,w->0}
\begin{aligned}
&\frac{{}_{\indhel}R^{\text{in,tra}}_{lm\omega}}{{}_{\indhel}R^{\text{in,inc}}_{lm\omega}} \sim \\
&\sim (r_{+}-r_{\_})^{2\indhel-1}I_{\tilde{\omega}}^*
\frac{\Gamma(1-\indhel+l)\Gamma(1+\indhel+l)\Gamma(1+l-iq)}{\Gamma(1+2l)\Gamma(2+2l)\Gamma(1-\indhel-iq)}(-2ik_p)^{1-\indhel+l} 
\\ & \qquad\qquad\qquad\qquad\qquad\qquad\qquad\qquad\qquad\qquad\qquad\qquad\qquad(k_p  \rightarrow 0)\\
&\frac{{}_{\indhel}R^{\text{in,ref}}_{lm\omega}}{{}_{\indhel}R^{\text{in,inc}}_{lm\omega}} \sim (-1)^{1+\indhel+l}(r_{+}-r_{\_})^{2\indhel}
\frac{\Gamma(1+\indhel+l)}{\Gamma(1-\indhel+l)}(2ik_p)^{-2\indhel}   
\\ & \qquad\qquad\qquad\qquad\qquad\qquad\qquad\qquad\qquad\qquad\qquad\qquad\qquad(k_p  \rightarrow 0)
\end{aligned}
\end{equation}

In particular, for helicity $-1$,
\begin{equation} \label{eq:R_1_in coeffs,w->0}
\begin{aligned}
&\frac{{}_{-1}R^{\text{in,tra}}_{lm\omega}}{{}_{-1}R^{\text{in,inc}}_{lm\omega}}\sim
\frac{I_{\tilde{\omega}}^*}{(r_{+}-r_{\_})^{3}}
\frac{\Gamma(2+l)\Gamma(l)\Gamma(1+l-iq)}{\Gamma(1+2l)\Gamma(2+2l)\Gamma(2-iq)}
(-2ik_p)^{2+l} \propto \omega^{2+l} \\
& \qquad\qquad\qquad\qquad\qquad\qquad \qquad\qquad\qquad \qquad\qquad\qquad\qquad\quad(k_p  \rightarrow 0)                 \\
&\frac{{}_{-1}R^{\text{in,ref}}_{lm\omega}}{{}_{-1}R^{\text{in,inc}}_{lm\omega}} \sim (-1)^l\frac{1}{(r_{+}-r_{\_})^{2}}
\frac{\Gamma(l)}{\Gamma(2+l)}(2ik_p)^2 \propto \omega^2 \quad \qquad \qquad\quad(k_p  \rightarrow 0)
\end{aligned}
\end{equation}

The asymptotic behaviour for small frequency of the `in' radial solution is obtained 
from equations (\ref{eq:sln approx low k_px eqA7Page'76}), (\ref{eq:C_in in terms of R_in coeffs}) 
and (\ref{eq:R_in coeffs,w->0}):
\begin{equation}  \label{eq:R_in,w->0}
\begin{aligned}
&\frac{{}_{\indhel}R^{\text{in}}_{lm\omega}}{{}_{\indhel}R^{\text{in,inc}}_{lm\omega}} \sim \\
&\sim (r_{+}-r_{\_})^{l+iq/2}(r-r_{+})^{-\indhel-iq/2}
\frac{\Gamma(1-\indhel+l)\Gamma(1+\indhel+l)\Gamma(1+l-iq)}{\Gamma(1+2l)\Gamma(2+2l)\Gamma(1-\indhel-iq)}\times
\\&
\times{}_2F_{1}\left(-l-\indhel ,l-\indhel+1;1-\indhel-iq;-\left(\frac{r-r_{+}}{r_{+}-r_{\_}}\right)\right)(-2i\omega )^{1-\indhel+l} \\
& \qquad\qquad\qquad\qquad\qquad\qquad\qquad\qquad\qquad\qquad\qquad\qquad (k_px<<l+1)
\end{aligned}
\end{equation}

In particular, for helicity $\pm 1$,
\begin{equation} \label{eq:R_1_in,w->0}
\begin{aligned}
&\frac{{}_{-1}R^{\text{in}}_{lm\omega}}{{}_{-1}R^{\text{in,inc}}_{lm\omega}} \sim \\
& \sim (r_{+}-r_{\_})^{l+iq/2}(r-r_{+})^{1-iq/2}
\frac{\Gamma(2+l)\Gamma(l)\Gamma(1+l-iq)}{\Gamma(1+2l)\Gamma(2+2l)\Gamma(2-iq)}\times
\\&
\times{}_2F_{1}\left(1-l,2+l;2-iq;-\left(\frac{r-r_{+}}{r_{+}-r_{\_}}\right)\right)(-2i\omega )^{2+l},  \quad (k_px<<l+1) \\
&
\frac{{}_{+1}R^{\text{in}}_{lm\omega}}{{}_{+1}R^{\text{in,inc}}_{lm\omega}} \sim \\
& \sim  (r_{+}-r_{\_})^{l+iq/2}(r-r_{+})^{-1-iq/2}
\frac{\Gamma(l)\Gamma(2+l)\Gamma(1+l-iq)}{\Gamma(1+2l)\Gamma(2+2l)\Gamma(-iq)}\times
\\&
\times{}_2F_{1}\left(-1-l,l;-i;-\left(\frac{r-r_{+}}{r_{+}-r_{\_}}\right)\right)(-2i\omega )^{l}, \quad\qquad\quad (k_px<<l+1)
\end{aligned}
\end{equation}

\item `up' solution:

Proceeding likewise for the `up' radial solution with WKB approximation (\ref{eq:R_up}), we find
\begin{equation} \label{eq:C_up in terms of R_up coeffs}
\begin{aligned}
C^{\text{up}}_1&=\frac{{}_{\indhel}R^{\text{up,ref}}_{lm\omega}I_{\tilde{\omega}}}{(r_{+}-r_{\_})^{2\indhel}}                   \\
C^{\text{up}}_2&={}_{\indhel}R^{\text{up,inc}}_{lm\omega}I_{\tilde{\omega}}^*
\end{aligned}
\end{equation}
and
\begin{equation} \label{eq:match R_2 and R_up for r->inf,a)}
\frac{\Gamma(2l+2)}{\Gamma(1+\indhel+l)}D^{\text{up}}_1(-2ik_p)^{-1+\indhel-l}+
\frac{\Gamma(-2l)}{\Gamma(+\indhel-l)}D^{\text{up}}_2(-2ik_p)^{\indhel+l}=
0                                     
\end{equation}
\begin{equation} \label{eq:match R_2 and R_up for r->inf,b)}
\begin{aligned}
\frac{\Gamma(2l+2)}{\Gamma(1-\indhel+l)}D^{\text{up}}_1(2ik_p)^{-1-\indhel-l}+
&\frac{\Gamma(-2l)}{\Gamma(-\indhel-l)}D^{\text{up}}_2(2ik_p)^{-\indhel+l}= \\
&\qquad \qquad=\frac{{}_{\indhel}R^{\text{up,tra}}_{lm\omega}}{(r_{+}-r_{\_})^(1+2\indhel)}
\end{aligned}
\end{equation}

Proceeding as for the `in' case, from equations (\ref{eq:C_up in terms of R_up coeffs}), (\ref{eq:rlns between C and D coeffs}) 
and (\ref{eq:match R_2 and R_up for r->inf,a)}) we find
\begin{equation} \label{eq:Rs_up_ref,w->0}
\begin{aligned}
&\frac{{}_{\indhel}R^{\text{up,ref}}_{lm\omega}}{{}_{\indhel}R^{\text{up,inc}}_{lm\omega}} \sim \\
&\sim (-1)^{\indhel+1}I_{\tilde{\omega}}^{* 2}
\frac{\Gamma(1+\indhel+iq)\Gamma(1-\indhel+l)\Gamma(1+l-iq)}
{\Gamma(1+l+iq)\Gamma(1+\indhel+l)\Gamma(1-\indhel-iq)}(r_{+}-r_{\_})^{2\indhel}, 
\quad (k_p  \rightarrow 0)
\end{aligned}
\end{equation}
which agrees with ~\cite{ar:J&McL&Ott'95} in the limit $Q=0=0$ and $\indhel=+2$.
\draft{do not know how to solve problem in paragraph below}

Proceeding the same way we are not able to find the asymptotic behaviour for ${}_{\indhel}R^{\text{up,tra}}_{lm\omega}$.
It follows from (\ref{eq:C_up in terms of R_up coeffs}) and (\ref{eq:rlns between C and D coeffs}), that in the expressions
for $D^{\text{up}}_1$ and $D^{\text{up}}_2$ in terms of ${}_{\indhel}R^{\text{up,inc}}_{lm\omega}$ and ${}_{\indhel}R^{\text{up,ref}}_{lm\omega}$ 
there appears no explicit $k_p$. This means that,
when these expressions are inserted in (\ref{eq:match R_2 and R_up for r->inf,a)}), then ${}_{\indhel}R^{\text{up,ref}}_{lm\omega}$ 
will be such that, to its 
lowest order in $k_p $, makes $D^{\text{up}}_1$ zero. This implies that, when the expressions are inserted in 
(\ref{eq:match R_2 and R_up for r->inf,b)}), to the lowest order in the calculations, ${}_{\indhel}R^{\text{up,tra}}_{lm\omega}$ 
will be zero. We can, however, find the asymptotic behaviour of ${}_{-1}R^{\text{up,tra}}_{lm\omega}/{}_{-1}R^{\text{up,inc}}_{lm\omega}$ 
by using the wronskian relations.
From (c) in Table \ref{table:radial wronsks} and equation (\ref{eq:R_1_in coeffs,w->0}) we obtain
\begin{equation} \label{eq:R_1_up_tra,w->0}
\begin{aligned}
&\frac{{}_{-1}R^{\text{up,tra}}_{lm\omega}}{{}_{-1}R^{\text{up,inc}}_{lm\omega}} \rightarrow 
-I_{\tilde{\omega}}^*\frac{\Gamma(2+l)\Gamma(l)\Gamma(1+l-iq)}
{\Gamma(2+2l)\Gamma(1+2l)\Gamma(1-iq)}\frac{(-2ik_p)^{l+1}}{(r_{+}-r_{\_})}
\qquad (k_p  \rightarrow 0)
\end{aligned}
\end{equation}

\end{itemize}

It is easy to check that all the coefficients with $\indhel =-1$ , given by (\ref{eq:R_1_in coeffs,w->0}),
(\ref{eq:Rs_up_ref,w->0}) and (\ref{eq:R_1_up_tra,w->0}), do satisfy the wronskian relations 
in Table \ref{table:radial wronsks}. For the coefficients with $\indhel =+1$, it can also be checked that 
${}_{+1}R^{\text{in,ref}}_{lm\omega}/{}_{-1}R^{\text{in,ref}}_{lm\omega}$, 
${}_{+1}R^{\text{in,tra}}_{lm\omega}/{}_{-1}R^{\text{in,tra}}_{lm\omega}$ and 
${}_{+1}R^{\text{up,ref}}_{lm\omega}/{}_{-1}R^{\text{up,ref}}_{lm\omega}$ satisfy the relations 
(\ref{eq:R1 coeffs from R_1's}).

\chapter{Spin-weighted spheroidal harmonics}  \label{ch:SWSH}

\draft{include graphs for SWSH and eigenvals. in ch.\ref{ch:SWSH} (apart from the ones in next chapter)?}

\section{Introduction}

The angular Teukolsky equation resulting from the separation of the Teukolsky equation 
and its solution are important for several reasons.
For one, the angular equation determines the eigenvalue which appears in the radial equation. 
The angular solution is of particular interest to us due to its central role in
the parity transformation $\{\theta\leftrightarrow \pi-\theta, \phi\leftrightarrow \phi+\pi\}$, 
an issue that we deal with in Chapter \ref{stress-energy tensor}. 
Finally, the asymptotic behaviour of the eigenvalues and the angular solution for large frequency is 
important in its own right and is the subject of Chapter \ref{ch:high freq. spher}.

In the remainder of this section we present a summary of the main results in the literature
relating to the angular Teukolsky equation and its solution in the various limits of values of
its parameters. 
In the next section, we present the asymptotic behaviour close to the 
boundary points of the angular solutions as well as other basic properties that we have obtained, 
all of which are needed for later calculations. In Section \ref{sec:num. method; SWSH} we describe
the numerical method, algorithms and their implementation in different Fortran90 programs
that we have used in order to numerically solve the angular differential equation. In the last section
of this chapter we display and analyze the numerical results that we have 
obtained and we compare them against previous results in the literature.

When expanding the field $\Omega_h$ in terms of the Fourier modes (\ref{eq:Fourier expansion for Omega_h}), the Teukolsky
equation (\ref{eq:Teuk.eq.}) becomes separable. The corresponding angular ordinary differential equation is
\begin{equation} \label{eq:ang. teuk. eq.}
\left[
\frac{d}{dx}\left((1-x^{2})\frac{d}{dx}\right)+c^{2}x^{2}-2\indhel cx-\frac{(m+\indhel x)^{2}}{1-x^{2}}+{}_{\indhel}\mathcal{A}_{lm\omega}+\indhel 
\right] {}_{\indhel}S_{lm\omega}(x)=0
\end{equation}
where the new variables $x\equiv \cos \theta$ and $c\equiv a\omega $ have been defined. The constant of separation between
the angular and radial equations is 
\begin{equation} \label{eq:lambda}
{}_{\indhel}\mathcal{A}_{lm\omega}\equiv {}_{\indhel}E_{lm\omega}-\indhel (\indhel+1)\equiv {}_{\indhel}\lambda_{lm\omega}-c^{2}+2mc
\end{equation}
It might be more logical to label the angular solutions and the eigenvalues by $c$ rather than $\omega$,
but following the convention of literature on the Teukolsky equation we label them by $\omega$.

The differential equation (\ref{eq:ang. teuk. eq.}) has two regular singular points at $x=\pm 1$ and one 
essential singularity \catdraft{(quina es la diff. amb 'irregular pt.'?)} at $x=\infty$.
We are only interested, however, in solutions for real values of the independent variable $x$ that lie
in the interval $x\in[-1,+1]$. 
We henceforth restrict $x$ to this range of validity and therefore we have only to consider the two regular singular points at $x=\pm 1$.
The differential equation (\ref{eq:ang. teuk. eq.}), together with the 
boundary condition that its solution ${}_{\indhel}S_{lm\omega}(x)$ is regular for $x\in[-1,+1]$,
is a parametric eigenvalue problem, the parameters being $c$, $m$ and $\indhel$. 
\catdraft{p.73BRW;p.69,70Stewart'75: no tinc clar quins son els parametres?!}
This differential equation is called the \define{spin-weighted spheroidal differential equation}; \catdraft{(check!)}
it reduces to the 
\define{spin-weighted spherical differential equation} \catdraft{(check!)}
when $c=0$ (but $\indhel\neq 0$),
it reduces to the \define{spheroidal differential equation} when $\indhel=0$  (but $c\neq 0$) and to the 
\define{spherical differential equation} 
when $c=0$ as well as $\indhel=0$. 
All four equations possess the same singularity structure described above.
The spheroidal differential equation also follows from the scalar wave equation in flat space-time 
separated in oblate spheroidal co-ordinates (see ~\cite{ar:Leaver86}).

The physical requirements of single-valuedness and of regularity at $x=\pm 1$ requires that $l$ and $m$ are integers
with $|m|\leq l$. 
\catdraft{1) quina cond. fa que hagi de ser $|m|\leq l$?, 2) a p.48Hartle\&Wilk'74 diu que $|m|<l$?}
In the particular case of integral $l$ and $m$, the spherical differential equation and its regular \ddraft{he afegit 'regular' jo pero ho he de comprovar!} 
solution are respectively called the \define{associated Legendre equation} 
and \define{associated Legendre functions of the first kind} $P^m_l(\theta)$. 
Similarly, for integral $l$ and $m$, the spheroidal differential equation and its regular solution are respectively
called the \define{oblate angular differential equation} and \define{oblate angle functions of the first kind} $S_{lm\omega}(\theta)$
(the other independent solutions of this equation have logarithmic singularities at $x=\pm1$ and are called 
\define{oblate angle functions of the second kind}).

When the regular \catdraft{segur?} solution of the spherical equation, the spin-weighted spherical equation, 
the spheroidal equation or the spin-weighted spheroidal equation is multiplied by $e^{im\phi}$, 
we then obtain
the 
\define{spherical harmonics} $Y_{lm}(\theta,\phi)$,
the \define{spin-weighted spherical harmonics} ${}_{\indhel}Y_{lm}(\theta,\phi)$,
the \define{spheroidal harmonics} $Z_{lm\omega}(\theta)$ or
the \define{spin-weighted spheroidal harmonics (SWSH)} ${}_{\indhel}Z_{lm\omega}(\theta)$
respectively.
With a slight abuse of terminology, common throughout the literature, we will also refer to the solutions
$S_{lm\omega}(\theta)$ and ${}_{\indhel}S_{lm\omega}(\theta)$ as spheroidal harmonics and
spin-weighted spheroidal harmonics (SWSH) respectively.
It will be clear from the context which of the two possible functions we are referring to.
See Table \ref{table:notation angular funcs.} for a summary of the notation of the different angular functions we have introduced.

\begin{table} 
\begin{center}
\begin{tabular}{c|c}
Values of parameters  & Name of solution of corresponding equation         \\
\hline
\hline
$c=0$, $\indhel=0$        & Spherical harmonics $Y_{lm}(\theta,\phi)$                         \\
\hline
$c\neq0$, $\indhel=0$     & Spheroidal harmonics $S_{lm\omega}(\theta)$, $Z_{lm\omega}(\theta,\phi)$                       \\
\hline
$c=0$, $\indhel\neq0$     & Spin-weighted spherical harmonics ${}_{\indhel}Y_{lm}(\theta,\phi)$      \\
\hline
$c\neq0$, $\indhel\neq0$  & Spin-weighted spheroidal harmonics ${}_{\indhel}S_{lm\omega}(\theta)$, ${}_{\indhel}Z_{lm\omega}(\theta,\phi)$     \\
\end{tabular}
\end{center}
\caption{Names and symbols of the regular \catdraft{segur?} solutions 
of the various differential equations that are
derived from (\ref{eq:ang. teuk. eq.}) by making zero none, one or two of the parameters $c=a\omega$ and $\indhel$.} \label{table:notation angular funcs.}
\end{table}

There exists in the literature much analytic and numerical work on the spheroidal harmonics and
substantial analytic, but not so much numerical, work on the spin-weighted spherical harmonics.
There exists little work, either analytic or numerical, on the spin-weighted spheroidal harmonics.
We will now present the main known properties of these various angular functions.

Flammer ~\cite{bk:Flammer} together with Abramowitz and Stegun ~\cite{bk:AS} 
are the most comprehensive works on the different properties, 
expansions and approximations for different limits of the parameters of the solutions and eigenvalues of the 
spheroidal differential equation. In addition, they both tabulate numerical results obtained for certain values of the parameters.
\catdraft{1) mention Erdelyi, Meixner\&Schaftzke, A\&S, Leaver, 2) dir quina cosa concreta vaig trobar a Meixner\&Schaftzke que no hi era als altres}

\catdraft{1) talk about conformal weight? 2) unify $\eth$ vs. $\vartheta$ denomination, 3) write $\eth$ in complex stereographic coords.?
3) include eq.3.19N\&P66 and property below it? and gral.eqs. in sec.2 of BRW? 
4) check sign in def. sphcal.harms. so that eq2.6Goldb.et al. coincs. with ours, 
5) in all N\&P'66,Goldberg et al.,Campbell'71 $m^{\mu}$ is ours for $r\rightarrow\infty$, i.e., in flat s-t, what's the story
(e.g., def. of spin,etc) for finite $r$?, 6)) a intro. BRW diu que spin-weighted spherical harmonics
els van introduir Gel'fand, Minlos \& Shapiro'63?, 7)Quantities defined only on a sphere.}

Newman and Penrose ~\cite{ar:N&P'66} introduced the spin-weighted spherical harmonics.
They first define \define{spin weight} $s$ in the same manner as helicity $h$ is defined in (\ref{eq:def. helicity}) 
for a wave travelling in the direction of $\vec{l}$ or $\vec{n}$. 
That is, they say that a quantity $\eta$ has helicity (spin-weight) $\indhel$ if it transforms as $\eta\to e^{\indhel i \vartheta}\eta$
under the transformation $m^{\mu}\to e^{i\vartheta}m^{\mu}$.
Since it was later proven by Campbell ~\cite{ar:Campbell'71} that the spin weight defined by
Newman and Penrose actually corresponds to a helicity, as we shall see later, 
we will use the term helicity even where Newman and Penrose used the term spin weight.
They define the operator $\eth$ (called \define{edth} or \define{thop}) and $\bar{\eth}$ 
acting on a quantity $\eta$ of helicity $\indhel$ defined on the $(\theta,\phi)$-sphere as   
\begin{equation}
\begin{aligned}
\eth \eta&=-\sin^{\indhel}\theta\left[\pardiff{}{\theta}+\frac{i}{\sin\theta}\pardiff{}{\phi}\right]\left(\left(\sin^{-\indhel}\theta\right)\eta\right)  \\
\bar{\eth} \eta&=-\sin^{-\indhel}\theta\left[\pardiff{}{\theta}-\frac{i}{\sin\theta}\pardiff{}{\phi}\right]\left(\left(\sin^{\indhel}\theta\right)\eta\right)
\end{aligned}
\end{equation}
If the $\phi$-dependence  of $\eta$ is $e^{im\phi}$, then $\eth$ and $\bar{\eth}$ are related to the operator 
defined in (\ref{eq:def. L_n}) in the manner:
\begin{equation} \label{eq:rln. between L and edth}
\mathcal{L}^{\topbott{\dagger}{}}_{\mp \indhel}\eta
=\left[-\topbott{\eth}{\bar{\eth}}\pm a\omega \sin\theta \right]\eta
\end{equation}

It can be proven that $\eth \eta$ is then a quantity of helicity $\indhel+1$, i.e., $\eta\to e^{(\indhel+1)i\vartheta}\eth \eta$ 
under $m^{\mu}\rightarrow e^{i\vartheta}m^{\mu}$, so that
$\eth$ is effectively a quantity of helicity unity. It is in this sense that we can say that $\eth$ raises the helicity by one unit.   
Similarly, $\bar{\eth}$ lowers the helicity by one unit.
The spin-weighted spherical harmonics satisfy equation (\ref{eq:ang. teuk. eq.}) 
with $c=0$, which can be re-written as
\begin{equation}
\bar{\eth}\eth{}_{\indhel}Y_{lm}=-(l-\indhel )(l+\indhel+1){}_{\indhel}Y_{lm}
\end{equation}
since in that case
\begin{equation} \label{eq:eigenval. for c=0}
{}_{\indhel}E_{lm}\equiv {}_{\indhel}E_{lm\omega=0}= l(l+1)-\indhel (\indhel+1).
\end{equation}
\catdraft{com es prova?}
Newman and Penrose proved that 
\begin{subequations}   \label{eq:{s+/-1}_Y in terms of s_Y}
\begin{align}
{}_{\indhel+1}Y_{lm}&=\left[(l-\indhel )(l+\indhel+1)\right]^{-1/2}\eth \ {}_{\indhel}Y_{lm} \label{eq:eq.B1bJ,McL,Ott'95} \\
{}_{\indhel-1}Y_{lm}&=-\left[(l+\indhel )(l-\indhel+1)\right]^{-1/2}\bar{\eth} \ {}_{\indhel}Y_{lm}  \label{eq:eq.B1cJ,McL,Ott'95}
\end{align}
\end{subequations}
so that by repeated application of (\ref{eq:{s+/-1}_Y in terms of s_Y}) the spin-weighted spherical harmonics 
can be expressed in terms of the spherical harmonics 
\begin{equation} \label{eq:sphcal.harms.}
{}_{\indhel=0}Y_{lm}\equiv Y_{lm}=\left[\frac{2l+1}{4\pi}\frac{(l-m)!}{(l+m)!}\right]^{1/2}P^{m}_l(\cos\theta)e^{im\phi} \\
\end{equation}
where the associated Legendre polynomials are given by
\begin{equation}  \label{eq:assoc.Leg.pols.}
P^{m}_l(x)=\frac{(-1)^m}{2^ll!}(1-x^2)^{m/2}\frac{\d^{l+m}}{\d x^{l+m}}(x^2-1)^l=(-1)^m\frac{(l+m)!}{(l-m)!}P^{-m}_l(x)
\end{equation}
It is clear from (\ref{eq:{s+/-1}_Y in terms of s_Y}) that the ${}_{\indhel}Y_{lm}$ vanish for $|\indhel|>l$.
It is easy to prove by induction on (\ref{eq:{s+/-1}_Y in terms of s_Y}) that the ${}_{\indhel}Y_{lm}$ 
form a set of orthonormal functions of helicity $\indhel$ on the $(\theta,\phi)$-sphere:
\begin{equation} \label{eq:orthogonality of {}_sY_{lm}}
\int\d\Omega\ {}_{\indhel}Y_{lm}\ {}_{\indhel}Y^*_{l'm'}=\delta_{ll'}\delta_{mm'}
\end{equation}
and that they are complete for helicity $\indhel$ quantities on the sphere, so that apart from 
(\ref{eq:orthogonality of {}_sY_{lm}}) they also satisfy:
\begin{equation}
\sum_{l=\indhel}^{\infty}\sum_{m=-l}^l{}_{\indhel}Y_{lm}(\theta,\phi){}_{\indhel}Y^*_{lm}(\theta',\phi')=\delta(\phi-\phi')\delta(\cos\theta-\cos\theta')
\end{equation}
Other properties satisfied by the spin-weighted spherical harmonics are (~\cite{ar:J&McL&Ott'91}) 
\begin{equation} \label{eq:eq.2aJ,McL,Ott'91}
\sum_{m=-l}^l {}_{-1}Y_{lm}(\theta,\phi){}_1Y_{lm}^*(\theta,\phi)=0
\end{equation}
and the ``addition theorem''  (~\cite{th:McL'90}) 
\begin{equation} \label{eq:eq.B6J,McL,Ott'95}
\sum_{m=-l}^l {}_{\indhel}Y_{lm}(\theta,\phi){}_{\indhel}Y^*_{lm}(\theta',\phi')=\frac{2l+1}{4\pi}P_l(\cos\gamma)
\end{equation}
where $\gamma$ is defined by $\cos\gamma\equiv \cos\theta\cos\theta'+\sin\theta\sin\theta'\cos(\phi-\phi')$.
The ``addition theorem'' is a consequence of the group multiplication law. 

\catdraft{If $s=0$, $\eth^*\eth$ is essentially the total angular momentum operator.(p.866N\&P'66): 
es de fet obvi per {eq:orb.ang.mom.op. related to edth}}

\catdraft{Mention next?:
Given any suitable regular $\eta$ on the sphere, of (integral) spin weight $s>0$, there exists $\xi$ of
spin weight zero for which $\eta=\eta_{\text{e}}+\eta_{\text{m}}$ with $\eta_{\text{e}}=\eth^{\indhel}\Re{\xi}$
and $\eta_{\text{m}}=\eth^{\indhel}\Im{\xi}$. $\eta_{\text{e}}$ and $\eta_{\text{m}}$ are respectively
called the \define{electric} and \define{magnetic} parts of $\eta$. The reason for this denomination
lies in its application to the spin coefficient $\sigma$. 
$\sigma(u,r,\theta,\phi)=\frac{\sigma^0(u,r,\theta,\phi)}{r^2}+O(r^{-4})=\sigma^0_{\text{e}}+\sigma^0_{\text{m}}$.
By analogy with the linear theory, we may suppose $\sigma^0_{\text{e}}$ to be associated with
``electric''-type radiation (e.g., arising from changes in the mass quadrupole) and $\sigma^0_{\text{m}}$
to be associated with ``magnetic''-type radiation (e.g., arising from changes in the angular momentum
quadrupole).}

\catdraft{talk about the fact that $\eth$ is covariant differentiation: $m^{\alpha}=\sqrt{2}P\delta^{\alpha}_{\zeta}$
and $\eth\eta=\sqrt{2}\eta_{(\alpha\dots\beta);\gamma}m^{\alpha}\dots m^{\beta}m^{\gamma}$? (p.2155,2156Gold.et al) Adrian:No}

The following results by Goldberg et al.\ ~\cite{ar:Gold.etal.} and by Campbell ~\cite{ar:Campbell'71} 
that we briefly present below were obtained in those papers in flat space-time, and they are therefore
valid asymptotically in the Kerr-Newman space-time. 
In particular, a tetrad can be written in the limit $r\rightarrow+\infty$ in terms of the usual unit polar vectors 
$\{\hat{\vec{e}}_t,\hat{\vec{e}}_r,\hat{\vec{e}}_{\theta},\hat{\vec{e}}_{\phi}\}$ 
in flat space-time. In this limit, the Kinnersley tetrad becomes: 
\begin{equation} \label{eq:Kinnersley tetrad, r->inf}
\begin{aligned}\vec{l}&\rightarrow -\hat{\vec{e}}_t+\hat{\vec{e}}_r & \qquad (r\rightarrow+\infty)\\
\vec{n}&\rightarrow -\frac{1}{2}\left(\hat{\vec{e}}_t+\hat{\vec{e}}_r\right) & (r\rightarrow+\infty) \\
\vec{m}&\rightarrow +\frac{1}{\sqrt{2}}\left(\hat{\vec{e}}_{\theta}+i\hat{\vec{e}}_{\phi}\right) & (r\rightarrow+\infty) \\
\end{aligned}
\end{equation}
and the vector $\vec{m}$ defined in those papers coincides with the Kinnersley tetrad vector $\vec{m}$.

In a subsequent paper to Newman and Penrose's, Goldberg et al. ~\cite{ar:Gold.etal.} further identify 
the spin-weighted spherical harmonics with the elements of the
matrices of the representation $D^{l}$ of the ordinary rotation group $R_3$ associated with total angular momentum $l$. 
If a spatial rotation $R(\phi,\theta,\gamma)$ of Euler angles $\phi,\theta,\gamma$ is composed of $\gamma$ about the OZ axis
followed by $\theta$ about OY and then $\phi$ about OZ and it transforms $x^i$ to $x^{' i}=R^{ij}x^j$,
then the matrix $D^l$ may be defined by its action
on spherical harmonics: $Y_{lm}(\vec{x}')=\sum_{m'}Y_{lm'}(\vec{x})D^l_{m'm}(R^{-1})$. Goldberg et al. prove that
\begin{equation}
{}_{\indhel}Y_{lm}(\theta,\phi)e^{-i\indhel \gamma}=\left[\frac{(2l+1)}{4\pi}\right]^{1/2}D^l_{-\indhel m}(\phi,\theta,\gamma)
\end{equation}
 and therefore
\begin{equation}
{}_{\indhel}Y_{lm}(\theta,\phi)=\left[\frac{(2l+1)}{4\pi}\right]^{1/2}D^l_{-\indhel m}(\phi,\theta,0)
\end{equation}
\catdraft{in eq.2.5.5/App.A ''Scatt.from b-h's'' there's a $(-1)^m$?}
The orthogonality and completeness relations for the spin-weighted spherical harmonics then follow directly
from the fact that the functions $D^l_{m'm}(\phi,\theta,\gamma)$ form a complete orthonormal basis for functions defined on $R_3$.

Goldberg et al. also relate $\eth$ to an ordinary angular-momentum raising operator. 
The well-known \define{angular momentum commutation relations} are
\begin{equation} \label{eq:ang. mom. commutation rlns.}
\begin{aligned}
\left[J_z,J_{\pm}\right]&=\pm J_{\pm} \\
\left[J_+,J_-\right]&=2J_{z}
\end{aligned}
\end{equation}
where $J_{\pm}\equiv J_x\pm iJ_y$. If $\ket{j,m_j}$ is a simultaneous eigenvector of $\vec{J}^2$ and $J_z$ with
\begin{equation} \label{eq:eigenvect. of J^2,J_z}
\begin{aligned}
\vec{J}^2 \ket{j,m_j}&=j(j+1)\ket{j,m_j} \\
J_z \ket{j,m_j}&=m_j\ket{j,m_j}
\end{aligned}
\end{equation}
then it is also an eigenvector of $J_{\pm}$ with 
\begin{equation}  \label{eq:eigenvect. of J_+/-}
J_{\pm} \ket{j,m_j}=\left[(j(j+1)-m_j(m_j\pm1)\right]^{1/2}\ket{j,m_j\pm 1}
\end{equation}
which is why $J_+$ and $J_-$ are respectively called angular-momentum \define{raising} and 
\define{lowering} operators.

It is well-known that the spherical harmonics $Y_{lm}$ are eigenvectors of the orbital angular 
momentum operator $\vec{L}=-i\vec{r}\times\nabla$ satisfying equations (\ref{eq:eigenvect. of J^2,J_z}) 
and (\ref{eq:eigenvect. of J_+/-}) with $\vec{J} \rightarrow \vec{L}$, $\ket{j,m_j}\rightarrow Y_{lm}$,
$j\rightarrow l$ and $m_j \rightarrow m$.  
\catdraft{aixo es nomes valid en flat s-t o no??}

We next give the relationship between the orbital angular momentum and the edth operators, 
which will be of use later on:
\begin{equation} \label{eq:orb.ang.mom.op. related to edth}
\topbott{\eth}{\bar{\eth}} \eta=\mp \sin^{\pm \indhel}\theta\left[\left(\vec{e}_{\theta}\pm i\vec{e}_{\phi}\right)\vec{L}\right](\sin^{\mp \indhel}\theta)\eta
\end{equation}
where $\hat{\vec{e}}_{\theta}$ and $\hat{\vec{e}}_{\phi}$ are the usual unit polar angular vectors in flat space
and $\eta$ is a quantity of helicity $\indhel$. 

It is also known that the operator $\vec{\mathtt{L}}$ defined as
\begin{equation}
\begin{aligned}
\mathtt{L}_z&\equiv-i\pardiff{}{\phi} \\
\mathtt{L}_{\pm}&\equiv \pm e^{\pm i\phi}\left(\pardiff{}{\theta}\pm i\cot\theta\pardiff{}{\phi}\pm i\csc\theta\pardiff{}{\gamma}\right)
\end{aligned}
\end{equation}
obeys the angular momentum commutation relations (\ref{eq:ang. mom. commutation rlns.}) and that $D^l_{-\indhel m}$ are its eigenvectors,
i.e., equations (\ref{eq:eigenvect. of J^2,J_z}) and (\ref{eq:eigenvect. of J_+/-}) are satisfied with 
$\vec{J} \rightarrow \vec{\mathtt{L}}$, $\ket{j,m_j}\rightarrow D^l_{-\indhel m}$, $j\rightarrow l$ and $m_j \rightarrow m$.
It is therefore immediate to show that the operator $\vec{\Lambda}$ such that $\Lambda_z\equiv L_z$ and 
$\Lambda_{\pm}\equiv L_{\pm}-\indhel \csc\theta e^{\pm i\phi}$, satisfies (\ref{eq:ang. mom. commutation rlns.}) and
that the spin-weighted spherical harmonics ${}_{\indhel}Y_{lm}$ are its eigenvectors with $\vec{J} \rightarrow \vec{\Lambda}$, 
$\ket{j,m_j}\rightarrow {}_{\indhel}Y_{lm}$, $j\rightarrow l$ and $m_j \rightarrow m$.                      

Based on the symmetry of
$D^l_{-\indhel m}(\phi,\theta,\gamma)$ with respect to $(m$, $\phi)$ on the one hand and $(\indhel$, $-\gamma)$ on the other,
Goldberg et al. define an angular-momentum operator $\vec{K}$, which commutes with $\mathtt{L}$, as:
\begin{equation}
\begin{aligned}
K_z&\equiv i\pardiff{}{\gamma} \\
K_{\pm}&\equiv \pm e^{\pm i\gamma}\left(\pardiff{}{\theta}\pm i\cot\theta\pardiff{}{\gamma}\pm i\csc\theta\pardiff{}{\phi}\right)
\end{aligned}
\end{equation}
It is then easy to see that it does indeed satisfy (\ref{eq:ang. mom. commutation rlns.}) as well as
equations (\ref{eq:eigenvect. of J^2,J_z}) and (\ref{eq:eigenvect. of J_+/-}) with 
$\vec{J} \rightarrow \vec{K}$, $\ket{j,m_j}\rightarrow D^l_{-\indhel m}$, $j\rightarrow l$ and now with
$m_j \rightarrow s$ (rather than $m$).        
The relationship between the operator $\eth$ and the angular-momentum raising differential operator $K_+$ 
is thus established as
\begin{equation} \label{eq:rln. between K_+ and eth}
\left[K_{+} D^l_{-\indhel m}\right]_{\gamma=0}=\eth D^l_{-\indhel m}(\phi,\theta,0)
\end{equation}
from which equations (\ref{eq:{s+/-1}_Y in terms of s_Y}) immediately follow given (\ref{eq:eigenvect. of J_+/-}).
Table \ref{table:ang. mom. op., eigenvects. and eigenvals.} summarizes the 
different angular momentum operators and corresponding eigenvectors and
eigenvalues we have looked at.

\catdraft{wrong sign in (\ref{eq:rln. between K_+ and eth})?}

\begin{table}
\begin{center}
\begin{tabular}{ccc}
$\vec{J}$           &  $\ket{j,m_j}$ & $(j,m_j)$  \\
\hline
\hline
$\vec{L}$           &  $Y_{lm}$      &   $(l,m)$    \\
$\vec{\mathtt{L}}$  &  $D^l_{-\indhel m}$   &   $(l,m)$    \\
$\vec{\Lambda}$     &  ${}_{\indhel}Y_{lm}$  &   $(l,m)$    \\
$\vec{K}$           &  $D^l_{-\indhel m}$   &   $(l,\indhel)$    \\
\hline
\end{tabular} 
\end{center}
\caption{Different angular momentum operators $\vec{J}$ and their corresponding eigenvectors $\ket{j,m_j}$ 
and parameters $(j,m_j)$ that make up the eigenvalues in (\ref{eq:eigenvect. of J^2,J_z}) 
and (\ref{eq:eigenvect. of J_+/-}).} 
\label{table:ang. mom. op., eigenvects. and eigenvals.}
\end{table}

By using vector harmonics, Campbell ~\cite{ar:Campbell'71} shows that $\indhel$ can be 
interpreted as a helicity.
The \define{vector harmonics} are defined as 
\begin{equation}
\begin{aligned}
T_i(+1,l,m;\vec{\hat{r}})&\equiv m_i\left[{}_{-1}Y_{lm}(\theta,\phi)\right] \\
T_i(0,l,m;\vec{\hat{r}})&\equiv r_i\left[{}_{0}Y_{lm}(\theta,\phi)\right] \\
T_i(-1,l,m;\vec{\hat{r}})&\equiv -m^*_i\left[{}_{+1}Y_{lm}(\theta,\phi)\right]
\end{aligned}
\end{equation}
where $m_i$ are the co-ordinates of the vector $\vec{m}$ in the limit of large $r$ 
in equation (\ref{eq:Kinnersley tetrad, r->inf}).  
It is easy to see that the vector harmonics form a complete set for vector functions of $\theta$ and $\phi$.
The total angular momentum operator for these harmonics, which is the generator of rotations for vector functions,
is given by
\begin{equation}
\begin{aligned}
\left(\mathcal{J}_k\right)_{ij}&=\delta_{ij}L_k+\left(S_k\right)_{ij} \\
\left(S_k\right)_{ij}&\equiv-i\epsilon_{ijk}
\end{aligned}
\end{equation}
where $S_k$ is the \define{spin operator} for cartesian 3-vectors.
Campbell then proves that
\begin{equation}
\begin{aligned}
\left(\mathcal{J}^2\right)^j_i T_j(\indhel,l,m;\vec{\hat{r}})&=l(l+1)T_i(\indhel,l,m;\vec{\hat{r}})     \\
\left(\mathcal{J}_z\right)^j_i T_j(\indhel,l,m;\vec{\hat{r}})&=mT_i(\indhel,l,m;\vec{\hat{r}})     \\
\left(\mathcal{J}_{\pm}\right)^j_i T_j(\indhel,l,m;\vec{\hat{r}})&=\left[l(l+1)-m(m\pm 1)\right]^{1/2}T_i(\indhel,l,m\pm 1;\vec{\hat{r}})     \\
\left(S^2\right)^j_i T_j(\indhel,l,m;\vec{\hat{r}})&=2T_i(\indhel,l,m;\vec{\hat{r}})     \\
\left(\vec{\hat{r}}\vec{S}\right)^j_i T_j(\indhel,l,m;\vec{\hat{r}})&=
\left(\vec{\hat{r}}\vec{\mathcal{J}}\right)^j_i T_j(\indhel,l,m;\vec{\hat{r}})=\indhel T_i(\indhel,l,m;\vec{\hat{r}})     \\
\end{aligned}
\end{equation}
We are therefore able to relate $l$ to a total angular momentum, $m$ to its $z$-projection, and $\indhel$ to the
radial component of the spin. If we think in terms of outgoing radiation, $\indhel$ can be thought of as a \define{helicity}.
Note that the value of the helicity of a vector harmonic is minus the value of the parameter $\indhel$ 
of the spin-weighted spherical harmonic used to construct the vector harmonic with.

\catdraft{1) veure discussio p.1763(darr),1764(darr)Campbell'71, 2) mirar de traslladar tant del que dic com sigui
possible a Kerr $\forall r$!, 3) incloure eqs.3.12-3.14Campbell'71}

As mentioned above, despite the large amount of research on both spheroidal harmonics and spin-weighted spherical harmonics, little
has been done on the angular funcions that concern us, the spin-weighted spheroidal harmonics. These functions were 
first introduced in 1973 by Teukolsky ~\cite{ar:Teuk'73} 
as a result of the separation of the Teukolsky equation for general spin, as we have seen. 
Shortly after, Press and Teukolsky ~\cite{ar:Press&Teuk'73}
used ordinary perturbation theory to obtain an expansion to second order in $c$ for the eigenvalue ${}_{\indhel}E_{lm\omega}$ and 
used a continuation technique for small, real $c$ to obtain solutions of (\ref{eq:ang. teuk. eq.}) in the form 
${}_{\indhel}S_{lm\omega}(\theta)=\sum_{l'} {}_{\indhel}A^m{}_{ll'}(\theta,c){}_{\indhel}Y_{lm}(\theta)$.                 
They (~\cite{ar:Press&Teuk'73} and ~\cite{ar:Teuk&Press'74}) tabulated their results for ${}_{\indhel}E_{lm\omega}$ 
for certain sets of $\{s,l,m\}$ and range of $c$, which we discuss later. Fackerell and Grossman ~\cite{ar:Fac&Gross'76} 
expressed ${}_{\indhel}S_{lm\omega}$ as a series involving
Jacobi polynomials to find a certain transcendental equation involving a continued fraction for the determination
of ${}_{\indhel}E_{lm\omega}$ as a power series in $c$ (particularly useful in the case of complex frequencies), which they 
evaluated (later corrected by Seidel ~\cite{ar:Seidel'89}) up to order 7. 
A similar geometrical interpretation of the SWSH to the one for spin-weighted spherical harmonics as eigenfunctions
of the Laplace operator on the unit sphere has not been found. As a matter of fact, the SWSH are not eigenfunctions
of the Laplace operator on a spheroid. \catdraft{aixo era aixi fins BRW->comprovar que no s'hagi trobat encara tal interpretac.!}  
They do, however, form a complete and orthonormal set of functions on a prolate spheroid (~\cite{ar:BRW}).  
Stewart ~\cite{ar:Stewart'75} showed that the SWSH form a strongly complete set if $c$ is real and he could only prove weak completeness if $c$ is complex.
\catdraft{1) provar/veure! donar eqs.corresponents!!-mirar Stewart'75!  
2) does this imply that for ex., $\int\d\Omega{}_{\indhel}Z_{lm}\ {}_{\indhel}Z^*_{l'm'}=\delta_{ll'}\delta_{mm'}$, as nobody seems to write it?}
Finally, some attempts 
(~\cite{ar:BRW}, ~\cite{ar:Breuer'75}) have been made
at finding the behaviour of ${}_{\indhel}E_{lm\omega}$ and ${}_{\indhel}S_{lm\omega}$ for large frequency, which we will discuss at length in
Chapter ~\ref{ch:high freq. spher}.

\draft{1) include refs. in Seidel'89, especially Leaver'95 and Starob\&Churi'74, 
2) BRW: analytic structure in Hartle and Wilkins'74 (!?)}


\section{General properties}

\draft{change terminology so that for ex. ${}_{\indhel}E_{lm\omega}$ (/${}_{\indhel}\mu_{lm\omega}$) represents val. 
which is not equal to eigenval. and ${}_{\indhel}E_{lm\omega}^{(c)}$ (/${}_{\indhel}E_{lm\omega}$)
when it is, to make phrasing easier?}

In this section we wish to establish some basic, useful properties of the solutions 
of the angular Teukolsky equation (\ref{eq:ang. teuk. eq.}).
 
The symmetries of the equation are immediate: the equation remains invariant under the change
in sign of two quantities among $[\indhel, (m,\omega), x]$, where we are considering that $(m,\omega)$
constitutes one single quantity.       
As a consequence, the SWSH satisfy the following symmetries, where the choice of signs will be justified
later on:
\begin{subequations} \label{eq: S symms}
\begin{align}
&{}_{\indhel}S_{lm\omega}(\theta)=(-1)^{l+m}{}_{-\indhel}S_{lm\omega}(\pi-\theta)   \label{eq:S symm.->pi-t,-s} \\
&{}_{\indhel}S_{lm\omega}(\theta)=(-1)^{l+\indhel }{}_{\indhel}S_{l-m-\omega}(\pi-\theta)  \label{eq:S symm.->pi-t,-m,-w}\\
&{}_{\indhel}S_{lm\omega}(\theta)=(-1)^{\indhel+m}{}_{-\indhel}S_{l-m-\omega}(\theta)     \label{eq:S symm.->-s,-m,-w}
\end{align}
\end{subequations}
where any one symmetry follows from the other two.
The eigenvalues must consequently also satisfy the symmetries:
\begin{subequations} \label{eq:eigenval. symms.}
\begin{align}
&{}_{\indhel}E_{lm\omega}={}_{-\indhel}E_{lm\omega}   \label{eq:eigenval. symm.->-s} \\
&{}_{\indhel}E_{lm\omega}={}_{\indhel}E_{l-m-\omega}  \label{eq:eigenval. symm.->-m,-w}
\end{align}
\end{subequations}

We give here some useful expressions for $\indhel=\pm1$, which may be easily obtained using the angular differential equation and 
the Teukolsky-Starobinski\u{\i} identities (\ref{eq:Teuk-Starob. ids.}):
\begin{subequations} \label{eq:Ss,LSs as func. of S_s,LS_s}
\begin{align}
{}_1B_{lm\omega}{}_{\pm1}S_{lm\omega}&=
\left[\mp 2\mathcal{Q}\mathcal{L}^{\topbott{\dagger}{}}_1-(\pm 2a\omega \cos\theta+{}_{-1}\lambda_{lm\omega})\right]{}_{\mp1}S_{lm\omega} 
\label{eq:Ss as func. of S_s,LS_s} \\
{}_1B_{lm\omega}\mathcal{L}^{\topbott{}{\dagger}}_1{}_{\pm1}S_{lm\omega}&=
\left[(\pm 2a\omega \cos\theta-{}_{-1}\lambda_{lm\omega})\mathcal{L}^{\topbott{\dagger}{}}_1\pm2a\omega \sin\theta\right]{}_{\mp1}S_{lm\omega} 
\label{eq:LSs as func. of S_s,LS_s} \\
\pm2\mathcal{Q}\mathcal{L}^{\topbott{}{\dagger}}_1{}_{\pm1}S_{lm\omega}&=
(\mp 2a\omega \cos\theta+{}_{-1}\lambda_{lm\omega}){}_{\pm1}S_{lm\omega}+{}_1B_{lm\omega}{}_{\mp1}S_{lm\omega} \label{eq:LSs as func. of Ss,S_s}
\end{align}
\end{subequations}

Combining the symmetry (\ref{eq:S symm.->pi-t,-s}) with the relation (\ref{eq:LSs as func. of S_s,LS_s}) evaluated
at $\theta=\pi/2$, the following relation at the equator can be immediately obtained:
\begin{equation}
\left.\diff{{}_{-1}S_{lm\omega}}{\theta}\right|_{\pi/2}=
\left[\frac{2a\omega }{({}_{-1}\lambda_{lm\omega}-(-1)^{l+m}{}_1B_{lm\omega})}-a\omega +m\right]{}_{-1}S_{lm\omega}(\pi/2)
\end{equation}
which we have verified to be satisfied by our numerical results for modes for several sets of $\{l,m,\omega \}$.

The differential equation (\ref{eq:ang. teuk. eq.}) has singular points at $x=\pm 1$. 
By using the Frobenius method it can be found that the solution that is regular at both 
boundary points $x=+1$ and $-1$ is given by                             
\begin{equation} \label{eq:asympt. S for x->+/-1}
{}_{\indhel}S_{lm\omega}(x)=(1-x)^{\alpha}(1+x)^{\beta}{}_{\indhel}y_{lm\omega}(x)
\end{equation}
where 
\begin{equation}
\begin{aligned}
\alpha &=\frac{|m+\indhel |}{2}, & \qquad \qquad \beta&=\frac{|m-\indhel |}{2}
\end{aligned}
\end{equation}
and the function 
${}_{\indhel}y_{lm\omega}(x)$ behaves close to the boundary points as
\begin{equation} \label{eq:asympt. y for x->+/-1}
{}_{\indhel}y_{lm\omega}(x)=\sum_{n=0}^{\infty} {{}_{\indhel}\topbott{a}{b}_{n,lm\omega}(1\mp x)^n}    \quad   \text{for} \quad x\to \pm1            
\end{equation}

On the other hand, the irregular solution at $x=\pm 1$ is given by
\begin{subequations} \label{eq:asympt. S irreg. for x->+/-1}
\begin{align}
{}_{\indhel}S^{\text{irreg}}_{lm\omega}&=(1-x)^{\alpha}(1+x)^{\beta}{}_{\indhel}y^{\text{irreg}}_{lm\omega}  \\
{}_{\indhel}y^{\text{irreg}}_{lm\omega}&=
\sum_{n=0}^{\infty}{}_{\indhel}\topbott{a}{b}^{\text{irreg}}_n(1\mp x)^{-2\topbott{\alpha}{\beta}+n} \quad   \text{for} \quad x\to \pm1 
\end{align}
\end{subequations}
It immediately follows from the above equations that
\begin{equation} \label{eq:y'^irreg as a func. of y^irreg}
\diff{{}_{\indhel}y^{\text{irreg}}_{lm\omega}}{x}=\frac{\pm |m\pm \indhel|}{(1\mp x)}{}_{\indhel}y^{\text{irreg}}_{lm\omega}, \quad x \rightarrow \pm 1
\end{equation}
for $m\neq \mp \indhel$, which will be useful for the numerical integration.

\draft{find for $m=\mp \indhel$!}

The function ${}_{\indhel}y_{lm\omega}(x)$ satisfies the differential equation
\begin{equation} \label{eq:ang. teuk. eq. for y}
\begin{aligned}
& \Bigg\{
 (1-x^{2})\ddiff{}{x}-2\left[\alpha-\beta+(\alpha+\beta+1)x\right]\diff{}{x}+   \\
&  +{}_{\indhel}E_{lm\omega}-(\alpha+\beta)(\alpha+\beta+1)+c^{2}x^{2}-2\indhel cx
\Bigg\} {}_{\indhel}y_{lm\omega}(x)=0
\end{aligned}
\end{equation}

We are interested in finding the behaviour of the regular solution 
${}_{\indhel}y_{lm\omega}$ at the boundary points.
We therefore substitute the expansion (\ref{eq:asympt. y for x->+/-1}) for ${}_{\indhel}y_{lm\omega}$
into the differential equation (\ref{eq:ang. teuk. eq. for y}) and obtain the recursive relation
\begin{equation} \label{eq:recursive rln. a_n}
\begin{aligned}
{}_{\indhel}\topbott{a}{b}_{n+1,lm\omega}&=\frac{1}{2(n+1)\left(n+1+2\topbott{\alpha}{\beta}\right)}
\Bigg\{\Big[2n(\alpha+\beta+1)- 
 \\
&
-\left({}_{\indhel}E_{lm\omega}-(\alpha+\beta)(\alpha+\beta+1)+c^2\mp2c\indhel\right)+n(n+1)\Big]{}_{\indhel}\topbott{a}{b}_{n,lm\omega}+ \\
&  +2c(c\mp \indhel){}_{\indhel}\topbott{a}{b}_{n-1,lm\omega}-c^2{}_{\indhel}\topbott{a}{b}_{n-2,lm\omega}\Bigg\}, \quad \forall n\in \mathbb{N}
\end{aligned}
\end{equation}
where is is understood that $\displaystyle {}_{\indhel}\topbott{a}{b}_{n,lm\omega}\equiv 0$ for $n<0$.

It is also useful to find the relationship between the asymptotic behaviours of ${}_{\indhel}S_{lm\omega}$ and 
${}_{-\indhel}S_{lm\omega}$ at the boundaries.
By inserting the asymptotic behaviour (\ref{eq:asympt. S for x->+/-1}) and (\ref{eq:asympt. y for x->+/-1})
for ${}_{\indhel}S_{lm\omega}$ into the Teukolsky-Starobinski\u{\i}
identities and making use of the relations (\ref{eq:recursive rln. a_n}), we obtain

\begin{equation} \label{eq:S1/S_1,x->1}
\begin{aligned}
\frac{{}_{+1}S_{lm\omega}(x)}{{}_{-1}S_{lm\omega}(x)}&\rightarrow
\left \{
\begin{array}{lll}
\displaystyle
{}_1B_{lm\omega}\frac{(1-x)}{2}\frac{1}{m(m+1)}                                                
&\displaystyle
=\frac{(1-x)}{(1+x)}\frac{{}_{+1}a_{0,lm\omega}}{{}_{-1}a_{0,lm\omega}}, & m\geq 1 \\
\displaystyle
-\left(\frac{{}_{-1}\lambda_{l,m=0,\omega }-2c}{{}_{-1}\lambda_{l,m=0,\omega }+2c}\right)^{1/2} 
&\displaystyle
=\frac{{}_{+1}a_{0,l,m=0,\omega}}{{}_{-1}a_{0,l,m=0,\omega}}, & m=0 \\
\displaystyle
\frac{1}{{}_1B_{lm\omega}}\frac{2}{(1-x)}m(m-1)                                                
&\displaystyle
=\frac{(1+x)}{(1-x)}\frac{{}_{+1}a_{0,lm\omega}}{{}_{-1}a_{0,lm\omega}}, & m\leq -1 
\end{array}
\right\}
, \\ &
\quad \qquad \qquad \qquad \qquad \qquad \qquad \qquad\qquad \qquad \qquad  \qquad  \qquad  (x \rightarrow +1) 
\end{aligned}
\end{equation}
Analogous relations for $x \rightarrow -1$ immediately follow from the ones above when the symmetry (\ref{eq:S symm.->pi-t,-s}) is used.
It is clear from (\ref{eq:S1/S_1,x->1}) that ${}_{+1}S_{lm\omega}(x)$ and ${}_{-1}S_{lm\omega}(x)$, 
where one is calculated from the other with the Teukolsky-Starobinski\u{\i} identities (\ref{eq:Teuk-Starob. ids.}), 
have the same sign for $x\rightarrow +1$ and for $x\rightarrow -1$ except when $m=0$.
This fact combined with the knowledge of the number of zeros that the SWSH have, given in Chapter \ref{ch:high freq. spher},
means that the sign taken in the symmetry relation (\ref{eq:S symm.->pi-t,-s}) is indeed the one that corresponds to the
Teukolsky-Starobinski\u{\i} identities (\ref{eq:Teuk-Starob. ids.}) that we are using.
\ddraft{but Chrzan. uses same Teukolsky-Starobinski\u{\i} identities and diff. sign in the symm.??}
The sign in (\ref{eq:S symm.->pi-t,-s}) also agrees with the corresponding sign for the associated Legendre polynomials in the case $c=0$
via equations (\ref{eq:{s+/-1}_Y in terms of s_Y}). 
The sign in the symmetry (\ref{eq:S symm.->-s,-m,-w}) has been chosen to coincide, in the case $c=0$, with the corresponding sign 
for the associated Legendre polynomials given by (\ref{eq:assoc.Leg.pols.}).

\section{Numerical method} \label{sec:num. method; SWSH}

\catdraft{1) justificar pq. faig servir ``shooting method'' enlloc de ``shooting to a fitting point'' (ch.17.2Num.Rec.) o ``relaxation'' (ch.17.3Num.Rec.; a p.747
 semla dir que ``relaxation'' es millor per quan hi ha ``extraneous slns.'', com deu ser-ho $y^{irreg}$?!),
 2) justificar pq. faig servir R-K integration (es aixi?) enlloc de,p.ex.,Sturlisch-Boer, 3) justificar pq. faig servir ``zbrent'' enlloc de Newton-Raphson
 i explicar metode de ``zbrent''
 4) definir quin tipus (i.e., 2n order, eigenval.,etc) de diff.eq. hem de solucionar i fer-ho tambe pel cas de la diff. radial eq.
 5) pq. es $l=0,1,2,...$ i $|m|\leq l$?}

The solution to the eigenvalue problem given by the second order differential equation (\ref{eq:ang. teuk. eq. for y}) 
that we wish to solve involves three unknowns: two constants of integration plus the eigenvalue. 
These three unknowns become determined by imposing one boundary condition
at each one of the end-points and one normalization condition. 
The boundary conditions we need to impose are for the solution ${}_{\indhel}y_{lm\omega}$ to be regular at $x=\pm 1$.
The normalization condition is the one in equation (\ref{eq:normalization SWSH}).

By means of the change of variables $y_1\equiv {}_{\indhel}y_{lm\omega}$, $y_2\equiv {}_{\indhel}y'_{lm\omega}$, $y_3\equiv {}_{\indhel}E_{lm\omega}$, 
the differential equation (\ref{eq:ang. teuk. eq. for y}) can be reduced to the
following system of first order differential equations 
\begin{equation} \label{eq:ang. prog. derivs.}
\begin{aligned}
y'_1&=y_2 \\
y'_2&=\frac{1}{(1-x^{2})}\Big\{2\left(\alpha-\beta+(\alpha+\beta+1)x\right)y_2-  \\
    &  -\left[y_3-(\alpha+\beta)(\alpha+\beta+1)+c^{2}x^{2}-2\indhel cx\right]y_1\Big\} \\
y'_3&=0
\end{aligned}
\end{equation}
The shooting method requires two initial, arbitrary values which, added to the boundary condition 
at one of the end-points, determine all three unknowns. The equation is then integrated from that end-point $x_1$ until 
the other one $x_2$ as an initial value problem. It is then assessed how well the values at $x_2$ of the solution obtained agree with  
a condition resulting from the boundary condition at that point. If the condition is not satisfied within the desired accuracy, 
one of the two initial, arbitrary values is modified and the integration starts again with the new value. The steps of 
modifying this value and integrating the differential equation are iterated until the condition
at the final end-point $x_2$ is met within the desired accuracy.
The other initial, arbitrary value is finally determined by imposing the normalization condition on the solution.

It is clear from (\ref{eq:ang. prog. derivs.}) that we cannot impose boundary conditions at exactly the end-points $x=\pm 1$. 
Instead, we impose them slightly away from the end-points, at $x_{\topbott{1}{2}}=\mp1 \pm \d{x}$, where $\d{x}\ll 1$, 
and we therefore use the asymptotic expansion (\ref{eq:asympt. S for x->+/-1}) to find the value of the solution at $x_{\topbott{1}{2}}$. 
We set $\d x=10^{-2}$ in the code.
We take initial, arbitrary values for ${}_{-1}a_{0,lm\omega}$ and for the eigenvalue, which we denote 
by ${}_{\indhel}\hat{E}_{lm\omega}$ to distinguish it from the actual eigenvalue ${}_{\indhel}E_{lm\omega}$.
The initial value ${}_{\indhel}\hat{E}_{lm\omega}$ is not entirely arbitrary but is chosen to be
close to the eigenvalue obtained for a value of $\omega$ slightly smaller
than the one in the present calculation.
The initial value for ${}_{-1}a_{0,lm\omega}$  plus the boundary condition (\ref{eq:asympt. S for x->+/-1}) 
of regularity at $x_1$ provide the initial values of $y_1(x_1)$ and $y_2(x_1)$ whereas the 
initial values ${}_{\indhel}\hat{E}_{lm\omega}$ provides the initial value $y_3(x_1)={}_{\indhel}\hat{E}_{lm\omega}$.
The arbitrary value ${}_{\indhel}\hat{E}_{lm\omega}$ will be modified appropriately so that $y_1(x_2)$ and $y_2(x_2)$ satisfy the boundary condition of
regularity at $x_2$ within the desired accuracy. Once the correct eigenvalue ${}_{\indhel}E_{lm\omega}$ and solution 
${}_{\indhel}y_{lm\omega}(x)$ and ${}_{\indhel}y'_{lm\omega}(x)$ are obtained,
the initially arbitrary value ${}_{-1}a_{0,lm\omega}$ is rescaled so that the normalization condition (\ref{eq:normalization SWSH}) is satisfied.

In general, for a value of ${}_{\indhel}\hat{E}_{lm\omega}$ different from the actual eigenvalue, the numerically integrated solution
is a combination of both the regular and the irregular solutions, i.e.,
\begin{equation}
{}_{\indhel}y^{\text{num}}_{lm\omega}=A({}_{\indhel}\hat{E}_{lm\omega}){}_{\indhel}y_{lm\omega}+A({}_{\indhel}\hat{E}_{lm\omega}){}_{\indhel}y^{\text{irreg}}_{lm\omega}
\end{equation}
where ${}_{\indhel}y^{\text{num}}_{lm\omega}$ is the numerically obtained value and 
${}_{\indhel}y_{lm\omega}$ is the analytic, regular value.
$A$ and $B$ are unknown functions of ${}_{\indhel}\hat{E}_{lm\omega}$. 
We need to modify the value of ${}_{\indhel}\hat{E}_{lm\omega}$ so that only the regular term $A{}_{\indhel}y_{lm\omega}$ is retained.
In the scalar case, the boundary condition at $x_2$ may be imposed by requiring that ${}_{\indhel=0}\hat{E}_{lm\omega}$ is a zero of the function 
$g({}_{\indhel=0}\hat{E}_{lm\omega})\equiv {}_{\indhel=0}y^{' \text{num}}_{lm\omega}(x_2)-{}_{\indhel=0}y'_{lm\omega}(x_2)$, \catdraft{or equivalently for $y(x_2)$?}
where the analytic value ${}_{\indhel}y'_{lm\omega}(x_2)$ is known for the scalar case because
${}_{\indhel=0}y'_{lm\omega}(x)\propto {}_{\indhel=0}y'_{lm\omega}(-x)$. 
The function $g({}_{\indhel=0}\hat{E}_{lm\omega})$ should tend to zero as ${}_{\indhel=0}\hat{E}_{lm\omega}$ approaches the
correct eigenvalue and should tend to infinity when it is far from it because of the behaviour (\ref{eq:y'^irreg as a func. of y^irreg}) of the irregular solution.
However, in general we have ${}_{\indhel}y'_{lm\omega}(x)\propto {}_{-\indhel}y'_{lm\omega}(-x)$, 
relating solutions of equations with different helicity when $h\neq 0$, and therefore we do
not know the analytic value ${}_{\indhel}y'_{lm\omega}(x_2)$ for a particular value $\indhel\neq 0$ of the helicity. 
Instead, we can look for a zero of the function 
\begin{equation}
g({}_{\indhel}\hat{E}_{lm\omega})\equiv 
\frac{{}_{\indhel}y^{' \text{num}}_{lm\omega}(x_2)}{{}_{\indhel}y^{\text{num}}_{lm\omega}(x_2)}-\frac{{}_{\indhel}y'_{lm\omega}(x_2)}{{}_{\indhel}y_{lm\omega}(x_2)}
\end{equation}
which does not require the knowledge of ${}_{\indhel}a_{0,lm\omega}$. 
The function $g({}_{\indhel}\hat{E}_{lm\omega})$ should also tend to zero as
${}_{\indhel}\hat{E}_{lm\omega}$ approaches the correct eigenvalue and to infinity when it is far from it. However, this function has the same
sign whether ${}_{\indhel}\hat{E}_{lm\omega}$ is greater or smaller than the actual eigenvalue, 
and therefore the zero of this function is also a minimum or a maximum point. 
This is a considerable drawback because looking for an 
extreme point of a function generally requires many more evaluations of the function 
than looking for a zero which is not an extreme point, and yet the accuracy is much smaller . 
Typically, an extreme point is only calculated up to the square root of the computer's floating-point precision. 
We therefore decided to 
find a zero of the function
\begin{equation}
g({}_{\indhel}\hat{E}_{lm\omega})\equiv {}_{\indhel}y^{' \text{num}}_{lm\omega}(x_2)-{}_{\indhel}y^{' \text{approx}}_{lm\omega}(x_2)
\end{equation}
instead, where ${}_{\indhel}y^{' \text{approx}}_{lm\omega}(x_2)$ is not the actual analytic value,
which we do not know, but an approximation to it:
\begin{equation}
{}_{\indhel}y_{lm\omega}^{' \text{approx}}(x_2)\simeq \frac{{}_{\indhel}y^{\text{num}}_{lm\omega}(x_2)}{{}_{\indhel}y_{lm\omega}(x_2)}{}_{\indhel}y'_{lm\omega}(x_2)
\end{equation}
We see numerically that $g({}_{\indhel}\hat{E}_{lm\omega})$ changes sign at the eigenvalue so that it is not an extreme point.

\catdraft{dir quins vals. de dx, epsilon,etc hem triat i quants termes hem pres en asympt. expansion i justificar-ho?}


As in the numerical integration of the radial equation in Section \ref{sec:num. method; radial func.}, 
we adapted the methods in ~\cite{bk:NumRec} to the particular problem we wish to solve.
In this section $Y$ denotes any of the numerically-calculated dependent variables $y_1$, $y_2$ or $y_3$, where
for clarity we omit the index that refers to a specific dependent variable.
The actual integration from $x_1$ until $x_2$ of the system of differential equations (\ref{eq:ang. prog. derivs.}) is done with the Runge-Kutta method.
We will use a fifth-order Runge-Kutta formula to calculate the value of $Y(x+h)$ 
by evaluating at six different points the right hand side of the 
differential equations (\ref{eq:ang. prog. derivs.}), which we denote by $f$. The advantage of the method is that 
a fourth-order Runge-Kutta formula is obtained with a different combination of the evaluations of $f$ at the same six points. 
By combining the fourth-order and the fifth-order formulae, not only do we obtain $Y(x+h)$ but also an estimate of the error being made,
both with only six evaluations of $f$. 
Having an actual estimate of the error allows us to adapt the stepsize (\define{adaptive stepsize}) to be taken at every step so that we can
obtain the solution within the required accuracy without taking too many steps. 
The sole knowledge of the order of the method does not provide an actual estimate of the error being made.
Efficient values of the constants involved in the two formulae were found by Cash and Karp and
we give them in Table \ref{table:C-K parameters for R-K method}.
The formula that calculates the value of the solution $Y_{n+1}\equiv Y(x+h)$ from the value $Y_{n}\equiv Y(x)$ 
is the fifth-order Runge-Kutta expression
\begin{equation} \label{eq:5th order R-K method}
\begin{aligned}
k_1&=hf(x_n,Y_n) \\
k_2&=hf(x_n+a_2h,Y_n+b_{21}k_1) \\
   & \vdots \\
k_6&=hf(x_n+a_6h,Y_n+b_{61}k_1+ \dots +b_{65}k_5) \\
Y_{n+1}&=Y_n+c_1k_1+c_2k_2+c_3k_3+c_4k_4+c_5k_5+c_6k_6+O(h^6)
\end{aligned}
\end{equation}
The fourth-order Runge-Kutta formula that we use to find an estimate of the error is
\begin{equation} \label{eq:embedded 4th order R-K method}
\bar{Y}_{n+1}=Y_n+\bar{c}_1k_1+\bar{c}_2k_2+\bar{c}_3k_3+\bar{c}_4k_4+\bar{c}_5k_5+\bar{c}_6k_6+O(h^5)
\end{equation}
An estimate of the error is thus given by
\begin{equation} \label{eq:error for 5th order R-K method}
Y^{\text{err}}_{n+1}=Y_{n+1}-\bar{Y}_{n+1}
\end{equation}
which is of order $O(h^5)$.

\catdraft{1) com pot ser que $Y^{\text{err}}_n\sim O(h^5)$ pero que de fet $Y_{n+1}$ es calculi amb formula amb error $O(h^6)$??!
2) que vol dir ``embedded''?}

\begin{table}
\begin{center}
\begin{tabular}{|c|c|ccccc|c|c|}
\hline
$i$& $a_i$         &                     &                  & $b_{ij}$           &                      &                  &$c_i$             & $\bar{c}_i$ \\
\hline
1  &               &                     &                  &                    &                      &                  &$\frac{37}{378}$  & $\frac{2825}{27648}$ \\
2  & $\frac{1}{5}$ & $\frac{1}{5}$       &                  &                    &                      &                  &0                 & 0\\
3  & $\frac{3}{10}$& $\frac{3}{40}$      & $\frac{9}{40}$   &                    &                      &                  &$\frac{250}{621}$ &$\frac{18575}{48384}$ \\
4  & $\frac{3}{5}$ & $\frac{3}{10}$      &-$\frac{9}{10}$   & $\frac{6}{5}$      &                      &                  &$\frac{125}{594}$ &$\frac{13525}{55296}$\\
5  &  1            &-$\frac{11}{54}$     & $\frac{5}{2}$    &-$\frac{70}{27}$    &$\frac{35}{27}$       &                  &0                 &$\frac{277}{14336}$ \\
6  & $\frac{7}{8}$ & $\frac{1631}{55296}$& $\frac{175}{512}$& $\frac{575}{13824}$&$\frac{44275}{110592}$&$\frac{253}{4096}$&$\frac{512}{1771}$&$\frac{1}{4}$ \\
\hline
\multicolumn{2}{|c}{$j=$}
                   & 1                   & 2                & 3                  & 4                    & 5                &      
\multicolumn{2}{c|}{} \\
\hline
\end{tabular} 
\end{center}
\caption{Cash-Karp parameters for embedded Runge-Kutta method} \label{table:C-K parameters for R-K method}
\end{table}

The driver routine \routine{odeint} is the same routine as the one used in the integration of the radial equation 
in Section \ref{sec:num. method; radial func.}. 
The quantity $Y^{\text{scal}}_{n+1}$, used to compare the absolute error $Y^{\text{err}}_{n+1}$ at the point $x_{n+1}$ with, 
is correspondingly given by equation (\ref{eq:def. yscal,odeint}). The stepper routine \routine{rkqs} obtains the numerical value $Y(x+h)$ of the solution  
and the error $Y^{\text{err}}_{n+1}$ of the method at this step by calling the algorithm routine \routine{rkck}. If the fractional error is
too large, i.e., 
$\epsilon_{n+1}\equiv \left|Y^{\text{err}}_{n+1}/Y^{\text{scal}}_{n+1}\right|>\epsilon$, 
the stepper routine then calls \routine{rkck} to try the integration again with
a new, reduced stepsize. Since the error of the method is of order $O(h^5)$, and the stepsize $h$ has resulted in a relative error
$\epsilon_{n+1}$, then  an estimate for a new stepsize that would produce a relative error  $\epsilon$ is given by 
$h\left(\epsilon_{n+1}/\epsilon\right)^{1/5}$. 
The factor 1/5 in the exponent is not exact since $Y^{\text{scal}}_{n+1}$
in (\ref{eq:def. yscal,odeint}) is in its turn also rescaled with the new stepsize. The new, reduced stepsize will thus be 
$Sh\left(\epsilon_{n+1}/\epsilon\right)^{1/4}$, 
where $S=0.9$ is a safety factor. If on the other hand, the error
of the integration is smaller than the minimum required, then \routine{rkqs} increases the stepsize to 
$Sh\left(\epsilon_{n+1}/\epsilon\right)^{1/5}$.
The algorithm routine \routine{rkck} straight-forwardly implements the fifth-order Cash-Karp Runge-Kutta method and returns
the numerical value $Y(x+h)$ obtained with (\ref{eq:5th order R-K method}) and an estimate of the error $Y^{\text{err}}_{n+1}$
given by (\ref{eq:error for 5th order R-K method}).

Because of the symmetry (\ref{eq:S symm.->pi-t,-s}) we only need to calculate the solution ${}_{\indhel}S_{lm\omega}(\theta)$ $\forall \theta \in [0,\pi]$ for one
particular value $\indhel$ of the helicity, which we choose to be $-1$ as that is the value of the 
helicity for which we have calculated the radial solution.

The Fortran90 program \program{\text{sphdrvKN mpi.f90}} calculates the eigenvalue to quadruple precision and the 
spherical function and its derivative to double precision.

\textbf{Parallel Programming}

To complete the explanation of the various methods and routines used for numerical calculations for this thesis, 
we shall briefly discuss a parallel algorithm which we used in the program \program{\text{sphdrvKN mpi.f90}}
as well as in other programs used in Chapters \ref{ch:high freq. spher} and \ref{stress-energy tensor}. In particular, this algorithm could have been used
in the radial program discussed in Chapter \ref{ch:radial sln.} or in any other program involving calculations
which must be performed in the exact same manner for different values of certain parameters.
In all the programs we developed, these parameters consisted in the pair $(l,m)$.
This algorithm only requires minor modifications for it to adapt to different programs of the characteristics mentioned.
Parallel programming is used in order to make the most out of several CPUs that may be available.
The parallel algorithm we implemented uses the Message-Passing Interface (MPI) ~\cite{bk:MPI} as the message-passing library.

The parallel algorithm we implemented is an instance of a `master'/`slave' application.
It consists in the initialization of \catdraft{creating/running/spawning?} one `master' process, which has knowledge of the various
values of the parameters for which the calculations must be performed, and a number $N$ of `slave' processes, which will actually
perform the calculations for particular values of the parameters.
The `master' process first reads the common data, if any, that is required by all the `slave' processes 
(e.g., the eigenvalues and tabulated data for the function $r_*=r_*(r)$ in the case of programs in Chapter \ref{stress-energy tensor}),
and it broadcasts this data to all the `slaves'.
The `master' process then reads the data that is necessary for the calculations to be performed for a pair of values of $(l,m)$.
It then sends sends to a `slave' the values of the parameters that it should perform the calculation with as well as the data that it requires.
This is done for the first $N$ number of pairs of values of $(l,m)$.
Each one of the `slave' processes receives the information from the `master' and then proceeds to perform 
the calculations for a particular pair of values of $(l,m)$ and range of values of $\omega$.
Once finished with the calculations, the `slaves' send the results back to the `master'.
When the `master' receives the results from one of the `slaves' it then either sends to this `slave' 
a new pair of values $(l,m)$ to perform the calculations with or, if all pairs of values have been completed, it tells the `slave' to finalize. 
The `master' process then stores the results in a file.
Since reading/writing data from/in a file is a very time-consuming process, it is an efficient procedure
that the `master' is in charge of these tasks and performs them while the `slaves' are performing the calculations.

The result of the implementation of this parallel algorithm is that the amount of time required to perform
the calculations a large number of times for various values of certain parameters is approximately reduced $N$ times.


\section{Numerical results} \label{sec:num. results; SWSH}

The results for the spherical function obtained with the program \program{\text{sphdrvKN mpi.f90}}
are compared in Table \ref{table:data S_1 a=0.95,l=2,m=-2,omega=-0.25;tableVIChandr} against Chandrasekhar's ~\cite{bk:Chandr}. 

\catdraft{some of them (especially, close to $x \sim 1$) do not coinc. to the 5th and 6th digits?!}

Chandrasekhar also displays results for the eigenvalue. Note that his eigenvalue, which we call $\lambda^{\text{chandr}}_{lm\omega}$, 
is actually ${}_{-1}\lambda_{lm\omega}$ for the spin-1 case and ${}_{+2}\lambda_{lm\omega}+4$ for the spin-2 case. 
Note also that his frequency $\sigma$ is equal to $-\omega$. Our numerical results for the eigenvalue do not coincide with his.
We believe that Chandrasekhar calculated $\lambda^{\text{chandr}}_{lm\omega}$ by using an approximation given previously by Teukolsky and Press, 
but Chandrasekhar used it for the wrong range of values for the frequency thus producing incorrect results.  
Teukolsky and Press (~\cite{ar:Press&Teuk'73} and ~\cite{ar:Teuk&Press'74}) obtained polynomial (in $\omega$)
approximations of the eigenvalues for several sets of $\{\indhel, l, m\}$ by applying a continuation method to a representation of spin-weighted spheroidal
harmonics in terms of spin-weighted spherical harmonics. 
Their approximations are valid only for $0\leq a\omega  \leq 3$ and are accurate to five digits. The results in Chandrasekhar's table are given for negative $\omega$
and coincide exactly with the ones produced by Teukolsky and Press's polynomials for that negative value of $\omega$, which is obviously not within the
range of validity $0\leq a\omega  \leq 3$. For Teukolsky and Press's polynomials to produce results for a certain $m$ and a certain negative $\omega$, 
the polynomial corresponding to $-m$ must be evaluated at $-\omega>0$ and then the symmetry (\ref{eq:eigenval. symm.->-m,-w}) must be used. 
All results (except for three of them discussed below) produced in this way coincide  with our numerical results up to the fifth digit. 
This is shown in Table \ref{table:data eigenval.s=+/-1,l=2,m=-2;tableVIIChandr}.

In view of the fact that for $a\omega=-0.2,-1.2,-1.4$ our numeric results for ${}_{1}E_{2,-2,\omega }$ do not coincide with those produced by
Teukolsky and Press's polynomial, we decided to check them by using two other methods. One is the relaxation method, which consists in replacing the
ordinary differential equation by a finite-difference equation and iteratively improve an initial guess for the solution at all grid points 
so that it ends up satisfying the finite-difference equation and the required boundary conditions. 
The other method we implemented is the one suggested by Sasaki and Nakamura
~\cite{ar:Sasa&Naka'82}, consisting in also replacing the differential equation by a finite-difference equation but then imposing for the determinant
of the matrix that represents the finite-difference equation to be zero. This method will be described in detail in the next chapter since
it was the main tool we used to obtain the eigenvalues for large frequency. Both the relaxation method and Sasaki and Nakamura's agree 
for the $a\omega=-0.2,-1.2,-1.4$ cases in Table \ref{table:data eigenval.s=+/-1,l=2,m=-2;tableVIIChandr} with our numerical results rather than
with Teukolsky and Press's. We therefore believe that the latter are not accurate to the fifth digit in these three instances.

\begin{table}
\begin{tabular}{c|cc}
$\cos \theta$& ${}_{-1}S^{\text{chandr}}_{2,-2,-0.25}$& ${}_{-1}S^{\text{num}}_{2,-2,-0.25}$\\
\hline
0.& 0.768015  & 0.768023791282718\\
0.04& 0.734306& 0.734314263501826\\
0.08& 0.699739& 0.699746529751899\\
0.12& 0.664460& 0.664467540117839\\
0.16& 0.628612& 0.628618596243942\\
0.20& 0.592331& 0.592336366304714\\
0.24& 0.555748& 0.555753833600832\\
0.28& 0.518996& 0.519001205617666\\
0.32& 0.482202& 0.482206809678456\\
0.36& 0.445494& 0.445498002401709\\
0.40& 0.408999& 0.409002123321547\\
0.44& 0.372844& 0.372847528852603\\
0.48& 0.337162& 0.337164752355464\\
0.52& 0.302086& 0.302087851246282\\
0.56& 0.267754& 0.267756026125980\\
0.60& 0.234314& 0.234315635527191\\
0.64& 0.201921& 0.201922793693838\\
0.68& 0.170746& 0.170746848326145\\
0.72& 0.140974& 0.140975232456899\\
0.76& 0.112820& 0.112820561950217\\
0.80& 0.086531& 0.086531629301585\\
0.84& 0.062411& 0.062411721362379\\
0.88& 0.040852& 0.040852325588169\\
0.92& 0.022405& 0.022405022463833\\
0.96& 0.007979& 0.007979518753337
\end{tabular} 
\caption{${}_{\indhel}S_{lm\omega}$ for 
$\indhel=-1$, $Q=0$, $a=0.95$, $l=2$, $m=-2$ and $\omega=-0.25$.${}_{-1}S^{\text{chandr}}_{2,-2,-0.25}$ is taken from Table VI in Chandrasekhar's ~\cite{bk:Chandr} 
Appendix and ${}_{-1}S^{\text{num}}_{2,-2,-0.25}$ has been calculated with Fortran90 program \program{sphdrvKN mpi.f90}}. 
\label{table:data S_1 a=0.95,l=2,m=-2,omega=-0.25;tableVIChandr}
\end{table}

\begin{table}
\begin{tabular}{c|cccc}
$a\omega $ & ${}_{1}E^{\text{chandr}}_{2,-2,\omega}$ & ${}_{1}E^{\text{TP}}_{2,-2,\omega}$ & ${}_{1}E^{\text{num}}_{2,-2,\omega}$ & ${}_{1}E^{\text{TP}}_{2,2,|\omega|}$\\
\hline
0.  & 6.     &  6.    & 6.                                      & 6.\\
-0.2&  5.8534&  5.8534&  5.8534429399013102317  & 5.8535\\
-0.4&  5.6789&  5.6789&  5.6790807066777195764  & 5.6791\\
-0.6&  5.4741&  5.4741&  5.4746409095537499953  & 5.4746\\
-0.8&  5.2362&  5.2362&  5.2375343807523856300  & 5.2375\\
-1.0&  4.9618&  4.9618&  4.9648622055126348357  & 4.9649\\
-1.2&  4.6472&  4.6472&  4.6534383466611133856  & 4.6535\\
-1.4&  4.2880&  4.2880&  4.2998313934573516407  & 4.2999\\
-1.6&  3.8792&  3.8792&  3.9004271601816162329  & 3.9004\\
-1.8&  3.4149&  3.4149&  3.4515108815270425522  & 3.4515\\
-2.0&  2.8886&  2.8886&  2.9493639889615692063  & 2.9494\\
-2.2&  2.2929&  2.2929&  2.3903668120214223888  & 2.3904\\
-2.4&  1.6194&  1.6194&  1.7710962418980976975  & 1.7711\\
-2.6&  0.8586&  0.8586&  1.0884074322264239240  & 1.0884\\
-2.8& -0.0002& -0.0002&  0.3394912459602174195  & 0.3395\\
-3.0& -0.9688& -0.9688& -0.4780963537474403940  &-0.4781
\end{tabular} 
\caption{${}_{\indhel}E_{lm\omega}$ for $\indhel=\pm 1$, $Q=0$, $l=2$ and $m=-2$. ${}_{1}E^{\text{chandr}}_{2,-2,\omega }$ 
is taken from Table VII in Chandrasekhar's ~\cite{bk:Chandr} 
Appendix;  ${}_{1}E^{\text{TP}}_{2,-2,\omega }$ and ${}_{1}E^{\text{TP}}_{2,2,|\omega|}$ 
are calculated with polynomial in Table 2 in Teukolsky and Press ~\cite{ar:Teuk&Press'74};
${}_1E^{\text{num}}_{2,-2,\omega }$ has been calculated with Fortran90 program \program{sphdrvKN mpi.f90}.} 
\label{table:data eigenval.s=+/-1,l=2,m=-2;tableVIIChandr}
\end{table}

\draft{describe 'zbrent', 'brent' routines?}






\chapter{High frequency asymptotics for the angular solution} \label{ch:high freq. spher}


\section{Introduction} \label{sec:Introduction}

\draft{is it right that the coef. ${}_{\indhel}A_{lm\omega}$ (same for B,C,D) have superscript $\omega $ or should it be omitted??}

Following standard conventions, in this thesis we refer to
`high frequency' in relation to the angular function and eigenvalues when
in fact what it is meant is large $c(=a\omega$).
The high frequency approximation of the spin-weighted spheroidal
equation is a particularly important subject that, nevertheless, has
been left unresolved thus far, except for the spin-0 case, due to its difficulty. 
This asymptotic study is
important when considering both classical and quantum perturbations. In the
classical case it is important, for example, when calculating
gravitational radiation emitted by a particle near the black hole
since the typical time-scale of the motion is short compared to the
scale set by the curvature of the black hole. In the quantum
case its importance lies in the fact that the high frequency
limit is at the root of the divergences that the expectation value
of the stress-energy tensor possesses. The correct subtraction of
the divergent terms from the expectation value of the stress-energy 
tensor is extremely troublesome in curved space-time,
particularly in one that is not spherically symmetric. 
Because the divergent terms arise from the high
frequency behaviour of the field, knowledge of this
behaviour is fundamental in such a subtraction.
This limit has also been recently considered in the Kerr background in the context of
quasinormal modes (see ~\cite{ar:Berti&Card&Yosh'04}).
Quasinormal frequencies with large imaginary part
have acquired great importance since Hod ~\cite{ar:Hod'98} established a correspondence between 
these frequencies and transitions in energy level of the quantum black hole.

The new results that we present in this
chapter contribute towards making this problem more tractable. 
However, we should note that the asymptotic study in this
paper is valid for fixed $m$ as $c$ tends to
infinity, a fuller understanding of the asymptotic behaviour of
the solution would require an anlysis uniform in $m$.
It is worthwhile remarking that the whole analysis in this chapter
has been done for general integral spin, so that it applies to the scalar, electromagnetic 
and, in particular, gravitational perturbations, which are of great interest in astrophysics.

In the remainder of this introductory section we discuss the
results for high frequency asymptotics of SWSH that have been
obtained in the literature up until now, show their shortcomings
and outline what our new results achieve. In the next section we
lay down the basic theory that we use in the following sections. In
Sections \ref{sec:Inner solutions}, \ref{sec:Outer solution},
\ref{sec:Matching the solutions} and \ref{sec:Evaluation of
gamma} we fully determine the aymptotic behaviour of the angular solution that is uniform in $x$ 
and the asymptotic behaviour of the eigenvalue. In
Section \ref{sec:num. method; high freq. sph.} we describe the
numerical method and programs used to obtain the numerical results,
which in the last section we show, analyze and compare to
numerical results in the literature.

Different authors have obtained high-frequency approximations to the solution and eigenvalues of the spheroidal
differential equation. Erd\'{e}lyi et al. ~\cite{bk:high_transc_funcs}, Flammer ~\cite{bk:Flammer} and Meixner and Sch\"{a}fke
~\cite{bk:Meixner&Schafke} have all done so using the fact that this differential equation becomes the
Laguerre differential equation in that limit. 

\draft{anybody tried for the s-weighted sphcal. harms.?}

Breuer ~\cite{ar:Breuer'75} was the first author to study the
high-frequency behaviour of the spin-weighted spheroidal
harmonics. Based on the work on the spin-0 case by the above
authors  he related the solution of a transformation of the
spin-weighted spheroidal equation for large $c$ and finite $m$ to the
generalized Laguerre polynomials. His work, however, was
fundamentally erroneous as it assumed that the solution was
either symmetric or antisymmetric under $x\rightarrow -x$, which is only true for $\indhel =0$.

Breuer, Ryan and Waller ~\cite{ar:BRW} (hereafter referred to as BRW) corrected this error and further developed this study by first
relating the SWSH to the confluent hypergeometric functions and then
reducing them to the generalized Laguerre polynomials
by imposing regularity far from the boundary points $x=\pm 1$. 
Unfortunately, their study of the high-frequency behaviour
was flawed and incomplete. 
The behaviour for high frequency of both the spherical functions and the eigenvalues obtained by BRW
depend critically on a certain parameter $\gamma$ which they left undetermined for the $\indhel \neq 0$ case. 
BRW did obtain the analytic value of $\gamma$ for the $\indhel =0$ case but for the $\indhel \neq 0$ case they could
only calculate it numerically for a handful of sets of values of
$\{\indhel,l,m\}$. BRW achieved this numerical calculation for the $\indhel \neq 0$ case 
by matching the high-frequency asymptotic expression for the eigenvalue that they obtained with the
expression for the eigenvalue given by Press and Teukolsky
~\cite{ar:Press&Teuk'73} valid for low frequency. Not only their
analytic expressions for both the spherical solution and the
eigenvalue for high frequency were thus left undetermined, but
also their expressions for the spherical solution are only valid
sufficiently close to the boundary points $x=1$ and $x=-1$, but
not for the region in-between them. This results in the
possibility that a zero of the solution near $x=0$, away
from $x=\pm1$, be overlooked. Furthermore, and crucially, their assumption that
the confluent hypergeometric functions should reduce to the
generalized Laguerre polynomials by imposing regularity far from
the boundary points is not correct. The reason why it is not correct is that in the
cases for which the confluent hypergeometric function diverges far
from one of the boundaries, the coefficient in front of it
decreases exponentially with $c$ so that the solution remains finite
in the whole region $x\in[-1,+1]$.  
We believe that the reason why they were not able to
analytically determine the value of the parameter $\gamma$
is because they ignored the behaviour of the
solution far from the boundaries, thus overlooking a possible
zero, and wrongly imposed regularity. 

The study of the behaviour of the solution and eigenvalues of the
spin-weighted spheroidal equation for high, real frequency and finite $m$ has not been
developed any further by these or any other authors and
therefore BRW's work is where this study stood until this thesis.

In this chapter we correct and complete BRW's study for high, real
frequency and finite $m$. We thus obtain an asymptotic solution for large, real
frequency to the spin-weighted spheroidal equation which is
uniformly valid everywhere within the range $x\in[-1,+1]$, not just near the boundaries. 
We also analyze the existence and location of a possible zero of the solution near $x=0$. 
We analytically determine the value of $\gamma$ by matching the
number of zeros that our asymptotic solution has with the number of zeros that the SWSH has. 
As a consequence, the
asymptotics of the eigenvalue in the same limit also become fully determined.
Finally, we have complemented all the analytic work with graphs
produced with numerically-obtained data. The graphs show the
behaviour of the eigenvalues for large frequency and how they
match with Press and Teukolsky's approximation for low frequency.
They also show the behaviour of the SWSH in this limit and the
location of its zeros.

\draft{1) fer esquema de en quines eqs. la s-weighted harm. eq. esdeve segons els diffs. lims. (e.g., w gran, w petita, etc)?
ja fet pels lims. $a=0$ i $\indhel =0$ pero? 2) parlar d'altres papers mes recents?}


\section{Boundary layer theory} \label{sec:boundary layer theory}

\draft{1) write a proper, general description of boundary layer theory and definitions of inner/outer solutions,
2) draw sketch of what solutions are valid for what range of $x$}

In the rest of this chapter we follow the approach to boundary
layer theory as presented by Bender and
Orszag~\cite{bk:Bender&Orszag}. The asymptotic solution that is
a valid approximation to the solution of the differential equation
from the boundary point $\pm 1$ until $x\sim \pm 1+O(c^{\delta})$,
where $-1\leq\delta<0$, is called the
\define{inner solution}. The region within which an inner solution
is valid is a
\define{boundary layer}. As we shall see, for the large frequency
approximation of the spin-weighted spheroidal equation, there are
two boundary layers within the region $x\in[-1,1]$, one close to
$x=-1$ and one close to $x=+1$. Close to the boundary points the
spin-weighted spheroidal function oscillates rather quickly in
$x$, and indeed it is there where all (with the possible exception
of one) the zeros of the function are located.

The asymptotic solution that is a valid approximation to the solution of the differential equation
in the range $-1+O(c^{-1})\ll x \ll +1-O(c^{-1})$, is called
the \define{outer solution}. This range comprises not only the region in between the two boundary layers but also a certain region
of both boundary layers. This region where both an inner solution and the outer solution are valid is called the
\define{overlap region}, and it is there that the outer and inner solutions are matched.

We shall see that in between the two boundary layers the function behaves rather smoothly, like a $\cosh x$
\catdraft{(solia dir-hi $\coth$ enlloc de $\cosh$?)} or a $\sinh x$, 
so that the SWSH may have at the most one zero close to $x=0$.
The behaviour of the outer solution is important despite its
smoothness because when matching it with the inner solutions it
will allow us to find an asymptotic solution which is uniformly
valid throughout the whole range of $x$. The outer solution is
also necessary in order to find out whether or not the uniform
solution has a zero close to $x=0$ and, if it does, to calculate
the analytic location of the zero.

This is a key feature that singles out the scalar case from the others: for the spin-$0$ case the differential equation
(\ref{eq:ang. teuk. eq.}) is clearly symmetric under $\{ x\leftrightarrow -x\}$ and therefore, depending
on its parity, it will have a zero at $x=0$ or not. On the other hand, for the general spin case, the differential equation does not
satisfy this symmetry but it does remain unchanged under the transformation
$\left \{ x\leftrightarrow -x, \indhel \leftrightarrow -\indhel \right \}$
instead. There is therefore no apparent reason why it should have
a zero near the origin. The outer solution is important for the
general spin case and not for spin-$0$ since, as we shall see, the
differential equation that the outer solution satisfies is symmetric under 
$\left \{ x\leftrightarrow -x, \indhel \leftrightarrow -\indhel \right \}$
to leading order in $c$.


\section{Inner solutions} \label{sec:Inner solutions}

BRW obtained an expression for the inner solution for general spin in terms of an undetermined parameter $\gamma$.
In this section we summarize and present their results in a compact way.

By making the variable substitution $u=2c(1-x)$, equation (\ref{eq:ang. teuk. eq. for y}) becomes
\begin{equation} \label{diff eq:y in u}
\begin{aligned}
&u\frac{d^{2}{}_{\indhel}y_{lm\omega}}{du^{2}}+(2\alpha+1)\frac{d{}_{\indhel}y_{lm\omega}}{du}-  \\
&\quad-\frac{1}{4}\left[u+2\indhel -\frac{1}{c}\left(c^{2}-(\alpha+\beta)(\alpha+\beta+1)+{}_{\indhel}E_{lm\omega}\right)
\right]{}_{\indhel}y_{lm\omega}-
\\
&\quad-\frac{1}{4c}\left[u^{2}\frac{d^{2}{}_{\indhel}y_{lm\omega}}{du^{2}}+2(\alpha+\beta+1)\frac{d{}_{\indhel}y_{lm\omega}}{du}-
\left(\frac{1}{4}u^{2}+\indhel u\right){}_{\indhel}y_{lm\omega}\right]=0
\end{aligned}
\end{equation}
\draft{Removed $\indhel (\indhel+1)-\indhel (\indhel+1)$ from middle term?} It is clear from
this equation that the leading order behaviour of ${}_{\indhel}E_{lm\omega}$
for large $c$ must be:
\begin{equation} \label{eq:series E for large w}
{}_{\indhel}E_{lm\omega}=-c^{2}+\gamma c+O(1)
\end{equation}
If its leading order were not $-c^2$, there would then be a leading
order term $+\frac{1}{4}c{}_{\indhel}y_{lm\omega}$ in the equation that it
could not be matched with any other term. Lower order
terms for ${}_{\indhel}E_{lm\omega}$ are given in BRW. It is crucial to know
the value of the parameter $\gamma$, as it will determine how the
angular function behaves asymptotically to leading order in $c$.
At this stage, $\gamma$ is an undetermined real number; \catdraft{pq. no complex?} 
we will determine its value later on.

Using the asymptotic behaviour (\ref{eq:series E for large w}) and letting $c\to \infty$, the terms in
(\ref{diff eq:y in u}) of order $O(c^{-1})$ can be ignored
with respect to the other ones and, to leading order in $c$, the function ${}_{\indhel}y_{lm\omega}$ satisfies
\begin{equation} \label{diff eq:y in u,1st order}
u\frac{d^{2}{}_{\indhel}y_{lm\omega}}{du^{2}}+(2\alpha+1)\frac{d{}_{\indhel}y_{lm\omega}}{du}-\frac{1}{4}\left(u+2\indhel -\gamma \right){}_{\indhel}y_{lm\omega}=0
\end{equation}
The solution of this differential equation that satisfies the boundary
condition of regularity at $x=+1$ is related to the confluent
hypergeometric function:
\begin{equation}
{}_{\indhel}y_{lm\omega}^{\text{inn},+1}={}_{\indhel}C_{lm\omega}e^{-u/2}{}_1F_{1}\Big((|m+\indhel |+\indhel+1)/2-\gamma/4,|m+\indhel |+1,u\Big)
\end{equation}
where ${}_{\indhel}C_{lm\omega}$ is a constant of integration.

\draft{no podria ser que l'altre sln., que no es regular a $x=+1$ tingues la irreg. compensada per coeff. (igual com
passa a $x=0$?}

Similarly, if we instead make a change of variable $u^{*}=2c(1+x)$ in equation (\ref{eq:ang. teuk. eq. for y}), due to the
$\left\{ x\leftrightarrow -x, \indhel\leftrightarrow -\indhel  \right \}$ symmetry we obtain
\begin{equation}
{}_{\indhel}y_{lm\omega}^{\text{inn},-1}={}_{\indhel}D_{lm\omega}e^{-u^{*}/2}{}_1F_{1}\Big((|m-\indhel |-\indhel+1)/2-\gamma/4,|m-\indhel |+1,u^{*}\Big)
\end{equation}
as the solution that is regular at $x=-1$.

We use the following obvious notation to refer to the solutions of the spin-weighted spheroidal
equation that correspond to the inner solutions of (\ref{diff eq:y in u,1st order}):
\begin{equation}
{}_{\indhel}S_{lm\omega}^{\text{inn},\pm 1}=(1-x)^{\alpha}(1+x)^{\beta}{}_{\indhel}y_{lm\omega}^{\text{inn},\pm 1}
\end{equation}
The inner solution ${}_{\indhel}S_{lm\omega}^{\text{inn},\pm 1}$ is only a valid
approximation in the region from the boundary point $\pm 1$ until 
a point $x\sim \pm 1\mp O(c^{\delta})$
with $-1\leq\delta<0$. The reason is that in the
step from (\ref{diff eq:y in u}) to (\ref{diff eq:y in u,1st
order}) we have ignored terms with $u^{\topbott{}{*}}/c$ with respect to terms of order
$O(1)$, and therefore the inner solution has been found for
$\pm 1 -x \sim u^{\topbott{}{*}}/c\ll O(1)$
and so we must have $\delta<0$. On the other hand, we are not
ignoring $u$ with respect to the $O(1)$ term $2(\indhel -q)$ in
equation (\ref{diff eq:y in u,1st order}), so that it must be
$u\sim O(c^{\delta+1})$ with $\delta+1\geq 0$.
From the fact that we are not ignoring $2(\indhel -q)$ with respect to $u$
it does not follow that $\delta+1\leq 0$, since the inner solution is valid at
the boundary point $x=+1$, where $u=0$.
That is, the term $2(\indhel -q)$ cannot be ignored with respect to $u$
everywhere in the region from $+1$ up to a point $x\sim +1-O(c^{\delta})$
even if $\delta+1\geq 0$.
A similar reasoning applies to $u^*$.

We therefore have one boundary layer comprising the region in $x$ from
$-1$ to  $(-1-x)\sim O(c^{\delta})$ and another boundary layer
from $(+1-x)\sim O(c^{\delta})$ to $+1$.

To leading order in $c$ the solution to the 
spin-weighted spheroidal equation which is valid within the two boundary layers is given by
\begin{equation} \label{eq: inner solution}
{}_{\indhel}S_{lm\omega}^{\text{inn}} =(1-x)^{\alpha}(1+x)^{\beta}
\begin{cases}
{}_{\indhel}C_{lm\omega}e^{-u/2}{}_1F_{1}(-p,2\alpha+1,u)         & \qquad x>0 \\
{}_{\indhel}D_{lm\omega}e^{-u^{*}/2}{}_1F_{1}(-p',2\beta+1,u^{*}) & \qquad x<0
\end{cases}
\end{equation}
where we have defined
\begin{equation} \label{eq:def pp'}
\begin{cases}
p\equiv -(|m+\indhel |+\indhel+1)/2+\gamma/4    \\
p'\equiv -(|m-\indhel |-\indhel+1)/2+\gamma/4
\end{cases}
\end{equation}

BRW then require that $p,p'\in \mathbb{Z}^{+}$ in order that
the inner solution ${}_{\indhel}S_{lm\omega}^{\text{inn}}$ is regular at $x=0$, where
$u, u^*\rightarrow \infty$. 
Correspondingly, they replace the
confluent hypergeometric functions ${}_1F_{1}(a,b,x)$ by the
generalized Laguerre polynomials $L_{-a}^{(b-1)}(x)$. 
As we shall see, this is erroneous:
$p,p'\in \mathbb{Z}^{+}$ is not a necessary condition for
regularity since in the cases for which this condition is not
satisfied, the coefficients ${}_{\indhel}C_{lm\omega}$ and ${}_{\indhel}D_{lm\omega}$
diminish exponentially for large $c$ in such a way that
${}_{\indhel}S_{lm\omega}^{\text{inn}}$ remains regular.


\section{Outer solution} \label{sec:Outer solution}

We now proceed to find the outer solution of (\ref{eq:ang. teuk.
eq.}). The analysis in this section is new as the outer solution has been overlooked by previous authors.
We first make the variable substitution
\begin{equation}
y(x)=g(x)\exp{\int
\frac{\alpha-\beta+(\alpha+\beta+1)x}{1-x^2}dx}=g(x)(1-x)^{-\frac{(2\alpha+1)}{2}}(1+x)^{-\frac{(2\beta+1)}{2}}
\end{equation}
which transforms equation (\ref{eq:ang. teuk. eq.}) into
\begin{equation} \label{eq:g}
g''(x)+f(x,c)g(x)=0
\end{equation}
where
\begin{equation}
\begin{aligned}
f(x,c)&=\frac{G(x,c)}{1-x^{2}}+ 
\frac{(\alpha+\beta+1)(1-x^{2})+2x\left[\alpha-\beta+(\alpha+\beta+1)x\right]}{(1-x^{2})^{2}}-  
\\&
-\frac{\left[\alpha-\beta+(\alpha+\beta+1)x\right]^{2}}{(1-x^{2})^{2}}
\end{aligned}
\end{equation}
and $G(x,c)$ is the coefficient of ${}_{\indhel}y_{lm\omega}$ in (\ref{eq:ang. teuk. eq. for y}), i.e.,
\begin{equation}
G(x,c)={}_{\indhel}E_{lm\omega}-(\alpha+\beta)(\alpha+\beta+1)+c^{2}x^{2}-2\indhel cx 
\end{equation}

We now perform a WKB-type expansion: $g(x)=e^{\mathcal{G}(x)}$. This change of variable converts equation (\ref{eq:g}) into
\begin{equation} \label{eq:phi}
\mathcal{G}''(x)+\mathcal{G}'(x)^{2}+f(x,c)=0
\end{equation}
Performing an asymptotic expansion of $f(x,c)$ in $c$:
\begin{equation}
f(x,c)=f_{0}(x)c^{2}+f_{1}(x)c+O(1),
\end{equation}
with
\begin{equation}
f_{0}(x)=-1, \qquad f_{1}(x)=\frac{2(q-\indhel x)}{1-x^{2}} ,
\end{equation}
where we have used the asymptotic expansion of ${}_{\indhel}E_{lm\omega}$ in
$c$ and we have also introduced the parameter $q \equiv
\gamma/2$. We will prove in Section \ref{sec:Evaluation of gamma} that $q$ must be an integer.
It is clear that to leading order in $c$ the outer solution is
symmetric under $\{ x\leftrightarrow -x\}$. 
We are avoiding any possible turning points \ddraft{define?}
by assuming that $f(x,c)\neq 0$ for $x$ values of interest. 
This condition is clearly satisfied if $c$ is large enough.

Next we perform an asymptotic expansion of $\mathcal{G}(x)$ in $c$.
We do not know what the leading order is, and  we will determine it
with the method of dominant balance. 
Let the expansion of $\mathcal{G}(x)$ for large $c$ be
$\mathcal{G}(x)=h_{0}(c)\mathcal{G}_{0}(x)+o(h_{0}(c))$. 
We use the small letter $o$ to indicate lower order than the order of its argument.
On substituting the asymptotic expansions for $f(x,c)$ and $\mathcal{G}(x)$ into
(\ref{eq:phi}) we obtain
\begin{equation}
h_{0}(c)\mathcal{G}''_{0}(x)+h_{0}(c)^2\left[\mathcal{G}'_{0}(x)\right]^{2}+c^{2}f_{0}(x)+o\big(h_{0}(c)^2\big)+o(c^{2})=0
\end{equation}
We could try and cancel out the $c^{2}f_{0}(x)$ term with
$h_{0}(c)\mathcal{G}''_{0}(x)$. That woud give $h_{0}=c^{2}$, but then
$h_{0}(c)\mathcal{G}''_{0}(x)$ would be subdominant
to $h_{0}^2(\mathcal{G}'_{0})^{2}$. The other option is to cancel the
$c^{2}f_{0}(x)$ term with  $h_{0}^2(\mathcal{G}'_{0})^{2}$ instead. This
gives $h_{0}=c$, which  works. We therefore find
$\mathcal{G}(x)=c\mathcal{G}_{0}(x)+\mathcal{G}_{1}(x)+O(c^{-1})$.

The resulting equation for the leading order term in $\mathcal{G}$ is
\begin{equation}
\left[\mathcal{G}'_{0}(x)\right]^{2}+f_{0}(x)=0,
\end{equation}
the solution of which is $\mathcal{G}_{0}=\pm(x-x_{0})$. The equation for
the next order in
 $c$ is
\begin{equation}
 \mathcal{G}''_{0}(x)+2\mathcal{G}'_{0}(x)\mathcal{G}'_{1}(x)+f_{1}(x)=0,
\end{equation}
 which gives
\begin{equation}
\mathcal{G}_{1}=\pm
\left[-\frac{\indhel}{2}\log(1-x^{2})-\frac{q}{2}\log\left(\frac{1+x}{1-x}\right)\right].
\end{equation}
The physical optics approximation for the outer solution is
therefore:
\begin{equation}\label{eq: outer solution}
\begin{aligned}
&{}_{\indhel}S_{lm\omega}^{\text{out}}(x)=(1-x)^{\alpha}(1+x)^{\beta}{}_{\indhel}y_{lm\omega}^{\text{out}}(x)
\\ 
&{}_{\indhel}y_{lm\omega}^{\text{out}}(x)=(1-x)^{-(2\alpha+1)/2}(1+x)^{-(2\beta+1)/2}
\times \\ & 
\qquad \qquad \qquad \qquad
\times 
\Big[{}_{\indhel}A_{lm\omega}(1-x)^{+(q-\indhel )/2}(1+x)^{-(q+\indhel )/2}e^{+cx}+  
\\&
\qquad \qquad\qquad \qquad \ 
+{}_{\indhel}B_{lm\omega}(1-x)^{-(q-\indhel )/2}(1+x)^{+(q+\indhel )/2}e^{-cx}\Big]
\end{aligned}
\end{equation}
where the constant $x_{0}$ has been absorbed within ${}_{\indhel}A_{lm\omega}$ and ${}_{\indhel}B_{lm\omega}$.

This solution is valid in the region $-1+O(c^{-1})\ll x \ll +1-O(c^{-1})$.

\catdraft{explicar el pq.}


\section{Matching the solutions} \label{sec:Matching the solutions}
We have found three different solutions. One of the two inner
solutions is valid in the region $-1 \leq x \lesssim
-1+O(c^{\delta})$ for any $\delta$ such that $-1\leq\delta <0$,
and the other one for $+1-O(c^{\delta}) \lesssim x \leq +1$. The
outer solution is valid for $-1+O(c^{-1})\ll x \ll +1-O(c^{-1})$.
Clearly all three solutions together span the whole physical region $-1\leq
x \leq +1$. There are also two regions of overlap, one close to -1
and one close to +1, where both the outer solution and one of the
inner solutions are valid. We can proceed to match the solutions
in these regions and we will do so only to leading order in $c$ as
matching to lower orders would not bring any more insight into the
behaviour of the SWSH. When the matching is completed to leading
order, the two overlap regions are given one by
the points $x$ satisfying $O(c^{-1})\ll 1+x \lesssim O(c^{\delta})$ and the other one by the points satisfying $O(c^{-1})\ll 1-x \lesssim O(c^{\delta})$.
For the overlap regions to exist it is therefore required that we choose a $\delta$ satisfying $-1<\delta <0$.

In order to obtain an expression for the inner solution in the overlap region, we expand the inner solution for $u,u^{*}\sim \infty$.
For that, we need to know how
the confluent hypergeometric functions behave when the independent variable is large. From ~\cite{bk:AS} we have
\begin{equation}
{}_1F_{1}(b,c,z) \rightarrow \frac{\Gamma(c)e^{+i\pi b}z^{-b}}{\Gamma(c-b)}+\frac{\Gamma(c)e^{z}z^{b-c}}{\Gamma(b)}, \qquad (|z|\rightarrow +\infty)
\end{equation}
when $z=|z |e^{i\vartheta}$ with $-\pi/2 < \vartheta < 3\pi/2$, which includes the case we are considering: $\vartheta=0$. 
This means that the inner solution valid close to $x=+1$ behaves like
\begin{equation} \label{eq: inner +1 asympt}
\begin{aligned}
&{}_{\indhel}y_{lm\omega}^{\text{inn},+1}\rightarrow
\\ & \rightarrow {}_{\indhel}C_{lm\omega}
\left \{
\begin{array}{ll}
\displaystyle\frac{\Gamma(|m+\indhel |+1)\left[2c(1-x)\right]^{(-p-|m+\indhel  |-1)}e^{+c(1-x)}}{\Gamma(-p)},        & p\notin \mathbb{Z}^{+}\cup\{0\} \\
\displaystyle\frac{\Gamma(|m+\indhel |+1)e^{-i\pi p}\left[2c(1-x)\right]^{p}e^{-c(1-x)}}{\Gamma(|m+\indhel  |+1+p)},  & p\in
\mathbb{Z}^{+}\cup\{0\}
\end{array}
\right \}
\\ & \qquad \qquad \qquad  \qquad \qquad \qquad  \qquad \qquad \qquad  \qquad
\qquad \qquad \qquad \qquad   , (|u|\rightarrow+\infty)
\end{aligned}
\end{equation}
The behaviour of the inner solution valid close to $x=-1$ is
similarly obtained by simultaneously replacing $x$ with $-x$,
$\indhel $ with $-\indhel $ (which also implies replacing $p$ by $p'$) 
and ${}_{\indhel}C_{lm\omega}$ with ${}_{\indhel}D_{lm\omega}$ above.

On the other hand, in order to obtain an expression for ${}_{\indhel}y_{lm\omega}^{\text{out}}$ valid in the overlap region
we perform a Taylor series expansion around $x=+1$ or $-1$ depending on where we are
doing the matching, and keep only the first order in the series:

\begin{itemize}

\item[a)] \textbf{Around $x=+1$}.

To first order in $(1-x)$:
\begin{equation} \label{eq: outer +1 asympt}
\begin{aligned}
{}_{\indhel}y_{lm\omega}^{\text{out}}(x)&\sim{}_{\indhel}A_{lm\omega}(1-x)^{\left[+(q-\indhel -1)/2-\alpha\right]}2^{\left[-(q+\indhel+1)/2-\beta\right]}e^{+cx}+ \\
&+{}_{\indhel}B_{lm\omega}(1-x)^{\left[-(q-\indhel+1)/2-\alpha\right]}2^{\left[+(q+\indhel -1)/2-\beta\right]}e^{-cx} \qquad (x\rightarrow +1)
\end{aligned}
\end{equation}
By matching the inner and outer solution in the overlap region $O(c^{-1})\ll 1-x \lesssim O(c^{\delta})$, i.e.,
by matching equations (\ref{eq: inner +1 asympt}) and (\ref{eq: outer +1 asympt}), we obtain the following relations
depending on the value of $p$:

\begin{itemize}
\item[a1)] if $p\notin \mathbb{Z}^{+}\cup\{0\}$:
\begin{equation} \label{eq: match a1}
\begin{cases}
{}_{\indhel}A_{lm\omega}=0  \\
\displaystyle
{}_{\indhel}B_{lm\omega}=2^{\left[-(q+\indhel -1)/2+\beta\right]}\frac{\Gamma(|m+\indhel |+1)}{\Gamma(-p)}(2c)^{\left[-p-|m+\indhel |-1\right]}e^{+c}{}_{\indhel}C_{lm\omega}
\end{cases}
\end{equation}

\item[a2)] if $p\in \mathbb{Z}^{+}\cup\{0\}$:
\begin{equation} \label{eq: match a2}
{}_{\indhel}A_{lm\omega}=2^{\left[+(q+\indhel+1)/2+\beta\right]}\frac{\Gamma(|m+\indhel |+1)}{\Gamma(|m+\indhel |+1+p)}e^{-i\pi p}(2c)^{p}e^{-c}{}_{\indhel}C_{lm\omega}
\end{equation}
\end{itemize}

\item[b)] \textbf{Around $x=-1$} (similar to the $x=+1$ case).

To first order in $(1+x)$:
\begin{equation}
\begin{aligned}
{}_{\indhel}y_{lm\omega}^{\text{out}}(x)&\sim{}_{\indhel}A_{lm\omega}(1+x)^{\left[-(q+\indhel+1)/2-\beta\right]}2^{\left[+(q-\indhel -1)/2-\alpha\right]}e^{+cx}+   \\
&+{}_{\indhel}B_{lm\omega}(1+x)^{\left[+(q+\indhel -1)/2-\beta\right]}2^{\left[-(q-\indhel+1)/2-\alpha\right]}e^{-cx} \qquad (x\rightarrow -1)
\end{aligned}
\end{equation}

\begin{itemize}
\item[b1)] if $p'\notin \mathbb{Z}^{+}\cup\{0\}$:
\begin{equation} \label{eq: match b1}
\begin{cases}
{}_{\indhel}B_{lm\omega}=0  \\
\displaystyle
{}_{\indhel}A_{lm\omega}=
2^{\left[-(q-\indhel -1)/2+\alpha\right]}\frac{\Gamma(|m-\indhel |+1)}{\Gamma(-p')}(2c)^{\left[-p'-|m-\indhel |-1\right]}e^{+c}{}_{\indhel}D_{lm\omega}
\end{cases}
\end{equation}

\item[b2)] if $p'\in \mathbb{Z}^{+}\cup\{0\}$:
\begin{equation} \label{eq: match b2}
{}_{\indhel}B_{lm\omega}=2^{\left[+(q-\indhel+1)/2+\alpha\right]}\frac{\Gamma(|m-\indhel |+1)}{\Gamma(|m-\indhel |+1+p')}e^{-i\pi p'}(2c)^{p'}e^{-c}{}_{\indhel}D_{lm\omega}
\end{equation}
\end{itemize}

\end{itemize}

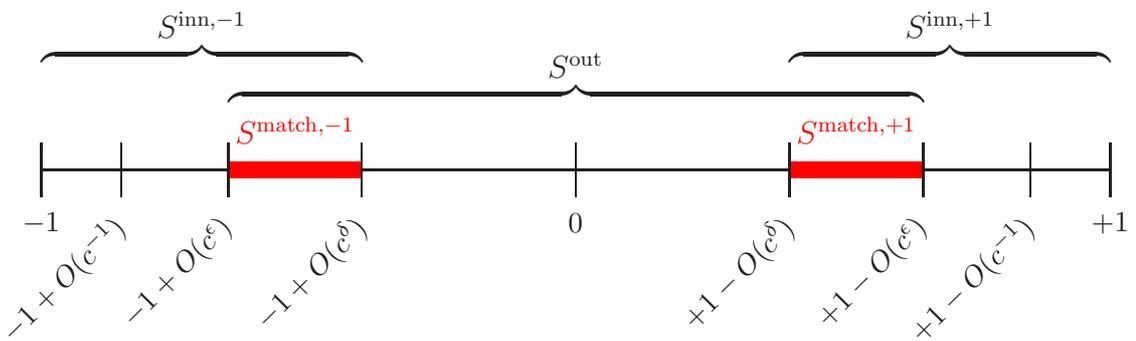
\begin{figure}
\begin{center}
\setlength{\unitlength}{1pt}
\begin{picture}(400,150)
\put(0,80){\line(1,0){400}}
\put(0,70){\line(0,1){20}}
\put(0,60){\rotatebox{0}{\makebox(0,0)[c]{\small$-1$}}}
\put(30,70){\line(0,1){20}}
\put(30,60){\rotatebox{45}{\makebox(0,0)[r]{\small$-1+O(c^{-1})$}}}
\put(70,80){\makebox(0,0)[l]{\colorbox{red}{\makebox[44pt][c]{}}}}
\put(70,60){\rotatebox{45}{\makebox(0,0)[r]{\small$-1+O(c^{\epsilon})$}}}
\put(70,70){\line(0,1){20}}
\put(120,70){\line(0,1){20}}
\put(120,60){\rotatebox{45}{\makebox(0,0)[r]{\small$-1+O(c^{\delta})$}}}
\put(200,70){\line(0,1){20}}
\put(200,60){\rotatebox{0}{\makebox(0,0)[c]{\small$0$}}}
\put(280,80){\makebox(0,0)[l]{\colorbox{red}{\makebox[44pt][c]{}}}}
\put(280,60){\rotatebox{45}{\makebox(0,0)[r]{\small$+1-O(c^{\delta})$}}}
\put(280,70){\line(0,1){20}}
\put(330,70){\line(0,1){20}}
\put(330,60){\rotatebox{45}{\makebox(0,0)[r]{\small$+1-O(c^{\epsilon})$}}}
\put(370,70){\line(0,1){20}}
\put(370,60){\rotatebox{45}{\makebox(0,0)[r]{\small$+1-O(c^{-1})$}}}
\put(400,70){\line(0,1){20}}
\put(400,60){\rotatebox{0}{\makebox(0,0)[c]{\small$+1$}}}
\put(95,90){\makebox(0,0)[b]{\color{red}$S^{\text{match},-1}$}}
\put(305,90){\makebox(0,0)[b]{\color{red}$S^{\text{match},+1}$}}
\put(200,105){\makebox(0,0)[b]{$\overbrace{\text{\makebox[260pt]{}}}$}}
\put(200,115){\makebox(0,0)[b]{$S^{\text{out}}$}}
\put(60,120){\makebox(0,0)[b]{$\overbrace{\text{\makebox[120pt]{}}}$}}
\put(60,130){\makebox(0,0)[b]{$S^{\text{inn},-1}$}}
\put(340,120){\makebox(0,0)[b]{$\overbrace{\text{\makebox[120pt]{}}}$}}
\put(340,130){\makebox(0,0)[b]{$S^{\text{inn},+1}$}}
\end{picture}
\caption{Regions of validity in the $x$ axis of the various approximations to the SWSH for large $c$.
It must be $-1<\epsilon<\delta<0$.
For clarity, the mode labels have been dropped. 
$S^{\text{match},\pm 1}$ refers to the asymptotic approximation valid in the overlap region (red) close to $x=\pm 1$.
The uniform solution is constructed as $S^{\text{unif}}=S^{\text{out}}+S^{\text{inn},+1}+S^{\text{inn},-1}-S^{\text{match},+1}-S^{\text{match},-1}$.}
\label{fig:regions of validity for asymptotic SWSH}
\end{center}
\end{figure}

From the above matching equations we can obtain a uniform asymptotic approximation to ${}_{\indhel}S_{lm\omega}$ valid
throughout the whole region $x\in [-1,+1]$ and also find out
where the zeros of the function are. 
Following ~\cite{bk:Bender&Orszag}, the uniform asymptotic approximation is obtained by adding the outer and the two inner solutions, and then subtracting the asymptotic
approximations in the two overlap regions since these have been included twice. 
Figure \ref{fig:regions of validity for asymptotic SWSH} depicts the region of validity of the various asymptotic solutions for large $c$
that we have obtained.

We can distinguish three cases:

\subsection*{$p,p'\notin\mathbb{Z}^{+}\cup\{0\}$}
 From equations (\ref{eq: match a1}) and
(\ref{eq: match b1}) it must be
${}_{\indhel}A_{lm\omega}={}_{\indhel}B_{lm\omega}=0={}_{\indhel}C_{lm\omega}={}_{\indhel}D_{lm\omega}$, so this case
is the trivial solution and we discard it.

\subsection*{$p\in \mathbb{Z}^{+}\cup\{0\}$ and $p'\notin
\mathbb{Z}^{+}\cup\{0\}$, or vice-versa}

Either ${}_{\indhel}A_{lm\omega}$ or ${}_{\indhel}B_{lm\omega}$ is equal to zero (but not
both), so that the function ${}_{\indhel}S_{lm\omega}$ cannot have a zero close
to $x=0$. All the zeros, if there are any, of ${}_{\indhel}S_{lm\omega}$ are
zeros
of the inner solutions and thus they are located inside the boundary layers, close to $x=\pm 1$. \\ \\
In this case we can already directly obtain the uniform asymptotic approximation, up to an overall normalization constant ${}_{\indhel}C_{lm\omega}$:
\begin{equation} \label{eq:unif S,p,p' not}
\begin{aligned}
&{}_{\indhel}S_{lm\omega}^{\text{unif}}={}_{\indhel}C_{lm\omega}(1-x)^{\alpha}(1+x)^{\beta}
\Bigg\{e^{-c(1-x)}{}_1F_{1}\Big(-p,2\alpha+1,2c(1-x)\Big) +
\\
&+\frac{\Gamma(2\alpha+1)}{\Gamma(2\alpha+1+p)}\frac{\Gamma(-p')}{\Gamma(2\beta+1)}e^{-i\pi p}(2c)^{p+p'+2\beta+1}e^{-2c}2^{(q+\beta-\alpha)}
e^{-c(1+x)} 
\times \\ &\times 
{}_1F_{1}\Big(-p',2\beta+1,2c(1+x)\Big)+
2^{\left[+(q+\indhel+1)/2+\beta\right]}\frac{\Gamma(2\alpha+1)}{\Gamma(2\alpha+1+p)}e^{-i\pi p}(2c)^{p}e^{-c}e^{+cx}
\times \\ &\times
\left[(1-x)^{+(q-\indhel -1)/2-\alpha}(1+x)^{-(q+\indhel+1)/2-\beta}-2^{\left[-(q+\indhel+1)/2-\beta\right]}(1-x)^{+(q-\indhel -1)/2-\alpha}- \right.
\\
&\left. -2^{\left[+(q-\indhel -1)/2-\alpha\right]}(1+x)^{-(q+\indhel+1)/2-\beta}\right]\Bigg\}
\qquad \qquad \text{when } p\in \mathbb{Z}^{+}\cup\{0\} \text{ and } p'\notin \mathbb{Z}^{+}\cup\{0\}
\end{aligned}
\end{equation}
The uniform approximation when $p\notin \mathbb{Z}^{+}\cup\{0\}$ and $p'\in \mathbb{Z}^{+}\cup\{0\}$ may be obtained by
making the substitutions $x\leftrightarrow -x$ and $\indhel \leftrightarrow -\indhel $ (which imply the substitutions
$\alpha\leftrightarrow \beta$ and $p\leftrightarrow p'$) in (\ref{eq:unif S,p,p' not}).

The irregularity arising from
$e^{-c(1+x)}{}_1F_{1}(-p',2\beta+1,2c(1+x))\sim e^{2c}$
(ignoring factors independent of $x$ and $c$) in the limit
$x\rightarrow +1$ and $c \rightarrow +\infty$ prompted BRW to
discard the case $p'\notin \mathbb{Z}^{+}\cup\{0\}$.
It is clear from (\ref{eq:unif S,p,p' not}), however, that 
this irregularity is nullified by the factor $e^{-2c}$ in front of it, 
brought in by the coefficient ${}_{\indhel}D_{lm\omega}$. 
Note that despite the factor $e^{-2c}$,
close to $x=-1$ this term (which is part of the inner solution
valid in the boundary layer there) is not dominated by the first
term in (\ref{eq:unif S,p,p' not}) (which is the inner solution
valid in the boundary layer near $x=+1$). The reason is that
$e^{-c(1+x)}{}_1F_{1}(-p',2\beta+1,2c(1+x))\sim e^{-2c}$ and
$e^{-c(1-x)}{}_1F_{1}(-p,2\alpha+1,2c(1-x)) \sim e^{-2c}$
where both limits are $x\rightarrow -1$ and $c \rightarrow +\infty$
and we have ignored factors independent of $x$ and $c$. 
In the boundary layer around $x=\pm 1$, the asymptotic approximation valid in the overlap region close to $x=\mp 1$
cancels out the inner solution ${}_{\indhel}S_{lm\omega}^{\text{inn},\mp 1}$ in expression (\ref{eq:unif S,p,p' not}). 
Similarly, in the same boundary layer, 
the asymptotic approximation valid in the overlap region close to $x=\pm 1$ 
cancels out the outer solution, so that only ${}_{\indhel}S_{lm\omega}^{\text{inn},\pm 1}$
contributes to the uniform approximation in that boundary layer.

A similar
reasoning can be applied to the case $p\notin
\mathbb{Z}^{+}\cup\{0\}$.

\subsection*{$p,p'\in \mathbb{Z}^{+}\cup\{0\}$}
 In this case, apart
from the overall normalization constant there is another unknown
constant. We are going to determine this extra unknown by imposing
the $\left\{ x\leftrightarrow -x, \indhel\leftrightarrow -\indhel  \right \}$
symmetry. Using the Teukolsky-Starobinski\u{\i} identities
(\ref{eq:Teuk-Starob. ids.}) and (\ref{eq:Teuk-Starob ids. for
spher.s=2}) together with the symmetry (\ref{eq:S symm.->pi-t,-s}) in the 
inner solution (\ref{eq: inner solution})
we obtain
\begin{equation} \label{eq: ratio D/C spin1}
\begin{aligned}
\frac{{}_{-1}D_{lm\omega}}{{}_{-1}C_{lm\omega}}&=\frac{{}_{+1}C_{lm\omega}}{{}_{+1}D_{lm\omega}}=(-1)^{(l+m)}\frac{{}_{+1}C_{lm\omega}}{{}_{-1}C_{lm\omega}}= \\
& =(-1)^{(l+m)}
\begin{cases}
\displaystyle
\frac{2 \sqrt{(m-q+1)(m-q-1)}}{m(m+1)}c  & \qquad \text{when} \quad m\geq +1     \\
\displaystyle
-\frac{\sqrt{q-1}}{\sqrt{q+1}}       &  \qquad\text{when} \quad m=0           \\
\displaystyle
\frac{m(m-1)}{2 \sqrt{(m-q+1)(m-q-1)}}\frac{1}{c}  & \qquad \text{when} \quad m\leq -1
\end{cases}
\end{aligned}
\end{equation}
\begin{equation} \label{eq: ratio D/C spin2}
\begin{aligned}
&\frac{{}_{-2}D_{lm\omega}}{{}_{-2}C_{lm\omega}}=\frac{{}_{+2}C_{lm\omega}}{{}_{+2}D_{lm\omega}}=(-1)^{(l+m)}\frac{{}_{+2}C_{lm\omega}}{{}_{-2}C_{lm\omega}}=  
(-1)^{(l+m)}
\times \\ &\times
\begin{cases}
\displaystyle
\frac{4 \sqrt{(m-q+1)(m-q-1)(m-q+3)(m-q-3)}}{(m+2)(m+1)m(m-1)}c^{2}  & \text{when} \quad m\geq +2     \\
\displaystyle
\frac{-\sqrt{q(q-2)(q-4)}}{3\sqrt{q+2}}c       & \text{when} \quad m=+1           \\
\displaystyle
\frac{\sqrt{(q-3)(q-1)}}{\sqrt{(q+3)(q+1)}}       & \text{when} \quad m=0           \\
\displaystyle
\frac{-3 \sqrt{q-2}}{\sqrt{q(q+2)(q+4)}}\frac{1}{c}       & \text{when} \quad m=-1           \\
\displaystyle
\frac{(m+1)m(m-1)(m-2)}{4 \sqrt{(m-q+1)(m-q-1)(m-q+3)(m-q-3)}}\frac{1}{c^{2}}  & \text{when} \quad m\geq -2
\end{cases}
\end{aligned}
\end{equation}
Equations (\ref{eq: ratio D/C spin1}) and (\ref{eq: ratio D/C spin2}) have been obtained without imposing
any restrictions on the values of $p$ or $p'$ and might therefore seem to
contradict the result from (\ref{eq: match a2}) and (\ref{eq: match b1}) [or (\ref{eq: match a1}) and (\ref{eq: match b2})]  giving an exponential behaviour with $c$
for the ratio ${}_{\indhel}D_{lm\omega}/{}_{\indhel}C_{lm\omega}$ for the case $p\in \mathbb{Z}^{+}\cup\{0\}$ and $p'\notin \mathbb{Z}^{+}\cup\{0\}$ [or viceversa]. 
We shall see in the next section, however, that equations (\ref{eq: ratio D/C spin1}) and (\ref{eq: ratio D/C spin2}) can only actually be applied to the case
$p,p'\in \mathbb{Z}^{+}\cup\{0\}$ so that there is no such contradiction.

We can already determine in what cases the outer solution has a zero. 
Clearly, from equations (\ref{eq: match a2}), (\ref{eq: match b2}), (\ref{eq: ratio D/C spin1}) and (\ref{eq: ratio D/C spin2}),
the ratio between the coefficients ${}_{\indhel}A_{lm\omega}$ and ${}_{\indhel}B_{lm\omega}$ is proportional 
to a power of $c$,
where the constant of proportionality does not depend on $c$. 
It then follows from the form (\ref{eq: outer solution}) of the outer solution that one exponential term will
dominate for positive $x$ and the other exponential term will dominate for negative $x$, when $c\rightarrow \infty$. 
Therefore the outer solution does not possess a zero far from $x=0$ for large $c$.
The outer solution has a zero if ${}_{\indhel}A_{lm\omega}$ 
and ${}_{\indhel}B_{lm\omega}$ have
different sign and it does not otherwise. From equations (\ref{eq:def pp'}), (\ref{eq: match a2}),
(\ref{eq: match b2}), (\ref{eq: ratio D/C spin1}) and
(\ref{eq: ratio D/C spin2}) we have:
\begin{equation}
sign\left(\frac{{}_{\indhel}A_{lm\omega}}{{}_{\indhel}B_{lm\omega}}\right)=
(-1)^{(p-p')}*sign\left(\frac{{}_{\indhel}C_{lm\omega}}{{}_{\indhel}D_{lm\omega}}\right)=(-1)^{(l+m)}
\end{equation}

Furthermore, we can calculate what the location of the zero of the outer solution is to leading order in $c$: by setting the outer solution
(\ref{eq: outer solution}) equal to zero and using (\ref{eq: match a2}) and (\ref{eq: match b2}) (since we have already seen that if $p$ and/or $p'$
$\notin \mathbb{Z}^{+}\cup\{0\}$ the outer solution does not have a zero) we obtain that for large frequency the zero is located at the following
value of $x$:
\begin{equation}
\begin{aligned}
x_0&=\frac{1}{2c}\log\left(-\frac{{}_{\indhel}B_{lm\omega}}{{}_{\indhel}A_{lm\omega}}\right)= \\
& =\frac{1}{2c}\log\left(-
2^{(-\indhel+\alpha-\beta)}\frac{\Gamma(|m-\indhel |+1)\Gamma(|m+\indhel |+1+p)}{\Gamma(|m+\indhel |+1)\Gamma(|m-\indhel |+1+p')}\right.\times\\
&\qquad\times\left.e^{-i\pi(p'-p)}(2c)^{(p'-p)}\frac{{}_{\indhel}D_{lm\omega}}{{}_{\indhel}C_{lm\omega}}\right)
\end{aligned}
\end{equation}
Clearly there is one zero in the region between the two boundary layers tending to the location $x=0$ as $c$ becomes large if ${}_{\indhel}B_{lm\omega}$ and
${}_{\indhel}A_{lm\omega}$ have different sign and there is not a zero if they have the same sign.

Finally, the uniform asymptotic approximation for this case is:
\begin{equation} \label{eq:unif S,p,p'}
\begin{aligned}
&{}_{\indhel}S_{lm\omega}^{\text{unif}}={}_{\indhel}C_{lm\omega}(1-x)^{\alpha}(1+x)^{\beta}\Bigg\{e^{-c(1-x)}{}_1F_{1}\Big(-p,2\alpha+1,2c(1-x)\Big)+
\\ &
+\frac{{}_{\indhel}D_{lm\omega}}{{}_{\indhel}C_{lm\omega}}e^{-c(1+x)}{}_1F_{1}\Big(-p',2\beta+1,2c(1+x)\Big)+
\\ & +
2^{\left[(q+\indhel+1)/2+\beta\right]}\frac{\Gamma(2\alpha+1)}{\Gamma(2\alpha+1+p)}e^{-i\pi p}(2c)^{p}e^{-c}e^{+cx}
\times \\ &\times
\left[(1-x)^{+(q-\indhel -1)/2-\alpha}(1+x)^{-(q+\indhel+1)/2-\beta}-2^{-\left[(q+\indhel+1)/2+\beta\right]}(1-x)^{+(q-\indhel -1)/2-\alpha}\right]+
\\ &
+\frac{{}_{\indhel}D_{lm\omega}}{{}_{\indhel}C_{lm\omega}}2^{\left[(q-\indhel+1)/2+\alpha\right]}
\frac{\Gamma(2\beta+1)}{\Gamma(2\beta+1+p')}e^{-i\pi p'}(2c)^{p'}e^{-c}e^{-cx}
\\ &
\left[(1+x)^{+(q+\indhel -1)/2-\beta}(1-x)^{-(q-\indhel+1)/2-\alpha}-2^{-\left[(q-\indhel+1)/2+\alpha\right]}(1+x)^{+(q+\indhel -1)/2-\beta}\right]\Bigg\}
\\ & \qquad \qquad \qquad \qquad \qquad \qquad \qquad \qquad \qquad \qquad \qquad\qquad \qquad \text{when } p,p'\in \mathbb{Z}^{+}\cup\{0\}
\end{aligned}
\end{equation}
where the ratio between ${}_{\indhel}D_{lm\omega}$ and ${}_{\indhel}C_{lm\omega}$ is given by 
(\ref{eq: ratio D/C spin1}) for $\indhel=\pm 1$ and by (\ref{eq: ratio D/C spin2}) for $\indhel=\pm 2$.

Similar cancelations to the ones for the case $p\in \mathbb{Z}^{+}\cup\{0\}$ and $p'\notin \mathbb{Z}^{+}\cup\{0\}$
occur in the present case for the uniform solution (\ref{eq:unif S,p,p'}). 
The only difference is that now, in the boundary layer around $x=\pm 1$, the asymptotic approximation valid in the overlap
region close to $x=\mp 1$ only cancels out part of the outer solution. The other part of the outer solution, however,
is exponentially negligible with respect to the inner solution ${}_{\indhel}S_{lm\omega}^{\text{inn},\pm 1}$.


\section{Calculation of $\gamma$} \label{sec:Evaluation of gamma}
To finally determine the value of $\gamma$ we only need to impose
that our asymptotic solution must have the correct number of
zeros. BRW give the number of zeros of the SWSH for non-negative
$m$ and $\indhel $. Straightforwardly generalizing their result for all
$m$ and $\indhel $ using the symmetries of the differential equation we
have:
\begin{theorem}\label{th: zeros S}
\textbf{Zeros of S: }
The number of zeros of ${}_{\indhel}S_{lm\omega}$ is independent of $c$ and for $x\in (-1,1)$ is equal to
\begin{equation}
\left\{
\begin{array}{ll}
l-|m| & \text{for} \quad |m|\geq |\indhel|\\
l-|\indhel| & \text{for} \quad |m|< |\indhel|
\end{array}
\right.
\end{equation}
\end{theorem}

\catdraft{dir si i quan te zeros a $x=\pm 1$}

The number of zeros of the confluent hypergeometric function is also needed, and that is given by Buchholz ~\cite{bk:Buchholz}:

The number of positive, real zeros of ${}_1F_{1}(-a,b,z)$ when $b>0$ is
\begin{equation}
\left\{
\begin{array}{ll}
-[-a] & \text{for} \quad  +\infty >a \geq 0 \\
0     & \text{for} \quad  0 \geq a >-\infty
\end{array}
\right.
\end{equation}
where $[n]$ means the largest integer $\leq n$.

Since the confluent hypergeometric functions are part of the inner
solutions and the region of validity of these solutions becomes
tighter to the boundary points as $c$ increases, the zeros of
${}_1F_{1}(-p,2\alpha+1,u)$ are grouped together close
to $x=+1$, and likewise for ${}_1F_{1}(-p',2\beta+1,u^{*})$ close
to $x=-1$. Apart from these zeros, for
large $c$, the function ${}_{\indhel}S_{lm\omega}$ may only have other zeros at $x=\pm 1$ and/or at
$x=x_0$. The possible one at $x=x_0$ is not due to the confluent
hypergeometric functions but to the outer solution. We define the
variable $z_{0}$ so that it has value $+1$ if ${}_{\indhel}S_{lm\omega}$ has a
zero at $x=x_0$ and value $0$ if it does not.

From equation (\ref{eq:def pp'}) we see that $p'=p+(|m+\indhel |+2\indhel -|m-\indhel |)/2$, 
and therefore if either $p$ or $p'$ is integer then the other one must be integer as
well. But, as we saw in Section \ref{sec:Matching the solutions}, at least one of $p$ and $p'$ 
(if not both) must be a positive integer or zero.
Therefore both $p$ and $p'$ must be integers and at least one of them is positive or zero. It also follows from (\ref{eq:def pp'}) that
\begin{equation}
\gamma=2(p+p')+2+|m+\indhel |+|m-\indhel |=2q
\end{equation}
where it is now clear that $q\in \mathbb{Z}$.

Requiring that the number of zeros of the asymptotic solution coincides with the number of zeros of the
SWSH results in the condition

\begin{equation} \label{eq:equal number of zeros}
\begin{aligned}
&\left\{
\begin{array}{ll}
-(|m+\indhel |+\indhel+1)/2+q/2 & \text{for} \quad q \geq |m+\indhel |+\indhel+1 \\
0                    & \text{for} \quad q < |m+\indhel |+\indhel+1
\end{array}
\right\}+ \\
+&\left\{
\begin{array}{ll}
-(|m-\indhel |-\indhel+1)/2+q/2 & \text{for} \quad q \geq |m-\indhel |-\indhel+1 \\
0                  & \text{for} \quad q < |m-\indhel |-\indhel+1
\end{array}
\right\}+ \\
+&
z_{0}= \left\{
\begin{array}{ll}
l-|m| & \text{for} \quad |m|\geq |\indhel|\\
l-|\indhel| & \text{for} \quad |m|< |\indhel|
\end{array}
\right\}
\end{aligned}
\end{equation}

From (\ref{eq:equal number of zeros}) and the fact that $z_{0}=0$ when either $p$ or $p'\notin \mathbb{Z}^{+}\cup\{0\}$ as seen in
Section \ref{sec:Matching the solutions}, we obtain the value of $q$ in all different cases:

\begin{subequations}  \label{eq:val. of q}
\begin{align}
\begin{split}
q&=\left\{
\begin{array}{ll}
l-|m| &\text{for} \quad |m|\geq |\indhel|\\
l-|\indhel| &\text{for} \quad |m|<|\indhel|
\end{array}
\right\}
+\frac{(|m+\indhel |+|m-\indhel |)}{2}+1-z_{0}
\quad\\
&\hspace{7cm} \text{if} \quad l \geq l_{1},l_{2} \ (\text{i.e., } p,p'\in \mathbb{Z}^{+}\cup\{0\}) \label{eq:1st q} \\
\end{split} \\ \begin{split}
q&=2\left\{
\begin{array}{ll}
l-|m| &\text{for} \quad |m|\geq |\indhel|\\
l-|\indhel| &\text{for} \quad |m|<|\indhel|
\end{array}
\right\}
+|m+\indhel |+\indhel+1
\quad \\
&\hspace{7cm}\text{if} \quad l<l_{2} \ (\text{i.e., } p\in,p'\notin \mathbb{Z}^{+}\cup\{0\})  \label{eq:2nd q} \\
\end{split} \\ \begin{split}
q&=2\left\{
\begin{array}{ll}
l-|m| &\text{for} \quad |m|\geq |\indhel|\\
l-|\indhel| &\text{for} \quad |m|<|\indhel|
\end{array}
\right\}
+|m-\indhel |-\indhel+1
\quad \\
&\hspace{7cm}\text{if} \quad l<l_{1} \ (\text{i.e., } p\notin,p'\in
\mathbb{Z}^{+}\cup\{0\})  \label{eq:3rd q}
\end{split}
\end{align}
\end{subequations}

where \\ \\
$l_{1}\equiv \left\{ \begin{array}{ll} |m| &\text{for} \quad |m|\geq |\indhel|\\ |\indhel| &\text{for} \quad |m|<|\indhel|\end{array} \right\}+
(|m+\indhel |-|m-\indhel |)/2+\indhel $ \\
$l_{2}\equiv \left\{ \begin{array}{ll} |m| &\text{for} \quad |m|\geq |\indhel|\\ |\indhel| &\text{for} \quad |m|<|\indhel| \end{array} \right\}+
(|m-\indhel |-|m+\indhel |)/2-\indhel $ \\

By requiring in (\ref{eq:1st q}) that $q$ must also satisfy (\ref{eq:def pp'}) and bearing in mind that $z_{0}$ can only have the values $0$ or $1$, it must
be
\begin{equation} \label{eq:z0=0,1 if S has zero at x=0 or not}
z_{0}=
\begin{cases}
0 & \text{for} \quad l-l_{1} \quad \text{even}   \\
1 & \text{for} \quad l-l_{1} \quad \text{odd}
\end{cases}
\end{equation}
where $l_{2}$ instead of $l_{1}$ could have been used, since one
is equal to the other one plus an even number.

It can be trivially seen that if $l_{1}$ has an allowed value,
i.e.,
\begin{equation}
l_{1}\geq \left\{ \begin{array}{ll} |m| &\text{for} \quad |m|\geq
|\indhel|\\ |\indhel| &\text{for} \quad |m|<|\indhel|\end{array} \right\},
\end{equation}
then $l_{2}$ does not, and vice-versa, so that cases (\ref{eq:2nd
q}) and (\ref{eq:3rd q}) are mutually exclusive.

Clearly, when $l<l_{1}$ or $l<l_{2}$, for fixed $\indhel $ and $m$, as $l$ is increased by $1$ the corresponding value of $q$ is also increased by $1$, so that two
different values of $l$ correspond to two different values of $q$. However, once the threshold $l \geq \max(l_{1},l_{2})$ is reached, every increase of $2$ in $l$
will involve the subtraction of an extra $1$ in (\ref{eq:1st q}) via $z_{0}$, so that its corresponding value of $q$ will be the same as for the previous $l$.
Therefore, in the region $l \geq \max(l_{1},l_{2})$, every value of $q$ will correspond to two consecutive, different $l$'s: the two corresponding SWSH's
will have the same number of zeros and behaviour close to the boundary points, but one will have a zero at $x=x_0$ and the other one will not.

%

Another feature that can be seen is that, for $\indhel = \pm 1$, the case
$l < \max(l_{1},l_{2})$ (i.e., $p_+$ and/or $p_-\notin \mathbb{Z}^{+}\cup\{0\}$) 
 implies $q-m=\pm 1$ or $q=+1$ when
$m\geq 1$ and $m=0$ respectively, so that the leading order behaviour given by
(\ref{eq: ratio D/C spin1}) vanishes for these cases. 
That is, in these cases we have not gone far enough in the asymptotic expansion (\ref{eq: ratio D/C spin1}).
When $m\leq -1$ it follows from (\ref{eq:2nd q})
and (\ref{eq:3rd q}) that 
$l < l_{1}(=l_{2})$ 
requires $l<|m|$,
which is not allowed.
Therefore, expression (\ref{eq: ratio D/C spin1}) is not applicable to the case $l < \max(l_{1},l_{2})$,
as already mentioned in the previous section.

Similarly, for $\indhel = \pm 2$, the case
$l < \max(l_{1},l_{2})$
implies
\begin{equation*}
\begin{cases}
m-q=\pm 1,\pm 3& \text{when} \quad m \geq 2   \\
q=0,2,4 & \text{when} \quad m=1 \\
q=1,3& \text{when} \quad m=0 \\
q=2& \text{when} \quad m=-1 \\
l<|m|& \text{when} \quad m\leq-2 \\
\end{cases}
\end{equation*}
so that the leading order behaviour given by (\ref{eq: ratio D/C spin2}) does then not apply.

Note that the scalar case is obtained from our formulae as a
particular case. Setting $\indhel =0$ in the equations above we have
$l_{1}=l_{2}=|m|$ and therefore $l$ will always be greater or
equal than both $l_{1}$ and $l_{2}$ so that (\ref{eq:1st q}) will
apply, and it gives $q=l+1-z_{0}$ with
\begin{equation*}
z_{0}=
\left\{
\begin{array}{ll} 0 &\text{for} \quad l-|m| \quad \text{even}   \\
1 &\text{for}  \quad l-|m| \quad \text{odd} \end{array} \right\}.
\end{equation*}
We also have $p=p'\in  \mathbb{Z}^{+}\cup\{0\}$ and $2\alpha
=2\beta =|m|$ and then the confluent hypergeometric functions are
just the generalized Laguerre polynomials:
${}_1F_{1}(-p,|m|+1,z)\propto L_{p}^{(|m|)}(z)$. Finally, because
of the existence of the $x\leftrightarrow -x$ symmetry in the
scalar case, we have that ${}_0B_{lm\omega}=\pm {}_0A_{lm\omega}$ in
(\ref{eq: outer solution}) and therefore the zero of the outer
solution, if it exists, will be located exactly at $x=0$. All
these results for the scalar case coincide with
~\cite{bk:high_transc_funcs},~\cite{bk:Flammer} and ~\cite{bk:Meixner&Schafke}.


\section{Numerical method} \label{sec:num. method; high freq. sph.}

Two different methods have been used to obtain the numerical data. One method is the one used by Sasaki and Nakamura~\cite{ar:Sasa&Naka'82}, consisting in
approximating the differential equation (\ref{eq:ang. teuk. eq. for y}) by a finite difference equation, 
and then finding the eigenvalue as the value of ${}_{\indhel}E_{lm\omega}$
that makes zero the determinant of the resulting (tri-diagonal) matricial equation. We have used this method to find the eigenvalues for several
large values of the frequency. However, we used the shooting method described in the previous chapter
to calculate the spin-weighted spheroidal function.

Sasaki and Nakamura's method, which they only develop explicitly for the case $\indhel =-2$ and $m=0$ solves the
angular differential equation (\ref{eq:ang. teuk. eq. for y}) re-written with derivatives with respect to $\theta$ rather than $x$:
\begin{equation} \label{eq:ang. teuk. eq. for y with theta-derivs.}
\begin{aligned}
& \Bigg\{
 \ddiff{}{\theta}+\frac{1}{\sin\theta}\left[2(\alpha-\beta)+2(\alpha+\beta)\cos\theta+\cos\theta\right]\diff{}{\theta}+  \\
&  +{}_{\indhel}E_{lm\omega}-(\alpha+\beta)(\alpha+\beta+1)+c^{2}\cos^{2}\theta-2\indhel c\cos\theta 
\Bigg\} {}_{\indhel}y_{lm\omega}(\theta)=0
\end{aligned}
\end{equation}
This equation is approximated by a finite-difference equation. Apart from at the boundaries, the derivatives are replaced with central differences.
At the boundary points, the regularity condition (\ref{eq:asympt. S for x->+/-1}) requires that $\left. \d{{}_{\indhel}y_{lm\omega}}/\d{\theta}\right|_{x=\pm 1}=0$ and
the first order derivative (which has a factor $1/\sin\theta$ in front) is approximated by a forward/backward difference at $x=+1/-1$ respectively.
The result is that equation
(\ref{eq:ang. teuk. eq. for y with theta-derivs.}) is approximated by
\begin{equation} \label {eq:finite-diff. ang. teuk. eq. for y with theta-derivs.}
\begin{aligned}
&
\frac{{}_{\indhel}y^{i+1}_{lm\omega}-2{}_{\indhel}y^{i}_{lm\omega}+{}_{\indhel}y^{i-1}_{lm\omega}}{(\Delta\theta)^2}+\\
&+\frac{1}{\sin\theta_i}\left[2(\alpha-\beta)+2(\alpha+\beta)\cos\theta_i+\cos\theta_i\right]
\frac{{}_{\indhel}y^{i+1}_{lm\omega}-{}_{\indhel}y^{i-1}_{lm\omega}}{2\Delta\theta}+ \\
&
+\left[{}_{\indhel}E_{lm\omega}-(\alpha+\beta)(\alpha+\beta+1)+c^{2}\cos^{2}\theta_i-2\indhel c\cos\theta_i
\right]{}_{\indhel}y^{i}_{lm\omega}=0,\\
&\hspace{10cm} \text{for\, } i=2,\dots,2N \\
& 2(1+2\alpha)\frac{2{}_{\indhel}y^{i+1}_{lm\omega}-2{}_{\indhel}y^{i}_{lm\omega}}{(\Delta\theta)^2}+ \\
& +\left[{}_{\indhel}E_{lm\omega}-(\alpha+\beta)(\alpha+\beta+1)+c^{2}-2\indhel c\right]{}_{\indhel}y^{i}_{lm\omega}=0,  \text{\, for\, } i=1 \, (\theta=0) \\
& -4\beta\frac{2{}_{\indhel}y^{i-1}_{lm\omega}-2{}_{\indhel}y^{i}_{lm\omega}}{(\Delta\theta)^2}+ \\
& +\left[{}_{\indhel}E_{lm\omega}-(\alpha+\beta)(\alpha+\beta+1)+c^{2}+2\indhel c\right]{}_{\indhel}y^{i}_{lm\omega}=0, \text{\, for\, } i=2N+1 \, (\theta=\pi)
\end{aligned}
\end{equation}
\draft{wrong factor of 2 (p.21J'')}
where
$\theta_i=\pi(i-1)/(2N)\equiv \Delta\theta(i-1)$ and
$i=1,2,\dots,2N+1$. Equation (\ref{eq:finite-diff. ang. teuk. eq.
for y with theta-derivs.}) can be represented as the product of a
square, tridiagonal matrix $A$ of dimension $(2N+1)\times(2N+1)$
and the vector of elements ${}_{\indhel}y^{i}_{lm\omega}$ equal to zero.
In order to find the eigenvalue we impose that the determinant of matrix $A$ is zero.

\draft{describe calculation of determinant with LU-factoritzacio (taken from ch.2.5Gerald\&Wheatley) followed by (Lapack?) routine 'dgdbi'?}

We found that, already with $N=100$, for most modes the values of ${}_{\indhel}E_{lm\omega}$ obtained to quadruple precision  actually provided
values of the determinant so large that were even greater than the machine's largest number. We therefore decided to use this method
(where the argument and the exponent of the value of the determinant must be passed on separately to the zero-finding routine \routine{zbrent})
only to find eigenvalues and use the program \program{sphdrvKN mpi.f90} (described in the previous chapter)
when we wish to find both eigenvalues and spherical functions.
In fact, Sasaki and Nakamura's method without finding the spherical funcion is so much faster than \program{sphdrvKN mpi.f90},
that it is the preferable method to use if we want to find eigenvalues far from any known eigenvalue (as we analytically do for $\omega =0$ for example).
This is why we used Sasaki and Nakamura's method to find the eigenvalues for large frequency and then used the resulting eigenvalue to find
the corresponding spherical function with \program{sphdrvKNlargew mpi.f90}, a variation of \program{sphdrvKN mpi.f90}.
\catdraft{\program{sphdrvKN mpi.f90 -r1.1.1.2} has been renamed to \program{sphdrvKNlargew mpi.f90 -r1.1.1.2} 
and \program{sphdrvDetZeroKN mine mpi.f90} has been renamed to \program{sphdrvDetZeroKN mpi.f90} in the thesis!->do likewise in CD!}

The Fortran90 program \program{sphdrvDetZeroKN mpi.f90}
implements Sasaki and Nakamura's method to find eigenvalues,
particularly adapted to the case of large frequency. It calculates
${}_{\indhel}\lambda_{lm\omega}$ rather than ${}_{\indhel}E_{lm\omega}$ since
${}_{\indhel}\lambda_{lm\omega}\sim O(c)$ for large $c$ whereas
${}_{\indhel}E_{lm\omega}\sim O(c^2)$. It starts with the known
value of ${}_{\indhel}\lambda_{l,m,\omega =0}$ (\ref{eq:eigenval. for c=0}) and
finds the eigenvalue ${}_{\indhel}\lambda_{lm\omega}$ for increasing frequency
by looking for a zero of the determinant of the matrix $A$. This
procedure is smooth no matter how large the frequency is if
$l<l_1$ or $l<l_2$. However, if $l\geq \max(l_1,l_2)$, for some large
value of the frequency, the eigenvalues for two consecutive values
of $l$ are so close (since they correspond to the same $q$ and
therefore their leading order term for large frequency is the
same) that the initial bracketing of the eigenvalue includes both
eigenvalues and therefore $\det A$ calculated with the values of
${}_{\indhel}\lambda_{lm\omega}$ at the two ends of the bracket has the same
sign. From this value of the frequency on, instead of looking for
a zero of the determinant the program just looks for the value
${}_{\indhel}\lambda_{lm\omega}$ that is an extreme of the determinant. The
reason is that this provides a point which is in between the two
actual eigenvalues and it is therefore useful both as an
approximation and as a bracket point for either of them. Instead
of using minimization/maximization routines, which are very costly
in terms of accuracy and time, in order to find an extreme of
$\det A$, the program \program{sphdrvDetZeroKN mpi.f90} looks
for a zero of the derivative of $\det A$, which can be calculated
to be
\begin{equation}
\diff{(\det A)}{{}_{\indhel}\lambda_{lm\omega}}=\mathop{\mathrm{trace}}
\left[(\det A) A^{-1}\right]
\end{equation}
and is very easy to evaluate.
The program \program{sphdrvDetZeroKN mpi.f90} provided the graphs of ${}_{\indhel}\lambda_{lm\omega}$
as a function of $\omega $ for large frequency and there is therefore no need for it to distinguish with accuracy between the two consecutive eigenvalues.

The extreme point of the determinant found by \program{sphdrvDetZeroKN mpi.f90} is used by the program
\program{sphdrvKNlargew mpi.f90} to bracket and determine the two close eigenvalues and their corresponding angular functions.
The program \program{sphdrvKNlargew mpi.f90} uses the shooting method and Runge-Kutta integration as used by \program{sphdrvKN mpi.f90} but the
main part of the program is adapted to look for eigenvalues for large frequency. It initially looks for a zero of the function
$g({}_{\indhel}E_{lm\omega})$ inside a bracket of the eigenvalue, and if it is not bracketed it
then assumes that it is because the frequency is large enough so that there are two eigenvalues inside the bracket corresponding to two different,
consecutive $l$'s. It then calls
the routine \routine{brent} to look for a minimum of $g({}_{\indhel}E_{lm\omega})$ (with a possible change of sign if there is a maximum instead) and uses that
minimum to find a zero to its right or to its left depending on which one corresponds to the $l$ we are interested in, according to (\ref{eq:val. of q}).
The program \program{sphdrvKNlargew mpi.f90} also finds the zero of the function ${}_{\indhel}S_{lm\omega}$ close to $x=0$ for large $\omega $
if it has one as indicated by (\ref{eq:z0=0,1 if S has zero at x=0 or not}), uses a smaller stepsize in $x$ close to $x=\pm 1$ to
cater for the rapid oscillations of the angular function there for large $\omega $ and makes use of equations (\ref{eq:val. of q})
and (\ref{eq:series E for large w}) to help bracket the eigenvalue.
The program \program{sphdrvKNlargew mpi.f90} provided the graphs of ${}_{\indhel}S_{lm\omega}(\theta)$ for large frequency.


\section{Numerical results} \label{sec:num. results; high freq. sph.}

\draft{describe S\&N's method in detail?}

All the numerical results and graphs in this section have been obtained setting $Q=0$, $a=0.95$ and $M=1$.

There is an obvious numerical problem when $p,p'\in
\mathbb{Z}^{+}\cup\{0\}$. In this case, as mentioned in Section
\ref{sec:Evaluation of gamma}, the eigenvalues for two different
values of $l$ (but same $\indhel ,m$) become exponentially close as
$c$ increases (~\cite{ar:BRW}). This means that for this case we are
not able to find the functions for very large values of the
frequency. For example, in the case below for $\indhel =-1$ and $m=1$, when
$\omega =25$ the eigenvalues for $l=3$ and $l=4$ only differ in their
14th digit.

\begin{figure}[p]
\rotatebox{90}
\centering
\includegraphics*[width=80mm,angle=270]{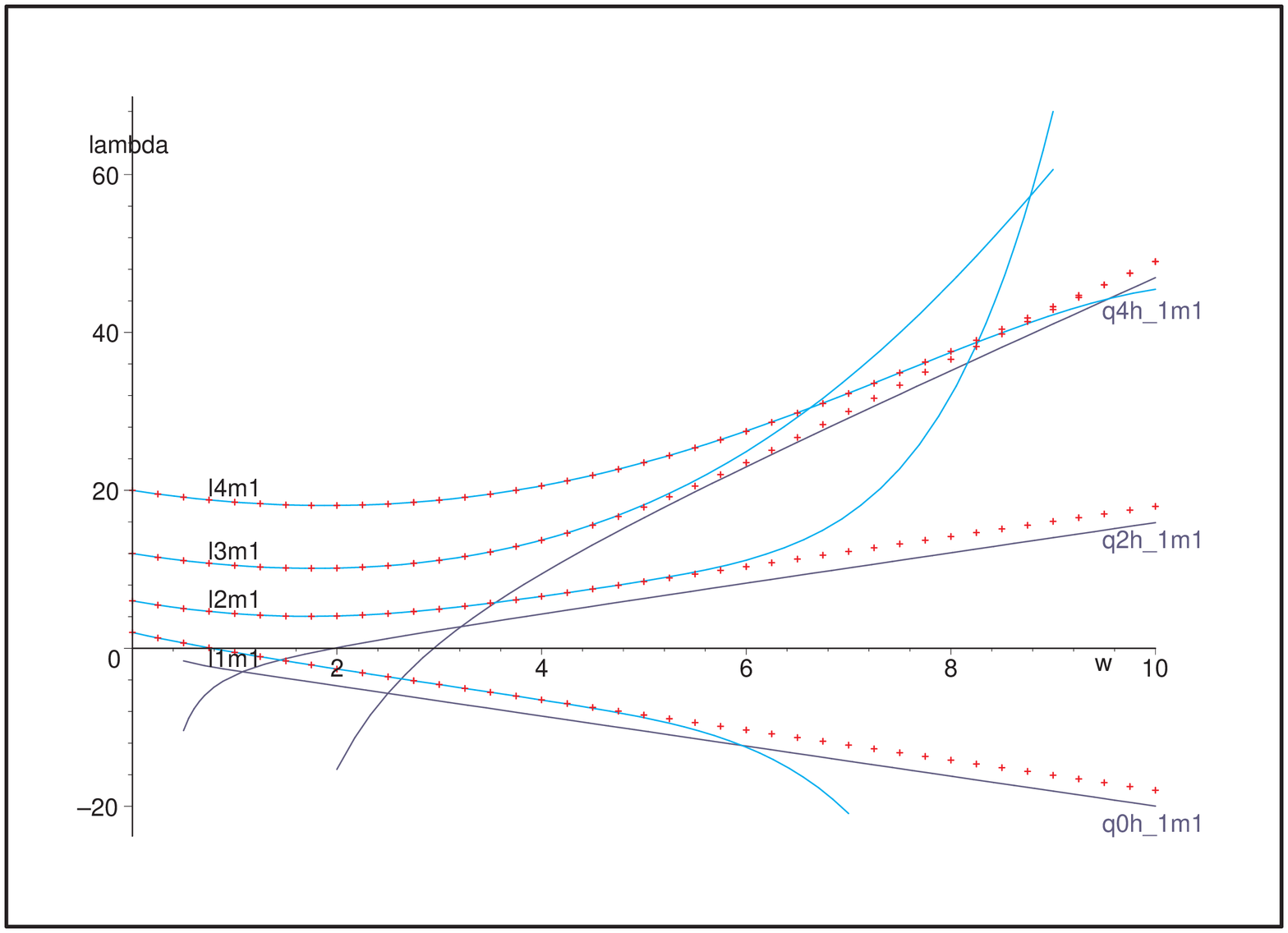}
\includegraphics*[width=80mm,angle=270]{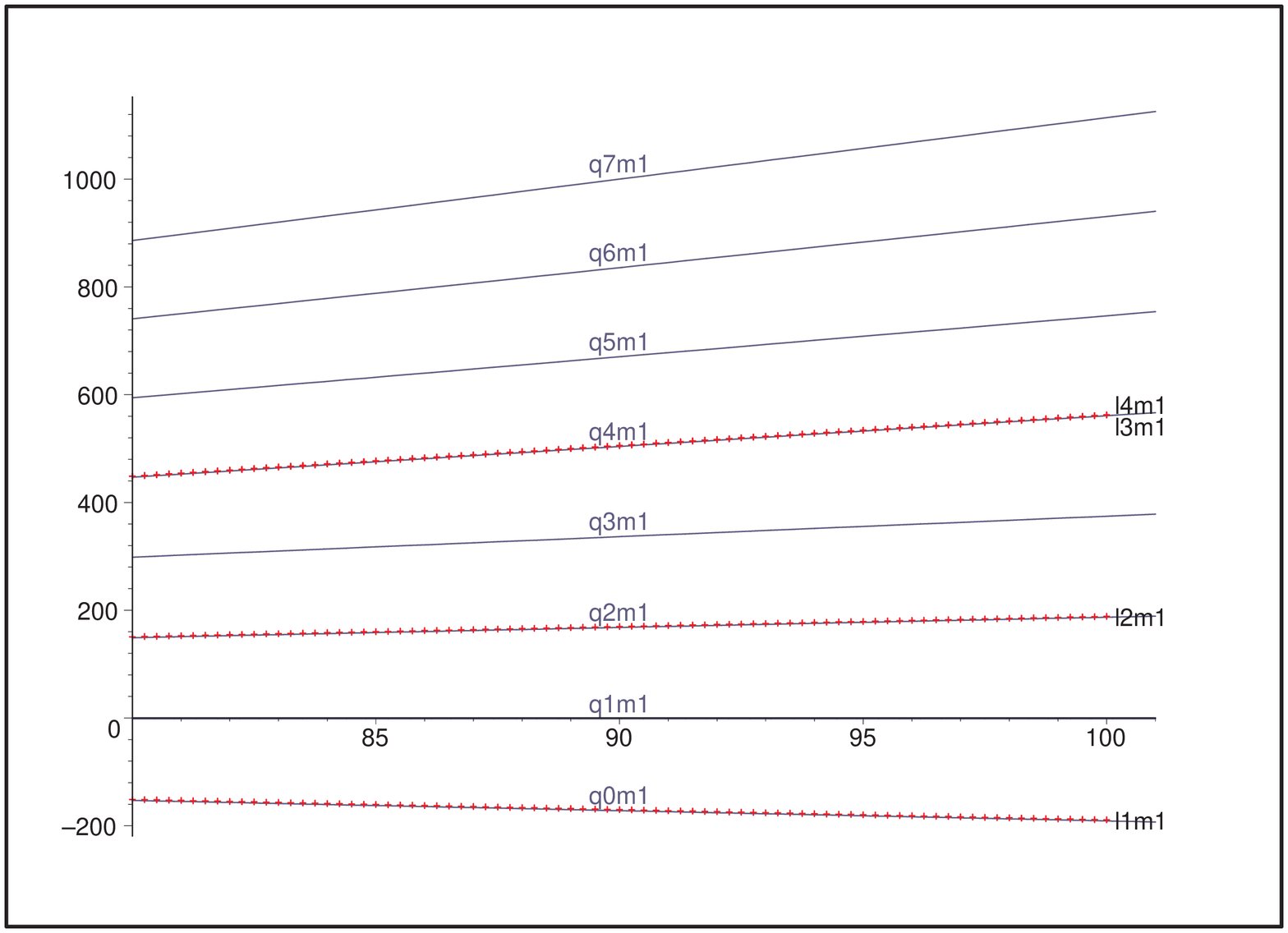}
\caption{${}_{-1}\lambda_{l,1,\omega}$ as a function of $\omega$ for several $l$ and $q$. 
The red crosses are the numerical data. 
The navy blue lines are using BRW's expansion for ${}_{\indhel}\lambda_{lm\omega}$
and the light blue lines are Press and Teukolsky's.
} \label{fig:lambda_s_1m1w0to100}
\end{figure}
\begin{figure}[p]
\rotatebox{90}
\centering
\includegraphics*[width=80mm,angle=270]{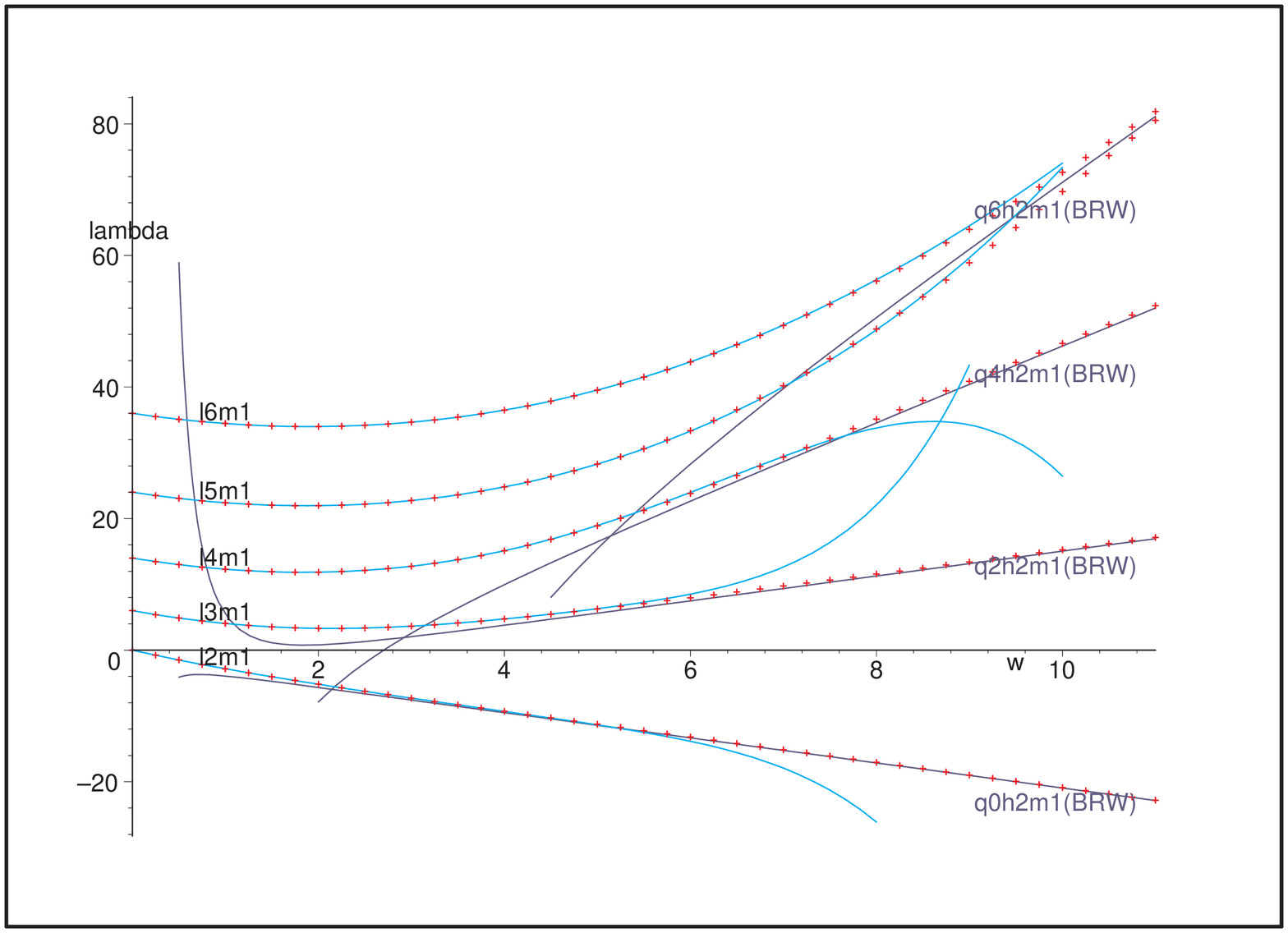}
\includegraphics*[width=80mm,angle=270]{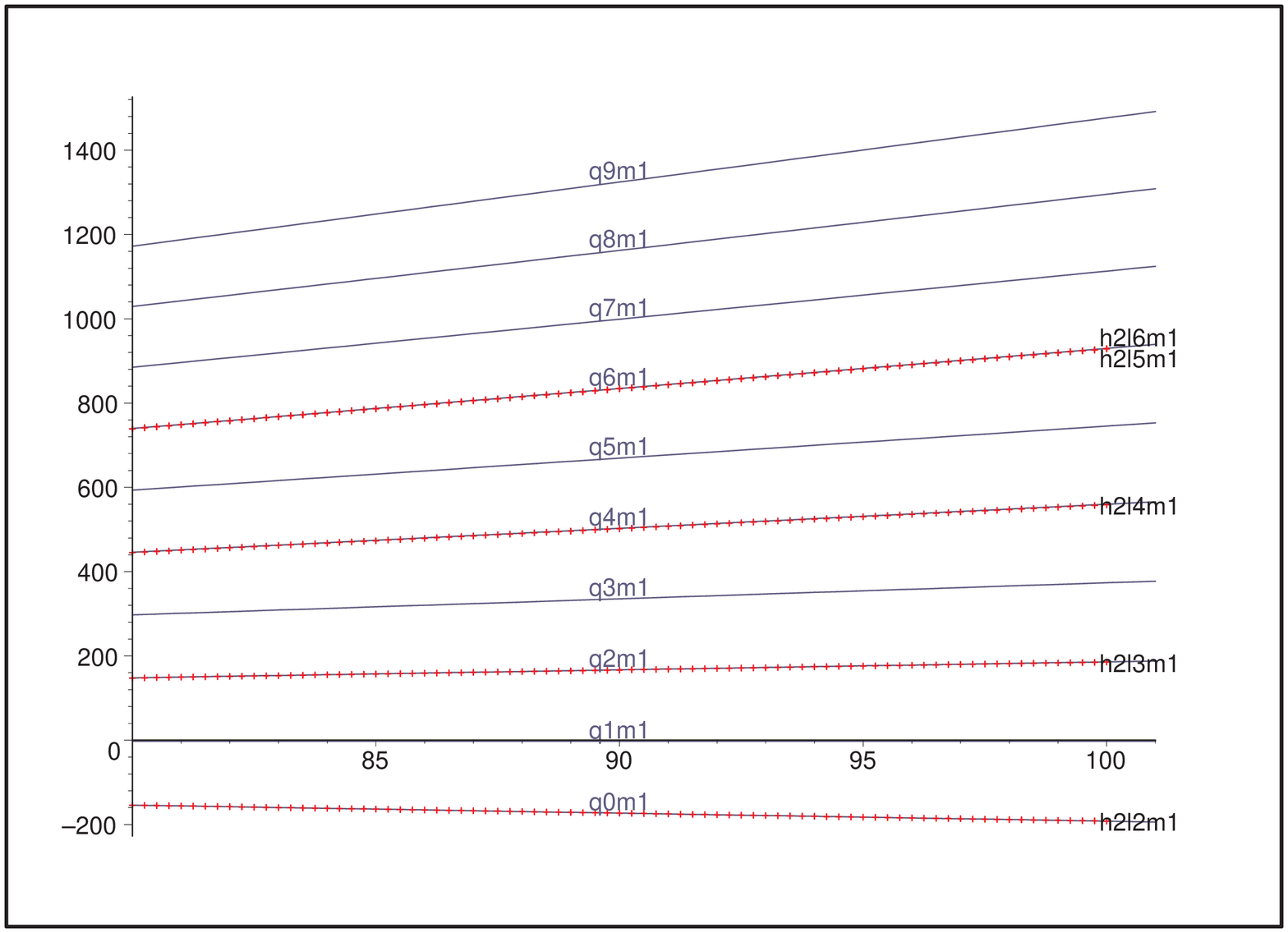}
\caption{${}_{+2}\lambda_{l,1,\omega}$ as a function of $\omega$ for several $l$ and $q$. 
The red crosses are the numerical data. 
The navy blue lines are using BRW's expansion for ${}_{\indhel}\lambda_{lm\omega}$
and the light blue lines are Press and Teukolsky's.
} \label{fig:lambda_s2m1w0to100}
\end{figure}

BRW do give the analytical value for $q$ for spin-0. For spin
different from zero, however, they try to numerically match their
large-frequency asymptotic expansion of the eigenvalue with the
expansion for small frequency given by Press and Teukolsky
(~\cite{ar:Press&Teuk'73} and ~\cite{ar:Teuk&Press'74}). As can
be seen in Figures 
\ref{fig:lambda_s_1m1w0to100}
and
\ref{fig:lambda_s2m1w0to100}, this matching at intermediate
values of the frequency might be good for certain cases,
especially for
 small $l$, but not for other ones. All eigenvalues start off for frequency zero at the value given by (\ref{eq:eigenval. for c=0}), as expected, and when
$l\geq l_{1}$ or $l_{2}$ the pairs of curves that share the same
value of $q$ become exponentially closer and closer to each other
 as the frequency increases. When the frequency is as large as $100$, the curves fully coincide in
the expected pairs for large frequency (given by
equation (\ref{eq:series E for large w}), and BRW for lower order terms) where $q$ comes in as
a parameter. From this, the corresponding value of $q$ for a certain set of values of $\{l,m,\indhel\}$ can be inferred, 
and this coincides with the one given by equations (\ref{eq:1st q}), (\ref{eq:2nd q}) and (\ref{eq:3rd q}).

\begin{figure}[p]
\rotatebox{90}
\centering
\includegraphics*[width=90mm,angle=270]{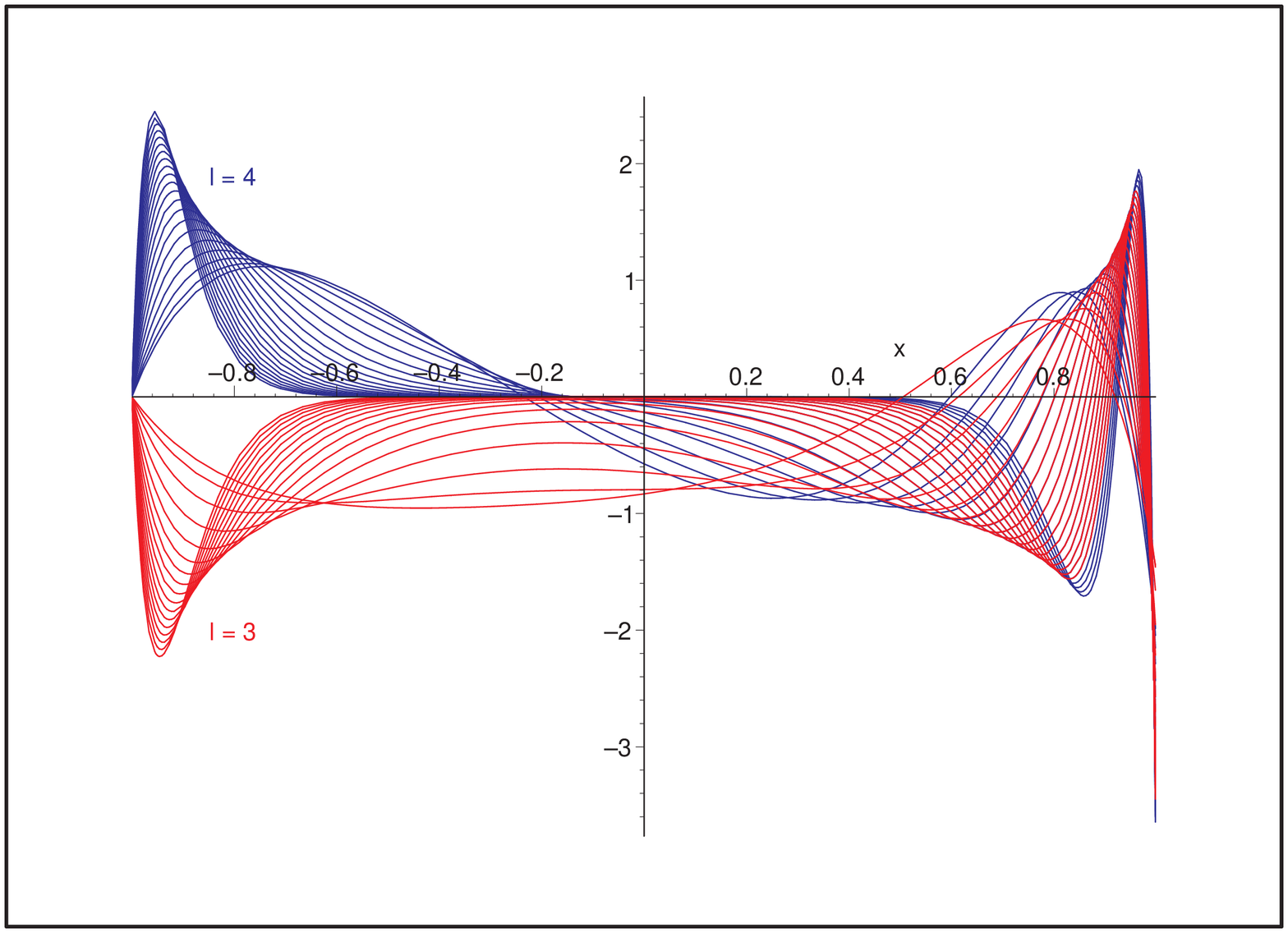}
\caption{${}_{-1}S_{l,1,\omega}$ for $l=3\ \&\ 4$, $\omega =5\rightarrow25$.
Blue lines correspond to $l=4$ and the red ones to $l=3$.
As $\omega$ increases the curves become increasingly flattened out in the region close to the origin.} \label{fig:sph_n3n4m1w5to25 first}
\includegraphics*[width=90mm,angle=270]{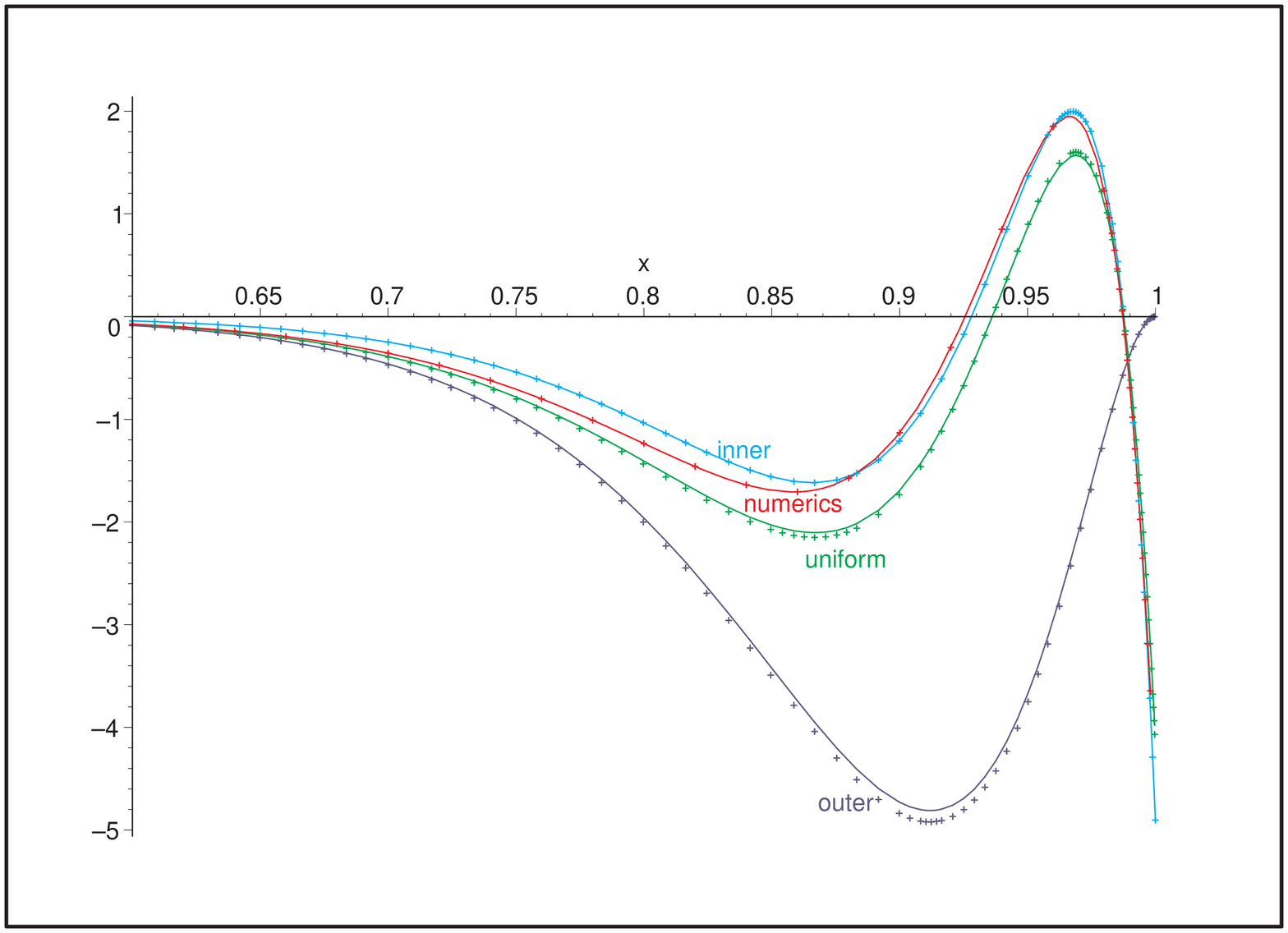}
\caption{${}_{-1}S_{l,1,25}$ for $l=3\ \&\ 4$.
Different solutions as labeled. 
The continuous lines correspond to $l=4$ and the dotted ones to $l=3$.} \label{fig:sph_n3n4m1w5to25 first bis}
\label{fig:sph_n3n4m1w5to25}
\end{figure}

We calculated and plotted in Figure \ref{fig:sph_n3n4m1w5to25 first} the
SWSH for $\indhel =-1$, $l=3\ \&\ 4$, $m=1$, where the value of $q$,
given by (\ref{eq:2nd q}), is the same for both of them: $q=4$
(this is a case where $p,p'\in \mathbb{Z}^{+}\cup\{0\}$). 
Several features can be seen. 
Firstly, as the frequency increases from $\omega =5$ to 25,
 the functions become flattened out in the middle region of $x$ and squeezed out towards the edges. Since the value of $q$ is the same for both cases, the inner solution is
the same for both of them, with the only exception of the relative sign between the inner solution for positive $x$ and negative $x$
(\ref{eq: ratio D/C spin1}). The function for $l=4$ has three zeros and the one for $l=3$ has two (see Theorem \ref{th: zeros S}). The inner solution provides
for the two zeros of $l=3$ and the corresponding two of $l=4$, and these become closer to the boundary point $x=+1$ as the frequency increases. The extra
zero of $l=4$ comes from the outer solution and becomes closer to $x=0$ with increasing frequency.

\begin{figure}[p]
\rotatebox{90}
\centering
\\
\includegraphics*[width=90mm,angle=270]{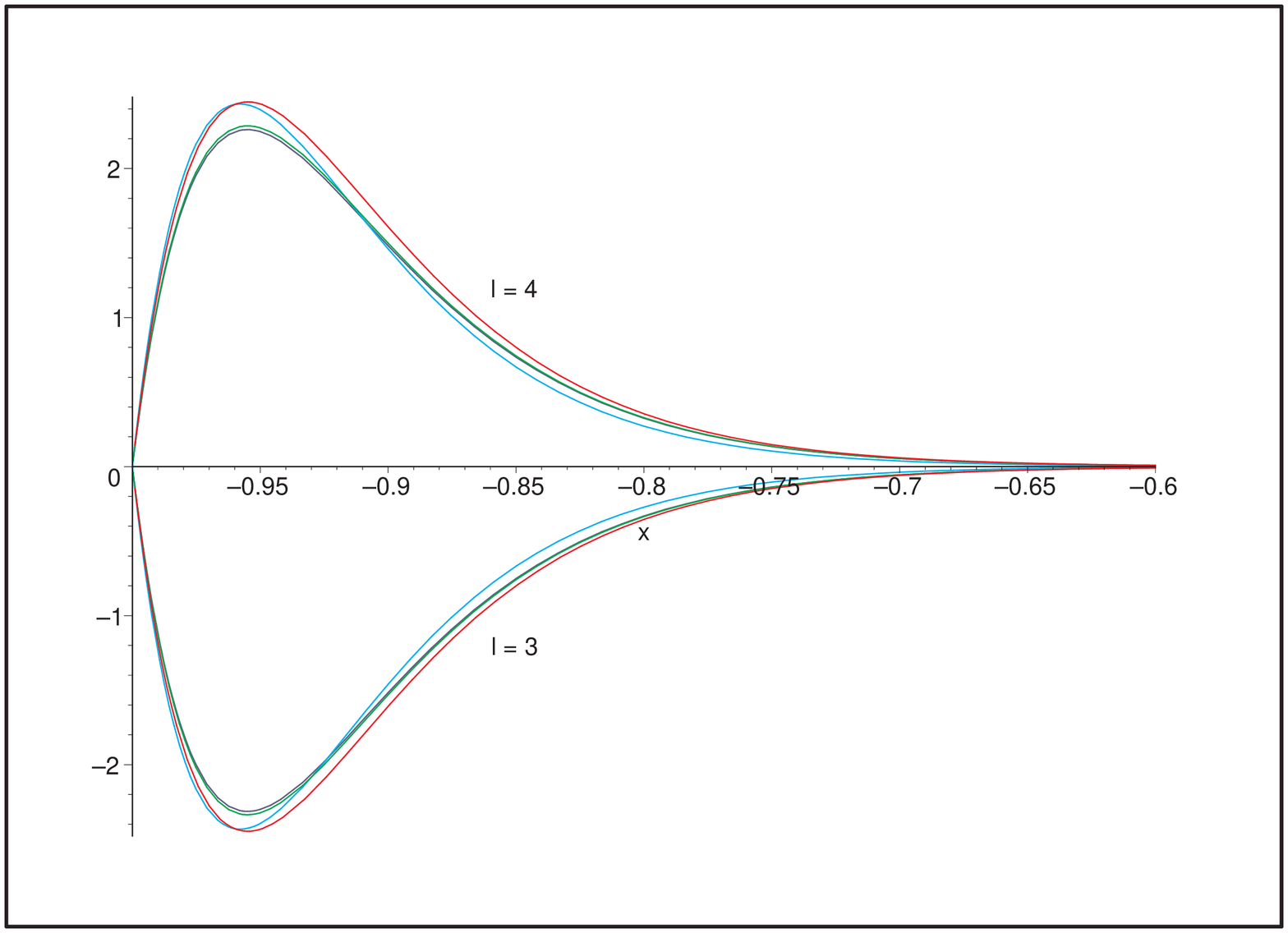}
\caption{${}_{-1}S_{l,1,25}$ for $l=3\ \&\ 4$.
The curves above the $x$-axis correspond to $l=4$ and below the axis to $l=3$. 
Correspondence between colours and solutions is the same as in Figure \ref{fig:sph_n3n4m1w5to25}.} 
\includegraphics*[width=90mm,angle=270]{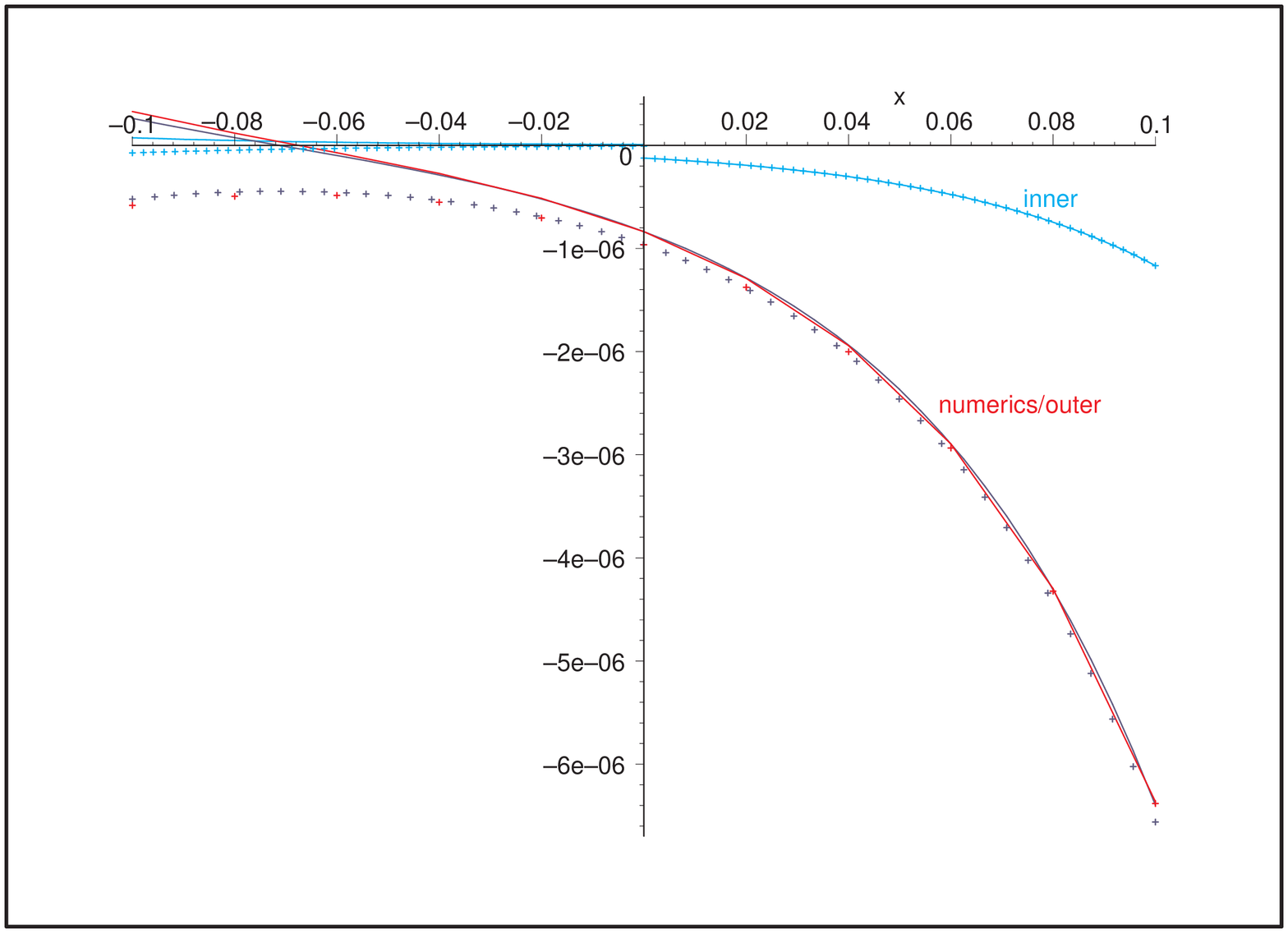}
\caption{${}_{-1}S_{l,1,25}$ for $l=3\ \&\ 4$.
The continuous lines correspond to $l=4$ and the dotted ones to $l=3$.
Correspondence between colours and solutions is the same as in Figure \ref{fig:sph_n3n4m1w5to25}.}  \label{fig:sph_n3n4m1w5to25 last}
\end{figure}

In Figures \ref{fig:sph_n3n4m1w5to25 first bis}--\ref{fig:sph_n3n4m1w5to25 last} the lines labelled as `inner' have been obtained with (\ref{eq: inner solution}), 
the ones labelled `outer' with (\ref{eq: outer solution}), the ones labelled `uniform' with (\ref{eq:unif S,p,p'}) and the ones
labelled `numerics' with the programs described in Section \ref{sec:num. method; high freq. sph.}.
These figures show that the outer (normalized to agree with the numerical data at $x=0$), inner (normalized to agree with the numerical data
at $x=\pm 0.96$) and uniform (also normalized to agree with the numerical data at $x=0$) solutions
approximate the numerical data for $\omega =25$ in the boundary layers and in the neighbourhood of $x=0$. The outer solution is valid until the boundary point $x=-1$ but not
until $x=+1$ since the function has two zeros close to it and the outer solution cannot cater for them, whereas the uniform solution is a valid approximation
for all $x$. The inner solutions, on the other hand, prove to be a good approximation in the boundary layers but not close to $x=0$.

\begin{figure}[p]
\rotatebox{90}
\centering
\includegraphics*[width=90mm,angle=270]{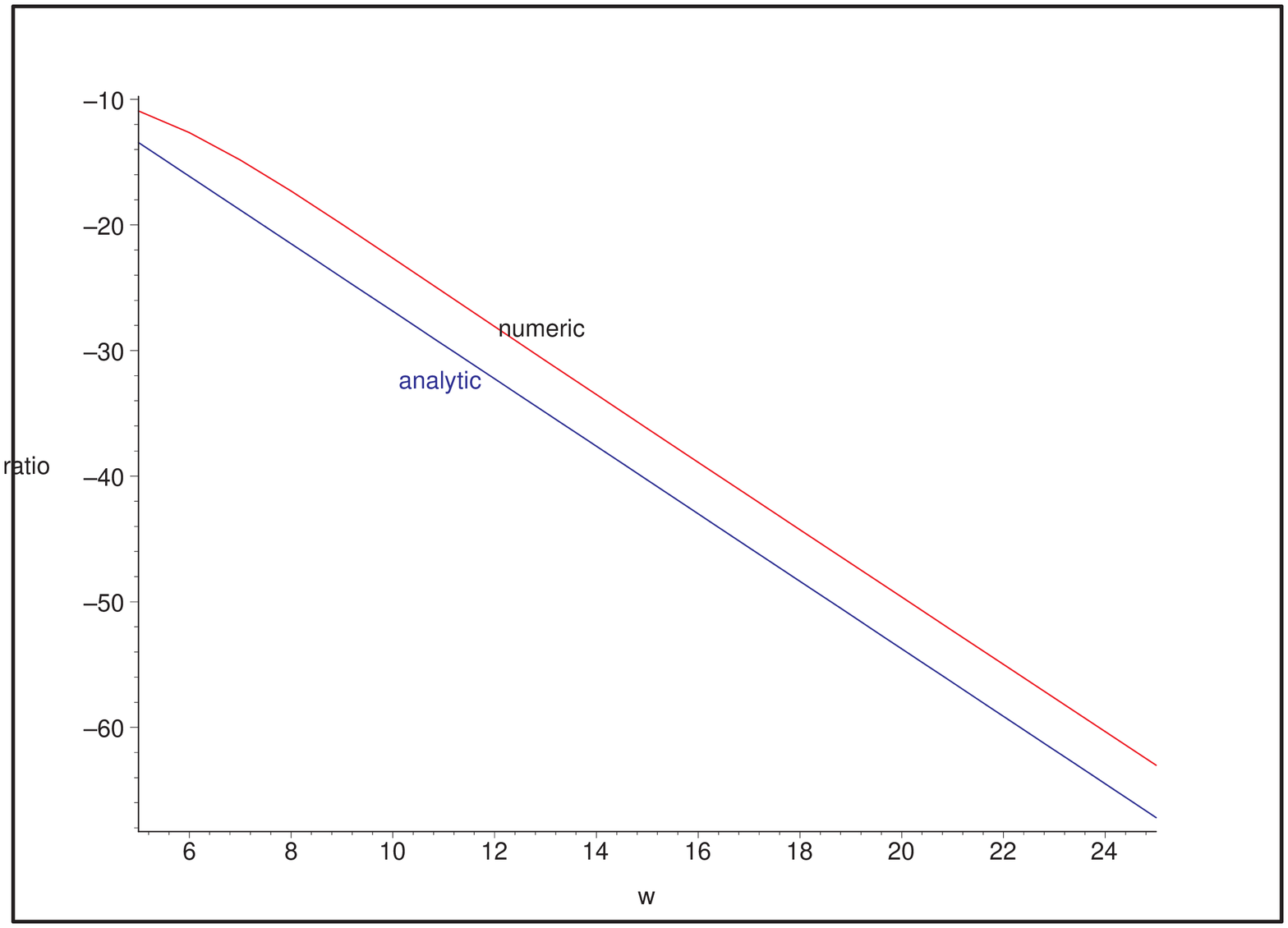}
\caption{$\frac{{}_{-1}D_{4,1,\omega}}{{}_{-1}C_{4,1,\omega}}$ for $\omega =5\rightarrow25$.
The slope of the analytic curve is given by the leading order behaviour (\ref{eq: ratio D/C spin1}).
The shift between the two curves is due to lower order, $O(1)$, terms.
} \label{fig:D_1divC_1_mge1_q4n4m1w5to25}
\includegraphics*[width=90mm,angle=270]{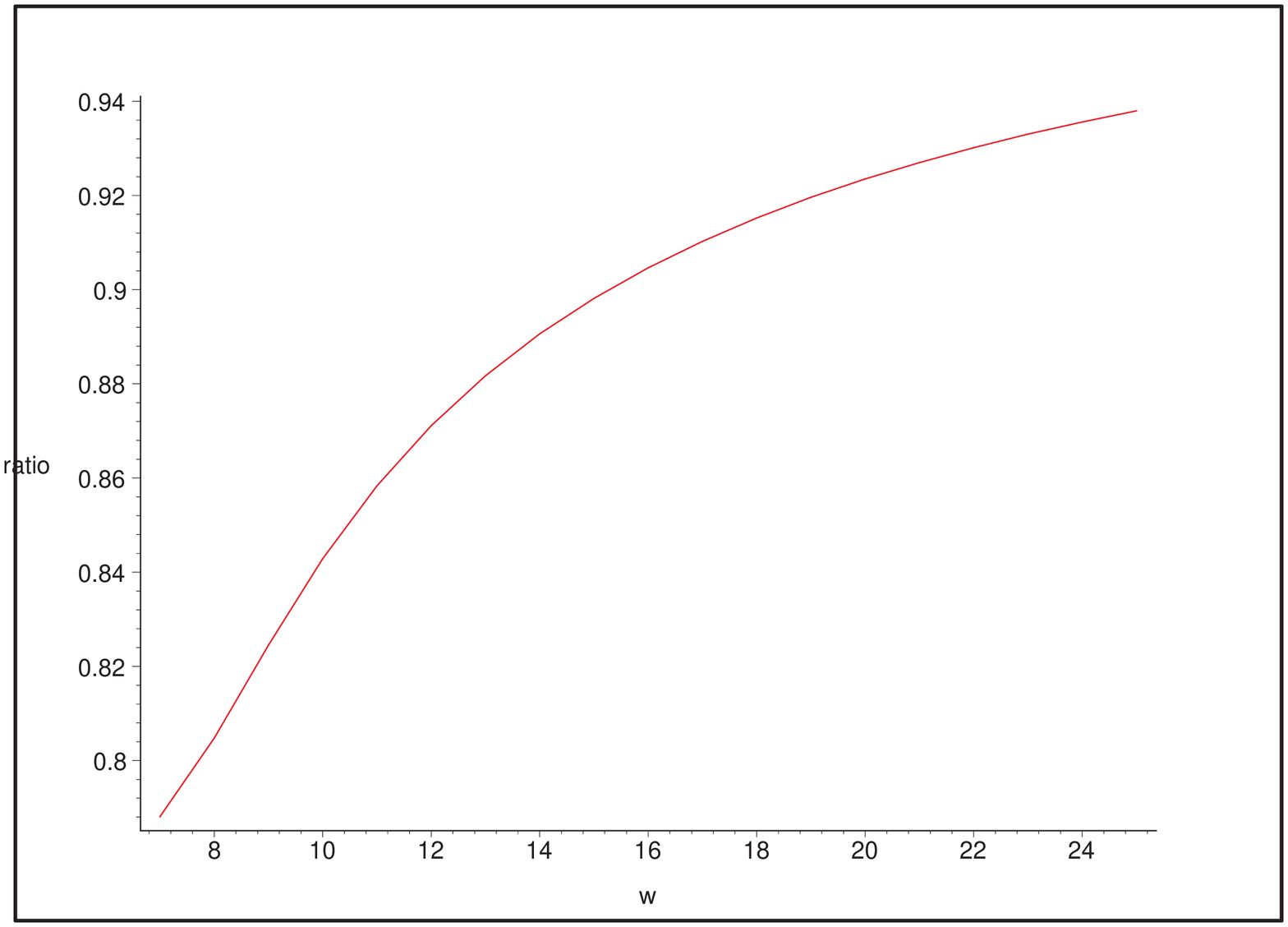}
\caption{Ratio between numeric and analytic values of $\frac{{}_{-1}D_{4,1,\omega}}{{}_{-1}C_{4,1,\omega}}$.
The analytic values have been obtained with (\ref{eq: ratio D/C spin1}).} \label{fig:ratio_D_1divC_1_mge1_q4n4m1w5to25}
\end{figure}

Figures \ref{fig:D_1divC_1_mge1_q4n4m1w5to25} and \ref{fig:ratio_D_1divC_1_mge1_q4n4m1w5to25} 
prove equation (\ref{eq: ratio D/C spin1}) to be correct for the case
$m\geq 1$: for the specific values $\indhel =-1$, $l=4$, $m=1$ and $q=4$ the inner solution (\ref{eq: inner solution})
has been normalized to match the numerical data at the points $x=\pm 0.998$ for different
values of the frequency from 5 to 25, in order to be able to calculate ${}_{-1}D_{4,1,\omega}$, ${}_{-1}C_{4,1,\omega}$ and ${}_{-1}D_{4,1,\omega}/{}_{-1}C_{4,1,\omega}$.
When plotting this numerical ratio together with the analytical result (\ref{eq: ratio D/C spin1}), the two lines are parallel and therefore agree to
highest order, and the ratio between the numerical and the analytical data tends to $1$.

\begin{figure}[p]
\rotatebox{90}
\centering
\includegraphics*[width=90mm,angle=270]{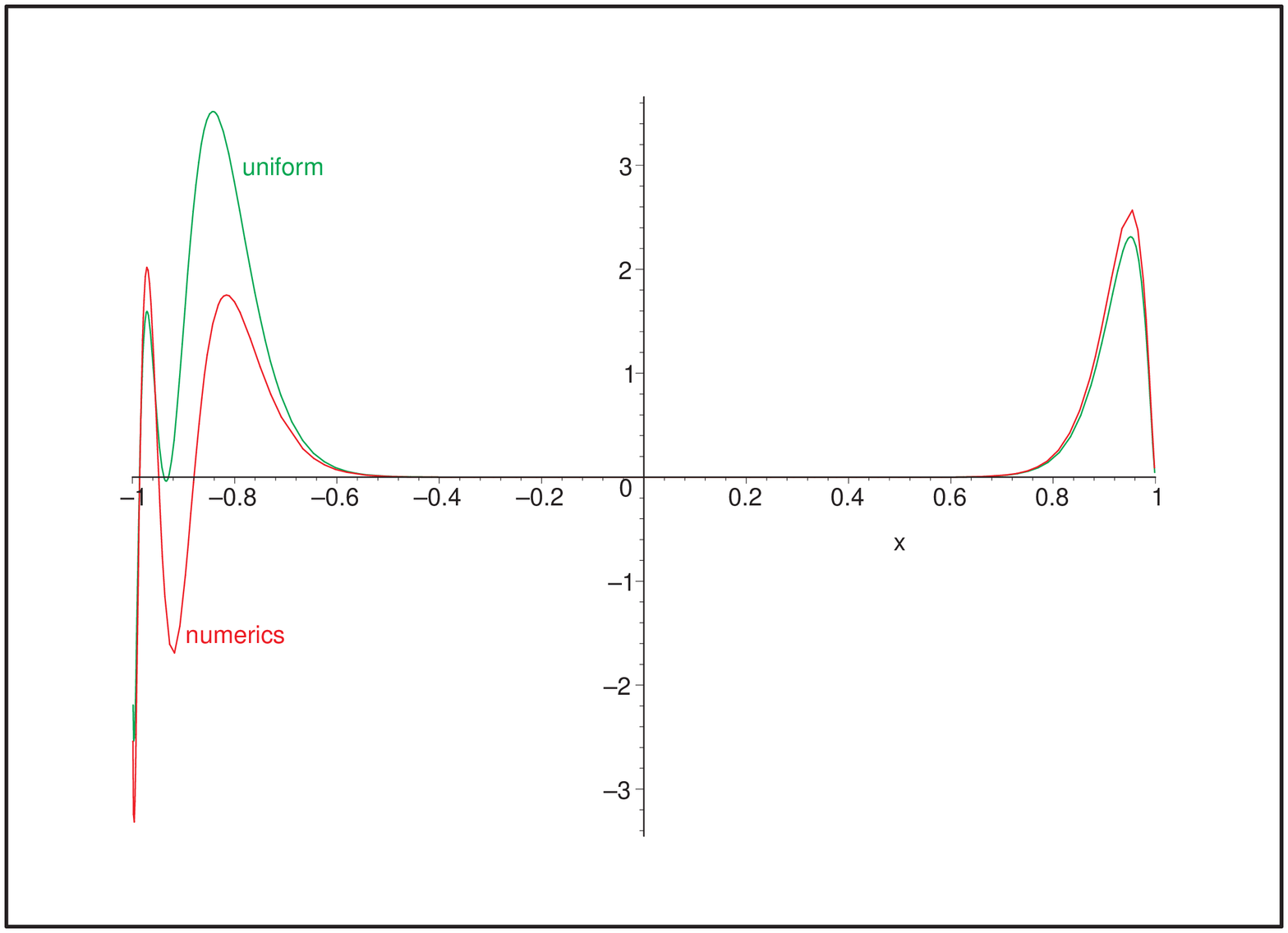}
\caption{${}_{+2}S_{5,1,35}$.
Green line corresponds to uniform solution (\ref{eq:unif S,p,p'}) and red line to numerics.} \label{fig:sph_s2n5m1w1to37_w35_x_1to1}
\includegraphics*[width=90mm,angle=270]{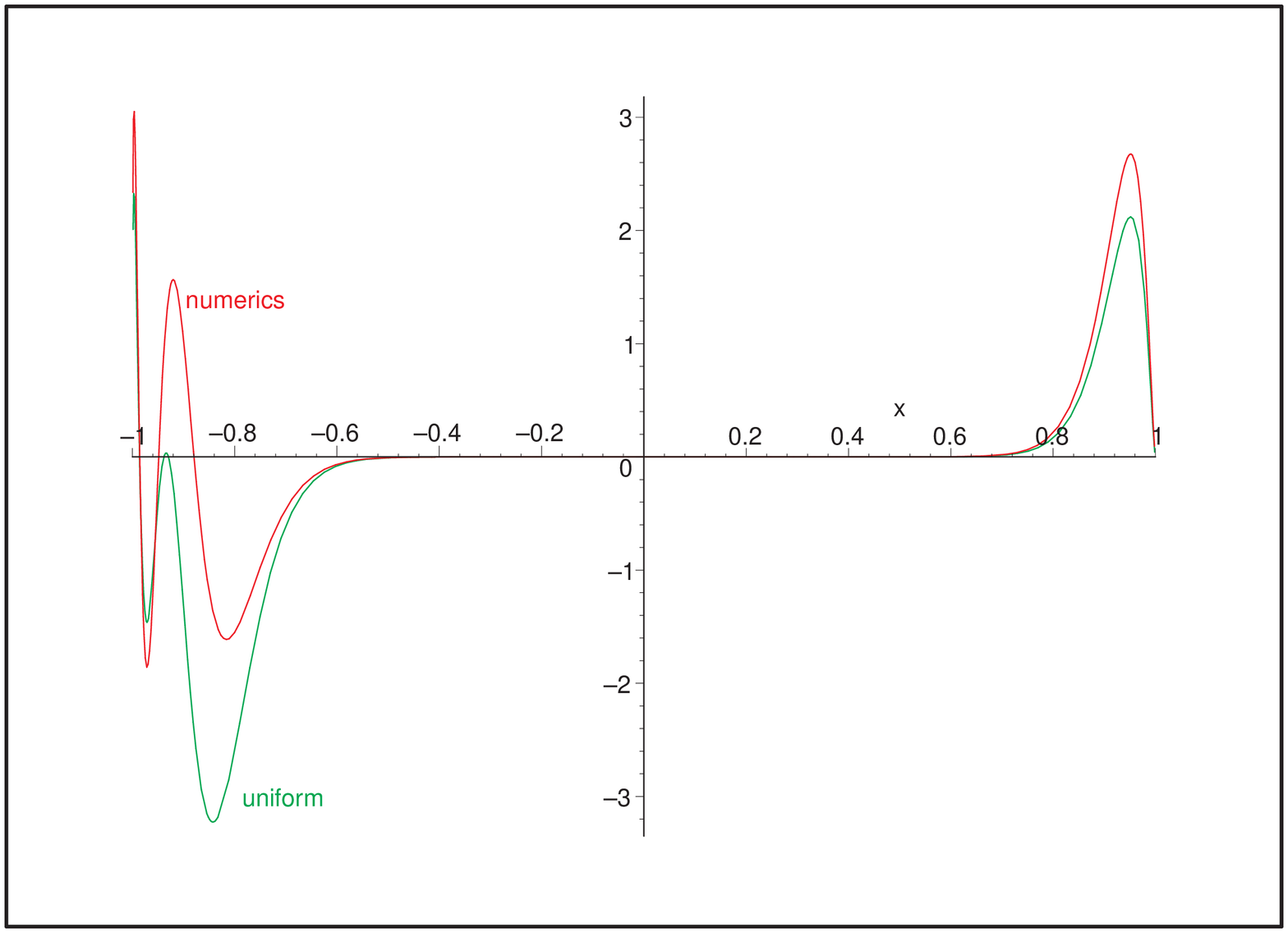}
\caption{${}_{+2}S_{6,1,35}$.
Green line corresponds to uniform solution (\ref{eq:unif S,p,p'}) and red line to numerics.}
\label{fig:sph_s2n5n6m1w1to35}
\end{figure}

Figures \ref{fig:sph_s2n5m1w1to37_w35_x_1to1}--\ref{fig:sph_s2n6m1w1to37_w35_x_0p6to_1}
correspond to modes with $h=+2$, $m=1$, $\omega=35$ and $l=5$ or $l=6$. The modes for both values of $l$ yield $q=6$. 
However, the mode with $l=5$ does not possess a zero at $x_0$ whereas the mode with $l=6$ does. 
The behaviour for positive $x$ is very similar for both values of $l$ but for negative $x$ the behaviours for the two modes differ by a sign. 

\begin{figure}[p]
\rotatebox{90}
\centering
\includegraphics*[width=90mm,angle=270]{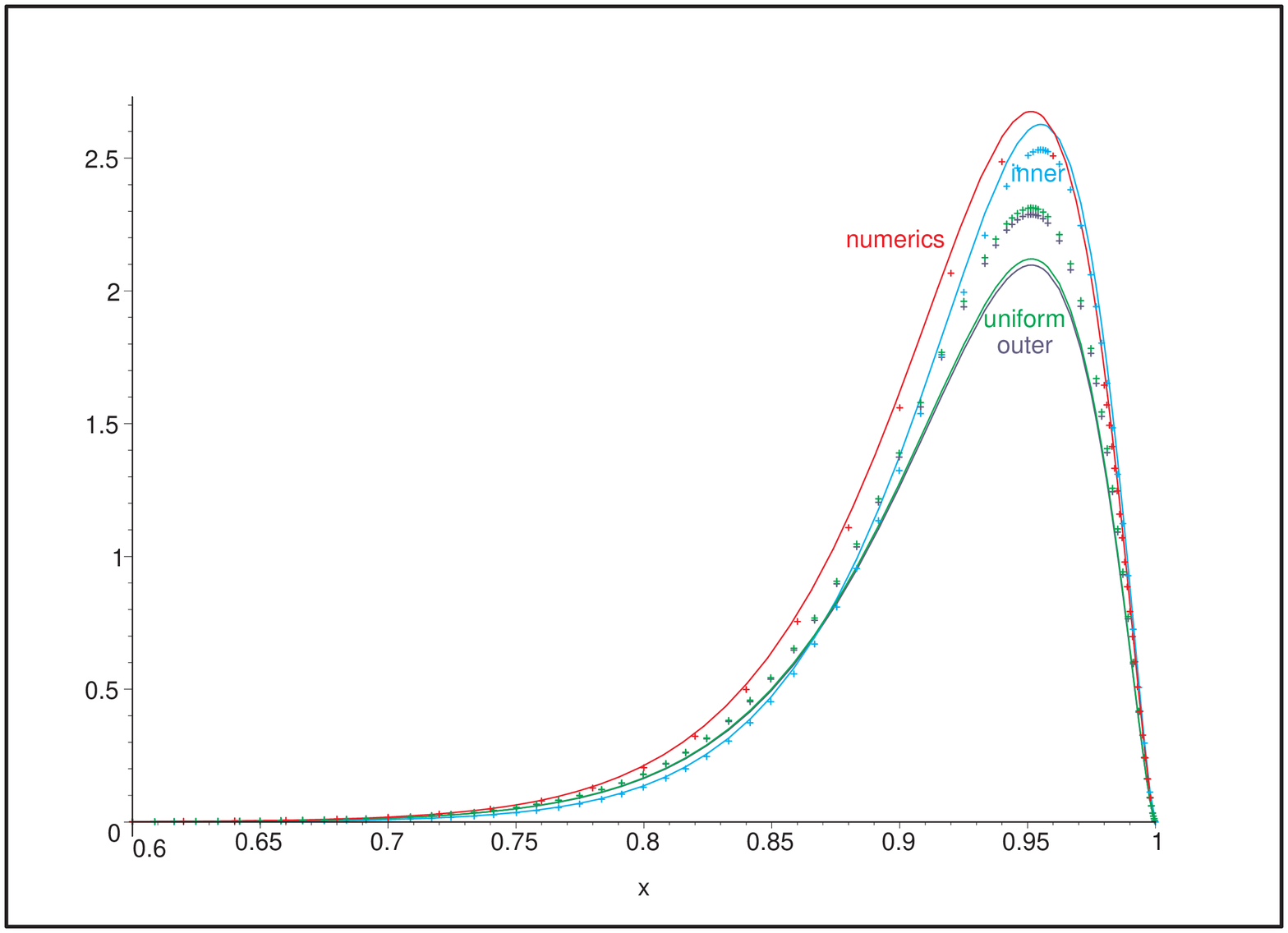}
\caption{${}_{+2}S_{l,1,35}$ for $l=5\ \&\ 6$.
The continuous lines correspond to $l=6$ and the dotted ones to $l=5$.
Correspondence between colours and solutions is the same as in Figure \ref{fig:sph_n3n4m1w5to25}.} 
\includegraphics*[width=90mm,angle=270]{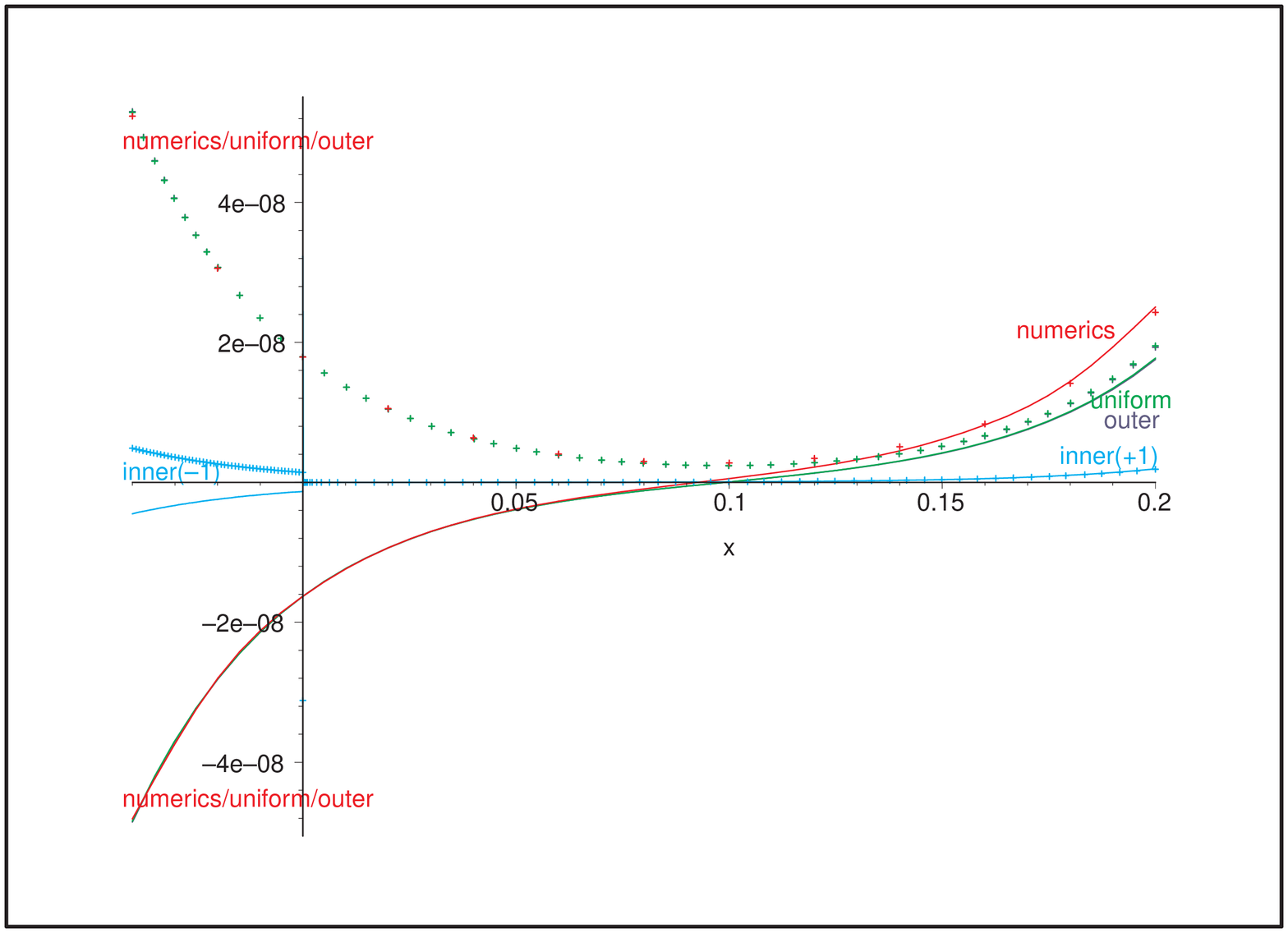}
\caption{${}_{+2}S_{l,1,35}$ for $l=5\ \&\ 6$.
The continuous lines correspond to $l=6$ and the dotted ones to $l=5$.
Correspondence between colours and solutions is the same as in Figure \ref{fig:sph_n3n4m1w5to25}.} 
\end{figure}

\begin{figure}[!p]
\rotatebox{90}
\centering
\includegraphics*[width=100mm,angle=270]{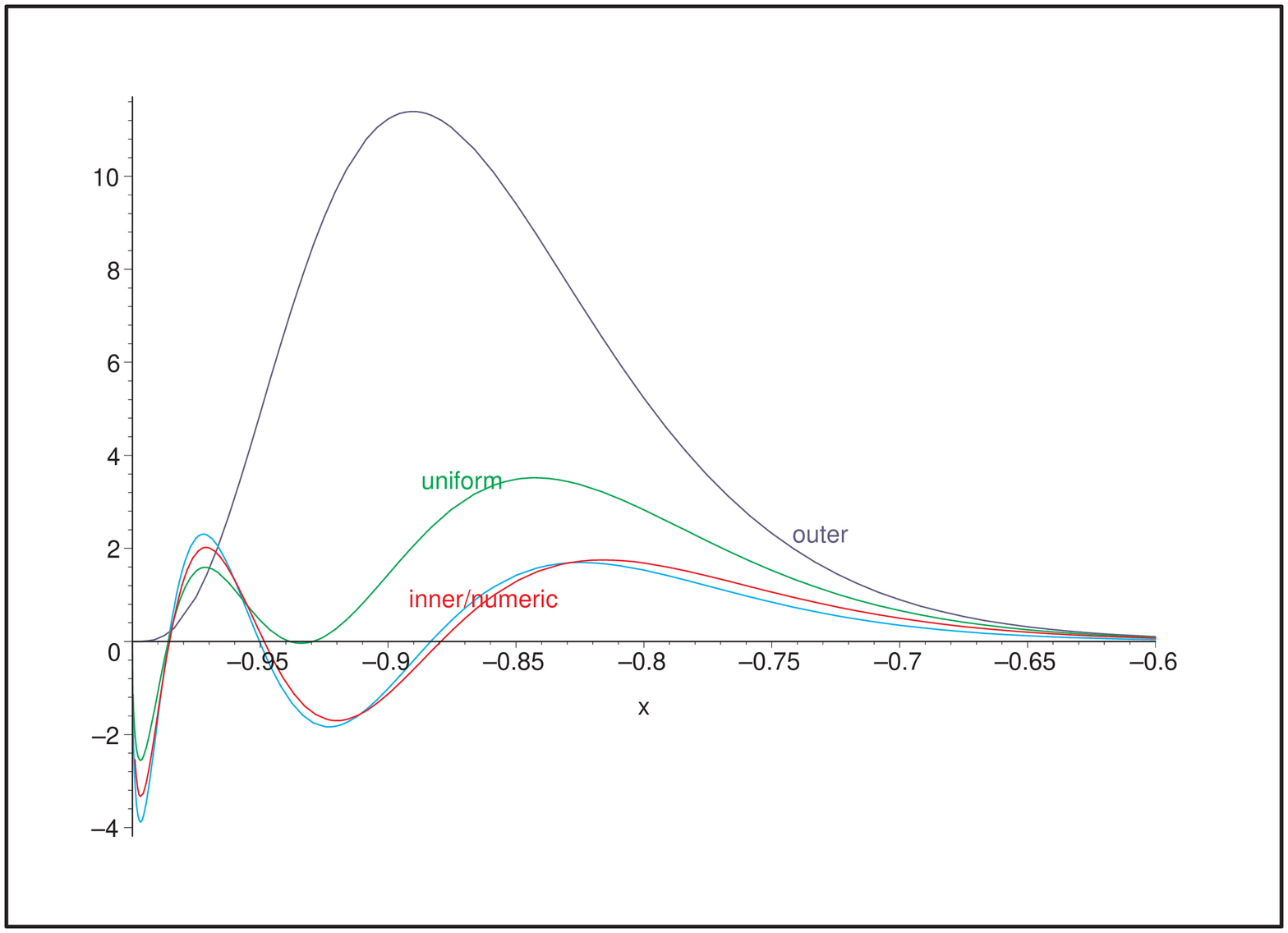}
\caption{${}_{+2}S_{5,1,35}$. Correspondence between colours and solutions is the same as in Figure \ref{fig:sph_n3n4m1w5to25}.}
\label{fig:sph_s2n5m1w1to37_w35_x_0p6to_1}
\includegraphics*[width=100mm,angle=270]{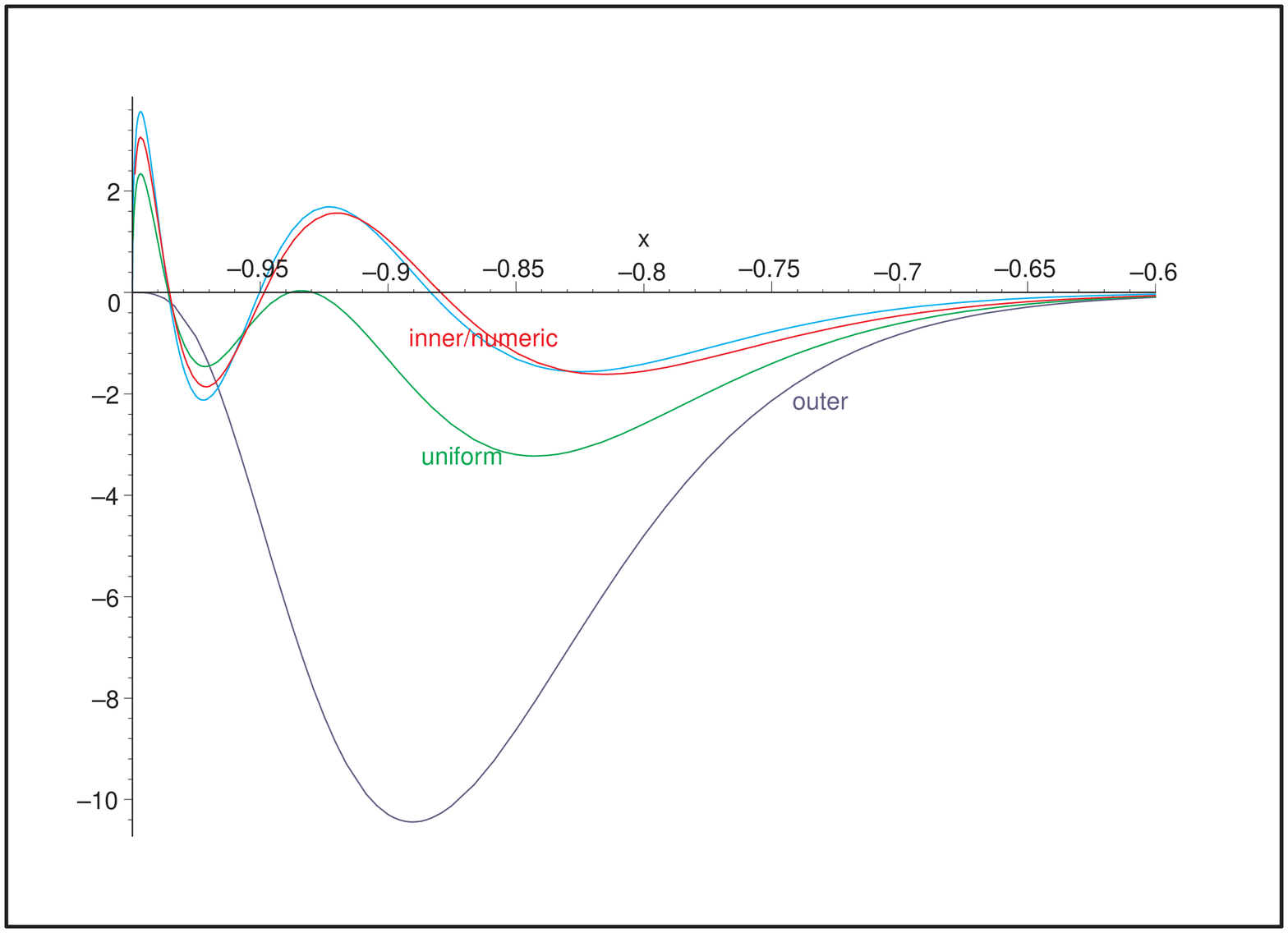}
\caption{${}_{+2}S_{6,1,35}$. Correspondence between colours and solutions is the same as in Figure \ref{fig:sph_n3n4m1w5to25}.}
\label{fig:sph_s2n6m1w1to37_w35_x_0p6to_1}
\end{figure}

\begin{figure}[!ht]
\rotatebox{90}
\centering
\includegraphics*[width=90mm,angle=270]{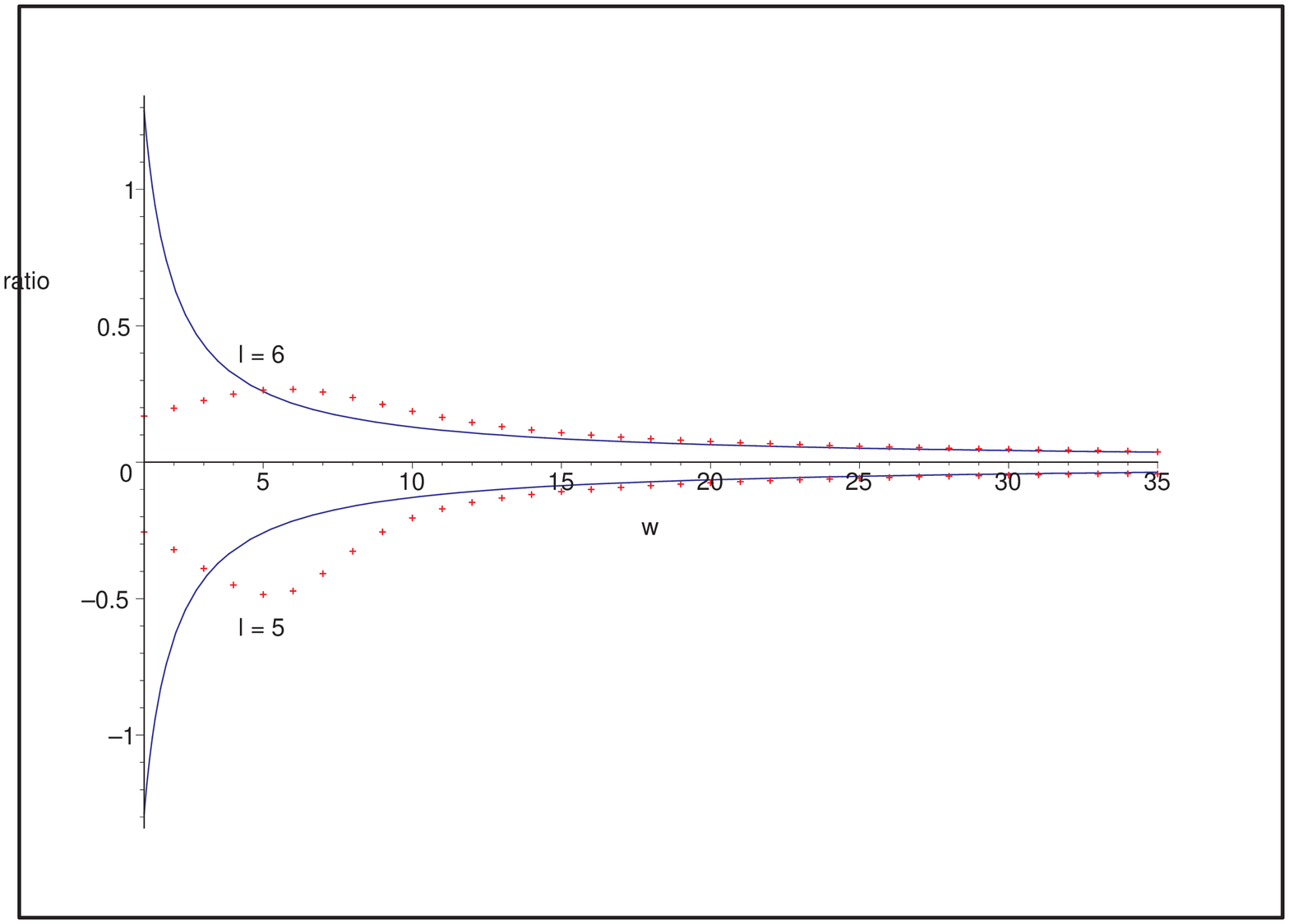}
\caption{$\frac{{}_{-1}D_{4,1,\omega}}{{}_{-1}C_{4,1,\omega}}$ for $\omega =1\rightarrow 35$.
The curves above the x-axis correspond to $l=6$ and below to $l=5$.
The continuous lines correspond to the analytic expression (\ref{eq: ratio D/C spin2}) and the dotted ones to the numerical data.}
\includegraphics*[width=90mm,angle=270]{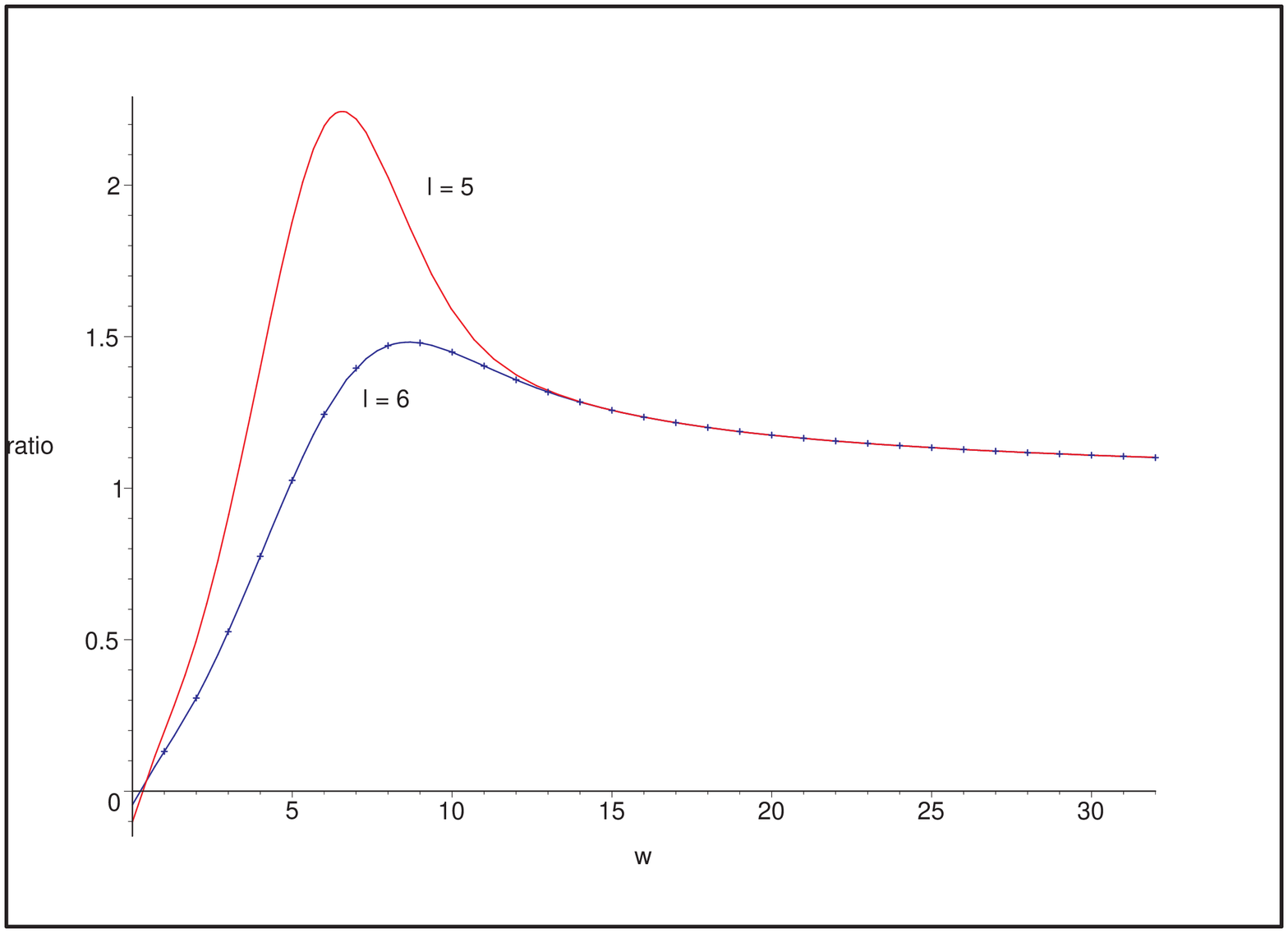}
\caption{Ratio between numeric and analytic values of $\frac{{}_{+2}D_{l,1,\omega}}{{}_{+2}C_{l,1,\omega}}$ for $\omega =1\rightarrow 34$.
Blue lines (plotted both continuous and dotted to show agreement with red line for large $\omega$) correspond to $l=6$ and
red line to $l=5$.
} \label{fig:ratio_D2divC2_m1_s2q6n5n6m1w1to34_x0p998}
\end{figure}

\begin{figure}[p]
\rotatebox{90}
\centering
\includegraphics*[width=90mm,angle=270]{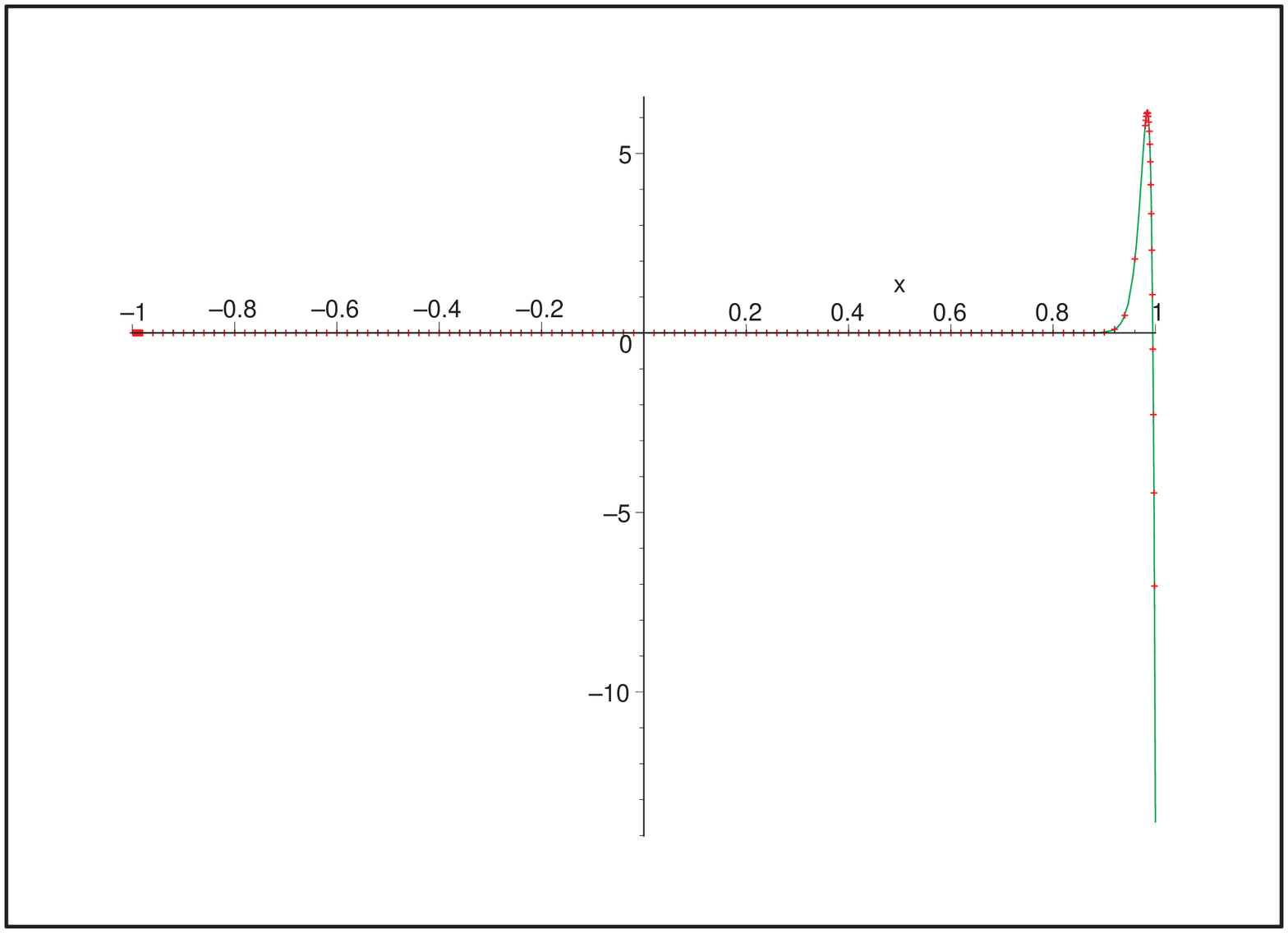} 
\caption{${}_{-1}S_{2,1,100}$.
The continuous, green line corresponds to the uniform solution (\ref{eq:unif S,p,p' not}) and the dotted, red one to the numerical data}
\includegraphics*[width=90mm,angle=270]{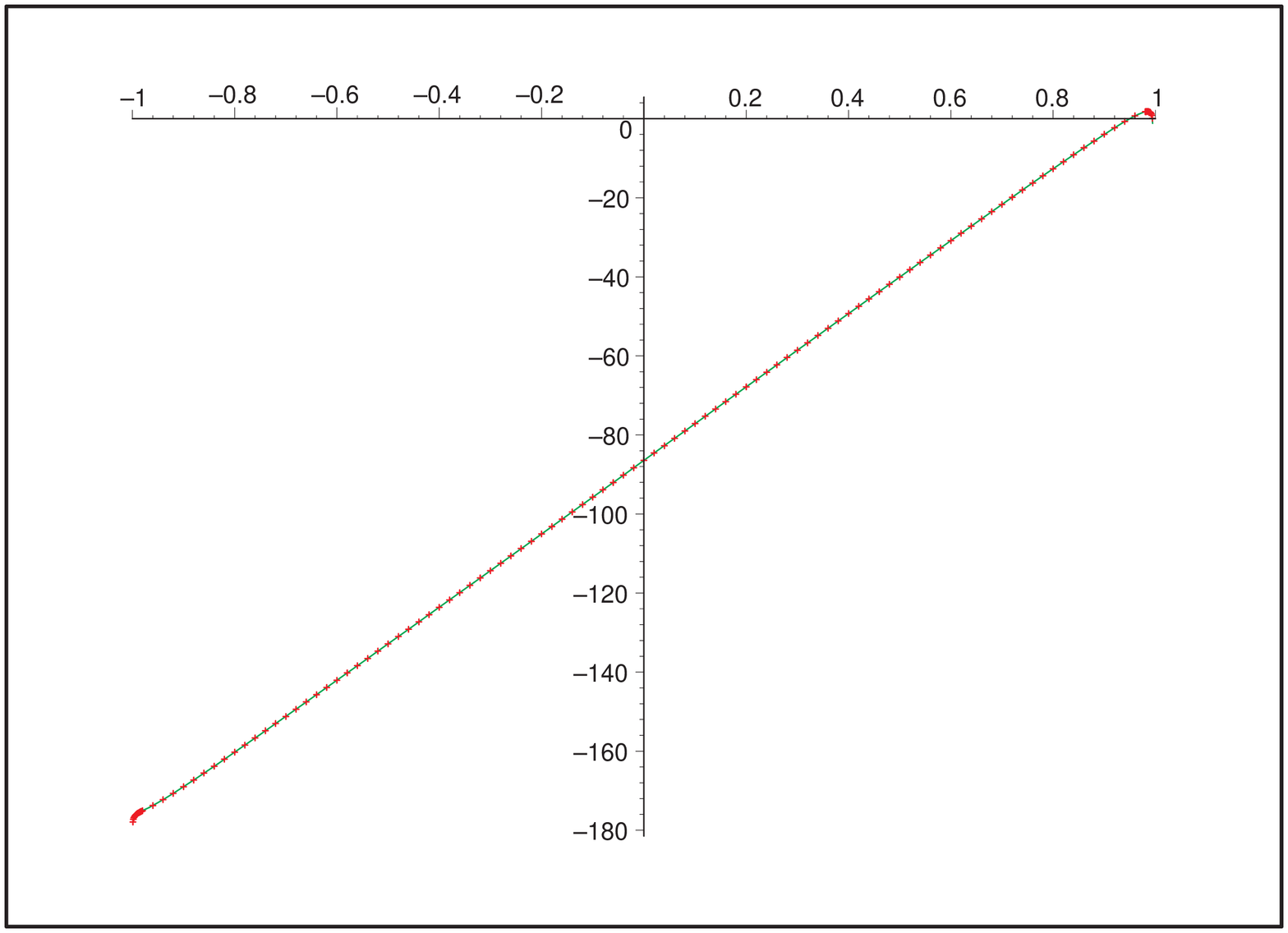} \label{fig:sph_n2m1w100_x_1to1}
\caption{$\log({}_{-1}S_{2,1,100})$.
The continuous, green line corresponds to the uniform solution (\ref{eq:unif S,p,p' not}) and the dotted, red one to the numerical data}
\label{fig:sph_n2m1w100}
\end{figure}

\begin{figure}[p]
\rotatebox{90}
\centering
\includegraphics*[width=90mm,angle=270]{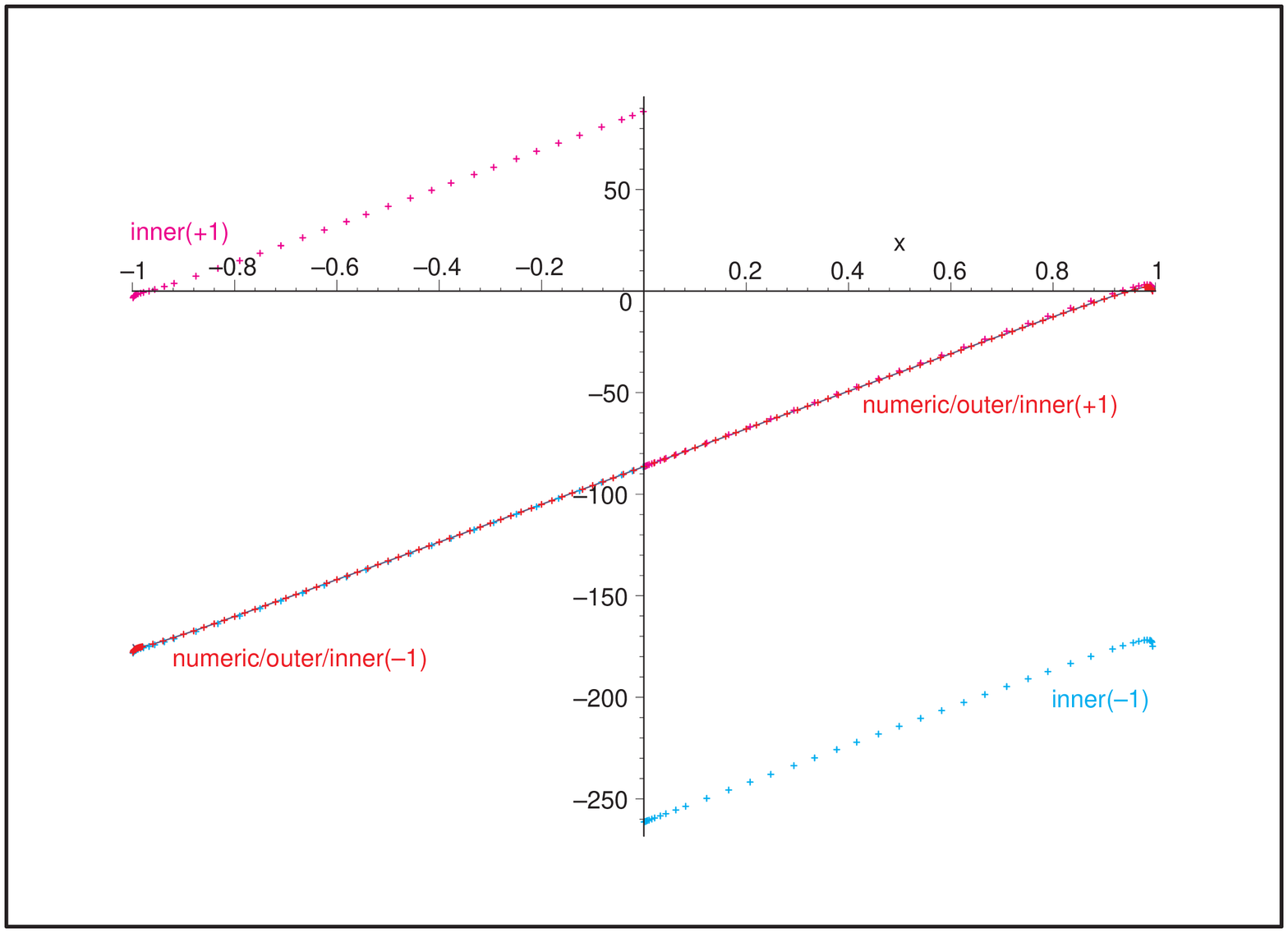}
\caption{$\log({}_{-1}S_{2,1,100})$.  
The red line (numerical data) overlaps with the navy line (outer solution). 
The light blue line (inner solution valid at  $x\sim -1$) 
and the magenta line (inner solution valid at  $x\sim +1$) overlap with the red/navy lines for negative and positive $x$ respectively.}
\label{fig:log_sph_n2m1w100_x_1to1}
\includegraphics*[width=90mm,angle=270]{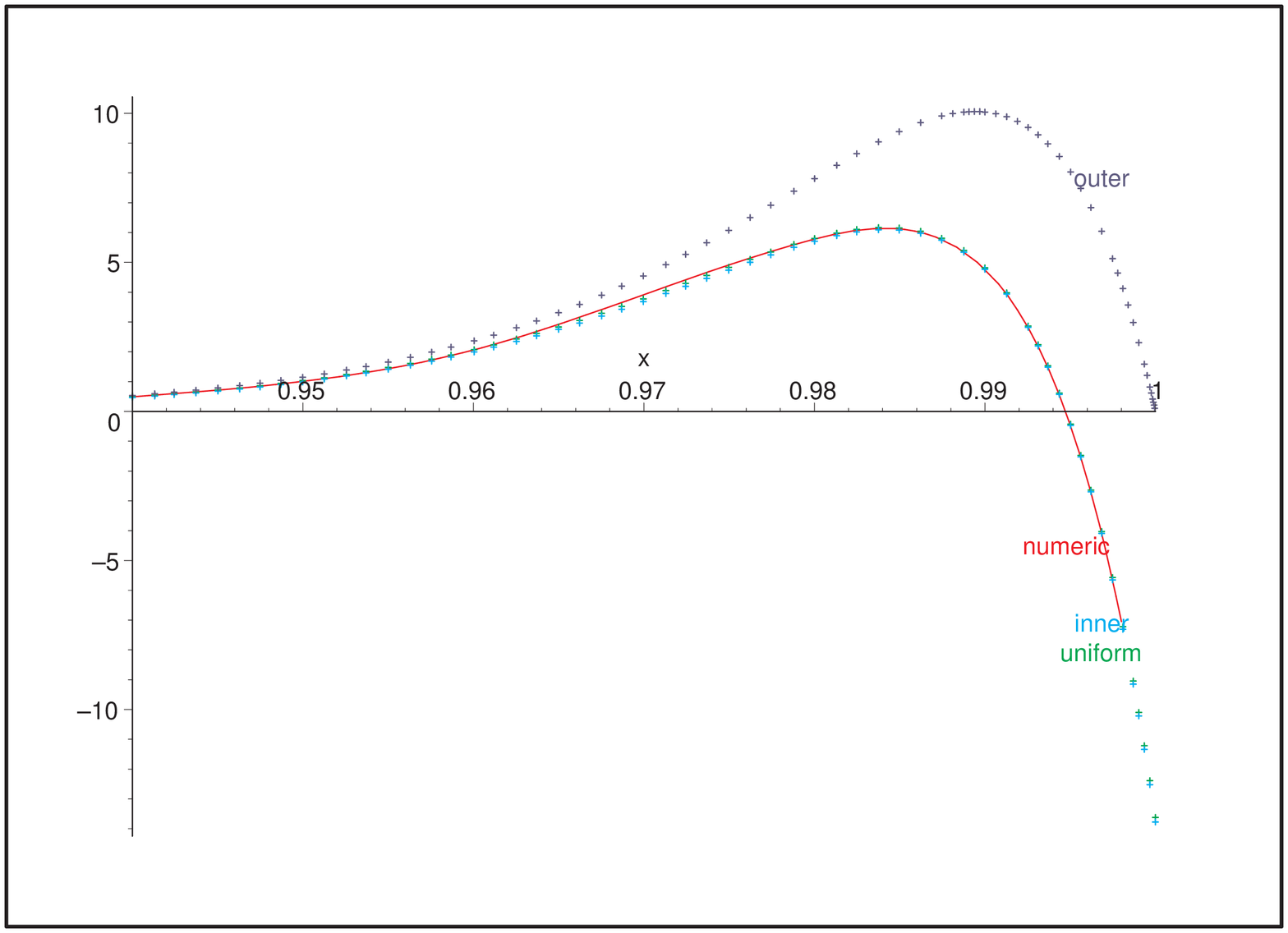}
\caption{${}_{-1}S_{2,1,100}$. The numeric (continuous, red), inner (light blue) and uniform (green) solutions overlap close to $x=+1$.} 
\label{fig:sph_unif_n2m1w100_x0p94to1}
\label{fig:sph_n2m1w100bis}
\end{figure}

For $\indhel =-1$, $l=2$, $m=1$ and $\omega =100$, the corresponding value of $q$ is $2$. 
This is a case where $p\in \mathbb{Z}^{+}\cup\{0\}$ and
$p'\notin \mathbb{Z}^{+}\cup\{0\}$. The numerical solution together with the uniform expansion (\ref{eq:unif S,p,p' not}) is plotted over the whole range
$x\in[-1,1]$ in Figures \ref{fig:sph_n2m1w100_x_1to1}--\ref{fig:log_sph_n2m1w100_x_1to1}.

As we have seen, in this case the function has an exponential
behaviour far from the boundary layers, so that a plot of the
$\log$ of the function allows us to see the behaviour over the
whole range of $x$. Both the uniform expansion and the outer
solution have been normalized so that they coincide with the
numerical value at $x=0$, and the inner solution has been
normalized once at $x=10^{-8}$ and once at $x=-10^{-8}$. The
uniform expansion agrees with the numerical solution for all
values of $x$. The outer solution agrees with the numerics
everywhere except very close to $x=\pm 1$, where it veers off.
The inner solutions are valid all the way from their respective
boundary layers until, and past, $x=0$, which is due to the
exponential nature of the function in the region between the
boundary layers. The inner solutions show a jump at $x=0$ due to
the different orders in $c$ of ${}_{\indhel}C_{lm\omega}$ and ${}_{\indhel}D_{lm\omega}$.

\begin{figure}[p]
\rotatebox{90}
\centering
\includegraphics*[width=90mm,angle=270]{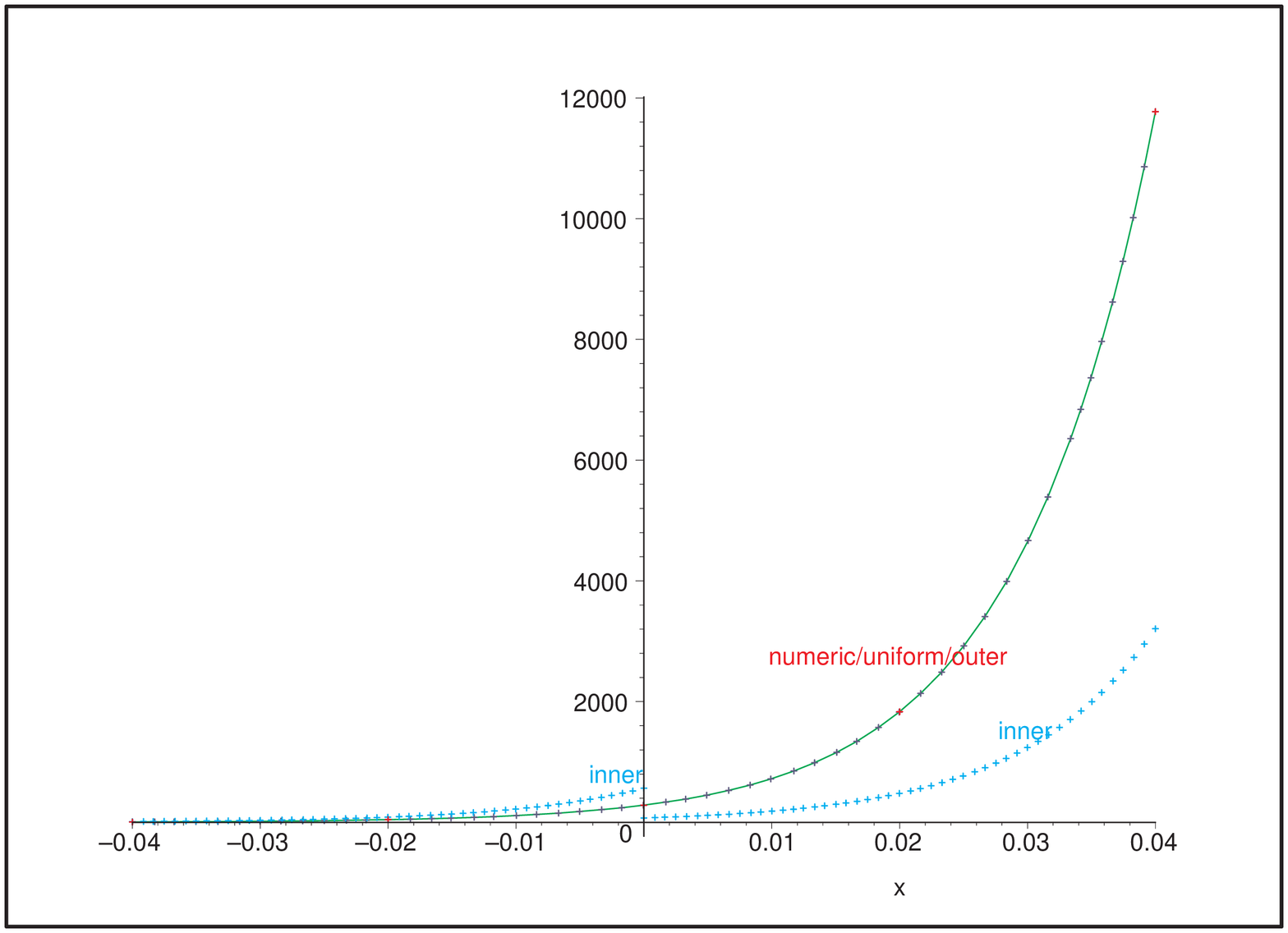}
\caption{$10^{40}{}_{-1}S_{2,1,100}$. The numeric (red), outer (navy) and uniform (continuous, green) solutions overlap close to $x=0$.
}
\includegraphics*[width=90mm,angle=270]{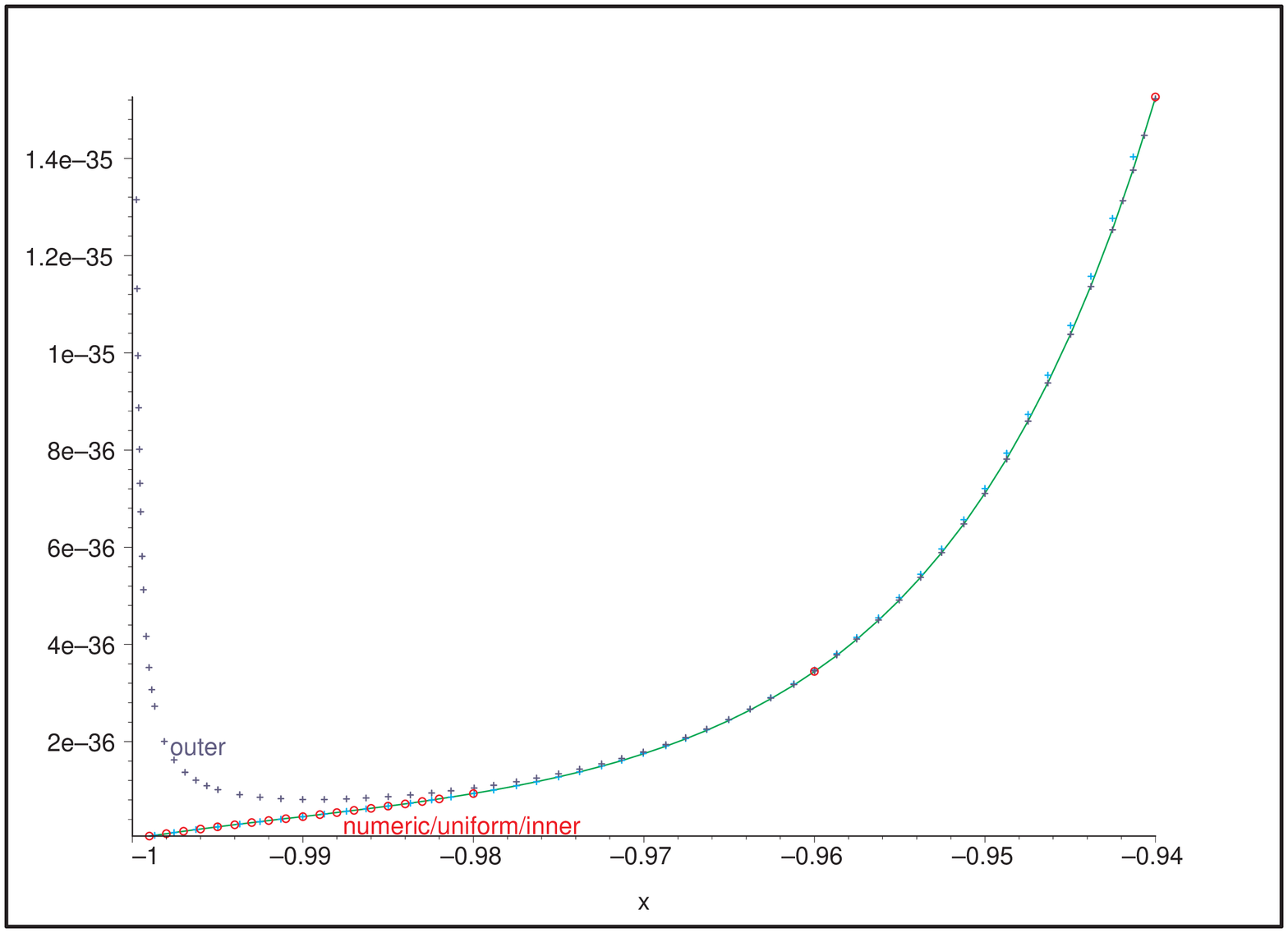}
\caption{$10^{40}{}_{-1}S_{2,1,100}$. The numeric (red circles), inner (light blue crosses) and uniform (continuous, green) solutions overlap close to $x=-1$.
} \label{fig:sph_unif_n2m1w100_x_0p04to_1}
\end{figure}

The above features can be seen in detail for $x$ close to $0$ and $\pm 1$ in
Figures \ref{fig:sph_unif_n2m1w100_x0p94to1}--\ref{fig:sph_unif_n2m1w100_x_0p04to_1} 
where they have been rescaled by $10^{40}$ for $x$ close to $0$ and $-1$.

\chapter{Stress-energy tensor}  \label{stress-energy tensor}


\draft{graphs for different polarizations in the last section must still be plotted}

\draft{1) where should the first graphs go?and where CCH's (\ref{eq:eq.3.7CCH;mine})? maybe both in 
section \ref{sec:luminosity} and then change its title?, 3) put section \ref{sec:RRO} as the last one so I can then
use/quote in it the results in section \ref{sec:theta->pi-theta Symmetry}?but section \ref{sec:theta->pi-theta Symmetry}
refers to asymptotics in section \ref{sec:RRO}?, 4) first graphs should be after \ref{sec:theta->pi-theta Symmetry}?
and maybe section \ref{sec:luminosity} as well?, 5) Frolov\&Bolashenko's paper is not being referenced...}

\draft{to be renamed to QFT? quantization of the field?....}


\section{Introduction} \label{sec:Intro. in stress-energy tensor}


This chapter is restricted to the uncharged case $Q=0$ for definiteness. 
Most of the results presented in this chapter do also apply to the Kerr-Newman space-time, which we will on occasion point out, and the programs
we have developed may perform the calculations in this space-time merely by setting $Q\neq 0$.
However, all results in the literature that we refer to in this chapter focus on the Kerr (or Schwarzschild) space-time;
most notably the construction of the various physical states as well as a paper that is central to this thesis, ~\cite{ar:CCH}.

This chapter is organized as follows.
In Section \ref{sec:CCH} we provide a canonical quantization of the electromagnetic potential and field. 

In the following three sections we give a description of the main physical states on the Schwarzschild and Kerr
space-times. Most of these descriptions relate to the scalar field.
We particularly focus on the various attempts in the literature to construct states
on the Kerr space-time with the same defining features as the Hartle-Hawking state on Schwarzschild.

In Section \ref{sec:CCH's expressions} we give expressions for the expectation value of the electromagnetic field
in the Kerr space-time, originally given in
Candelas, Chrzanowski and Howard ~\cite{ar:CCH}, hereafter referred to as CCH.
We also present the results from numerical calculations
of differences between two states of the renormalized expectation value of the stress-energy tensor (abbreviated as RSET).

In the subsequent section we endeavor to calculate the luminosity of the Kerr black hole in the past Boulware and past
Unruh states for the spin-1 case. We discuss the difficulties in the calculation and the differences with respect 
to the scalar case.

In Section \ref{sec:RRO} we study the form of the RSET close to the horizon when the field is in the past Boulware state.
CCH show a form of this RSET which is not exactly (minus) thermal.
We rederive CCH's analytic result, show why it is incorrect and compare it against our numerical results.
We also study the rate of rotation of this RSET, for which there is no unanimous consensus in the literature.

We initially used expressions given by CCH for the expectation value of the stress-energy tensor when the field
is in various states in order to calculate differences between two states of the RSET. 
The results, both analytically and numerically were not 
symmetric under the parity operation $\mathcal{P}:(\theta,\phi)\to(\pi-\theta,\phi+\pi)$. This is the topic of the last section. 
It is split into three subsections. In the first one we show that this lack of symmetry is indeed present. In the second
subsection we see that the reason for its presence are incorrect expressions given by CCH. 
We derive the correct expressions for the expectation value of the stress-energy tensor when the electromagnetic field
is in the main physical states in the Kerr space-time. 
In the last subsection we give a physical interpretation of the various sets of terms appearing in 
the expectation value of the stress-energy tensor when the field is in different states.

All figures in this chapter have been obtained for the values: $Q=0$, $a=0.95M$ and $M=1$.
Note that the Boyer-Lindquist radius of the event horizon for such values of the black hole
parameters is $r_+\simeq 1.3122$.
 
\catdraft{es stress-energy tensor o energy-momentum (com diu p.117GavPhD)?}


Before finishing this introductory section, we will derive the classical stress-energy tensor of the theory.
Einstein's field equations, which describe the generation of space-time curvature by energy, are
\begin{equation} \label{eq:Einstein eqs.}
R_{\mu\nu}-\frac{1}{2}Rg_{\mu\nu}+\Lambda g_{\mu\nu}=+8\pi T_{\mu\nu}
\end{equation}
where $\Lambda$ is the cosmological constant.
We may construct the total action
\begin{equation}
S=S_g+S_m
\end{equation}
where $S_g$ refers to the gravitational action and $S_m$ includes the contribution from the matter fields.
If the gravitational action is given by
\begin{equation}
S_g=\frac{1}{16\pi}\int_{\mathcal{M}}(R-2\Lambda)\sqrt{-g} \d^4{x}
\end{equation}
where $\mathcal{M}$ is a fixed space-time, and the action $S_m$ is related to the stress-energy tensor by
\begin{equation} \label{eq:stress tensor from action}
\frac{2}{\sqrt{-g}}\frac{\delta S_m}{\delta g^{\mu\nu}}=T_{\mu\nu}
\end{equation}
then Einstein's field equations may be derived by imposing that the total action $S$ satisfies the condition
\begin{equation}
\frac{2}{\sqrt{-g}}\frac{\delta S}{\delta g^{\mu\nu}}=0
\end{equation}
It is clear that, except for a factor, the imposition of this condition to only the gravitational action $S_g$ instead of $S$
yields the left hand side of (\ref{eq:Einstein eqs.}), whereas imposing it to only the action $S_m$ yields the right hand side of the equation.

In this thesis we are interested in the case where the only matter field present is the electromagnetic field.
Therefore, in this thesis, the action $S_m$ is equal to the electromagnetic action $S_{em}$.
The electromagnetic action must yield the Maxwell field equations (\ref{eq:Maxwell eqs. with potential}) 
via the Euler-Lagrange equations, which result from requiring the electromagnetic 
action to be stationary under an infinitesimal variation of the fields:
\begin{equation}
\frac{\delta S_{em}}{\delta A_{\mu}(x)}=0
\end{equation}
The simplest electromagnetic action which leads to the
Maxwell field equations via the Euler-Lagrange equations is
\begin{equation} \label{eq:emag. action}
S_{em}=\int_{\mathcal{M}} \mathcal{L}_{em}(x)\d^4{x}
\end{equation}
with the electromagnetic Lagrangian
\begin{equation} \label{eq:emag. lagrangian}
\mathcal{L}_{em}(x)=\left(-\frac{1}{4}F_{\alpha\beta}F^{\alpha\beta}-J_{\alpha}A^{\alpha}\right)\sqrt{-g}
\end{equation}
\catdraft{diff. with ref.21Unruh'74 in c.c. (and sign)? Adrian:probably conventions}
Although in the absence of a charge current the Maxwell lagrangian $\mathcal{L}_{em}$ is gauge-invariant,
in the presence of a charge current it is not: under a gauge transformation (\ref{eq:gauge transf.}) it acquires a new, pure-divergence term, 
which does not alter the field equations by virtue of the law of current conservation. 

We will restrict ourselves to the case of absence of a charge current.
The electromagnetic stress-energy tensor $T^{\mu\nu}$ is calculated from the action (\ref{eq:emag. action}) 
using (\ref{eq:stress tensor from action}), and the result is
\begin{equation} \label{eq:stress tensor, spin 1, in terms of F}
T^{\mu\nu}=\frac{1}{4}g^{\mu\nu}F_{\alpha\beta}F^{\alpha\beta}-F^{\mu\alpha}F^{}_{\alpha}{}^{\nu}
\end{equation}
in the absence of sources.
This form for the stress-energy tensor is
gauge-invariant, conserved, symmetric and traceless.

By virtue of the Maxwell field equations, the stress-energy tensor 
(\ref{eq:stress tensor, spin 1, in terms of F})
satisfies the following \define{conservation equation}   
\begin{equation} \label{eq:conservation law for emag. stress tensor}
T^{\mu\nu}{}_{;\mu}=0
\end{equation}

The classical, electromagnetic stress-energy tensor (\ref{eq:stress tensor, spin 1, in terms of F}) can be expressed
in terms of the NP Maxwell scalars as
\begin{equation} \label{eq:stress tensor, spin 1}
\begin{aligned}
T_{\mu\nu}&=\left\{\phi_{-1}\phi_{-1}^*n_{\mu}n_{\nu}+2\phi_0\phi_0^*\left[l_{(\mu}n_{\nu)}+
m_{(\mu}m_{\nu)}^*\right]+\phi_{+1}\phi_{+1}^*l_{\mu}l_{\nu}-
\right.
\\ 
&
\left.
-4\phi_0\phi_{-1}^*n_{(\mu}m_{\nu)}-4\phi_{+1}\phi_0^*l_{(\mu}m_{\nu)}+2\phi_{+1}\phi_{-1}^*m_{\mu}m_{\nu}\right\}+c.c.
\end{aligned}
\end{equation}
Note that it follows from (\ref{eq:parity op. on NP objs.}) that all the pairs of null tetrad vectors  
appearing in the different terms in (\ref{eq:stress tensor, spin 1}) remain invariant under
the parity operation, except for the ones that have a factor containing $\phi_0$ together with either $\phi_{-1}$ or $\phi_{+1}$, which
change sign. That is, a pair of null 
vectors $\vec{e}_{(a)}\vec{e}_{(b)}$ 
appearing in (\ref{eq:stress tensor, spin 1}) with a factor
$\phi_{\indhel}\phi_{\indhel'}^*$ changes under the parity operation as
\begin{equation}  \label{eq:pairs of vects. in stress tensor under parity}
\mathcal{P}\left(\vec{e}_{(a)}\vec{e}_{(b)}\right)=(-1)^{\indhel+\indhel'}\vec{e}_{(a)}\vec{e}_{(b)}
\end{equation}

\section{Quantization of the electromagnetic potential/field}  \label{sec:CCH}

The abundance in the literature of the quantization of the scalar field in a curved background is in sharp contrast
with the scarce treatment of the quantization of the electromagnetic -or gravitational- field in such a background.
In particular, the definitions of the various states that we shall give in the following three sections have all been done in relation to the scalar case.
CCH did quantize both the electromagnetic and the
gravitational fields in the Kerr background. They used a canonical quantization method, which is the one
we have chosen to use in this thesis.

The terminology we will use is the following.
As until now, a bullet (and a primed bullet) superscript indicates either `in' or `up' modes.
Correspondingly, the symbol $\omega^{\bullet}$ is defined as being equal to $\omega$ when it is part of an expression
containing `in' modes and it is equal to $\tilde{\omega}$ when the expression contains `up' modes.
We will refer to the `ingoing' and `upgoing' gauge potentials with the corresponding superscript, rather than with
the notation in Chapter \ref{ch:field eqs.} of a $\pm 1$ subindex.
The variable $\Gamma$ refers to either the potential components $A_{\mu}$ or the NP scalars $\phi_{\indhel}$.

Notice first that the symmetry property
\begin{equation}
\mathcal{P}{}_{lm\omega}A_{\mu}^{\bullet}=(-1)^{l+m}{}_{l-m-\omega}A_{\mu}^{\bullet *}
\end{equation}
which is easily obtained from (\ref{eq:R symm.->cc,-m,-w}), (\ref{eq:S symm.->-s,-m,-w}) and (\ref{eq:potential as a func. of RS}),
leads to the following expression for the potential modes (\ref{eq:def. of lmwPA}):
\begin{equation} \label{eq: def. of lmwPA_mu}
{}_{lm\omega P}A_{\mu}^{\bullet}={}_{lm\omega}A_{\mu}^{\bullet}+P\mathcal{P}{}_{lm\omega}A_{\mu}^{\bullet}
\end{equation}
This was actually Chrzanowski's ~\cite{ar:Chrzan'75} starting point for the derivation of the potential modes, 
as referred to in Section \ref{sec:Wald in Kerr}.
The decomposition of the potential into eigenstates of the parity operator $\mathcal{P}$ is the natural choice because of the 
invariance of the Kerr metric under this operation. Note that this also applies to the Kerr-Newman solution, as both
its metric and the Maxwell's equations are invariant under $\mathcal{P}$.

With the above definitions the Fourier series expansion for either the potential components or the NP Maxwell scalars may be expressed as
\begin{equation} \label{eq:Fourier series of Gamma field}
\Gamma^{\bullet}= \sum_{lmP} \int_{-\infty}^{+\infty}\d{\omega^{\bullet}}{}_{lm\omega P}a^{\bullet}{}_{lm\omega P}\Gamma^{\bullet}
\end{equation}
where ${}_{lm\omega P}a^{\bullet}$ are the coefficients of the Fourier series.
This Fourier series may be re-arranged as
\begin{equation} \label{eq:Fourier series of Gamma field for pos.freq.}
\Gamma^{\bullet}=\sum_{lmP} \int_{0}^{+\infty}\d{\omega^{\bullet}}\left({}_{lm\omega P}a^{\bullet}{}_{lm\omega P}\Gamma^{\bullet}+
(-1)^{l+m}P{}_{lm\omega P}a^{\bullet *}{}_{l-m-\omega P}\Gamma^{\bullet}\right)
\end{equation}
We can now use the symmetry relations
\begin{subequations} \label{eq: symms. of lmwA_mu and lmwphi_i}
\begin{align}
\mathcal{P}{}_{lm\omega P}A_{\mu}^{\bullet}&=(-1)^{l+m}{}_{l-m-\omega P}A_{\mu}^{\bullet *}=
P{}_{lm\omega P}A_{\mu}^{\bullet} \label{eq: symm. of lmwA_mu}
\\
\mathcal{P}{}_{lm\omega}\phi_{\indhel}^{\bullet}&=(-1)^{l+m+1+\indhel}{}_{l-m-\omega}\phi_{\indhel}^{\bullet *} \label{eq: symm. of lmwphi_i}
\end{align}
\end{subequations}

where the property that $\mathcal{P}^2$ is the identity operator was used in (\ref{eq: symm. of lmwA_mu}).
The equations above for $\Gamma$ are equally valid for $A_{\mu}$ and $\phi_{\indhel}$. In particular, we only need to 
apply the operator $K^{(a)}_{\indhel}e^{\mu}_{(a)}$ (which is explicitly given in (\ref{eq:op. K})) to an equation for $A_{\mu}$ in order to obtain the 
corresponding equation for $\phi_{\indhel}$. However the last step in (\ref{eq: symm. of lmwA_mu}) has no 
equivalent for  $\phi_{\indhel}$ in (\ref{eq: symm. of lmwphi_i}).   
The reason is that
\begin{equation}
\left(K^{(a)}_{\indhel}e^{\mu}_{(a)}\right)\mathcal{P}{}_{lm\omega}A_{\mu}^{\bullet} \propto 
\left(K^{(a)}_{\indhel}e^{\mu}_{(a)}\right) {}_{l-m-\omega}A_{\mu}^{\bullet *} \equiv 0
\end{equation}
as can be checked; that is, this term is pure gauge. 
Hence the fact that
${}_{lm\omega}\phi_{\indhel}^{\bullet}=\left(K^{(a)}_{\indhel}e^{\mu}_{(a)}\right){}_{lm\omega P}A_{\mu}^{\bullet}$ 
does not actually depend on $P$ and that is why $P$ is not a subindex of the NP scalar modes. 
The potential is real whereas the field 
components are not, as seen in equations (\ref{eq:classical mode expansion for A(in)_mu}) 
and (\ref{eq:classical mode expansion for phi(in)_i}) below. 
From (\ref{eq:Fourier series of Gamma field for pos.freq.}) and using (\ref{eq: symm. of lmwA_mu}) and 
(\ref{eq: symm. of lmwphi_i}) for the potential and the field respectively, we have
\begin{equation} \label{eq:classical mode expansion for A(in)_mu}
\begin{aligned}
A_{\mu}^{\bullet}
&=\sum_{lmP} \int_{0}^{+\infty}\d{\omega^{\bullet}}\left({}_{lm\omega P}a^{\bullet}{}_{lm\omega P}A_{\mu}^{\bullet}+
P{}_{lm\omega P}a^{\bullet *}\mathcal{P}{}_{lm\omega P}A_{\mu}^{\bullet *}\right)=            \\    
&=\sum_{lmP} \int_{0}^{+\infty}\d{\omega^{\bullet}}\left({}_{lm\omega P}a^{\bullet}{}_{lm\omega P}A_{\mu}^{\bullet}+
{}_{lm\omega P}a^{\bullet *}{}_{lm\omega P}A_{\mu}^{\bullet *}\right)
\end{aligned}
\end{equation}
\draft{link (\ref{eq:classical mode expansion for A(in)_mu}) with (\ref{eq:make potential real})!}

and
\begin{equation} \label{eq:classical mode expansion for phi(in)_i}
\phi_{\indhel}^{\bullet}=\sum_{lmP} \int_{0}^{+\infty}\d{\omega^{\bullet}}\left({}_{lm\omega P}a^{\bullet}{}_{lm\omega}\phi_{\indhel}^{\bullet}+
(-1)^{\indhel+1}P{}_{lm\omega P}a^{\bullet *}\mathcal{P}{}_{lm\omega}\phi_{\indhel}^{\bullet *}\right)
\end{equation}

We now quantize the field by promoting ${}_{lm\omega P}a^{\bullet}$ and ${}_{lm\omega P}a^{\bullet *}$ to 
operators ${}_{lm\omega P}\hat{a}^{\bullet}$ and ${}_{lm\omega P}\hat{a}^{\bullet \dagger}$ respectively.
It became apparent in Chapter \ref{ch:field eqs.} that the whole theory may be expressed in terms of one
single NP complex scalar, which represents the two radiative degrees of freedom of the electromagnetic perturbations. 
If we introduce expansion (\ref{eq:classical mode expansion for phi(in)_i}) 
\ddraft{should be expanded in terms of complete set formed by `in' plus `up' modes?}
for the NP scalars into $T^{00}$ given by (\ref{eq:stress tensor, spin 1}), we then obtain a hamiltonian which is a superposition of 
independent harmonic oscillator hamiltonians, one for each mode of the electromagnetic field. 
From the standard quantization of the harmonic oscillator, we know that the operators 
${}_{lm\omega P}\hat{a}^{\bullet}$ and ${}_{lm\omega P}\hat{a}^{\bullet \dagger}$ must
satisfy the commutation relations:
\begin{equation} \label{eq:commut. rlns. for a,a_dagger}
\begin{aligned}
\left[\hat{a}_{lm\omega P}^{\bullet},\hat{a}_{l'm'w'P'}^{\bullet \dagger}\right]=\delta(\omega-\omega')\delta_{ll'}\delta_{mm'}\delta_{PP'} \\
\Big[\hat{a}_{lm\omega P}^{\bullet},\hat{a}_{l'm'w'P'}^{\bullet}\Big]=\left[\hat{a}_{lm\omega P}^{\bullet \dagger},\hat{a}_{l'm'w'P'}^{\bullet \dagger}\right]=0
\end{aligned}
\end{equation}

These commutation relations are satisfied provided that the orthonormality conditions
\begin{equation} \label{eq:orthonormality conds. for potential}
\begin{aligned}
\left\langle {}_{lm\omega P}A^{\bullet}_{\alpha},{}_{l'm'w'P'}A^{\bullet'}_{\alpha}\right\rangle_{\mathcal{S}}&=
\delta_{\bullet \bullet'}\delta_{ll'}\delta_{mm'}\delta(\omega-\omega')\delta_{PP'} \\
\left\langle {}_{lm\omega P}A^{\bullet *}_{\alpha},{}_{l'm'w'P'}A^{\bullet'}_{\alpha}\right\rangle_{\mathcal{S}}&=0
\end{aligned}
\end{equation}
are satisfied, where $\mathcal{S}$ is any complete Cauchy hypersurface for the outer region of the space-time 
and where the \define{Klein-Gordon inner product} is taken as  
\begin{equation} \label{eq:def. inner prod.}
\left\langle \psi_{\alpha},\varphi_{\alpha}\right\rangle_{\mathcal{S}}=
i\int_{\mathcal{S}}\d^3{\Sigma}^{\mu}\left(\psi^{\alpha *}\nabla_{\mu}\varphi_{\alpha}-\varphi^{\alpha}\nabla_{\mu}\psi^*_{\alpha}+
\varphi_{\mu}\nabla_{\alpha}\psi^{\alpha *}-\psi^*_{\mu}\nabla_{\alpha}\varphi^{\alpha}\right)
\end{equation}
\catdraft{definir $\varphi_{\mu}$, $\psi_{\alpha}$}
The inner product (\ref{eq:def. inner prod.}) has the same form as the one taken by CCH. 
However, CCH give an expression for the stress-energy tensor which includes a factor $4\pi$ in (\ref{eq:stress tensor, spin 1}),
corresponding to unrationalized units. If unrationalized units are used, then a factor $4\pi$ should also be included in the
inner product (\ref{eq:def. inner prod.}). 
We believe that despite the fact that CCH give an expression of the stress-energy tensor in unrationalized units, they 
calculate it in rationalized units, as corresponds to (\ref{eq:def. inner prod.}). 

Note that the electromagnetic inner product is gauge-independent     
\begin{equation} \label{eq:gauge-indep inner prod.}
\left\langle {}_{lm\omega P}A^{\bullet}_{\mu}+\alpha_{, \mu},{}_{l'm'w'P'}A^{\bullet'}_{\nu}\right\rangle_{\mathcal{S}}=
\left\langle {}_{lm\omega P}A^{\bullet}_{\mu},{}_{l'm'w'P'}A^{\bullet'}_{\nu}\right\rangle_{\mathcal{S}}
\end{equation}
if the electromagnetic field is source-free.
\catdraft{1) ref.21Unruh'74:potser tambe cal que background s-t sigui source-free?, 2)definir $\alpha$}

Constants of normalization are to be included in front of the radial functions so that the potential modes (\ref{eq:tableIChrzan.}) satisfy
the orthonormality conditions (\ref{eq:orthonormality conds. for potential}) given the asymptotic behaviour of the radial functions in (\ref{eq:R_in/up}).
In order to find the constants of normalization we use the potentials in (\ref{eq:tableIChrzan.}): $A_{-1}{}_{\mu}$ is
used when `in' boundary conditions are taken and $A_{+1}{}_{\mu}$ when `up' boundary conditions are taken. 
Using the `in'/`up' radial functions as determined by (\ref{eq:R_in/up}) we find that:
\begin{subequations}  \label{eq:normalization consts.}
\begin{align}
|N_{-1}^{\text{in}}|^2&=\frac{1}{2^5\omega^3\pi} \\
|N_{+1}^{\text{up}}|^2&=\frac{1}{2^3\pi|\EuFrak{N}|^2\tilde{\omega}(r_+^2+a^2)} \\
|N_{-1}^{\text{up}}|^2&=|N_{+1}^{\text{up}}|^2\left|\frac{{}_{-1}R^{\text{up,inc}}_{lm\omega}}{{}_{+1}R^{\text{up,inc}}_{lm\omega}}\right|^2=
\frac{\tilde{\omega}(r_+^2+a^2)}{2\pi {}_1B_{lm\omega}^4}
\end{align}
\end{subequations}
where $\EuFrak{N}$ is given in (\ref{eq: def. EuFrak{N}_s}).
We have chosen ${}_{-1}R^{\text{in,inc}}_{lm\omega}=1$ and ${}_{+1}R^{\text{up,inc}}_{lm\omega}=1$ respectively for the first two equations.
The constant of normalization $|N_{-1}^{\text{up}}|$ is calculated as indicated with the use of (\ref{eq:R1 coeffs from R_1's}). It is 
therefore the constant of normalization that corresponds to using the radial function (\ref{eq:R_up}) when setting ${}_{-1}R^{\text{up,inc}}_{lm\omega}=1$,
which is the actual normalization we have used in the numerical calculation of the `up' solutions.
The NP scalars are therefore assumed to include the constants of normalization (\ref{eq:normalization consts.}). 
That is, the NP scalar modes ${}_{lm\omega}\phi_{\indhel}^{\bullet}$ are to be calculated from expressions 
(\ref{eq:phi_0/2(in/up)}) and (\ref{eq:phi0(ch)}) with the inclusion of the appropriate constant of normalization (\ref{eq:normalization consts.}),
while the radial functions remain unaltered.

\draft{1) check signs/factors in the normalization consts.!, 2) check/show surface elements (see p.15'Nin/up(darrr))}

It can be checked that the set of modes $\{{}_{lm\omega P}A_{\mu}^{\text{in}},{}_{lm\omega P}A_{\mu}^{\text{up}}\}$
forms a complete set of orthonormal solutions to the Maxwell equations in the outer region of the Kerr space-time. 
Similarly, it can be checked that $\{{}_{lm\omega P}A_{\mu}^{\text{out}},{}_{lm\omega P}A_{\mu}^{\text{down}}\}$ also
form a complete set.
We may expand the electromagnetic potential by using the
complete set of solutions $\{{}_{lm\omega P}A_{\mu}^{\text{in}},{}_{lm\omega P}A_{\mu}^{\text{up}}\}$ and then quantize it as:
\begin{equation} \label{eq:quantum mode expansion for A in in/up modes}
\begin{aligned}
\hat{A}_{\mu}&
=\sum_{lmP} \int_{0}^{+\infty}\d{\omega}
\left({}_{lm\omega P}\hat{a}^{\text{in}}{}_{lm\omega P}A_{\mu}^{\text{in}}+
{}_{lm\omega P}\hat{a}^{\text{in} \dagger}{}_{lm\omega P}A_{\mu}^{\text{in} *}\right)+\\ 
&+\sum_{lmP} \int_{0}^{+\infty}\d{\tilde{\omega}}
\left({}_{lm\omega P}\hat{a}^{\text{up}}{}_{lm\omega P}A_{\mu}^{\text{up}}+
{}_{lm\omega P}\hat{a}^{\text{up} \dagger}{}_{lm\omega P}A_{\mu}^{\text{up} *}\right)
\end{aligned}
\end{equation}

Alternatively, we could proceed exactly in the same manner but using the complete set of solutions
$\{{}_{lm\omega P}A_{\mu}^{\text{out}},{}_{lm\omega P}A_{\mu}^{\text{down}}\}$ instead.
The result is then:
\begin{equation} \label{eq:quantum mode expansion for A in out/dn modes}
\begin{aligned}
\hat{A}_{\mu}&=
\sum_{lmP} \int_{0}^{+\infty}\d{\omega}\left({}_{lm\omega P}\hat{a}^{\text{out}}{}_{lm\omega P}A_{\mu}^{\text{out}}+
{}_{lm\omega P}\hat{a}^{\text{out} \dagger}{}_{lm\omega P}A_{\mu}^{\text{out} *}\right)+\\ 
&+\sum_{lmP} \int_{0}^{+\infty}\d{\tilde{\omega}}\left({}_{lm\omega P}\hat{a}^{\text{down}}{}_{lm\omega P}A_{\mu}^{\text{down}}+
{}_{lm\omega P}\hat{a}^{\text{down} \dagger}{}_{lm\omega P}A_{\mu}^{\text{down} *}\right)
\end{aligned}
\end{equation}

The asymptotic behaviour in terms of the advanced and retarded time co-ordinates of the electromagnetic
potential and NP scalars for the `in' and `up' modes is the same as the one exhibited by the modes (\ref{eq:X_in/up as func. of u,v}).
The same applies to the asymptotic behaviour of the 
`out' and `down' modes exhibited in (\ref{eq:X_out/down as func. of u,v}).
Accordingly, the operators ${}_{lm\omega P}\hat{a}^{\text{in} \dagger}$, ${}_{lm\omega P}\hat{a}^{\text{up} \dagger}$, 
${}_{lm\omega P}\hat{a}^{\text{out} \dagger}$ and ${}_{lm\omega P}\hat{a}^{\text{down} \dagger}$ are creation operators of particles incident
from $\mathcal{I}^-$, $\mathcal{H}^-$, $\mathcal{I}^+$ and $\mathcal{H}^+$ respectively.  

\draft{is this true, despite of the fact of the existence of factors depending on $r$ and $\theta$ which
are different at each asymptotic region (in particular, does the interpretation of unit flux remain the same)??}

Since the `in' and `out' modes are only defined for $\omega$ non-negative, they have non-negative energy as measured by
an observer following the integral curve of $\vec{\xi}$, by virtue of (\ref{eq:hamiltonians on in/up modes}).
Similarly, the `up' and `down' modes, defined for $\tilde{\omega}$ non-negative, have non-negative energy with respect
to observers following the integral curve of $\vec{\chi}$. 

We may now construct the stress-energy tensor operator from either the potential operator 
(\ref{eq:quantum mode expansion for A in in/up modes}) or (\ref{eq:quantum mode expansion for A in out/dn modes}).
It is well-known that the stress-energy tensor as an operator does not have a well-defined meaning. It suffers from ultra-violet divergences 
and its expectation value when the field is in a certain state $\ket{\Psi}$ must be renormalized. 
There are several techniques for renormalization.
The \define{point-splitting technique} consists in temporarily displacing the point where one field in 
every quadratic term in the stress-energy tensor is evaluated, thus forming the object $\vac{\hat{T}_{\alpha\beta}(x,x')}{\Psi}$. 
This object is finite. 
Specific divergent terms, gathered in the bitensor $T^{\text{div}}_{\alpha\beta}(x,x')$, 
which are purely geometric and thus independent of the quantum state, are then subtracted from $\vac{\hat{T}_{\alpha\beta}(x,x')}{\Psi}$.
The end result is obtained by finally bringing the separated points together:
\begin{equation}
\vac[ren]{\hat{T}_{\alpha\beta}(x)}{\Psi}=\lim_{x'\to x}\left(\vac{\hat{T}_{\alpha\beta}(x,x')}{\Psi}-T^{\text{div}}_{\alpha\beta}(x,x')\right)
\end{equation}
It is this renormalized expectation value of the stress-energy tensor (RSET)
that is the source in Einstein's field equations (\ref{eq:Einstein eqs.}) in the semiclassical theory.
Christensen ~\cite{ar:Christ'78} has explicitly calculated the divergent terms $T^{\text{div}}_{\alpha\beta}$ by using covariant geodesic point separation. 
Jensen, McLaughlin and Ottewill ~\cite{ar:J&McL&Ott'88} calculated a linearly divergent term for the spin-1 case, 
which was not explicitly given by Christensen. The reason being that this term does not have to be included when an average 
is taken over the covariant derivative of the biscalar of geodetic interval $\sigma^{\mu}$ and $-\sigma^{\mu}$, 
as performed by Christensen.

Before we start a description of the various physical states of the field, we give an important result found by 
Unruh ~\cite{ar:Unruh'76} and further established by ~\cite{ar:Brown&Ott'83} and ~\cite{ar:Grove&Ott'83}. The result is that a `particle detector'
will react to states of the field which have positive frequency with respect to the detector's proper time. This means that a certain
observer will see as a vacuum state the one that has been defined with positive frequency modes with respect to the 
4-velocity of the observer.   
That is, if a certain observer $\mathcal{A}$ makes measurements relative to a certain vacuum state $\ket{\Xi}$, then 
he or she will measure a stress tensor 
\begin{equation}   \label{eq:stress for gral. obs.&vac.}
\vac[\mathcal{A}]{\hat{T}_{\alpha\beta}}{\Psi}=\vac{\hat{T}_{\alpha\beta}}{\Psi}-\vac{\hat{T}_{\alpha\beta}}{\Xi}
\end{equation}
when the field is in a certain state $\ket{\Psi}$.
In flat space-time, this means that an inertial observer makes measurements
relative to the Minkowski vacuum $\ket{M}$ and a Rindler observer (RO) relative to the Fulling vacuum $\ket{F}$.


\section{Boulware vacuum} \label{sec:B vac.}

\textbf{Schwarzschild space-time}

The \define{Boulware vacuum state}, denoted by $\ket{B}$, is defined in the Schwarzschild space-time as the vacuum that corresponds 
to quantizing the field with normal modes that have all positive frequency with respect to the space-time's hypersurface-orthogonal
timelike killing vector $\vec{\xi}$. 
This state respects the isometries of the Schwarzschild space-time.
Since it is the static observers SO the ones that move along integral curves
of $\vec{\xi}$, from Unruh's result stated in the previous section it follows that these observers will make
measurements relative to the Boulware vacuum $\ket{B}$. That is, a SO will see the vacuum $\ket{B}$ as empty.
Candelas ~\cite{ar:Candelas'80}, based on conjectures made previously by Christensen and Fulling ~\cite{ar:Christ&Fulling'77},
has found that the RSET when the scalar field is in the Boulware vacuum is zero at both $\mathcal{I}^-$ and $\mathcal{I}^+$
in the Schwarzschild space-time. 
Candelas also found that the RSET, close to the horizon, when the field is in the Boulware vacuum diverges and corresponds to the
absence from the vacuum of black-body radiation at the black hole temperature. 
The Boulware vacuum is therefore irregular at 
$\mathcal{H}^-$ and $\mathcal{H}^+$. The Boulware vacuum models a cold star with a Boyer-Lindquist radius slightly larger than
its Schwarzschild radius.   \catdraft{pq.?pq. si fos b-h, un obs. real podria apropar-se a l'horitzo i no seria fisic pq.
mesuraria quantitat div.; en canvi, si es estrella, un obs. no pot arribar a $r_+$(?)}

\textbf{Kerr space-time}

In the Schwarzschild space-time the Boulware vacuum is associated to the field expansion in terms of either the
`in' and `up' modes or the `out' and `down', both pairs of sets of complete modes defining the same vacuum $\ket{B}$.
We can perform a similar expansion for the electromagnetic potential in the Kerr space-time.
The \define{past Boulware state} is defined by
\begin{equation} \label{eq:a^in/up on B}
\begin{aligned}
{}_{lm\omega P}\hat{a}^{\text{in}}\ket{B^-}&=0  \\
{}_{lm\omega P}\hat{a}^{\text{up}}\ket{B^-}&=0
\end{aligned}
\end{equation}
corresponding to an absence of particles at $\mathcal{H}^-$ and $\mathcal{I}^-$. 
This is the state mentioned in Section \ref{sec:superrad.}, which exhibits the Starobinski\u{\i}-Unruh effect.
Due to this effect, the past Boulware state is not empty at $\mathcal{I}^+$.

We can also define the \define{future Boulware state}, as that state which is empty at $\mathcal{I}^+$ and $\mathcal{H}^+$:  
\begin{equation} \label{eq:a^out/down on B}
\begin{aligned}
{}_{lm\omega P}\hat{a}^{\text{out}}\ket{B^+}&=0  \\
{}_{lm\omega P}\hat{a}^{\text{down}}\ket{B^+}&=0
\end{aligned}
\end{equation}
The Bogolubov transformation between the pair of operators ${}_{lm\omega P}\hat{a}^{\text{in}}$ and ${}_{lm\omega P}\hat{a}^{\text{up}}$
and the pair ${}_{lm\omega P}\hat{a}^{\text{down}}$ and ${}_{lm\omega P}\hat{a}^{\text{out}}$ is non-trivial: 
the expression for ${}_{lm\omega P}\hat{a}^{\text{in} \dagger}\left[{}_{lm\omega P}\hat{a}^{\text{up} \dagger}\right]$ 
in terms of `out' and `down' operators contains ${}_{l,-m,-\omega,P}\hat{a}^{\text{down} \dagger}\left[{}_{l,-m,-\omega,P}\hat{a}^{\text{out} \dagger}\right]$
for modes in the superradiant regime.    
\catdraft{segur que es aixi per spin-1?}
This implies that the past Boulware state contains both outgoing and downgoing superradiant particles, 
and is therefore not empty at $\mathcal{I}^+$ and $\mathcal{H}^+$.
This flux of particles out to $\mathcal{I}^+$ corresponds to the Starobinski\u{\i}-Unruh effect.  
Similarly, the future Boulware state contains ingoing and upgoing superradiant particles, and is therefore not empty 
at $\mathcal{I}^-$ and $\mathcal{H}^-$. 
Because of the fact that ${}_{lm\omega P}\hat{a}^{\text{up}}$, when expressed in terms of `out' and `down' operators, contains 
the creator operator ${}_{l,-m,-\omega,P}\hat{a}^{\text{out} \dagger}$, it is not possible to construct a state which is empty at both 
$\mathcal{I}^-$ and $\mathcal{I}^+$, like the Boulware vacuum in the Schwarzschild space-time.

From the definitions (\ref{eq:a^in/up on B}) and (\ref{eq:a^out/down on B}) together with the relations (\ref{NP scalars in/up->out/down}),
the past and future Boulware states are obtainable
one from the other under the transformation $(t,\phi)\to (-t,-\phi)$. 
Because the two states are not equivalent, it follows that neither is invariant under this symmetry of the Kerr (and Kerr-Newman) space-time. 
\draft{to what extent are relations (\ref{NP scalars in/up->out/down}) sufficient to deduct the above, when there's a change in the sign
of $m$ and $\omega$ and $\phi_{-1}$ goes to $\phi_{+1}$ plus factors?
is it correct to say something like ${}_{lm\omega P}\hat{a}^{\text{in}}\to {}_{l-m-\omega P}\hat{a}^{\text{out}} \text{under} (t,\phi)\to (-t,-\phi)$?}

\draft{find and give Bogolubov transfs. for spin-1 case}

Several attempts have been made to construct a state which is stable and is also empty at both $\mathcal{I}^+$ and
$\mathcal{I}^-$. Matacz, Davies and Ottewill ~\cite{ar:Mat&Dav&Ott'93} considered a highly relativistic rotating star 
by assuming that the space-time outside the star at a radius $r_*=x$ is given by the Kerr metric and
requiring the scalar modes to vanish at the surface $r_*=x$ of the star. The radius $r_*=x$ is outside the horizon 
but close enough to it that there is an ergosphere. They then find that the state empty at $\mathcal{I}^-$ is related
to the state empty at $\mathcal{I}^+$ by a trivial Bogolubov transformation, so that the state is indeed empty
at both $\mathcal{I}^-$ and $\mathcal{I}^+$. It is also invariant under the symmetry $(t,\phi)\to (-t,-\phi)$ of the space-time.
This result, however, was proved under the explicit assumption of the absence of solutions with complex frequencies.
They say that if solutions with complex frequencies were to exist, then their result would not be valid.
Furthermore, Friedman ~\cite{ar:Friedman'78} 
proved that any stationary and asymptotically flat space-time
which has an ergosphere but no event horizon is classically unstable to scalar perturbations. 
It is straightforward to show that Friedman's result is applicable to the model of a star in ~\cite{ar:Mat&Dav&Ott'93}
and therefore solutions with complex frequencies do exist. Kang ~\cite{ar:Kang'97} has shown that the response function of an Unruh
`particle detector' is unstable if modes with complex frequency exist. 
The state state defined by ~\cite{ar:Mat&Dav&Ott'93} is therefore unstable.

Winstanley ~\cite{th:WinstMSc} has constructed a state in Kerr with the defining features of the Boulware vacuum in Schwarzschild
by using a variant of the $\eta$ formalism (which employs non-standard commutation relations for the creation and annihilation operators)
introduced by Frolov and Thorne ~\cite{ar:F&T'89}.
This state $\ket{B_{\mathcal{I}}}$ is defined as the one satisfying
\begin{equation} \label{eq:a^in/up on B_I}
\begin{aligned}
{}_{lm\omega P}\hat{a}^{\text{in}}\ket{B_{\mathcal{I}}}&=0  & \\
{}_{lm\omega P}\hat{a}^{\text{up}}\ket{B_{\mathcal{I}}}&=0 & \text{for} \quad \omega>0 \\
{}_{lm\omega P}\hat{a}^{\text{up} \dagger}\ket{B_{\mathcal{I}}}&=0 & \text{for} \quad \omega<0
\end{aligned}
\end{equation}
\catdraft{i cas $\omega=0$?}
Defining the state as in (\ref{eq:a^in/up on B_I}) is equivalent to defining it as 
\begin{equation} \label{eq:a^in/out on B_I}
\begin{aligned}
{}_{lm\omega P}\hat{a}^{\text{in}}\ket{B_{\mathcal{I}}}&=0  \\
{}_{lm\omega P}\hat{a}^{\text{out}}\ket{B_{\mathcal{I}}}&=0 
\end{aligned}
\end{equation}
The state $\ket{B_{\mathcal{I}}}$ is therefore empty at both $\mathcal{I}^+$ and $\mathcal{I}^-$ and it is invariant
under $(t,\phi)$ reversal. Winstanley has also proved 
that the RSET when the field is in this state goes asymptotically to zero as $O(r^{-3})$ for $r\rightarrow +\infty$. 
Therefore $\ket{B_{\mathcal{I}}}$ possesses some of the characteristic properties that the Boulware state possesses in
the Schwarzschild space-time. 
It still remains to be checked that this state is regular everywhere.
The formalism in ~\cite{ar:F&T'89} which Winstanley used to construct this state, however, introduces irregularities in the Hartle-Hawking case,
as we shall see in the next section.

Finally, one can similarly define a state as the one that satisfies
\begin{equation} \label{eq:a^up/down on B_I}
\begin{aligned}
{}_{lm\omega P}\hat{a}^{\text{up}}\ket{B_{\mathcal{H}}}&=0  \\
{}_{lm\omega P}\hat{a}^{\text{down}}\ket{B_{\mathcal{H}}}&=0 
\end{aligned}
\end{equation}
This state is then empty at both $\mathcal{H}^+$ and $\mathcal{H}^-$ and is invariant under $(t,\phi)$ reversal.

\draft{remove subindex P everywhere for scalar case}


\section{Hartle-Hawking state} \label{sec:H-H vac.}

\draft{for HH and U vacs. I can't write expressions for potential in terms of pos.freq.modes w.r.t. to $\pardiff{}{U}$
and $\pardiff{}{V}$ as I should find the corresponding eqs. to 4.3.1->4.3.4GavPhD for spin-1, plus I can't write them
either in terms of coth's in front of `in' and `up' modes as that is what I deduce in the last section?!->put them
(incl. for Boulware) for scalar field? find my two-point function (i.e., eq.4.3.15GavPhD)}

The defining features of a Hartle-Hawking state is that it possesses the symmetries
of the space-time and that it is regular everywhere, including on both the past and the future event horizons. 
Kay and Wald ~\cite{ar:Kay&Wald'91} have proven that for any globally hyperbolic space-time which has a Killing field with a bifurcate Killing 
horizon there can be at most one state with the above features. 
Kay and Wald have further shown that for the Kerr space-time this state does not exist. 
The Rindler and the Schwarzschild space-times are covered by Kay and Wald's theorem. 
In the Rindler space-time this state is clearly the Minkowski vacuum.  

\textbf{Schwarzschild space-time}

In the Schwarzschild space-time the state $\ket{H}$ is defined as the one that corresponds to quantizing the
field with upgoing normal modes which on $\mathcal{H}^-$ have positive frequency with respect to the Kruskal co-ordinate $U\equiv -e^{-\kappa_+u}$
and with ingoing normal modes which on $\mathcal{H}^+$ have positive frequency with respect to the Kruskal co-ordinate $V\equiv e^{\kappa_+v}$. 

\catdraft{why? where's the proof? why are the $\mu^{up}$,etc modes constructed like that? aren't the $u^{up}$,etc
equally eigenfuncs. of $\pardiff{}{u}$ at $\mathcal{H^-}$ (p.111GavPhD)??}

The state $\ket{H}$ is then the one that has the mentioned Hartle-Hawking features in the Schwarzschild space-time.
Indeed, Candelas ~\cite{ar:Candelas'80} showed that the state defined in the above manner is regular on both the past and 
future horizons. He also found that the RSET at infinity when the field is in the $\ket{H}$ state corresponds 
to that of a bath of black body radiation at the black hole temperature $T_H$. The Hartle-Hawking state models a black
hole in unstable thermal equilibrium with an infinite destribution at the Hawking temperature. 

From the above results and from the previous section we know that $\vac{\hat{T}_{\alpha\beta}}{H-B}$ is thermal both for
$r\rightarrow r_+$ and for $r\rightarrow +\infty$. Christensen and Fulling conjecture that this is the case everywhere.
However, Jensen, McLaughlin and Ottewill \cite{ar:J&McL&Ott'92} numerically show that $\vac{\hat{T}_{\alpha\beta}}{H-B}$
deviates from isotropic, thermal form as one moves away from the horizon.

We can establish a direct correspondence between observers and states in different space-times.
Candelas showed that the RSET close to the horizon when the field is in the $\ket{B}$ state diverges
like minus the stress tensor of black body radiation at the black hole temperature, and that it must also be equal 
to $-\vac[SO]{\hat{T}_{\alpha\beta}}{H}$, due to (\ref{eq:stress for gral. obs.&vac.}) and to the regularity of $\ket{H}$ 
\ddraft{and irregularity of $\ket{B}$ close to the horizon?}.  
Analogously, Unruh ~\cite{ar:Unruh'76} showed that in flat space-time and when the field is in the Minkowski vacuum, 
a Rindler observer RO will also see a bath of black body radiation at the Hawking temperature of a black hole 
with surface gravity $\kappa_+=a\alpha$, where $a$ is the RO's acceleration and $\alpha$ is the lapse function in Rindler space.   
We can therefore establish a correspondence RO$\leftrightarrow$SO and also $\ket{M}\leftrightarrow\ket{H}$; see also ~\cite{ar:Israel'76}.    
\ddraft{not quite: the relationship between RO and SO comes from the fact that they are accelerated in the same way=>change}      
The correspondence between $\ket{M}$ and $\ket{H}$ is related to the fact that they are both regular close to the horizon
of their respective space-times. We can then also establish a correspondence between the RO's own vacuum, $\ket{F}$,
and the SO's own vacuum, $\ket{B}$, related to the fact that they are both divergent close to the horizon of their respective space-times. 
\catdraft{is $\ket{F}$ div.?} Finally, from the $\ket{M}\leftrightarrow\ket{H}$ correspondence we can also establish a
correspondence between the inertial observers in flat space-time and the freely-falling observers in Schwarzschild.  


\textbf{Kerr space-time}

With the variant of the $\eta$ formalism mentioned in the previous section, Frolov and Thorne ~\cite{ar:F&T'89}
define a new ``Hartle-Hawking'' state $\ket{FT}$ invariant under the symmetries of the Kerr space-time. They go on to
prove that the RSET when the field is in the $\ket{FT}$ state is finite at the horizon but that, at least
for arbitrarily slow rotation, it is equal to the stress tensor of a thermal distribution at the Hawking
temperature rigidly rotating with the horizon. This suggests that it becomes irregular wherever $\vec{\chi}$ is not timelike,
that is on and outside the speed-of-light surface.          
Ottewill and Winstanley ~\cite{ar:Ott&Winst'00}, however, proved that although $\ket{FT}$ has a Feynman propagator
with the correct properties for regularity on the horizons, its two-point function is actually pathological almost everywhere, not just outside the
speed-of-light surface. Only at the axis of symmetry, where all the modes in the two-point function for the scalar field 
are evaluated for $\tilde{\omega}=\omega$ (i.e., $m=0$), it does not suffer from this pathology. 

Frolov and Thorne claim that close to the horizon ZAMOs make measurements relative to an unspecified Boulware vacuum.
They also claim that, when the field is in the state $\ket{FT}$, ZAMOs measure close to the horizon 
a stress tensor equal to that of a thermal distribution at the Hawking temperature rigidly rotating with the horizon. 

\draft{add to it the fact that the unspecified Boulware vac. must be $\ket{B_{\mathcal{H}}}$ 
and therefore it should apply to RROs instead of ZAMOs?}   

Duffy ~\cite{th:GavPhD} modified the Kerr space-time by introducing a mirror and constructed a state $\ket{H_{\mathcal{M}}}$
for the scalar field that is invariant under the isometries of the modified space-time.
He then showed that $\ket{H_{\mathcal{M}}}$ is regular everywhere in the modified space-time if, and only if, the mirror
removes the region outside the speed-of-light surface.  
He constructed another state $\ket{B_{\mathcal{M}}}$ invariant under the isometries of the modified space-time and empty on both the past
and future horizons. 
\ddraft{pq. horizons? en tot cas hauria de ser empty a past and future infinity?! el Gav. no sembla que digui ni l'una cosa ni l'altra en el seu PhD}
This is the state that RROs make measurements relative to in the modified space-time. 
He also numerically showed that when the field is in the $\ket{H_{\mathcal{M}}}$ state the stress tensor measured by a RRO
is, close to the horizon, that of a thermal distribution at the Hawking temperature rigidly rotating with the horizon. 

Finally, CCH defined a new Hartle-Hawking-type state, which we will hereafter denote by $\ket{CCH^-}$. 
This state is obtained by thermalizing the `in' and `up' modes with respect to their natural energy. 
Ottewill and Winstanley ~\cite{ar:Ott&Winst'00} showed that this state is, however, not invariant under the symmetry transformation 
$(t,\phi)\to (-t,-\phi)$ of the space-time.
They further argued that the RSET when the scalar field is in the $\ket{CCH^-}$ state is regular on the future horizon but irregular on the past horizon.
We must note that these results were derived in ~\cite{ar:Ott&Winst'00} based on a stress-energy tensor for which the $t\theta$- and $\phi\theta$-components
are identically zero. We shall see in Section \ref{sec:luminosity} that although this is indeed the case for the scalar field, which is the case
they considered, this is most probably not true for the electromagnetic field.
\draft{say that my numerical results, however, do agree with their conclusions, or do they?}
A similar state could be constructed by applying the transformation $(t,\phi)\to (-t,-\phi)$ to the state $\ket{CCH^-}$. 
This state, suitably named $\ket{CCH^+}$, would then be irregular on the future horizon and regular on the past horizon.
In Section \ref{sec:RRO} we will investigate
the form close to the horizon of $\vac[ren]{\hat{T}^{\mu}{}_{\nu}}{CCH^--B^-}$ for electromagnetism.


\section{Unruh state} \label{sec:U vac.}

\textbf{Schwarzschild space-time}

Unruh ~\cite{ar:Unruh'76} constructed a state $\ket{U^-}$ in the Schwarzschild space-time 
that would model the state of a black hole at late times. With this purpose he replaced the stellar collapse 
by certain boundary conditions at $\mathcal{H^-}$. He then expanded the
scalar field in modes that are positive frequency with respect to the proper time $t$ of 
inertial observers in $\mathcal{I^-}$ and modes that are positive frequency with respect to
the proper time of inertial observers close to $\mathcal{H^-}$.            
Unruh showed that 
an inertial observer falling into $\mathcal{H^-}$ \ddraft{$\mathcal{H^+}$ according to p.140GavPhD??} will see no
particles flowing out of the black hole but he will see particles flowing in, when the field is
in the $\ket{U^-}$ state. \ddraft{sec.5.2GavPhD says that observer will see no particles at all??!!} 

\catdraft{posar p.141(dalt) GavPhD!!}

Unruh and later Candelas ~\cite{ar:Candelas'80} (aided by the regularity of $\ket{H}$ at the horizon) showed that $\ket{U^-}$
is regular on $\mathcal{H^+}$, but is irregular on $\mathcal{H^-}$.                  
Candelas also showed that the RSET when the field is in the $\ket{U^-}$ state corresponds to 
a flux of thermal radiation at the Hawking temperature outgoing at $\mathcal{I^+}$.  
This radiation is the Hawking radiation discussed in Section \ref{sec:b-h thermodynamics}.
\ddraft{then Candelas's result is only a corroboration of Hawking's original results?}

\textbf{Kerr space-time}

It is possible to construct a state $\ket{U^-}$ in Kerr with the same positive-frequency mode definitions
as for the Unruh state in Schwarzschild. This state, like $\ket{U^-}$ in Schwarzschild, is empty at $\mathcal{I^-}$
but
is thermally populated at the Hawking temperature at $\mathcal{I^+}$, 
corresponding to the Hawking radiation.
\catdraft{1) but if $\vac[ren]{T}{U^-}$ includes both Hawking rad. and Starob-Unruh rad. then previous statement is incorrect/loose?
Adrian: Starob-Unruh rad. is considered as part of the Hawking rad.
2) $\vac[ren]{T}{U^-}$ is the same as $\vac{T}{U^--B^+}$ at $\mathcal{I^+}$ and the latter is said in Ott\&Winst'00 to correspond
to the Hawking effect, but from 1) should it not really be $\vac{T}{U^--B^-}$ that at $\mathcal{I^+}$ corresponds to the Hawking effect?
sln: same as for 1)}
 It is clearly not invariant under $(t,\phi)\to (-t,-\phi)$, like $\ket{U^-}$ in Schwarzschild.   
\catdraft{hauria de dir que $\mathcal{I^-}\leftrightarrow \mathcal{I^+}$ i $\mathcal{H^-}\leftrightarrow \mathcal{H^+}$
sota $(t,\phi) \leftrightarrow (-t,-\phi)$! es aixi??!}
The state $\ket{U^-}$ is referred to as the \define{past Unruh state}.
However, since in the Kerr space-time there is no Hartle-Hawking-type state regular on both
the past and future horizons, it is a difficult task to prove any properties of the state $\ket{U^-}$ near the horizon.   
Duffy ~\cite{th:GavPhD}, however, obtained numerical results which indicate that this state is
regular on the future horizon.      

\draft{mention Ott\&Winst'00 and Punsley's results that $\ket{U^-}$ 'seems to' be regular on $\mathcal{H^+}$ and irregular on $\mathcal{H^-}$?}

The expressions for the expectation value of the stress-energy tensor as measured by various observers that we have 
discussed in the present and in the two previous sections are summarized in Table \ref{stress tensor by diff. obs. for r->r_+}. 
This table also summarizes
the thermal behaviour that some of these expectation values possess proven so far in the literature.

\begin{table}
\begin{center}
\renewcommand{\arraystretch}{1.4}
\begin{tabular}{|c|l|}
\hline
Space-time    &   \\
\hline
\hline
Flat          & $\vac[RO]{\hat{T}_{\alpha\beta}}{\Psi}=\vac[ren]{\hat{T}_{\alpha\beta}}{\Psi-F}$     \\
              & $\vac[RO]{\hat{T}_{\alpha\beta}}{M}=T^{\text{th}}_{\alpha\beta}$            \\ [0.5ex]
\hline                                                                   
Schwarzschild & $\vac[SO]{\hat{T}_{\alpha\beta}}{\Psi}=\vac[ren]{\hat{T}_{\alpha\beta}}{\Psi-B}$      \\
              & $\vac[SO]{\hat{T}_{\alpha\beta}}{H}\rightarrow -\vac[ren]{\hat{T}_{\alpha\beta}}{B}\rightarrow T^{\text{th}}_{\alpha\beta}$  
                \quad $(r\rightarrow r_+)$ \\   [0.5ex]      
\hline                                                                   
Kerr          & $\vac[RRO]{\hat{T}_{\alpha\beta}}{\Psi}=\vac[ren]{\hat{T}_{\alpha\beta}}{\Psi-B_\mathcal{H}}$       \\[0.5ex]
\hline                                                                   
Modified Kerr & $\vac[RRO]{\hat{T}_{\alpha\beta}}{\Psi_\mathcal{M}}=\vac[ren]{\hat{T}_{\alpha\beta}}{\Psi-B_\mathcal{M}}$ \\
              & 
          $\vac[RRO]{\hat{T}_{\alpha\beta}}{H_{\mathcal{M}}}\rightarrow -\vac[ren]{\hat{T}_{\alpha\beta}}{B_{\mathcal{M}}}\rightarrow T^{\text{(th,RR)}}_{\alpha\beta}$  
\quad $(r\rightarrow r_+)$ \\[0.5ex]
\hline
\end{tabular}
\end{center}
\caption{Expectation value of the stress-energy tensor as measured by various observers in different space-times and their thermal behaviour
in certain states. The states $\ket{\Psi}$ and $\ket{\Psi_\mathcal{M}}$ represent any state in the corresponding space-times.
The stress-energy tensor $T^{\text{(th,RR)}}_{\alpha\beta}$ corresponds to a rigidly-rotating thermal distribution at the Hawking
temperature and its form is given in (\ref{eq:thermal RR stress tensor}).
}
\label{stress tensor by diff. obs. for r->r_+}
\end{table}

\draft{
1) should regular corrections term be added to 
$\vac[RRO]{\hat{T}_{\alpha\beta}}{H_{\mathcal{M}}}\rightarrow -\vac[ren]{\hat{T}_{\alpha\beta}}{B_{\mathcal{M}}}$?,
2) is $\vac[RRO]{\hat{T}_{\alpha\beta}}{\Psi}=\vac[ren]{\hat{T}_{\alpha\beta}}{\Psi-B_\mathcal{H}}$ valid everywhere or only at horizon?
3) what are the observers that see the $\ket{M}$ and the $\ket{H}$ states as their own? add them to table?
4) include in the table F\&T's wrong results? 5) include Gavin's numerics indicating $\vac[ren]{\hat{T}_{\alpha\beta}}{U^--B^-}$ being thermal
in the diagonal in Kerr (since this is the sort of thing I prove in my graphs)?}



\section{Expectation value of the stress tensor} \label{sec:CCH's expressions}

In this section we give the expectation values of a quadratic operator and of the stress-energy tensor when the field is in
various physical states, as given by CCH. 
To our knowledge, these important expressions for the electromagnetic field in the Kerr space-time have only been given so far by CCH.
These are the expressions that we initially used in our numerical calculations.
We felt forced to review them, however, and that is addressed later in Section \ref{sec:theta->pi-theta Symmetry}.

The following expressions are given by CCH where $\hat{Q}$ is any quadratic operator in the field and its derivatives
and $Q$ is its classical counterpart:
\begin{subequations}  \label{eq:quadratic op. for s=1 on vacua}
\begin{align}
\begin{split}
&\expct{\hat{Q}}{B^-}=\\
&=\sum_{lmP}\left\{
\int_0^{\infty}\d{\tilde{\omega}}\, 
Q\left[{}_{lm\omega}\phi_{\indhel}^{\text{up}},{}_{lm\omega}\phi_{\indhel}^{\text{up} *}\right]+ 
\int_0^{\infty}\d{\omega}\, 
Q\left[{}_{lm\omega}\phi_{\indhel}^{\text{in}},{}_{lm\omega}\phi_{\indhel}^{\text{in} *}\right]
\right\} 
\end{split}
\label{eq:quadratic op. for s=1 on B-} \\ 
\begin{split}
&\expct{\hat{Q}}{U^-}=\\
&=\sum_{lmP}\left\{
\int_0^{\infty}\d{\tilde{\omega}}\coth{\left(\frac{\pi\tilde{\omega}}{\kappa}\right)}
Q\left[{}_{lm\omega}\phi_{\indhel}^{\text{up}},{}_{lm\omega}\phi_{\indhel}^{\text{up} *}\right]+ 
\int_0^{\infty}\d{\omega}
Q\left[{}_{lm\omega}\phi_{\indhel}^{\text{in}},{}_{lm\omega}\phi_{\indhel}^{\text{in} *}\right]
\right\} 
\end{split}
\label{eq:quadratic op. for s=1 on U-} \\ 
\begin{split}
&\expct{\hat{Q}}{CCH^-}=\sum_{lmP}\left\{
\int_0^{\infty}\d{\tilde{\omega}}\coth{\left(\frac{\pi\tilde{\omega}}{\kappa}\right)}
Q\left[{}_{lm\omega}\phi_{\indhel}^{\text{up}},{}_{lm\omega}\phi_{\indhel}^{\text{up} *}\right]+ 
\right. \\ &\left.
+\int_0^{\infty}\d{\omega}\coth{\left(\frac{\pi \omega}{\kappa}\right)}
Q\left[{}_{lm\omega}\phi_{\indhel}^{\text{in}},{}_{lm\omega}\phi_{\indhel}^{\text{in} *}\right]
\right\} \label{eq:quadratic op. for s=1 on CCH}
\end{split}
\end{align}
\end{subequations}
From the above equations (\ref{eq:quadratic op. for s=1 on vacua}), the expressions for the expectation value 
of the stress-energy tensor when the field is in different states follow:
\begin{subequations} \label{eq:stress tensor for s=1 on all vac.}
\begin{align}
\begin{split}
&\expct{\hat{T}_{\mu\nu}}{B^-}=\\
&=\sum_{lmP}\left\{
\int_0^{\infty}\d{\tilde{\omega}}\, T_{\mu\nu}
\left[{}_{lm\omega}\phi_{\indhel}^{\text{up}},{}_{lm\omega}\phi_{\indhel}^{\text{up} *}\right]+
\int_0^{\infty}\d{\omega}\, T_{\mu\nu}
\left[{}_{lm\omega}\phi_{\indhel}^{\text{in}},{}_{lm\omega}\phi_{\indhel}^{\text{in} *}\right]
\right\}   
\end{split}
\label{eq:stress tensor for s=1 on B-} \\
\begin{split}
&\expct{\hat{T}_{\mu\nu}}{U^-}=\\
&=\sum_{lmP}\left\{
\int_0^{\infty}\d{\tilde{\omega}}\coth{\left(\frac{\pi\tilde{\omega}}{\kappa}\right)}
T_{\mu\nu}
\left[{}_{lm\omega}\phi_{\indhel}^{\text{up}},{}_{lm\omega}\phi_{\indhel}^{\text{up} *}\right]+
\int_0^{\infty}\d{\omega}T_{\mu\nu}
\left[{}_{lm\omega}\phi_{\indhel}^{\text{in}},{}_{lm\omega}\phi_{\indhel}^{\text{in} *}\right]
\right\}   
\end{split}
\label{eq:stress tensor for s=1 on U-}  \\
\begin{split}
&\expct{\hat{T}_{\mu\nu}}{CCH^-}=\sum_{lmP}\left\{
\int_0^{\infty}\d{\tilde{\omega}}\coth{\left(\frac{\pi\tilde{\omega}}{\kappa}\right)}
T_{\mu\nu}
\left[{}_{lm\omega}\phi_{\indhel}^{\text{up}},{}_{lm\omega}\phi_{\indhel}^{\text{up} *}\right]+
\right. \\  & \left.+
\int_0^{\infty}\d{\omega}\coth{\left(\frac{\pi \omega}{\kappa}\right)}
T_{\mu\nu}
\left[{}_{lm\omega}\phi_{\indhel}^{\text{in}},{}_{lm\omega}\phi_{\indhel}^{\text{in} *}\right]
\right\}   
\end{split}
\label{eq:stress tensor for s=1 on CCH}  
\end{align}
\end{subequations}

We use the obvious notation that $Q\left[{}_{lm\omega}\phi_{\indhel}^{\bullet},{}_{lm\omega}\phi_{\indhel}^{\bullet *}\right]$
and $T_{\mu\nu}\left[{}_{lm\omega}\phi_{\indhel}^{\bullet},{}_{lm\omega}\phi_{\indhel}^{\bullet *}\right]$ denote the general
expressions for $Q$ and the stress-energy tensor (\ref{eq:stress tensor, spin 1}) respectively, where the scalars 
$\phi_{\indhel}$ have been replaced by the modes ${}_{lm\omega}\phi_{\indhel}^{\bullet}$. 
We will also use the symbol ${}_{lm\omega}T^{\bullet}_{\mu\nu}$ to refer to 
$T_{\mu\nu}\left[{}_{lm\omega}\phi_{\indhel}^{\bullet},{}_{lm\omega}\phi_{\indhel}^{\bullet *}\right]$.

CCH's original expressions contained the symbols $Q\left[u^{\bullet}_{\omega lmP},u^{\bullet *}_{\omega lmP}\right]$
and $T_{\mu\nu}\left[u^{\bullet}_{\omega lmP},u^{\bullet *}_{\omega lmP}\right]$, which we have respectively replaced in the expressions
(\ref{eq:quadratic op. for s=1 on vacua}) and (\ref{eq:stress tensor for s=1 on all vac.}) above by 
$Q\left[{}_{lm\omega}\phi_{\indhel}^{\bullet},{}_{lm\omega}\phi_{\indhel}^{\bullet *}\right]$ and 
$T_{\mu\nu}\left[{}_{lm\omega}\phi_{\indhel}^{\bullet},{}_{lm\omega}\phi_{\indhel}^{\bullet *}\right]$.
Note that the above expressions (\ref{eq:quadratic op. for s=1 on vacua}) and 
(\ref{eq:stress tensor for s=1 on all vac.}) without these replacements are indeed valid for the scalar field case.

CCH did not give an expression for the expectation value of the stress tensor when the field is in the state $\ket{FT}$,
as this state was only defined later.
We give here an expression for this expectation value obtained by direct generalization from the corresponding
one for the scalar field as in the case of the states in (\ref{eq:stress tensor for s=1 on all vac.}):
\begin{equation} \label{eq:stress tensor for s=1 on FT}
\begin{aligned}
&\expct{\hat{T}_{\mu\nu}}{FT}=\sum_{lmP}\left\{
\int_0^{\infty}\d{\tilde{\omega}}\coth{\left(\frac{\pi\tilde{\omega}}{\kappa}\right)}
T_{\mu\nu}\left[{}_{lm\omega}\phi_{\indhel}^{\text{up}},{}_{lm\omega}\phi_{\indhel}^{\text{up} *}\right]
\right. \\ &\left. +
\int_0^{\infty}\d{\omega}\coth{\left(\frac{\pi \tilde{\omega}}{\kappa}\right)}
T_{\mu\nu}\left[{}_{lm\omega}\phi_{\indhel}^{\text{in}},{}_{lm\omega}\phi_{\indhel}^{\text{in} *}\right]
\right\}  
\end{aligned}
\end{equation}
We will only use this expression in Section \ref{subsec:lack of symmetry}.
We shall show in that section that CCH's expressions (\ref{eq:stress tensor for s=1 on all vac.}) are incorrect.
We obtain the corrected expressions and give them in (\ref{eq:corrected stress tensor for s=1 on B-,U-}).

\begin{figure}[p]
\rotatebox{90}
\centering
\includegraphics*[width=70mm,angle=270]{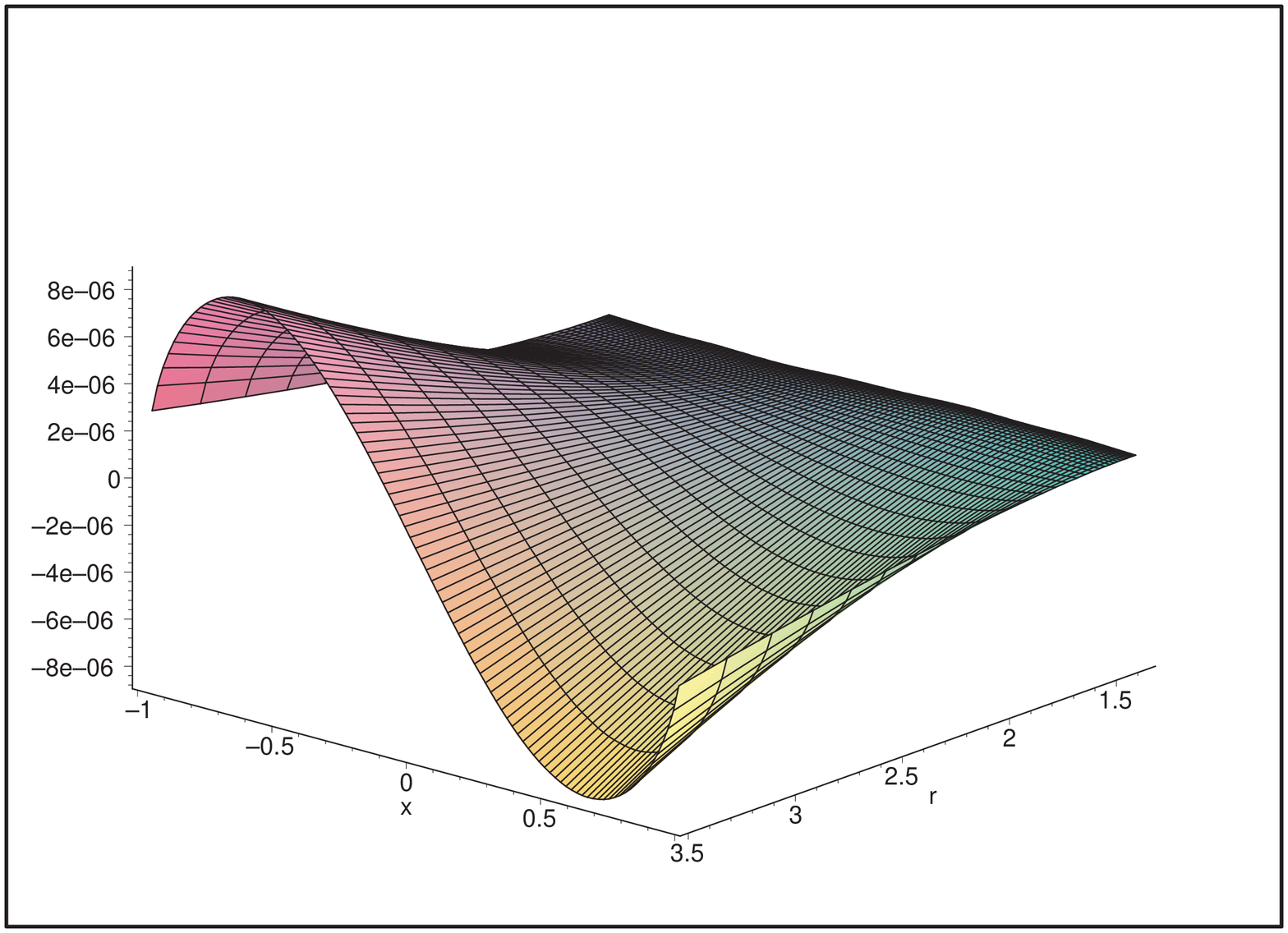}
\caption{$\frac{1}{4\pi}\Delta^2\vac{\hat{T}_{r\theta}}{U^--B^-}$}
\label{fig:delta2SymmTrtheta_u_b}
\includegraphics*[width=70mm,angle=270]{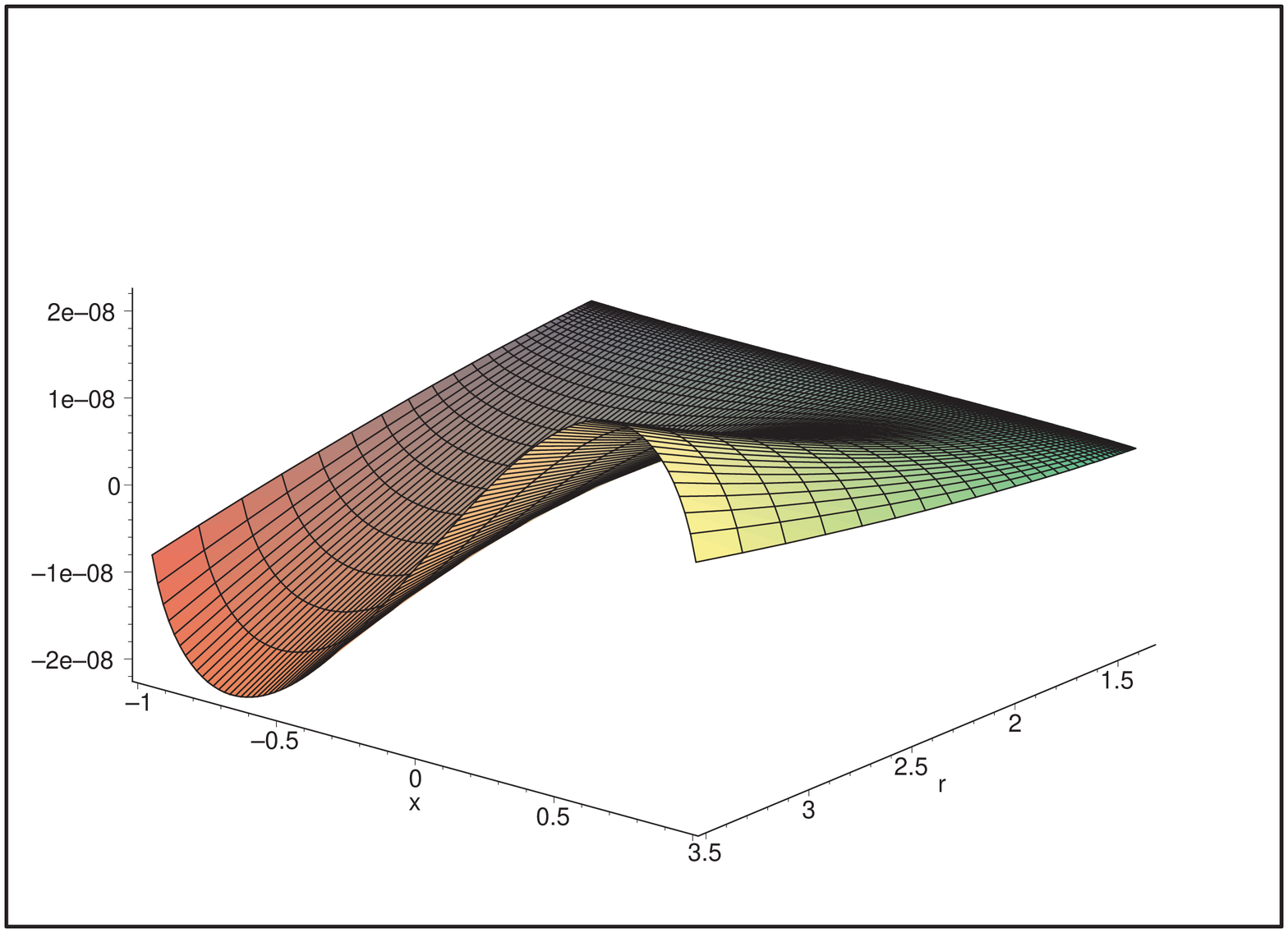}
\caption{$\frac{1}{4\pi}\Delta^2\vac{\hat{T}_{r\theta}}{CCH^--U^-}$}
\label{fig:delta2SymmTrtheta_cch_u_past}
\end{figure}

\begin{figure}[p]
\rotatebox{90}
\centering
\includegraphics*[width=70mm,angle=270]{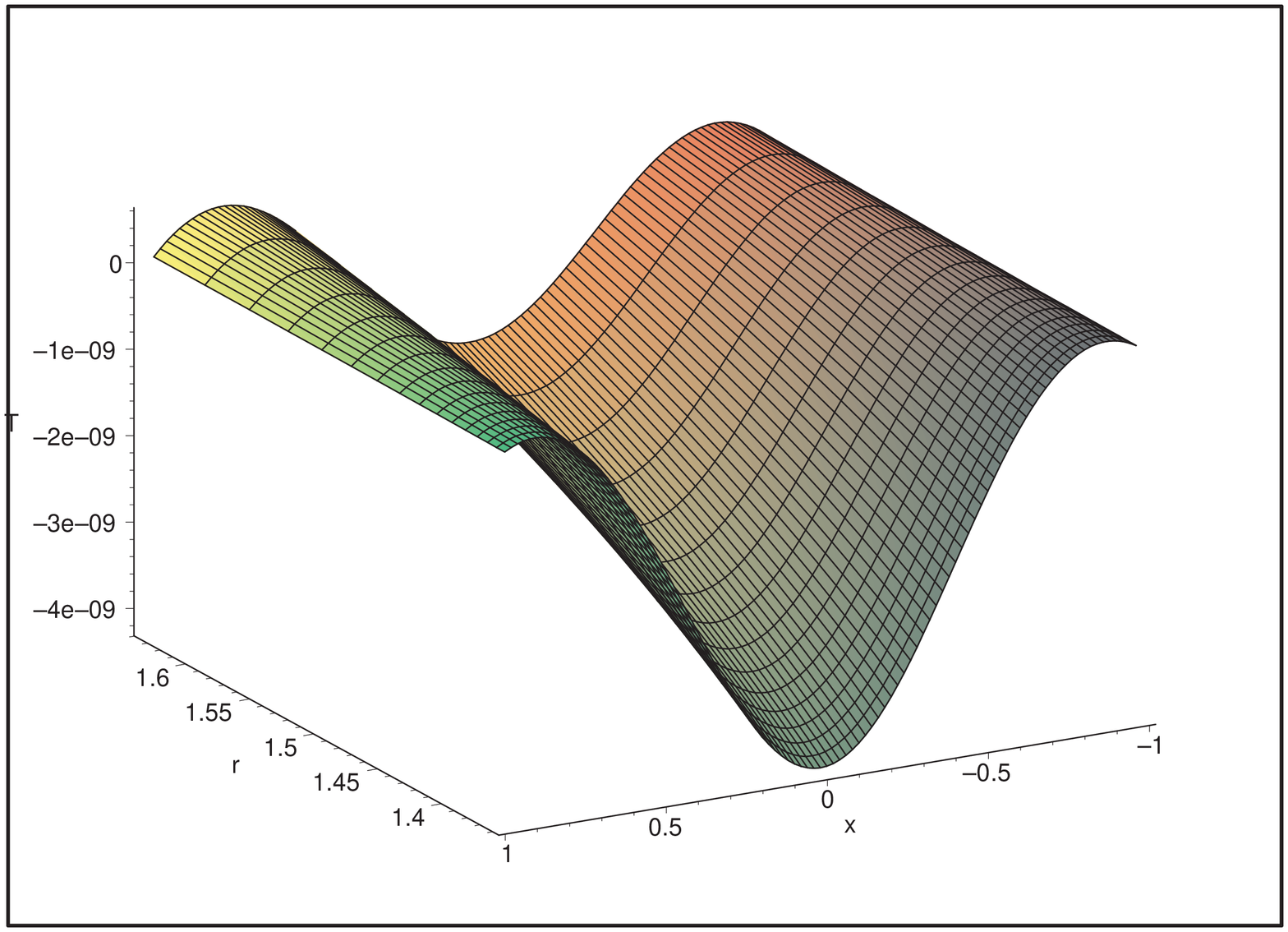} 
\caption{$\frac{1}{4\pi}\vac{\hat{T}_{t\phi}}{CCH^--U^-}$}
\label{fig:SymmTtphi_cch_u_past}
\end{figure}

Graphs \ref{fig:delta2SymmTrtheta_u_b}--\ref{fig:SymmTtphi_cch_u_past}
have been obtained with these corrected expressions.
We plot the $r\theta$-component of the stress tensor for the difference between the $\ket{U^-}$ and the $\ket{B^-}$ states.
This is the only component that is not plotted in later sections.
A plot of the same component for the difference between the $\ket{CCH^-}$ and the $\ket{B^-}$ states is identical to 
Figure \ref{fig:delta2SymmTrtheta_u_b} for the range of $r$ displayed.
The components for the difference
between $\ket{CCH^-}$ and $\ket{U^-}$ are trivially obtained by subtracting the previous two differences between states.
We have included the $r\theta$- and $t\phi$-components for the difference between $\ket{CCH^-}$ and $\ket{U^-}$.

\draft{include the components that are not included in Section \ref{}, i.e., the comps. for which the thermal RR tensor
is identically zero. how do we know their asymptotic behaviour close to the horizon and at infinity? could $T_{tr}$ be
compared against anything?}

\catdraft{1) parlar del fet que $T_{t\theta}$ no es zero (p.71K->74K), 2) dir si $T_{tr}^{div}$ es zero o no (p.76K,etc), 3) parlar de conserv.eqs. i trace-freeness
i que han estat comprovades numericament (p.79K,88K,etc)}


\section{Luminosity} \label{sec:luminosity}

\catdraft{parlar/descriure/explicar superradiance en aquesta seccio!}

The total energy flux at infinity per unit solid angle is given by
\begin{equation} \label{eq:dE/dtdOmega}
\frac{\d{E}}{\d{t} \d{\Omega}}=
\lim_{r\rightarrow +\infty}r^2 T^r{}_t
\end{equation}
\catdraft{pq.?}
Let $E^{\text{(inc)}}$ and $E^{\text{(ref)}}$ denote, respectively, the energy incident and the energy reflected
by the black hole at infinity. 
The corresponding incident and reflected energy fluxes at infinity per unit solid angle are then calculated with (\ref{eq:dE/dtdOmega}) including in the 
NP scalars only the ingoing (i.e., incident) and outgoing (i.e., reflected) parts, respectively, 
of the radial functions ${}_{\indhel}R^{\text{in}}$.
The expressions at radial infinity of the corresponding NP scalars are denoted by $\phi^{\text{(in,inc)}}_{\indhel}$ 
and $\phi^{\text{(in,ref)}}_{\indhel}$ respectively.
The following expressions can be immediately checked: 
\begin{subequations} \label{eq:dE(inc/out)/dtdOmega}
\begin{align}
\frac{\d{E^{\text{(inc)}}}}{\d{t} \d{\Omega}}&=\frac{r^2}{8\pi}\left|\phi^{\text{(in,inc)}}_{-1}\right|^2 \label{eq:dE(inc)/dtdOmega} \\
\frac{\d{E^{\text{(ref)}}}}{\d{t} \d{\Omega}}&=\frac{r^2}{2\pi}\left|\phi^{\text{(in,ref)}}_{+1}\right|^2 \label{eq:dE(out)/dtdOmega}
\end{align}
\end{subequations}
Let $E^{\text{(tra)}}$ denote the energy going down across the event-horizon of the black hole. 
If there is a flux of energy across the 2-surface element formed by the intersection of an element of the horizon with two
surfaces of constant $v$ separated by $\d{v}$, then the change in energy of the black hole is 
\begin{equation} \label{eq:dE}
\d{E}^{\text{(tra)}}=T_{\mu}{}^{\nu}\xi^{\mu}\d^3{\Sigma_{\nu}}
\end{equation}
where $\d^3{\Sigma_{\nu}}$ is the $3$-surface element of the horizon, normal to the inward radial direction 
of the Kerr system $\{v,r,\theta,\bar{\phi}\}$. 
It can be checked that the corresponding flux of energy per unit solid angle is
\begin{equation}\label{eq:dEtra/dtdOmega}
\frac{\d{E^{\text{(tra)}}}}{\d{t} \d{\Omega}}=\frac{\Delta^2}{8\pi(r_+^2+a^2)}\frac{\omega}{\tilde{\omega}}\left|\phi^{\text{(in,tra)}}_{-1}\right|^2
\end{equation}
where $\phi^{\text{(in,tra)}}_{-1}$ refers to the inclusion in the NP scalar of only the transmitted part 
close to the horizon of the radial function ${}_{+1}R^{\text{in}}$.

The wronskian relations in Table \ref{table:radial wronsks} relate the above energy fluxes.
Indeed these wronskian relations correspond to the conservation of energy law that equates the net flux of energy coming in
from infinity to the net flux of energy going down into the black hole:
\begin{equation} \label{eq:conserv. of energy law}
\diff{E^{\text{(inc)}}}{t}-\diff{E^{\text{(ref)}}}{t}=\diff{E^{\text{(tra)}}}{t}
\end{equation}
The \define{reflection coefficient} $\mathbb{R}_{lm\omega}$  and the \define{transmission coefficient} $\mathbb{T}_{lm\omega}$ 
of an incoming wave mode are defined as the following flux ratios:
\begin{subequations}
\begin{align}
\mathbb{R}_{lm\omega}&\equiv \frac{\d E^{\text{(ref)}}_{lm\omega}/\d{t}}{\d E^{\text{(inc)}}_{lm\omega}/\d{t}}\\
\mathbb{T}_{lm\omega}&\equiv \frac{\d E^{\text{(tra)}}_{lm\omega}/\d{t}}{\d E^{\text{(inc)}}_{lm\omega}/\d{t}}
\end{align}
\end{subequations}

The transmission coefficient $\mathbb{T}_{lm\omega}$ is also commonly interpreted as the absorption probability 
for an incoming wave mode and is then denoted by $\Gamma_{lm\omega}$.
The conservation of energy law (\ref{eq:conserv. of energy law}) can then be re-expressed as
\begin{equation} \label{eq:def. coeff. absorption}
\mathbb{R}_{lm\omega}=1-\mathbb{T}_{lm\omega}=
4\frac{\left|{}_{lm\omega}\phi^{\text{(in,ref)}}_{+1}\right|^2}{\left|{}_{lm\omega}\phi^{\text{(in,inc)}}_{-1}\right|^2}
\end{equation}
where we have made use of expressions (\ref{eq:dE(inc/out)/dtdOmega}).

By using the asymptotic expressions for the NP scalars together with the relations (\ref{eq:R1 coeffs from R_1's}) and (\ref{eq:R_1 coeffs from X's})
and the wronskians in Table \ref{table:radial wronsks} we immediately find various, equivalent expressions for the reflection
and absorption coefficients:
\begin{equation} \label{eq:val. of trans. coeff.}
\mathbb{T}_{lm\omega}=1-\mathbb{R}_{lm\omega}=1-\left|A^{\text{in}}_{lm\omega}\right|^2=\frac{-i}{2^4\omega^3}W[Y_{+1}^{\text{in}},Y_{-1}^{\text{in} *}]_{lm\omega}
\end{equation}
We use the obvious notation that subindices outside the square brackets of the wronskian apply to the radial functions inside the brackets.
It is clear that the coefficients of the radial function $X_{lm\omega}$, rather than those of ${}_{\indhel}R_{lm\omega}$, are
the natural ones in the description of the scattering of wave modes.
As anticipated in Section \ref{sec:short-range potentials}, $|A^{\text{in}}_{lm\omega}|^2$ is the 
fractional gain or loss of energy of an incoming wave mode. 
We know that for superradiant wave modes this quantity is greater than one and therefore 
the reflection coefficient $\mathbb{R}_{lm\omega}$ is also greater than one while
the transmission coefficient $\mathbb{T}_{lm\omega}$ is negative for these modes.

\draft{should coeffs. $\mathbb{T}_{lm\omega}$ and $\mathbb{R}_{lm\omega}$ have subindex $P$? it doesn't seem like they should but
Page'76 does include them. Then coeffs. $A^{\text{in/up}}_{lm\omega}$ and ${}_{\indhel}R^{\text{in/up},tra/ref/inc}_{lm\omega}$ should
have subindex $P$ as well?}

The conservation equations $\nabla_{\nu}T_{\mu}{}^{\nu}=0$ can alternatively be written ~\cite{bk:Dirac} as
\begin{equation} \label{eq:compact conserv.eqs.}
\partial_{\nu}\left(T_{\mu}{}^{\nu}\sqrt{-g}\right)=\frac{1}{2}\sqrt{-g}\left(\partial_{\mu}g_{\alpha\beta}\right)T^{\alpha\beta}
\end{equation}
Assuming that the stress-energy tensor is independent of $t$ and $\phi$, like the Kerr metric (and the Kerr-Newman solution), the $\mu=t$ and 
$\mu=\phi$ components of equations (\ref{eq:compact conserv.eqs.}) become
\begin{subequations}
\begin{align}
\partial_r\left(\Sigma\sin\theta T_t{}^{r}\right)+\partial_{\theta}\left(\Sigma\sin\theta T_t{}^{\theta}\right)&=0 \\
\partial_r\left(\Sigma\sin\theta T_{\phi}{}^{r}\right)+\partial_{\theta}\left(\Sigma\sin\theta T_{\phi}{}^{\theta}\right)&=0 
\end{align}
\end{subequations}
After integrating these equations over $r$ the result is:
\begin{subequations} \label{eq:integrated conserv.eqs., T_tr,Ttphi}
\begin{align}
T_{tr}&=\frac{K(\theta)}{\Delta}-\frac{1}{\Delta\sin\theta}\partial_{\theta}\left(\sin\theta\int_{r_+}^r\d{r'}T_{t\theta}\right) 
\label{eq:integrated conserv.eqs., T_tr} \\
T_{\phi r}&=\frac{L(\theta)}{\Delta}-\frac{1}{\Delta\sin\theta}\partial_{\theta}\left(\sin\theta\int_{r_+}^r\d{r'}T_{\phi\theta}\right) 
\label{eq:integrated conserv.eqs., Ttphi}
\end{align}
\end{subequations}
where $K(\theta)$ and $L(\theta)$ are arbitrary functions.
The function $K(\theta)$ is related to the \define{luminosity}, 
which is defined as the instantaneous flux of energy per unit time. 
The luminosity when the field is in the state $\ket{\Psi}$ is given by
\begin{equation} \label{eq:def. dM/dt}
\diff{M}{t}=\Delta\int_S\d{\Omega}\vac[ren]{\hat{T}_{tr}}{\Psi}
\end{equation}
where the surface $S$ can be any surface of constant $t$ and $r$. 
A non-zero value for the luminosity when the field is in the past Boulware state in a background
possessing an ergosphere is a manifestation of the Starobinski\u{\i}-Unruh effect.
Analogously, a non-zero value for the luminosity when the field is in the past Unruh state in such a 
background is a manifestation of the Hawking radiation.
\catdraft{plus Starob-Unruh rad.? no, see sln. in ATENCIO in Section \ref{sec:U vac.}}
In the forthcoming the subindex $A$ refers to either $t$ or $\phi$ and the subindex $X$ to either $r$ or $\theta$.

In order to compare some spin-1 results with the corresponding spin-0 results we shall briefly outline the latter. 
Consequently, the present and following paragraphs only apply to the scalar case.
On the one hand, it can be proved that, for spin-0, it is $T_{A\theta}(x,x')=0$ for any points $x$ and $x'$.  
On the other hand, Frolov and Thorne ~\cite{ar:F&T'89} prove that $T_{AX}^{\text{div}}=0$ for spin-0
whenever the separation between the points $x$ and $x'$ is the particular choice that they both lie in the same two-dimensional surface
$\Sigma\equiv\{t,\phi\}$, i.e., $x=(t,r_*,\theta,\phi)$ and $x'=(t'=t,r'_*,\theta',\phi'=\phi)$. 
Combining both results it follows that $\vac[ren]{\hat{T}_{A\theta}}{\Psi} =0$ for spin-0, where $\ket{\Psi}$ is any state
among $\ket{B^\pm}$, $\ket{U^-}$, $\ket{CCH^-}$ or $\ket{FT}$. 

Equations (\ref{eq:integrated conserv.eqs., T_tr,Ttphi}) give
\begin{equation} \label{eq:T_Ar for s=0}
\left.
\begin{aligned}
T_{tr}&=\frac{K(\theta)}{\Delta} \\
T_{\phi r}&=\frac{L(\theta)}{\Delta} 
\end{aligned}
\right\}  
\quad \text{if} \quad T_{A\theta}=0
\end{equation}
so that, in particular, equations (\ref{eq:T_Ar for s=0}) apply to the RSET when the scalar field is in any of the above-mentioned states.
It may indeed be calculated directly from the expression for the spin-0 stress-energy tensor 
that all the radial dependence of $\Delta {}_{lm\omega}T_{tr}$ can be expressed as a radial wronskian. 
It can also be checked that ${}_{lm\omega}T_{tr}^{\text{in}}=-{}_{lm\omega}T_{tr}^{\text{up}}$ for spin-0 so that 
the only contribution to the luminosity in the past Boulware vacuum comes from the superradiant modes:
\begin{equation}
\Delta \vac[ren]{\hat{T}_{tr}}{B^-}=-2\sum_{l=1}^{\infty}\sum_{m=1}^{l}\int_0^{m\Omega_+}\d{\omega}\Delta {}_{lm\omega}T_{tr}^{\text{up}} \qquad \text{for} \quad s=0
\end{equation}  

\catdraft{1) quin sentit te el $T_{AX}^{\text{div}}$ de dalt, calculat prenent $x=x'$ (idem $\Delta {}_{lm\omega}T_{tr}$)?
o es que val $\forall x,x'$? 
2) wrong sign in $\vac[ren]{T_{tr}}{B-}$?}

It is immediately apparent that for the spin-1 case the task to prove analytically 
whether $\Delta T_{tr}$ is constant in $r$ or not is much more arduous than for spin-0. 
In the expression for ${}_{lm\omega}T_{tr}$
both terms with ${}_{-1}S_{lm\omega}^2$ and other terms with ${}_{+1}S_{lm\omega}^2$ appear in it. 
As a matter of fact, when evaluated at the axis of symmetry $\theta=0$ or $\pi$, only one term of the first type and one
of the second type appear. 
Furthermore, neither of these two terms is constant in $r$.
It is therefore apparent that if we wish to prove that $\Delta T_{tr}$ is constant in $r$, or otherwise, 
we must then somehow relate ${}_{-1}S_{lm\omega}^2$ to ${}_{+1}S_{lm\omega}^2$.
It follows from the symmetries (\ref{eq: S symms}) that we can only relate one spherical function to the other at the same 
point by applying the transformation $(m,\omega)\to (-m,-\omega)$ to one
of them. The change in sign of $m$ can be overturned due to the symmetric sum in $m$ in the Fourier sums 
(\ref{eq:stress tensor for s=1 on all vac.}). The change in sign of $\omega$, however, is a problem when trying to relate a term with 
${}_{-1}S_{lm\omega}^2$ to a term with ${}_{+1}S_{lm\omega}^2$
due to the non-symmetric nature under $(m,\omega)\to (-m,-\omega)$ of the integrals over $\omega$ or $\tilde{\omega}$ 
for all states involved in (\ref{eq:stress tensor for s=1 on all vac.}). 
\ddraft{maybe text above is unclear?: so far done for stress tensor in gral., not for expectation val. in those states}
The use of (\ref{eq:Ss as func. of S_s,LS_s}) does not help
either since it introduces an undesirable derivative in $\theta$, 
which could not be cancelled out.

We encounter a similar problem when trying to prove
whether $T_{t\theta}$ is zero or not. Since we do not know whether it is zero or not we cannot
see either  from (\ref{eq:integrated conserv.eqs., T_tr}) that $\Delta T_{tr}$ is constant in $r$, as we
did for spin-0.

In the case $a=0$, since the spin-weighted spherical harmonics do not depend on $\omega$, we only need
a change in the sign of $m$ to relate the two types of terms in $\Delta {}_{lm\omega}T_{tr}$. 
Indeed, use of (\ref{eq:eq.B6J,McL,Ott'95}) allows us
to prove that $\sum_m \Delta {}_{lm\omega}T_{tr}$ is constant in $r$ and that 
$\sum_m {}_{lm\omega}T_{tr}^{\text{up}}=-\sum_m {}_{lm\omega}T_{tr}^{\text{in}}$ in the Schwarzschild background.   

The solution to this deadlock for the spin-1 case in the Kerr background consists in integrating over the solid angle. 
This allows us to relate a term with
$\int\d{\Omega}{}_{-1}S_{lm\omega}^2$ to a term with $\int\d{\Omega}{}_{+1}S_{lm\omega}^2$, 
when both types of terms appear in $\int\d{\Omega}\Delta {}_{lm\omega}T_{tr}$ .
This is in accord with the fact that if we integrate the conservation
equation (\ref{eq:integrated conserv.eqs., T_tr}) over the solid angle we immediately obtain that
\begin{equation}
\int\d{\Omega}\Delta T_{tr}=\int\d{\Omega} K(\theta)=const.
\end{equation}
Indeed, we analytically calculated $\int\d{\Omega} \Delta \vac{\hat{T}_{tr}}{U^-}$ in the above manner and
found that it is constant and in agreement with Page's ~\cite{ar:PageII'76} expression. Since the calculations were
not immediate, we will give here a brief outline.  
For each term in (\ref{eq:stress tensor, spin 1}) for ${}_{lm\omega}T_{tr}$ that contains ${}_{lm\omega}\phi_0$ we 
choose to use one particular expression for this NP scalar out of the two in (\ref{eq:phi0(ch)}). 
The expression we choose for ${}_{lm\omega}\phi_0$ in each term is the one that uses the same ${}_{h}S_{lm\omega}$ 
and different ${}_{h}R_{lm\omega}$ from those, as given by (\ref{eq:phi_0/2(in/up)}), appearing in the other NP scalar in that term. 
The reason for this choice is two-fold. One, so that we can later easily identify the wronskian expressions (\ref{eq:gral. radial wronsk.,R_+/-1}). 
Secondly, so that we can directly compare these terms containing ${}_{lm\omega}\phi_0$ with the ones containing 
$\left|{}_{lm\omega}\phi_{\pm1}\right|^2$, since then both types of terms involve ${}_{\mp1}S_{lm\omega}^2$. 
For each term in ${}_{lm\omega}T_{tr}$ we then factor out parts which are functions of $r$ only. We 
identify and group terms such that their factorized parts containing $\theta$
are equal (bar a sign) after being integrated over the solid angle 
and the symmetry (\ref{eq:S symm.->pi-t,-s}) is used. The result after also including the
complex conjugate part is that  all terms can be grouped together with common factor either 
$\left[{}_{-1}R^*_{lm\omega}\mathcal{D}_0^{\dagger}\left(\Delta{}_{+1}R_{lm\omega}\right)-\Delta{}_{+1}R_{lm\omega}\mathcal{D}_0^{\dagger}{}_{-1}R^*_{lm\omega}\right]+c.c.$, 
which is zero from (\ref{eq:gral. radial wronsk.,R_+/-1}), or else 
$\left[{}_{-1}R^*_{lm\omega}\mathcal{D}_0^{\dagger}\left(\Delta{}_{+1}R_{lm\omega}\right)-\Delta{}_{+1}R_{lm\omega}\mathcal{D}_0^{\dagger}{}_{-1}R^*_{lm\omega}\right]-c.c.$.
The factor multiplying the latter can be simplified to eventually yield the desired result:
\begin{equation} \label{eq:DeltaTtr integrated over solid angle}
\int\d{\Omega} \Delta {}_{lm\omega}T_{tr}^{\text{up}}=-\int\d{\Omega} \Delta {}_{lm\omega}T_{tr}^{\text{in}}=\frac{-1}{4\pi}\omega\mathbb{T}_{lm\omega}
\end{equation}
where we have included the constants of normalization (\ref{eq:normalization consts.}).

We can now give simple expressions for the luminosity when the electromagnetic field is in the past Boulware state and in the past Unruh state:
\begin{subequations} \label{eq:dM/dt for B-,U-}
\begin{align}
\left.\diff{M}{t}\right|_{B^-}&=
\frac{1}{2\pi}\sum_{l=1}^{\infty}\sum_{m=1}^{+l}\sum_{P=\pm1}\int_0^{m\Omega_+}\d{\omega}\omega\mathbb{T}_{lm\omega} \label{eq:dM/dt for B-}
\\
\left.\diff{M}{t}\right|_{U^-}&=
\frac{1}{2\pi}\sum_{l=1}^{\infty}\sum_{m=-l}^{+l}\sum_{P=\pm1}\int_0^{\infty}\d{\omega}\frac{\omega\mathbb{T}_{lm\omega}}{e^{2\pi\tilde{\omega}/\kappa}-1}
\label{eq:dM/dt for U-}
\end{align}
\end{subequations}
The former corresponds to the Starobinski\u{\i}-Unruh radiation and the latter to the Hawking radiation.
Since only superradiant modes are being included in the Starobinski\u{\i}-Unruh radiation (\ref{eq:dM/dt for B-}) and 
the transmission coefficient $\mathbb{T}_{lm\omega}$ is negative for these modes, there is a constant outflow of energy from the black hole
when the field is in the past Boulware state. 

\draft{wrong $-2$ in (\ref{eq:dM/dt for U-}) and wrong $2$ in (\ref{eq:dM/dt for B-})?! have arbitrarily changed both factors!!}

We numerically evaluated (\ref{eq:dM/dt for B-,U-}) for the case $Q=0$ and $a=0.95M$.
The results, compared against values in the literature are:
\begin{subequations} \label{eq:dM/dt for B- for a=0.95,M=1}
\begin{align}
M^2\left.\diff{M}{t}\right|_{B^-}&=-4.750*10^{-4}    && (\text{spin-1})          \label{eq:num. dM/dt in B-, s=1}\\
M^2\left.\diff{M}{t}\right|_{B^-}&=-5.01*10^{-5}      && (\text{spin-0, Duffy})   \label{eq:Gav. num. dM/dt in B-, s=0} 
\end{align}
\end{subequations}
in the past Boulware state, and
\begin{subequations} \label{eq:dM/dt for U- for a=0.95,M=1}
\begin{align}
M^2\left.\diff{M}{t}\right|_{U^-}&=-1.1714*10^{-3}    && (\text{spin-1})                \label{eq:num. dM/dt in U-, s=1} \\
M^2\left.\diff{M}{t}\right|_{U^-}&=-1.18*10^{-3}      && (\text{spin-1, Page})   \label{eq:Page num. dM/dt in U-, s=1} 
\end{align}
\end{subequations}
in the past Unruh state.
The value (\ref{eq:Gav. num. dM/dt in B-, s=0}) for the scalar field is calculated by Duffy ~\cite{th:GavPhD} and 
we have calculated (\ref{eq:Page num. dM/dt in U-, s=1}) from splining Page's ~\cite{ar:PageII'76} numerical results. Both
of them have also been calculated for $Q=0$ and $a=0.95M$.

\catdraft{encara no entenc les unitats de (\ref{eq:dM/dt for B- for a=0.95,M=1}) i (\ref{eq:dM/dt for U- for a=0.95,M=1});
check vals. against p.319DeWitt; see p.179McLPhD and p.3264,fig.1Page'76II?!!;val. (\ref{eq:num. dM/dt in U-, s=1}) includes factor $2$ correction}

The above results for the expectation value of the stress-energy tensor for a spin-1 field have been obtained using
CCH's expressions (\ref{eq:stress tensor for s=1 on all vac.}).
We now investigate what effect it has in these results the use of the correction 
(\ref{eq:corrected stress tensor for s=1 on B-,U-}) to CCH's expressions.
The difficulty we encountered above when trying to see whether $T_{t\theta}$ is zero and
whether $\Delta T_{tr}$ is constant in $r$ does not exist when calculating the expectation value of 
these components in the past Boulware and past Unruh states using expressions 
(\ref{eq:corrected stress tensor for s=1 on B-,U-}). Indeed, due to the symmetry 
(\ref{eq:S symm.->pi-t,-s}) we can now relate terms in each of these components that contain
${}_{-1}S_{lm\omega}^2+\mathcal{P}{}_{-1}S_{lm\omega}^2$ to terms that contain ${}_{+1}S_{lm\omega}^2+\mathcal{P}{}_{+1}S_{lm\omega}^2$.
The calculation of $\Delta \left({}_{lm\omega}T_{tr}+\mathcal{P}{}_{lm\omega}T_{tr}\right)$ follows through in a very
similar manner to the calculation of $\int\d{\Omega} \Delta {}_{lm\omega}T_{tr}$ described above and the result is:
\begin{equation} \label{eq:Delta(Ttr+Ttr(t->pi-t))}
\begin{aligned}
&\Delta \left({}_{lm\omega}T_{tr}^{\text{up}}+\mathcal{P}{}_{lm\omega}T_{tr}^{\text{up}}\right)=
\\&=
\frac{\mathbb{T}_{lm\omega}}{4\pi^2\Sigma}\Big\{-\omega\Sigma\left({}_{-1}S_{lm\omega}^2+{}_{+1}S_{lm\omega}^2\right)+ 
\frac{a^3\cos\theta\sin^2\theta}{\Sigma}\left({}_{-1}S_{lm\omega}^2-{}_{+1}S_{lm\omega}^2\right)+
\\ &+
a\sin\theta\left({}_{-1}S_{lm\omega}\partial_{\theta}{}_{-1}S_{lm\omega}-{}_{+1}S_{lm\omega}\partial_{\theta}{}_{+1}S_{lm\omega}\right)\Big\}
\end{aligned}
\end{equation}
The corresponding result for the `in' modes is equal to (\ref{eq:Delta(Ttr+Ttr(t->pi-t))}) with a change of sign, by virtue of
(\ref{eq:val. of trans. coeff.}) and the property (\ref{eq: wronskian in=-up}).

Even if we used equation (\ref{eq:LSs as func. of Ss,S_s}) to rid of the derivatives in (\ref{eq:Delta(Ttr+Ttr(t->pi-t))}), 
we would not be able to express its last term in terms of ${}_{-1}S_{lm\omega}^2$ and ${}_{+1}S_{lm\omega}^2$ only. 
It follows that $\Delta \left({}_{lm\omega}T_{tr}+\mathcal{P}{}_{lm\omega}T_{tr}\right)$ is not
constant in $r$ and therefore neither is $\Delta\vac[ren]{\hat{T}_{tr}}{B^-}$ nor $\Delta\vac[ren]{\hat{T}_{tr}}{U^-}$. 
Similarly, a calculation of $\left({}_{lm\omega}T_{t\theta}+\mathcal{P}{}_{lm\omega}T_{t\theta}\right)$
shows that it is not zero and therefore neither $\vac[ren]{\hat{T}_{t\theta}}{B^-}$ nor $\vac[ren]{\hat{T}_{t\theta}}{U^-}$
are zero. 
However, since $\int\d{\Omega} \mathcal{P}f(\theta)=\int\d{\Omega} f(\theta)$, the results obtained above 
involving integration over the solid angle, i.e., equations (\ref{eq:dM/dt for B-,U-})--(\ref{eq:dM/dt for U- for a=0.95,M=1}),
remain unaltered by the correction (\ref{eq:corrected stress tensor for s=1 on B-,U-}).   

\begin{figure}[p]
\rotatebox{90}
\centering
\includegraphics*[width=70mm,angle=270]{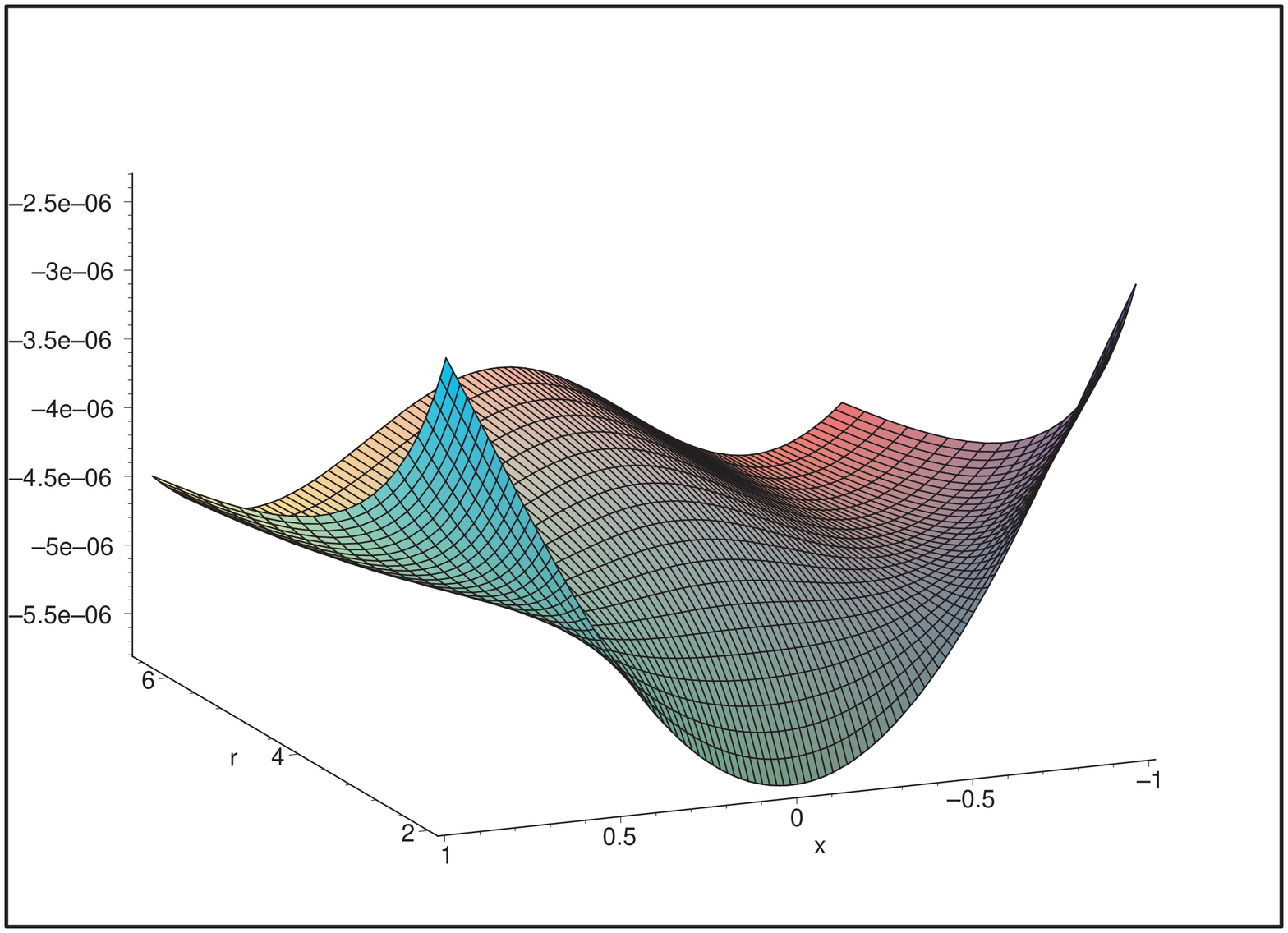} 
\caption{$\frac{1}{4\pi}\Delta\vac{\hat{T}_{tr}}{U^--B^-}$}    \label{fig:deltaSymmTtr_u_b}
\includegraphics*[width=70mm,angle=270]{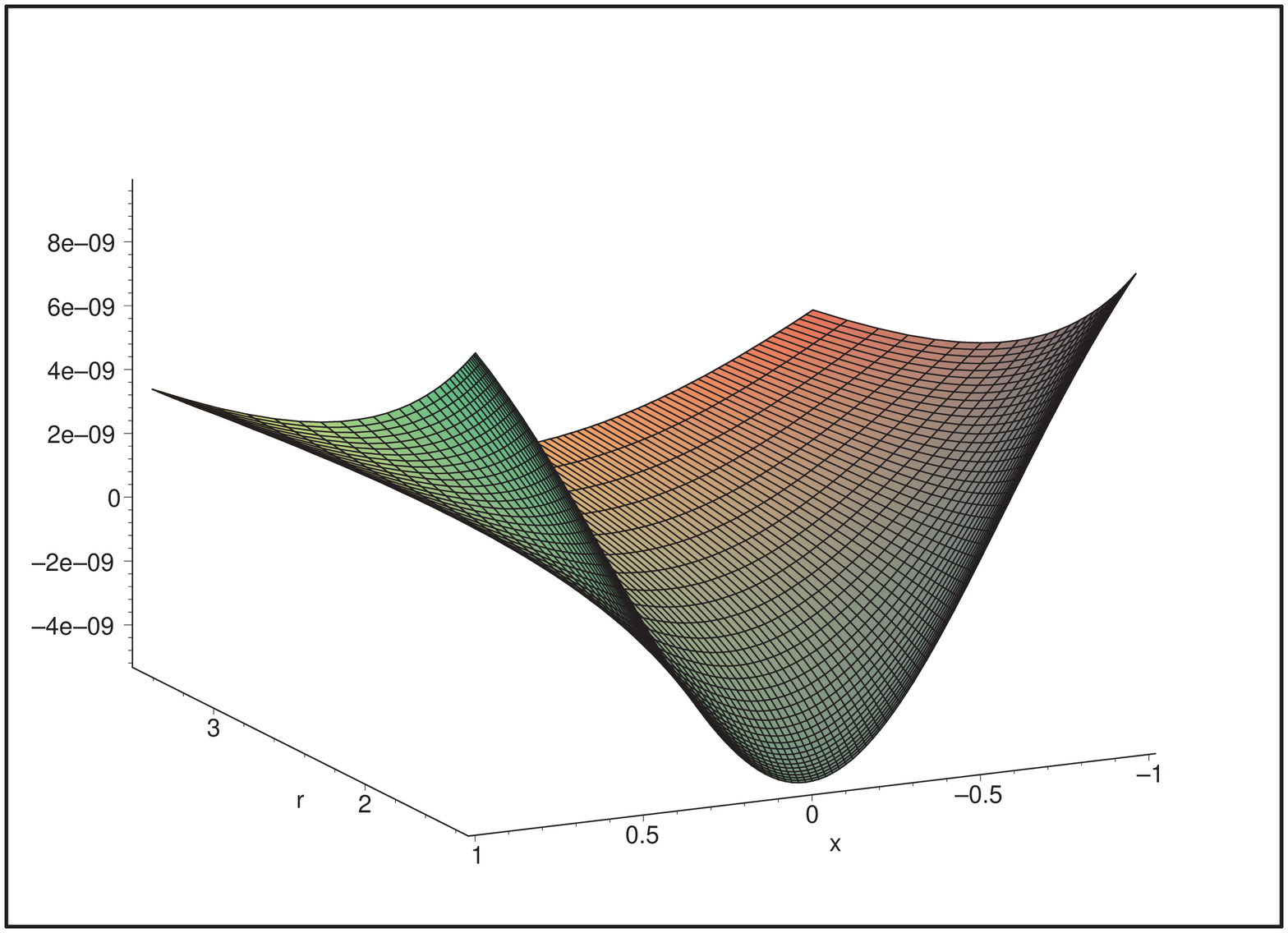} 
\caption{$\frac{1}{4\pi}\Delta\vac{\hat{T}_{tr}}{CCH^--U^-}$}    \label{fig:deltaSymmTtr_cch_u_past}
\end{figure}

\begin{figure}[p]
\rotatebox{90}
\centering
\includegraphics*[width=70mm,angle=270]{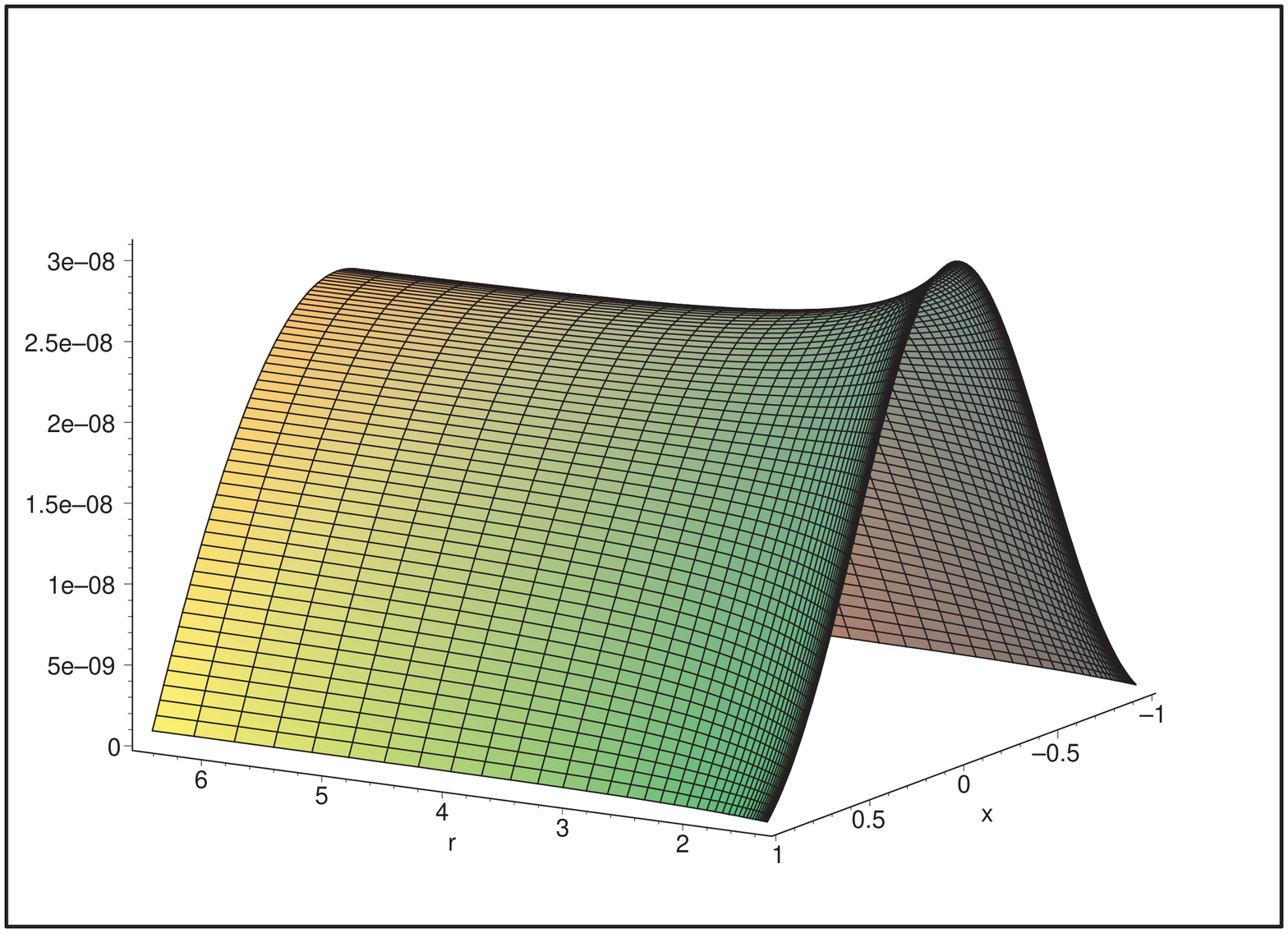} 
\caption{$\frac{1}{4\pi}\Delta\vac{\hat{T}_{r\phi}}{CCH^--U^-}$}    \label{fig:deltaSymmTrphi_cch_u_past}
\includegraphics*[width=70mm,angle=270]{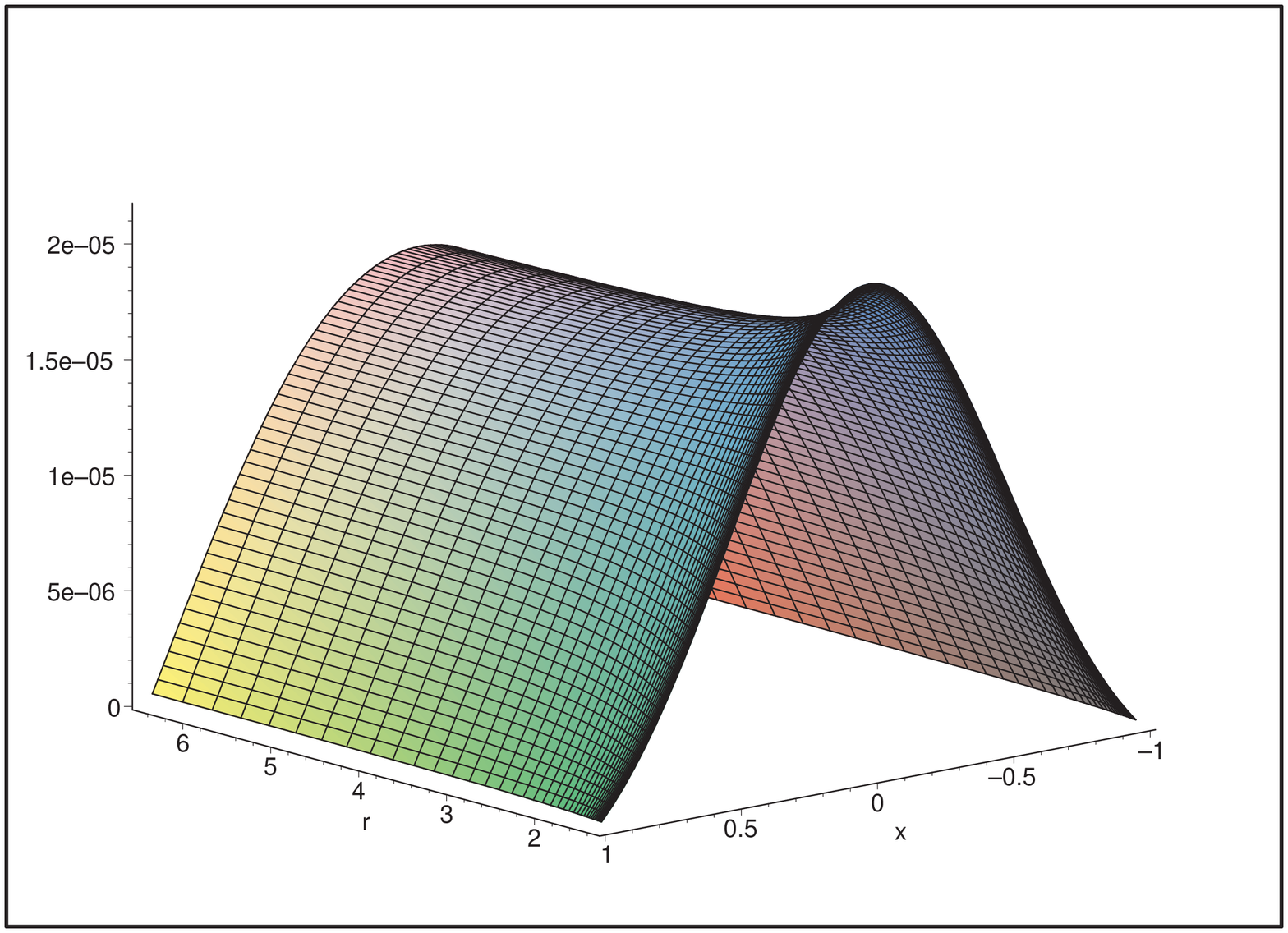} 
\caption{$\frac{1}{4\pi}\Delta\vac{\hat{T}_{r\phi}}{U^--B^-}$}    \label{fig:deltaSymmTrphi_u_b}
\end{figure}

\begin{figure}[p]
\rotatebox{90}
\centering
\includegraphics*[width=70mm,angle=270]{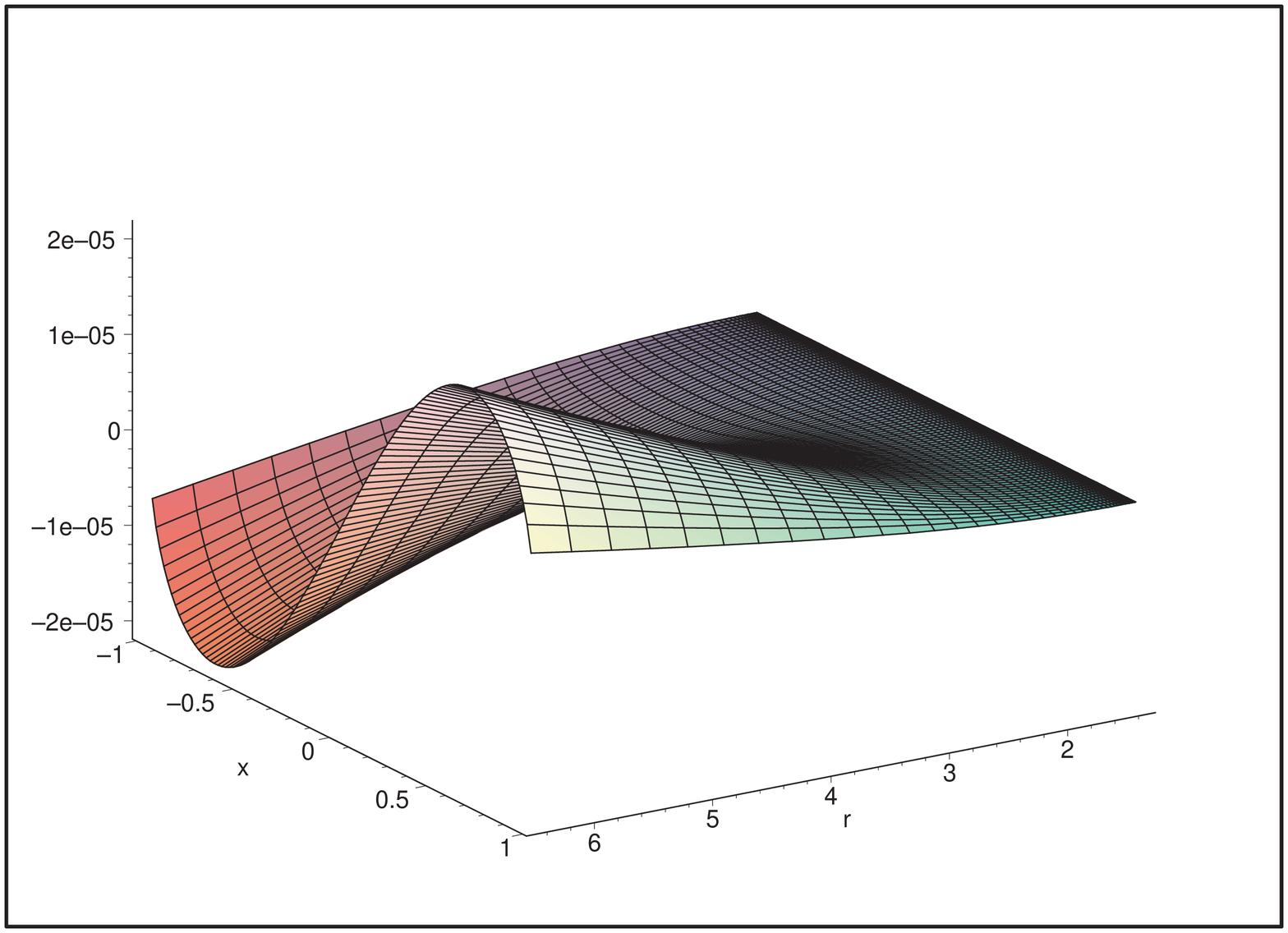} 
\caption{$\frac{1}{4\pi}\Delta^2\vac{\hat{T}_{t\theta}}{U^--B^-}$.
Note that the viewing angle in this and the following graphs is different from that of the previous graphs 
in order to make more visible the region far from the horizon.}    \label{fig:delta2SymmTttheta_u_b}
\includegraphics*[width=70mm,angle=270]{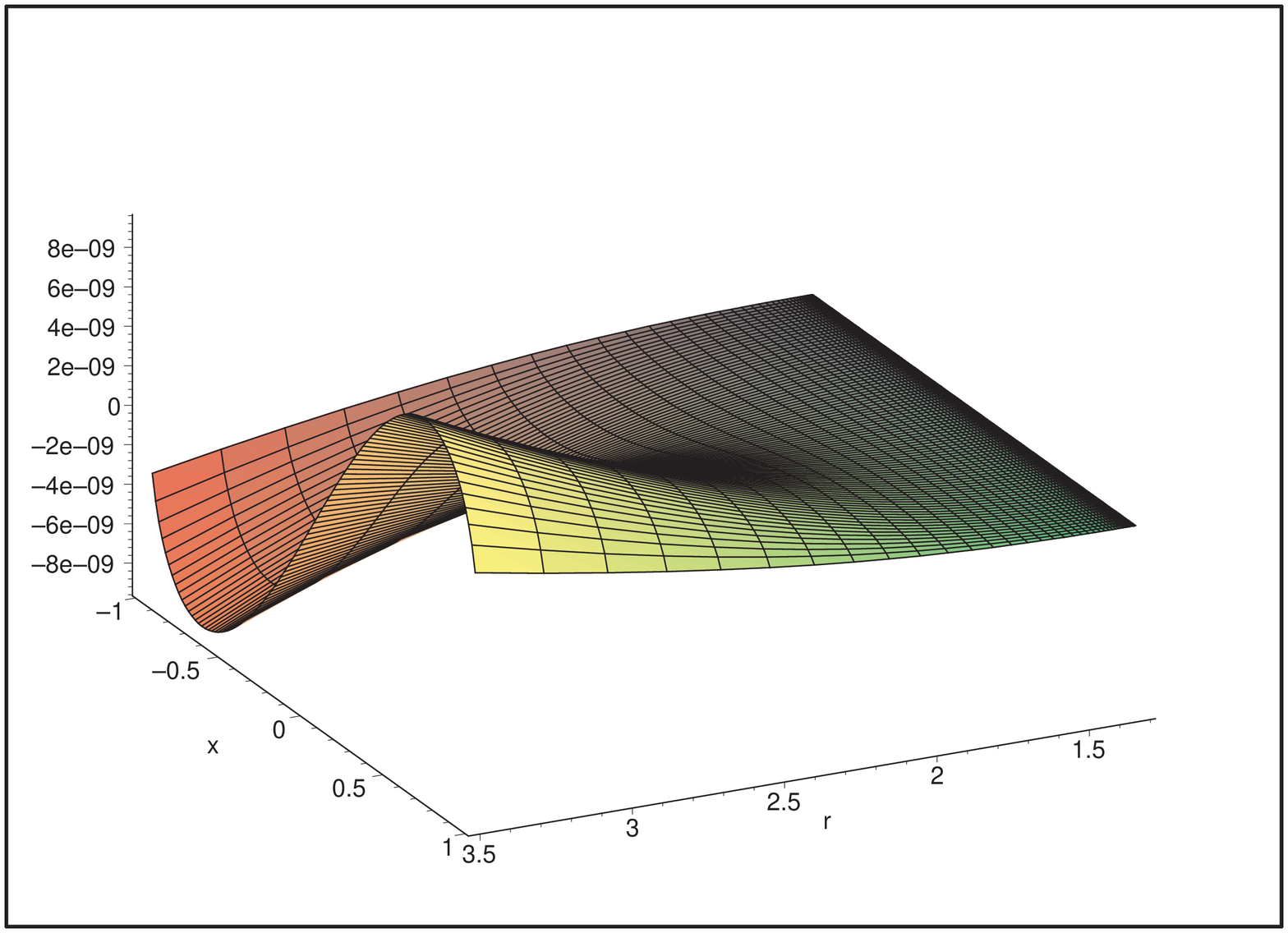} 
\caption{$\frac{1}{4\pi}\Delta^2\vac{\hat{T}_{t\theta}}{CCH^--U^-}$}    \label{fig:delta2SymmTttheta_cch_u_past}
\end{figure}

\begin{figure}[p]
\rotatebox{90}
\centering
\includegraphics*[width=70mm,angle=270]{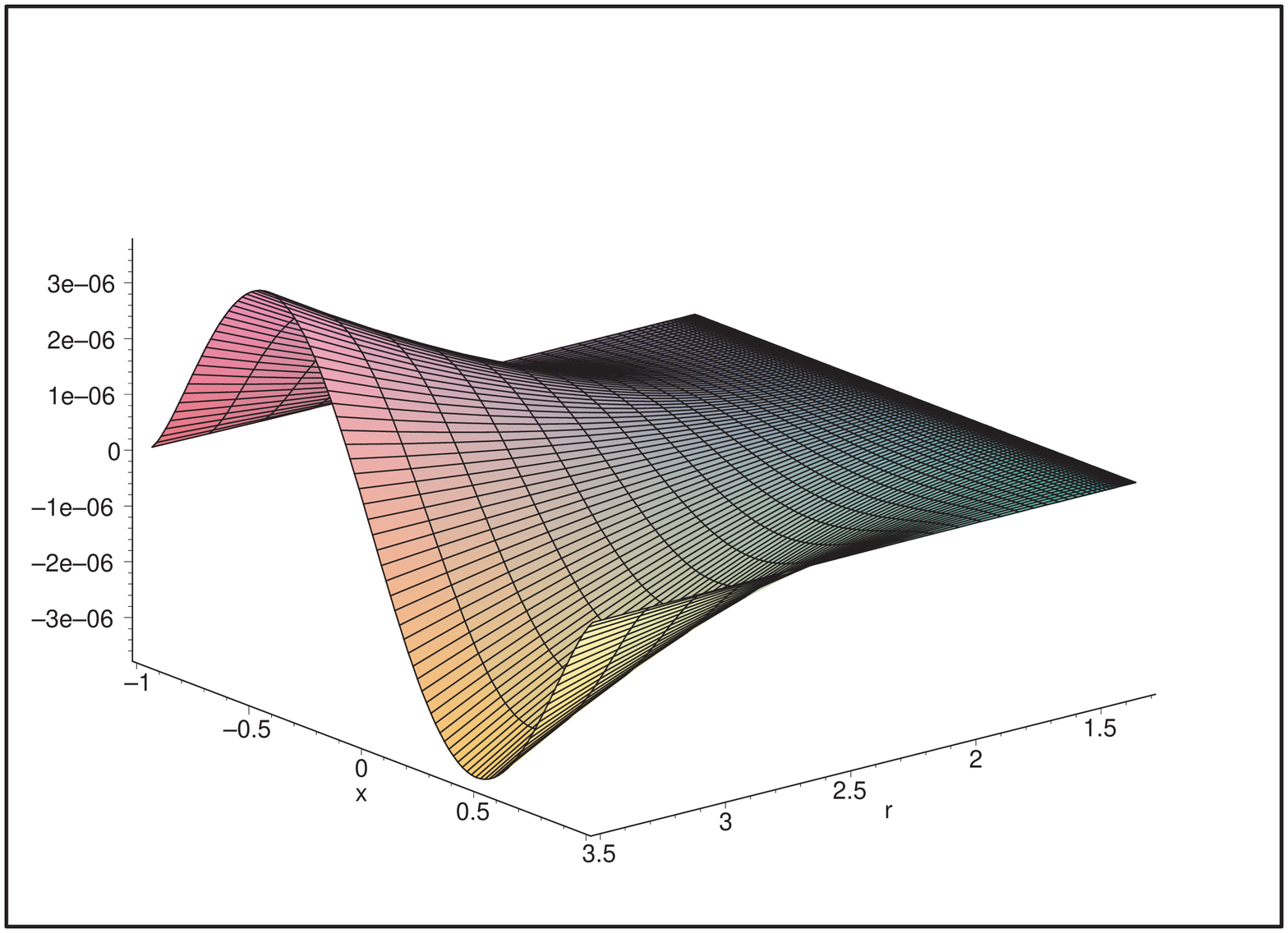} 
\caption{$\frac{1}{4\pi}\Delta^2\vac{\hat{T}_{\theta\phi}}{U^--B^-}$}    \label{fig:delta2SymmTthetaphi_u_b}
\includegraphics*[width=70mm,angle=270]{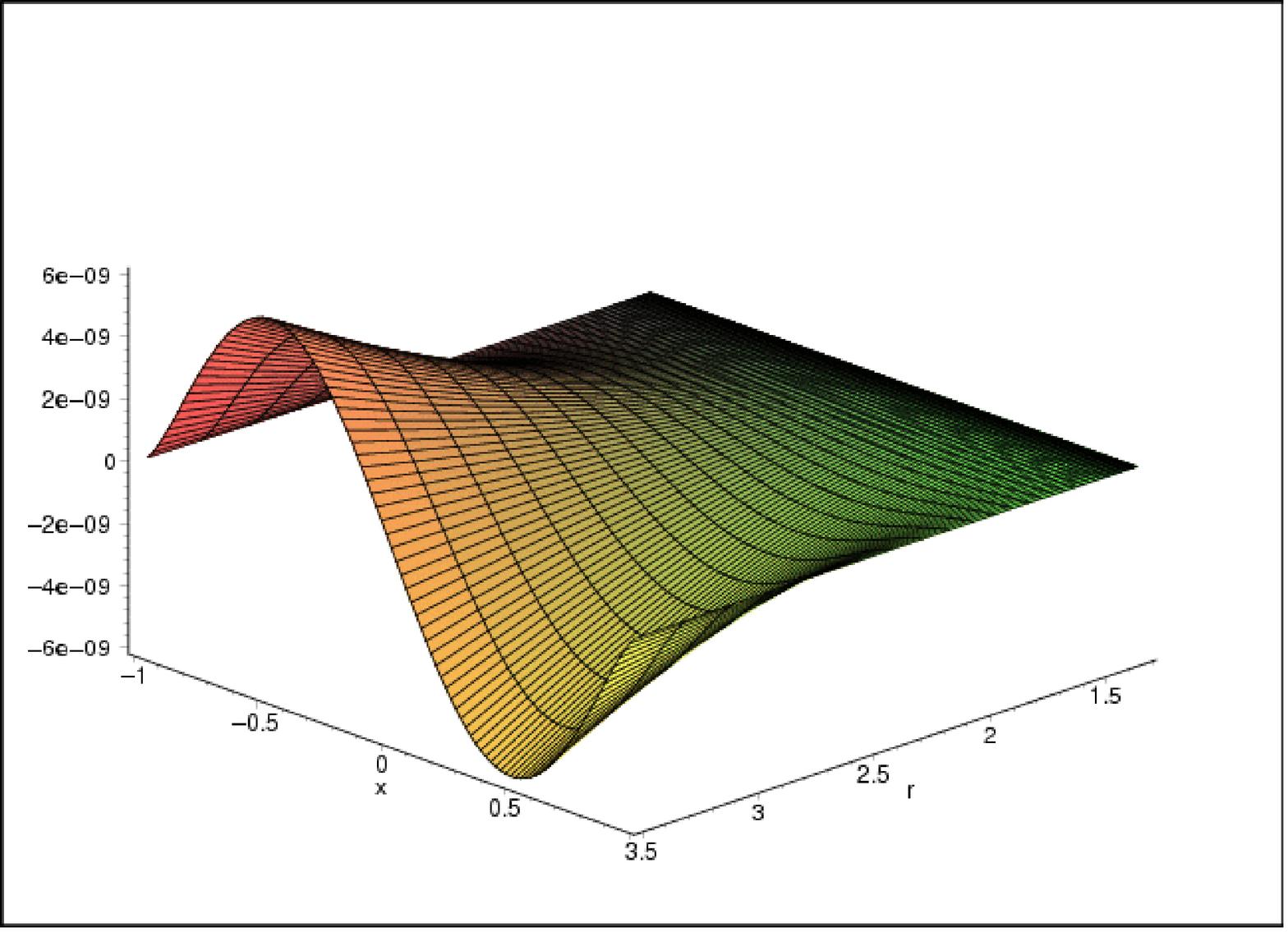} 
\caption{$\frac{1}{4\pi}\Delta^2\vac{\hat{T}_{\theta\phi}}{CCH^--U^-}$}    \label{fig:delta2SymmTthetaphi_cch_u_past}
\end{figure}

Indeed, Graphs \ref{fig:deltaSymmTtr_u_b}--\ref{fig:delta2SymmTthetaphi_cch_u_past} numerically corroborate the above 
conclusions.
Graphs \ref{fig:deltaSymmTtr_u_b} and \ref{fig:deltaSymmTtr_cch_u_past} show that neither 
$\Delta\vac{\hat{T}_{tr}}{U^--B^-}$ nor $\Delta\vac{\hat{T}_{tr}}{CCH^--U^-}$ are constant in $r$.
Graphs \ref{fig:delta2SymmTttheta_u_b}--\ref{fig:delta2SymmTthetaphi_cch_u_past} show that neither 
$\vac[ren]{\hat{T}_{A\theta}}{U^-B^-}$ nor $\vac[ren]{\hat{T}_{A\theta}}{CCH^--U^-}$ are zero.
Graphs \ref{fig:deltaSymmTrphi_cch_u_past}--\ref{fig:deltaSymmTrphi_u_b}, however, seem to indicate that
both $\Delta\vac{\hat{T}_{r\phi}}{CCH^--U^-}$ and $\Delta\vac{\hat{T}_{r\phi}}{U^--B^-}$ might actually be constant in $r$.
\draft{1) this is strange since $\vac[ren]{\hat{T}_{\phi\theta}}{U^-B^-}$ and $\vac[ren]{\hat{T}_{\phi\theta}}{CCH^--U^-}$ are clearly non-zero??!,
2) plot $\phi\theta$ graphs starting from $r$ further from $r_+$ to show constanct in $r$ more clearly?}

\draft{1) (p.19Ttr) surface $\int\d{\Omega}$ is only for large $r$ but $\forall r$ it should be a diff. gral. one!
2) it's not Page's expression but Hawking's?,
3) wrong factor $-2$ throughout, 5) (\ref{eq:DeltaTtr integrated over solid angle}) only applies to `up',
4) should these graphs be placed together with graphs in next section instead?,
5) include $\vac[ren]{\hat{T}_{A\theta}}{CCH^---B^-}$ graphs?}


\section{RSET close to the horizon in the Boulware vacuum} \label{sec:RRO}

Candelas and Deutsch ~\cite{ar:Cand&Deutsch'77} consider flat space-time in the presence of an accelerating barrier
with acceleration $a_B^{-1}$. They then calculate the spin-1 RSET in the tetrad of an accelerating
observer RO with local acceleration $\xi^{-1}$. In the limit $\xi/a_B\rightarrow \infty$ 
the vacuum state above the accelerating mirror approximates the Fulling vacuum $\ket{F}$. 
The result is
\begin{equation} \label{eq:stress tensor in Fulling in RO tetrad}
\begin{aligned}
\vac[ren]{T^{\bar{\mu}}{}_{\bar{\nu}}}{F}
\sim& -\frac{1}{\pi^2\xi^4}\int_0^{\infty}\d{x}
\frac{x^3+x}{e^{2\pi x}-1}\text{diag}\left(-1,\frac{1}{3},\frac{1}{3},\frac{1}{3}\right)= \\
&=\frac{-11}{240\pi^2\xi^4}\text{diag}\left(-1,\frac{1}{3},\frac{1}{3},\frac{1}{3}\right)  \qquad (\xi/a_B\rightarrow \infty)
\end{aligned}
\end{equation}
where the bars on the indices indicate RO tetrad.  
\ddraft{in Section \ref{sec:H-H vac.} the comparison with flat space was due to Unruh dealing with flat space 
without a mirror, slightly different to the comparison with flat space done here?...}
Expression (\ref{eq:stress tensor in Fulling in RO tetrad}) is equivalent to minus the stress-energy tensor for thermal radiation
at a temperature of $(2\pi\xi)^{-1}$.
We saw in Section \ref{sec:H-H vac.} that in Schwarzschild space-time, analogously to (\ref{eq:stress tensor in Fulling in RO tetrad}) in flat space,
the RSET close to the horizon when the field is in the Boulware vacuum diverges like minus the stress tensor of black body radiation at the black hole temperature. 
It is therefore reasonable to expect that if there existed a
state in Kerr with the defining features that the Boulware vacuum possesses in Schwarzschild, then
the RSET close to the horizon
when the field were in this vacuum, would 
diverge like minus the stress tensor of black body radiation at the black hole temperature rotating with the horizon.
The past Boulware vacuum is not invariant under $(t,\phi)$ reversal because of the existence of the Starobinski\u{\i}-Unruh radiation.
However, the stress tensor components $tr$ and $r\phi$, which correspond to the Starobinski\u{\i}-Unruh radiation, are expected 
(from Section \ref{sec:luminosity}) to have a divergence of one lower leading order than that of the diagonal components as the horizon is approached.
It is with this understanding that we say that a state is isotropic at the horizon and that, in particular, the past Boulware vacuum might be isotropic.
It is obvious that to next order in $\Delta$ the past Boulware vacuum cannot be isotropic, 
\ddraft{why not? it might not be in Boy-Lindq coords. but it might be in some other frame?}
but $\vac[ren]{\hat{T}^{\mu}{}_{\nu}}{CCH^--B^-}$
might be since $\ket{CCH^-}$ is not invariant under $(t,\phi)$ reversal either. 

CCH claim that the RSET of the electromagnetic field in the past Boulware vacuum close to the horizon 
differs from that of minus the stress-energy tensor of a thermal distribution rotating at the angular 
velocity of a Carter observer by a factor which is a function of $\theta$.
In the present section, we will show that CCH's result is due to a flawed assumption in the asymptotic behaviour of the SWSH. 
We will show this by re-calculating their result using the assumptions we believe they used. 
The numerical results back up the fact that the mentioned RSET is (minus) thermal at the horizon.

We also saw in Section \ref{sec:H-H vac.} that Frolov and Thorne claim that close to the horizon ZAMOs measure a thermal 
stress tensor which is rigidly rotating with the horizon when the field is in the $\ket{FT}$ state.
That is, they argue that $\vac[ren]{\hat{T}^{\mu}{}_{\nu}}{FT-B}$ (where $\ket{B}$ is an unspecified Boulware-type state)
is thermal close to the horizon, and isotropic in the frame of a RRO.
Duffy, in turn, shows that close to the horizon RROs measure a thermal state which is rigidly
rotating with the horizon when the field is in the $\ket{H_{\mathcal{M}}}$ state in the Kerr space-time 
modified with the introduction of a mirror. 
That is to say, close to the horizon $\vac[ren]{\hat{T}^{\mu}{}_{\nu}}{H_{\mathcal{M}}-B_{\mathcal{M}}}$ is 
thermal and isotropic in the frame of a RRO.
Finally, as mentioned above, CCH claim that $\vac[ren]{\hat{T}^{\mu}{}_{\nu}}{CCH^--B^-}$ is, bar a factor, thermal at the horizon and
isotropic in the Carter tetrad. 
Of course, the angular velocity at the horizon of a Carter observer, a RRO and a ZAMO is $\omega=\Omega_+$ for them all,
so that CCH's result
does not actually distinguish between these observers.

Ottewill and Winstanley ~\cite{ar:Ott&Winst'00Lett} have proved that if a certain stress-energy tensor is thermal and rigidly-rotating with
the horizon everywhere, then it is divergent on the speed-of-light surface in the Boyer-Lindquist co-ordinates,
which are regular on this surface. \ddraft{is this obvious and thus not actually `discovered' by ~\cite{ar:Ott&Winst'00Lett}?}
This implies that if $\vac[ren]{\hat{T}^{\mu}{}_{\nu}}{CCH^--B^-}$ were thermal and
rigidly-rotating with the horizon everywhere then the state $\ket{CCH^-}$ would have to be irregular on the speed-of-light surface.
In the present section we will numerically investigate the rate of rotation of the thermal distribution in question.


\draft{1) so everybody agrees that it's RRO to 1st order then (even F\&T)?!, 2) see p.1064F\&Z'84!!!!!!}

The stress-energy tensor of a spin-1 thermal distribution at the Hawking temperature rigidly rotating with the horizon is given by
\begin{equation} \label{eq:thermal RR stress tensor}
T^{\text{(th,RR)} \mu}{}_{\nu}=\frac{11T^4\pi^2}{45}\left[\delta^{\mu}_{\nu}-4\frac{\chi^{\mu}\chi_{\nu}}
{\chi^{\rho}\chi_{\rho}}\right]
\end{equation}
where
\begin{equation} \label{eq:def. T}
T \equiv \frac{\kappa_+}{2\pi}\frac{1}{\sqrt{-\chi^{\rho}\chi_{\rho}}}
\end{equation}
is the local temperature. Note that this stress-energy tensor is obviously isotropic in the frame of a RRO, but it is not
in the rigidly-rotating co-ordinate system $\{t_+,r,\theta,\phi_+\}$, which is not adapted to a RRO.

\catdraft{check sign in (\ref{eq:thermal RR stress tensor}) and (\ref{eq:thermal RR stress tensor in RRO coords.})}
\draft{give thermal tetrad for general obs. $\vec{e}_{(t)}$ instead of only for RRO? (\ref{eq:thermal RR stress tensor in RRO coords.})
would actually be the same just that in comps. in gral. frame}

In primed co-ordinates, which are adapted to a RRO, the rigidly-rotating thermal stress tensor becomes
\begin{equation} \label{eq:thermal RR stress tensor in RRO coords.}
T^{\text{(th,RR)} \mu'}{}_{\nu'}=\frac{11(r_+-r_-)^4}{2^8\cdot 3^2\cdot 5\pi^2}\frac{1}{\Delta^2\Sigma^2}\text{diag}\left(-1,\frac{1}{3},\frac{1}{3},\frac{1}{3}\right)
\end{equation}
in the Kerr space-time.

CCH calculate an expression for the RSET close to the horizon when the electromagnetic field is in the 
past Boulware state. They make the assumption that the RSET close to the horizon when the field 
is in this state is more irregular than when it is in the $\ket{CCH^-}$ state and therefore approximate
\begin{equation} \label{eq:RSET in B^- at horizon}
\vac[ren]{\hat{T}^{\mu}{}_{\nu}}{B^-}\sim\vac[ren]{\hat{T}^{\mu}{}_{\nu}}{B^-}-\vac[ren]{\hat{T}^{\mu}{}_{\nu}}{CCH^-}=
\vac{\hat{T}^{\mu}{}_{\nu}}{B^-}-\vac{\hat{T}^{\mu}{}_{\nu}}{CCH^-} \qquad (r\rightarrow r_+)
\end{equation}
They can then use their expressions (\ref{eq:stress tensor for s=1 on all vac.}), and it is clear that only the
`up' modes are involved in the calculation.
Their result, when the components of the stress tensor are put in the Carter orthonormal
tetrad (\ref{eq:def. Carter ortho. tetrad}) is:
\begin{equation}   \label{eq:CCH's RSET in B^- in Carter, for r->r_+}
\begin{aligned}
\vac[ren]{\hat{T}^{\hat{\mu}}{}_{\hat{\nu}}}{B^-} &\sim
-\frac{8M^3r_+}{\pi^2\Delta^2\Sigma}\int_0^{\infty}\d{\tilde{\omega}}
\frac{\tilde{\omega}\left(\tilde{\omega}^2+\kappa^2\right)}{e^{2\pi\tilde{\omega}/\kappa}-1}\text{diag}\left(-1,\frac{1}{3},\frac{1}{3},\frac{1}{3}\right)=\\
&=-\frac{1}{r_+}\frac{11(r_+-r_-)^4}{2^8\cdot 3^2\cdot 5\pi^2}\frac{1}{\Delta^2\Sigma (2Mr_+)}\text{diag}\left(-1,\frac{1}{3},\frac{1}{3},\frac{1}{3}\right)
\end{aligned}
\end{equation}
where the hats on the indices indicate adaptation to the Carter orthonormal tetrad.
This expression and the expected result, minus 
(\ref{eq:thermal RR stress tensor in RRO coords.}), differ in a factor of $r_+(2Mr_+)/\Sigma$.
\catdraft{should they coincide when one is rotating like a RRO and the other like a Carter observer (even though
at the horizon they are undistinguishable but the off-diagonal components depend on the next order terms, which is why $t_+\phi_+$-component
allows to distinguish between the 3 rotating observers)? yes, they should coincide because we're only comparing diag. comps. 
(the off-diag. comps. were ``eaten up'' by CCH)}
We proceed to reproduce CCH's expression to explain
this disagreement.

We believe that \ddraft{add that this was recently ratified by Candelas in person?} 
CCH followed Candelas ~\cite{ar:Candelas'80} method for spin-0 to obtain asymptotic expansions 
for the radial solutions for spin-1 close to the horizon. This is the method that we developed in Section 
\ref{sec:asympts. close to r_+}. Armed with the asymptotics of that section, we can proceed to calculate 
the different components of the stress-energy tensor. In order to do that,
we are first going to separately calculate the asymptotic expressions for the various terms that occur in 
the classical stress-energy tensor (\ref{eq:stress tensor, spin 1}).

As mentioned in Section \ref{sec:asympts. close to r_+}, for the asymptotic behaviour we are seeking here 
we can replace the spin-weighted spheroidal harmonics ${}_{\indhel}S_{lm}$ by the spin-weighted spherical 
harmonics ${}_{\indhel}Y_{lm}$. We can then make use of
(\ref{eq:eq.2aJ,McL,Ott'91}), which immediately leads to
\begin{equation} \label{eq:phi_iphi_j,i<>j,approx l->inf,r->rplus}
\sum_{m=-l}^l {}_{lm\omega}\phi_{\indhel}^{\text{up}}{}_{lm\omega}\phi_{h'}^{up *} 
\rightarrow 0 
\qquad (l\rightarrow +\infty, r \rightarrow r_+) 
\qquad \text{when } h\neq h'
\end{equation}

\draft{still to be proven that $\sum_{m=-l}^l {}_{0}Y_{lm}{}_{\pm1}Y_{lm}^*=0$!!! and clarify
that Y here is Z(a=0) rather than S(a=0)}

The asymptotic calculation of the term $\abs{{}_{lm\omega}\phi_{0}^{\text{up}}}^2$ requires a more careful treatment.
We observed in Section \ref{sec:asympts. close to r_+} that the large-$l$ modes dominate the Fourier series for the
`up' radial solution close to the horizon.
Using (\ref{eq:phi0(ch)}) and replacing $\sum_{l=0}^{\infty}$ with $\int_{0}^{\infty} \d{l}$ 
as we are only interested in the behaviour for $l\rightarrow +\infty$, we have
\begin{equation} \label{eq:phi1^2 approx l->inf,r->rplus;1st step}
\begin{aligned}
&\sum_{l,m,P}\abs{{}_{lm\omega}\phi_{0}^{\text{up}}}^2 \sim
\int_{0}^{\infty} \d{l}
\frac{|N^{\text{up}}_{+1}|^2}{2^2\Sigma^2}\left[
\abs{\left(\rho^{-1}\mathcal{D}_0^{\dagger}+1\right)(\Delta{}_{+1}R^{\text{up}}_{lm\omega})}^2 
\sum_{m=-l}^l \abs{\mathcal{L}^{\dagger}_{1} {}_{-1}Y_{lm}}^2 +
\right.
\\
& 
+a^2\sin^2\theta \abs{\mathcal{D}_0^{\dagger}(\Delta{}_{+1}R^{\text{up}}_{lm\omega})}^2 \sum_{m=-l}^l \abs{{}_{-1}Y_{lm}}^2+
\\ & \left.+
\text{terms with} \sum_{m=-l}^l\left({}_{0}Y_{lm}{}_{-1}Y_{lm}^*+{}_{-1}Y_{lm}{}_{0}Y_{lm}^*\right)
\right]  \qquad \quad (l\rightarrow +\infty, r \rightarrow r_+) 
\end{aligned}
\end{equation}

Using equations (\ref{eq:rln. between L and edth}), (\ref{eq:eq.B1bJ,McL,Ott'95}), (\ref{eq:eq.B6J,McL,Ott'95}) and 
 (\ref{eq:eq.2aJ,McL,Ott'91}) and the fact that of the two independent variables $\tilde{\omega}$ and $m$,
${}_{\indhel}R^{\text{up}}_{lm\omega}$ depends only on $\tilde{\omega}$ in the limit $(l\rightarrow +\infty, r \rightarrow r_+)$,
whereas ${}_{\indhel}Y_{lm}$ depends only on $m$, equation (\ref{eq:phi1^2 approx l->inf,r->rplus;1st step}) can 
be simplified to
\begin{equation} \label{eq:phi1^2 approx l->inf,r->rplus;2nd step}
\sum_{l,m,P}\abs{{}_{lm\omega}\phi_{0}^{\text{up}}}^2 \sim
\frac{1}{2^3\pi\Sigma}
\int_{0}^{\infty} \d{l}
l^3
|N^{\text{up}}_{+1}|^2 \abs{\mathcal{D}_0^{\dagger}(\Delta{}_{+1}R^{\text{up}}_{lm\omega})}^2
\qquad \quad  (l\rightarrow +\infty, r \rightarrow r_+) 
\end{equation}
where we have also used the fact that 
\begin{equation}
|N^{\text{up}}_{+1}|^2\mathcal{D}_0^{\dagger}(\Delta{}_{+1}R^{\text{up}}_{lm\omega}) \gg |N^{\text{up}}_{+1}|^2\Delta{}_{+1}R^{\text{up}}_{lm\omega}
\qquad \quad (l\rightarrow +\infty, r \rightarrow r_+) 
\end{equation}
\catdraft{hauria de ser $(l\rightarrow +\infty,r \rightarrow r_+,lx^{1/2}\ \text{finite})$ com a (\ref{eq:Ddagger Delta R1 `up' approx l->inf,r->rplus}) ,etc?}
as can be seen from 
(\ref{eq:R1 `up' approx l->inf,r->rplus;compact version}) and 
(\ref{eq:Ddagger Delta R1 `up' approx l->inf,r->rplus}).
We then substitute (\ref{eq:def.A_s}), (\ref{eq:def.D}), and 
(\ref{eq:Ddagger Delta R+1 `up' approx l->inf,r->rplus})
in the above equation and approximate ${}_1B_{lm\omega} \sim l^2$. The next integral, found in ~\cite{bk:GR}, is needed:
\begin{equation} \label{eq:eq.6.576(4)G&R,lambda=-3}
\int_{0}^{\infty} \d{l} l^3 K_{iq}^2(2lx^{1/2})=\frac{q^2(1+q^2)\abs{\Gamma(iq)}^2}{3\cdot 2^4x^2}
\end{equation}

We finally obtain 
\begin{equation} \label{eq:phi1^2 approx l->inf,r->rplus}
\begin{aligned}
\sum_{l,m,P}\abs{{}_{lm\omega}\phi_{0}^{\text{up}}}^2 \sim
\frac{Mr_+\tilde{\omega}\abs{\EuFrak{N}}^2}{6\pi^2\Sigma \Delta^2}
\qquad \quad  \text{($l\rightarrow +\infty$, $r \rightarrow r_+$)} 
\end{aligned}
\end{equation}

The other terms in the expression for the stress-energy tensor can be obtained in a similar manner,
but they are easier to calculate. We will therefore only give the final results:
\begin{subequations} \label{eq:phi0^2,phi2^2 approx l->inf,r->rplus}
\begin{align}
\sum_{l,m,P}\abs{{}_{lm\omega}\phi_{-1}^{\text{up}}}^2 & \sim
\frac{2Mr_+\tilde{\omega}\abs{\EuFrak{N}}^2}{3\pi^2 \Delta^3}
&
(l\rightarrow +\infty, r \rightarrow r_+)
\label{eq:phi0^2 approx l->inf,r->rplus} \\ 
\sum_{l,m,P}\abs{{}_{lm\omega}\phi_{+1}^{\text{up}}}^2 & \sim
\frac{Mr_+\tilde{\omega}\abs{\EuFrak{N}}^2}{6\pi^2 \Sigma^2 \Delta}
&
(l\rightarrow +\infty, r \rightarrow r_+)  \label{eq:phi2^2 approx l->inf,r->rplus}
\end{align}
\end{subequations}

We can now use equations (\ref{eq:phi_iphi_j,i<>j,approx l->inf,r->rplus}), (\ref{eq:phi1^2 approx l->inf,r->rplus}) 
and (\ref{eq:phi0^2,phi2^2 approx l->inf,r->rplus}) together with the quantum expressions (\ref{eq:RSET in B^- at horizon})
and (\ref{eq:stress tensor for s=1 on all vac.}) to reproduce equation 3.7 in CCH. We obtain
\begin{equation} \label{eq:eq.3.7CCH;mine}
\begin{aligned}
&\vac[ren]{\hat{T}^{\mu}{}_{\nu}}{B^-} \sim \vac{\hat{T}^{\mu}{}_{\nu}}{B^--CCH^-} \sim
\frac{-8M^3r_+^3}{3\pi^2\Delta^2\Sigma^2}\int_0^\infty 
\frac{\d{\tilde{\omega}}\tilde{\omega}(\tilde{\omega}^2+\kappa^2)}{e^{2\pi\tilde{\omega}/\kappa}-1}
\times
\\
&
\times
\left(
\begin{array}{cccc}
-3(r_+^2+a^2)-a^2\sin^2\theta & 0 & 0 & 4a\sin^2\theta(r_+^2+a^2)  \\
0 & \Sigma & 0 & 0 \\
0 & 0 & \Sigma & 0 \\
-4a & 0 & 0 & (r_+^2+a^2)+3a^2\sin^2\theta
\end{array}
\right)
\qquad \quad 
(r \rightarrow r_+)
\end{aligned}
\end{equation}
in Boyer-Lindquist co-ordinates.
\draft{wrong factor $-2$ throughout!!!!!!!!!}

This is exactly equation 3.7 in CCH except for the fact that (\ref{eq:eq.3.7CCH;mine}) contains a factor $r_+^3$ instead
of a $r_+$ in CCH. We believe that the discrepancy is due to a typographical error in CCH 
since otherwise the stress-energy tensor would not have the correct units.
We have also checked that equation (\ref{eq:eq.3.7CCH;mine}), when the tensor indices are adapted
to the Carter orthonormal tetrad, produces the result (\ref{eq:CCH's RSET in B^- in Carter, for r->r_+}) above.
Again, the discrepancy with respect to (\ref{eq:CCH's RSET in B^- in Carter, for r->r_+}) is only 
in the power of $r_+$.
It seems that, despite the dicrepancy in the power of $r_+$, this is the method that CCH used to calculate
their expression (\ref{eq:CCH's RSET in B^- in Carter, for r->r_+}).
However, as we pointed out in Section \ref{sec:asympts. close to r_+}, this asymptotic analysis is only valid when both $\omega$ and $m$ are kept bounded
since otherwise we would not be able to replace the spin-weighted spheroidal harmonics by the spin-weighted
spherical harmonics. In the analysis we have just carried out $\omega$ and $m$ do not both remain bounded in general.
The only points in the Kerr space-time where both remain bounded are the points along the axis $\theta=0$ or $\pi$ since, 
there, the Newman-Penrose scalars ${}_{lm\omega}\phi_{\indhel}$ are only non-zero for $m=\pm1,0$ and thus $m$ is bounded. 
The frequency $\omega$ is then also kept bounded because the factor in the integrand diminishes
exponentially with $\tilde{\omega}$ and thus the contribution is only important when $\tilde{\omega}$ is bounded.
Equations (\ref{eq:eq.3.7CCH;mine}) and (\ref{eq:CCH's RSET in B^- in Carter, for r->r_+}) are therefore
only valid at the axis. An asymptotic behaviour of the `up' radial solutions uniform both in $l$
and $\tilde{\omega}$ is required.

Another issue is the fact that the state $\ket{CCH^-}$ has been used in (\ref{eq:RSET in B^- at horizon}) as a Hartle-Hawking
state, regular on both the past and future horizons. 
We know from Kay and Wald's work that there exists no such state on the Kerr space-time satisfying its isommetries.
Since $\ket{CCH^-}$ is not invariant under $(t,\phi)$ reversal it is not covered by Kay and Wald's result and thus it might
be regular on both $\mathcal{H^-}$ and $\mathcal{H^+}$.
We saw that Ottewill and Winstanley ~\cite{ar:Ott&Winst'00} argued that in the scalar case this state is 
 irregular on $\mathcal{H^-}$ and regular on $\mathcal{H^+}$.
Even if that were also the case for spin-1, using $\ket{CCH^-}$ in the preceding calculation could still be acceptable if the divergence
of $\ket{CCH^-}$ close to $r_+$ is of a smaller order than that of $\ket{B^-}$. 
In the Schwarzschild background Candelas has shown that the Unruh state is irregular on $\mathcal{H^-}$, regular on $\mathcal{H^+}$ and
that the order of its divergence close to $r_+$ is smaller than that of the Boulware state. 
It is therefore reasonable to expect that the order of the divergence of $\ket{CCH^-}$ close to $r_+$ in the Kerr background
is smaller than that of the past Boulware state.
Indeed, our numerical data indicate that the approximation in (\ref{eq:RSET in B^- at horizon}) is correct.

\begin{figure}[p]
\rotatebox{90}
\centering
\includegraphics*[width=70mm,angle=270]{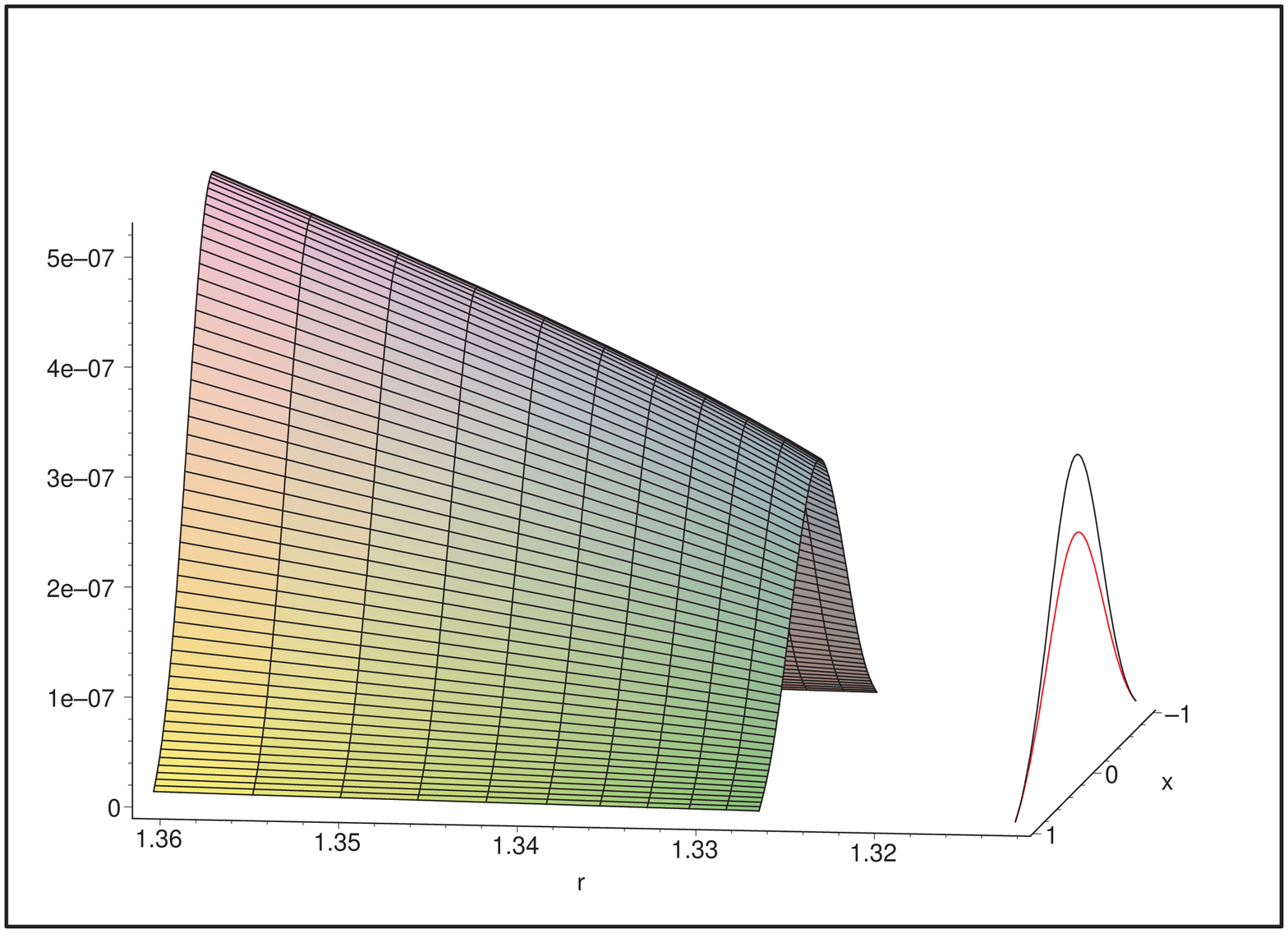} 
\caption{$\frac{1}{4\pi}\Delta^2\vac{\hat{T}_{tt}}{CCH^--B^-}$}    \label{fig:delta2Ttt_cch_b_past}
\includegraphics*[width=70mm,angle=270]{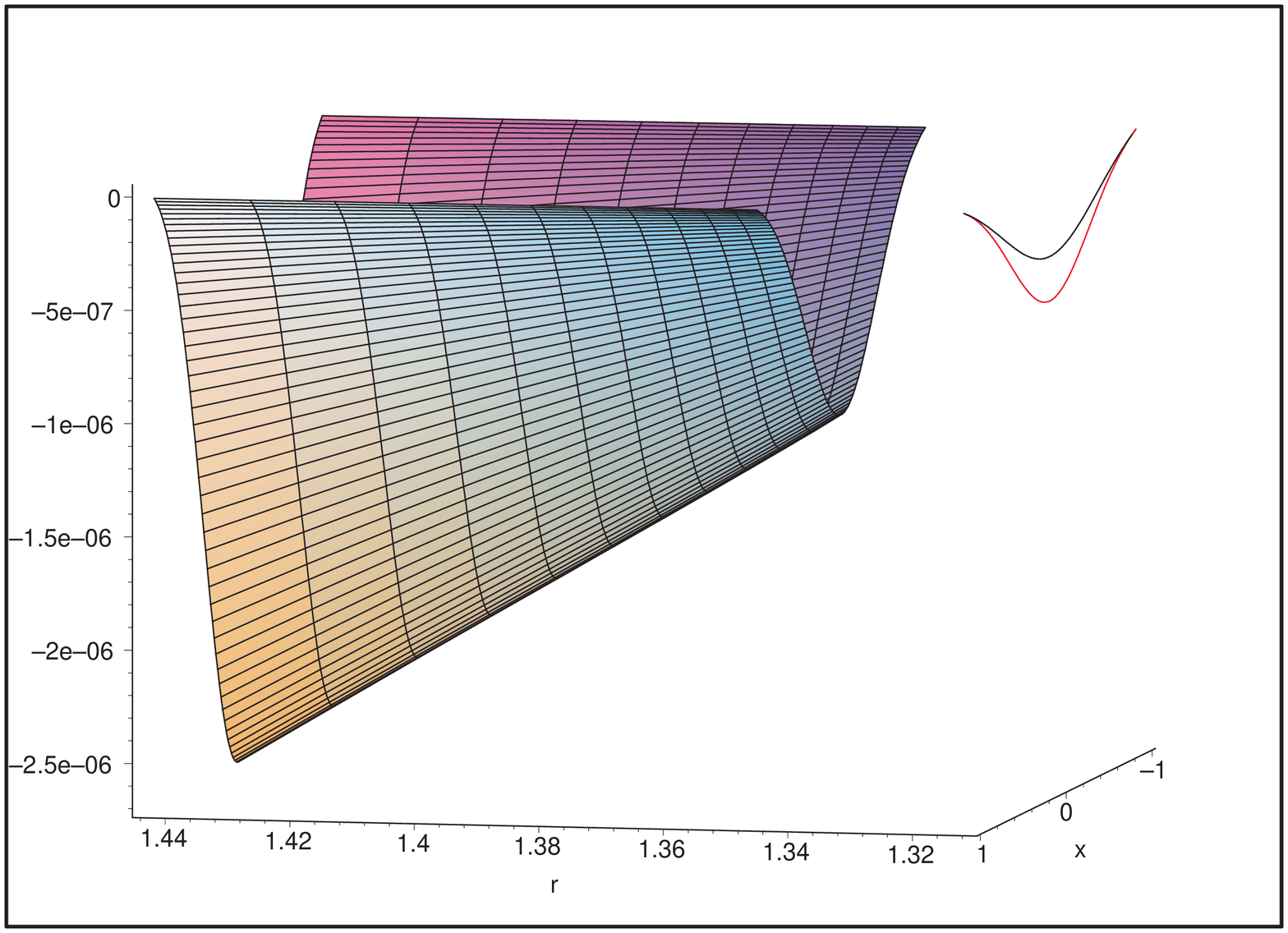}
\caption{$\frac{1}{4\pi}\Delta^2\vac{\hat{T}_{t\phi}}{CCH^--B^-}$}             \label{fig:delta2Ttphi_cch_b_past}
\end{figure}

\begin{figure}[p]
\rotatebox{90}
\centering
\includegraphics*[width=70mm,angle=270]{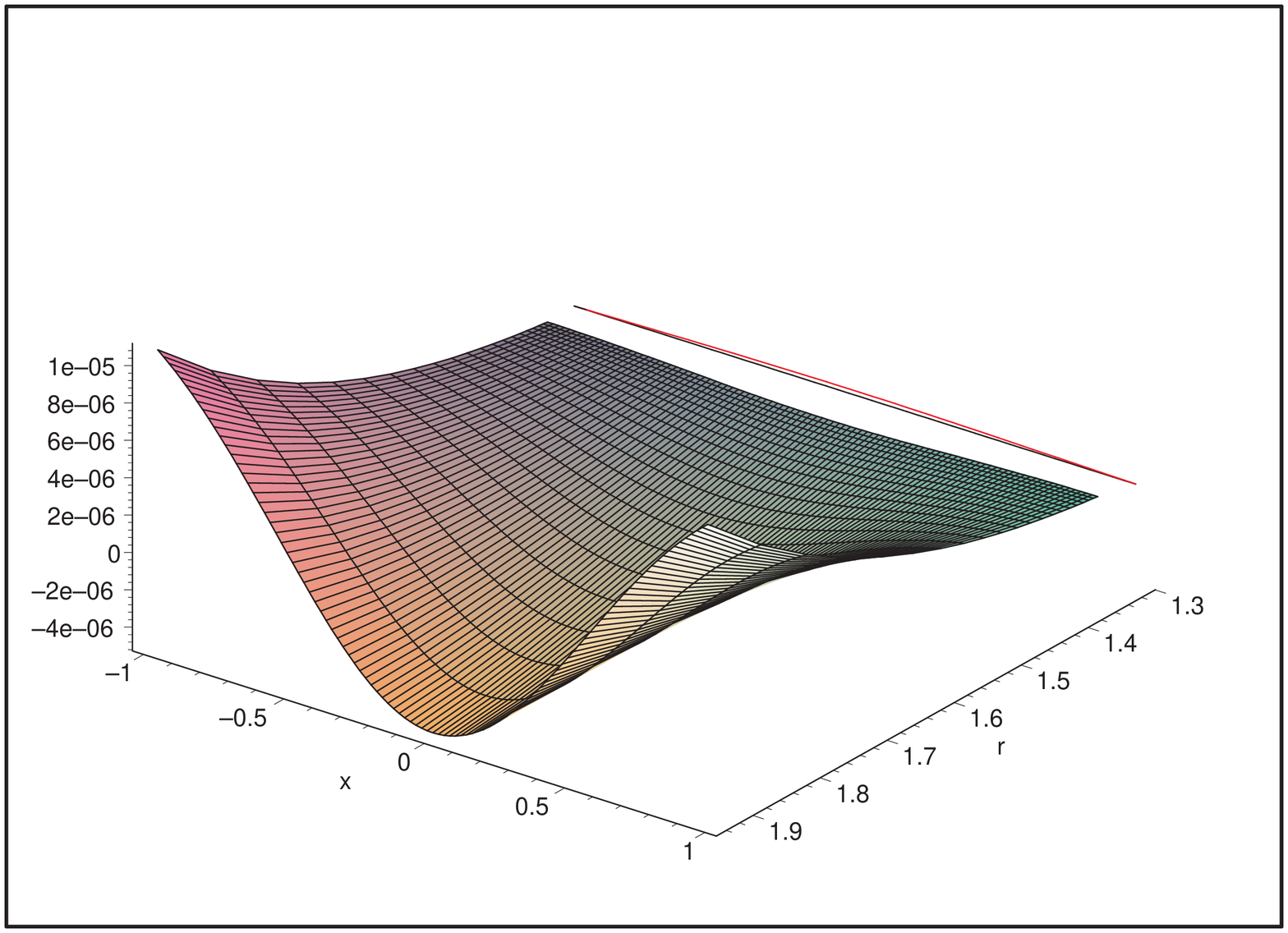} 
\caption{$\frac{1}{4\pi}\Delta^3\vac{\hat{T}_{rr}}{CCH^--B^-}$}                \label{fig:delta3Trr_cch_b_past}
\end{figure}

\begin{figure}[p]
\rotatebox{90}
\centering
\includegraphics*[width=70mm,angle=270]{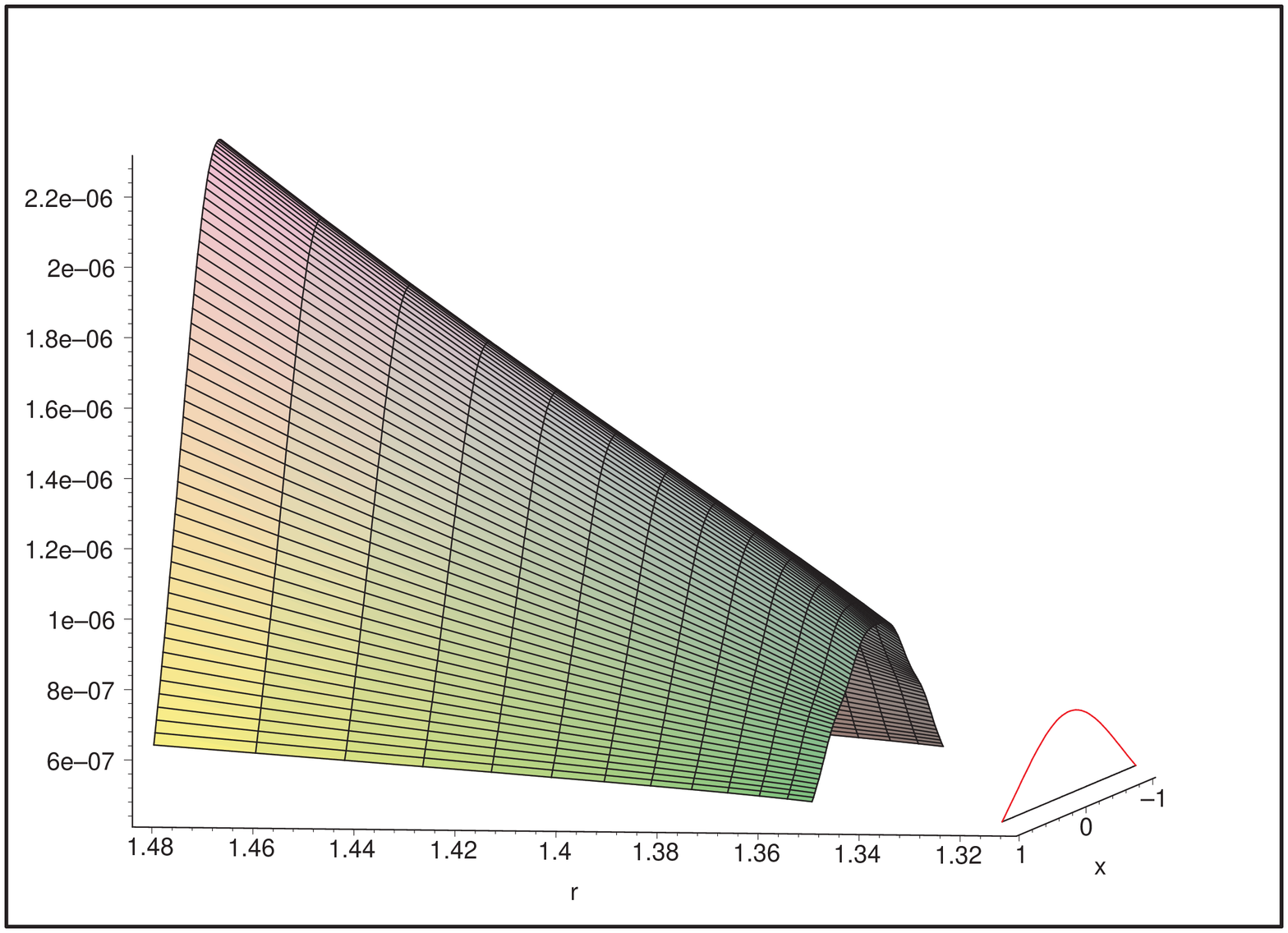}
\caption{$\frac{1}{4\pi}\Delta^2\vac{\hat{T}_{\theta\theta}}{CCH^--B^-}$}      \label{fig:delta2Tthetatheta_cch_b_past_bis}
\includegraphics*[width=70mm,angle=270]{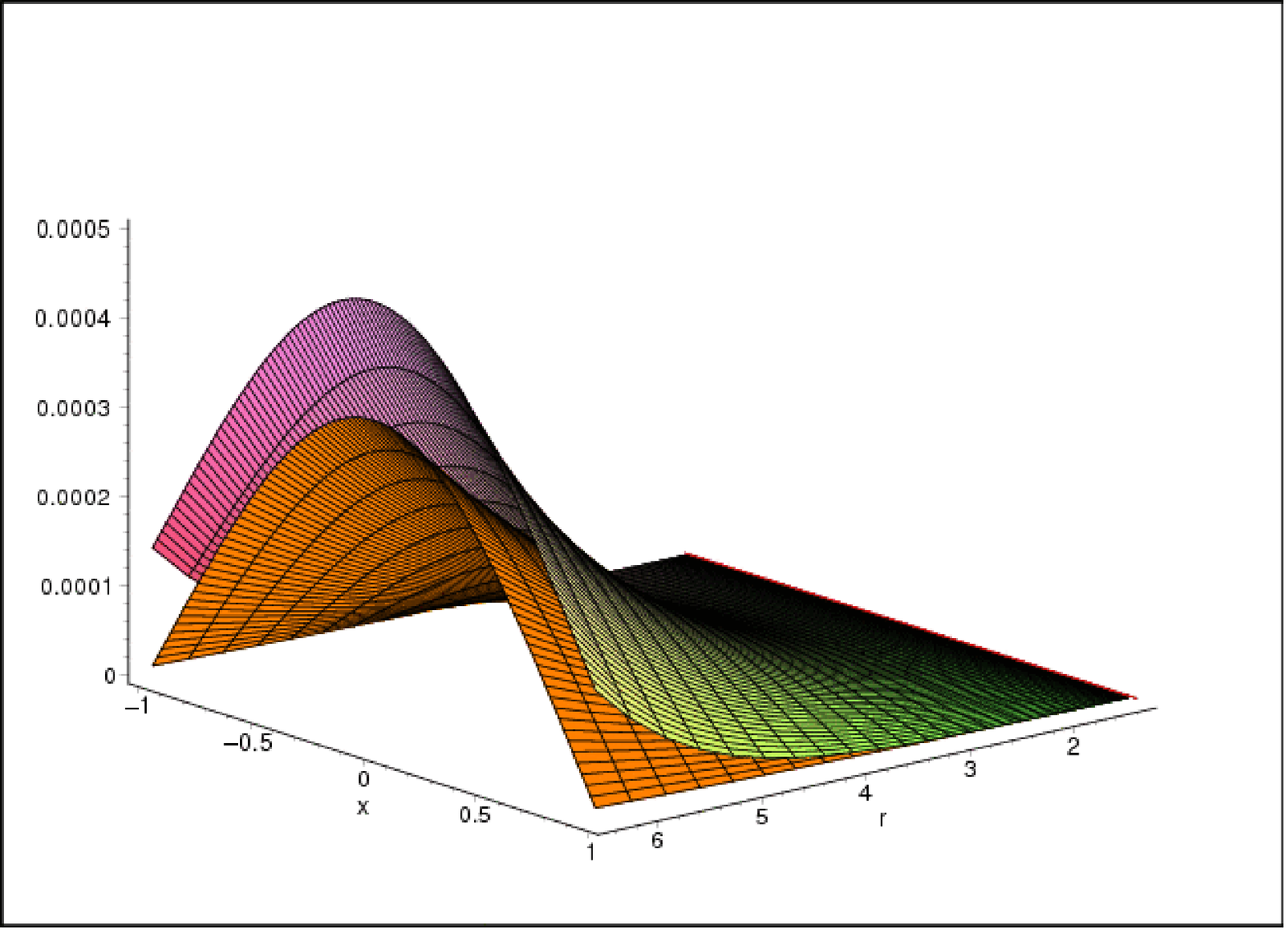} 
\caption{$\frac{1}{4\pi}\Delta^2\vac{\hat{T}_{\theta\theta}}{CCH^--B^-}$ and 
$\frac{1}{4\pi}\Delta^2\vac{\hat{T}_{\theta\theta}}{U^--B^-}$(orange)}    \label{fig:delta2Tthetatheta_u_b_cch_b_past}
\end{figure}

\begin{figure}[p]
\rotatebox{90}
\centering
\includegraphics*[width=70mm,angle=270]{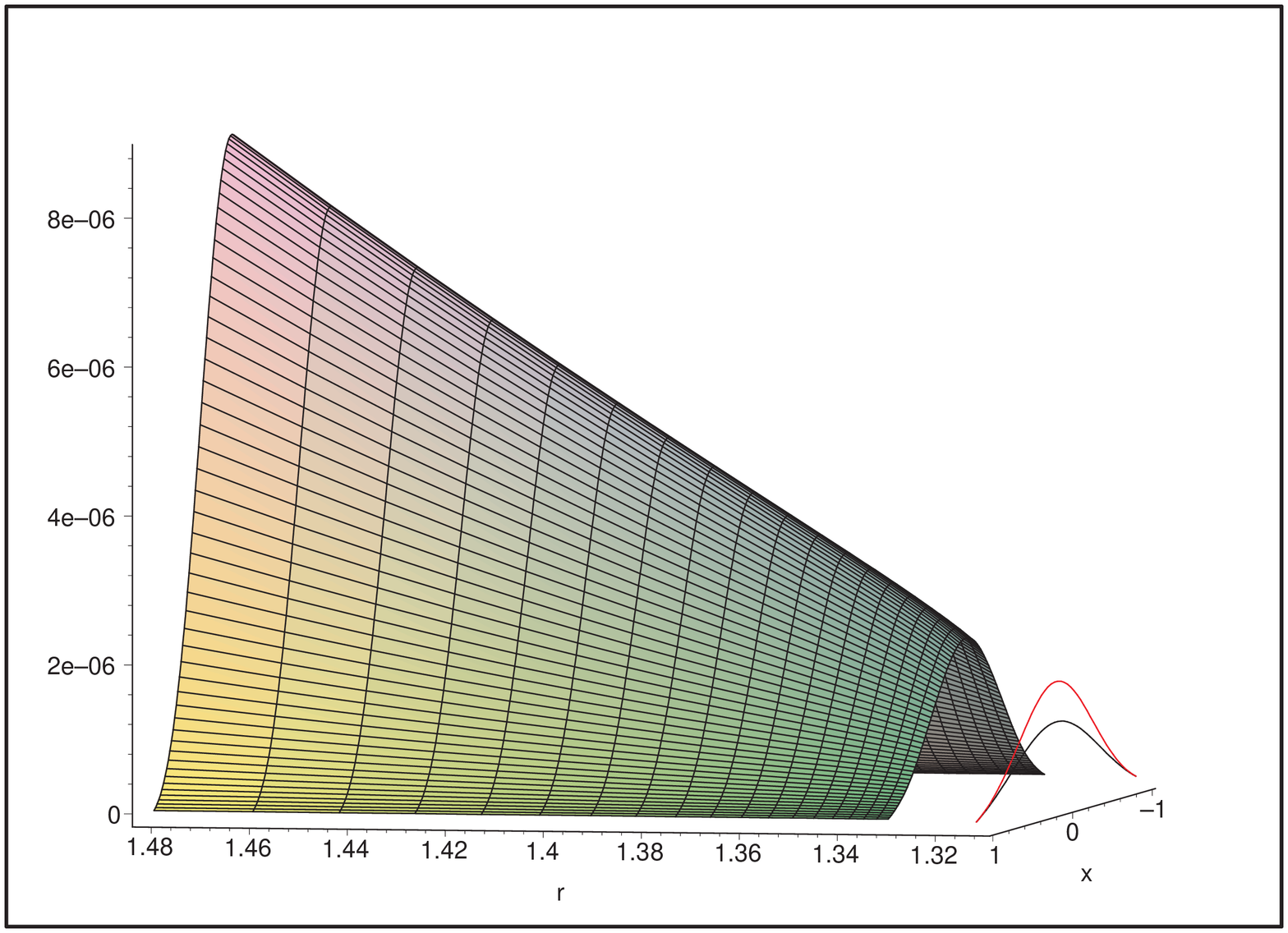} 
\caption{$\frac{1}{4\pi}\Delta^2\vac{\hat{T}_{\phi\phi}}{CCH^--B^-}$}          \label{fig:delta2Tphiphi_cch_b_past}
\includegraphics*[width=70mm,angle=270]{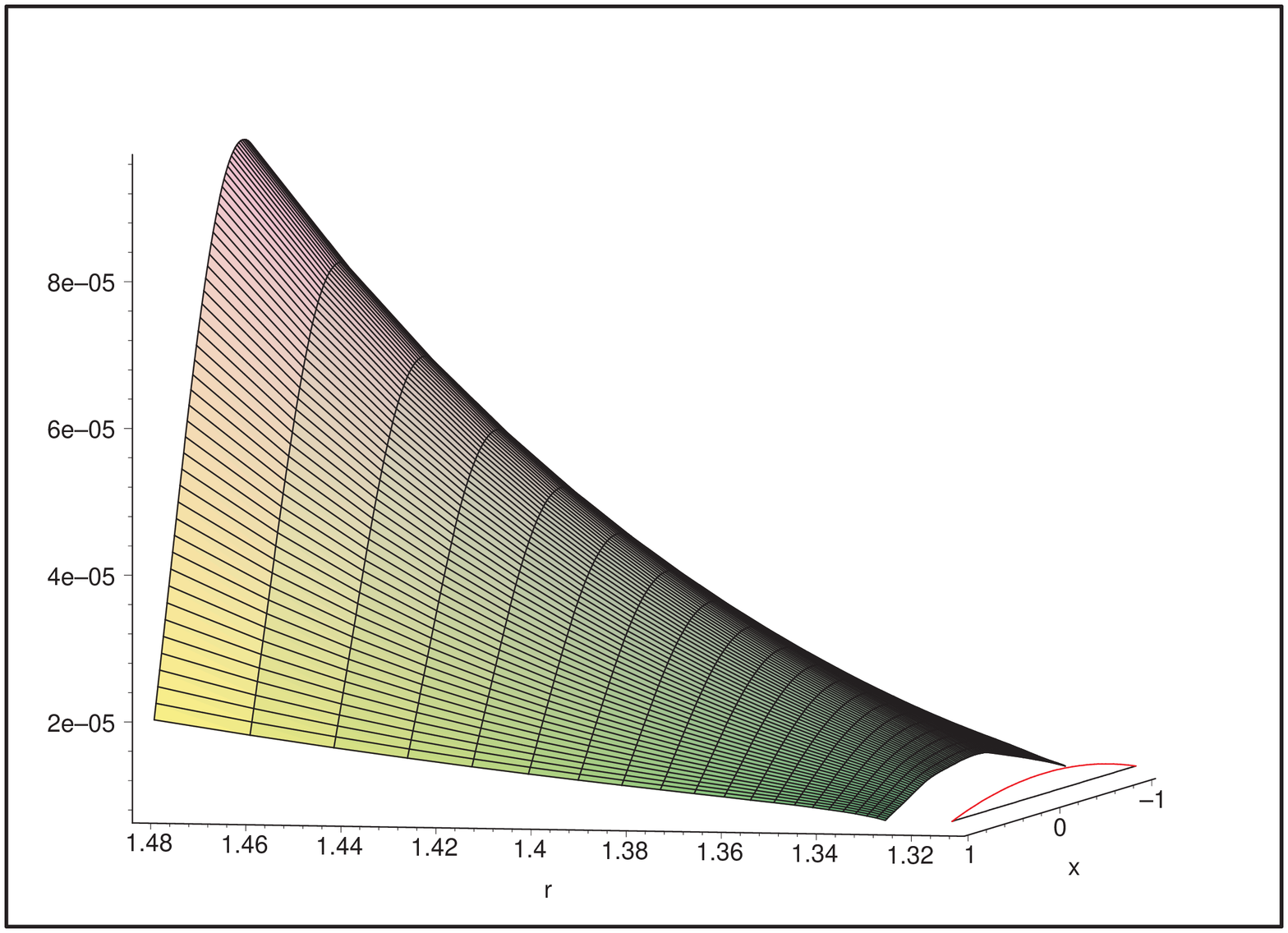}
\caption{$\frac{\Sigma\Delta^2}{4\pi\sin^2\theta}\vac{\hat{T}_{\phi\phi}}{CCH^--B^-}+\frac{11(r_+-r_-)^4Mr_+a^2\sin^2\theta}{2^9\cdot 3^2\cdot 5\pi^3}$
(so that CCH's expression is constant in $\theta$)}
\label{fig:CCHconstdelta2Tphiphi_cch_b_past_bis}
\end{figure}

Graphs \ref{fig:delta2Ttt_cch_b_past}--\ref{fig:CCHconstdelta2Tphiphi_cch_b_past_bis}
show that the RSET when the field is in the past Boulware vacuum approaches a thermal distribution rotating
with the horizon rather than CCH's result (\ref{eq:eq.3.7CCH;mine}). 
The red lines in the graphs correspond to the thermal stress tensor (\ref{eq:thermal RR stress tensor}) 
rotating with the horizon evaluated at $r=r_+\simeq 1.3122$.
The black lines are also located at $r=r_+$ and correspond to CCH's result (\ref{eq:eq.3.7CCH;mine}).
It can be seen in the graphs that as $r$ becomes closer to the horizon, $\vac[ren]{\hat{T}{}_{\mu\nu}}{CCH^--B^-}$ approaches 
the thermal stress tensor (\ref{eq:thermal RR stress tensor}) (red line) rather than CCH's 
corrected equation (\ref{eq:eq.3.7CCH;mine}) (black line). At the poles, however, it can be straight-forwardly checked
analytically that the two coincide, as expected. Only for the $rr$-component, which is the only component that diverges like
$O(\Delta^{-3})$ close to the horizon, we did not seem to be able to obtain a clear plot. 

Within the range of $r$ considered in Graphs \ref{fig:delta2Ttt_cch_b_past}--\ref{fig:CCHconstdelta2Tphiphi_cch_b_past_bis}
(except \ref{fig:delta2Tthetatheta_u_b_cch_b_past})
for the difference between the states $\ket{CCH^-}$ and $\ket{B^-}$ of the various expectation values, 
the corresponding plots for the difference between the states $\ket{U^-}$ and $\ket{B^-}$ are identical. 
This is the expected behaviour since for small radius $r$ the `up' modes dominate in these RSETs.
Graph \ref{fig:delta2Tthetatheta_u_b_cch_b_past} includes the two differences 
for the $\theta\theta$-component of the stress-energy tensor up to a value of $r$ large enough so that the two differences become clearly distinct.

\begin{figure}[p]
\rotatebox{90}
\centering
\includegraphics*[width=80mm,angle=270]{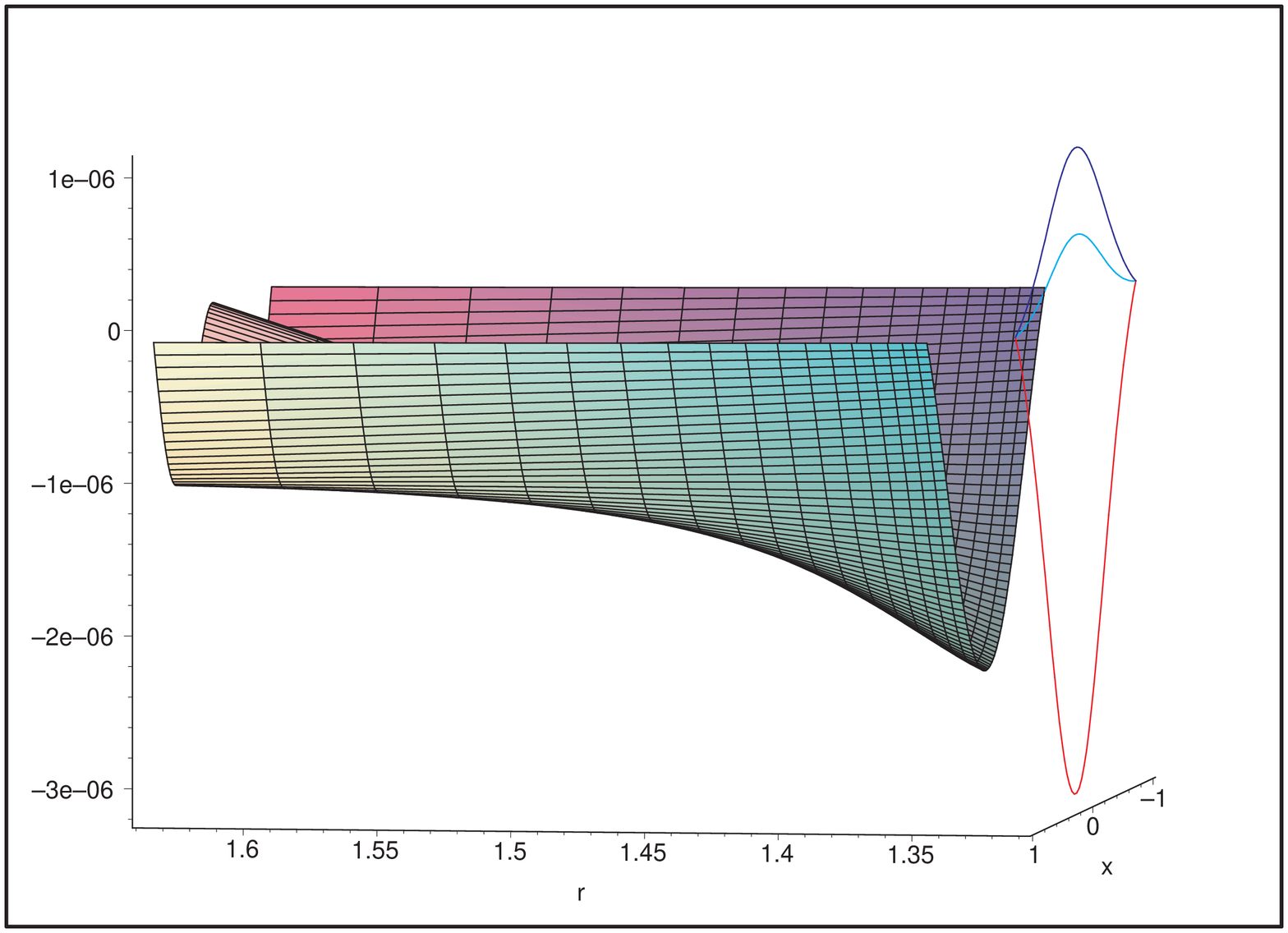} 
\caption{$\frac{1}{4\pi}\Delta\vac{\hat{T}_{t_+\phi_+}}{CCH^--B^-}$, $\frac{1}{4\pi}\Delta T^{\text{(th,RR)}}_{t_+\phi_+}$ (red), 
$\frac{1}{4\pi}\Delta T^{\text{(th,ZAMO)}}_{t_+\phi_+}$ (blue)
and $\frac{1}{4\pi}\Delta T^{\text{(th,Carter)}}_{t_+\phi_+}$ (cyan).}           \label{fig:deltaTtplusphiplus_cch_b_past}
\includegraphics*[width=80mm,,angle=270]{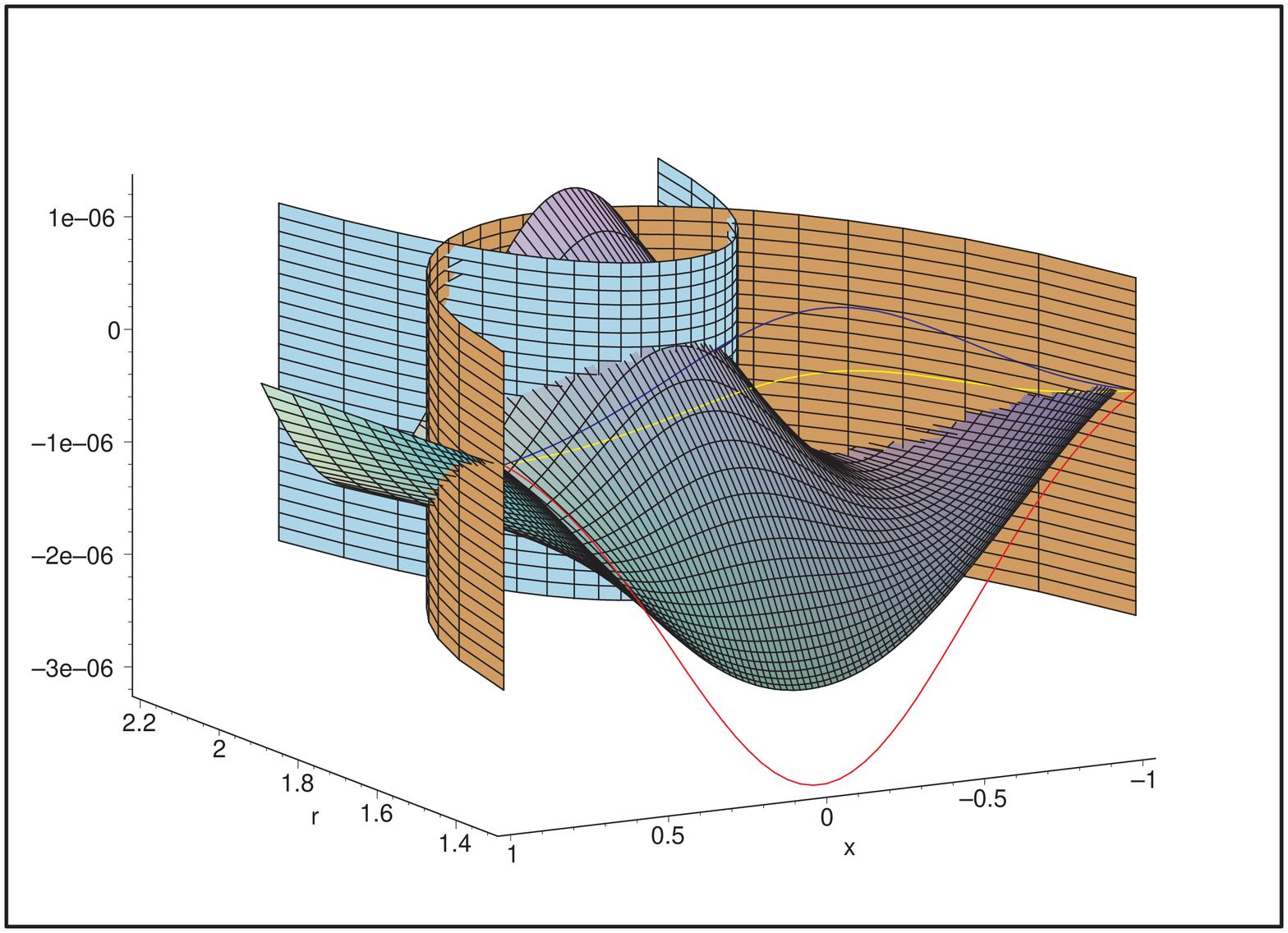} 
\caption{$\frac{1}{4\pi}\Delta\vac{\hat{T}_{t_+\phi_+}}{CCH^--B^-}$, $\frac{1}{4\pi}\Delta T^{\text{(th,RR)}}_{t_+\phi_+}$ (red), 
$\frac{1}{4\pi}\Delta T^{\text{(th,ZAMO)}}_{t_+\phi_+}$ (blue)
and $\frac{1}{4\pi}\Delta T^{\text{(th,Carter)}}_{t_+\phi_+}$ (yellow). The light blue and brown surfaces correspond to the speed-of-light and 
the static limit surfaces respectively.}                     \label{fig:deltaTtplusphiplus_cch_b_past_surfs}
\end{figure}

In following with the notation used in (\ref{eq:tetrad of stationary obs.}) and the one used so far for tensor components 
in Boyer-Lindquist co-ordinates, we use the obvious
notation of `$(\alpha\beta)$-component' to refer to the stress-energy tensor component $T{}_{\mu\nu}e_{(\alpha)}{}^{\mu}e_{(\beta)}{}^{\nu}$
in the tetrad of a stationary observer.
Since the angular velocities of a RRO, ZAMO
and Carter observer all equal $\Omega_+$ at the horizon, each one of the diagonal components of a stress tensor for a thermal
distribution will be the same 
in any of the three tetrads adapted to these observers.
The $(r\theta)$-component will also be the same 
in any of the three tetrads
since the tetrad vectors $\vec{e}_{(r)}$ and $\vec{e}_{(\theta)}$ do not depend
on the rate of rotation. The $(t\phi)$-component,
however, vanishes to leading order for the radial functions as $r\rightarrow r_+$.
To the next leading order for the radial functions this component does depend on the rate of rotation of 
the stationary observer that the tetrad is adapted to.
\ddraft{1) why is it not the same for $tr$,$t\theta$,$r\phi$,$\theta\phi$ components?, 2) maybe to the next leading order
the approximation $\vac[ren]{\hat{T}^{\mu}{}_{\nu}}{B^-}\sim -\vac{\hat{T}^{\mu}{}_{\nu}}{CCH^--B^-}$ is not valid anymore?}
Graphs \ref{fig:deltaTtplusphiplus_cch_b_past}--\ref{fig:deltaTtplusphiplus_cch_b_past_surfs} for $\vac[ren]{\hat{T}^{}_{t_+\phi_+}}{B^-}$ 
show that the rate of rotation of the thermal distribution
approaches, to next order in $\Delta$, that of a RRO, rather than that of a ZAMO or a Carter observer.
This result tallies with Duffy~\cite{th:GavPhD}'s results for the spin-0 case in the Kerr space-time 
modified with a mirror when the field is in the $\ket{H_{\mathcal{M}}}$ state. 
He also numerically shows that $\vac{\hat{T}{}_{\mu\nu}}{U^--B^-}$ is, close to the horizon and for the scalar field, 
thermal and rotating at the rate of a RRO to $O(\Delta)$ in the angular frequency. 
We calculated and plotted $\vac[ren]{\hat{T}_{t_+\phi_+}}{U^--B^-}$ and 
it fully coincided with $\vac[ren]{\hat{T}_{t_+\phi_+}}{CCH^--B^-}$ in the region of Graphs 
\ref{fig:deltaTtplusphiplus_cch_b_past}--\ref{fig:deltaTtplusphiplus_cch_b_past_surfs}, which is why we do not include them.
We conclude that the rate of rotation close to the horizon for the difference between the states $\ket{U^-}$ and $\ket{B^-}$ is also
that of a RRO, in agreement with Duffy's results.

\begin{figure}[p]
\centering
\includegraphics*[width=70mm,angle=270]{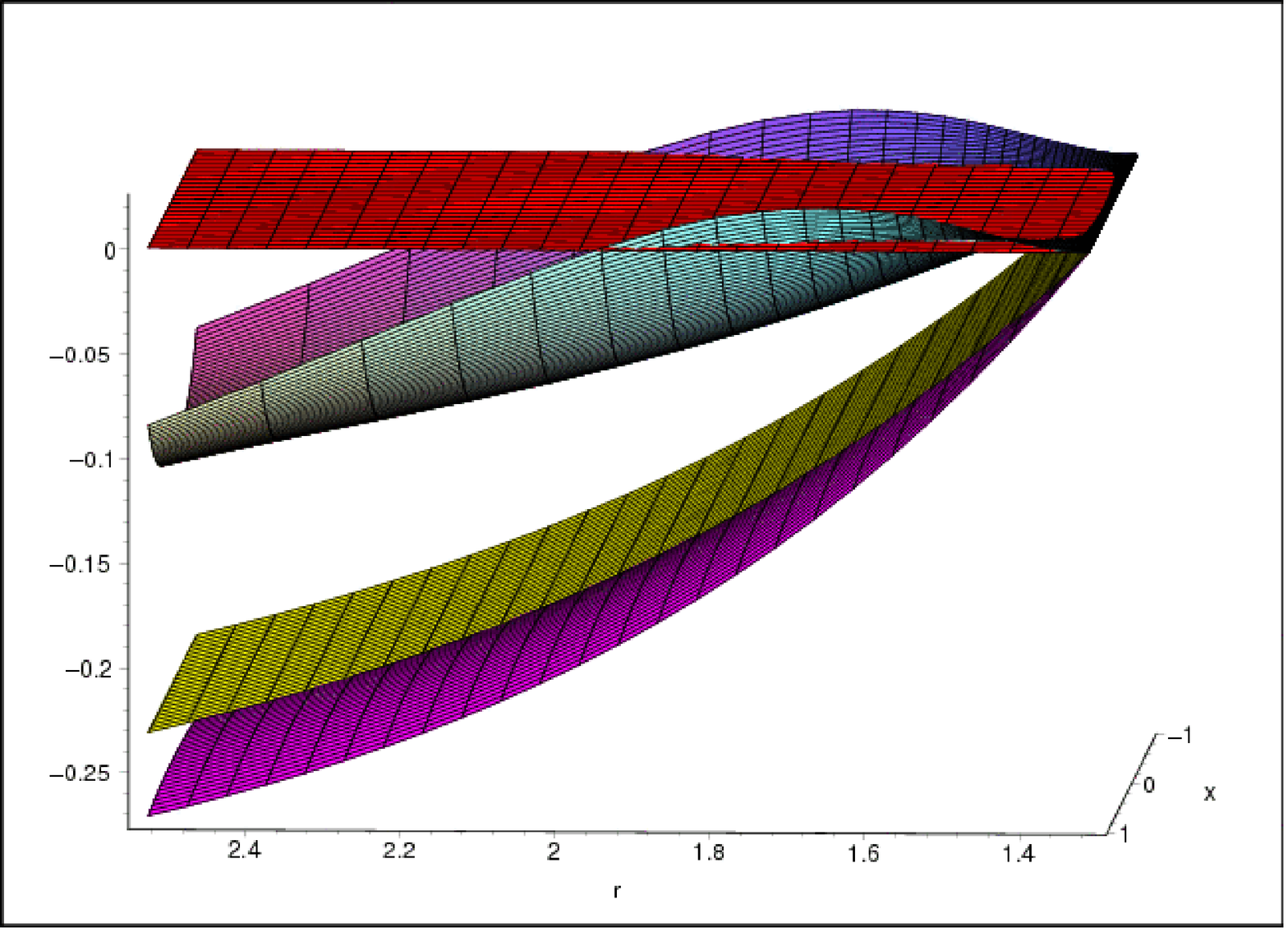} \\
$(\omega_{\text{ZEFO}}-\Omega_+)$ \\
\includegraphics*[width=70mm,angle=270]{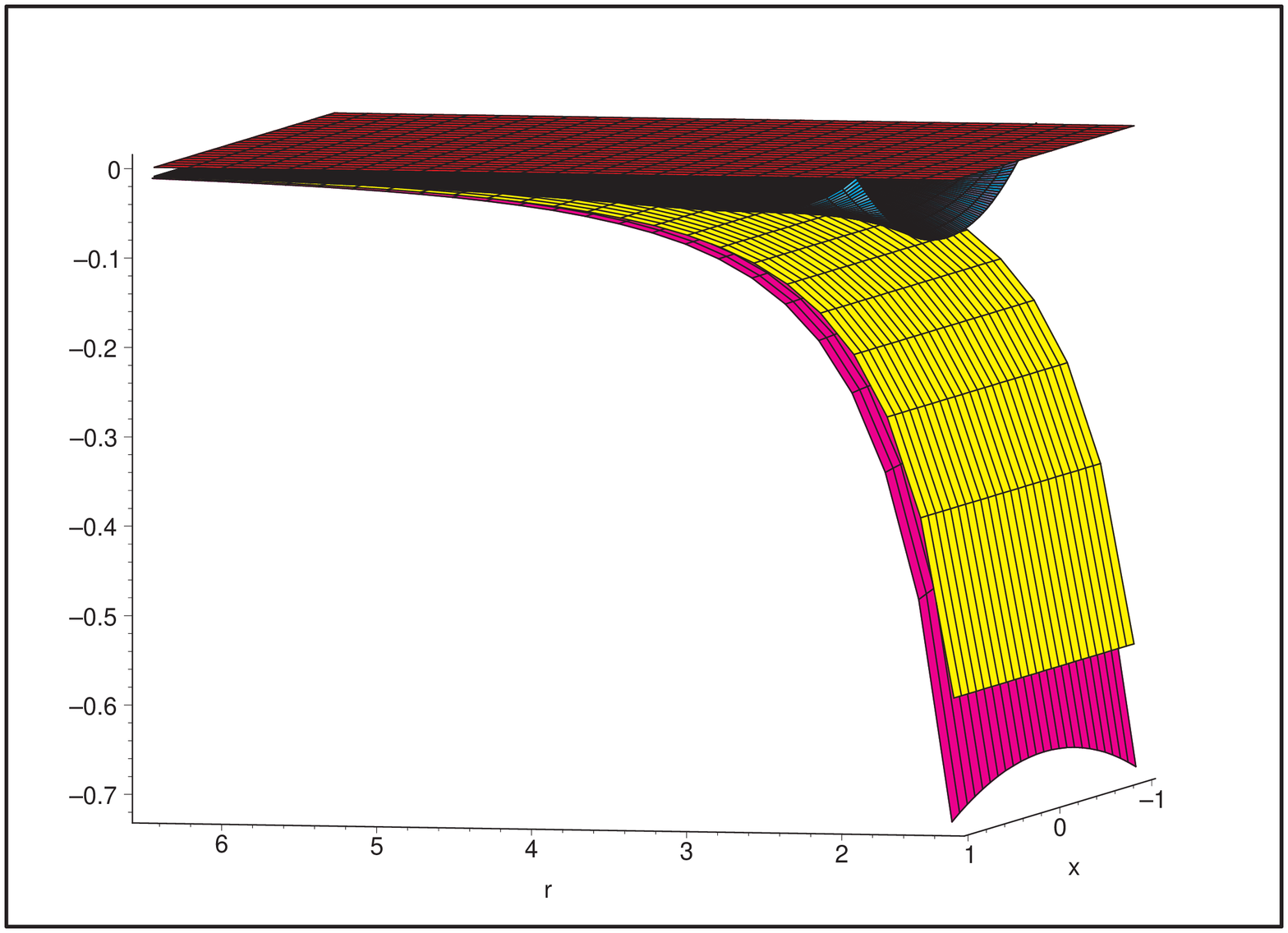} \\
$\left(\omega_{\text{ZEFO}}-\Omega_+\right)/\Delta$ \\
\caption{Plots of $(\omega_{\text{ZEFO}}-\Omega_+)$ and  $\left(\omega_{\text{ZEFO}}-\Omega_+\right)/\Delta$ (dark surfaces) where 
$T_{\mu\nu}$ is replaced by $\vac[ren]{\hat{T}^{}_{\mu\nu}}{CCH^--B^-}$ in (\ref{eq:omega_ZEFO}), together with the corresponding
plots with the angular velocities of a RRO (red), ZAMO (magenta) and Carter observer (yellow).
}
\label{fig:freqCarterZAMO_omega_hor_cch_b_past}
\end{figure}

An alternative technique for investigating what is the rate of rotation of the thermal distribution at the horizon
is as follows. We find what is the frequency $\omega=\omega_{\text{ZEFO}}$ of rotation of the tetrad frame 
(\ref{eq:tetrad of stationary obs.}) such that $T_{(t\phi)}=0$, where
the term ZEFO stands for \define{zero energy flux observer}. The answer is
\begin{equation} \label{eq:omega_ZEFO}
\omega_{\text{ZEFO}}=\frac{-2C}{B+\sqrt{B^2-4AC}}
\end{equation}
where
\begin{equation} \label{eq:def. of A,B,C for ZEFO}
\begin{aligned}
A&=g_{\phi\phi}T_{t\phi}-g_{t\phi}T_{\phi\phi} \\
B&=g_{\phi\phi}T_{tt}-g_{tt}T_{\phi\phi}\\
C&=g_{t\phi}T_{tt}-g_{tt}T_{t\phi}
\end{aligned}
\end{equation}
We then plot $\omega_{\text{ZEFO}}$ where $T_{\mu\nu}$ is replaced by 
$\vac[ren]{\hat{T}_{\mu\nu}}{CCH^--B^-}$ in (\ref{eq:def. of A,B,C for ZEFO}). This plot is compared against that of the angular velocities 
of a RRO, ZAMO and Carter observer in Figure \ref{fig:freqCarterZAMO_omega_hor_cch_b_past}.
We also plotted $\omega_{\text{ZEFO}}$ where $T_{\mu\nu}$ is replaced by 
$\vac[ren]{\hat{T}_{\mu\nu}}{U^--B^-}$ and it fully coincided with the corresponding one for $\vac[ren]{\hat{T}_{\mu\nu}}{CCH^--B^-}$
in the region of Figure \ref{fig:freqCarterZAMO_omega_hor_cch_b_past}.

\begin{figure}[p]
\centering
\begin{tabular}{cc}
\includegraphics*[height=65mm,width=75mm,angle=270]{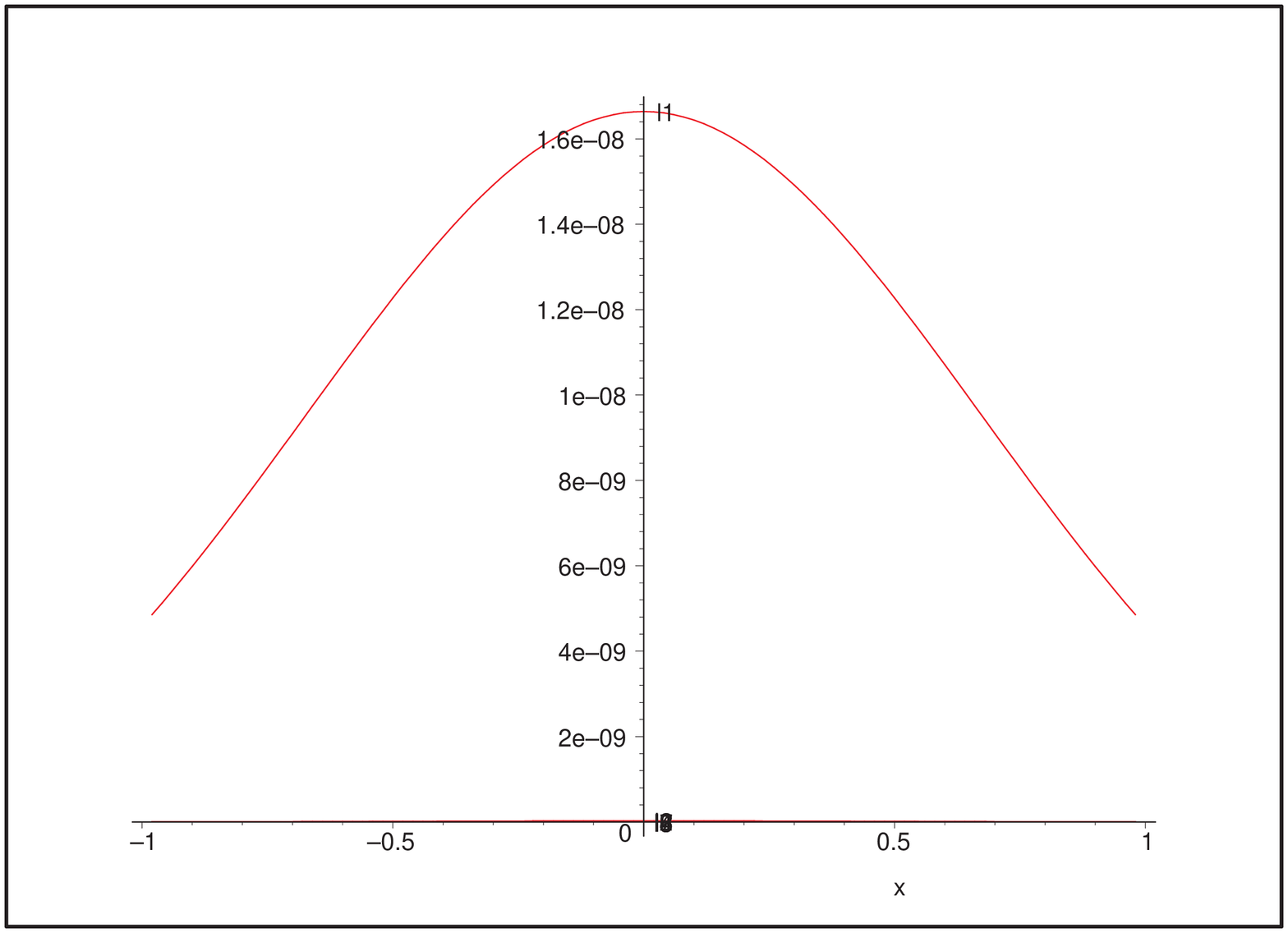} &
\includegraphics*[height=65mm,width=75mm,angle=270]{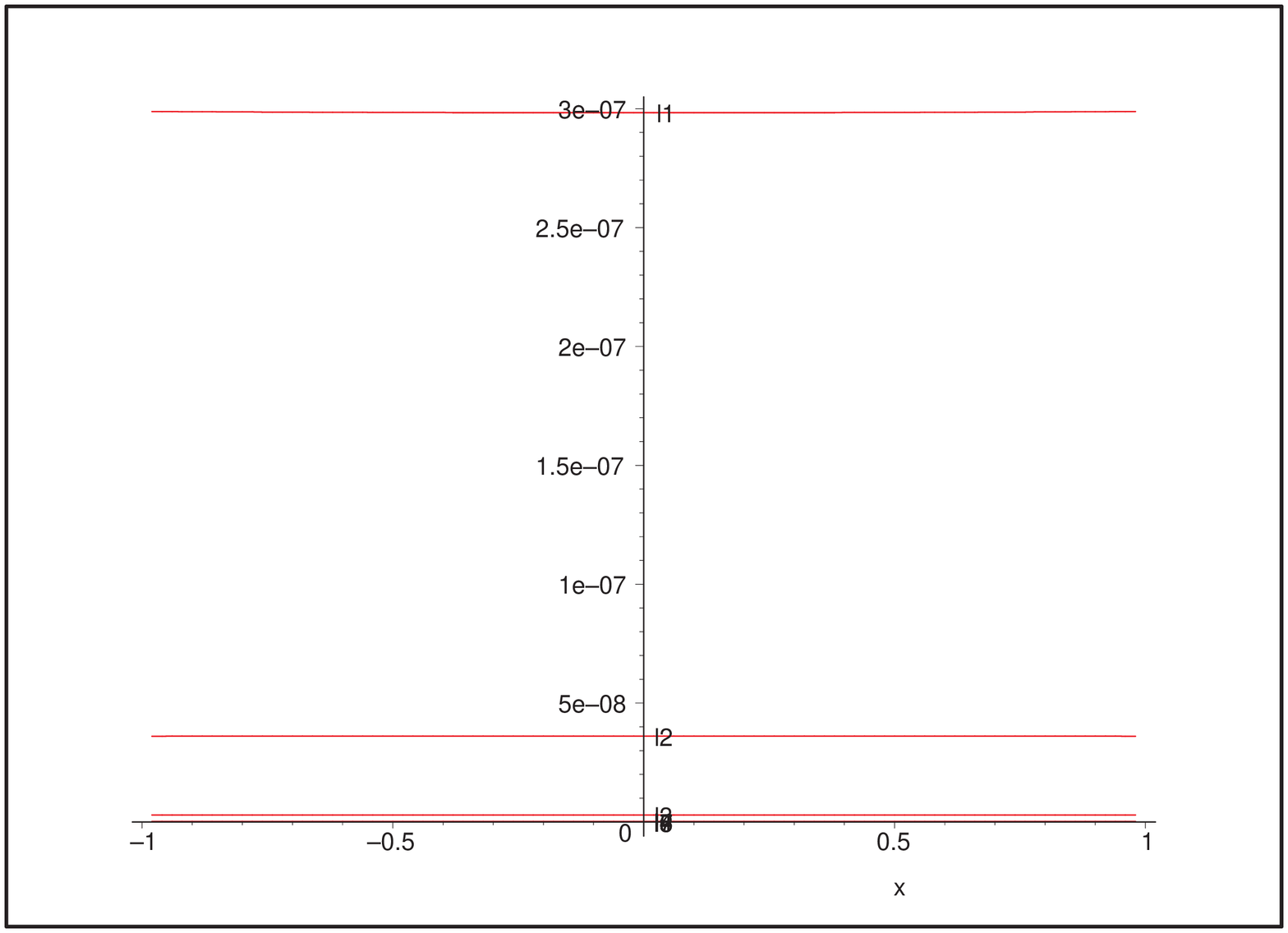} \\
$\frac{1}{4\pi}\vac{\hat{T}_{\theta\theta}(r \simeq 1.36,\theta)}{CCH^--U^-}$ & $\frac{1}{4\pi}\vac{\hat{T}_{\theta\theta}(r \simeq 10.26,\theta)}{CCH^--U^-}$\\
\includegraphics*[height=65mm,width=75mm,angle=270]{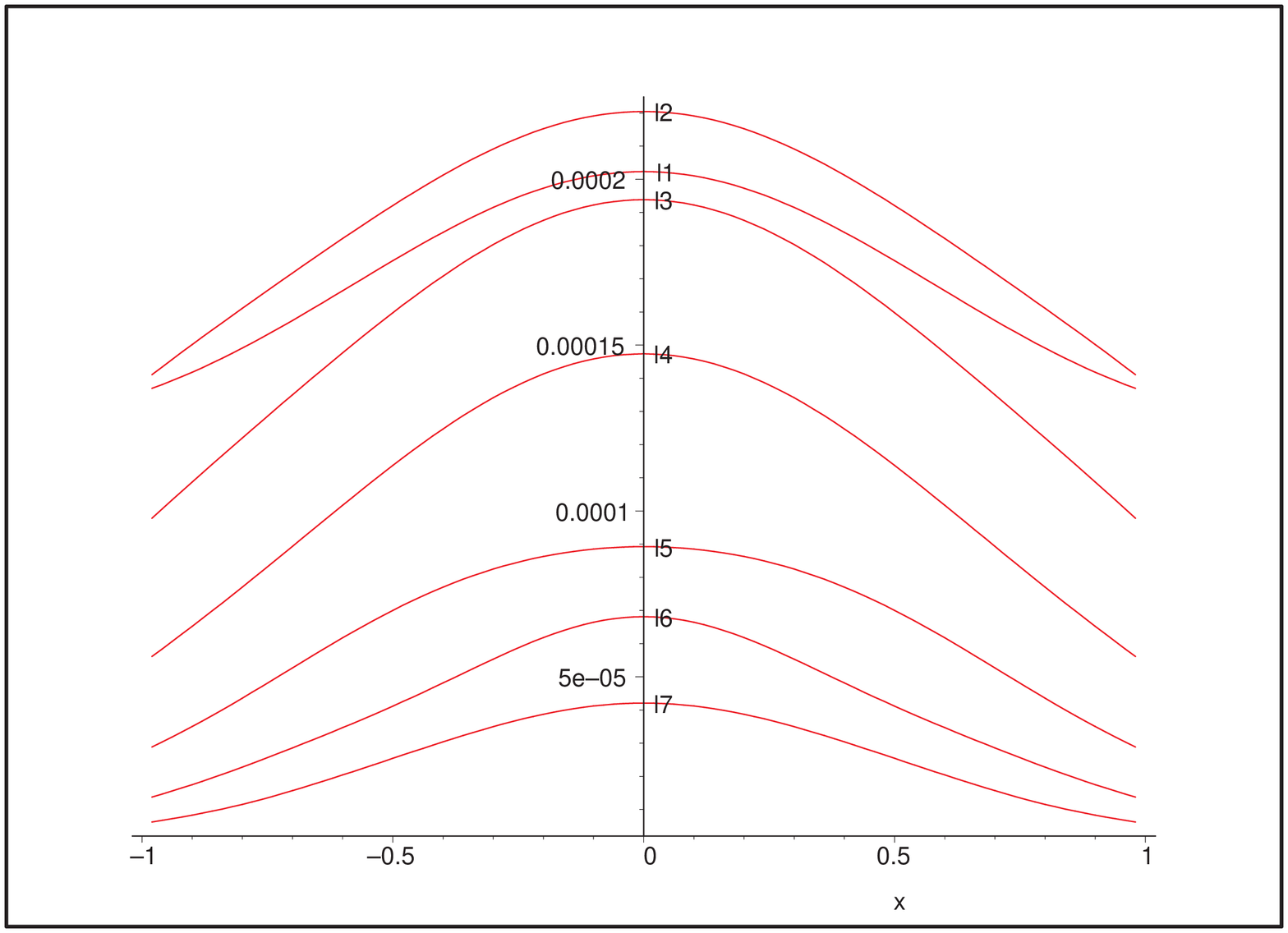} &
\includegraphics*[height=65mm,width=75mm,angle=270]{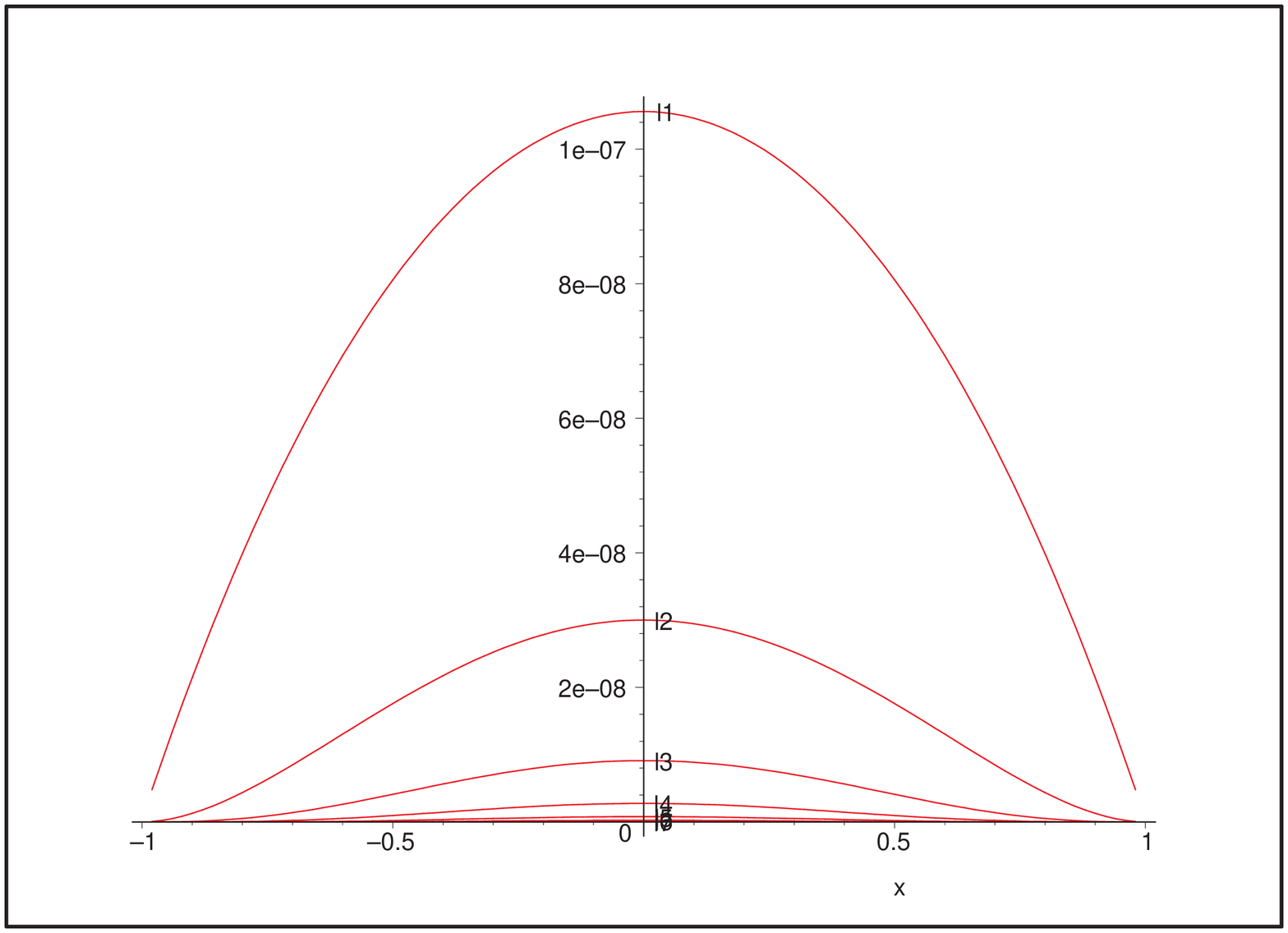} \\
$\frac{1}{4\pi}\vac{\hat{T}_{\theta\theta}(r \simeq 1.36,\theta)}{U^-B^-}$    & $\frac{1}{4\pi}\vac{\hat{T}_{\theta\theta}(r \simeq 10.26,\theta)}{U^-B^-}$\\
\end{tabular}
\caption{
The sum over $l$ has not been performed. For each value of $l$ the sum $\sum_{m=-l}^{l}$ has been performed.
In the case of $\vac{\hat{T}_{\theta\theta}}{CCH^--U^-}$ the low-$l$ modes clearly dominate close to the horizon. 
On the other hand, the high-$l$ modes dominate close to the horizon in the case of $\vac{\hat{T}_{\theta\theta}}{U^--B^-}$.
}
\label{fig:Tthetatheta_cch_u_past_u_b_spher_last_r100r150_n}
\end{figure}

\begin{figure}[p]
\rotatebox{90}
\centering
\begin{tabular}{c}
\includegraphics*[width=70mm,angle=270]{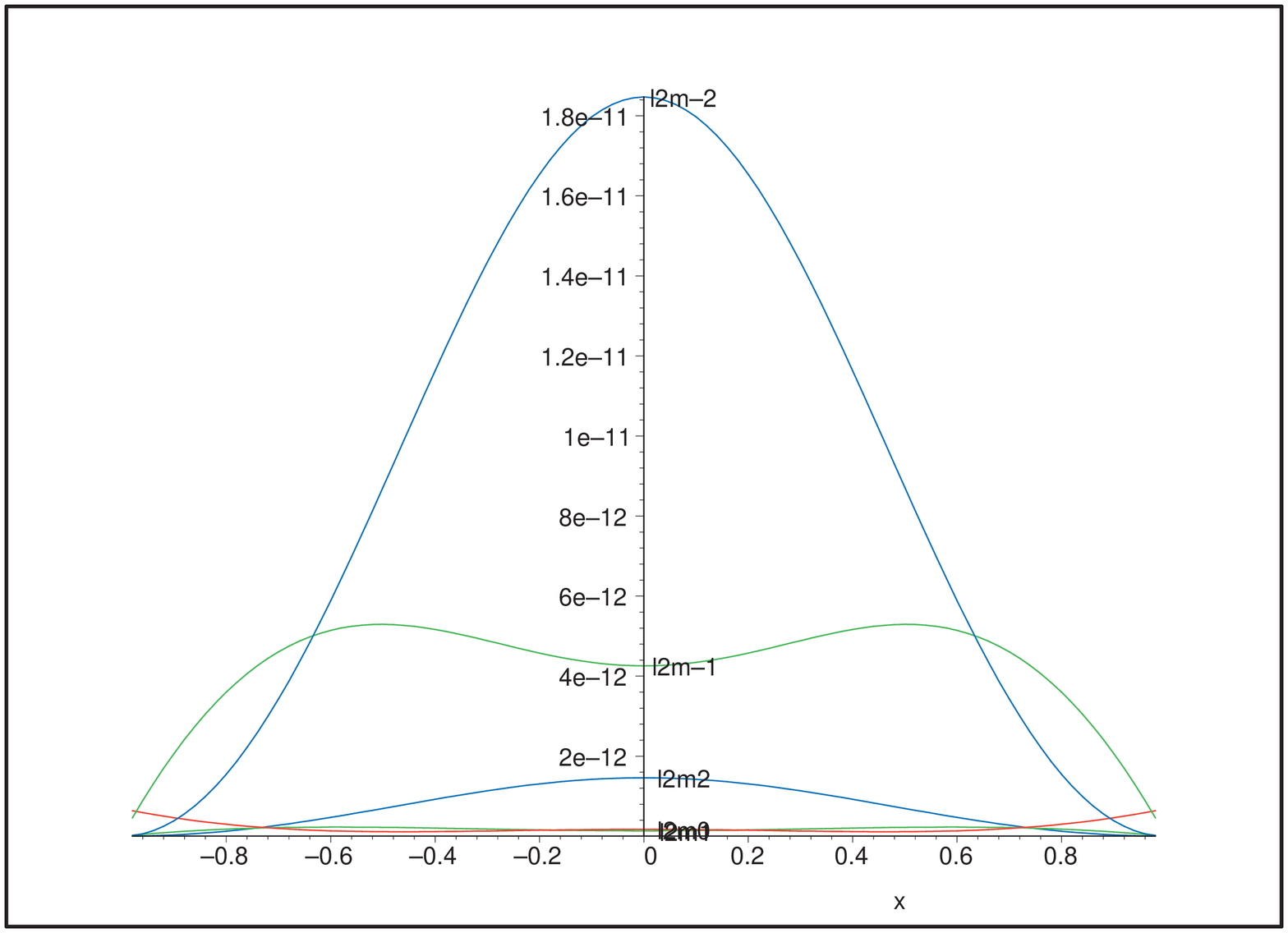} \\
$l=2$ at  $r \simeq 1.36$ \\
\includegraphics*[width=70mm,angle=270]{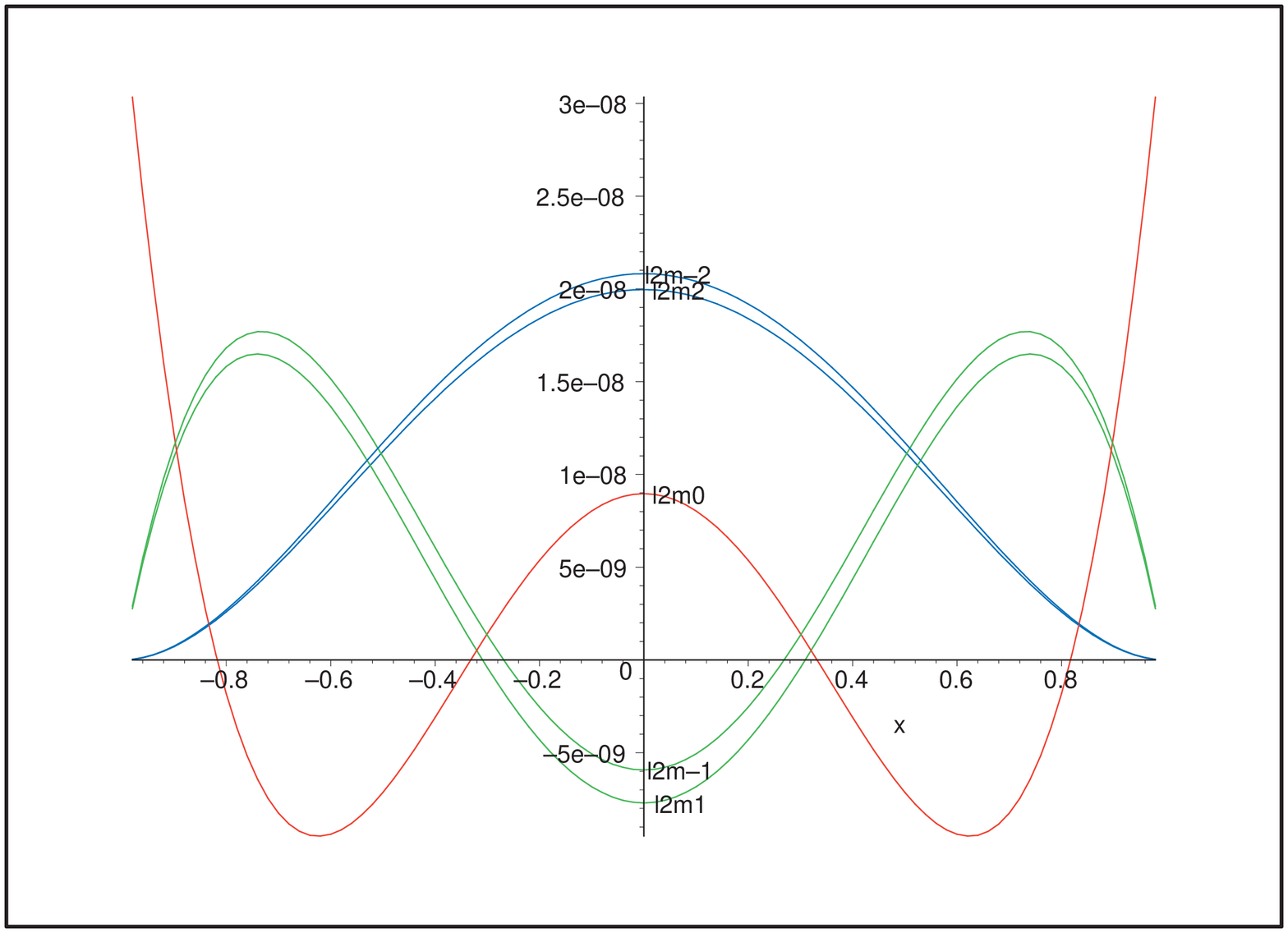} \\
$l=2$ at  $r \simeq 10.26$\\
\end{tabular}
\caption{
$\frac{1}{4\pi}\vac{\hat{T}_{\theta\theta}}{CCH^--U^-}$ where the sums over $l$ and $m$ have not been performed.
The graphs for the modes with $(l,|m|)$ and with $(l,-|m|)$ are very similar in shape. 
Close to the horizon the ones with $(l,-|m|)$ dominate.
}
\label{fig:Tthetatheta_cch_u_past_spher_last_r100r150_n2m}
\end{figure}

\begin{figure}[p]
\rotatebox{90}
\centering
\begin{tabular}{c}
\includegraphics*[width=70mm,angle=270]{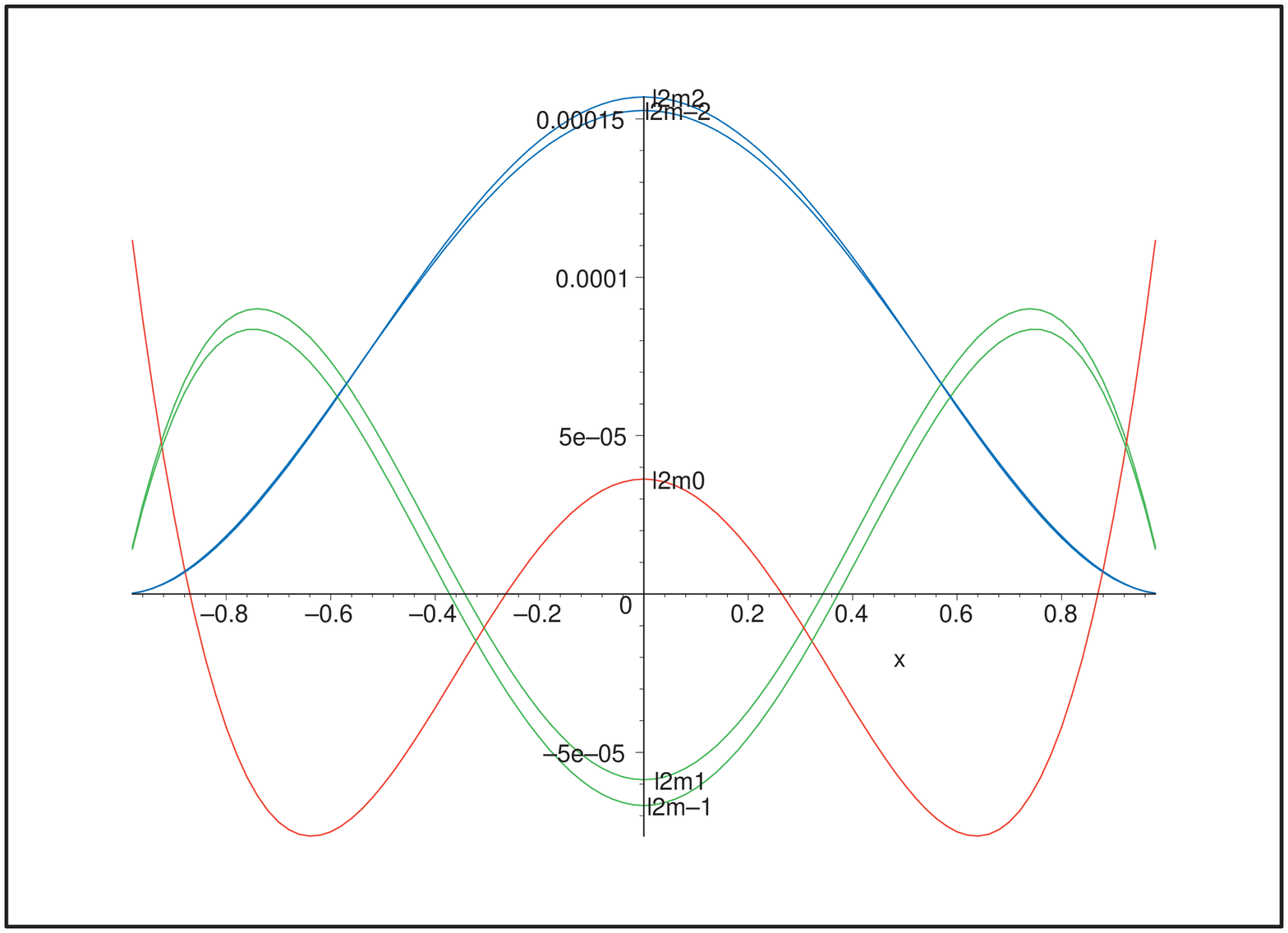} \\
$l=2$ at $r \simeq 1.36$ \\
\includegraphics*[width=70mm,angle=270]{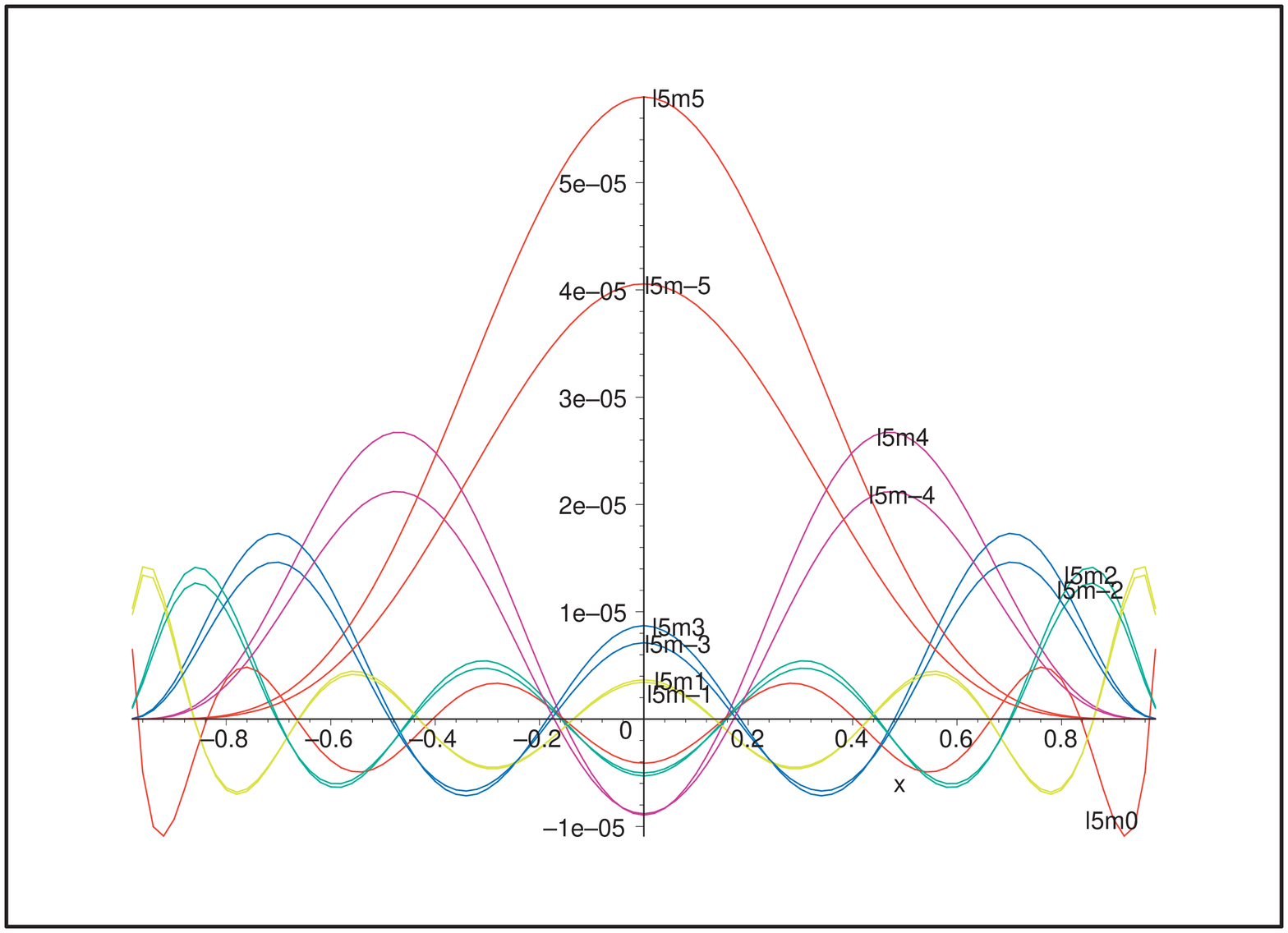} \\
$l=5$ at $r \simeq 1.36$\\
\end{tabular}
\caption{$\frac{1}{4\pi}\vac{\hat{T}_{\theta\theta}}{U^--B^-}$ where the sums over $l$ and $m$ have not been performed.
The graphs for the modes with $(l,|m|)$ and with $(l,-|m|)$ are very similar in shape and magnitude, both close
and far from the horizon.}
\label{fig:Tthetatheta_u_b_spher_last_r100_n2n5m}
\end{figure}

\begin{figure}[p]
\rotatebox{90}
\centering
\includegraphics*[width=70mm,angle=270]{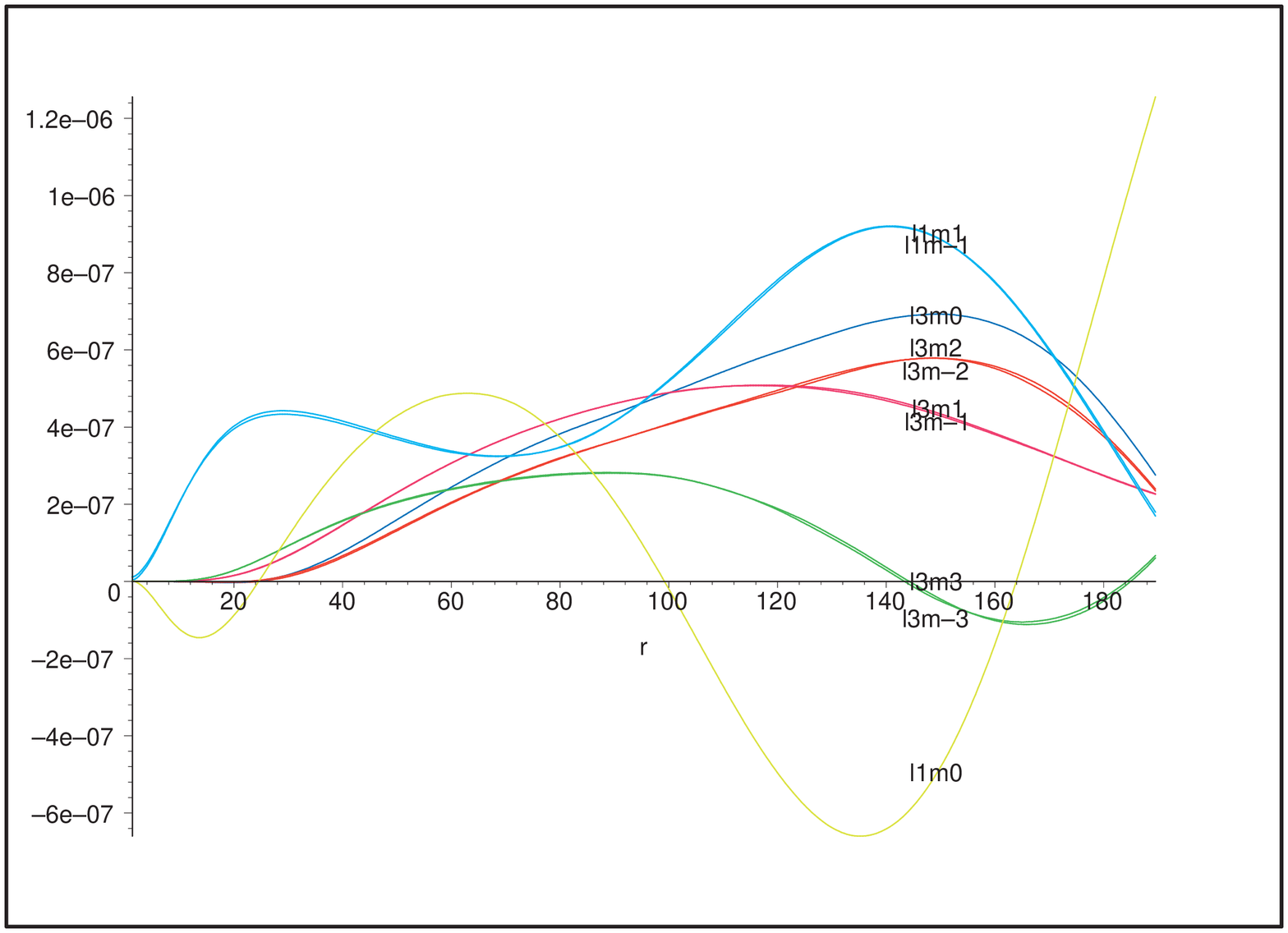} 
\caption{
$\frac{1}{4\pi}\vac{\hat{T}_{\theta\theta}(r,\theta=\pi/2)}{CCH^--U^-}$ for $l=1,3$ and $m=-l\to l$.
The modes $(l,|m|)$ and $(l,-|m|)$ are paired up far from the horizon, 
where they intertwine. One set does not dominate over the other in that region. 
Low-$l$ modes dominate over high-$l$ ones close to the horizon but neither set dominates far from the horizon. }
\includegraphics*[width=70mm,angle=270]{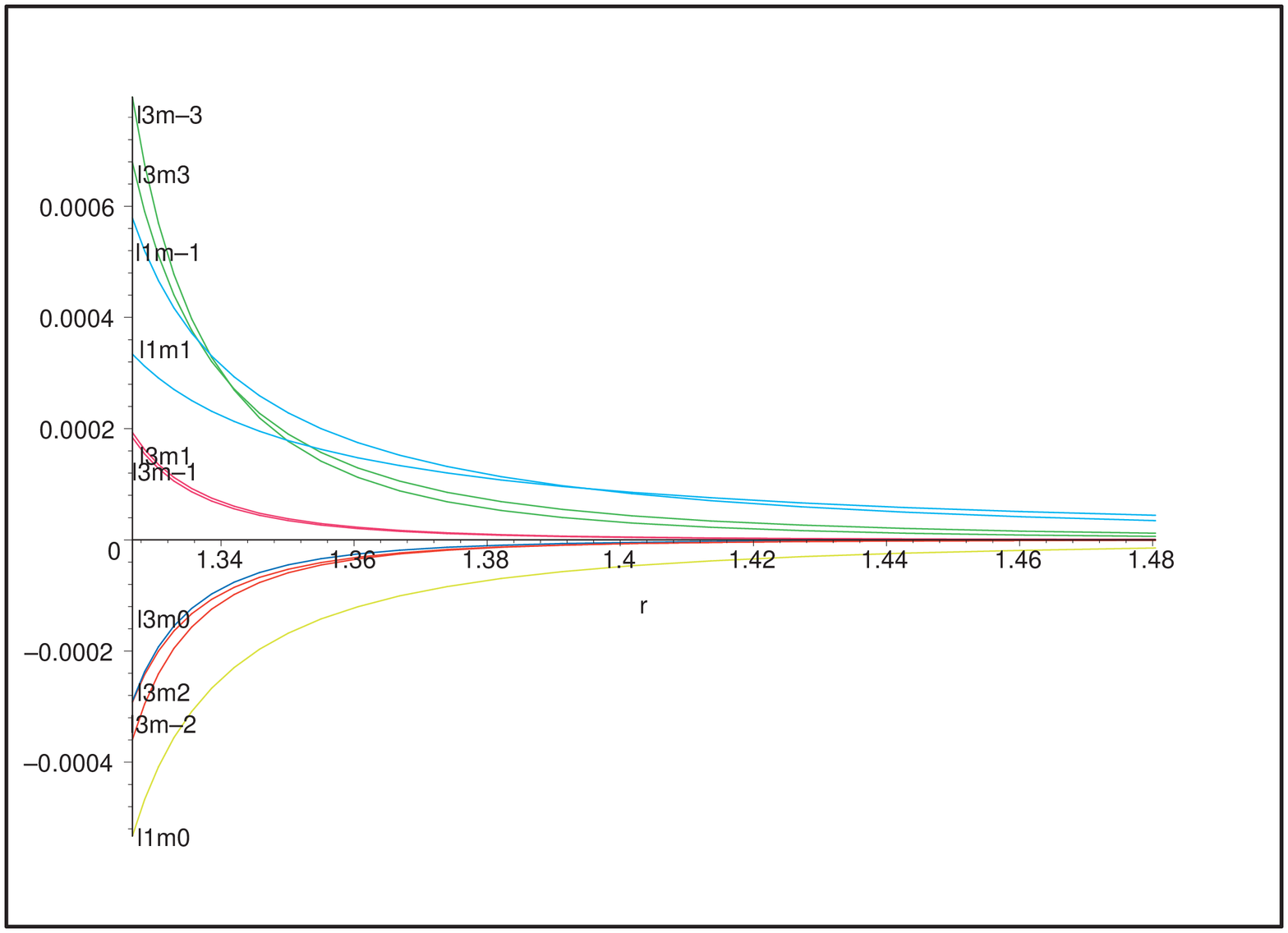}
\caption{
$\frac{1}{4\pi}\vac{\hat{T}_{\theta\theta}(r,\theta=\pi/2)}{U^--B^-}$ for $l=1,3$ and $m=-l\to l$.
Modes $(l,-|m|)$ dominate over $(l,-|m|)$ much closer to the the horizon than they do for the
`in' modes case. High-$l$ modes dominate over low-$l$ ones close to the horizon.
}
\label{fig:Tthetatheta_u_b_spher_last_t50_n1n3m}
\end{figure}

Graphs \ref{fig:Tthetatheta_cch_u_past_u_b_spher_last_r100r150_n}--\ref{fig:Tthetatheta_u_b_spher_last_t50_n1n3m}
show the behaviour of the various modes as the horizon is approached. 
Most of the features described in their captions are explained by the 
horizon asymptotics developed in Section \ref{sec:asympts. close to r_+}.


\draft{1) change in sign of $T_{tt}$ with CCH's expression must be checked out,
2) plot $\omega_{ZEFO}$ $\forall r$ and show that it goes to zero for large $r$!}

\catdraft{p.36eq.3.7CCH: sembla que hauria de posar comps. en RR coords., ja que $T_{tt}$, $T_{t\phi}$ i $T_{\phi\phi}$
tendeixen totes al mateix (tret de factor): $T_{t_+\phi_+}$??! de fet $T_{\phi_+\phi_+}=T_{\phi\phi}$ i $T_{t_+\phi_+}$
ja les poso=> la unica que no poso es $T_{t_+t_+}$, que incorriria en el mateix tipus d'extra round-off
error que $T_{t_+\phi_+}$}
\catdraft{hauria d'estar dividint per r amb que div. a infinit pq. aixi pendent es mes petit, graphs serien mes smooth, i
podria dibuixar fins a vals. de $r$ mes llunyans?!->com puc saber amb quin ordre de $r$ van si se que no van com thermal stress tensor?}
\catdraft{Adrian: figs. \ref{fig:delta2Ttphi_cch_b_past} and \ref{fig:delta2Tphiphi_cch_b_past} son sorprenentment lineals}



\section{Symmetry $(\theta,\phi)\to (\pi-\theta,\phi+\pi)$ } \label{sec:theta->pi-theta Symmetry}

When we initially used the expressions (\ref{eq:stress tensor for s=1 on all vac.}) given by CCH for the calculation of the
difference in the RSET when the field is in two different states, we found to our surprise that the
results were not symmetric under the parity operation $\mathcal{P}:(\theta,\phi)\to (\pi-\theta,\phi+\pi)$.
Analytically, there was a strong indication that the results were not symmetric under $\mathcal{P}$, although 
it was hard to prove the lack of symmetry. Numerically, the results obtained were clearly not symmetric
under the parity operation.

In the first subsection of this last section we analytically prove that the results obtained
using CCH's expressions (\ref{eq:stress tensor for s=1 on all vac.}) are not symmetric under $\mathcal{P}$.
In the second subsection we find that the reason for this lack of symmetry is that expressions 
(\ref{eq:stress tensor for s=1 on all vac.}) are not correct, and we find the correct expressions.
In the third and last subsection we give a physical interpretation of the different sets of terms
in the expressions for the expectation value of the stress-energy tensor in various states.

Even though the full parity operation involves a transformation in both angular variables $\theta$ and $\phi$,
it is clear that we only need to consider the transformation in $\theta$ as regards to the stress-energy tensor.


\subsection{Lack of symmetry} \label{subsec:lack of symmetry}

The invariance under the transformation $(\theta \to \pi -\theta)$ is straight-forwardly satisfied on the Schwarzschild 
background by the RSET of a field of any spin when the field is in any of the states in expressions (\ref{eq:stress tensor for s=1 on all vac.}).
The reason for this invariance in the Schwarzschild background is that when $a=0$, the angular function does not depend on $\omega$ 
and the radial function does not depend on $m$. Therefore, applying the 
transformation $(\theta \to \pi -\theta)$ on any mode in (\ref{eq:stress tensor for s=1 on all vac.}) 
is equivalent to performing $(m \to -m)$ only, by virtue of the relations (\ref{eq:R symm.->cc,-m,-w})
and (\ref{eq:S symm.->pi-t,-m,-w}) and the reality of the stress tensor.
The sum over $m$ appearing in (\ref{eq:stress tensor for s=1 on all vac.})
then guarantees the invariance of those expressions under $(\theta \to \pi -\theta)$, and therefore under $\mathcal{P}$.

Such a straight-forward reasoning does not follow in the Kerr background because applying the parity operation on the
expressions (\ref{eq:stress tensor for s=1 on all vac.}) implies a change in the sign of $\omega$ as well as in the sign of $m$.
Whereas the sum over $m$ is symmetric with respect to $m=0$, the integration over the frequency is not symmetric
with respect to $\omega=0$. In order to investigate the symmetry, or otherwise, under $\mathcal{P}$ of the expectation values 
(\ref{eq:stress tensor for s=1 on all vac.}) on the Kerr background, we will calculate these expectation values
evaluated at the point $(r,\theta)$ minus their value at the point $(r,\pi-\theta)$.
This procedure is obviously not useful for those components of the stress-energy tensor such that one index is 
$\theta$ and the other one is not. For these components, the symmetry under parity should be investigated by adding
the value of the expectation value at $(r,\theta)$ to that at $(r,\pi-\theta)$.

The classical stress-energy tensor (\ref{eq:stress tensor, spin 1}) is made up of the sum of various terms that are quadratic in the field.
We conveniently gather these terms into groups that appear when expressing the stress-energy tensor in
Boyer-Lindquist co-ordinates.
We then calculate the difference between the value at $(r,\theta)$ and the value at $(r,\pi-\theta)$ of these groups of terms, mode by mode. 
The following are useful expressions for some of such differences:

\begin{subequations}  \label{eq:stress tensor terms t->pi-t}
\begin{align}
\begin{split}
&\frac{\abs{{}_{lm\omega}\phi_{-1}(r,\theta)}^2}{4}+\frac{\abs{{}_{lm\omega}\phi_{+1}(r,\theta)}^2}{\Delta^2\Sigma^2}
-\tpit= \\
&\quad\quad\quad =\frac{-iK}{2B\Delta^2}\left({}_{+1}S_{lm\omega}^2-{}_{-1}S_{lm\omega}^2\right)W[Y_{+1},Y_{-1}^{*}]_{lm\omega} 
\end{split} \label{eq: phi0^2+phi2^2 t->pi-t}  \\
\begin{split}
&\frac{|{}_{lm\omega}\phi_{-1}(r,\theta)|^2\Delta}{4\Sigma}-\frac{|{}_{lm\omega}\phi_{+1}(r,\theta)|^2\Sigma}
{\Delta}-\tpit=\\
&\quad\quad\quad =\frac{\Delta}{4\Sigma}\left({}_{+1}S_{lm\omega}^2-{}_{-1}S_{lm\omega}^2\right)
\left(|{}_{+1}R_{lm\omega}|^2+\frac{\Sigma^4}{\Delta^2}|{}_{-1}R_{lm\omega}|^2\right)
\end{split} \label{eq: phi0^2-phi2^2 t->pi-t}  \\
\begin{split}
&|{}_{lm\omega}\phi_{0}(r,\theta)|^2-\tpit=\\ 
&\quad\quad\quad =\frac{-iaW[Y_{+1},Y_{-1}^{*}]_{lm\omega}}
{2\Sigma^2{}_1B_{lm\omega}^2}
\left[
2\cos \theta (\mathcal{L}^{\dagger}_{1}{}_{-1}S_{lm\omega})(\mathcal{L}_{1}{}_{+1}S_{lm\omega})+
\right. \\ 
&\quad\quad\quad\left.
+\sin \theta (
{}_{+1}S_{lm\omega}(\mathcal{L}^{\dagger}_{1}{}_{-1}S_{lm\omega})+
{}_{-1}S_{lm\omega}(\mathcal{L}^{\dagger}_{1}{}_{+1}S_{lm\omega})
)
\right]
\end{split}\label{eq: phi1^2 t->pi-t}  \\
&{}_{lm\omega}\phi_{+1}(r,\theta){}_{lm\omega}\phi_{-1}^*(r,\theta)\frac{1}{\rho^2}-\tpit=0 \label{eq: phi2phi0 t->pi-t}
\end{align}
\end{subequations}
We are adopting the notation that the symbol $\tpit$ at the end of an expression 
represents all the previous terms in that expression being evaluated at $\pi-\theta$ instead of $\theta$.
Note that we are not using any particular boundary conditions for the radial funcions.
We therefore do not include any constants of normalization and will only include them when we wish to calculate the
expectation value of the stress-energy tensor.

An outline of two useful properties that some of these groups of terms possess is given in Table \ref{table:individual terms t->pi-t}.
When the stress-energy tensor is expressed in Boyer-Lindquist co-ordinates and its value at $(r,\pi-\theta)$ is subtracted from the one at $(r,\theta)$, 
there exist two other groups of terms apart from those in (\ref{eq:stress tensor terms t->pi-t})
which we have included in Table \ref{table:individual terms t->pi-t} but which do not possess any of the two properties in question.
Table \ref{table:comps. contain what individual terms t->pi-t} shows which groups of terms appear for each component of the stress tensor. 

\begin{table}[p]
\centering
\rotatebox{90}{
\begin{tabular}{cc||c|c}
&
& has $\left({}_{+1}S_{lm\omega}^2-{}_{-1}S_{lm\omega}^2\right)$? & has $W[Y_{+1},Y_{-1}^{*}]_{lm\omega}$? \\
\hline                                                                   
\hline                                                                   
(a)& $\frac{\abs{{}_{lm\omega}\phi_{-1}}^2}{4}+\frac{\abs{{}_{lm\omega}\phi_{+1}}^2}{\Delta^2\Sigma^2}
-\tpit$ & Y & Y    \\
\hline                                                                   
(b)& $\frac{|{}_{lm\omega}\phi_{-1}|^2\Delta}{4\Sigma}-\frac{|{}_{lm\omega}\phi_{+1}|^2\Sigma}{\Delta}
-\tpit$    &  Y & N \\         
\hline                                                                   
(c)& $|{}_{lm\omega}\phi_{0}|^2-\tpit$    &     N & Y  \\
\hline                                                                   
(d)& ${}_{lm\omega}\phi_{+1}\ {}_{lm\omega}\phi_{-1}^*\frac{1}{\rho^2}-\tpit$ & 
\multicolumn{2}{c}{0} 
\\  
\hline                                                                   
(e)& ${}_{lm\omega}\phi_0\ {}_{lm\omega}\phi_{-1}^*\frac{\Delta}{\Sigma}+
2{}_{lm\omega}\phi_{+1}\ {}_{lm\omega}\phi_0^*-\tpit$ & 
N&N
\\
\hline                                                                   
(f)& $-{}_{lm\omega}\phi_0\ {}_{lm\omega}\phi_{-1}^*\frac{\Delta}{\Sigma}+
2{}_{lm\omega}\phi_{+1}\ {}_{lm\omega}\phi_0^*-\tpit$ & 
N&N
\end{tabular}}
\caption{
This table indicates which of the groups of terms appearing in ${}_{lm\omega}T_{\mu\nu}(r,\theta)-\tpit$
in Boyer-Lindquist co-ordinates contain the factors $\left({}_{+1}S_{lm\omega}^2-{}_{-1}S_{lm\omega}^2\right)$ 
and $W[Y_{+1},Y_{-1}^{*}]_{lm\omega}$} \label{table:individual terms t->pi-t}
\end{table}

\begin{table}
\begin{center}
\begin{tabular}{rl}
$T_{\mu\nu}$ & \\
\hline
$t\phi$, $tt$, $\phi\phi$    & :(a),(c),(d),(e) \\
$tr$, $r\phi$                & :(b),(f) \\
$t\theta$, $\theta\phi$      & :(d),(e) \\
$rr$                         & :(a),(c) \\
$\theta\theta$               & :(c),(d) \\
$r\theta$                    & :(f)
\end{tabular}
\end{center}
\caption{
Groups of terms in Table \ref{table:individual terms t->pi-t} that appear in the expression for ${}_{lm\omega}T_{\mu\nu}(r,\theta)-\tpit$ for
each one of the components in Boyer-Lindquist co-ordinates.}
\label{table:comps. contain what individual terms t->pi-t}
\end{table}

We calculate in this subsection the expectation value of any quadratic operator when the field is in a certain state 
of interest with CCH's expressions (\ref{eq:quadratic op. for s=1 on vacua}).
In order to evaluate 
the difference in the RSET in a particular state between the points 
$(r,\theta)$ and $(r,\pi-\theta)$, we also need to know how $T^{\text{div}}_{\mu\nu}$ behaves under 
$(\theta \to \pi-\theta)$. 
Since $T^{\text{div}}_{\mu\nu}$ is a purely geometrical object and the metric is invariant under $\mathcal{P}$, this
divergent stress tensor must also be invariant under $\mathcal{P}$.  
This invariance implies that 
\begin{equation}
\vac{\hat{T}_{\mu\nu}(r,\theta)}{\Psi}-(-1)^{\vartheta}\vac{\hat{T}_{\mu\nu}(r,\pi-\theta)}{\Psi}=
\vac[ren]{\hat{T}_{\mu\nu}(r,\theta)}{\Psi}-(-1)^{\vartheta}\vac[ren]{\hat{T}_{\mu\nu}(r,\pi-\theta)}{\Psi}
\end{equation}
 where $\Psi$ represents any state. 
The variable $\vartheta$ is defined so that $(-1)^{\vartheta}$ is equal to -1 if one
index of the component of the stress tensor is $\theta$ and the other one is not, and it is equal to +1 otherwise. 
\draft{maybe easier to give expression for $\vac{\hat{\vec{T}}}{\Psi}$ instead?}

By virtue of the symmetries (\ref{eq: symm. of lmwphi_i}) and the property (\ref{eq: wronskian in=-up}),
the expectation value in certain states of operators corresponding to the groups of terms for which the 
radial functions only appear as part of a wronskian will adopt a particularly simple form.
This is the case for the groups of terms (a) and (c) in Table \ref{table:individual terms t->pi-t}.
Table \ref{table:terms t->pi-t with wronsk. in vacua} shows the form of the
expectation value of such operators when the electromagnetic field is in various states when expressions
(\ref{eq:stress tensor for s=1 on all vac.}) and (\ref{eq:stress tensor for s=1 on FT}) are used.

\begin{table}
\begin{tabular}{c|c}
$\ket{\Psi}$ & $\vac{\hat{Q}}{\Psi}$\\
\hline                                                                   
\hline                                                                   
$\ket{B^-}$ 
& 0                   \\
\hline                                                                   
$\ket{FT}$ & 
$-2\sum_{m=1}^l {\int_{0}^{m\Omega_+}\d{\omega}\coth(\frac{\pi\tilde{\omega}}{\kappa})}$   \\
\hline                                                                   
$\ket{CCH^-}$ & 
$-2\sum_{m=1}^l {\int_{0}^{m\Omega_+}\d{\omega}\coth(\frac{\pi\tilde{\omega}}{\kappa})}+
\sum_{m=-l}^l {\int_{0}^{\infty}\d{\omega}\left[\coth(\frac{\pi\tilde{\omega}}{\kappa})-
\coth(\frac{\pi \omega}{\kappa})\right]}$   \\
\hline
$\ket{U^-}$ & 
$-2\sum_{m=1}^l 
{\int_{0}^{m\Omega_+}\d{\omega}\coth(\frac{\pi\tilde{\omega}}{\kappa})}+
\sum_{m=-l}^l {\int_{0}^{\infty}\d{\omega}\left[\coth(\frac{\pi\tilde{\omega}}{\kappa})-1\right]}$              
\end{tabular}
\caption{
$\hat{Q}$ is any quadratic operator in the field and its derivatives such that all the radial
functions content of its classical counterpart 
$Q\left[{}_{lm\omega}\phi_{\indhel}^{\bullet},{}_{lm\omega}\phi_{\indhel}^{\bullet *}\right]$
can be expressed as a wronskian.
For clarity, $\sum_{lP}$ has been omitted from the sums and 
$Q\left[{}_{lm\omega}\phi_{\indhel}^{\text{up}},{}_{lm\omega}\phi_{\indhel}^{\text{up} *}\right]$
from all integrands.}
\label{table:terms t->pi-t with wronsk. in vacua}              
\end{table}



We now proceed to analytically prove the lack of symmetry under parity of the RSET for the easiest case.
Tables \ref{table:individual terms t->pi-t} and \ref{table:comps. contain what individual terms t->pi-t} 
reveal that the easiest component of the RSET for which to investigate the symmetry under parity is the $\theta\theta$-component.
The easiest point where to evaluate the stress-energy tensor is at the axis of symmetry. 
We are therefore going to calculate ${}_{lm\omega}T_{\theta\theta}(r,\theta=0)-{}_{lm\omega}T_{\theta\theta}(r,\theta=\pi)$,
which only involves term (c) in Table \ref{table:individual terms t->pi-t}.

From equations (\ref{eq:asympt. y for x->+/-1}) and (\ref{eq:def. L_n}) it follows that
\begin{subequations}
\begin{align}
\mathcal{L}_{n}^{\topbott{}{\dagger}}{}_{\indhel}S_{lm\omega} 
&\sim
2^{(\beta-1/2)}{}_{\indhel}a_{n=0,lm\omega}\left(2\alpha\pm m+n\right)(1-x)^{(\alpha-1/2)} & (x \rightarrow +1)  
\label{eq:L_1 S, x->+1} \\
\mathcal{L}_{n}^{\topbott{}{\dagger}}{}_{\indhel}S_{lm\omega} 
&\sim
2^{\left(\alpha-1/2\right)}
{}_{\indhel}b_{n=0,lm\omega}\left(-2\beta\pm m-n\right)(1+x)^{(\beta-1/2)}                 & (x \rightarrow -1)
\label{eq:L_1 S, x->-1}
\end{align}
\end{subequations}
We insert these equations into the two equivalent expressions for ${}_{lm\omega}\phi_{0}$ given in (\ref{eq:phi0(ch)}).
These expressions refer to either `in' or `up' modes and we will temporarily follow the same normalization
as in ~\cite{bk:Chandr}. This normalization is obtained by replacing ${}_{+1}R^{\text{in}}_{lm\omega}$ by 
$-2\sqrt{2\pi}{}_{+1}R_{lm\omega}$ and ${}_{-1}R^{\text{in}}_{lm\omega}$ by 
$-\sqrt{2\pi}{}_{-1}R_{lm\omega}/{}_1B_{lm\omega}$. Alternatively, this normalization can be obtained from an expression for `up' modes
by replacing ${}_{+1}R^{\text{up}}_{lm\omega}$ by $-2\sqrt{2\pi}{}_{+1}R_{lm\omega}/{}_1B_{lm\omega}$ and ${}_{+1}R^{\text{up}}_{lm\omega}$ by 
$-\sqrt{2\pi}{}_{+1}R_{lm\omega}/{}_1B_{lm\omega}^2$.  
We will later restore the appropriate factors.
We can then deduce the behaviour at the axis of ${}_{lm\omega}\phi_{0}$ in this normalization:
\begin{equation} \label{eq:phi_1, x->+/-1}
\begin{aligned}
&{}_{lm\omega}\phi_{0}\left(r,\theta=\topbott{0}{\pi}\right)
\sim
 \\ &  
\sim
\frac{\left(\pm \abs{m\pm 1}+m\pm 1\right)2^{\left(\abs{m\mp 1
}/2-1\right)}{}_{+1}\topbott{a}{b}_{n=0,lm\omega}\left(1\mp x \right)^{\left(\abs{m\pm 1
}-1\right)/2}}{\left(r\mp ia\right)^2{}_1B_{lm\omega}}\times \\
& \qquad \qquad \qquad \qquad \qquad \qquad \qquad \times\left[\left(r\mp ia\right)\mathcal{D}_0-1\right]{}_{-1}R_{lm\omega}= 
 \\ & =
\frac{-\left(\pm\abs{m\mp 1}-m\pm1\right)2^{\left(\abs{m\pm
1}/2-1\right)}{}_{-1}\topbott{a}{b}_{n=0,lm\omega}\left(1\mp x\right)^{\left(\abs{m\mp 1
}-1\right)/2}}{\left(r\mp ia\right)^2{}_1B_{lm\omega}}\times  \\
& \qquad \qquad \qquad \qquad \qquad \qquad \qquad \times\left[\left(r\mp ia\right)\mathcal{D}_0^{\dagger}-1\right]\left(\Delta{}_{+1}R_{lm\omega}\right)
\qquad \left(x \rightarrow \pm 1\right)
\end{aligned}
\end{equation}
\catdraft{1) comprovar eq. 1's que he afegit a exponents a (\ref{eq:phi_1, x->+/-1}) siguin correctes, 2) aquesta
expressio es `ch' pero obtinguda a partir d'expressio per `in' i `up', explicar com!!}

By looking at the coefficient and the exponent of $\left(1 \mp x\right)$ in (\ref{eq:phi_1, x->+/-1}) 
we can see that ${}_{lm\omega}\phi_{0}$ is only non-zero at $\theta=0,\pi$ if $m=0$.
The value of this NP scalar in that case is
\begin{equation} \label{eq:phi_1, x-=+/-1}
\begin{aligned}
{}_{l,m=0,\omega}\phi_{0}\left(r,\theta=\topbott{0}{\pi}\right)
&=\frac{\pm2{}_{+1}\topbott{a}{b}_{n=0,l,m=0,\omega}}{\sqrt{2}\left(r\mp ia\right)^2{}_1B_{lm\omega}}
\left[\left(r\mp ia\right)\mathcal{D}-1\right]{}_{-1}R_{l,m=0,\omega}= \\
&=\frac{\mp 2{}_{-1}\topbott{a}{b}_{n=0,l,m=0,\omega}}{\sqrt{2}\left(r\mp ia\right)^2{}_1B_{lm\omega}}
\left[\left(r\mp ia\right)\mathcal{D}^{\dagger}-1\right]\left(\Delta{}_{+1}R_{l,m=0,\omega}\right)
\end{aligned}
\end{equation}

Proceeding now similarly to the way we did to obtain (\ref{eq: phi1^2 t->pi-t}), we have that at the axis of symmetry
\begin{equation} \label{eq: phi1^2 0->pi}  
\begin{aligned}
&\abs{{}_{l,m=0,\omega}\phi_{0}(r,\theta=0)}^2-\abs{{}_{l,m=0,\omega}\phi_{0}(r,\theta=\pi)}^2= \\
&=\frac{-4iaW[Y_{+1},Y_{-1}^{*}]_{l,m=0,\omega}({}_{-1}a_{n=0,l,m=0,\omega}\ {}_{+1}a_{n=0,l,m=0,\omega})}
{(r^2+a^2)^2{}_1B_{l,m=0,\omega}^2}
\end{aligned}
\end{equation}
is satisfied.

As we have seen, Table \ref{table:terms t->pi-t with wronsk. in vacua} applies to the 
term $\abs{{}_{lm\omega}\phi_{0}(r,\theta=0)}^2-\abs{{}_{lm\omega}\phi_{0}(r,\theta=\pi)}^2$,
and thus to ${}_{lm\omega}T_{\theta\theta}(r,\theta=0)-{}_{lm\omega}T_{\theta\theta}(r,\theta=\pi)$. 
Since this term is zero at the axis for $m\neq 0$, this table shows that 
the $\theta\theta$-component of the RSET when the field is in the states $\ket{FT}$ or $\ket{CCH^-}$
(as well, of course, as in the state $\ket{B^-}$) is invariant under parity at the axis.
Only in the state $\ket{U^-}$ this component might not be invariant at the axis.
From (\ref{eq:stress tensor, spin 1}), (\ref{eq: phi1^2 0->pi}) and Table \ref{table:terms t->pi-t with wronsk. in vacua} 
we can finally find a simple expression for the difference in the $\theta\theta$-component of the RSET in the past Unruh state 
evaluated at $\theta=0$ and at $\theta=\pi$. 
This expression, where we now include the constant of normalization and we restore the appropriate factors for the `in' and `up' modes, is:
\begin{equation} \label{eq:T_thetatheta 0->pi,past Unruh}
\begin{aligned}
&\vac[ren]{\hat{T}_{\theta\theta}(r,\theta=0)}{U^-}-\vac[ren]{\hat{T}_{\theta\theta}(r,\theta=\pi)}{U^-}= \\
&=\sum_{l=0}^{\infty}
\int_{0}^{\infty}\d{\omega}\left[\coth\left(\frac{\pi \omega}{\kappa}\right)-1\right]
\times \\ & \qquad \quad\times
\frac{-2^3ia|N^{\text{up}}_{+1}|^2W[Y_{+1}^{\text{up}},Y_{-1}^{\text{up} *}]_{l,m=0,\omega}({}_{-1}a_{n=0,l,m=0,\omega}\ {}_{+1}a_{n=0,l,m=0,\omega})}
{(r^2+a^2)^2{}_1B_{l,m=0,\omega}^2}=      \\
&=\sum_{l=0}^{\infty}
\int_{0}^{\infty}\d{\omega}\left[\coth\left(\frac{\pi \omega}{\kappa}\right)-1\right]
\times \\ & \qquad \quad \times
\frac{2^3ia|N^{\text{up}}_{+1}|^2W[Y_{+1}^{\text{up}},Y_{-1}^{\text{up} *}]_{l,m=0,\omega}\ {}_{+1}a_{n=0,l,m=0,\omega}^2}
{(r^2+a^2)^2{}_1B_{l,m=0,\omega}^2}\frac{\sqrt{{}_{-1}\lambda_{l,m=0,\omega}+2a\omega}}{\sqrt{{}_{-1}\lambda_{l,m=0,\omega}-2a\omega}}
\end{aligned}
\end{equation}
where in the last step we have made use of (\ref{eq:S1/S_1,x->1}).

From Table \ref{table:radial wronsks} we can see that 
$iW[Y_{+1}^{\text{up}},Y_{-1}^{\text{up} *}]_{l,m=0,\omega}=
\frac{{}_1B^2_{l,m=0,\omega}}{\omega}\abs{{}_{-1}R^{\text{up,tra}}_{l,m=0,\omega}}^2 \ge 0$ 
as long as 
$\omega \ge 0$, and therefore
the integrand in (\ref{eq:T_thetatheta 0->pi,past Unruh}) is non-negative for $\omega \ge 0$.

Proceeding similarly for the other groups of terms in Table \ref{table:individual terms t->pi-t}, we
can see that (e) and (f) will be zero at the axis after summing over $m$ since 
${}_{lm\omega}\phi_{-1}$, ${}_{lm\omega}\phi_{0}$ and ${}_{lm\omega}\phi_{+1}$ are only non-zero at $\theta=\topbott{0}{\pi}$ 
when $m=\topbott{-1}{+1}$, $\topbott{0}{0}$ and $\topbott{+1}{-1}$ respectively. 
This result is obviously equally valid if the expectation value of the quadratic term at the point $\pi-\theta$
is added, rather than subtracted, to that at the point $\theta$, so that this result is also useful for those components of the
stress tensor such that one index is $\theta$ and the other one is not.

On the other hand, we are not able to prove whether the groups of terms (a) and (b) are zero or not at the axis.
The reason is the presence of the factor $\left({}_{+1}S_{lm\omega}^2-{}_{-1}S_{lm\omega}^2\right)$:
after the summation over $m$ we have two separate terms,
one for the mode $m=+1$ and the other for the mode $m=-1$. It is not possible to combine together these two modes using the symmetries
(\ref{eq: S symms}) of the angular function unless the transformation $(\omega \to -\omega)$ is also applied.

A summary of the analytical results relating to the symmetry under $\mathcal{P}$ of the Boyer-Lindquist components of
the RSET in various states of interest when expressions (\ref{eq:stress tensor for s=1 on all vac.}) 
and (\ref{eq:stress tensor for s=1 on FT}) are used, is as follows:
\begin{itemize}
\item $\vac[ren]{\hat{T}_{rr}}{B^-}$ and $\vac[ren]{\hat{T}_{\theta\theta}}{B^-}$ are both symmetric under $\mathcal{P}$ everywhere, and \\
      $\vac[ren]{\hat{T}_{tt}}{B^-}$, $\vac[ren]{\hat{T}_{t\phi}}{B^-}$ and $\vac[ren]{\hat{T}_{\phi\phi}}{B^-}$ 
are symmetric under $\mathcal{P}$ at the axis.
\item $\vac[ren]{\hat{T}_{r\theta}}{\Psi}$,
$\vac[ren]{\hat{T}_{t\theta}}{\Psi}$, $\vac[ren]{\hat{T}_{\phi\theta}}{\Psi}$ where $\ket{\Psi}$ may be any one state among 
$\ket{B^-}$, $\ket{FT}$, $\ket{CCH^-}$ ,$\ket{U^-}$ are all symmetric under $\mathcal{P}$ at the axis.
\item $\vac[ren]{\hat{T}_{\theta\theta}}{\Psi}$ where $\Psi$ may be any one state among $\ket{B^-}$, $\ket{FT}$ ,$\ket{CCH^-}$ 
are all symmetric under $\mathcal{P}$ at the axis.
\item $\vac[ren]{\hat{T}_{\theta\theta}}{U^-}$ is not symmetric under $\mathcal{P}$ at the axis.
\end{itemize}

To conclude this subsection, we consider two analytic results in the literature for which the electromagnetic 
RSET on the Kerr background exhibits an invariance under parity. 


One result is, of course, CCH's asymptotic result (\ref{eq:eq.3.7CCH;mine}), which we know is only valid at the poles.
We have already seen in Section \ref{sec:RRO}
that the replacement of spheroidal functions by spherical functions
together with the use of the asymptotic behaviour (\ref{eq:approx eqA6Cand'80}) at the horizon for the radial functions leads to
stress-energy tensor components for the `up' modes that are symmetric under $\mathcal{P}$ to leading order.
Indeed, if the wronskian is calculated with the asymptotics at the horizon for the `up' radial functions, its leading
order behaviour is zero. Therefore, groups of terms (a) and (c) in Table \ref{table:individual terms t->pi-t} 
for the `up' modes are zero to leading order at the horizon.
The group of terms (b) is zero to leading order at the poles for the `up' modes because the spheroidal functions 
can be replaced by spherical functions. All the groups of terms in Table \ref{table:individual terms t->pi-t}  
for the `up' modes are therefore either identically zero or zero to leading order at the poles. 
It follows that the leading order at the poles of all components of the stress-energy tensor for the `up' modes 
are symmetric under parity.

This means that $\vac[ren]{\hat{T}_{\theta\theta}}{U^-}$ is symmetric under $\mathcal{P}$ at the poles.
However, since there is no divergence along the axis off the horizon, modes other than those with
$l\rightarrow +\infty$ contribute to (\ref{eq:T_thetatheta 0->pi,past Unruh}), making it non-zero.
Even though $\vac[ren]{\hat{T}_{\theta\theta}}{U^-}$ is symmetric under $\mathcal{P}$ at the poles, 
it is not symmetric along the axis off the horizon.

The other result we wish to mention was obtained by Frolov and Zel'nikov ~\cite{ar:F&Z'85}. 
They calculated the electromagnetic RSET at the pole $(r=r_+,\theta=0)$ when the field is in the state $\ket{FT}$. 
We used their method to obtain the same result at the other pole, $(r=r_+,\theta=\pi)$.
The electromagnetic RSET when the field is in the state $\ket{FT}$ is therefore symmetric under parity at the poles.
The result they find at the pole is finite and we therefore cannot apply the same reasoning as above. 
We have seen that the $r\theta$, $t\theta$, $\phi\theta$ and $\theta\theta$ components should all be symmetric
at the axis when the field is in the state $\ket{FT}$ but unfortunately we do not have an explanation 
for the symmetry of the other components at the axis.

\catdraft{$S_s$ can be replaced by $Y_s$ only for bounded m (therefore only at the axis) and bounded w so
explanation for $B^-$ is only valid at the axis?!->I'm not able to explain symmetry for $B^-$ at the horizon
off the axis the same way CCH were not able to find analytic asymptotics in such region}

\catdraft{$\vac[ren]{T^{\mu}_{\nu}}{CCH^-}$ in  ~\cite{ar:F&Z'85} at the poles is not irregular$=>$probably
the $l->\infty$ reasoning is not valid for this case?! how come that it is regular in ~\cite{ar:F&Z'85}
but it is not supposed to be (Winst\&Ott explanation relies on there being only non-superradiant modes
at the axis but there are $m=\pm 1$ ones)?!}

\catdraft{CCH is not supposed to be regular at the horizon but 
$\vac[ren]{T^{\mu}_{\nu}}{U^-} \sim \vac{T^{\mu}_{\nu}}{U^-}-\vac{T^{\mu}_{\nu}}{CCH^-}$ is still valid 
because its divergence is probably smaller than that of U's, but this has not been proved??no, order of irregularity is the same-
Not according to p.14Ott\&Winst'00??!}

\begin{figure}[p]
\rotatebox{90}
\centering
\includegraphics*[width=70mm,angle=270]{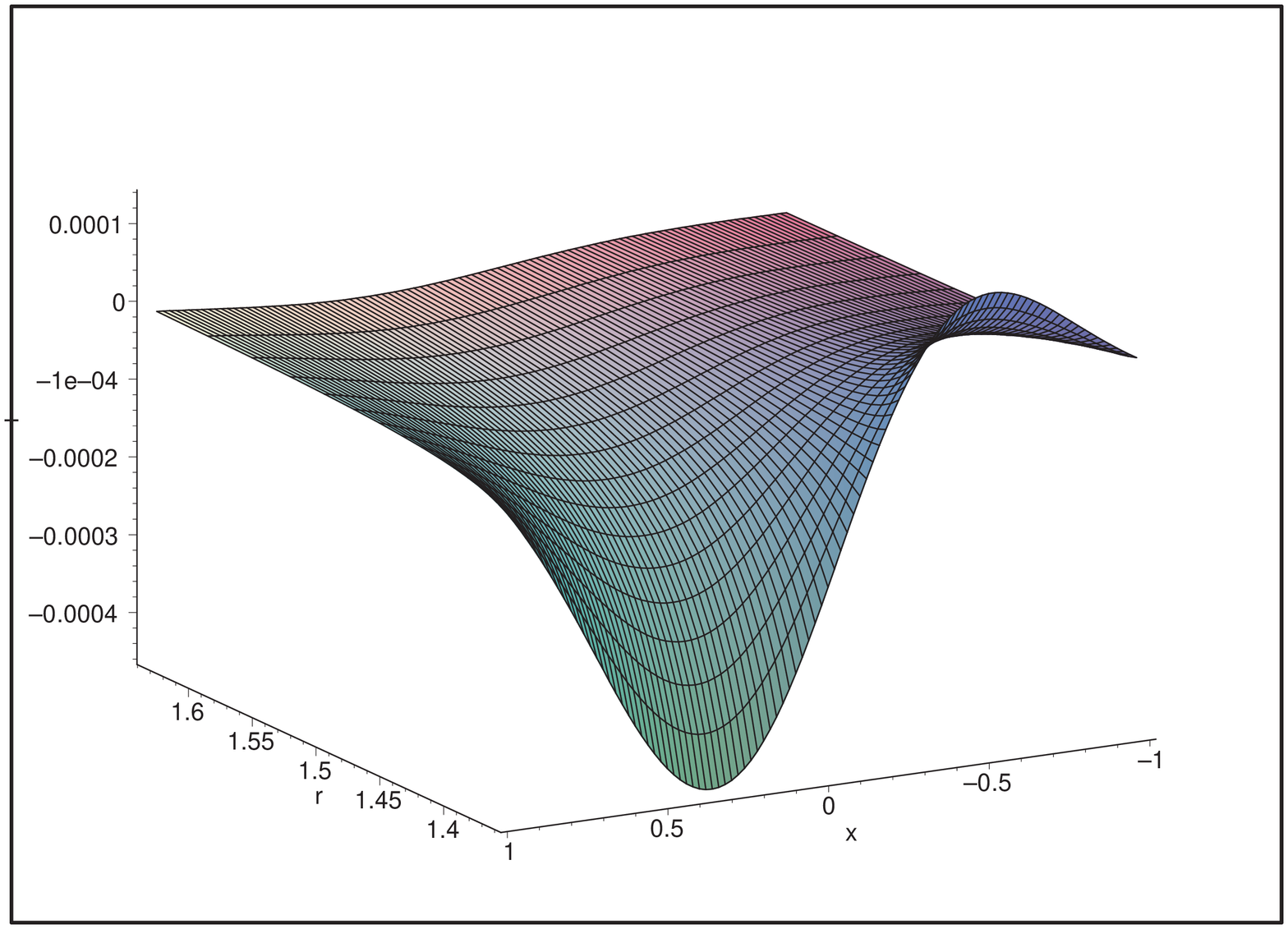}
\caption{$\frac{1}{4\pi}\vac{\hat{T}_{tr}}{U^--B^-}$}
\label{fig:Ttr_u_b_spher_last_r100to115}
\end{figure}

We include one graph for one of the components, the $tr$-component, for the difference in the RSET between the 
past Unruh and past Boulware states as an example of the clear lack of symmetry under the parity operation for most components.
Graph \ref{fig:Ttr_u_b_spher_last_r100to115} has been obtained using the expressions in (\ref{eq:stress tensor for s=1 on all vac.}).

\subsection{New expressions for the quantization of the field}

Having proved that CCH's expressions (\ref{eq:stress tensor for s=1 on all vac.}) lead to expectation values
of the stress-energy tensor which are not invariant under the parity operation, in the present subsection we will find
the reason for this asymmetry.

Note that, from (\ref{eq: symm. of lmwA_mu}), the potential mode
${}_{lm\omega P}A_{\mu}^{\bullet}$ is indeed an
eigenfunction of the parity operator. However, the general solution 
$A_{\mu}^{\bullet}$, for which no boundary conditions have been specified, is not:
\begin{equation}
\mathcal{P}A_{\mu}^{\bullet}=\sum_{lmP} \int_{-\infty}^{+\infty}\d{\omega^{\bullet}}P{}_{lm\omega P}a^{\bullet}{}_{lm\omega P}A_{\mu}^{\bullet}\neq
\pm\sum_{lmP} \int_{-\infty}^{+\infty}\d{\omega^{\bullet}}{}_{lm\omega P}a^{\bullet}{}_{lm\omega P}A_{\mu}^{\bullet}=\pm A_{\mu}^{\bullet}
\end{equation}    
Similarly, from (\ref{eq: symm. of lmwphi_i}), one NP scalar mode is complex-conjugated (and $m$ and $\omega$ change sign)
under the parity operation but the general solution is not:
\begin{equation}
\begin{aligned}
\mathcal{P}\phi_{\indhel}^{\bullet}&=
(-1)^{\indhel+1}\sum_{lmP} \int_{-\infty}^{+\infty}\d{\omega^{\bullet}}P{}_{lm\omega P}a^{\bullet *}{}_{lm\omega}\phi_{\indhel}^{\bullet *}\neq  
\\ & \neq
\pm \sum_{lmP} \int_{-\infty}^{+\infty}\d{\omega^{\bullet}}{}_{lm\omega P}a^{\bullet *}{}_{lm\omega}\phi_{\indhel}^{\bullet *}=
\pm\phi_{\indhel}^{\bullet *}
\end{aligned}
\end{equation}  

We will now call ${}_{lm\omega P}\Phi_{\indhel}^{\bullet}$ the integrand in 
(\ref{eq:classical mode expansion for phi(in)_i}). We immediately have that
\begin{equation}
\begin{array}{l} \label{eq: symm. of lmwP_phi_i}
\mathcal{P} {}_{lm\omega P}\Phi_{\indhel}^{\bullet}=(-1)^{\indhel+1}P{}_{lm\omega P}\Phi_{\indhel}^{\bullet *} \\
 \mathcal{P} \left({}_{lm\omega P}\Phi_{\indhel}^{\bullet}\ {}_{l'm'w'P'}\Phi_{\indhel'}^{\bullet *}+c.c.\right)=
(-1)^{\indhel+\indhel'}PP'\left({}_{lm\omega P}\Phi_{\indhel}^{\bullet *}\ {}_{l'm'w'P'}\Phi_{\indhel'}^{\bullet}+c.c.\right)
\end{array}
\end{equation}
and therefore the latter is an eigenfunction of the parity operator.

We look next at the quantized expressions, obtained by promoting the coefficients ${}_{lm\omega P}a^{\bullet}$ and
${}_{lm\omega P}a^{\bullet *}$ to operators.
There is, however, an operator-ordering ambiguity in the transition.
The classical term $\left(\phi_{\indhel}\phi_{\indhel'}^*+c.c.\right)$ should be quantized to
the symmetrized form 
$\left(\hat{\phi}_{\indhel}\hat{\phi}_{\indhel'}^{\dagger}+\hat{\phi}_{\indhel'}^{\dagger}\hat{\phi}_{\indhel}\right)/2+h.c.$ ,
where the symbol $`h.c.'$ stands for hermitian conjugate. 
There is no physical reason why the
classical term $\left(\phi_{\indhel}\phi_{\indhel'}^*+c.c.\right)$ should be quantized
to one particular choice between 
$\left(\hat{\phi}_{\indhel}\hat{\phi}_{\indhel'}^{\dagger}+h.c.\right)$
and
$\left(\hat{\phi}_{\indhel'}^{\dagger}\hat{\phi}_{\indhel}+h.c.\right)$.
However, in the expression that CCH give for the electromagnetic stress-energy tensor operator,
the choice of one option over the other seems to have been arbitrarily taken for each of the various quadratic terms 
appearing in it- the correct, symmetrized form was not used for any of the terms.
Furthermore, we shall see that CCH have used
the first option over the second one and over the symmetrized form
when calculating the expression for the expectation value
of a general quadratic operator in the field in the past Boulware state.
In the analogous expression for the past Unruh state neither of the
two options nor the symmetrized form was used.

We look at what is the result of 
using separately each one of the two options, i.e., each one of the two quadratic terms in the symmetrized form.
It follows from (\ref{eq: symm. of lmwP_phi_i}) that
\begin{equation} \label{eq:symm. of op. lmwP_phi_i}
\begin{array}{l}
\mathcal{P} {}_{lm\omega P}\hat{\Phi}_{\indhel}^{\bullet}=(-1)^{\indhel+1}P{}_{lm\omega P}\hat{\Phi}_{\indhel}^{\bullet \dagger} \\
\mathcal{P} \left({}_{lm\omega P}\hat{\Phi}_{\indhel}^{\bullet}\ {}_{l'm'w'P'}\hat{\Phi}_{\indhel'}^{\bullet \dagger}+h.c.\right)
=\mathcal{P} \left({}_{lm\omega P}\hat{\Phi}_{\indhel}^{\bullet}\ {}_{l'm'w'P'}\hat{\Phi}_{\indhel'}^{\bullet \dagger}+
 {}_{l'm'w'P'}\hat{\Phi}_{\indhel'}^{\bullet}\ {}_{lm\omega P}\hat{\Phi}_{\indhel}^{\bullet \dagger}\right)=    \\
=(-1)^{\indhel+\indhel'}PP'\left({}_{lm\omega P}\hat{\Phi}_{\indhel}^{\bullet \dagger}\ {}_{l'm'w'P'}\hat{\Phi}_{\indhel'}^{\bullet}+
{}_{l'm'w'P'}\hat{\Phi}_{\indhel'}^{\bullet \dagger}\ {}_{lm\omega P}\hat{\Phi}_{\indhel}^{\bullet}\right)
\end{array}
\end{equation}
In the past Boulware state we have
\begin{equation} \label{eq:phi(in)_iphi(in,dagger)_j on vac.0}
\bra{B^-}
\left(\hat{\phi}_{\indhel}^{\bullet} \hat{\phi}_{\indhel'}^{\bullet \dagger}+h.c.\right)
\ket{B^-}
=\sum_{lmP}\int_{0}^{+\infty}\d{\omega^{\bullet}}\left({}_{lm\omega}\phi^{\bullet}_{\indhel}\ {}_{lm\omega}\phi_{\indhel'}^{\bullet *}+c.c.\right)
\end{equation}
whereas
\begin{equation} \label{eq:phi(in,dagger)_jphi(in)_i on vac.0}
\begin{aligned}
\bra{B^-}
\left(\hat{\phi}_{\indhel'}^{\bullet \dagger} \hat{\phi}_{\indhel}^{\bullet}+h.c.\right) 
\ket{B^-}
&=(-1)^{\indhel+\indhel'}\sum_{lmP}\int_{0}^{+\infty}\d{\omega^{\bullet}}
\Big[\mathcal{P}\left({}_{lm\omega}\phi^{\bullet}_{\indhel'}{}_{lm\omega}\phi_{\indhel}^{\bullet *}\right)+c.c.\Big]=  \\
& =(-1)^{\indhel+\indhel'}\mathcal{P}\left(
\bra{B^-}
\left(\hat{\phi}_{\indhel}^{\bullet} \hat{\phi}_{\indhel'}^{\bullet \dagger}+h.c.\right)
\ket{B^-}
\right)
\end{aligned}
\end{equation}
\catdraft{te sentit posar b.c. `up'/`in' en operador (e.g., $\bra{B^-}\hat{\phi}_j^{\bullet \dagger} \hat{\phi}_{\indhel}^{\bullet}\ket{B^-}$)?????}
We can see from (\ref{eq: symm. of lmwphi_i}) that the two options 
(\ref{eq:phi(in)_iphi(in,dagger)_j on vac.0})
and (\ref{eq:phi(in,dagger)_jphi(in)_i on vac.0}) will in principle give different results.
It is the first option, (\ref{eq:phi(in)_iphi(in,dagger)_j on vac.0}), that CCH used to obtain their expression (\ref{eq:quadratic op. for s=1 on B-}).
We have seen, however, that in the Schwarzschild space-time the two coincide (except for a possible different sign) since 
the transformation $(\omega \to -\omega)$ is not required in the symmetry (\ref{eq: symm. of lmwphi_i}). 

The expectation value in the state $\ket{B^-}$ of one of the two terms $\hat{\phi}_{\indhel}^{\bullet}\hat{\phi}_{\indhel'}^{\bullet \dagger}$
and $\hat{\phi}_{\indhel'}^{\bullet \dagger}\hat{\phi}_{\indhel}^{\bullet}$ is derived from 
that of the other term by operating with $\mathcal{P}$ and multiplying by $(-1)^{\indhel+\indhel'}$, but only if each term is added to its own hermitian conjugate.
Therefore the quantum-mechanical symmetrization guarantees 
that the expectation value in the state $\ket{B^-}$ of a hermitian, quadratic operator will be invariant (bar a sign) under parity. 
The sign $(-1)^{\indhel+\indhel'}$ is precisely the same sign appearing in (\ref{eq:pairs of vects. in stress tensor under parity}). 
This implies that if the quadratic terms in the expression (\ref{eq:stress tensor, spin 1})
are quantum-mechanically symmetrized when promoting the NP scalars to operators, 
then the expectation value in the state $\ket{B^-}$ of the stress-energy tensor will be invariant under parity.

\catdraft{1) QM-symmetrized form is symm. under parity regardless of the vac. only because it's 
QM-symmetrized form? or does the vac. def. $\hat{a}\ket{0}=0$ imply symm. under parity of $\ket{0}$  ??
2) 'quantum-mechanical' expression vs. 'quantum-field-theoretical' (e.g., for QM stress tensor)?
3) do a table with the scheme of the relationships between the different
quantum quadratic operators? 
}

In order to calculate the expectation value of the quadratic terms 
$\hat{\phi}_{\indhel}^{\bullet}\hat{\phi}_{\indhel'}^{\bullet \dagger}$ and
$\hat{\phi}_{\indhel'}^{\bullet \dagger}\hat{\phi}_{\indhel}^{\bullet}$ in the past Unruh state we are going to
make use of the expression calculated in ~\cite{ar:F&T'89} which gives the past Unruh state in terms of the past Boulware state:
\begin{equation} \label{eq:U in terms of B}
\ket{U^-}=\prod_{lm\tilde{\omega}P}C_{lm\omega P}\exp\left(e^{-\pi\tilde{\omega}/\kappa_+}{}_{lm\omega P}\hat{a}^{\text{up} \dagger}
{}_{lm\omega P}\hat{a}^{\text{up'} \dagger}\right)\ket{B^-}
\end{equation}
where $C_{lm\omega P}$ are normalization constants and ${}_{lm\omega P}\hat{a}^{\text{up'} \dagger}$ are
creation operators in region $I^*$ of the extended Kerr space-time. 

We will also make use of the following expression in ~\cite{ar:Schum&Caves'85}:
\begin{equation} \label{eq:eq3.66Schum&Caves'85}
S(r,\phi)=(\cosh r)^{-1}e^{-\hat{a}^{\dagger}_+\hat{a}^{\dagger}_-e^{2i\phi}\tanh r}
e^{-(\hat{a}^{\dagger}_+\hat{a}_{+}+\hat{a}^{\dagger}_-\hat{a}_-)\ln(\cosh r)}e^{\hat{a}_+\hat{a}_{-}e^{-2i\phi}\tanh r}
\end{equation}
where $S(r,\phi)$ is the two-mode squeeze operator
\begin{equation} \label{eq:S op.} 
S(r,\phi)=e^{r(\hat{a}_+\hat{a}_{-}e^{-2i\phi}-\hat{a}^{\dagger}_+\hat{a}^{\dagger}_-e^{2i\phi})}
\end{equation}
and the independent operators $\hat{a}_+$ and $\hat{a}_{-}$ satisfy the standard commutation relations.
By using (\ref{eq:eq3.66Schum&Caves'85}) and (\ref{eq:S op.}) we can re-express (\ref{eq:U in terms of B}) as
\begin{equation}
\ket{U^-}=\exp{\left\{\sum_{lmP}\int_0^{\infty}\d{\tilde{\omega}}\Big[\ln C_{lm\omega P}+\ln(\cosh r_{\tilde{\omega}})\Big]\right\}}e^{-\hat{A}}\ket{B^-}
\end{equation}
with
\begin{equation}
\hat{A}\equiv\sum_{lmP}\int_0^{\infty}\d{\tilde{\omega}}\  
r_{\tilde{\omega}}\left({}_{lm\omega P}\hat{a}^{\text{up} \dagger}{}_{lm\omega P}\hat{a}^{\text{up'} \dagger}-
{}_{lm\omega P}\hat{a}^{\text{up}}{}_{lm\omega P}\hat{a}^{\text{up'}}\right)
\end{equation}
and 
\begin{equation}
r_{\tilde{\omega}}\equiv -\tanh^{-1} \left(e^{-\pi\tilde{\omega}/\kappa_+}\right)
\end{equation}
Since $\left(e^{\hat{A}}\right)^{\dagger}=e^{-\hat{A}}$, 
the normalization $\left\langle\right.U^- \left|\right.U^-\left.\right\rangle=1$ implies 
\begin{equation}
\exp{\left\{\sum_{lmP}\int_0^{\infty}\d{\tilde{\omega}}\Big[\ln C_{lm\omega P}+\ln C^*_{lm\omega P}+2\ln(\cosh r_{\tilde{\omega}})\Big]\right\}}=1
\end{equation}
\draft{include two-mode squeeze op. rln. between $\ket{B^-}$ and $\ket{B^+}$ found by Mat,Dav\&Ott'93?}

Using now the Baker-Campbell-Hausdorff equation (~\cite{bk:Louisell})
\begin{equation} \label{eq:BCK}
e^{\xi \hat{P}}\hat{Q}e^{-\xi \hat{P}}=\hat{Q}+\xi [\hat{P},\hat{Q}]+\frac{\xi^2}{2!}[\hat{P},[\hat{P},\hat{Q}]]+\frac{\xi^3}{3!}[\hat{P},[\hat{P},[\hat{P},\hat{Q}]]]+\ldots
\end{equation}
where $\hat{P}$ and $\hat{Q}$ are any two operators and $\xi$ is a parameter, we can find that
\begin{equation} \label{}
e^{\hat{A}}{}_{lm\omega P}\hat{a}^{\text{up}}e^{-\hat{A}}=
{}_{lm\omega P}\hat{a}^{\text{up}}\cosh r_{\tilde{\omega}}+{}_{lm\omega P}\hat{a}^{\text{up'} \dagger}\sinh r_{\tilde{\omega}}
\end{equation}
and finally
\begin{subequations} \label{eq:a_dagger*a on unruh}
\begin{align}
\bra{U^-}{}_{lm\omega P}\hat{a}^{\text{up} \dagger}{}_{l'm'w'P'}\hat{a}^{\text{up}}\ket{U^-}&=
\frac{1}{2}\left[\coth \left(\frac{\pi\tilde{\omega}}{\kappa_+}\right)-1\right]\delta(\omega-\omega')\delta_{ll'}\delta_{mm'}\delta_{PP'} \\
\bra{U^-}{}_{lm\omega P}\hat{a}^{\text{up} \dagger}{}_{l'm'w'P'}\hat{a}^{\text{up} \dagger}\ket{U^-}&=0\\
\bra{U^-}{}_{lm\omega P}\hat{a}^{\text{up}}{}_{l'm'w'P'}\hat{a}^{\text{up}}\ket{U^-}&=0
\end{align}
\end{subequations}
With the above results we find that
\begin{equation}  \label{eq:phi*phi^dagger `up' on U-}
\begin{aligned}
&\bra{U^-}\hat{\phi}_{\indhel}^{\text{up}}\hat{\phi}_{\indhel'}^{\text{up} \dagger}\ket{U^-}=  
\\ &=
\frac{1}{2}\sum_{lmP}\int_0^{\infty}\d{\tilde{\omega}}
\begin{aligned}[t]
\bigg\{
&
\left[{}_{lm\omega}\phi_{\indhel}^{\text{up}}{}_{lm\omega}\phi_{\indhel'}^{\text{up} *}
+(-1)^{\indhel+\indhel'}\mathcal{P}({}_{lm\omega}\phi_{\indhel}^{\text{up} *}{}_{lm\omega}\phi_{\indhel'}^{\text{up}})\right]
\coth\left(\frac{\pi\tilde{\omega}}{\kappa_+}\right)+
\\ +&
\left[{}_{lm\omega}\phi_{\indhel}^{\text{up}}{}_{lm\omega}\phi_{\indhel'}^{\text{up} *}-
(-1)^{\indhel+\indhel'}\mathcal{P}({}_{lm\omega}\phi_{\indhel}^{\text{up} *}{}_{lm\omega}\phi_{\indhel'}^{\text{up}})\right]
\bigg\}
= 
\end{aligned}
\\ & =
(-1)^{\indhel+\indhel'}\mathcal{P}\left(\bra{U^-}\hat{\phi}_{\indhel}^{\text{up} \dagger}\hat{\phi}_{\indhel'}^{\text{up}}\ket{U^-}\right)
\end{aligned}
\end{equation}
Note the minus sign in the second term in (\ref{eq:phi*phi^dagger `up' on U-}). Its presence may seem a bit surprising at first but,
as we shall now see, it is precisely this sign that causes the expectation value of the stress-energy tensor in the
past Unruh state to adopt a more familiar form by having all `up' terms multiplied by a $\coth$ factor. 
This is already clear from looking at (\ref{eq:phi*phi^dagger `up' on U-}) and realizing that
when quantum-symmetrizing the classic expression $\phi_{\indhel}^{\text{up}}\phi_{\indhel'}^{\text{up} *}$ 
the terms without a $\coth$ factor will cancel out.

For the `in' modes we have
\begin{equation}
\bra{U^-}\hat{\phi}_{\indhel}^{\text{in}}\hat{\phi}_{\indhel'}^{\text{in} \dagger}\ket{U^-}=
\sum_{lmP}\int_0^{\infty}\d{\omega}{}_{lm\omega}\phi_{\indhel}^{\text{in}}{}_{lm\omega}\phi_{\indhel'}^{\text{in} *}=
(-1)^{\indhel+\indhel'}\mathcal{P}\left(\bra{U^-}\hat{\phi}_{\indhel}^{\text{in} \dagger}\hat{\phi}_{\indhel'}^{\text{in}}\ket{U^-}\right)
\end{equation}

The following identities are therefore immediately satisfied
\begin{equation}
\begin{aligned}
\vac{\left[\hat{\phi}_{\indhel}^{\bullet},\hat{\phi}_{\indhel'}^{\bullet \dagger}\right]}{U^-}&=
\vac{\left[\hat{\phi}_{\indhel}^{\bullet},\hat{\phi}_{\indhel'}^{\bullet \dagger}\right]}{B^-}=\\
&=\sum_{lmP}\int_0^{\infty}\d{\omega^{\bullet}}
\left[{}_{lm\omega}\phi_{\indhel}^{\bullet}{}_{lm\omega}\phi_{\indhel'}^{\bullet *}
-(-1)^{\indhel+\indhel'}\mathcal{P}\left({}_{lm\omega}\phi_{\indhel}^{\bullet *}{}_{lm\omega}\phi_{\indhel'}^{\bullet}\right)\right]
\end{aligned}
\end{equation}
and therefore
\begin{equation}
\vac{\left[\hat{\phi}_{\indhel}^{\bullet},\hat{\phi}_{\indhel'}^{\bullet \dagger}\right]}{U^--B^-}=0
\end{equation}
as it should be.

When the classical term $\left(\phi_{\indhel}\phi_{\indhel'}^*+c.c.\right)$ is quantized to
the symmetrized term $\left(\hat{\phi}_{\indhel}\hat{\phi}_{\indhel'}^{\dagger}+\hat{\phi}_{\indhel'}^{\dagger}\hat{\phi}_{\indhel}\right)/2+h.c.$.
it gives the following real, parity-invariant expressions in the past Boulware and past Unruh states:
\begin{subequations}
\begin{align}
\begin{split}
&\vac{\frac{\hat{\phi}_{\indhel}\hat{\phi}_{\indhel'}^{\dagger}+\hat{\phi}_{\indhel'}^{\dagger}\hat{\phi}_{\indhel}}{2}+h.c.}{B^-}= 
\\ &=
\frac{1}{2}\sum_{lmP}\bigg(\int_0^{\infty}\d{\tilde{\omega}}
\left[
%
{}_{lm\omega}\phi_{\indhel}^{\text{up}}{}_{lm\omega}\phi_{\indhel'}^{\text{up} *}+
(-1)^{\indhel+\indhel'}\mathcal{P}({}_{lm\omega}\phi_{\indhel}^{\text{up}}{}_{lm\omega}\phi_{\indhel'}^{\text{up} *})\right]+
\\ &
\qquad \quad\ \
+
\int_0^{\infty}\d{\omega}
\left[{}_{lm\omega}\phi_{\indhel}^{\text{in}}{}_{lm\omega}\phi_{\indhel'}^{\text{in} *}+
(-1)^{\indhel+\indhel'}\mathcal{P}({}_{lm\omega}\phi_{\indhel}^{\text{in}}{}_{lm\omega}\phi_{\indhel'}^{\text{in} *})\right]
\bigg)+c.c.
\end{split}\\
\begin{split}
&\vac{\frac{\hat{\phi}_{\indhel}\hat{\phi}_{\indhel'}^{\dagger}+\hat{\phi}_{\indhel'}^{\dagger}\hat{\phi}_{\indhel}}{2}+h.c.}{U^-}= 
\\ &=
\frac{1}{2}\sum_{lmP}\bigg(\int_0^{\infty}\d{\tilde{\omega}}
\left[{}_{lm\omega}\phi_{\indhel}^{\text{up}}{}_{lm\omega}\phi_{\indhel'}^{\text{up} *}
+(-1)^{\indhel+\indhel'}\mathcal{P}({}_{lm\omega}\phi_{\indhel}^{\text{up}}{}_{lm\omega}\phi_{\indhel'}^{\text{up} *})\right]
\coth\left(\frac{\pi\tilde{\omega}}{\kappa}\right)+
\\ &
\qquad \quad\ \
+
\int_0^{\infty}\d{\omega}
\left[{}_{lm\omega}\phi_{\indhel}^{\text{in}}{}_{lm\omega}\phi_{\indhel'}^{\text{in} *}+
(-1)^{\indhel+\indhel'}\mathcal{P}({}_{lm\omega}\phi_{\indhel}^{\text{in}}{}_{lm\omega}\phi_{\indhel'}^{\text{in} *})\right]
\bigg)+c.c.
\end{split}
\end{align}
\end{subequations}

Note that in the above expressions we have been able to complex conjugate the mode functions that are operated on by $\mathcal{P}$
because of the existence of the $+c.c.$ terms.
This immediately leads to the following real, parity-invariant expressions for the stress-energy tensor in
the past Boulware and past Unruh states:
\begin{subequations} \label{eq:corrected stress tensor for s=1 on B-,U-}
\begin{align}
\begin{split}
&\expct{\hat{T}_{\mu\nu}}{B^-}= \\
&=\frac{1}{2}\sum_{lmP}\left(
\int_0^{\infty}\d{\tilde{\omega}}\, 
\Big\{T_{\mu\nu}
\left[{}_{lm\omega}\phi_{\indhel}^{\text{up}},{}_{lm\omega}\phi_{\indhel}^{\text{up} *}\right]
+(-1)^{\vartheta}\mathcal{P}\left(T_{\mu\nu}
\left[{}_{lm\omega}\phi_{\indhel}^{\text{up}},{}_{lm\omega}\phi_{\indhel}^{\text{up} *}\right]
\right)\Big\}
+  \right. \\
& 
\qquad \quad \ \
\left. +\int_0^{\infty}\d{\omega}\, 
\Big\{T_{\mu\nu}
\left[{}_{lm\omega}\phi_{\indhel}^{\text{in}},{}_{lm\omega}\phi_{\indhel}^{\text{in} *}\right]
+(-1)^{\vartheta}\mathcal{P}\left(T_{\mu\nu}
\left[{}_{lm\omega}\phi_{\indhel}^{\text{in}},{}_{lm\omega}\phi_{\indhel}^{\text{in} *}\right]
\right)\Big\}
\right)   
\end{split} \label{eq:corrected stress tensor for s=1 on B-} \\ \begin{split}
&\expct{\hat{T}_{\mu\nu}}{U^-}= 
\frac{1}{2}\sum_{lmP}
\\&
\left(
\int_0^{\infty}\d{\tilde{\omega}}\, 
\coth\left(\frac{\pi\tilde{\omega}}{\kappa_+}\right)
\Big\{T_{\mu\nu}
\left[{}_{lm\omega}\phi_{\indhel}^{\text{up}},{}_{lm\omega}\phi_{\indhel}^{\text{up} *}\right]
+(-1)^{\vartheta}\mathcal{P}\left(T_{\mu\nu}
\left[{}_{lm\omega}\phi_{\indhel}^{\text{up}},{}_{lm\omega}\phi_{\indhel}^{\text{up} *}\right]
\right)\Big\}
+  \right. \\
& 
\qquad \qquad \quad
\left. +\int_0^{\infty}\d{\omega}\, 
\Big\{T_{\mu\nu}
\left[{}_{lm\omega}\phi_{\indhel}^{\text{in}},{}_{lm\omega}\phi_{\indhel}^{\text{in} *}\right]
+(-1)^{\vartheta}\mathcal{P}\left(T_{\mu\nu}
\left[{}_{lm\omega}\phi_{\indhel}^{\text{in}},{}_{lm\omega}\phi_{\indhel}^{\text{in} *}\right]
\right)\Big\}
\right) \end{split} \label{eq:corrected stress tensor for s=1 on U-}
\end{align}
\end{subequations}
Note that the sign $(-1)^{\vartheta}$ appears in the above expressions instead of $(-1)^{\indhel+\indhel'}$ by virtue of the change under the parity operation 
of the coefficients of the quadratic field operators that appear in the expression for the stress-energy tensor, as seen in
(\ref{eq:pairs of vects. in stress tensor under parity}).
\draft{maybe easier to give expression for $\expct{\hat{\vec{T}}}{U^-}$ instead?}

This is in sharp contrast with the expressions given by CCH. The expressions for the stress-energy tensor 
given by CCH, i.e., 
(\ref{eq:stress tensor for s=1 on B-}) and (\ref{eq:stress tensor for s=1 on U-}), are not invariant under parity.
They are equivalent to (\ref{eq:corrected stress tensor for s=1 on B-,U-}) if 
$\mathcal{P} \left({}_{lm\omega}\phi_{\indhel}^{\bullet}{}_{lm\omega}\phi_{\indhel'}^{\bullet *}\right)=
(-1)^{\indhel+\indhel'}{}_{lm\omega}\phi_{\indhel}^{\bullet}{}_{lm\omega}\phi_{\indhel'}^{\bullet *}$
$\forall \indhel,\indhel'$,  
which we have proved in this chapter that it is not the case.
\draft{mention scalar case?}
The general expressions (\ref{eq:quadratic op. for s=1 on vacua}) for a quadratic operator $\hat{Q}$ 
given by CCH, when applied to the cases $\hat{\phi}_{\indhel}\hat{\phi}_{\indhel'}^{\dagger}$
and $\hat{\phi}^{\dagger}_{\indhel}\hat{\phi}_{\indhel'}$, yield the surprising result: 
\begin{equation}
\begin{aligned}
&\vac{\left[\hat{\phi}_{\indhel}^{\bullet},\hat{\phi}_{\indhel'}^{\bullet \dagger}\right]}{U^--B^-}= \\
&=\sum_{lmP}\int_0^{\infty}\d{\tilde{\omega}}\,
\left\{ 
\left[\coth\left(\frac{\pi\tilde{\omega}}{\kappa_+}\right)-1\right]
\left[{}_{lm\omega}\phi_{\indhel}^{\text{up}}{}_{lm\omega}\phi_{\indhel'}^{\text{up} *}-
\mathcal{P}({}_{lm\omega}\phi_{\indhel}^{\text{up}}{}_{lm\omega}\phi_{\indhel'}^{\text{up} *})\right]\right\}
\end{aligned}
\end{equation}
which is not generally zero, as proved in the previous subsection.

To our knowledge, this is the first time that the expressions (\ref{eq:corrected stress tensor for s=1 on B-,U-})
for the expectation value of the stress-energy tensor when the electromagnetic field is in the past Boulware 
and past Unruh states have been given.

By comparing the expectation values in (\ref{eq:stress tensor for s=1 on all vac.}) with their symmetrized 
versions in (\ref{eq:corrected stress tensor for s=1 on B-,U-}) for the past Boulware and past Unruh states, 
we can give an analogous symmetrized version for the state $\ket{CCH^-}$:
\begin{equation} \label{eq:corrected stress tensor for s=1 on CCH^-}
\begin{aligned}
&\expct{\hat{T}_{\mu\nu}}{CCH^-}= 
\frac{1}{2}\sum_{lmP}
\\ &
\left(
\int_0^{\infty}\d{\tilde{\omega}}\, 
\coth\left(\frac{\pi\tilde{\omega}}{\kappa_+}\right)
\Big\{T_{\mu\nu}
\left[{}_{lm\omega}\phi_{\indhel}^{\text{up}},{}_{lm\omega}\phi_{\indhel}^{\text{up} *}\right]
+(-1)^{\vartheta}\mathcal{P}\left(T_{\mu\nu}
\left[{}_{lm\omega}\phi_{\indhel}^{\text{up}},{}_{lm\omega}\phi_{\indhel}^{\text{up} *}\right]
\right)\Big\}
+ \right.  \\
& \left. +\int_0^{\infty}\d{\omega}\, 
\coth\left(\frac{\pi\omega}{\kappa_+}\right)
\Big\{T_{\mu\nu}
\left[{}_{lm\omega}\phi_{\indhel}^{\text{in}},{}_{lm\omega}\phi_{\indhel}^{\text{in} *}\right]
+(-1)^{\vartheta}\mathcal{P}\left(T_{\mu\nu}
\left[{}_{lm\omega}\phi_{\indhel}^{\text{in}},{}_{lm\omega}\phi_{\indhel}^{\text{in} *}\right]
\right)\Big\}
\right) 
\end{aligned}
\end{equation}


\subsection{Polarization} \label{sec:polarization}

In this last subsection we will give a physical interpretation of the non-parity term and the parity term appearing in the
expressions (\ref{eq:corrected stress tensor for s=1 on B-,U-}) and (\ref{eq:corrected stress tensor for s=1 on CCH^-})
for the expectation value of the stress-energy tensor in different states.
We denote by parity term in a certain expression a term that explicitly contains the parity operator $\mathcal{P}$, and by
non-parity term one in the same expression that does not explicitly contain this operator.

It is clear from the classical expression (\ref{eq:A1 as a func. of psi_j^*}) for the `upgoing gauge' potential ${}_{lm\omega}A^{\text{up} \mu}$ 
that this potential only contains the null vectors $\vec{n}$ and $\vec{m}$.
The parity term $P\mathcal{P}{}_{lm\omega}A^{\text{up} \mu}$ in (\ref{eq: def. of lmwPA_mu}), 
because of the transformations (\ref{eq:parity op. on NP objs.}) of the null base under parity, 
contains the vectors $\vec{n}$ and $\vec{m}^*$. 
The potential ${}_{lm\omega P}A^{\text{up} \mu}$ therefore contains the vectors $\vec{n}$, $\vec{m}$ and $\vec{m}^*$.
However, as we saw in Section \ref{sec:gauge inv.}, only two of them are physically significant. Indeed, we know that the parity term is
a pure gauge and therefore the contribution to the physical quantities from the term with $\vec{m}^*$ in the potential 
is zero. Only the terms with $\vec{n}$ and $\vec{m}$ in the `upgoing' potential contribute to physical quantities. 

\draft{this is wrong since term with $\vec{m}^*$ does contribute to electric and magnetic fields $\vec{E}$ and $\vec{B}$??
even though the parity term should not contribute to NP scalars and might contribute to $\vec{E}$ and $\vec{B}$, this means the
link with Section \ref{sec:gauge inv.} cannot be established and it seems that $\vec{E}$ and $\vec{B}$ depends on three null vectors
(and two polar vectors for $r\to\infty$ or four $\forall r$), contradicting Section \ref{sec:gauge inv.} which says there are only two
physical vectors?? sln.:fields $\vec{E}$ and $\vec{B}$ must be made real and then term with $\vec{m}^*$ does not contribute to them?-check
sln (p.68(bis1119,bis6->bis7)t->pi-t): 
yes, both $\vec{m}$ and $\vec{m}^*$ contribute to $\vec{E}$ and $\vec{B}$ but the neutral pol. coming from $\vec{l}$ and $\vec{n}$
does not contribute to them so that the 2 physically significant vects. are $\vec{m}$ and $\vec{m}^*$ rather than $\vec{n}$ and $\vec{m}$!
only one between $\vec{m}$ and $\vec{m}^*$ contributes to $T^{\text{up}}$ but the other one does to $T^{\text{in}}$ so that both end up
contributing to $\vac{\hat{T}}{\psi}$.}

In particular, it is immediate from expressions (\ref{eq:phi as func. of potential with K}) 
for the Maxwell scalars that only the term in the potential ${}_{lm\omega P}A^{\text{up} \mu}$ that contains the vector
$\vec{m}$ contributes to ${}_{lm\omega}\phi^{\text{up}}_{+1}$ whereas only the term with $\vec{n}$ contributes
to ${}_{lm\omega}\phi^{\text{up}}_{-1}$. Both, terms with $\vec{n}$ and terms with $\vec{m}$, contribute to ${}_{lm\omega}\phi^{\text{up}}_{0}$.

It is in the limit for large $r$ that the physical meaning of the various vectors becomes clear.
We know that in flat space-time an electric field mode of positive frequency that is proportional to the vector 
$\left(\hat{\vec{e}}_{\theta}+i\hat{\vec{e}}_{\phi}\right)[\left(\hat{\vec{e}}_{\theta}-i\hat{\vec{e}}_{\phi}\right)]$
possesses a positive[negative] angular momentum and we thus say that it is positively[negatively] polarized.
If the mode is instead of negative frequency, the sign of the angular momentum changes and then an electric field mode
proportional to $\left(\hat{\vec{e}}_{\theta}+i\hat{\vec{e}}_{\phi}\right)[\left(\hat{\vec{e}}_{\theta}-i\hat{\vec{e}}_{\phi}\right)]$
is said to be negatively[positively] polarized.
\draft{valid for both classical and QM theory?only for outgoing wave?only if $t-$dependence is $e^{-i\omega t}$?
according to eq.16.65Jackson the sign of the ang.mom.depends on the sign of $m$, which I'm ignoring here?->probl. that
calculation of ang. mom. seems to require lower order in $r$;our electric\&mag. fields are similar, but not exactly equal,
to those of a multipole expansion in flat s-t}
Therefore, according to (\ref{eq:Kinnersley tetrad, r->inf}), 
an electric and a magnetic field modes of positive frequency that are proportional to the vector $\vec{m}[\vec{m}^*]$ correspond, 
in the flat space limit, to a positive[negative] polarization, whereas the vectors $\vec{l}$ and $\vec{n}$ correspond both to neutral polarization.
This implies that the positive-frequency modes ${}_{lm\omega}\phi^{\text{up}}_{-1}$, ${}_{lm\omega}\phi^{\text{up}}_{+1}$ and 
${}_{lm\omega}\phi^{\text{up}}_{0}$ are obtained from terms in the potential that, in the flat space limit, are
neutrally-, positively- and both neutrally- and positively- polarized respectively. 
As we have seen, the parity term in the potential in (\ref{eq: def. of lmwPA_mu}), 
which is of the opposite polarization to that of the non-parity term, 
does not contribute to any NP scalar mode because it is pure gauge. 
This does not mean that the opposite polarization to that of the non-parity term in the potential does not contribute to the NP scalars. 
Indeed, it does contribute through the negative-frequency
modes when the integration is over all frequencies, as in (\ref{eq:Fourier series of Gamma field}). 
In the expression (\ref{eq:classical mode expansion for A(in)_mu}) for the potential
or (\ref{eq:classical mode expansion for phi(in)_i}) for the NP scalars, 
in which we have rid of the negative-frequency modes, the opposite polarization appears
via the complex-conjugate term or the parity-term respectively.

Even though all three Maxwell scalars appear in the classical expression for the electromagnetic stress tensor, 
due to their different asymptotic behaviour (\ref{eq:peeling th}) for large $r$, 
the terms in the stress tensor (\ref{eq:stress tensor, spin 1}) with ${}_{lm\omega}\phi^{\text{up}}_{+1}$ predominate in this limit.
That is, the radiation field components of the stress tensor $T^{\text{up} \mu\nu}$ are calculated in the flat space limit 
from modes in the potential (\ref{eq:Fourier series of Gamma field}) which for positive[negative] frequency 
correspond to a positive[negative] polarization. 
Note that the complex-conjugation of NP scalars in the stress tensor does not change the polarization 
of the field since it is merely a consequence of the fact that the null tetrad contains complex vectors, and does
not imply the complex-conjugation of the tensor field components $F_{\mu\nu}$.

\catdraft{the above is not true, for ex., for $T_{\theta\theta}$, which only contains $|{}_{lm\omega}\phi_{0}|^2$
and ${}_{lm\omega}\phi_{-1}{}_{lm\omega}\phi^*_{+1}$ terms??!although it looks like for standard radially outgoing
classical waves in flat space the neutrally-polarized terms also contribute to $T_{\theta\theta}$?
both $T_{\theta\theta}$ and ang. mom. vanish to leading order (also for flat s-t).
to what extend can it be said that they're pos.pol. when to lower orders they contain neutral pol.? to what extend
can pos./negat./neutral pol. be defined to lower orders, for which s-t is not flat? to what extend can pos./negat./neutral
pol. be defined even to leading order in the flat s-t sense, when the ang.mom. vanishes to leading order (even for flat s-t)?
mirar sln. a p.68(bis117)(bis118)t->pi-t}

We give here expressions for the `upgoing' potential modes in the limit for large $r$.
We wish, however, to obtain expressions for the fields that are real mode by mode. 
We will therefore not calculate the potential modes ${}_{lm\omega P}A^{\text{up} \mu}$
from expressions (\ref{eq:tableIChrzan.}), since the transformation $(m,\omega)\to (-m,-\omega)$ has been applied to
the complex-conjugated term in (\ref{eq:def. of lmwPA}). 
Even though the potential is obviously real, the potential modes ${}_{lm\omega P}A^{\text{up} \mu}$ are not.
Instead, we will calculate potential modes that are real mode by mode by applying equation (\ref{eq:make potential real}) mode by mode.
We denote these modes by ${}_{lm\omega P}A^{' \text{up} \mu}$
From equations (\ref{eq: def. of lmwPA_mu}), (\ref{eq:A1 as a func. of psi_j^*}) and (\ref{eq:R_up}) it then follows that
\begin{equation}   \label{eq:A_up, r->inf}
\begin{aligned}
&{}_{lm\omega P}A^{' \text{up} \mu}\equiv \left(\Pi^{\dagger}_j{}^{\alpha}{}_{lm\omega}\NPadj_j\right)^* + \Pi^{\dagger}_j{}^{\alpha}{}_{lm\omega}\NPadj_j\rightarrow 
-\frac{\omega i |N^{\text{up}}_{+1}|}{\sqrt{2}r}
\times \\  &\times
\left[{}_{-1}Y_{lm\omega}{}_{+1}R^{\text{up,tra}}_{lm\omega}e^{-i\omega (t-r)}\left(\hat{\vec{e}}_{\theta}+i\hat{\vec{e}}_{\phi}\right)-
{}_{-1}Y_{lm\omega}^*{}_{+1}R^{\text{up,tra} *}_{lm\omega}e^{+i\omega (t-r)}\left(\hat{\vec{e}}_{\theta}-i\hat{\vec{e}}_{\phi}\right)\right]
\\ & \qquad \qquad \qquad \qquad \qquad \qquad \qquad \qquad \qquad \qquad \qquad  \qquad \qquad(r\rightarrow+\infty)
\end{aligned}
\end{equation}
where we have used the fact that in flat space we can replace ${}_{\indhel}Z_{lm\omega}$ by ${}_{\indhel}Y_{lm\omega}$
\draft{how is that justified when in the expression for $Z$ in terms of $Y$, $r$ does not appear?}
The difference between ${}_{lm\omega P}A^{\text{up} \mu}$ and ${}_{lm\omega P}A^{' \text{up} \mu}$ lies only on a sign factor and
a change of sign in $(m,\omega)$ on the second term.
We know, however, that the second term is pure gauge mode by mode and it therefore does not contribute to the NP scalars.
The electric and magnetic fields that corresond to the above potential follow through trivially when $a=0$:
\catdraft{no tinc clar quins resultats per $r\to\infty$ valen per $a\neq 0$ i quins nomes per $a=0$??(mirar dalt p.68(bis74)t->pi-t)}
\begin{equation}  \label{eq:E_up,B_up, r->inf}
\begin{aligned}
&{}_{lm\omega P}\vec{E}^{\text{up}}=-\nabla {}_{lm\omega P}A^{' \text{up} 0}-\pardiff{{}_{lm\omega P}\vec{A}^{' \text{up}}}{t}=
\frac{\omega^2 |N^{\text{up}}_{+1}|}{\sqrt{2}r}
\times \\&\times
\left[{}_{-1}Y_{lm\omega}{}_{+1}R^{\text{up,tra}}_{lm\omega}e^{-i\omega (t-r)}\left(\hat{\vec{e}}_{\theta}+i\hat{\vec{e}}_{\phi}\right)+
{}_{-1}Y_{lm\omega}^*{}_{+1}R^{\text{up,tra} *}_{lm\omega}e^{+i\omega (t-r)}\left(\hat{\vec{e}}_{\theta}-i\hat{\vec{e}}_{\phi}\right)\right]
\\
&{}_{lm\omega P}\vec{B}^{\text{up}}=\nabla\times{}_{lm\omega P}\vec{A}^{' \text{up}}=
\frac{-\omega^2 i|N^{\text{up}}_{+1}|}{\sqrt{2}r}
\times \\&\times
\left[{}_{-1}Y_{lm\omega}{}_{+1}R^{\text{up,tra}}_{lm\omega}e^{-i\omega (t-r)}\left(\hat{\vec{e}}_{\theta}+i\hat{\vec{e}}_{\phi}\right)-
{}_{-1}Y_{lm\omega}^*{}_{+1}R^{\text{up,tra} *}_{lm\omega}e^{+i\omega (t-r)}\left(\hat{\vec{e}}_{\theta}-i\hat{\vec{e}}_{\phi}\right)\right]
\end{aligned}
\end{equation}
where ${}_{lm\omega P}A^{' \text{up}}=\left({}_{lm\omega P}A^{' \text{up} 0},{}_{lm\omega P}\vec{A}^{' \text{up}}\right)$.
It is worth noting that even though the second term in either (\ref{eq:make potential real}) or (\ref{eq:def. of lmwPA}) 
does not contribute to the NP scalars it does contribute to the fields. 
The reason is that the NP scalars must be calculated from real fields. 
It does not make physical sense to consider the contribution to the NP scalars from a non-real field, such as the second
term in the above expressions for the field. We say that this term is `pure gauge' in the sense that it does not contribute
to the NP scalars even if it does contribute to the physical fields so as to make them real. 
\catdraft{continua semblant contradiccio?? posar igualment eqs. per ${}_{lm\omega P}A^{\text{up} \mu}$? 3) canviar eqs. ${eq:E_up,B_up, r->inf}$,
trure subindex $P$ a ${}_{lm\omega P}A^{' \text{up} \mu}$?}
The angular funcions can be expressed in terms of the orbital angular momentum acting on the spherical harmonics as
\begin{equation}
{}_{\pm1}Y_{lm\omega}=-\left[l(l+1)\right]^{-1/2}\left(\hat{\vec{e}}_{\theta}\pm i\hat{\vec{e}}_{\phi}\right)\vec{L}Y_{lm\omega}
\end{equation}
where we have made use of the relationships (\ref{eq:{s+/-1}_Y in terms of s_Y}) and (\ref{eq:orb.ang.mom.op. related to edth}).
The large-$r$ asymptotics for the NP Maxwell scalars in terms of the electric and magnetic fields are easily obtained:
\begin{equation}  \label{eq:phi_i, r->inf}
\begin{aligned}
\phi_{-1}&\rightarrow -\frac{1}{\sqrt{2}}\left(\vec{E}+\vec{B}i\right)\left(\hat{\vec{e}}_{\theta}+i\hat{\vec{e}}_{\phi}\right)  & (r\rightarrow+\infty)\\
\phi_{0}&\rightarrow \frac{1}{2}\left(\vec{E}+\vec{B}i\right)\hat{\vec{e}}_{r}  & (r\rightarrow+\infty) \\
\phi_{+1}&\rightarrow \frac{1}{2\sqrt{2}}\left(\vec{E}+\vec{B}i\right)\left(\hat{\vec{e}}_{\theta}-i\hat{\vec{e}}_{\phi}\right)  & (r\rightarrow+\infty)
\end{aligned}
\end{equation}
It is clear that the parity and non-parity terms correspond to opposite polarizations both 
for the potential (\ref{eq:A_up, r->inf}) and the fields (\ref{eq:E_up,B_up, r->inf}). 
It is also clear that the only contribution to $\phi^{\text{up}}_{+1}$ 
from the `upgoing' electric and magnetic fields comes from the non-parity term. 
To leading order in $r$ for the electric and magnetic fields both $\phi^{\text{up}}_{-1}$ and $\phi^{\text{up}}_{0}$ vanish, 
in agreement with (\ref{eq:peeling th}).
To next order in $r$, expressions (\ref{eq:E_up,B_up, r->inf}) and (\ref{eq:phi_i, r->inf}) must be calculated to include lower
order terms and it is therefore not valid to conclude from them that the only contribution to $\phi^{\text{up}}_{-1}$ and $\phi^{\text{up}}_{0}$ 
comes from negatively- and neutrally- polarized terms respectively. 
We have indeed seen in the beginning of this subsection that this is not the case. 
Finally, it is also manifest from the above asymptotic expressions that the positive- and the negative- frequency modes of the potential 
(\ref{eq:Fourier series of Gamma field}) have opposite polarization. 
Indeed, since the negative-frequency ones correspond to the positive-frequency, complex-conjugated term 
in (\ref{eq:classical mode expansion for A(in)_mu}), 
the term with $\left(\hat{\vec{e}}_{\theta}+i\hat{\vec{e}}_{\phi}\right)$ in the fields (\ref{eq:E_up,B_up, r->inf}),
which is the only term that contributes to the NP scalars and which is positively-polarized, is complex-conjugated to
a negatively-polarized term with $\left(\hat{\vec{e}}_{\theta}-i\hat{\vec{e}}_{\phi}\right)$.

The reasoning used so far for the `upgoing gauge' potential can be applied in the same manner to the `ingoing gauge' potential ${}_{lm\omega}A^{\text{in} \mu}$.
In this case, the potential contains one term with the vector $\vec{l}$, which is the only one that contributes to
${}_{lm\omega}\phi^{\text{in}}_{+1}$, and one term with the vector $\vec{m}^*$, which is the only one that contributes to
${}_{lm\omega}\phi^{\text{in}}_{-1}$. Like in the `upgoing gauge' case, both terms contribute to ${}_{lm\omega}\phi^{\text{up}}_{0}$,
and the parity term (containing $\vec{l}$ and $\vec{m}$) does not contribute to any of the NP scalars.
The scalar ${}_{lm\omega}\phi^{\text{in}}_{-1}$ is the one that diminishes more slowly in the limit for large $r$.
The radiative components of the classical stress tensor $T^{\text{in} \mu\nu}$ is thus calculated in the flat space limit from modes in the potential 
(\ref{eq:Fourier series of Gamma field}) that for positive[negative] frequency correspond to negative[positive] polarization.

It is clear that the two terms in the asymptotic expression (\ref{eq:A_up, r->inf}) for the `upgoing' potential are both derived from the modes
of the null tetrad component $A_m^{\text{up}}$. Similarly, the only asymptotic contribution to the `ingoing' potential comes from $A_{m^*}^{\text{in}}$.
We can thus say that $A_m$ and $A_{m^*}$ are the asymptotically gauge independent parts of $A_{\mu}$.

So far in this subsection we have looked at the physical meaning of the different terms in classical expressions only.
We are now in a position to understand the physical meaning of the terms in the quantum field theory expressions.
The positive frequency modes in (\ref{eq:Fourier series of Gamma field}) correspond to the non-parity term
in the expression (\ref{eq:classical mode expansion for phi(in)_i}) for the NP scalar and, ultimately, 
give rise to the non-parity term in the expectation value of the stress tensor (\ref{eq:corrected stress tensor for s=1 on B-,U-}).
Similarly, the negative frequency modes in (\ref{eq:Fourier series of Gamma field}) give rise to the 
parity term in the NP scalars and the parity term in the expectation value of the stress tensor.
We therefore reach the conclusion that the non-parity terms in expressions (\ref{eq:corrected stress tensor for s=1 on B-,U-})
for the expectation value of the stress tensor \ddraft{this should be said instead for a single quadratic term?}
correspond in the flat space limit to one specific polarization (positive in the `up' case and negative in the `in' case)
and that the corresponding parity terms in the same expressions correspond to the opposite polarization.
Both the contribution from the positive-polarization terms and from the negative-polarization terms are
separately real, as it should be.
We also know, from the beginning of Subsection \ref{subsec:lack of symmetry}, that in the spherically-symmetrical case $a=0$, 
the contribution to the expectation value of the stress tensor from the positive-polarization terms 
is identical to the one from the negative-polarization terms, as one would expect.

The notable exception to this picture are the `up' superradiant modes. 
Indeed, these modes have a sign of $\omega$ opposite to the non-superradiant modes in the same term in the expectation
value, whether the parity term or the non-parity term. The polarization of the `up' superradiant modes
is therefore the opposite to the non-superradiant modes in the same term, that is, it is negative if part of the non-parity
term and positive if part of the parity term. Note, however, that the `in' superradiant modes
have the same sign of $\omega$ (positive), and therefore the same polarization, as the non-superradiant modes
in the same term in the expectation value.

When CCH only include non-parity terms in their expressions (\ref{eq:stress tensor for s=1 on all vac.}) 
they are only including one polarization and leaving out the other one for the `in' modes. For the `up' modes, 
they are only including one polarization for the non-superradiant modes and the opposite polarization for the superradiant modes.
In particular, when subtracting the expectation value of the stress tensor 
in the past Boulware state from the one in the past Unruh state, only `up' modes are needed.
Neglecting the parity terms is in this case equivalent to neglecting negative polarization non-superradiant modes
as well as positive polarization superradiant modes. 
That is the case in the calculation of $\vac[ren]{\hat{T}^{\mu}{}_{\nu}}{B^-}$ close to the horizon in Section \ref{sec:RRO} 
but,
as explained in Subsection \ref{subsec:lack of symmetry}, in this limit
the non-parity and the parity terms coincide. \ddraft{is there a physical explanation for this coincidence?}

\draft{1) p.68(bis115)t->pi-t:we have linked the two opposite polarizations directly to pos. vs. negat. freqs. but in classical flat space
it looks as if there're always two polarizations and always integration for all freqs., so that this link does not seem
to exist?,}

\begin{figure}[p]
\rotatebox{90}
\centering
\includegraphics*[width=70mm,angle=270]{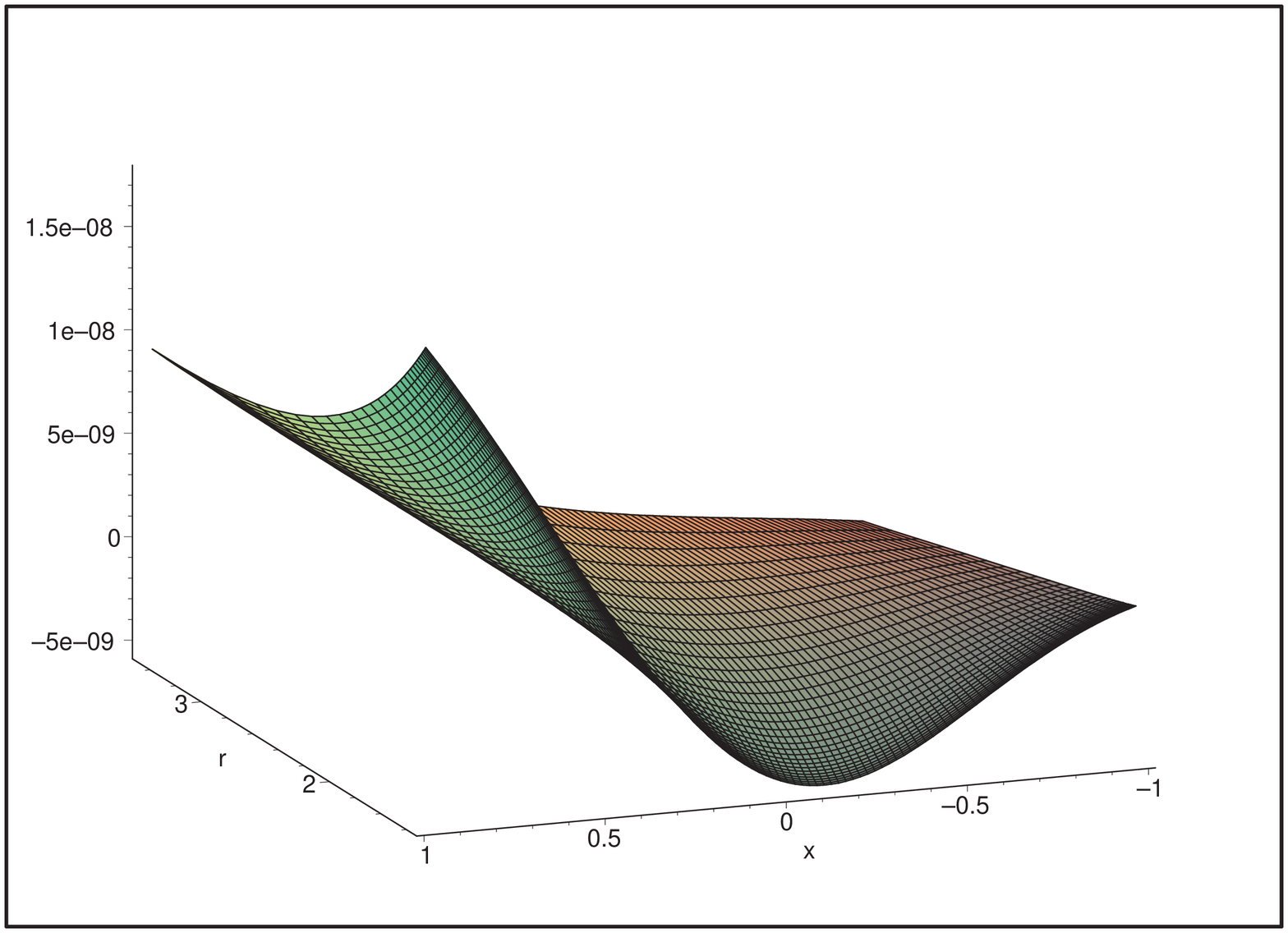}
\caption{Positive polarization terms of $\frac{1}{4\pi}\Delta\vac{\hat{T}_{tr}}{CCH^--U^-}$}
\label{fig:deltaTtr_cch_u_past}
\includegraphics*[width=70mm,angle=270]{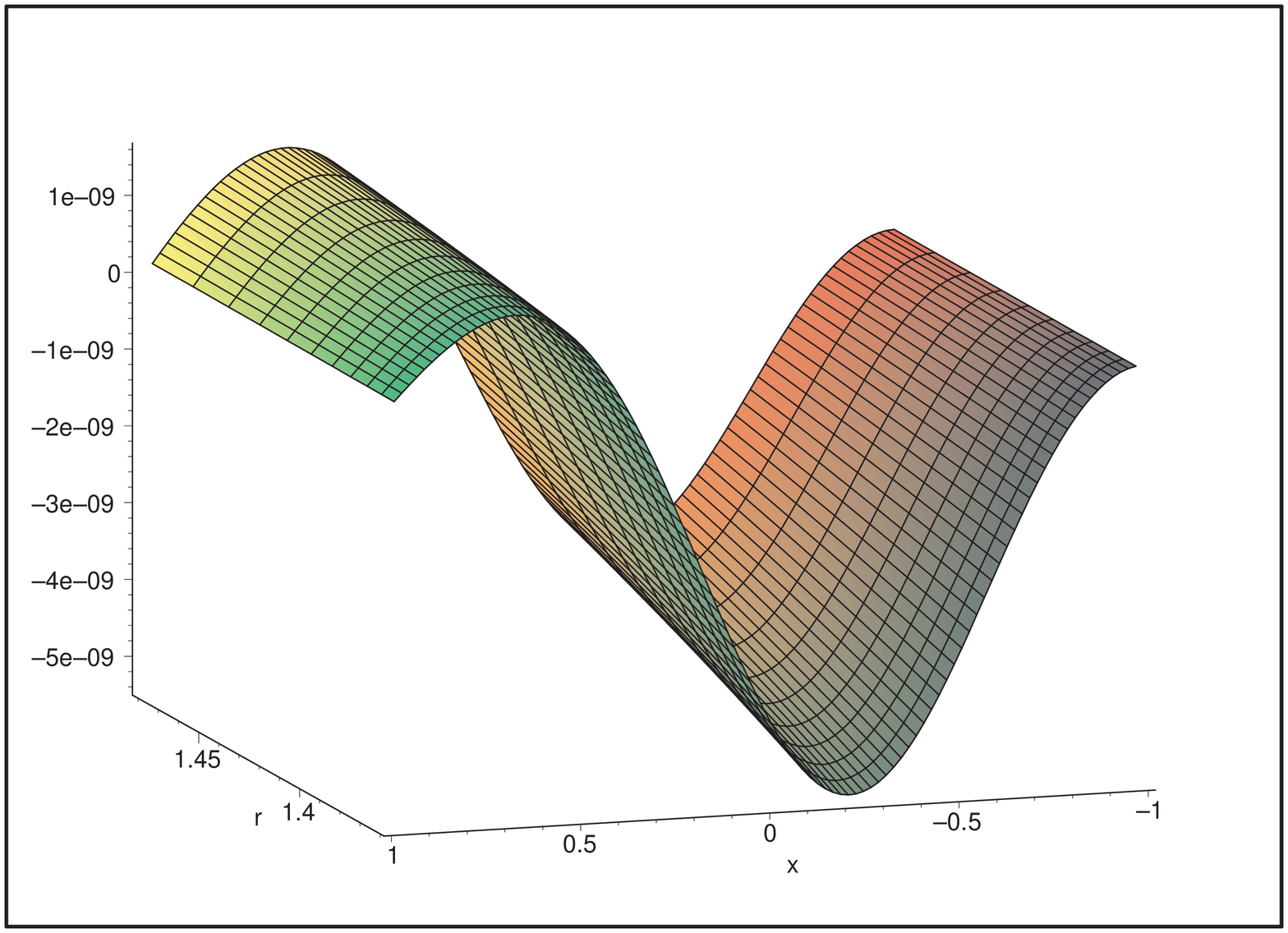}
\caption{Positive polarization terms of $\frac{1}{4\pi}\vac{\hat{T}_{t\phi}}{CCH^--U^-}$}
\label{fig:Ttphi_cch_u_past}
\end{figure}

\begin{figure}[p]
\rotatebox{90}
\centering
\includegraphics*[width=70mm,angle=270]{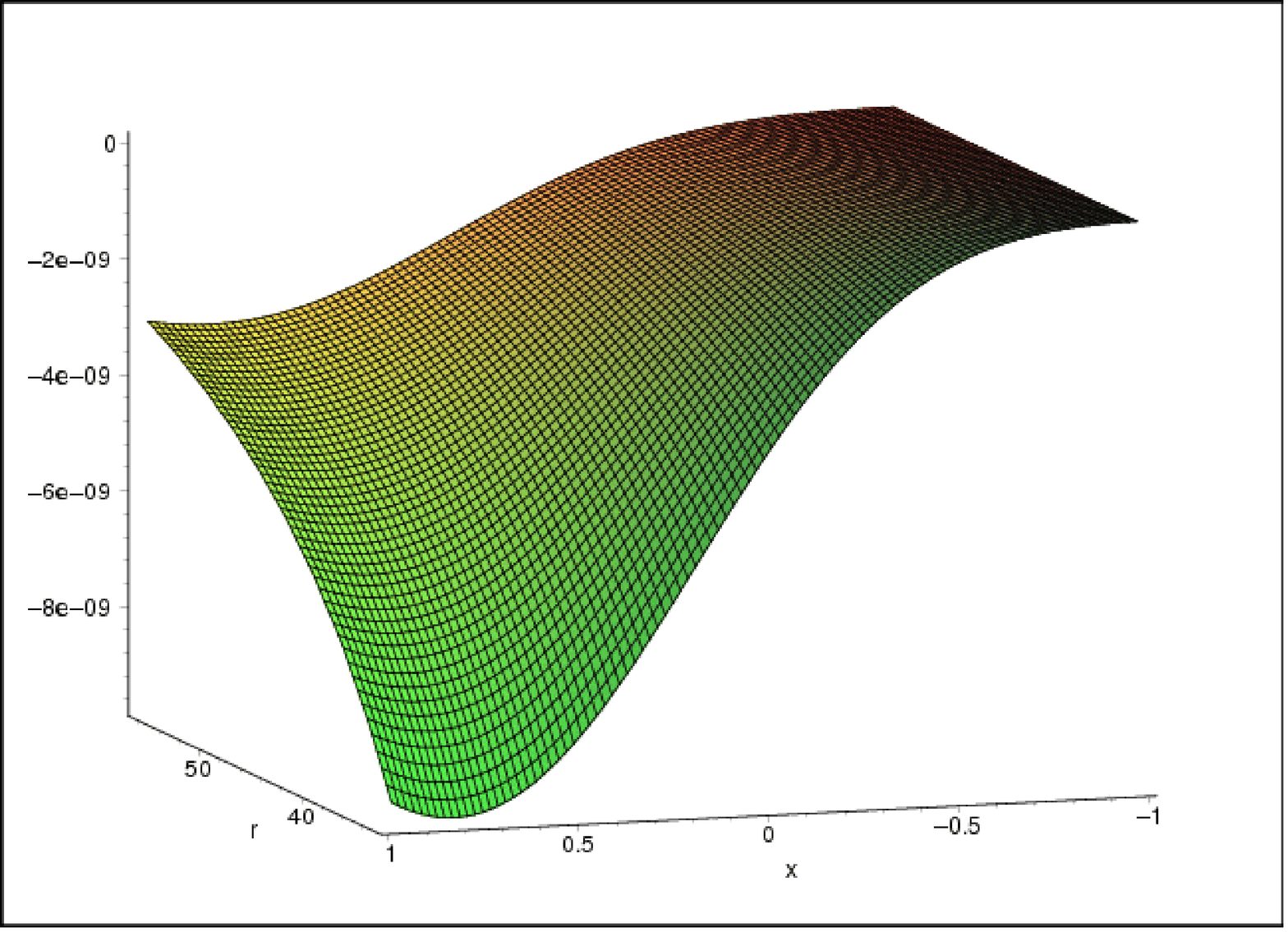}
\caption{Positive polarization terms of $\frac{1}{4\pi}\vac{\hat{T}_{tr}}{U^--B^-}$}
\label{fig:Ttr_polpos_u_b}
\includegraphics*[width=70mm,angle=270]{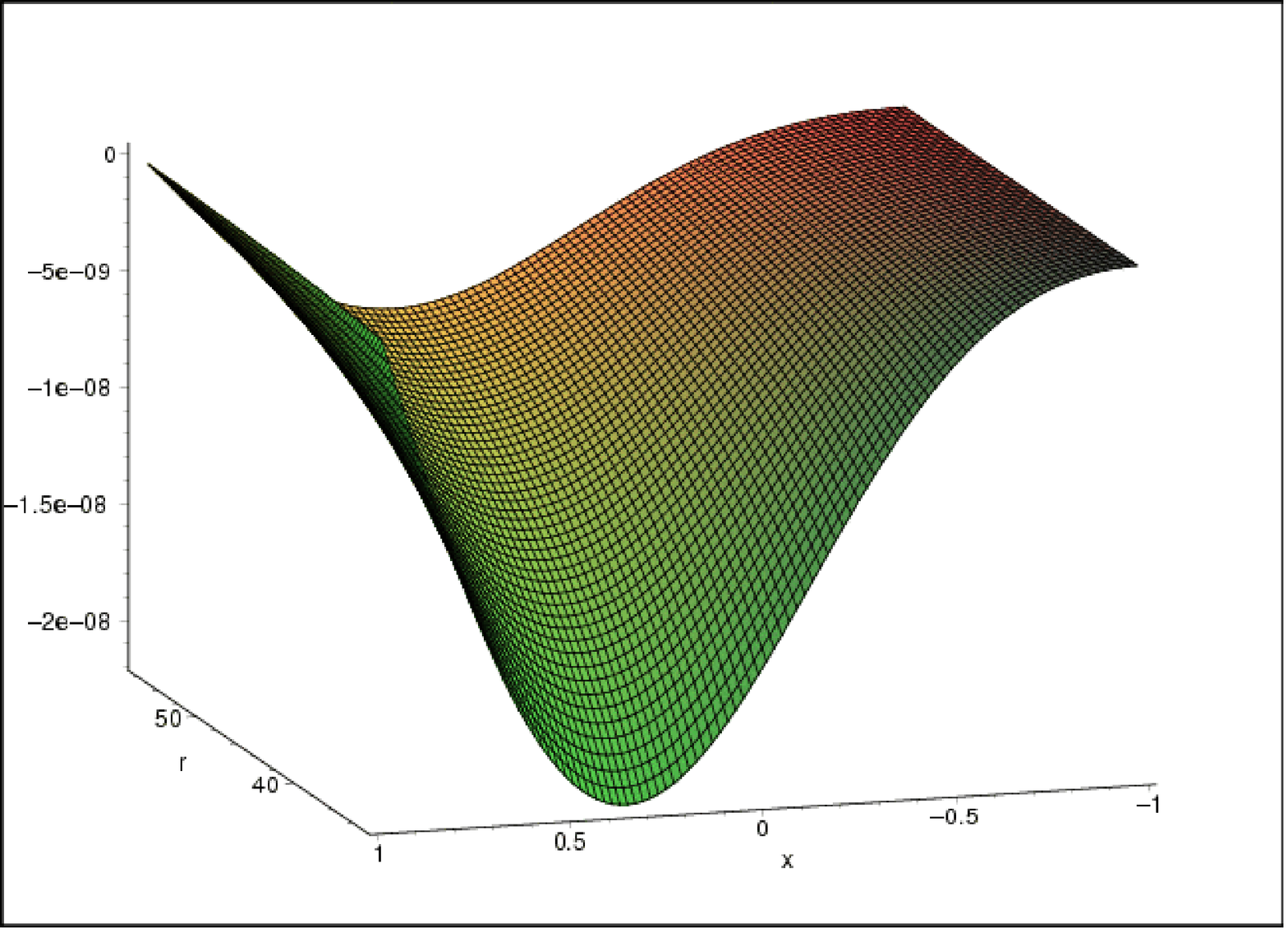}
\caption{Positive polarization terms of $\frac{1}{4\pi}\vac{\hat{T}_{t\phi}}{U^--B^-}$}
\label{fig:Ttphi_polpos_u_b}
\end{figure}

It is interesting to group the terms with the same polarization in the expectation value of the stress energy tensor.
Of course, in the case of the difference between the states $\ket{CCH^-}$ and $\ket{U^-}$, which only has contribution from the `in' modes,
the sum of the positive polarization terms coincides with the direct evaluation of CCH's expressions (\ref{eq:stress tensor for s=1 on all vac.}).
The negative polarization contribution can be obtained by applying the transformation $x\to -x$.
We include the plots of the tensor components corresponding to the fluxes of energy and angular momentum 
from the positive polarization terms in Figures \ref{fig:deltaTtr_cch_u_past}--\ref{fig:Ttphi_cch_u_past}.
The evaluation of the positive polarization contribution to the difference in the expectation value of the stress energy tensor 
between the states $\ket{U^-}$ and $\ket{B^-}$ requires carefully adding the contribution of the superradiant modes to the appropriate
polarization. We calculated the positive polarization contribution to these differences of 
expectation values and plot them in Figures \ref{fig:Ttr_polpos_u_b}--\ref{fig:Ttphi_polpos_u_b}.
The corresponding negative polarization contribution is, again, obtained by applying the transformation $x\to -x$.
The interest of these graphs lies in the region far from the horizon. Close to the horizon 
the irregularity of the state $\ket{B^-}$ dominates and thermality guarantees symmetry with respect to the equator.

\chapter*{Conclusions}  \label{ch:conclusions}
\addcontentsline{toc}{chapter}{Conclusions}
\markboth{}{Conclusions}

\subsection*{Results}

In this thesis we have aimed to give a precise and complete account of the quantum theory of linear spin-1 
perturbations of the Kerr and Kerr-Newman space-times.
This is a scarce subject in the literature compared to the volume of work on the scalar field in the Kerr
space-time or on the electromagnetic field in the Schwarzschild space-time, precisely because it is
considerably more difficult to deal with.

In Chapter \ref{ch:field eqs.} we gave a full account of the classical theory on a Type-D background 
where $\kappa=\sigma=\nu=\lambda=0$, based on the elegant and compact formalism introduced by Wald. 
We also showed that the ingoing and upgoing gauge electromagnetic potentials can both be naturally expressed 
in terms of one single Newman-Penrose Maxwell scalar, $\phi_0$. 
It is therefore possible
to reduce the quantization of the electromagnetic theory to that of a simpler, complex scalar theory. Unfortunately,
we showed that this was not viable in the Kerr space-time when using either the Kinnersley or the Carter
null tetrads since the field equation for $\phi_0$ is not separable in either case. It is however possible to do so in the 
Reissner-Nordstr\"{o}m space-time.

In Chapter \ref{ch:radial sln.} we studied the solution to the radial Teukolsky equation, corresponding to the decoupling
of the field equations for the other two Newman-Penrose scalars, $\phi_{+1}$ and $\phi_{-1}$, in the Kerr-Newman space-time.
The radial Teukolsky equation has a long-range potential, behaving as $1/r$ for large $r$. 
Its solution cannot be expressed in terms of any standard
functions and must be solved numerically. We considered the various alternative methods that convert the radial equation 
into one with a short-range potential. 
We numerically integrated the equation and compared our numerical results against those in the literature. 
We also completed an analysis of the behaviour of the general-spin radial solution close to the horizon 
following a study by Candelas that he only developed for spin-0. 
The chapter ended with a study of the asymptotics for small frequency based on a method presented by Page.

The solution of the angular Teukolsky equation was the topic of the following chapter. 
We presented the background research on these solutions and on their limiting cases of either spin-0 in Kerr space-time 
or else general spin in the Schwarzschild space-time. 
We numerically solved the angular equation for the spin-1 case in the Kerr-Newman space-time and presented the results.

Chapter \ref{ch:high freq. spher} was dedicated to the asymptotic analysis of the angular solution in the limit
of large frequency and fixed $m$. The study was based on a paper by Breuer, Ryan and Waller. Their analysis, however was incomplete and
partly flawed. They wrongly imposed a regularity condition on the solution and they also ignored the asymptotic solution 
that is valid in the region far from the boundary points. 
These are the reasons why they could not determine the parameter $\gamma$ on which the asymptotic
behaviour crucially depends. In this chapter we made the appropriate corrections to their paper and gave a complete 
account of the large frequency asymptotics with fixed $m$ of the eigenvalue and the angular solution for general spin in the
Kerr-Newman space-time. Such an account has not been presented in the literature before.

The last chapter undertook the quantization of the electromagnetic field on the Kerr background. It starts with a revision
of the results in the literature related to the construction of physical states in the Kerr background that have the same
defining features as the Boulware, Hartle-Hawking or Unruh states in the Schwarzschild background. 
We quantized the electromagnetic field following a canonical quantization method presented by Candelas, Chrzanowski and Howard (CCH)
which they used for the electromagnetic and gravitational fields. We calculated, both analytically and numerically, the luminosity
of a black hole when the electromagnetic field is in the past Boulware vacuum and when it is in the past Unruh state. 
We compared the results against related ones in the literature and exposed some of the algebraic complications existing in the calculation
of stress-energy tensor components for spin-1. We also calculated the expectation value of the 
renormalized stress-energy tensor (RSET) close to the horizon when the field is in the past Boulware vacuum. This calculation was
prompted by a result in CCH which did not agree with the expected result that it should correspond to minus the stress-energy
tensor of a thermal distribution at the Hawking temperature rigidly rotating with the horizon. 
Our numerical calculations agree with the latter rather than with CCH's result. 
We showed that the error in their calculation was caused by the fact that their asymptotic
approximation of the radial and angular functions close to the horizon was not uniform in $\tilde{\omega}$.
We further showed that the rate of rotation close to the horizon of the mentioned thermal distribution approaches that of a RRO
rather than that of a ZAMO or a Carter observer.

We initially used expressions in CCH for the expectation value of the stress-energy tensor when the field is in the $\ket{B^-}$,
$\ket{CCH^-}$ or $\ket{U^-}$ states. Both analytically and numerically they led to the surprising result that the difference of the 
RSET between two of the previous states was not invariant under the parity transformation $(\theta,\phi)\to(\pi-\theta,\phi+\pi)$.
We found that the reason for this asymmetry was the non-symmetrization of the quantum operators. 
We obtained the correct expressions, which are invariant under the parity transformation.
We finally showed that the non-parity and the parity terms appearing in these corrected expressions
correspond to two opposite polarizations, except for the case of the `up' superradiant modes which have opposite
polarization to the `up' non-superradiant modes in the same term.

\subsection*{Future work}

We have seen that the reduction in the Kerr or Kerr-Newman backgrounds of the quantization of the 
electromagnetic theory to a complex scalar theory is not possible using the Kinnersley or Carter 
null tetrads due to the non-separability of the equations for $\phi_0$. 
However, by making full use of the three classes of rotation of the NP frame it might be possible to find another 
null tetrad for which the corresponding equation for $\phi_0$ is separable.
Alternatively, it might be possible to prove either that such a tetrad does exist or else that it does not; we are not
aware of the existence of such a theorem. In any case, it is possible to separate the equation for $\phi_0$ in the
Reissner-Nordstr\"{o}m space-time. 
We intend to develop the quantization of the electromagnetic field in this space-time treating it as a complex scalar field.

We have obtained an asymptotic analysis of the radial solution close to the horizon and of the angular solution
for large frequency and fixed $m$. The asymptotic analysis of the angular solution that would allow us to obtain asymptotic
results for the RSET in different physical states close to the horizon is one that is uniform in $m$. 
A possible approach for obtaining this asymptotic analysis uniform in $m$ consists in performing an asymptotic analysis of 
the angular solution for large frequency and large $m$ and
then matching this analysis with the one for large frequency and fixed $m$. This is still an open problem.

In this thesis we have described the algebraic difficulties that calculations for the spin-1 field imply
in relation to those for the spin-0 field.
We believe, however, that with the insight we have gained into these calculations for the spin-1 field, a similar
analysis of the stess-energy tensor to that carried out in ~\cite{ar:Ott&Winst'00} for the spin-0 field
in terms of general physical principles is ready to be performed for spin-1.
Such an analysis would be very interesting in order to further our knowledge of the properties -regularity and symmetries in particular- 
of the different physical states in the Kerr-Newman space-time.

Finally, one of our initial aims was to investigate the extreme charged Kerr-Newman black hole. 
Even though we gradually diverted from this aim as we encountered various challenges, 
our analysis and programs only require minor modifications to produce results for the  extreme charged Kerr-Newman black hole.
This black hole has recently acquired relevant importance, particularly in light of the result (~\cite{ar:Bard&Horo'99}) that its geometry
close to the horizon has similar properties to the $A\d S_2\times S^2$ geometry. 
Some asymptotically anti-de-Sitter space-times have the interesting property that they can be in stable equilibrium with a thermal distribution 
(~\cite{ar:Hawking&Page'83}, ~\cite{ar:HawkHunt&Rob'99}).
The space-time corresponding to the geometry of a Kerr-Newman black hole embedded in the anti-de-Sitter universe is an example of these space-times.
These space-times are recently of huge interest because of a conjectured correspondence between gravity in the
anti-de-Sitter universe and conformal field theory on its boundary- the AdS/CFT correspondence.
We believe that a great part of our analysis and our programs can be adapted for investigation of such space-times.




\appendix
\chapter{Radial numerics} \label{ch:App.A}

The following are the values of the coefficients ${}_ic_{lm\omega}$ that appear in the asymptotic expansion 
(\ref{eq:X_up/B_up, r->inf}) of the radial function $X^{\text{up} *}_{lm\omega}/B^{\text{up} *}_{lm\omega}$ 
in the limit $r_* \rightarrow +\infty$:

\begin{subequations}
\begin{align}
{}_1c_{lm\omega}&=-\frac{i\left({}_{-1}\lambda_{lm\omega}+2am\omega\right)}{2\omega}
\\
{}_2c_{lm\omega}&=\frac{\left({}_{-1}\lambda_{lm\omega}+2am\omega\right)}{4\omega^2}-iamM
\\
\begin{split}
{}_3c_{lm\omega}&=-\frac{M{}_{-1}\lambda_{lm\omega}}{2\omega^2}-
\frac{i}{24\omega^3}
\left({}_{-1}\lambda_{lm\omega}^2+4a\omega m{}_{-1}\lambda_{lm\omega}+4a\omega m+32am\omega^3M^2- \right.\\&\left.
-8a^3\omega^3m-3\kappa-4a^2\omega^2-4\omega^2{}_{-1}\lambda_{lm\omega}a^2-8a\omega^3Q^2m\right)
\end{split}
\\
\begin{split}
{}_4c_{lm\omega}&=
\frac{1}{16\omega^4}\left(2{}_{-1}\lambda_{lm\omega}^2+4{}_{-1}\lambda_{lm\omega}\omega^2Q^2+4m^2a^2\omega^2+8{}_{-1}\lambda_{lm\omega}ma\omega-8a^3\omega^3m-
\right.\\&\left.
-4a^2\omega^2-4a^2\omega^2{}_{-1}\lambda_{lm\omega}+4\omega am-3\kappa\right)
-\frac{i}{16\omega^4}\left(-M\omega {}_{-1}\lambda_{lm\omega}^2-\right.\\&
-16\omega^4MQ^2am+4\omega^3Ma^2-16Ma^3\omega^4m+32amM^3\omega^4-4amM\omega^2+\\&\left.
+10M\omega \kappa\right)
\end{split}
\displaybreak[0]
\\
\begin{split}
{}_5c_{lm\omega}&=
\frac{1}{80\omega^5}\left(-80\omega^2M{}_{-1}\lambda_{lm\omega}am+80\omega^3{}_{-1}\lambda_{lm\omega}a^2M+80\omega^3a^2M+130M\omega \kappa- \right.\\&\left.
-40M{}_{-1}\lambda_{lm\omega}^2\omega-80M\omega^2am\right)
-\frac{i}{80\omega^5}\left(-8a^2\omega^2+2{}_{-1}\lambda_{lm\omega}^2+30\kappa-\right.\\&
-4\omega^2{}_{-1}\lambda_{lm\omega}a^2+16a^4\omega^4-15\kappa{}_{-1}\lambda_{lm\omega}-40a^3\omega^3m-8a^2\omega^4Q^2+\\&
+24a^2\omega^2m^2+20\omega^2\kappa a^2-30\omega^2\kappa Q^2+8a\omega^3Q^2m+4a\omega m{}_{-1}\lambda_{lm\omega}-\\&
-50a\omega m\kappa+16a^5\omega^5m+8a^4\omega^4{}_{-1}\lambda_{lm\omega}-4a^2\omega^2{}_{-1}\lambda_{lm\omega}^2+2\omega^2{}_{-1}\lambda_{lm\omega}^2Q^2+\\&
+{}_{-1}\lambda_{lm\omega}^3-16a^3\omega^3m{}_{-1}\lambda_{lm\omega}+8a^2\omega^2m^2{}_{-1}\lambda_{lm\omega}+6a\omega m{}_{-1}\lambda_{lm\omega}^2+\\&
+32a^3\omega^5Q^2m-192M^2a^3\omega^5m+8a\omega m+16\omega^5Q^4am-\\&\left.
-192Q^2\omega^5M^2am+256M^4\omega^5am-60\omega^2M^2\kappa\right)
\end{split}
\\
\begin{split}
{}_6c_{lm\omega}&=-\frac{1}{48\omega^6}\left(24Q^2\omega^4{}_{-1}\lambda_{lm\omega}a^2+54\omega^2Q^2\kappa-30\omega^2\kappa a^2+24a^4\omega^4m^2+\right.\\&
+48\omega^4M^2a^2-24\omega^2M^2{}_{-1}\lambda_{lm\omega}^2-24a\omega m{}_{-1}\lambda_{lm\omega}^2+48a^3\omega^3m{}_{-1}\lambda_{lm\omega}+\\&+
84\omega am\kappa+210\omega^2M^2\kappa-8a^3\omega^3m^3-36a^2\omega^2m^2{}_{-1}\lambda_{lm\omega}+12a^2\omega^2{}_{-1}\lambda_{lm\omega}^2-\\&-
12a^4\omega^4{}_{-1}\lambda_{lm\omega}-12\omega^2{}_{-1}\lambda_{lm\omega}^2Q^2-24a^5\omega^5m+24a^2\omega^4Q^2-24a^4\omega^4-\\&-
24Q^2\omega^3{}_{-1}\lambda_{lm\omega}am+27\kappa{}_{-1}\lambda_{lm\omega}-12{}_{-1}\lambda_{lm\omega}a\omega m-4{}_{-1}\lambda_{lm\omega}^3-\\&-
24\omega^3Q^2am-48\omega^3M^2am-48\omega^2a^2m^2-12a\omega m+12a^2\omega^2+72\omega^3a^3m+\\&\left. +
12\omega^2{}_{-1}\lambda_{lm\omega}a^2-45\kappa-3{}_{-1}\lambda_{lm\omega}^2\right)+\frac{i}{48\omega^6}\left(256a^3\omega^6M^3m-256\omega^6M^5am- \right.\\&-
138M\omega^2am\kappa-63M\omega \kappa{}_{-1}\lambda_{lm\omega}+96M\omega^3\kappa a^2-42M\omega^3\kappa Q^2+\\&+
32M\omega^2am-6Ma^2{}_{-1}\lambda_{lm\omega}^2\omega^3+6ma{}_{-1}\lambda_{lm\omega}^2M\omega^2+2{}_{-1}\lambda_{lm\omega}^3M\omega +\\&+
231M\omega \kappa+24a^4M\omega^5+24M\omega^3a^2m^2+256Q^2\omega^6M^3am-\\&-
96Q^2\omega^6Ma^3m-48Q^4\omega^6Mam-48a^5M\omega^6m-8M{}_{-1}\lambda_{lm\omega}a^2\omega^3+\\&\left.+
8M{}_{-1}\lambda_{lm\omega}a\omega^2m-32a^2\omega^3M-48Ma^3\omega^4m+8M{}_{-1}\lambda_{lm\omega}^2\omega \right)
\end{split}
\end{align}
\end{subequations}

The following table shows the values of 
the radial functions ${}_{-1}R^{\text{chandr}}_{2,-2,-0.5}$ from TableV in Chandrasekhar's ~\cite{bk:Chandr} Appendix 
and ${}_{-1}R^{\text{sym,num}}_{2,-2,-0.5}$ calculated with Fortran90 
program \program{raddrv2KN.f} described in Section \ref{sec:num. method; radial func.}.

\newpage

\tablecaption{
Radial functions and their derivatives for $\indhel=-1$, $Q=0$, $a=0.95$, $l=2$, $m=-2$, $\omega=-0.5$.
Within each cell for the radial functions and for the derivatives the top value corresponds to the real part and
the bottom value to the imaginary part.} 
\begin{supertabular}{c|cccc}
\tabletail{\hline}
$r/M$& ${}_{-1}R^{\text{chandr}}_{2,-2,-0.5}$           & ${}_{-1}R^{\text{sym,num}}_{2,-2,-0.5}$ 
     & $\diff{{}_{-1}R^{\text{chandr}}_{2,-2,-0.5}}{r}$ & $\diff{{}_{-1}R^{\text{sym,num}}_{2,-2,-0.5}}{r}$\\
\hline
\hline
2.1& 1.2003& 1.20026785136567&2.18133& 2.18334724475591 \\  &
-0.10873& -0.10875922190824& -1.2468&-1.24677817342713 \\ \hline
2.2& 1.4350& 1.43492663358771&2.5017& 2.50176744330508 \\  &
-0.24999& -0.25002380500959& -1.5818&-1.58179707611184 \\ \hline 
2.3& 1.6992& 1.69915736492414&2.7757& 2.77575249647072 \\  &
-0.42574& -0.42576456994030& -1.9361&-1.93613390496899 \\ \hline 
2.4& 1.9887& 1.98869418055624&3.0079& 3.00797480302144 \\  &
-0.63781& -0.63783865284620& -2.3080&-2.30801570052304 \\ \hline 
2.5& 2.2994& 2.29938556170459&3.1991& 3.19916057586282 \\  &
-0.88783& -0.88785789689252& -2.6946&-2.69460457858789 \\ \hline 
2.6& 2.6272& 2.62716311795229&3.3492& 3.34919957529641 \\  &
-1.1771& -1.17714459161815& -3.0926&-3.09264287195368 \\ \hline 
2.7& 2.9679& 2.96782782649048&3.4575& 3.45746844788633 \\  &
-1.5066& -1.50664472608102& -3.4985&-3.49853230609708 \\ \hline 
2.8& 3.3173& 3.31722704239298&3.5234& 3.52337153886738 \\  &
-1.8770& -1.87698636533703& -3.9087&-3.90869011633273 \\ \hline 
2.9& 3.6711& 3.67107176062917&3.5463& 3.54629329952196 \\  &
-2.2884& -2.28840304104407& -4.3194&-4.31941466954318 \\ \hline 
3.0& 4.0251& 4.02503739941419&3.5258& 3.52580194116473 \\  &
-2.7407& -2.74076500607478& -4.7270&-4.72703002192608 \\ \hline 
3.1& 4.3748& 4.37477566784955&3.4617& 3.46169488540498 \\  &
-3.2336& -3.23358278453457& -5.1279&-5.12790900693526 \\ \hline 
3.2& 4.7160& 4.71592778158095&3.3540& 3.35402047306941 \\  &
-3.7660& -3.76600886397982& -5.5185&-5.51848547182964 \\ \hline 
3.3& 5.0442& 5.04413646655539&3.2031& 3.20308181562904 \\  &
-4.3368& -4.33681855987963& -5.8953&-5.89525158089140 \\ \hline 
3.4& 5.3551& 5.35511788785783&3.0095& 3.00948363393847 \\  &
-4.9445& -4.94448127866285& -6.2549&-6.25483895346162 \\ \hline 
3.5& 5.6447& 5.64463977090707&2.7741& 2.77408878295112 \\  &
-5.5871& -5.58710647199081& -6.5940&-6.59396595505323 \\ \hline 
3.6& 5.9086& 5.90858014436306&2.4980& 2.49803814126054 \\  &
-6.2625& -6.26248554733942& -6.9095&-6.90949152985754 \\ \hline 
3.7& 6.1430& 6.14293415562655&2.1827& 2.18273401664141 \\  &
-6.9681& -6.96812791317567& -7.1984&-7.19842063374193 \\ \hline 
3.8& 6.3439& 6.34387247839626&1.8298& 1.82983839651524 \\  &
-7.7012& -7.70119148771223& -7.4579&-7.45791354146316 \\ \hline
3.9& 6.5077& 6.50771674910113&1.4412& 1.44124730666313 \\  &
-8.4586& -8.45862800399240& -7.6853&-7.68530442839311 \\
 \hline 
4.0& 6.6310& 6.6310&1.0191& 1.01908734960253 \\  &
-9.2371& -9.2371& -7.8781&-7.87810703311225 \\ \hline 
\goodbreak
4.1& 6.7105& 6.71049567442221&0.56571& 0.56570603328326 \\  &
-10.0330& -10.03301997548350& -8.0341&-8.03403834407662 \\ \hline 
4.2& 6.7432& 6.74317769234848&0.83637& 0.08363318452101 \\  &
-10.8430& -10.84260964545399& -8.1510&-8.15099433940105 \\ \hline 
4.3& 6.7264& 6.72632534827829&-0.042439& -0.42440357766359 \\  &
-11.6620& -11.66187058988782& -8.2272&-8.22710903213187 \\ \hline 
4.4& 6.6576& 6.65754159377316&-0.09555& -0.95549606758222 \\  &
-12.4870& -12.48661417641590& -8.2608&-8.26076519166212 \\ \hline 
4.5& 6.5346& 6.53458927651571&-1.5066& -1.50663575355877 \\  &
-13.3130& -13.31255141698046& -8.2505&-8.25048650875395 \\ \hline 
4.6& 6.3557& 6.35565169996788&-2.0746& -2.07464130742694 \\  &
-14.1350& -14.13520779834626& -8.1951&-8.19509763948480 \\ \hline 
4.7& 6.1192& 6.11911479670302&-2.6562& -2.65625398486783 \\  &
-14.9500& -14.95002770851276& -8.0937&-8.09358159395761 \\ \hline 
4.8& 5.8241& 5.82397789741934&-3.2480& -3.24805124849254 \\  &
-15.7520& -15.75237325602916& -7.9455&-7.94536770095228 \\ \hline 
4.9& 5.4694& 5.46938483164315&-3.8466& -3.84658353999649 \\  &
-16.5380& -16.53754452954476& -7.7500&-7.74998554852685 \\ \hline 
5.0& 5.0547& 5.05467081851123&-4.4484& -4.44837518344008 \\  &
-17.3010& -17.30079952288854& -7.5071&-7.50709894810016 \\ \hline 
5.1& 4.5797& 4.57969767206908&-5.0499& -5.04986329143664 \\  &
-18.0370& -18.03737403247108& -7.2168&-7.21674248154373 \\ \hline 
5.2& 4.0448& 4.04462210157875&-5.6475& -5.64747476292679 \\  &
-18.7430& -18.74251756497168& -6.8793&-6.87915515030467 \\ \hline 
5.3& 3.4505& 3.45026879397524&-6.2376& -6.23760299462583 \\  &
-19.4120& -19.41162505323660& -6.4952&-6.49506253045761 \\ \hline 
5.4& 2.7977& 2.79759552270645&-6.8167& -6.81668061192606 \\  &
-20.0400& -20.04007705958522& -6.0653&-6.06525657188396 \\ \hline 
5.5& 2.0876& 2.08760103455425&-7.3812& -7.38118691076062 \\  &
-20.6230& -20.62321598203533& -5.5906&-5.59052672884331 \\ \hline 
5.6& 1.3220& 1.32190457349643&-7.9277& -7.92764415522192 \\  &
-21.1570& -21.15664442715797& -5.0722&-5.07211926348503 \\ \hline 
5.7& 0.50282& 0.50271763626229&-8.4527& -8.45263822831099 \\  &
-21.6360& -21.63621445472158& -4.5118&-4.51170165964859 \\ \hline 
5.8& -0.36768& -0.36768764471486&-8.9528& -8.95282258003868 \\  &
-22.0580& -22.05775138133475& -3.9110&-3.91093638972109 \\ \hline 
5.9& -1.2868& -1.28682498342566&-9.4250& -9.42493652904060 \\  &
-22.4170& -22.41718655405400& -3.2716&-3.27161266174137 \\ \hline 
6.0& -2.2516& -2.25163946019482&-9.8659& -9.86586697577774 \\  &
-22.7110& -22.71085893051983& -2.5960&-2.59593849391648 \\ \hline 
6.1& -3.2589& -3.25887087136051&-10.2730& -10.27260628469677 \\  &
-22.9350& -22.93522595577702& -1.8863&-1.88623285460110 \\ \hline 
6.2& -4.3049& -4.30492749584150&-10.6420& -10.64231072724808 \\  &
-23.0870& -23.08705554175912& -1.1451&-1.14504283409244 \\ \hline 
6.3& -5.3860& -5.38600209115882&-10.9720& -10.97226665323438 \\  &
-23.1630& -23.16328981815513& -0.37505&-0.37503743621995 \\ \hline 
6.4& -6.4980& -6.49797039906948&-11.2600& -11.25994137512047 \\  &
-23.1610& -23.16120229611491& 0.42089&0.42089987073883 \\ \hline 
6.5& -7.6365& -7.63650873314380&-11.5030& -11.50295464488002 \\  &
-23.0780& -23.07828061195214& 1.2397&1.23977308060567 \\ \hline 
6.6& -8.7970& -8.79701255583228&-11.6990& -11.69920334879844 \\  &
-22.9130& -22.91253679020954& 2.0784&2.07837074709545 \\ \hline 
6.7& -9.9747& -9.97471998568154&-11.8470& -11.84673549848341 \\  &
-22.6620& -22.66217107792423& 2.9334&2.93340696991069 \\ \hline 
6.8& -11.1650& -11.16467396352343&-11.9440& -11.94373627176464 \\  &
-22.3260& -22.32556159484567& 3.8015&3.80149794211117 \\ \hline 
6.9& -12.3620& -12.36170540720160&-11.9890& -11.98854519165635 \\  &
-21.9020& -21.90131337097084& 4.6791&4.67914339708598 \\ \hline 
7.0& -13.5610& -13.56055044001096&-11.9800& -11.97988874356997 \\  &
-21.3900& -21.38872477349637& 5.5627&5.56273274110761\\
\end{supertabular} 
\label{table:data R_1 a=0.95,l=2,m=-2,w=-0.5;tableVChandr}

\bibliographystyle{plain}

\begin{thebibliography}{10}

\bibitem{bk:AS}
Milton Abramowitz and Irene~A. Stegun.
\newblock {\em Handbook of Mathematical Functions}.
\newblock Dover Publications,Inc., New York, USA, ninth edition, 1965.

\bibitem{ar:B&Carter&Hawk'73}
James~M. Bardeen, B.~Carter, and S.W. Hawking.
\newblock The four laws of black hole mechanics.
\newblock {\em Commun.\ Math.\ Phys}, 31:161--170, 1973.

\bibitem{ar:Bard&Horo'99}
James~M. Bardeen and Gary~T. Horowitz.
\newblock The extreme {K}err throat geometry: A vacuum analog of
  ${A}d{S}_2\times {S}^2$.
\newblock {\em Phys.\ Rev.\ D}, 60:104030, 1999.

\bibitem{bk:Bender&Orszag}
Carl~M. Bender and Steven~A. Orszag.
\newblock {\em Advanced mathematical methods for scientists and engineers}.
\newblock McGraw-Hill, 1978.

\bibitem{ar:Berti&Card&Yosh'04}
Emanuele Berti, Vitor Cardoso, and Shijun Yoshida.
\newblock Highly damped quasinormal modes of {K}err black holes: A complete
  numerical investigation.
\newblock {\em Phys.\ Rev.\ D}, 69:124018, 2004.

\bibitem{ar:Bose'75}
S.K. Bose.
\newblock Studies in the {K}err-{N}ewman metric.
\newblock {\em J.\ Math.\ Phys.}, 16(4):772--775, 1975.

\bibitem{ar:Boyer&Lind'67}
Robert~H. Boyer and Richard~W. Lindquist.
\newblock Maximal analytic extension of the {K}err metric.
\newblock {\em J.\ Math.\ Phys.}, 8(2):265--281, 1967.

\bibitem{ar:BRW}
R.A. Breuer, M.P. Ryan~Jr, and S.~Waller.
\newblock Some properties of spin-weighted spheroidal harmonics.
\newblock {\em Proc.\ R.\ Soc.\ Lond. A}, 358:71--86, 1977.

\bibitem{ar:Breuer'75}
Reinhard~A. Breuer.
\newblock Gravitational perturbation theory and synchrotron radiation.
\newblock {\em Lecture Notes in Physics}, 44, 1975.

\bibitem{ar:Brown&Ott'83}
M.R. Brown and Adrian~C. Ottewill.
\newblock The energy momentum operator in curved space-time.
\newblock {\em Proc.\ R.\ Soc.\ Lond. A}, 389:379--403, 1983.

\bibitem{bk:Buchholz}
H.~Buchholz.
\newblock {\em The Confluent Hypergeometric Function}.
\newblock Springer Tracts in Natural Philosophy, 1969.

\bibitem{ar:Campbell'71}
William~B. Campbell.
\newblock Tensor and spinor spherical harmonics and the spin-$s$ harmonics
  ${}_sy_{lm}(\theta,\phi)$.
\newblock {\em J.\ Math.\ Phys.}, 12(8):1763--1770, 1971.

\bibitem{ar:Candelas'80}
Philip Candelas.
\newblock Vacuum polarization in {S}chwarzschild spacetime.
\newblock {\em Phys.\ Rev.\ D}, 21(8):2185--2202, 1980.

\bibitem{ar:CCH}
Philip Candelas, P.~Chrzanowski, and K.~W. Howard.
\newblock Quantization of electromagnetic and gravitational perturbations of a
  {K}err black hole.
\newblock {\em Phys.\ Rev.\ D}, 24(2):297--304, 1981.

\bibitem{ar:Cand&Deutsch'77}
Philip Candelas and David Deutsch.
\newblock On the vacuum stress induced by uniform acceleration or supporting
  the ether.
\newblock {\em Proc.\ R.\ Soc.\ Lond. A}, 354:79--99, 1977.

\bibitem{ar:Carter'68b}
B.~Carter.
\newblock {H}amilton-{J}acobi and {S}chrodinger separable solutions of
  {E}instein's equations.
\newblock {\em Commun.\ Math.\ Phys.}, 10:280--210, 1968.

\bibitem{bk:Carter-DeWittDeWitt}
B.~Carter.
\newblock In B.~DeWitt and C.~DeWitt, editors, {\em Black Holes}, New York,
  USA, 1973. Gordon and Breach.

\bibitem{bk:Carter-Cargese'86}
B.~Carter.
\newblock Mathematical foundations of the theory of relativistic stellar and
  black hole configurations.
\newblock In B.~Carter and J.B. Hartle, editors, {\em Gravitation in
  Astrophysics. Carg\`{e}se 1986}, New York, USA, 1987. Plenum Press and NATO
  Scientific Affairs Division.

\bibitem{ar:Carter'68a}
Brandon Carter.
\newblock Global structure of the {K}err family of gravitational fields.
\newblock {\em Phys.\ Rev.}, 174(5):1559--1571, 1968.

\bibitem{bk:Chandr}
S.~Chandrasekhar.
\newblock {\em The Mathematical Theory of Black Holes}.
\newblock Oxford University Press, Oxford, UK, second edition, 1992.

\bibitem{ar:Christ'78}
S.~M. Christensen.
\newblock Regularization, renormalization, and the covariant geodesic point
  separation.
\newblock {\em Phys.\ Rev.\ D}, 17(4):946--963, 1978.

\bibitem{ar:Christ&Fulling'77}
S.~M. Christensen and S.~A. Fulling.
\newblock Trace anomalies and the {H}awking effect.
\newblock {\em Phys.\ Rev.\ D}, 15(8):2088--2104, 1977.

\bibitem{ar:Christo'70}
D.~Christodoulou.
\newblock Reversible and irreversible transformations in black hole physics.
\newblock {\em Phys.\ Rev.\ Lett.}, 25:1596--1597, 1970.

\bibitem{ar:Christo&Ruff'71}
D.~Christodoulou and R.~Ruffini.
\newblock Reversible transformations of a charged black hole.
\newblock {\em Phys.\ Rev.\ D}, 4:3552--3555, 1971.

\bibitem{ar:Chrzan'75}
Paul~L. Chrzanowski.
\newblock Vector potential and metric perturbations of a rotating black hole.
\newblock {\em Phys.\ Rev.\ D}, 11(8):2042--2062, 1975.

\bibitem{ar:CMSR}
Paul~L. Chrzanowski, Richard~A. Matzner, Vernon~D. Sandberg, and Michael~P.
  Ryan~Jr.
\newblock Zero-mass plane waves in nonzero gravitational backgrounds.
\newblock {\em Phys.\ Rev.\ D}, 14(2):317--326, 1976.

\bibitem{ar:Chrzan&Misner'75}
Paul~L. Chrzanowski and C.W. Misner.
\newblock Geodesic synchroton radiation in the {K}err geometry by the method of
  asymptotically factorized {G}reen's functions.
\newblock {\em Phys.\ Rev.\ D}, 10(6):1701--1721, 1974.

\bibitem{ar:Coh&Keg'74}
Jeffrey~M. Cohen and Lawrence~S. Kegeles.
\newblock Electromagnetic fields in curved spaces: A constructive procedure.
\newblock {\em Phys.\ Rev.\ D}, 10(4):1070--1084, 1974.

\bibitem{ar:Detw'76}
Stephen~L. Detweiler.
\newblock On the equations governing the electromagnetic perturbations of the
  {K}err black hole.
\newblock {\em Proc.\ R.\ Soc.\ Lond. A}, 349:217--230, 1976.

\bibitem{ar:Detw'77}
Stephen~L. Detweiler.
\newblock On resonant oscillations of a rapidly rotating black hole.
\newblock {\em Proc.\ R.\ Soc.\ Lond.\ A}, 352:381--395, 1977.

\bibitem{bk:Dirac}
P.A.M. Dirac.
\newblock {\em General Theory of Relativity}.
\newblock Wiley, New York, 1975.

\bibitem{th:GavPhD}
Gavin Duffy.
\newblock {\em Scalar Quantum Field Theory on the {K}err Black Hole
  Background}.
\newblock PhD thesis, University College Dublin, 2002.

\bibitem{bk:high_transc_funcs}
A.~Erd\'{e}lyi, W.~Magnus, F.~Oberhettinger, and F.G. Tricomi.
\newblock {\em Higher Transcendental Functions}.
\newblock Bateman Manuscript Project, 1953.

\bibitem{ar:Fac&Gross'76}
Edward~D. Fackerell and Robert~G. Grossman.
\newblock Spin-weighted angular spheroidal functions.
\newblock {\em J.\ Math.\ Phys.}, 18(9):1849--1854, 1977.

\bibitem{bk:Flammer}
C.~Flammer.
\newblock {\em Spheroidal Wave Functions}.
\newblock Stanford University Press, 1957.

\bibitem{ar:Friedman'78}
John~L. Friedman.
\newblock Ergosphere instability.
\newblock {\em Commun.\ Math.\ Phys}, 63:243--255, 1978.

\bibitem{ar:F&T'89}
Valery~P. Frolov and Kip~S. Thorne.
\newblock Renormalized stress-energy tensor near the horizon of a slowly
  evolving rotating black hole.
\newblock {\em Phys.\ Rev.\ D}, 39(8):2125--2154, 1989.

\bibitem{ar:F&Z'85}
Valery~P. Frolov and A.~I. Zel'nikov.
\newblock Vacuum polarization of the electromagnetic field near a rotating
  black hole.
\newblock {\em Phys.\ Rev.\ D}, 32(12):3150--3163, 1985.

\bibitem{ar:Gold.etal.}
J.N. Goldberg, A.J. MacFarlane, E.T. Newman, F.~Rohrlich, and E.C.G. Sudarshan.
\newblock Spin-$s$ spherical harmonics and $\eth$.
\newblock {\em J.\ Math.\ Phys.}, 8(11):2155--2161, 1967.

\bibitem{bk:GR}
I.~S. Gradshteyn and I.~M. Ryzhik.
\newblock {\em Table of Integrals, Series and Products}.
\newblock Academic Press, San Diego, fifth edition, 1995.

\bibitem{bk:MPI}
W.~Gropp, E.~Lusk, and A.~Skjellum.
\newblock {\em Using {MPI}. Portable Parallel Programming with the {M}essage
  {P}assing {I}nterface}.
\newblock The MIT Press, Cambridge, Massachusetts, second edition, 1999.

\bibitem{ar:Grove&Ott'83}
P.G. Grove and Adrian~C. Ottewill.
\newblock Notes on `particle detectors'.
\newblock {\em J.\ Phys.\ A}, 16:3905--3920, 1983.

\bibitem{ar:Hawking&Page'83}
S.~W. Hawking and Don~N. Page.
\newblock Thermodynamics of black holes in anti-de {S}itter space.
\newblock {\em Commun.\ Math.\ Phys.}, 87:577--587, 1983.

\bibitem{ar:Hawking'71b}
Stephen~W. Hawking.
\newblock Gravitational radiation from colliding black holes.
\newblock {\em Phys.\ Rev.\ Lett.}, 26:1344--1346, 1971.

\bibitem{ar:Hawking'75}
Stephen~W. Hawking.
\newblock Particle creation by black holes.
\newblock {\em Commun.\ Math.\ Phys.}, 43:199--220, 1975.

\bibitem{ar:HawkHunt&Rob'99}
Stephen~W. Hawking, C.J. Hunter, and M.M. Taylor-Robinson.
\newblock Rotation and the {A}d{S}/{CFT} correspondence.
\newblock {\em Phys.\ Rev.\ D}, 59:064005, 1999.

\bibitem{ar:Hod'98}
S.~Hod.
\newblock {B}ohr's correspondence principle and the area spectrum of quantum
  black holes.
\newblock {\em Phys.\ Rev.\ Lett.}, 81:4293--4296, 1998.

\bibitem{ar:Ipser'71}
James~R. Ipser.
\newblock Electromagnetic test fields around a {K}err-metric black hole.
\newblock {\em Phys.\ Rev.\ Lett.}, 27(8):529--531, 1971.

\bibitem{ar:Israel'76}
Werner Israel.
\newblock Thermo-field dynamics of black holes.
\newblock {\em Phys.\ Lett.\ A}, 57(2):107--110, 1976.

\bibitem{ar:Israel'86}
Werner Israel.
\newblock Third law of black hole dynamics: A formulation and proof.
\newblock {\em Phys.\ Rev.\ Lett.}, 57:397--399, 1986.

\bibitem{ar:J&McL&Ott'91}
Bruce~P. Jensen, John~G. McLaughlin, and Adrian~C. Ottewill.
\newblock Renormalized electromagnetic stress tensor for an evaporating black
  hole.
\newblock {\em Phys.\ Rev.\ D}, 43(12):4142--4144, 1991.

\bibitem{ar:J&McL&Ott'92}
Bruce~P. Jensen, John~G. McLaughlin, and Adrian~C. Ottewill.
\newblock Anisotropy of the quantum thermal state in {S}chwarzschild
  space-time.
\newblock {\em Phys.\ Rev.\ D}, 45(8):3002--3005, 1992.

\bibitem{ar:J&McL&Ott'95}
Bruce~P. Jensen, John~G. McLaughlin, and Adrian~C. Ottewill.
\newblock One-loop quantum gravity in {S}chwarzschild space-time.
\newblock {\em Phys.\ Rev.\ D}, 51(10):5676--5697, 1995.

\bibitem{ar:J&McL&Ott'88}
Bruce~P. Jensen, John~G. McLaughlin, and Adrian~C. Ottewill.
\newblock Renormalized electromagnetic energy density on the horizon of a
  {K}err black hole.
\newblock {\em Class.\ Quantum.\ Grav.}, 5:L187--L189, 1999.

\bibitem{ar:Kang'97}
Gungwon Kang.
\newblock Quantum aspects of ergoregion instability.
\newblock {\em Phys.\ Rev.\ D}, 55:7563--7573, 1997.

\bibitem{ar:Kay&Wald'91}
Bernard~S. Kay and Robert~M. Wald.
\newblock Theorems on the uniqueness and thermal properties of stationary,
  nonsingular, quasifree states on spacetimes with a bifurcate {K}illing
  horizon.
\newblock {\em Physics Reports}, 207(2):49--136, 1991.

\bibitem{co:Kerr&Schild'65}
Roy.~P. Kerr and A.~Schild.
\newblock A new class of vacuum solutions of the {E}instein field equations.
\newblock In {\em Proceedings of the {G}alileo {G}alilei Centenary Meeting on
  General Relativity, Problems of Energy and Gravitational Waves}, pages
  222--233, Florence, 1965. Comitato Nazionale per le Manifestazione
  Celebrative.

\bibitem{ar:Kinnersley'69}
William Kinnersley.
\newblock Type {D} vacuum metrics.
\newblock {\em J.\ Math.\ Phys.}, 10:1195--1203, 1969.

\bibitem{ar:Kundt&Thomp'62}
W.~Kundt and A.~Thompson.
\newblock Le tenseur de {Weyl} et une congruence associe\'{e} de
  g\'{e}od\'{e}siques isotropes sans distorsion.
\newblock {\em C.R.\ Acad.\ Sci.\ Paris}, 254:4257--4259, 1962.

\bibitem{ar:Kundt&Trump'62}
W.~Kundt and M.~Tr\"{u}mper.
\newblock Beitr\"{a}ge zur theorie der gravitations-strahlungsfelder.
\newblock {\em Akad.\ Wiss.\ Lit.\ Mainz, Abhandl.\ Math.-Nat.\ Kl.},
  12:965--1000, 1962.

\bibitem{ar:Leaver86}
E.~W. Leaver.
\newblock Solutions to a generalized spheroidal wave equation: {T}eukolsky's
  equations in general relativity, and the two-center problem in molecular
  quantum mechanics.
\newblock {\em J.\ Math.\ Phys.}, 27(5):1238--1265, 1986.

\bibitem{bk:Louisell}
William~H. Louisell.
\newblock {\em Quantum Statistical Properties of Radiation}.
\newblock Wiley-Interscience, 1990.

\bibitem{ar:Mat&Dav&Ott'93}
A.~L. Matacz, P.~C.~W. Davies, and Adrian~C. Ottewill.
\newblock Quantum vacuum instability near rotating stars.
\newblock {\em Phys.\ Rev.\ D}, 47(4):1557--1562, 1993.

\bibitem{th:McL'90}
John~Gerard McLaughlin.
\newblock {\em Renormalisation of the energy-momentum stress tensor for quantum
  fields on a curved background}.
\newblock PhD thesis, University of Oxford, 1990.

\bibitem{bk:Meixner&Schafke}
Josef Meixner and Friedrich~Wilhelm Sch\"{a}fke.
\newblock {\em Mathieusche Funktionen und Sph\"{a}roidfunktionen mit
  Anwendungen auf Physikalische und Technische Probleme}.
\newblock Springer-Verlag, 1954.

\bibitem{bk:M&T&W}
Charles~W. Misner, Kip~S. Thorne, and John~Archibald Wheeler.
\newblock {\em Gravitation}.
\newblock W.H. Freeman, San Francisco, 1973.

\bibitem{ar:Newmanetal'65}
E.T. Newman, E.~Couch, K.~Chinnapared, A.~Exton, A.~Prakash, and R.~Torrence.
\newblock Metric of a rotating, charged mass.
\newblock {\em J.\ Math.\ Phys.}, 6(6):918--919, 1965.

\bibitem{ar:Newman&Janis'65}
Ezra~T. Newman and A.I. Janis.
\newblock Note on the {K}err spinning-particle metric.
\newblock {\em J.\ Math.\ Phys.}, 6(6):915--917, 1966.

\bibitem{ar:N&P'62}
Ezra~T. Newman and Roger Penrose.
\newblock An approach to gravitational radiation by a method of spin
  coefficients.
\newblock {\em J.\ Math.\ Phys.}, 3(3):566--578, 1962.

\bibitem{ar:N&P'66}
Ezra~T. Newman and Roger Penrose.
\newblock Note on the {B}ondi-{M}etzner-{S}achs group.
\newblock {\em J.\ Math.\ Phys.}, 7:863--870, 1966.

\bibitem{ar:Ott&Winst'00Lett}
Adrian~C. Ottewill and Elizabeth Winstanley.
\newblock Divergence of a quantum thermal state on {K}err space-time.
\newblock {\em Phys.\ Lett.\ A}, 273:149--152, 2000.

\bibitem{ar:Ott&Winst'00}
Adrian~C. Ottewill and Elizabeth Winstanley.
\newblock Renormalized stress tensor in {K}err space-time: General results.
\newblock {\em Phys.\ Rev.\ D}, 62(8):084018, 2000.

\bibitem{ar:PageII'76}
Don~N. Page.
\newblock Particle emission rates from a black hole: {II}. massless particles
  from a rotating hole.
\newblock {\em Phys.\ Rev.\ D}, 14(12):3260--3273, 1976.

\bibitem{ar:PageI'76}
Don~N. Page.
\newblock Particle emission rates from a black hole: Massless particles from an
  uncharged, nonrotating hole.
\newblock {\em Phys.\ Rev.\ D}, 13(2):198--205, 1976.

\bibitem{ar:Penrose'69}
R.~Penrose.
\newblock Gravitational collapse: The role of general relativity.
\newblock {\em Rev.\ del Nuovo Cimento}, 1:252--276, 1969.

\bibitem{ar:Press&Teuk'73}
William~H. Press and Saul~A. Teukolsky.
\newblock Perturbations of a rotating black hole. {II}. {D}ynamical stability
  of the {K}err metric.
\newblock {\em The Astrophysical Journal}, 185:649--673, 1973.

\bibitem{bk:NumRec}
William~H. Press, Saul~A. Teukolsky, William~T. Vetterling, and Brian~P.
  Flannery.
\newblock {\em Numerical Recipes in Fortran}.
\newblock Cambridge University Press, Cambridge, second edition, 1992.

\bibitem{ar:Price'72}
Richard~H. Price.
\newblock Nonspherical perturbations of relativistic gravitational collapse.
  {II}. integer-spin, zero-rest-mass fields.
\newblock {\em Phys.\ Rev.\ D}, 5:2439--2454, 1972.

\bibitem{ar:Robin&Schild'63}
I.~Robinson and A.~Schild.
\newblock Generalization of a theorem by {G}oldberg and {S}achs.
\newblock {\em J.\ Math.\ Phys.}, 4:484--489, 1963.

\bibitem{ar:Sasa&Naka'82}
Misao Sasaki and Takashi Nakamura.
\newblock Gravitational radiation from a {K}err black hole. {I}.
\newblock {\em Progress of Theoretical Physics}, 67(6):1788--1809, 1982.

\bibitem{ar:Schum&Caves'85}
Bonny~L. Schumaker and Carlton~M. Caves.
\newblock New formalism for two-photon quantum optics. {II}. {M}athematical
  foundation and compact notation.
\newblock {\em Phys.\ Rev.\ A}, 31(5):3093--3111, 1985.

\bibitem{ar:Seidel'89}
Edward Seidel.
\newblock A comment on the eigenvalues of the spin-weighted spheroidal
  functions.
\newblock {\em Class.\ Quantum.\ Grav.}, 6:1057--1062, 1989.

\bibitem{ar:Starob'73a}
A.~A. Starobinski\u{\i}.
\newblock Amplification of waves during reflection from a rotating black hole.
\newblock {\em Zh.\ Eksp.\ Theor.\ Fiz.}, 64:48--57, 1973.

\bibitem{ar:Starob'73b}
A.~A. Starobinski\u{\i} and S.~M. Churilov.
\newblock Amplification of electromagnetic and gravitational waves scattered by
  a rotating black hole.
\newblock {\em Zh.\ Eksp.\ Theor.\ Fiz.}, 65:3--11, 1973.

\bibitem{ar:Stewart'75}
J.M. Stewart.
\newblock On the stability of {K}err's space-time.
\newblock {\em Proc.\ R.\ Soc.\ Lond. A}, 344:65--79, 1975.

\bibitem{ar:Teuk'72}
Saul~A. Teukolsky.
\newblock Rotating black holes: {S}eparable wave equations for gravitational
  and electromagnetic perturbations.
\newblock {\em Phys.\ Rev.\ Lett.}, 29(16):1114--1118, 1972.

\bibitem{ar:Teuk'73}
Saul~A. Teukolsky.
\newblock Perturbations of a rotating black hole. {I}. {F}undamental equations
  for gravitational, electromagnetic, and neutrino-field perturbations.
\newblock {\em The Astrophysical Journal}, 185:635--647, 1973.

\bibitem{ar:Teuk&Press'74}
Saul~A. Teukolsky and William~H. Press.
\newblock Perturbations of a rotating black hole. {III}. {I}nteraction of the
  hole with gravitational and electromagnetic radiation.
\newblock {\em The Astrophysical Journal}, 193:443--461, 1974.

\bibitem{ar:Unruh'74}
William~G. Unruh.
\newblock Second quantization in the {K}err metric.
\newblock {\em Phys.\ Rev.\ D}, 10(10):3194--3205, 1974.

\bibitem{ar:Unruh'76}
William~G. Unruh.
\newblock Notes on black-hole evaporation.
\newblock {\em Phys.\ Rev.\ D}, 14(4):870--891, 1976.

\bibitem{ar:Wald'78}
Robert~M. Wald.
\newblock Construction of solutions of gravitational, electromagnetic, or other
  perturbation equations from solutions of decoupled equations.
\newblock {\em Phys.\ Rev.\ Letters}, 41(4):203--206, 1978.

\bibitem{bk:Wald'84}
Robert~M. Wald.
\newblock {\em General Relativity}.
\newblock The University of Chicago Press, Chicago and London, 1984.

\bibitem{th:WinstMSc}
Elizabeth Winstanley.
\newblock Quantum field theory on black hole backgrounds.
\newblock First year dissertation, Oxford University, 1993.

\bibitem{ar:Zel'71}
Ya.~B. Zel'dovich.
\newblock The generation of waves by a rotating body.
\newblock {\em ZhETF Pis. Red.}, 14:270--, 1971.

\end{thebibliography}

\chapter*{Errata}
\addcontentsline{toc}{chapter}{Errata}
\markboth{this is markboth 1st}{Errata}




The following are corrections to the 
printed
version of the Ph.D.~thesis.

\setlength{\leftmargini}{0mm}

\begin{itemize}
\item
In Eqs. (\ref{eq:norm Aup/out/in/dn}), (\ref{eq:phi_j as funcs. of R_1S1}),
(\ref{eq:phi_j as funcs. of R1S_1}), (\ref{eq:commut. rlns. for a,a_dagger})
(\ref{eq:orthonormality conds. for potential}), (\ref{eq:gauge-indep inner prod.})
(\ref{eq: symm. of lmwP_phi_i}), (\ref{eq:symm. of op. lmwP_phi_i})
and (\ref{eq:a_dagger*a on unruh})
the index $w$ should be $\omega$.
\end{itemize}

{\bf Chapter \ref{ch:intro}} 

\begin{itemize}
\item
On Pages 18, 20 (twice) and 21, ``Schwarzschild'' has been mispelt as ``Schwarzchild''.

\end{itemize}
\noindent

{\bf Chapter \ref{ch:field eqs.}} 

\begin{itemize}

\item
The factor $\cos^2\theta$ in the Teukolsky equation (\ref{eq:Teuk.eq.})
should be $\cot^2\theta$ instead.

\item
The power of $\rho^*$ in Eq. (\ref{eq:Chrzan. potential as a func. of RS}) should be $(\pm 1-1)$, so
that the equation should read:
\begin{equation*} 
{}_{lm\omega}A_{\beta}=-\Pi_{\mp 1 \beta}{}^{\dagger *}\rho^{* (\pm 1-1)}{}_{\mp 1}R_{lm\omega}\ {}_{\pm 1}Z_{lm\omega}e^{-i\omega t}
\end{equation*}

\end{itemize}

{\bf Chapter \ref{ch:radial sln.}} 
\begin{itemize}

\item
There are two wrong signs and a power of two missing in the values of 
$a_D$ and $b_D$ in Eq. (\ref{eq:def a_D,b_D}). It should read:
\begin{equation*}
\begin{aligned}
a_D&=-\frac{2}{\Delta^2}\left[2K^2-\Delta(iK'+{}_{-1}\lambda_{lm\omega})\right] \\
b_D&=-\frac{4iK}{\Delta}
\end{aligned}
\end{equation*}

\item
There is a minus sign missing in the potential (\ref{eq:potential -1mathcalU}), in the
first term of the right hand side on the second line.
It should read:
\begin{equation*}
\begin{aligned}
&
{}_{-1}\mathcal{U}=\\
&-\frac{\left[-\omega (r^2+a^2)+am\right]^2}{(r^2+a^2)^2}+
\frac{\Delta{}_{-1}\lambda_{lm\omega}}{(r^2+a^2)^2}-
\frac{\Delta(\Delta r^2+4Ma^2r-Q^2(a^2-r^2))}{(r^2+a^2)^4}-                                     \\
&-\frac{\Delta\left[\Delta(10r^2+2\nu^2)-(r^2+\nu^2)(11r^2-10rM+\nu^2)\right]}{(r^2+a^2)^2\left[(r^2+\nu^2)^2+\eta\Delta\right]}+ \\
&+\frac{12\Delta r(r^2+\nu^2)^2\left[\Delta r-(r^2+\nu^2)(r-M)\right]}{(r^2+a^2)^2\left[(r^2+\nu^2)^2+\eta\Delta\right]^2}
-\frac{\Delta(r-M)^2\eta \left[2(r^2+\nu^2)^2-\eta\Delta \right]}{(r^2+a^2)^2\left[(r^2+\nu^2)^2+\eta\Delta\right]^2}
\end{aligned}
\end{equation*}

\item
The variable $\tilde{\omega}$ used for the first time on Page 74 is not defined anywhere.
For an uncharged matter field in a Kerr-Newman black hole it is  defined as $\tilde{\omega}\equiv \omega-m \Omega_+$.

\item
The normalization used in Table \ref{table:radial wronsks} has not been indicated.
\\
It is: ${}_{-1}R^{\text{in,inc}}_{lm\omega}={}_{-1}R^{\text{up,inc}}_{lm\omega}=1$.

\item
On Page 95 it should say ${}_{\indhel}Z_{lm\omega} \rightarrow {}_{\indhel}Y_{lm}$ rather than
${}_{\indhel}S_{lm} \rightarrow {}_{\indhel}Y_{lm}$.

\item
Eq. (\ref{eq:I_wtilde}) should read:
\begin{equation*} 
I_{\tilde{\omega}}\equiv e^{-i\tilde{\omega}r_+}\left(4M\kappa_+\right)^{-\frac{i\tilde{\omega}}{2\kappa_+}}\left(-4M\kappa_-\right)^{-\frac{i\tilde{\omega}}{2\kappa_-}}
\end{equation*}

\item
There is a missing subscript in the normalization constant in Eq. (\ref{eq:R1 `up' approx l->inf,r->rplus;compact version}):
\begin{equation*}
|N^{\text{up}}_{-1}|{}_{\indhel}R^{\text{up}}_{lm\omega} \rightarrow 
A_{\indhel}Nx^{-1/2}K_{\indhel+iq}(2lx^{1/2}) \qquad (l\rightarrow +\infty,r \rightarrow r_+,lx^{1/2}\ \text{finite})
\end{equation*}

\item
There is a factor $|N^{\text{up}}_{-1}|$ missing in Eqs. (\ref{eq:Ddagger Delta R1 `up' approx l->inf,r->rplus})
and (\ref{eq:Ddagger Delta R+1 `up' approx l->inf,r->rplus}). They should respectively read:
\begin{equation*} 
\begin{aligned}
&\frac{|N^{\text{up}}_{-1}|}{A_{\indhel}N}\mathcal{D}_0^{\dagger}\left(\Delta{}_{\indhel}R^{\text{up}}_{lm\omega}\right) \rightarrow \\
&\rightarrow (r_+-r_-)x^{-\indhel /2} \left[\left(-\frac{\indhel}{2}+1+i\frac{q}{2}\right)K_{\indhel+iq}(2lx^{1/2})+
lx^{1/2}K'_{\indhel+iq}(2lx^{1/2})
\right]  \\
& \qquad \qquad \qquad  \qquad \qquad\qquad \qquad\qquad(l\rightarrow +\infty,r \rightarrow r_+,lx^{1/2}\ \text{finite})
\end{aligned}
\end{equation*}
and
\begin{equation*} 
\frac{|N^{\text{up}}_{-1}|}{A_{\indhel}N}\mathcal{D}^{\dagger}_0\left(\Delta{}_{+1}R^{\text{up}}_{lm\omega}\right) \rightarrow 
-\frac{(r_+-r_-)}{2x^{1/2}} K_{iq} \qquad (l\rightarrow +\infty,r \rightarrow r_+)
\end{equation*}

\end{itemize}

{\bf Chapter \ref{ch:SWSH}} 

\begin{itemize}

\item
Eq. (\ref{eq:eigenval. for c=0}) should read:
\begin{equation*}
{}_{\indhel}E_{lm}\equiv {}_{\indhel}E_{lm\omega=0}= l(l+1).
\end{equation*}

\item
There is a typo in Eq. (\ref{eq:recursive rln. a_n}).
The term $n(n+1)$ should be replaced by $n(n-1)$
so that the equation reads
\begin{equation*} 
\begin{aligned}
{}_{\indhel}\topbott{a}{b}_{n+1,lm\omega}&=\frac{1}{2(n+1)\left(n+1+2\topbott{\alpha}{\beta}\right)}
\Bigg\{\Big[2n(\alpha+\beta+1)- 
 \\
&
-\left({}_{\indhel}E_{lm\omega}-(\alpha+\beta)(\alpha+\beta+1)+c^2\mp2c\indhel\right)+n(n-1)\Big]{}_{\indhel}\topbott{a}{b}_{n,lm\omega}+ \\
&  +2c(c\mp \indhel){}_{\indhel}\topbott{a}{b}_{n-1,lm\omega}-c^2{}_{\indhel}\topbott{a}{b}_{n-2,lm\omega}\Bigg\}, \quad \forall n\in \mathbb{N}
\end{aligned}
\end{equation*}

\end{itemize}

{\bf Chapter \ref{ch:high freq. spher}} 

\begin{itemize}

\item
There is a factor  $u$ missing in the third line of Eq. (\ref{diff eq:y in u}). It should read
\begin{equation*}
\begin{aligned}
&u\frac{d^{2}{}_{\indhel}y_{lm\omega}}{du^{2}}+(2\alpha+1)\frac{d{}_{\indhel}y_{lm\omega}}{du}-  \\
&\quad-\frac{1}{4}\left[u+2\indhel -\frac{1}{c}\left(c^{2}-(\alpha+\beta)(\alpha+\beta+1)+{}_{\indhel}E_{lm\omega}\right)
\right]{}_{\indhel}y_{lm\omega}-
\\
&\quad-\frac{1}{4c}\left[u^{2}\frac{d^{2}{}_{\indhel}y_{lm\omega}}{du^{2}}+2(\alpha+\beta+1)u\frac{d{}_{\indhel}y_{lm\omega}}{du}-
\left(\frac{1}{4}u^{2}+\indhel u\right){}_{\indhel}y_{lm\omega}\right]=0
\end{aligned}
\end{equation*}

\end{itemize}

{\bf Chapter \ref{stress-energy tensor}} 

\begin{itemize}

\item
We are using rationalized units (i.e., the Maxwell equations are given by Eq.~(\ref{eq:Maxwell eqs. with potential})
and the stress-energy tensor by Eq.~(\ref{eq:stress tensor, spin 1})).
There is therefore an incorrect factor of $4\pi$ in Eqs.~(\ref{eq:dE(inc/out)/dtdOmega})
and (\ref{eq:dEtra/dtdOmega}). 
Furthermore, the derivative sign $\d$ in the numerator of the left hand side of these equations
should be a second derivative sign $\d^2$.
These equations should respectively read
\begin{equation*} 
\begin{aligned}
\frac{\d^2{E^{\text{(inc)}}}}{\d{t} \d{\Omega}}&=\frac{r^2}{2}\left|\phi^{\text{(in,inc)}}_{-1}\right|^2 \\
\frac{\d^2{E^{\text{(ref)}}}}{\d{t} \d{\Omega}}&=2r^2\left|\phi^{\text{(in,ref)}}_{+1}\right|^2 
\end{aligned}
\end{equation*}
and
\begin{equation*}
\frac{\d^2{E^{\text{(tra)}}}}{\d{t} \d{\Omega}}=\frac{\Delta^2}{2(r_+^2+a^2)}\frac{\omega}{\tilde{\omega}}\left|\phi^{\text{(in,tra)}}_{-1}\right|^2
\end{equation*}

\item
Above Eq.~(\ref{eq:phi_iphi_j,i<>j,approx l->inf,r->rplus}) the spin-weighted spheroidal harmonics should be denoted
by ${}_{\indhel}Z_{lm\omega}$, not by ${}_{\indhel}S_{lm}$ .

\item
On Page 214, where it says "Graphs \ref{fig:deltaTtplusphiplus_cch_b_past}--\ref{fig:deltaTtplusphiplus_cch_b_past_surfs} for $\vac[ren]{\hat{T}^{}_{t_+\phi_+}}{B^-}$\dots"
it should instead say
"Graphs \ref{fig:deltaTtplusphiplus_cch_b_past}--\ref{fig:deltaTtplusphiplus_cch_b_past_surfs} for $\vac[ren]{\hat{T}^{}_{t_+\phi_+}}{CCH^--B^-}$\dots".

\item
The variable $B$ in Eq.~(\ref{eq: phi0^2+phi2^2 t->pi-t}) 
should be ${}_1B_{lm\omega}$.

\item
The terms `positive' and `negative' referring to the polarization should be swopped 
everywhere in Section \ref{sec:polarization}, except for in the very last paragraph
and in its corresponding Figs.~(\ref{fig:deltaTtr_cch_u_past})--(\ref{fig:Ttphi_polpos_u_b}).
As an example, the fourth paragraph of Section \ref{sec:polarization} should start as:
``It is in the limit for large $r$ that the physical meaning of the various vectors becomes clear.
We know that in flat space-time an electric field mode of positive frequency that is proportional to the vector 
$\left(\hat{\vec{e}}_{\theta}+i\hat{\vec{e}}_{\phi}\right)[\left(\hat{\vec{e}}_{\theta}-i\hat{\vec{e}}_{\phi}\right)]$
possesses a negative[positive] angular momentum and we thus say that it is negatively[positively] polarized.''.
This is because a time dependence $e^{+i\omega t}$ is assumed.
The reason for this confusion arose because of the non-standard notational
change in the sign of $(m,\omega)$ in Eq.~(\ref{eq:potential as a func. of RS}).
The sign of $(m,\omega)$ in the right hand side of
Eqs.~(\ref{eq:A_up, r->inf}) and (\ref{eq:E_up,B_up, r->inf}) should be changed
so that they read
\begin{equation*}   
\begin{aligned}
&{}_{lm\omega P}A^{' \text{up} \mu}\equiv \left(\Pi^{\dagger}_j{}^{\alpha}{}_{lm\omega}\NPadj_j\right)^* + \Pi^{\dagger}_j{}^{\alpha}{}_{lm\omega}\NPadj_j\rightarrow 
-\frac{\omega i |N^{\text{up}}_{+1}|}{\sqrt{2}r}
\times \\  &\times
\left[{}_{-1}Y_{l-m-\omega}{}_{+1}R^{\text{up,tra}}_{l-m-\omega}e^{+i\omega (t-r)}\left(\hat{\vec{e}}_{\theta}+i\hat{\vec{e}}_{\phi}\right)-
{}_{-1}Y_{l-m-\omega}^*{}_{+1}R^{\text{up,tra} *}_{l-m-\omega}e^{-i\omega (t-r)}\left(\hat{\vec{e}}_{\theta}-i\hat{\vec{e}}_{\phi}\right)\right]
\\ & \qquad \qquad \qquad \qquad \qquad \qquad \qquad \qquad \qquad \qquad \qquad  \qquad \qquad(r\rightarrow+\infty)
\end{aligned}
\end{equation*}
and
\begin{equation*} 
\begin{aligned}
&{}_{lm\omega P}\vec{E}^{\text{up}}=-\nabla {}_{lm\omega P}A^{' \text{up} 0}-\pardiff{{}_{lm\omega P}\vec{A}^{' \text{up}}}{t}=
\frac{\omega^2 |N^{\text{up}}_{+1}|}{\sqrt{2}r}
\times \\&\times
\left[{}_{-1}Y_{l-m-\omega}{}_{+1}R^{\text{up,tra}}_{l-m-\omega}e^{+i\omega (t-r)}\left(\hat{\vec{e}}_{\theta}+i\hat{\vec{e}}_{\phi}\right)+
{}_{-1}Y_{l-m-\omega}^*{}_{+1}R^{\text{up,tra} *}_{l-m-\omega}e^{-i\omega (t-r)}\left(\hat{\vec{e}}_{\theta}-i\hat{\vec{e}}_{\phi}\right)\right]
\\
&{}_{lm\omega P}\vec{B}^{\text{up}}=\nabla\times{}_{lm\omega P}\vec{A}^{' \text{up}}=
\frac{-\omega^2 i|N^{\text{up}}_{+1}|}{\sqrt{2}r}
\times \\&\times
\left[{}_{-1}Y_{l-m-\omega}{}_{+1}R^{\text{up,tra}}_{l-m-\omega}e^{+i\omega (t-r)}\left(\hat{\vec{e}}_{\theta}+i\hat{\vec{e}}_{\phi}\right)-
{}_{-1}Y_{l-m-\omega}^*{}_{+1}R^{\text{up,tra} *}_{l-m-\omega}e^{-i\omega (t-r)}\left(\hat{\vec{e}}_{\theta}-i\hat{\vec{e}}_{\phi}\right)\right]
\end{aligned}
\end{equation*}
respectively.

\end{itemize}

{\bf Bibiliography}
\begin{itemize}

\item
The year in~\cite{ar:J&McL&Ott'88} should be 1988, not 1999.

\end{itemize}





\end{document}